\documentclass[ 
		oneside,
		openright,  
		titlepage,
		numbers=noenddot,
		headinclude, 
	 	footinclude=true, 
	 	cleardoublepage=empty,
		dottedtoc, 
		BCOR=5mm,paper=a4,fontsize=11pt, 
		french
		]{scrreprt} 
		
\def\format{oneside}
\def\cover{true}
\def\showcolors{1}

\if 0\showcolors
\usepackage[usenames,dvipsnames]{xcolor} 

\definecolor{carnelian}{rgb}{0.7, 0.11, 0.11}
\definecolor{Prune}{RGB}{99,0,60}

\definecolor{burntorange}{rgb}{0.8, 0.33, 0.0}
\definecolor{mahogany}{rgb}{0.75, 0.25, 0.0}

\definecolor{deepcarminepink}{rgb}{0.94, 0.19, 0.22}
\definecolor{dogwoodrose}{rgb}{0.84, 0.09, 0.41}
\definecolor{byzantine}{rgb}{0.74, 0.2, 0.64}

\definecolor{carmine}{rgb}{0.59, 0.0, 0.09}

\definecolor{amber}{rgb}{1.0, 0.75, 0.0}
\definecolor{mikadoyellow}{rgb}{1.0, 0.77, 0.05}
\definecolor{tangerineyellow}{rgb}{1.0, 0.8, 0.0}
\definecolor{mikadoyellow}{rgb}{1.0, 0.77, 0.05}

\definecolor{blue(munsell)}{rgb}{0.0, 0.5, 0.69}
\definecolor{yaleblue}{rgb}{0.06, 0.3, 0.57}
\definecolor{richelectricblue}{rgb}{0.03, 0.57, 0.82}
\definecolor{frenchblue}{rgb}{0.0, 0.45, 0.73}
\definecolor{steelblue}{rgb}{0.27, 0.51, 0.71}
\definecolor{skobeloff}{rgb}{0.0, 0.48, 0.45}

\definecolor{yellow-green}{rgb}{0.6, 0.8, 0.2}
\definecolor{applegreen}{rgb}{0.55, 0.71, 0.0}
\definecolor{ballblue}{rgb}{0.13, 0.67, 0.8}

\definecolor{indigo(dye)}{rgb}{0.0, 0.25, 0.42}
\definecolor{prussianblue}{rgb}{0.0, 0.19, 0.33}
\definecolor{darkmidnightblue}{rgb}{0.0, 0.2, 0.4}

\definecolor{tealgreen}{rgb}{0.0, 0.51, 0.5}
\definecolor{tealgreenperso}{rgb}{0.0, 0.5, 0.5}
\definecolor{teal}{rgb}{0.0, 0.5, 0.5}

\definecolor{arsenic}{rgb}{0.23, 0.27, 0.29}
\definecolor{coolblack}{rgb}{0.0, 0.18, 0.39}
\definecolor{gray}{gray}{0.55}

\newcommand{\colortitle}{mahogany}
\newcommand{\colorsubtitle}{indigo(dye)}

\newcommand{\colorchapter}{black}
\newcommand{\colorpartnumber}{black}
\newcommand{\colorchapternumber}{black}
\newcommand{\colorpart}{black}
\newcommand{\colorsubpart}{black}

\newcommand{\colorsection}{black}
\newcommand{\colorsubsection}{black}

\newcommand{\colorurl}{black}
\newcommand{\colorcite}{black}
\newcommand{\colorlink}{black}

\newcommand{\colorsubsubsection}{black}
\newcommand{\colorparagraph}{black}
\newcommand{\colorkeyword}{black}

\newcommand{\colorcomment}{black}

\else
\usepackage[usenames,dvipsnames]{xcolor}

\definecolor{carnelian}{rgb}{0.7, 0.11, 0.11}
\definecolor{Prune}{RGB}{99,0,60}

\definecolor{burntorange}{rgb}{0.8, 0.33, 0.0}
\definecolor{mahogany}{rgb}{0.75, 0.25, 0.0}

\definecolor{deepcarminepink}{rgb}{0.94, 0.19, 0.22}
\definecolor{dogwoodrose}{rgb}{0.84, 0.09, 0.41}
\definecolor{byzantine}{rgb}{0.74, 0.2, 0.64}

\definecolor{carmine}{rgb}{0.59, 0.0, 0.09}

\definecolor{amber}{rgb}{1.0, 0.75, 0.0}
\definecolor{mikadoyellow}{rgb}{1.0, 0.77, 0.05}
\definecolor{tangerineyellow}{rgb}{1.0, 0.8, 0.0}
\definecolor{mikadoyellow}{rgb}{1.0, 0.77, 0.05}

\definecolor{blue(munsell)}{rgb}{0.0, 0.5, 0.69}
\definecolor{yaleblue}{rgb}{0.06, 0.3, 0.57}
\definecolor{richelectricblue}{rgb}{0.03, 0.57, 0.82}
\definecolor{frenchblue}{rgb}{0.0, 0.45, 0.73}
\definecolor{steelblue}{rgb}{0.27, 0.51, 0.71}
\definecolor{skobeloff}{rgb}{0.0, 0.48, 0.45}

\definecolor{yellow-green}{rgb}{0.6, 0.8, 0.2}
\definecolor{applegreen}{rgb}{0.55, 0.71, 0.0}
\definecolor{ballblue}{rgb}{0.13, 0.67, 0.8}

\definecolor{indigo(dye)}{rgb}{0.0, 0.25, 0.42}
\definecolor{prussianblue}{rgb}{0.0, 0.19, 0.33}
\definecolor{darkmidnightblue}{rgb}{0.0, 0.2, 0.4}

\definecolor{tealgreen}{rgb}{0.0, 0.51, 0.5}
\definecolor{tealgreenperso}{rgb}{0.0, 0.5, 0.5}
\definecolor{teal}{rgb}{0.0, 0.5, 0.5}

\definecolor{arsenic}{rgb}{0.23, 0.27, 0.29}
\definecolor{coolblack}{rgb}{0.0, 0.18, 0.39}
\definecolor{gray}{gray}{0.55}

\newcommand{\colortitle}{mahogany}
\newcommand{\colorchapternumber}{\colorchapter}
\newcommand{\colorpartnumber}{\colorchapter}
\newcommand{\colorpart}{\colortitle}
\newcommand{\colorsubpart}{\colortitle}

\newcommand{\colorchapter}{indigo(dye)}
\newcommand{\colorsubtitle}{\colorchapter}
\newcommand{\colorsection}{\colorchapter!92!white}
\newcommand{\colorsubsection}{\colorchapter!82!white}

\newcommand{\colorurl}{byzantine!95!black}
\newcommand{\colorcite}{tealgreenperso!95!black}
\newcommand{\colorlink}{tangerineyellow!85!black}

\newcommand{\colorsubsubsection}{black}
\newcommand{\colorparagraph}{black}
\newcommand{\colorkeyword}{black}

\newcommand{\colorcomment}{black}

\fi	
	
\usepackage{geometry}
\PassOptionsToPackage{
	eulerchapternumbers, 
	listings, 
	pdfspacing, 
	subfig, 
	beramono, 
	eulermath, 
	parts,
	}{classicthesis}

\newcommand{\myTitlesmall}{
Mean-field methods and algorithmic perspectives for high-dimensional
machine learning
}

\newcommand{\myName}{Benjamin Aubin\xspace}

\newcommand{\myDepartment}{Institut de Physique Th\'eorique\xspace}
\newcommand{\myFaculty}{CEA \& Universit\'e Paris-Saclay\xspace}
\newcommand{\myUni}{CEA \& Universit\'e Paris-Saclay}
\newcommand{\myLocation}{Saclay, France \xspace}
\newcommand{\myTime}{December 16, 2020 \xspace}
\newcommand{\myVersion}{version 1\xspace}
\newcounter{dummy}
\providecommand{\mLyX}{L\kern-.1667em\lower.25em\hbox{Y}\kern-.125emX\@}
\usepackage{Config/classicthesiscustom}

\SetSymbolFont{operators}   {normal}{OT1}{cmr} {m}{n}
\SetSymbolFont{letters}     {normal}{OML}{cmm} {m}{it}
\SetSymbolFont{symbols}     {normal}{OMS}{cmsy}{m}{n}
\SetSymbolFont{largesymbols}{normal}{OMX}{cmex}{m}{n}
\SetSymbolFont{operators}   {bold}  {OT1}{cmr} {bx}{n}
\SetSymbolFont{letters}     {bold}  {OML}{cmm} {b}{it}
\SetSymbolFont{symbols}     {bold}  {OMS}{cmsy}{b}{n}
\SetSymbolFont{largesymbols}{bold}  {OMX}{cmex}{m}{n}
\SetMathAlphabet{\mathbf}{normal}{OT1}{cmr}{bx}{n}
\SetMathAlphabet{\mathsf}{normal}{OT1}{cmss}{m}{n}
\SetMathAlphabet{\mathit}{normal}{OT1}{cmr}{m}{it}
\SetMathAlphabet{\mathtt}{normal}{OT1}{cmtt}{m}{n}
\SetMathAlphabet{\mathbf}{bold}  {OT1}{cmr}{bx}{n}
\SetMathAlphabet{\mathsf}{bold}  {OT1}{cmss}{bx}{n}
\SetMathAlphabet{\mathit}{bold}  {OT1}{cmr}{bx}{it}
\SetMathAlphabet{\mathtt}{bold}  {OT1}{cmtt}{m}{n}

\renewcommand{\thesubsubsection}{\thesubsection.\alph{subsubsection}}
\renewcommand{\subparagraph}[1]{\vspace{0.5cm} $\bullet$ \textit{#1} \newline}

\newcommand{\acrotarget}[1]{\hypertarget{#1}{}}
\newcommand{\aclink}[1]{\hyperlink{#1}{\ac{#1}}}

\AtBeginDocument{\renewcommand{\thepart}{\Roman{part}}}

\newcommand{\subpartpage}[2]{
	\addcontentsline{toc}{part}{{\scshape \color{\colorpart}{#1. ~ #2}}}
	\null\thispagestyle{empty}
	\vfill
    \begin{center}
    \vspace{-3cm}
    \begin{spacing}{2}
    {\bfseries\color{\colorpartnumber} \huge Part #1.\\}
    \vspace{0.5cm}
    {\bfseries\color{\colorsubpart} \huge \textls[160]{\MakeTextUppercase{#2}}}
    \end{spacing}
	\end{center}
	\vfill
}

\usepackage{import} 

\PassOptionsToPackage{utf8}{inputenc}
\usepackage{inputenc}
\PassOptionsToPackage{french,american,english,british}{babel} 
\usepackage{babel}
\usepackage{mathptmx}
\usepackage{tocloft}

\PassOptionsToPackage{T1}{fontenc} 
\usepackage{fontenc}
\usepackage{textcomp} 
\usepackage{scrhack} 
\usepackage{xspace} 
\usepackage{mparhack} 
\usepackage{fixltx2e} 
\PassOptionsToPackage{smaller}{acronym} 
\usepackage[printonlyused]{acronym}

\usepackage[osf]{libertine}

\PassOptionsToPackage{fleqn}{amsmath} 
\usepackage{amsmath,amsfonts,amssymb,amsthm,mathrsfs, bm}
\usepackage{amscd,amsbsy,dsfont}
\usepackage{nicefrac}
\usepackage{microtype}      
\usepackage{mathtools}
\usepackage{nicefrac}       
\usepackage{stmaryrd} 

\usepackage[intoc,english]{nomencl} 
\makenomenclature
\usepackage{etoolbox}
\usepackage{physics}
\usepackage{tensor}
\usepackage{empheq}

\usepackage{algorithmic}
\usepackage{algorithm2e}
	
\PassOptionsToPackage{pdftex}{graphicx}
\usepackage{graphicx} 
\usepackage{epstopdf}

\usepackage{mdwlist}
\usepackage{tabularx} 
\usepackage{booktabs} 
\usepackage{caption}
\usepackage{subcaption}
\usepackage{float}
\usepackage{cases}
\usepackage{pdfpages} 
\usepackage{bookmark}
\usepackage{adjustbox}
\usepackage{blkarray}
\usepackage{nomencl}
\makenomenclature

\usepackage{listings} 
\usepackage{imakeidx}

\PassOptionsToPackage{pdftex,hyperfootnotes=false,pdfpagelabels}{hyperref}
\usepackage{hyperref}
\hypersetup{
colorlinks=true, linktocpage=true, pdfstartpage=3, pdfstartview=FitV,
breaklinks=true, pdfpagemode=UseNone, pageanchor=true, pdfpagemode=UseOutlines,
plainpages=false, bookmarksnumbered, bookmarksopen=true, bookmarksopenlevel=1,
hypertexnames=true, pdfhighlight=/O,
urlcolor=\colorurl, linkcolor=\colorlink, citecolor=\colorcite,
pdftitle={Mean-field methods and algorithmic perspectives for high-dimensional machine learning},
pdfauthor={\textcopyright Benjamin Aubin, Institut de Physique Th\'eorique, CEA \& Universit\'e Paris-Saclay},
pdfsubject={},
pdfkeywords={statistical physics, machine learning, neural networks, statistical estimation, message-passing algorithms, replica method},
pdfcreator={pdfLaTeX},
pdfproducer={LaTeX}
}
\usepackage{url}
\PassOptionsToPackage{hyphens}{url}\usepackage{hyperref}
\usepackage{xr} 
\usepackage{dirtytalk} 

\makeatletter
\@ifpackageloaded{babel}
{
\addto\extrasamerican{

}
}{\relax}
\makeatother

\usepackage{setspace} 
\usepackage{afterpage}

\usepackage[most]{tcolorbox}

\usepackage{tikz}
\usetikzlibrary{arrows}
\tikzset{middlearrow/.style={decoration={markings,mark= at position 0.5 with {\arrow{#1}} ,},postaction={decorate}}}

\usetikzlibrary{decorations.markings}
\usetikzlibrary{fit}
\usetikzlibrary{bayesnet}
\usepackage{pgfplots}
\usetikzlibrary{calc}
\usepackage{ifthen}
\usetikzlibrary{decorations.text}
\usetikzlibrary{decorations.pathmorphing}
\usetikzlibrary{decorations.markings}
\usetikzlibrary{shapes.geometric}
\usetikzlibrary{shapes}
\usepackage{pgfplotstable}
\usetikzlibrary{fadings,shapes.arrows,shadows}   
\usetikzlibrary{calc}

\usetikzlibrary{external}
\tikzexternaldisable

\usepackage{mdframed}
\usepackage{multirow} 
\usepackage{multicol} 
\usepackage{scrextend} 
\usepackage[absolute]{textpos} 
\usepackage{colortbl}
\usepackage{array}

\newcommand{\ie}{i.\,e.~}
\newcommand{\eg}{e.\,g.~}
\newcommand{\Sec}{Sec.~}
\newcommand{\Part}{Part~}
\newcommand{\Chap}{Chap.~}
\newcommand{\App}{Appendix.~}
\newcommand{\Alg}{Algo.~}
\newcommand{\Fig}{Fig.~}

\newcommand{\Left}{(\textbf{Left})~}
\newcommand{\Right}{(\textbf{Right})~}
\newcommand{\Center}{(\textbf{Center})~}
\newcommand{\Leftn}{(\textbf{Left})}
\newcommand{\Rightn}{(\textbf{Right})}

\def \({\left(}
\def \){\right)}
\def \[{\left[}
\def \]{\right]}

\newcommand{\tbf}[1]{{\textbf{#1}}}

\newcommand{\bu}{{\textbf {u}}}
\newcommand{\bV}{{\mathbf {V}}}

\newcommand{\bh}{{\textbf {h}}}

\newcommand{\bv}{{\textbf {v}}}
\newcommand{\bA}{{\textbf {A}}}

\newcommand{\bx}{{\textbf {x}}}

\newcommand{\bz}{{\textbf {z}}}

\newcommand{\bxi}{{\boldsymbol{\xi}}}
\newcommand{\bgamma}{{\boldsymbol{\gamma}}}
\newcommand{\bGamma}{{\boldsymbol{\Gamma}}}
\newcommand{\bomega}{{\boldsymbol{\omega}}}
\newcommand{\boldeta}{{\boldsymbol{\eta}}}
\newcommand{\btheta}{{\boldsymbol{\theta}}}
\newcommand{\bphi}{{\boldsymbol{\phi}}}
\newcommand{\bchi}{{\boldsymbol{\chi}}}
\newcommand{\brho}{{\boldsymbol{\rho}}}
\newcommand{\bsigma}{{\boldsymbol{\sigma}}}
\newcommand{\bSigma}{{\boldsymbol{\Sigma}}}

\newcommand{\bLambda}{{\boldsymbol{\Lambda}}}

\newcommand{\bzero}{{\textbf{0}}}
\newcommand{\bmu}{{\boldsymbol{\mu}}}
\newcommand{\bepsilon}{{\boldsymbol{\epsilon}}}

\newcommand{\cC}{{\mathcal{C}}}

\renewcommand{\d}{\mathrm{d}}
\newcommand{\D}{\mathrm{D}}
\newcommand{\e}{\text {e}}

\newcommand{\eq}{\text{ eq}.~}

\newcommand{\be}{\begin{equation}}
\newcommand{\ee}{\end{equation}}
\newcommand{\beqa}{\begin{eqnarray}}
\newcommand{\eeqa}{\end{eqnarray}}

\newcommand{\bea}{\begin{align}}
\newcommand{\eea}{\end{align}}

\newtheorem{theorem}{Theorem}[section]
\newtheorem{lemma}[theorem]{\textbf{Lemma}}
\newtheorem{thm}[theorem]{\textbf{Theorem}}
\newtheorem{remark}[theorem]{\textbf{Remark}}
\newtheorem{proposition}[theorem]{\textbf{Proposition}}
\newtheorem{corollary}[theorem]{\textbf{Corollary}}
\newtheorem{definition}[theorem]{\textbf{Definition}}

\DeclareMathOperator{\atanh}{atanh}
\DeclareMathAlphabet{\varmathbb}{U}{bbold}{m}{n}
\newcommand{\id}{\mathds{1}}
\newcommand{\EE}{\mathbb{E}}
\newcommand{\bbR}{\mathbb{R}}
\newcommand{\bbX}{\mathbb{X}}
\newcommand{\bbH}{\mathbb{H}}
\newcommand{\bbK}{\mathbb{K}}
\newcommand{\bbY}{\mathbb{Y}}
\newcommand{\bbD}{\mathbb{D}}
\newcommand{\bbS}{\mathbb{S}}

\newcommand{\bbP}{\mathbb{P}}
\newcommand{\bbZ}{\mathbb{Z}}
\newcommand{\bbN}{\mathbb{N}}
\newcommand{\bbM}{\mathbb{M}}
\newcommand{\bbA}{\mathbb{A}}
\newcommand{\bbB}{\mathbb{B}}

\newcommand{\bbC}{\mathbb{C}}

\renewcommand{\d}{{\mathrm d}}
\renewcommand{\P}{{\mathrm P}}

\newcommand{\mZ}{\mathcal{Z}}
\newcommand{\mH}{\mathcal{H}}

\newcommand{\mI}{\mathcal{I}}
\newcommand{\mM}{\mathcal{M}}
\newcommand{\mN}{\mathcal{N}}
\newcommand{\mS}{\mathcal{S}}
\newcommand{\mC}{\mathcal{C}}
\newcommand{\mD}{\mathcal{D}}

\newcommand{\mL}{\mathcal{L}}
\newcommand{\mA}{\mathcal{A}}

\newcommand{\mR}{\mathcal{R}}
\newcommand{\mF}{\mathcal{F}}
\newcommand{\mG}{\mathcal{G}}
\newcommand{\mP}{\mathcal{P}}

\newcommand{\mE}{\mathcal{E}}
\newcommand{\mT}{\mathcal{T}}
	
\renewcommand{\tr}[1]{\textrm{Tr}\(#1\)}
\renewcommand{\det}[1]{\textrm{det}\(#1\)}
\newcommand{\diag}[1]{\textrm{diag}\(#1\)}
\newcommand{\td}[1]{{\tilde{#1}}}

\newcommand{\spacecase}[0]{\vspace{0.3cm} \\}
\newcommand{\Spacecase}[0]{\vspace{0.5cm} \\}
\newcommand{\hhspace}[0]{\hspace{0.3cm}}

\def\E{\mathbb{E}}

\def\cE{\mathcal E}
\newcommand{\Var}[1]{\textrm{Var}(#1)}

\newcommand{\andcase}[0]{\hspace{ 0.2cm }\textrm{ and }\hspace{ 0.2cm }}

\renewcommand{\vec}[1]{{\mathbf{#1}}}
\newcommand{\mat}[1]{{\mathbf{#1}}}

\newcommand{\sign}{{\textrm{sign}}}

\newcommand{\Z}{\mat{Z}}
\newcommand{\mW}[0]{{\mat{W}}}
\newcommand{\X}[0]{{\mat{X}}}

\newcommand{\m}[0]{\mat{m}}

\newcommand{\iid}[0]{{\textrm{iid}~}}
\newcommand{\out}[0]{\mathrm{out}}
\newcommand{\w}[0]{{\mathrm{w}}}
\renewcommand{\u}[0]{{\mathrm{u}}}
\renewcommand{\v}[0]{{\mathrm{v}}}
\newcommand{\z}[0]{{\mathrm{z}}}
\newcommand{\x}[0]{{\mathrm{x}}}
\newcommand{\y}[0]{{\mathrm{y}}}

\newcommand{\rI}[0]{\mat{I}}

\newcommand{\rC}[0]{{\mathrm{C}}}
\newcommand{\rH}[0]{{\mathrm{H}}}
\newcommand{\rU}[0]{{\mathrm{U}}}

\newcommand{\rP}[0]{{\mathrm{P}}}

\newcommand{\rQ}[0]{{\mathrm{Q}}}
\newcommand{\rE}[0]{{\mathrm{E}}}
\newcommand{\rV}[0]{{\mathrm{V}}}
\newcommand{\rF}[0]{{\mathrm{F}}}
\newcommand{\bg }[0]{{\textbf{g}}}
\newcommand{\rp}[0]{{\mathrm{p}}}
\renewcommand{\rq}[0]{{\mathrm{q}}}
\newcommand{\rg}[0]{{\mathrm{g}}}
\newcommand{\rd}[0]{{\mathrm{d}}}
\newcommand{\rX}[0]{{\mathrm{X}}}
\newcommand{\rY}[0]{{\mathrm{Y}}}

\newcommand{\underlim}[2]{\underset{#1 \to #2}{\longrightarrow}}
\newcommand{\extr}{{\textbf{extr}}}
\newcommand{\Diff}{\mathrm{D}}
\newcommand{\argmin}{\mathrm{argmin}}
\newcommand{\argmax}{\mathrm{argmax}}
\newcommand{\train}{\mathrm{train}}
\newcommand{\test}{\mathrm{test}}
\newcommand{\MSE}{\mathrm{MSE}}
\newcommand{\gs}{\mathrm{gs}}
\newcommand{\bayes}{\mathrm{b}}
\newcommand{\gibbs}{\mathrm{gibbs}}
\newcommand{\vc}{\mathrm{vc}}
\newcommand{\mmse}{\mathrm{mmse}}
\newcommand{\map}{\mathrm{map}}
\newcommand{\mle}{\mathrm{mle}}
\newcommand{\rL}[0]{{\ell}}
\newcommand{\gen}{\mathrm{gen}}
\newcommand{\rs}{\mathrm{rs}}

\definecolor{green}{RGB}{0, 153, 0}
\definecolor{light_blue}{RGB}{51, 153, 255}
\definecolor{orange}{RGB}{255, 204, 0}
\definecolor{bg}{RGB}{0, 153, 153}
\definecolor{blue}{RGB}{0, 102, 204}
\definecolor{red}{RGB}{204, 0, 0}
\definecolor{lg}{RGB}{214, 214, 214}

\definecolor{codegreen}{rgb}{0,0.6,0}
\definecolor{codegray}{rgb}{0.5,0.5,0.5}
\definecolor{codepurple}{rgb}{0.58,0,0.82}
\definecolor{backcolour}{rgb}{0.95,0.95,0.92}
\lstdefinestyle{mystyle}{
    backgroundcolor=\color{backcolour},   
    stringstyle=\color{codepurple},
    commentstyle=\color{codegreen},
    numberstyle=\tiny\color{codegray},
    basicstyle=\ttfamily\footnotesize,
    numbers=left,      
}
\newcommand*{\colorboxed}{}
\def\colorboxed#1#{%
  \colorboxedAux{#1}%
}
\newcommand*{\colorboxedAux}[3]{%
  \begingroup
    \colorlet{cb@saved}{.}%
    \color#1{#2}%
       \boxed{%
      \color{cb@saved}%
      #3%
    }%
  \endgroup
}

\renewcommand{\boxed}[1]{
\tikz[baseline={([yshift=-1ex]current bounding box.center)}] \node [rectangle,line width=1.25, minimum width=1ex,draw, color=red] {\normalcolor $\displaystyle#1$};}
 \makeatother
  
\newtheorem{prop}[theorem]{Proposition}
\newtheorem{conjecture}[theorem]{Conjecture}
\newtheorem{hypothesis}[theorem]{Hypothesis}

\theoremstyle{remark}
\newtheorem*{remark*}{\textbf{Remark}}

\def\E{\mathbb{E}}

\def\cE{\mathcal E}

\def\out{\text{out}}
\def\x{\text{x}}

\def\dd{\text{d}}
\def\MSE{\text{MSE}}

\def\IT{\text{IT}}
\def\alg{\text{alg}}
\def\sign{\text{sign}}

\def\extr{{  \textbf{extr}}}

\newcommand{\KL}{\textrm{KL}}
\newcommand{\tramp}{\textsf{tramp}\xspace}

\newcommand{\rdbrs}[1]{\left( #1 \right)}	
\newcommand{\sqbrs}[1]{\left\lbrack #1 \right\rbrack}

\def\dd{\text{d}}
\def\MSE{\text{MSE}}

\newcommand{\lb}{\llbracket}
\newcommand{\rb}{\rrbracket}

\newcommand\scalemath[2]{\scalebox{#1}{\mbox{\ensuremath{\displaystyle #2}}}}

\usepackage{csquotes}
\PassOptionsToPackage{
backend=bibtex8, 
bibencoding=ascii,
language=auto,
style=authoryear,
citestyle=authoryear,
url=false,
doi=false,
isbn=false, 
arxiv=false,
eprint=true,
maxbibnames=6,
maxcitenames=1,
backref=true, 
natbib=true
}{biblatex}
\usepackage{biblatex}

\newcommand{
\citepublication}[3]{\textit{\citetitle{#1}}. \citet*{#1} \\
Presented in \Chap\ref{#3}.\vspace{0.3cm}\\
\textbf{Summary}: {#2}\\
}

\newcommand{
\citepublicationfrench}[3]{\textit{\citetitle{#1}}. \citet*{#1} \\
Presenté dans le \Chap\ref{#3}.\vspace{0.3cm}\\
\textbf{Résumé}: {#2}\\
}

\newcommand{
\citepublicationnoref}[3]{\textit{\citetitle{#1}}. \citet*{#1}\vspace{0.3cm}\\
\textbf{Summary}: {#2}\\
}
\newcommand{
\citepublicationnoreffrench}[3]{\textit{\citetitle{#1}}. \citet*{#1}\vspace{0.3cm}\\
\textbf{Résumé}: {#2}\\
}

\DeclareCiteCommand{\cite}[\mkbibparens]
  {\usebibmacro{prenote}}
  {\usebibmacro{citeindex}%
   \printtext[bibhyperref]{\usebibmacro{cite}}}
  {\multicitedelim}
  {\usebibmacro{postnote}}

\DeclareCiteCommand*{\cite}[\mkbibparens]
  {\usebibmacro{prenote}}
  {\usebibmacro{citeindex}%
   \printtext[bibhyperref]{\usebibmacro{citeyear}}}
  {\multicitedelim}
  {\usebibmacro{postnote}}
  
\DeclareFieldFormat{citehyperref}{%
  \DeclareFieldAlias{bibhyperref}{noformat}
  \bibhyperref{#1}}

\DeclareFieldFormat{textcitehyperref}{%
  \DeclareFieldAlias{bibhyperref}{noformat}
  \bibhyperref{%
    #1%
    \ifbool{cbx:parens}
      {\bibcloseparen\global\boolfalse{cbx:parens}}
      {}}}

\savebibmacro{cite}
\savebibmacro{textcite}

\renewbibmacro*{cite}{%
  \printtext[citehyperref]{%
    \restorebibmacro{cite}%
    \usebibmacro{cite}}}

\renewbibmacro*{textcite}{%
  \ifboolexpr{
    ( not test {\iffieldundef{prenote}} and
      test {\ifnumequal{\value{citecount}}{1}} )
    or
    ( not test {\iffieldundef{postnote}} and
      test {\ifnumequal{\value{citecount}}{\value{citetotal}}} )
  }
    {\DeclareFieldAlias{textcitehyperref}{noformat}}
    {}%
  \printtext[textcitehyperref]{%
    \restorebibmacro{textcite}%
    \usebibmacro{textcite}}}

\newcommand{\ndim}{d}
\newcommand{\nsamples}{n}
\newcommand{\indsamples}{\mu}

\makeindex

\usetikzlibrary{decorations.text}
\tikzstyle{var}=[circle, draw, very thick, minimum size=12pt, inner sep=0pt, fill=burntorange, font=\small]
\tikzstyle{inter}=[rectangle, draw, very thick, minimum size=12pt, inner sep=0pt, fill=teal, font=\small]
\tikzstyle{field}=[rectangle, draw, very thick, minimum size=12pt, inner sep=0pt, fill=amber, font=\small]
\tikzstyle{fun}=[rectangle, draw, very thick, minimum size=15pt, inner sep=0pt, fill=teal]
\tikzstyle{edge} = [draw, thick, -]
\newcommand{\AxisRotator}[1][rotate=0]{%
    \tikz [x=0.25cm,y=0.60cm,line width=.2ex,-stealth,#1, dashed] \draw (0,0) arc (-150:150:1 and 1);%
}
\newcommand{\arcarrow}[8]{
  \pgfmathsetmacro{\rin}{#1}
  \pgfmathsetmacro{\rmid}{#2}
  \pgfmathsetmacro{\rout}{#3}
  \pgfmathsetmacro{\astart}{#4}
  \pgfmathsetmacro{\aend}{#5}
  \pgfmathsetmacro{\atip}{#6}
  \fill[#7] (\astart:\rin) arc (\astart:\aend:\rin)
       -- (\aend+\atip:\rmid) -- (\aend:\rout) arc (\aend:\astart:\rout)
       -- (\astart+\atip:\rmid) -- cycle;
  \path[decoration = {text along path, text = {#8},
    text align = {align = center}, raise = -0.5ex}, decorate]
    (\astart+\atip:\rmid) arc (\astart+\atip:\aend+\atip:\rmid);
}

\tikzset{My Arrow Style/.style={single arrow, fill=red!50, anchor=base, align=center,text width=2.8cm}}

\begin{document}

\frenchspacing 
\raggedbottom 
\selectlanguage{american} 
\pagenumbering{roman} 
\pagestyle{plain} 

\ifthenelse{\equal{\cover}{true}}
	{
\begin{titlepage}
\newgeometry{left=7.5cm,bottom=1cm, top=1cm, right=1cm}
\tikz[remember picture,overlay] \node[opacity=1,inner sep=0pt] at (-28mm,-135mm){\includegraphics{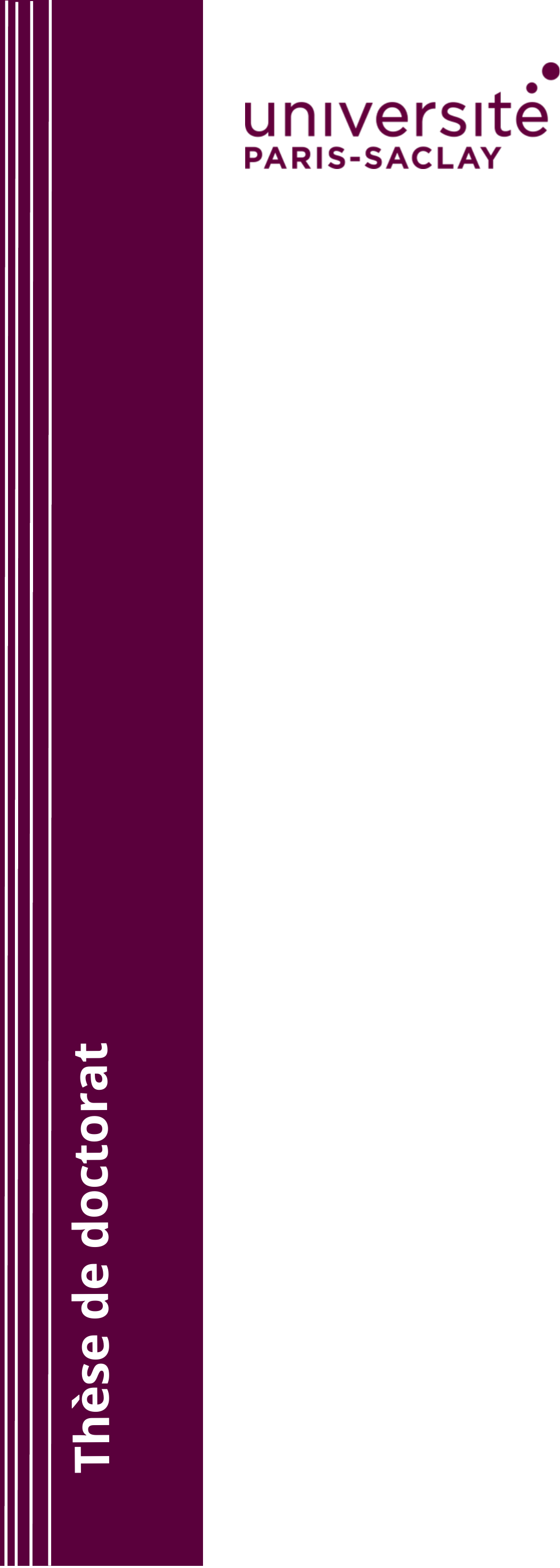}};

\fontfamily{fvs}\fontseries{m}\selectfont

\color{white}
\begin{picture}(0,0)
\put(-150,-735){\rotatebox{90}{NNT: 2020UPASP083}}
\end{picture}
 
\flushright
\vspace{15mm} 
\color{Prune}
\fontfamily{fvs}\fontseries{m}\fontsize{22}{26}\selectfont
\begin{minipage}[r]{0.95\linewidth}
	\myTitlesmall
\end{minipage}


\normalsize
\vspace{1cm}

\color{black}
\textbf{Thèse de doctorat de l'université Paris-Saclay}

\vspace{1cm}

École doctorale n$^{\circ}$ 564\\
École Doctorale Physique en Île-de-France (EDPIF) \\
\small Spécialité de doctorat: Physique\\
\footnotesize Unité de recherche: Université Paris-Saclay, CNRS, CEA\\
Institut de physique théorique\\
91191, Gif-sur-Yvette, France.\\
\footnotesize Référent: Faculté des sciences d’Orsay
\vspace{1cm}

\textbf{Thèse présentée et soutenue en visioconférence totale, le 16/12/2020, par}\\
\bigskip
\Large {\color{Prune} \textbf{Benjamin AUBIN}}

\vspace{0.5cm}


\flushleft \small \textbf{Composition du jury:}
\bigskip

\scriptsize
\begin{tabular}{|p{8cm}l}
\arrayrulecolor{Prune}
\textbf{Romain COUILLET} &  Président \\
Professeur, Centrale-Supélec & \\
Université Paris-Saclay & \\
\textbf{Sundeep RANGAN} &   Rapporteur \& Examinateur\\ 
Professeur, directeur associé & \\
NYU Wireless  & \\
\textbf{David SAAD} &  Rapporteur \& Examinateur \\ 
Professeur &\\
Aston University &\\
\textbf{Marc MEZARD} &  Examinateur \\ 
Directeur de recherche CNRS    &   \\ 
École Normale Supérieure & \\
\textbf{Alberto ROSSO} &  Examinateur \\ 
Directeur de recherche CNRS & \\
Université Paris-Saclay & \\
\end{tabular} 

\medskip
\begin{tabular}{p{8cm}l}\arrayrulecolor{white}
\textbf{Lenka ZDEBOROVA} &   Directrice de thèse\\ 
Directrice de recherche CNRS  & \\
EPFL & \\
\textbf{Florent KRZAKALA} &  Invité\\ 
Professeur & \\
EPFL & \\
\end{tabular} 

\end{titlepage}
} 
	{\newpage}

\ifthenelse{\equal{\format}{oneside}}
	{\clearpage\null\thispagestyle{empty}\newpage}
	{\cleardoublepage}
	
\begin{titlepage}
\afterpage{
\newgeometry{top=1.5cm, bottom=1.25cm, left=2cm, right=2cm}

\begin{center}
\large
\hfill
\vfill
\begingroup
\begin{spacing}{2}
\color{\colortitle}\spacedallcaps{\huge \bfseries 
\myTitlesmall
}
\end{spacing}
\bigskip \bigskip
\endgroup

\vspace{1cm}
{\color{\colorsubtitle}\spacedallcaps{ \bfseries \myName}}\\ \medskip 
\myDepartment \\
\myFaculty \\
\bigskip

\vfill
\myTime

\vfill
\end{center}

\clearpage
\restoregeometry
}
\end{titlepage}  
\ifthenelse{\equal{\format}{oneside}}
	{\clearpage\null\thispagestyle{empty}\newpage}
	{\cleardoublepage}
	
\thispagestyle{empty}
\refstepcounter{dummy}


\hspace{0pt}
\vfill
\begin{center}
The constructionist hypothesis breaks down when confronted with the twin difficulties of scale and complexity. The behavior of large and complex aggregates of elementary particules, it turns out, is not to be understood in terms of a simple extrapolation of the properties of a few particles. Instead, at each level of complexity entirely new properties appear, and the understanding of the new behaviors requires research which I think is as fundamental in its nature as any other.\\ \medskip
\emph{More is different}--- P. W. Anderson (1972)   
\end{center}
\vfill
\hspace{0pt}
\ifthenelse{\equal{\format}{oneside}}
	{\clearpage\null\thispagestyle{empty}}
	{\cleardoublepage}
	
\begingroup
\chapter*{Acknowledgements}
\addcontentsline{toc}{chapter}{Acknowledgements}
First of all, I would like to warmly thank Sundeep Rangan, David Saad, Romain Couillet, Marc Mézard and Alberto Rosso for accepting to read and review this Ph.D manuscript. \\
\vspace{-0.3cm}

Ensuite, je souhaite remercier chaleureusement Lenka et Florent pour m'avoir donné la chance de passer ces trois années à leur contact et l'occasion de voyager aux quatre coins du monde. Membre à part entière de leur grande famille, ils m'auront donné goût à la recherche fondamentale et énormément appris autour de nombreuses discussions et moments de convivialité.
Naturellement je remercie tous les membres des groupes de recherche \emph{Sphinx} \& \emph{Smile} pour leur bonne humeur et enthousiasme: Alia, Marylou, Antoine M, Stefano et Francesca avec qui j'ai partagé (à l'occasion) mon bureau à l'IPhT, et tous les autres:
 Federica, Jonathan, Cédric, Ruben, Hugo, Maria, Paula, Luca, Gabriele, Laura et Alejandro pour leur contact quotidien. Et surtout, un immense merci à Sebastian, Bruno, Christian, Antoine B, Stéphane et Pierre pour la relecture de ce manuscrit, et à Alaa, Thibault, André et Levent pour leurs nombreux conseils avisés.
Je souhaite évidemment remercier Jean, Nicolas, Will et Yue pour avoir contribué à démontrer certains résultats de ce manuscript. 
Une pensée particulière à Felix pour avoir partagé notre bureau, nos repas, nos débats socio-politiques et pour avoir bravé les vagues au milieu des "requins"; et enfin à Marco qui a su nous faire profiter d'une éclipse surréaliste au milieu de la Death Valley.
Merci à Tristan, Clément, Louise, José et Dhruv pour avoir égayé les couloirs de l'ENS; à Léo et Alexandre pour notre expérience de crypto-trading; à Samuel Kindermann et Velten Doering pour leurs suggestions musicales; et à Stefano et Riccardo pour m'avoir appris quelques "rudiments" d'italien. 
Merci à Pierfrancesco pour avoir répondu à toutes mes questions techniques, à Guilhem pour sa gentillesse et son humour détonant et à Giulio pour ses conseils et pour m'avoir laissé la possibilité d'étudier la physique tout en pratiquant la planche à voile et le catamaran, et en suivant la coupe du monde de football.
Enfin un grand merci à Laure, Sylvie et Carine pour leur gentillesse, leur sens du détail et le travail administratif colossal qu'elles m'auront aidé à surmonter.
Une attention toute particulière à G. Montambaux et J. P. Bouchaud qui m'ont introduit et enseigné la physique statistique des systèmes complexes à l'X, et sans qui mon parcours aurait été probablement très différent. 
Merci à Marc Goerbig pour son amitié et son accueil chaleureux au LPS d'Orsay et à Léon Bottou pour sa bienveillance au sein de FAIR.\\
\vspace{-0.3cm}

Enfin merci à Albane pour son soutien dans les moments de doute et surtout pour avoir toléré mon ordinateur allumé lors de nos soirées films! Pour finir, j'éprouve une reconnaissance toute particulière envers mes parents sans qui je ne serais jamais arrivé aussi loin.
\endgroup 
\ifthenelse{\equal{\format}{oneside}}
	{\clearpage\null\thispagestyle{empty}}
	{\cleardoublepage}
\pagestyle{scrheadings} 



\refstepcounter{dummy}
\renewcommand{\contentsname}{Table of contents}
\addcontentsline{toc}{chapter}{Table of contents}
\setcounter{tocdepth}{2} 
\setcounter{secnumdepth}{3} 
\manualmark
\markboth{\spacedlowsmallcaps{\contentsname}}{\spacedlowsmallcaps{\contentsname}}
\tableofcontents 
\automark[section]{chapter}
\renewcommand{\chaptermark}[1]{\markboth{\spacedlowsmallcaps{#1}}{\spacedlowsmallcaps{#1}}}
\renewcommand{\sectionmark}[1]{\markright{\thesection\enspace\spacedlowsmallcaps{#1}}}

\refstepcounter{dummy}
\begingroup 


\markboth{\spacedlowsmallcaps{Abbreviations and symbols}}{\spacedlowsmallcaps{Abbreviations and symbols}}
\phantomsection
\chapter*{List of abbreviations}
\addcontentsline{toc}{chapter}{List of abbreviations}

\acrodef{AMP}{Approximate Message Passing}
\acrotarget{AMP}
\acrodef{ML}{Multi-layer}
\acrotarget{ML}
\acrodef{SE}{State Evolution}
\acrotarget{SE}
\acrodef{SP}{Saddle Point}
\acrotarget{SP}
\acrodef{VAE}{Variational Auto-Encoder}
\acrotarget{VAE}
\acrodef{AE}{Auto-Encoder}
\acrotarget{AE}
\acrodef{GAN}{Generative Adversarial Network}
\acrotarget{GAN}
\acrodef{KL}{Kullback-Leibler}
\acrotarget{KL}
\acrodef{MLE}{Maximum Likelihood Estimator}
\acrotarget{MLE}
\acrodef{MAP}{Maximum A Posteriori}
\acrotarget{MAP}
\acrodef{MSE}{Mean Squared Error}
\acrotarget{MSE}
\acrodef{MMSE}{Minimum Mean Squared Error}
\acrotarget{MMSE}
\acrodef{RV}{Random Variable}
\acrotarget{RV}
\acrodef{PAC}{Probably Approximately Correct}
\acrotarget{PAC}
\acrodef{RS}{Replica Symmetry}
\acrotarget{RS}
\acrodef{RSB}{Replica Symmetry Breaking}
\acrotarget{RSB}
\acrodef{dAT}{de Almeida Thouless}
\acrotarget{dAT}
\acrodef{GD}{Gradient-Descent}
\acrotarget{GD}
\acrodef{SGD}{Stochastic Gradient-Descent}
\acrotarget{SGD}
\acrodef{JPD}{Joint Probability Distribution}
\acrotarget{JPD}
\acrodef{DAG}{Directed Acyclic Graphs}
\acrotarget{DAG}
\acrodef{MRF}{Markov Random Field}
\acrotarget{MRF}
\acrodef{CSP}{Constraints Satisfaction Problem}
\acrotarget{CSP}
\acrodef{rCSP}{random Constraints Satisfaction Problem}
\acrotarget{rCSP}
\acrodef{BP}{Belief Propagation}
\acrotarget{BP}
\acrodef{GLM}{Generalized Linear Model}
\acrotarget{GLM}
\acrodef{MF}{Mean-Field}
\acrotarget{MF}
\acrodef{i.i.d}{{\textrm{i.i.d}~}}
\acrotarget{i.i.d}
\acrodef{CLT}{Central Limit Theorem}
\acrotarget{CLT}
\acrodef{RFIM}{Random Field Ising Model}
\acrotarget{RFIM}
\acrodef{SK}{Sherrington-Kirkpatrick}
\acrotarget{SK}
\acrodef{EA}{Edwards-Anderson}
\acrotarget{EA}
\acrodef{MCMC}{Markov-Chain Monte-Carlo}
\acrotarget{MCMC}
\acrodef{MC}{Monte-Carlo}
\acrotarget{MC}
\acrodef{AI}{Artificial Intelligence}
\acrotarget{AI}
\acrodef{ML}{Machine Learning}
\acrotarget{ML}
\acrodef{DL}{Deep Learning}
\acrotarget{DL}
\acrodef{SVM}{Support Vector Machines}
\acrotarget{SVM}
\acrodef{GPU}{Graphics Processing Units}
\acrotarget{GPU}
\acrodef{CPU}{Central Processing Units}
\acrotarget{CPU}
\acrodef{VC}{Vapnik-Chervonenkis}
\acrotarget{VC}
\acrodef{PAC}{Probably Approximately Correct}
\acrotarget{PAC}
\acrodef{T-S}{Teacher-Student}
\acrotarget{T-S}
\acrodef{CS}{Compressed Sensing}
\acrotarget{CS}
\acrodef{PR}{Phase Retrieval}
\acrotarget{PR}
\acrodef{SM}{Statistical Mechanics}
\acrotarget{SM}
\acrodef{IT}{Information Theory}
\acrotarget{IT}
\acrodef{SG}{Spin Glass}
\acrotarget{SG}
\acrodef{SI}{Statistical Inference}
\acrotarget{SI}
\acrodef{ANN}{Artificial Neural Networks}
\acrotarget{ANN}
\acrodef{DNN}{Deep Neural Networks}
\acrotarget{DNN}
\acrodef{CNN}{Convolutional Neural Networks}
\acrotarget{CNN}
\acrodef{LSTM}{Long Short-Term Memory}
\acrotarget{LSTM}
\acrodef{RNN}{Recurrent Neural Network}
\acrotarget{RNN}
\acrodef{NLP}{Natural Language Processing}
\acrotarget{NLP}
\acrodef{RL}{Reinforcement Learning}
\acrotarget{RL}
\acrodef{PCA}{Principal Component Analysis}
\acrotarget{PCA}
\acrodef{SVD}{Singular Value Decomposition}
\acrotarget{SVD}
\acrodef{ERM}{Empirical Risk Minimization}
\acrotarget{ERM}
\acrodef{rBP}{relaxed Belief Propagation}
\acrotarget{rBP}
\acrodef{GAMP}{Generalized Approximate Message Passing}
\acrotarget{GAMP}
\acrodef{TAP}{Thouless-Anderson-Palmer}
\acrotarget{TAP}
\acrodef{ReLU}{REcitfied Linear Unit}
\acrotarget{ReLU}
\acrodef{EP}{Expectation Propagation}
\acrotarget{EP}
\acrodef{EC}{Expectation Consistency}
\acrotarget{EC}
\acrodef{f1RSB}{1-Step Replica Symmetry Breaking}
\acrotarget{1fRSB}
\acrodef{FRSB}{Full Replica Symmetry Breaking}
\acrotarget{FRSB}
\acrodef{1RSB}{One-step Replica Symmetry Breaking}
\acrotarget{1RSB}
\acrodef{2RSB}{Two-steps Replica Symmetry Breaking}
\acrotarget{2RSB}
\acrodef{RI}{Rotationally Invariant}
\acrotarget{RI}
\acrodef{LAMP}{Linearized Approximate Message Passing}
\acrotarget{LAMP}
\acrodef{BBP}{Baik, Ben Arous and P\'ech\'e}
\acrotarget{BBP}


\begin{acronym}[UML]\itemsep0pt
\acro{AI}{Artificial Intelligence}
\acro{AMP}{Approximate Message Passing}
\acro{ANN}{Artificial Neural Networks}

\acro{BBP}{Baik, Ben Arous and P\'ech\'e}
\acro{BP}{Belief Propagation}

\acro{CLT}{Central Limit Theorem}
\acro{CNN}{Convolutional Neural Networks}
\acro{CPU}{Central Processing Units}
\acro{CS}{Compressed Sensing}
\acro{CSP}{Constraints Satisfaction Problem}

\acro{DAG}{Directed Acyclic Graphs}
\acro{dAT}{de Almeida Thouless}
\acro{DL}{Deep Learning}
\acro{DNN}{Deep Neural Networks}

\acro{EA}{Edwards-Anderson}
\acro{EC}{Expectation Consistency}
\acro{EP}{Expectation Propagation}

\acro{f1RSB}{frozen 1-step Replica Symmetry Breaking}
\acro{FRSB}{Full Replica Symmetry Breaking}

\acro{GAMP}{Generalized Approximate Message Passing}
\acro{GAN}{Generative Adversarial Network}
\acro{GD}{Gradient-Descent}
\acro{GLM}{Generalized Linear Model}
\acro{GPU}{Graphics Processing Units}

\acro{i.i.d}{independent and identically distributed}
\acro{IT}{Information Theory}

\acro{JPD}{Joint Probability Distribution}

\acro{KL}{Kullback-Leibler}

\acro{LAMP}{Linearized Approximate Message Passing}
\acro{LSTM}{Long Short-Term Memory}

\acro{MAP}{Maximum A Posteriori}
\acro{MC}{Monte-Carlo}
\acro{MCMC}{Markov-Chain Monte-Carlo}
\acro{ML}{Machine Learning}
\acro{MLE}{Maximum Likelihood Estimator}
\acro{MMSE}{Minimum Mean Squared Error}
\acro{MRF}{Markov Random Field}
\acro{MSE}{Mean Squared Error}

\acro{NLP}{Natural Language Processing}

\acro{PAC}{Probably Approximately Correct}
\acro{PCA}{Principal Component Analysis}
\acro{PR}{Phase Retrieval}

\acro{r-BP}{relaxed Belief Propagation}
\acro{rCSP}{random Constraints Satisfaction Problem}
\acro{ReLU}{Recitfied Linear Unit}
\acro{RFIM}{Random Field Ising Model}
\acro{RL}{Reinforcement Learning}
\acro{RNN}{Recurrent Neural Network}
\acro{RS}{Replica Symmetry}
\acro{RSB}{Replica Symmetry Breaking}
\acro{1RSB, 2RSB}{One-step / Two-steps Replica Symmetry Breaking}
\acro{RI}{Rotationally Invariant}
\acro{RV}{Random Variable}

\acro{SE}{State Evolution}
\acro{SG}{Spin Glass}
\acro{SGD}{Stochastic Gradient-Descent}
\acro{SI}{Statistical Inference}
\acro{SK}{Sherrington-Kirkpatrick}
\acro{SP}{Saddle Point}
\acro{SVD}{Singular Value Decomposition}
\acro{SVM}{Support Vector Machines}

\acro{T-S}{Teacher-Student}
\acro{TAP}{Thouless-Anderson-Palmer}

\acro{VAE}{Variational Auto-Encoder}
\acro{VAMP}{Vector Approximate Message Passing}
\acro{VC}{Vapnik-Chervonenkis}

\end{acronym}  

\newpage

\phantomsection
\renewcommand{\nomname}{List of symbols}

\renewcommand\nomgroup[1]{%
  \item[\bfseries
  \ifstrequal{#1}{A}{Algebra}{%
  \ifstrequal{#1}{B}{Probabilities}{%
  \ifstrequal{#1}{C}{Calculus and functions}{%
  \ifstrequal{#1}{D}{Sets, graphs and indexing}{%
  \ifstrequal{#1}{E}{Physics}{}}}}}%
]}

\nomenclature[A, 01]{$a_i$}{A scalar}
\nomenclature[A, 02]{$\vec{a}$}{A vector}
\nomenclature[A, 03]{$\mat{A}$}{A matrix}
\nomenclature[A, 04]{$\vec{a} \cdot \vec{b}$}{Scalar product of the vectors $\vec{a}$ and $\vec{b}$}
\nomenclature[A, 05]{$\mat{A}^\intercal$}{Transpose of the matrix $\mat{A}$}
\nomenclature[A, 06]{$\mat{A}\otimes\mat{B}$}{Tensorial product of $\mat{A}$ and $\mat{B}$}
\nomenclature[A, 07]{$\mat{A} \times \mat{B}$}{Hadamard product of $\mat{A}$ and $\mat{B}$}
\nomenclature[A, 08]{$\mat{I}_\ndim$}{Identity matrix of size $\ndim \times \ndim$} 
\nomenclature[A, 09]{$\mat{J}_\ndim$}{Matrix full of ones of size $\ndim \times \ndim$}
\nomenclature[A, 10]{$\vec{1}_\ndim$}{Vector of ones of size $\ndim$} 
\nomenclature[A, 11]{$\det{\mat{A}}$}{Determinant of $\mat{A}$} 
\nomenclature[A, 12]{$\tr{\mat{A}}$}{Trace of $\mat{A}$} 

\nomenclature[B, 01]{$\rX \vert \rY$}{The random variable $\rX$ knowing the variable $\rY$}
\nomenclature[B, 02]{$\bbP(\rX)$}{The probability of the random variable $\rX$, shorthand for $\bbP(\rX=x)$}
\nomenclature[B, 03]{$\rX \sim \rP_\x(.)$}{$\rX$ is distributed according to the distribution $\rP_\x$}
\nomenclature[B, 04]{$\rp_\x(.)$}{Density of $\rP_\x(.)$: $\d \rP_\x(x) = \rp(x) ~ \d x $}
\nomenclature[B, 05]{$\EE_{\rX}[x]$}{Expectation of the random variable $\rX$}
\nomenclature[B, 06]{$\Var{\rX}$}{Variance of the random variable $\rX$}
\nomenclature[B, 07]{$\mN_{\vec{x}}\(\bmu, \bSigma\)$}{Gaussian distribution of the vector $\vec{x}$ with mean $\bmu$ and covariance matrix $\bSigma$}
\nomenclature[B, 08]{$\D \vec{x}$}{Gaussian measure: $\D \vec{x} = \d \vec{x} ~ \mN_{\vec{x}}\(\vec{0}, \mat{I}\)$}

\nomenclature[C, 01]{$\log$}{Natural logarithm}
\nomenclature[C, 02]{$\equiv$}{Defined as}
\nomenclature[C, 03]{$\simeq$}{Equal to, up to negligible terms}
\nomenclature[C, 04]{$f:\bbA \mapsto \bbB$}{A function $f$ with domain $\bbA$ and range $\bbB$}
\nomenclature[C, 05]{$f(\vec{x};\btheta)$}{A function of $\vec{x}$ parametrized by $\btheta$}
\nomenclature[C, 06]{$f \circ g$}{Composition of the functions $f$ and $g$}
\nomenclature[C, 08]{$\id[x]$}{Indicator function equals to $1$ if $x$ is true and $0$ otherwise}
\nomenclature[C, 07]{$\|.\|_p$}{$L^p$ norm}

\nomenclature[D, 01]{$\bbA$}{A set}
\nomenclature[D, 02]{$\bbA\setminus\bbB$}{A set containing the elements of $\bbA$ that are not in $\bbB$}
\nomenclature[D, 03]{$\bbR, \bbZ, \bbN$}{The sets of real numbers, integers and positive integers}
\nomenclature[D, 06]{$\lb n \rb $}{Integers interval between $1$ and $n$}
\nomenclature[D, 07]{$[ a ; b ]$}{Reals interval including $a$ and $b$}
\nomenclature[D, 08]{$\{x_i\}_{i=1}^n = \{x_1, \cdots, x_n \}$}{A set containing the elements $x_i$ for $i \in \lb n \rb $}
\nomenclature[D, 09]{$<ij>$}{All neighbouring pairs ($i, j$)}
\nomenclature[D, 10]{$\partial_i$}{All neighbours of the node $i$}
\nomenclature[D, 11]{$\partial_i \setminus j$}{$\partial_i$ except the node $j$}

\nomenclature[E, 01]{$\mH_\ndim$, $\mZ_\ndim$}{Hamiltonian and partition function of $d$ variables}
\nomenclature[E, 02]{$\Phi, \varphi$}{Free entropy and free energy}
\nomenclature[E, 03]{$\rH$, $\mI$}{Entropy and mutual information}
\nomenclature[E, 04]{$\beta$}{Inverse temperature}
\nomenclature[E, 05]{$\langle . \rangle_\beta$}{Gibbs/Boltzmann average at inverse temperature $\beta$}

\printnomenclature[1.5cm]
	
\endgroup

%
%

%
%

%
%
\ifthenelse{\equal{\format}{oneside}}
	{\clearpage\null\thispagestyle{empty}}
	{\clearpage\null\thispagestyle{empty}
	\clearpage\null\thispagestyle{empty}}
	
\phantomsection
\chapter*{Foreword}
\addcontentsline{toc}{chapter}{Foreword}
\markboth{\spacedlowsmallcaps{Foreword}}{\spacedlowsmallcaps{Foreword}}

At a time when the use of data has reached an unprecedented level, the access to large datasets precipitated
their intense use to train machine learning models. The corresponding algorithms essentially aim to detect 
and make use of structured informations within excessively large datasets. Specifically, after many twists and turns, 
the celebrated, now ubiquitous, deep-learning models, based on artificial neural networks architectures, brought important numerical progresses in this direction. 
Overtaking other existing models from the mid-2000s, they became, in just a few years, indispensable in many industrial 
applications such as image classification, speech recognition, text mining, \cite{LeCun15} or time series prediction, 
object detection for face recognition, natural language processing, medical diagnosis, etc.

However, understanding most of the practical \emph{gradient-based} algorithms used to train these oversized 
and complex networks, which contain up to millions of parameters, remains empirical and challenging to analyze theoretically. 
The main issue arising with \emph{deep} and \emph{wide} neural network architectures lies essentially in the succession of numerous layers through non-linear operations that make the space of optimization very \emph{high-dimensional} and \emph{complex}.
Handling and visualizing this large collection of parameters is the central mathematical difficulty in most of 
state-of-the-art machine learning models and algorithms.
This lack of theoretical understanding raises many questions about their efficiency and potential risks in many areas. 
As a result, establishing theoretical foundations on simple models and providing numerical prescriptions on which to base and explain empirical observations have become one of the fundamental challenges of the research community.\\
 
In this manuscript, we investigate these burning questions, arising in machine learning, through the lens of statistical physics of disordered systems. This singular transversal approach to computer science problems has a long and rich history \cite{engel2001statistical,mezard2009information,grassberger2012statistical, zdeborova2016statistical, advani2017high}, that we revisit in the \emph{high-dimensional regime} by focusing especially on modern algorithmic considerations and rigorous justifications.
Specifically in the context of oversized neural networks, for which the number of parameters explodes, exact analytical solutions are unknown most of the time and numerical computations are ruled out. Techniques from statistical physics have been precisely designed to infer the macroscopic behavior of such a large collection of \emph{particles} from the microscopic description of their elementary interactions. 
They offer a suitable set of approximations, called \emph{mean-fields methods}, that are simple enough to be computationally tractable and rich enough to capture and reproduce interesting features of the system. Moreover, in this \emph{thermodynamic limit}, physicists experienced that macroscopic behaviors are typically described correctly by a set of a few \emph{order parameters}. 

Applied to machine learning theory, which precisely lacks such techniques, we believe that statistical physics insights may contribute in identifying the set of relevant observables that control the large-scale properties of the system, and provide a powerful framework to analyze such complex artificial neural networks. 
Unfortunately, even though very powerful and believed to lead to the correct result in many situations, these techniques were derived historically without rigorous foundations. Therefore, this work is part of the current momentum of the mathematical physics community that focuses on proving former results obtained heuristically in the 90's. Additionally, while these former statistical analysis were not discussing computational perspectives, we revisit this approach by focusing on the potential \emph{algorithmic phase transitions}.\\ 

 At the heart of this work, we strongly capitalize on a probabilistic Bayesian reasoning, which contrasts with the traditional optimization approach. Moreover, we make an intense use of the deep connection between the \emph{replica method} and \emph{approximate message passing} algorithms to elicit the phase diagrams of simple theoretical models, which reveal nonetheless interesting features. By revisiting the \emph{teacher-student} paradigm, that allows to create synthetic, but tractable, tasks, we focus our attention on emphasizing the potential gaps between \emph{statistical} and \emph{computational} thresholds.

We illustrate the efficiency of these mean-field methods on various poorly understood machine learning models.
We essentially focus on synthetic tasks and data generated in the \emph{teacher-student} paradigm, and we contribute to their understanding by describing their rich phase diagrams.
First, we start by presenting the \emph{Bayes-optimal analysis} of committee machines that reveals the existence of large computational-to-statistical gaps. Next, in a \emph{worst-case} analysis, we bring to light a strong connection between the Rademacher generalization bound from statistical learning theory, and the storage capacity and ground state energies from the statistical physics literature, which allows us to explicitly compute the Rademacher complexity of perceptrons. We finally complete the picture by analyzing the intensively used \emph{empirical risk minimization} of generalized linear models and we compare it to the previous \emph{Bayes-optimal} and \emph{worst-case analysis}.
In another research direction, we define a general procedure to combine elementary models already analyzed to build up more complex and structured architectures.
In this way, we develop a framework that overcomes in particular the standard \emph{separable} prior assumption and makes possible to analyze estimation models, such as low-rank matrix estimation, phase retrieval or compressed sensing, with deep generative priors based on random weights.

\phantomsection
\section*{Organization of the manuscript}
\addcontentsline{toc}{section}{Organization of the manuscript}
\markboth{\spacedlowsmallcaps{Organization of the manuscript}}{\spacedlowsmallcaps{Organization of the manuscript}}

As my work of Ph.D lies at the crossroads of machine learning and statistical physics of disordered systems, in \Part\ref{part:introduction} I take the opportunity to pedagogically present the basic, yet essential, theoretical concepts to follow the rest of the manuscript. 
In \Chap\ref{chap:review_ml}, I propose a high-level overview of the field of machine learning with a focus on its tortuous history, basic concepts and current challenges. \Chap\ref{chap:statistical_physics} covers the basic tools of statistical physics that are relevant to understand the original approach we employ to tackle machine learning problems. These two first chapters are devoted to readers unfamiliar with one or the other background and can be skipped by experts.
In \Chap\ref{chap:phys_ml_together}, we provide a selection of important historical references to understand how these two fields are intertwined for over thirty years. It is also the occasion to review a selection of the current research axes of the statistical physics approach in artificial neural networks. Finally, we introduce the crucial Bayesian probabilistic framework and its crucial connection with statistical physics. This constitutes the cornerstone of our approach which allows us to analyze simultaneously statistical inference and random constraint satisfaction problems.
In \Chap\ref{main:chap:mean_field}, we propose a methodological review of selected fundamental mean-field inference methods, originally motivated in the spin glass literature \cite{mezard1987spin}, that are mainly used in the second part of the dissertation. Specifically, we remind the details of the derivations of the replica method and message passing algorithms on the class of \emph{generalized linear models}, as a core example throughout this manuscript. Moreover, by highlighting their complementarities, we attempt to clarify how the methods are related and allow to reveal rich statistical and algorithmic phase transitions. \\
      
The \Part\ref{part:contribution} of this manuscript is devoted to cover the works I have contributed as a Ph.D student from October 2017 to December 2020, at \emph{Institut de Physique Théorique in CEA-Saclay} under the supervision of Lenka Zdeborov\'{a} and Florent Krzakala. 
The contents of the articles have already been published in a series of works which can be found online in their original format. They have been revised in order to standardize the notations of this manuscript.
In particular, for the sake of clarity and conciseness, some of the lengthy proofs and calculations to which I have not directly contributed are not reported in this manuscript and can be found in the original publications listed in \Sec\ref{contributions}.

\phantomsection
\section*{Contributions}	
\addcontentsline{toc}{section}{Contributions}
\label{contributions}
\markboth{\spacedlowsmallcaps{Contributions}}{\spacedlowsmallcaps{Contributions}}

\Part\ref{part:contribution}, which brings together my main contributions, is separated in two sub-parts corresponding to parallel axes of research. In order to best reflect my work, I will detail my personal contributions to the various co-signed articles in which I participated.

In \Part\ref{part:contribution}.~A, we discuss the complementary analysis of the \emph{Bayes-optimal} and \emph{worst-case} scenarios and \emph{empirical risk minimization} of simple feed-forward neural networks with separable prior distributions. 
In \Chap\ref{chap:committee_machine}, we first present the \emph{Bayes-optimal} approach on committee machines, that provides an information theoretical lower-bound perspective. 
Next, we describe the analysis of the storage capacity problem in \Chap\ref{chap:binary_perceptron} and related ground state energies, within a generic random constraint satisfaction problem framework. In \Chap\ref{chap:rademacher}, we show that these quantities turn out to be closely related to the \emph{worst-case} Rademacher complexity generalization error upper bound. Finally in \Chap\ref{chap:erm}, we investigate the \emph{practical} case with the analysis of \emph{empirical risk minimization} which is performed in practice with gradient-descent algorithms. 

 \begin{enumerate}
	\item \citepublication{Aubin2018}
	{
		Heuristic tools from statistical physics have been used in the past 
		  to locate the phase transitions and compute the optimal learning and generalization errors in the
		  teacher-student scenario in multi-layer neural networks.  
		  In this contribution, we provide a rigorous justification of these approaches
		   for a two-layers neural network model called the committee machine, under a technical assumption. 
		  We also introduce a version of the approximate
		  message passing (AMP) algorithm for the committee machine, that allows to perform optimal learning in polynomial time for a large set of
		  parameters. 
		  We find that there are regimes in which a low
		  generalization error is information-theoretically achievable while
		  the AMP algorithm fails to deliver it; strongly suggesting that no
		  efficient algorithm exists for those cases, and unveiling
		  a large computational gap.
	}{chap:committee_machine}
	\textbf{Personal contributions}: I have developed and implemented the AMP algorithm and its state evolution to depict the corresponding phase diagrams.\\
	\item \citepublication{Aubin_2019}
	{We study the problem of determining the capacity of the binary perceptron for two
		variants of the problem where the corresponding constraint is symmetric. We call these variants the rectangle-binary-perceptron (RPB) and the $u-$function-binary-perceptron (UBP). We show that, unlike for the
		  usual step-function-binary-perceptron, the critical capacity in these symmetric cases is given by the annealed computation in a large region of parameter space, for all rectangular constraints and for narrow enough $u-$function constraints,
		 $K<K^*$.  We prove this result, under two natural assumptions, using the first and second moment
		methods.  We further use the second moment method to conjecture that solutions of the symmetric binary perceptrons are organized in a so-called
		frozen-1RSB structure, without using the replica method. We then use
		the replica method to estimate the capacity threshold for the UBP case
		when the $u-$function is wide $K>K^*$. We conclude that
		full-step-replica-symmetry breaking would have to be evaluated in
		order to obtain the exact capacity in this case.
	}{chap:binary_perceptron}
	\textbf{Personal contributions}: I have analyzed the RSB Ansätze of the replica free entropy to study the Gardner capacity and the configuration space geometry. I also contributed to the first and second moments proofs.  
	\item \citepublication{abbara2020rademacher}
	{
	Statistical learning theory provides bounds of the generalization gap, using in particular the Vapnik-Chervonenkis dimension and the Rademacher complexity. An alternative approach, mainly studied in the statistical physics literature, is the study of generalization in simple synthetic-data models. Here we discuss the connections between these approaches and focus on the link between the Rademacher complexity in statistical learning and the theories of generalization for \emph{typical-case} synthetic models from statistical physics, involving quantities known as \emph{Gardner capacity} and \emph{ground state energy}. We show that in these models the Rademacher complexity is closely related to the ground state energy computed by replica theories. Using this connection, one may reinterpret many results of the literature as rigorous Rademacher bounds in a variety of models in the high-dimensional statistics limit. Somewhat surprisingly, we also show that statistical learning theory provides predictions for the behavior of the ground-state energies in some full replica symmetry breaking models.
	}{chap:rademacher}
	\textbf{Personal contributions}: I derived and evaluated the ground state energies and draw the connection with the Rademacher complexity.
	\item \citepublication{aubin2020generalization}
	{  
			We consider a commonly studied supervised classification of
		  a synthetic dataset whose labels are generated by feeding a one-layer neural network with random \iid inputs. 
		We study the generalization performances of standard
		  classifiers in the high-dimensional regime
		  where $\alpha={\nsamples}/{\ndim}$ is kept finite in the limit
		  of a high dimension $\ndim$ and number of samples $\nsamples$. 
		Our contribution is three-fold: First, we prove
		  a formula for the generalization error achieved by $\rL_2-$regularized
		  classifiers that minimize a convex loss.
		  This formula was first
		  obtained by the heuristic replica
		  method of statistical physics.
		 Secondly, focusing on commonly used loss functions and optimizing
		 the $\rL_2$ regularization strength, we
		  observe that while ridge regression performance is
		  poor, logistic and hinge regression are surprisingly able to
		  approach the Bayes-optimal generalization error extremely closely. 
		  As $\alpha \to \infty$ they lead to Bayes-optimal rates, a fact that does
		  not follow from predictions of margin-based generalization error bounds. 
		Third, we design an optimal loss and regularizer that provably leads to Bayes-optimal
		  generalization error.
	}{chap:erm}
	\textbf{Personal contributions}: I conducted the theoretical analysis and the numerical evaluations. I also contributed to the proofs based on the Gordon min-max theorem.
\suspend{enumerate}

In \Part\ref{part:contribution}.~B, we present a line of research conducted in parallel that investigates different kinds of prior informations for estimation problems, such as the spiked matrix model presented in \Chap\ref{chap:generative_spiked} or compressed sensing and phase retrieval detailed in \Chap\ref{chap:generative_phase}. Specifically, we compare the statistical-to-algorithmic gaps for sparse separable priors and structured deep generative priors with random weights.

\resume{enumerate}
	\item \citepublication{Aubin2019c}
	{
	Using a low-dimensional parametrization of signals is a generic and powerful way to enhance performance in signal processing and statistical inference. A very popular and widely explored type of dimensionality reduction is sparsity; another
	type is generative modeling of signal distributions. Generative models based on neural networks, such as GANs or variational auto-encoders,
	are particularly performant and are gaining on applicability. In this paper we study spiked matrix models, where a low-rank matrix is observed through a noisy
	channel. This problem with sparse structure of the spikes has attracted
	broad attention in the past literature. Here, we replace the sparsity assumption by
	generative modelling, and investigate the consequences on statistical and
	algorithmic properties. We analyze the Bayes-optimal
	performance under specific generative models for the spike. In contrast with
	the sparsity assumption, we do not observe regions of parameters where
	statistical performance is superior to the best known algorithmic
	performance. We show that in the analyzed cases the approximate
	message passing algorithm is able to reach optimal performance. We also design
	enhanced spectral algorithms and analyze their performance and
	thresholds using random matrix theory, which was performed by collaborators, showing their superiority to the
	classical principal component analysis. We complement our theoretical
	results by illustrating the performance of the spectral algorithms when the spikes come from real datasets.
	}{chap:generative_spiked}
	\textbf{Personal contributions}: I co-developed the plug-in framework to combine the replica free entropies and the AMP algorithms of sub-models, in close collaboration with B.~Loureiro. I also co-developed the LAMP spectral method and conducted the numerical implementations.
	\item \citepublication{aubin2020exact}{
	We consider the problem of compressed sensing and of (real-valued) phase retrieval with random measurement matrix.
	We derive sharp asymptotics for the information-theoretically optimal performance and for the best known polynomial algorithm for an ensemble of generative priors consisting of fully connected deep neural networks with random weight matrices and arbitrary activations. 
	We compare the performance to sparse separable priors and conclude that in all cases analyzed generative priors have a smaller statistical-to-algorithmic gap than sparse priors, giving theoretical support to previous experimental observations that generative priors might be advantageous in terms of algorithmic performance.
	In particular, while sparsity does not allow to perform compressive phase retrieval efficiently close to its information-theoretic limit, it is found that under the random generative prior compressed phase retrieval becomes tractable.
	}
	{chap:generative_phase}
	\textbf{Personal contributions}: I conducted the numerical analysis and evaluation of the phase transitions.	
\suspend{enumerate}

Finally, I have contributed to an additional work, not covered in this dissertation, that introduces a modular python implementation of compositional inference on tree-structured inference models. However, in line with previous works, in \Sec\ref{chap:generative_phase:applications_real} we present simple applications of the algorithm to estimation problems with generative priors trained on real datasets.

\resume{enumerate}
	\item \citepublicationnoref{baker2020tramp}{
	We introduce \textsf{tramp}, standing for \emph{TRee Approximate Message Passing}, a python package for compositional inference in
	high-dimensional tree-structured models. The package provides an unifying framework to study several 
	approximate message passing algorithms previously derived for a variety of machine learning tasks such as generalized linear models, inference in multi-layer networks, matrix factorization, and reconstruction using non-separable penalties. For some models, the asymptotic performance of the algorithm can be theoretically predicted by the state evolution, and the measurements entropy estimated by the free entropy formalism. The implementation is modular by design: each module, which implements a factor, can be composed at will with other modules to solve complex inference tasks. The user only needs to declare the factor graph of the model: the inference algorithm, state evolution and entropy estimation are fully automated. The source code is publicly available at \href{https://github.com/sphinxteam/tramp}{\url{https://github.com/sphinxteam/tramp}} and the documentation is accessible at
	\href{https://sphinxteam.github.io/tramp.docs}{\url{https://sphinxteam.github.io/tramp.docs}}.\newline
	\textbf{Personal contributions}: I contributed to implement some parts of the source code and developed entirely the online documentation. I mainly investigated the reconstruction performances of the package for generative priors with weights trained with different GAN and VAE architectures.
	}
	\item
\end{enumerate}

\clearpage\null\thispagestyle{empty}
\clearpage\null\thispagestyle{empty}
\phantomsection
\chapter*{Avant-propos}
\addcontentsline{toc}{chapter}{Avant-propos}
\markboth{\spacedlowsmallcaps{Avant-propos}}{\spacedlowsmallcaps{Avant-propos}}

À une époque où l'utilisation des données a atteint un niveau sans précédent, l'accès à ce grand nombre de données a précipité leur utilisation intensive afin d’entrainer des modèles d'apprentissage automatique. Les algorithmes correspondants visent essentiellement à détecter et à utiliser des informations structurées au sein d'ensembles de données extrêmement volumineux. 
Plus précisément, après de nombreux rebondissements, les modèles d'apprentissage profond, basés sur des architectures de réseaux de neurones artificiels, ont apporté d'importants progrès numériques dans cette direction et sont désormais omniprésents. Leurs performances dépassant de loin celles des autres modèles existants, ils sont devenus à partir du milieu des années 2000, et en quelques années à peine, indispensables dans de nombreuses applications industrielles telles que la classification d'images, la reconnaissance vocale, l'analyse de texte \cite{LeCun15} ou la prédiction de séries temporelles, la détection d'objets et la reconnaissance faciale, le traitement du langage naturel, le diagnostic médical, la robotique, etc.
Cependant, la compréhension de la plupart des algorithmes, basés sur la descente de gradient d'une fonction de coût, utilisés en pratique pour entraîner des réseaux surdimensionnés et complexes, qui contiennent jusqu'à des millions de paramètres, reste essentiellement empirique et difficile à analyser en théorie. Le principal problème qui se pose avec les architectures de réseaux de neurones profonds réside essentiellement dans la succession de nombreuses couches constituées d’opérations non-linéaires qui rendent l'espace d'optimisation très complexe et de haute dimension. L’analyse et la visualisation de cette vaste collection de paramètres constituent la principale difficulté mathématique dans la plupart des modèles et algorithmes d'apprentissage automatique de pointe. Ce manque de compréhension théorique soulève de nombreuses questions sur leur efficacité et les risques potentiels dans de nombreux domaines d'application. En conséquence, établir des fondements théoriques sur des modèles simples et fournir des prescriptions numériques sur lesquelles fonder et expliquer les observations empiriques sont devenus l'un des défis fondamentaux de la communauté scientifique.\\

Dans ce manuscrit, nous étudions ces questions d'envergure, soulevées par la récente utilisation intensive de l'apprentissage automatique, à travers le prisme de la physique statistique des systèmes désordonnés. Transverse et singulière, cette approche des problèmes d'informatique par la physique a une longue et riche histoire \cite{engel1993statistical, mezard2009information, grassberger2012statistical, zdeborova_statistical_2016,advani2017high}, que nous revisitons dans le régime de haute dimension, en nous concentrant essentiellement sur des considérations algorithmiques modernes confortées par des preuves rigoureuses. 
Spécifiquement, dans le contexte de réseaux de neurones surdimensionnés, pour lesquels le nombre de paramètres explose, les solutions analytiques exactes sont la plupart du temps inconnues et les simulations numériques, quant à elles, très coûteuses. La plupart des techniques issues de la physique statistique ont été précisément conçues pour déduire le comportement macroscopique d'une aussi grande collection de particules à partir de la description microscopique de leurs interactions élémentaires. Ainsi, elles forment un ensemble d'approximations de choix, appelées méthodes à champ moyen, qui sont suffisamment simples pour être calculables et suffisamment riches pour décrire et reproduire les caractéristiques intéressantes du système. De plus, dans cette limite thermodynamique, les physiciens ont constaté que les comportements macroscopiques sont typiquement décrits correctement par seulement quelques paramètres d'ordre.
Appliquées à la théorie de l'apprentissage automatique, qui manque cruellement de telles techniques, nous pensons que les connaissances et techniques de la physique statistique peuvent contribuer à identifier cet ensemble d'observables pertinentes qui contrôlent les propriétés à grande échelle du système et fournissent un cadre puissant pour analyser ces réseaux de neurones artificiels complexes. Malheureusement, même si elles sont très puissantes et supposées conduire à des résultat corrects dans de nombreuses situations, ces techniques ont été utilisées historiquement sans fondement rigoureux. Par conséquent, ce travail fait partie de la dynamique actuelle de la communauté de physique-mathématique à démonter d'anciens résultats obtenus de manière heuristique dans les années 90. De plus, alors que ces analyses statistiques antérieures ne discutaient pas des considérations algorithmiques, nous revisitons cette approche en nous concentrant principalement sur ces potentielles transitions de phase algorithmiques.\\

Au coeur de ce travail, nous capitalisons fortement sur un raisonnement probabiliste Bayésien, qui contraste avec l'approche d'optimisation traditionnelle. De plus, nous utilisons intensément la connexion profonde entre la méthode des répliques et les algorithmes de passage de messages pour obtenir les diagrammes de phase de modèles théoriques simplifiés, qui révèlent néanmoins des caractéristiques intéressantes. En revisitant le paradigme \emph{enseignant-élève}, qui permet de créer des tâches synthétiques et analysables théoriquement, nous concentrons notre attention sur la mise en évidence des écarts potentiels entre les seuils statistiques et algorithmiques.
Nous illustrons l'efficacité de ces méthodes à champ moyen sur divers modèles d'apprentissage automatique qui restent mal compris. Nous nous intéressons essentiellement à des tâches synthétiques avec des données générées dans le paradigme enseignant-élève, et nous contribuons à leur compréhension en décrivant leurs riches diagrammes de phases. Tout d'abord, nous commençons par présenter l'analyse Bayes-optimale dans des machines à comité qui révèle l'existence de grandes lacunes algorithmiques par rapports aux seuils statistiques. Ensuite, dans une analyse pessimiste du pire scénario possible, nous mettons en évidence un lien fort entre la complexité de Rademacher, qui fournit une borne supérieure de l’erreur de généralisation et est liée à la théorie de l'apprentissage statistique, et la capacité de stockage et l’énergie de l'état fondamental abordés dans la littérature de physique statistique. Cela nous permet en particulier de calculer explicitement la complexité de Rademacher dans le cas des perceptrons. Nous complétons enfin le tableau en analysant la minimisation du risque empirique dans le cas des modèles linéaires généralisés, qui est intensivement utilisée en pratique, et nous la comparons aux précédentes analyses Bayes-optimales et du pire scénario. 
Dans une autre direction de recherche, nous définissons une procédure générale pour combiner des modèles élémentaires déjà analysés, afin de construire des architectures plus complexes et structurées. De cette manière, nous développons un cadre qui surmonte en particulier l'hypothèse standard de séparabilité et permet d'analyser des modèles d'estimation, tels que la factorisation matricielle avec un faible rang, la récupération de phase ou la détection compressée, avec des informations à priori fournies par des réseaux génératifs profonds avec des poids aléatoires.

\phantomsection
\section*{Organisation du manuscrit}
\addcontentsline{toc}{section}{Organisation du manuscrit}
\markboth{\spacedlowsmallcaps{Organisation du manuscrit}}{\spacedlowsmallcaps{Organisation du manuscrit}}

Comme mon travail de doctorat se situe au croisement de l'apprentissage automatique et de la physique statistique des systèmes désordonnés, je profite de l'occasion pour présenter pédagogiquement dans la Partie \ref{part:introduction} les concepts théoriques de base, mais essentiels pour suivre le reste du manuscrit. Dans le \Chap\ref{chap:review_ml}, je propose une vue d'ensemble du domaine de l'apprentissage automatique en mettant l'accent sur son histoire tortueuse, ses concepts de base et ses défis actuels. 
Le \Chap\ref{chap:statistical_physics} couvre les outils de base de la physique statistique qui sont pertinents pour comprendre l'approche originale que nous employons pour résoudre les problèmes d'apprentissage automatique. Ces deux premiers chapitres sont consacrés aux lecteurs qui ne connaissent pas l'un des deux domaines et peuvent être donc ignorés par les experts. 
Dans le \Chap\ref{chap:phys_ml_together}, nous proposons une sélection de références historiques importantes pour comprendre comment ces deux domaines sont liés depuis plus de trente ans. C'est aussi l'occasion de passer en revue une sélection des axes de recherche actuels auxquels s'intéresse la communauté de physique statistique des réseaux de neurones artificiels. Enfin, nous introduisons le cadre probabiliste Bayésien et son lien crucial avec la physique statistique. Ceci constitue la pierre angulaire de notre approche qui nous permet d'analyser simultanément les problèmes d'inférence statistique et de satisfaction de contraintes aléatoires. 
Dans le \Chap\ref{main:chap:mean_field}, nous proposons une revue méthodologique de certaines méthodes fondamentales d'inférence à champ moyen, motivées à l'origine dans la littérature des verres de spin \cite{mezard1987spin}, qui sont principalement utilisées dans la seconde partie de la thèse. Plus précisément, nous rappelons en détails la méthode des répliques et les algorithmes de passage de messages que nous présentons et illustrons sur la classe des \emph{modèles linéaires généralisés}, qui sert d'exemple de base tout au long de ce manuscrit. De plus, en mettant en évidence leurs complémentarités, nous tentons de clarifier comment les méthodes sont étroitement liées et permettent de révéler de riches transitions de phase statistiques et algorithmiques.\\

La partie \ref{part:contribution} de ce manuscrit est consacrée à couvrir les travaux auxquels j'ai contribué en tant que doctorant d'Octobre 2017 à Décembre 2020, à l'\emph{Institut de Physique Théorique du CEA-Saclay} sous la direction de Lenka Zdeborová et Florent Krzakala. Le contenu des articles a déjà été publié dans une série d'ouvrages qui peuvent être trouvés en ligne dans leur format original. Ils ont été révisés afin d'uniformiser les notations de ce manuscrit. En particulier, dans un souci de clarté, certaines des longues preuves et calculs auxquels je n'ai pas directement contribués ne sont pas rapportés dans ce manuscrit et peuvent être trouvés dans les publications originales.

\phantomsection
\section*{Contributions}	
\addcontentsline{toc}{section}{Contributions}
\label{contributions}
\markboth{\spacedlowsmallcaps{Contributions}}{\spacedlowsmallcaps{Contributions}}

La partie \ref{part:contribution}, qui rassemble mes principales contributions, est séparée en deux sous-parties correspondant à des axes de recherche menés en parallèle. Dans la partie \ref{part:contribution}.~A, nous discutons de l'analyse complémentaire des scénarios \emph{Bayes-optimal}, du \emph{pire cas}, et de la \emph{minimisation du risque empirique} dans le cadre de réseaux de neurones simples avec des distributions à priori séparables. 
Dans le \Chap\ref{chap:committee_machine}, nous présentons tout d'abord l'approche \emph{Bayes-optimale} sur les machines à comité, qui fournit une analyse des bornes inférieures d’un point de vue de la théorie de l'information. 
Ensuite, dans le \Chap\ref{chap:binary_perceptron} nous décrivons l'analyse du problème de la capacité de stockage et des énergies de l'état fondamental associées, dans le cadre générique des problèmes de satisfaction de contraintes aléatoires. 
Dans le \Chap\ref{chap:rademacher}, nous montrons que ces quantités s'avèrent être étroitement liées à la complexité de Rademacher, connue pour être une borne supérieure de l'erreur de généralisation. 
Dans le \Chap\ref{chap:erm}, nous étudions le cas le plus utilisé en pratique avec l'analyse de la minimisation du risque empirique qui est souvent réalisée grâce à des algorithmes de descente de gradient.

\begin{refsection}
 \begin{enumerate}
	\item \citepublicationfrench{Aubin2018}
	{
	Des outils heuristiques issus de la physique statistique ont été utilisés dans le passé
	pour localiser les transitions de phase et calculer les erreurs d'apprentissage et de 
	généralisation optimales de réseaux de neurones multicouches dans le scénario enseignant-élève. Dans cette contribution nous fournissons, sous une hypothèse technique, une 
	justification rigoureuse de ces approches pour un modèle de réseau de neurones à deux couches, 
	appelé machine à comité.
	Nous introduisons également une version de l'algorithme de passage de messages approximatifs 
	(AMP) pour la machine à comité, qui permet d'effectuer un apprentissage optimal en temps 
	polynomial pour une grande région de paramètres.
	Nous constatons cependant qu'il existe des régimes dans lesquels une faible erreur de généralisation est théoriquement réalisable alors que l'algorithme AMP ne parvient pas à l'atteindre; 
	suggérant fortement qu'aucun algorithme efficace n'existe dans cette région, ce qui met en évidence un grand écart entre seuils statistique et algorithmique.
	}{chap:committee_machine}
	\textbf{Contributions personnelles} : J'ai développé et implémenté l'algorithme AMP et son évolution d'état afin de représenter et analyser les diagrammes de phase correspondants.\\
	\item \citepublicationfrench{Aubin_2019}
	{		
	Nous étudions le problème du calcul de la capacité de stockage du perceptron binaire pour deux
	variantes du problème, dans lesquelles la contrainte correspondante est symétrique. Nous appelons ces variantes le perceptron binaire rectangulaire (RPB) et le perceptron binaire $u$ (UBP). Nous montrons que, contrairement au perceptron binaire habituel avec une fonction marche, la capacité de stockage dans ces alternatives symétriques est donnée par le calcul recuit dans une grande région d'espace de paramètres, \ie pour toutes les contraintes rectangulaires et pour des contraintes de fonction $u$ assez étroites pour $K<K^*$. 
	Nous prouvons ce résultat, sous deux hypothèses naturelles, en utilisant la méthode des premier et second moments. Nous utilisons en outre la méthode du second moment pour conjecturer que les solutions des perceptrons binaires symétriques sont organisées dans une configuration gelée dite 1RSB, et ce sans utiliser la méthode des répliques. Nous utilisons ensuite cette méthode des répliques pour estimer la capacité de stockage dans le cas UBP lorsque la fonction $u-$ est large avec $K> K^* $. Finalement, nous concluons que dans ce cas la rupture totale de la symétrie des répliques devrait être évaluée pour obtenir la capacité exacte.
	}{chap:binary_perceptron}
	\textbf{Contributions personnelles} : J'ai analysé l'entropie des modèles sous différents Ansätze pour étudier la capacité de stockage de Gardner et la géométrie de l'espace de configuration. J'ai également contribué aux preuves en utilisant la méthode des premier et second moments.	\\
	\item \citepublicationfrench{abbara2020rademacher}
	{
	La théorie de l'apprentissage statistique fournit des bornes sur l'erreur de généralisation en utilisant en particulier la dimension de Vapnik-Chervonenkis et la complexité de Rademacher. 
	Une approche alternative, principalement étudiée dans la littérature de physique statistique, est l'étude de la généralisation dans des modèles de données synthétiques simples. Nous discutons donc des liens entre ces approches et nous nous concentrons sur le lien entre la complexité de Rademacher en apprentissage statistique et la théorie de la généralisation pour des modèles synthétiques dans le \emph{cas typique} étudié en physique statistique. Cela implique notamment des quantités connues sous le nom de \emph{capacité de stockage de Gardner} et \emph{de l'énergie de l'état fondamental} du modèle. Nous montrons que dans ces modèles, la complexité de Rademacher est étroitement liée à l'énergie de l'état fondamental calculée par la méthode des répliques. En utilisant cette connexion, on peut dès lors réinterpréter de nombreux résultats de la littérature comme des bornes de Rademacher rigoureuses dans une variété de modèles et dans le régime de haute dimension. De manière assez surprenante, nous montrons également que la théorie de l'apprentissage statistique fournit des prédictions sur le comportement des énergies de l'état fondamental dans certains modèles présentant une rupture totale de la symétrie des répliques.
	}{chap:rademacher}
	\textbf{Contributions personnelles} : J'ai calculé et évalué les énergies de l'état fondamental et établi leur lien avec la complexité de Rademacher.
	\item \citepublicationfrench{aubin2020generalization}
	{  
	Nous considérons une tâche de classification supervisée d'un ensemble de données synthétiques dont les étiquettes sont générées en alimentant un réseau de neurone à une couche avec des entrées \iid aléatoires. Nous étudions les performances de généralisation de 
	classificateurs standards dans le régime de haute dimension dans lequel $ \alpha = {\nsamples} / {\ndim} $ est maintenu fini dans la limite d'une dimension $ \ndim $ et d'un nombre d'échantillons $ \nsamples $ infinis. Notre contribution est triple : Premièrement, nous prouvons une formule donnant l'erreur de généralisation obtenue par des
	classificateurs qui minimisent une fonction de coût convexe avec un terme de régularisation $ \rL_2$. Cette formule a été obtenue initialement et de façon heuristique par la méthode des répliques de la physique statistique. Deuxièmement, en nous concentrant sur des fonctions de coût couramment utilisées et en optimisant l'amplitude de la régularisation $ \rL_2 $, nous
	observons que même si les performances de la régression Ridge sont
	médiocres, en outre les régressions logistique et Hinge sont étonnamment capables d'approcher de très près l'erreur de généralisation Bayes-optimale.
	Dans le régime où $ \alpha \to \infty $, ces régressions conduisent à des taux de généralisation Bayes-optimaux, ce qui, cependant, ne découle pas des prédictions asymptotiques de l'erreur de généralisation basées sur les marges.
	Troisièmement, nous concevons une fonction de coût et un terme de régularisation optimaux qui conduisent de manière asymptotique et rigoureuse à l'erreur de généralisation Bayes-optimale.
	}{chap:erm}
	\textbf{Contributions personnelles} : J'ai réalisé l'analyse théorique et les évaluations numériques. J'ai également contribué aux preuves basées sur le théorème de Gordon.
\suspend{enumerate}

Dans la partie \ref{part:contribution}.~B , nous présentons une ligne de recherche menée en parallèle qui étudie plusieurs types d'informations à priori pour différents problèmes d'estimation, comme le modèle de factorisation de matrice présenté dans le \Chap\ref{chap:generative_spiked} ou la détection compressée et la récupération de phase, détaillées dans le \Chap\ref{chap:generative_phase}. Plus précisément, nous comparons les écarts statistiques et algorithmiques entre d'une part des informations à priori séparables et parcimonieuses, et d'autre part des informations à priori produites par des modèles génératifs profonds et structurés avec des poids aléatoires.

\resume{enumerate}
	\item \citepublicationfrench{Aubin2019c}
	{
	L'utilisation d'une paramétrisation de faible dimension des signaux est un moyen générique
	et puissant pour améliorer les performances de traitement du signal et d'inférence
	statistique. Un type de réduction de dimension très populaire et largement exploré est la 
	parcimonie; une autre méthode plus récente est la modélisation générative de la distribution de signaux. Les modèles génératifs basés sur des réseaux de neurones, tels que les GAN ou les auto-encodeurs variationnels (VAE), sont particulièrement performants et gagnent notamment en applicabilité.
	Dans cette contribution, nous étudions les modèles matriciels à pics, où une matrice de faible rang est observée à travers un canal potentiellement bruité. L'étude de ce problème avec une structure parcimonieuse a attiré une large attention dans la littérature. Ici, nous remplaçons l'hypothèse de parcimonie par un modèle génératif, et nous étudions les conséquences sur les propriétés statistiques et algorithmiques. Nous analysons les
	performances Bayes-optimales sous l'hypothèse spécifique de modèles génératifs pour les pics. En contraste avec l'hypothèse de parcimonie, nous n'observons pas de régions de paramètres où
	les performances statistiques sont supérieures aux performances algorithmiques du meilleur algorithme connu. Nous montrons que dans les cas analysés, l'algorithme de passage de messages est capable d'atteindre ces performances optimales. Nous concevons également
	de nouveaux algorithmes spectraux et analysons leurs performances et leurs
	seuils statistiques en utilisant la théorie des matrices aléatoires, qui a été réalisée par mes collaborateurs. Nous montrons leur supériorité par rapport à l'analyse classique de la composante principale (PCA). Nous complétons nos résultats théoriques avec l'illustration des performances des algorithmes spectraux dans le cas où les pics sont générés par des données réelles.	
}{chap:generative_spiked}
	\textbf{Contributions personnelles} : En étroite collaboration avec B.~Loureiro, j'ai co-développé le cadre théorique pour combiner les entropies des répliques et les algorithmes de passage de messages AMP de sous-modèles. J'ai également co-développé la méthode spectrale LAMP et réalisé les implémentations numériques.
	\item \citepublicationfrench{aubin2020exact}{
	Nous considérons le problème de détection compressée et de récupération de phase (à valeurs réelles) pour une matrice de mesure aléatoire.
	Nous calculons précisément le comportement asymptotique des performances optimales, au sens de la théorie de l'information, et celles du meilleur algorithme polynomial connu, dans le cas d'informations à priori génératives provenant de réseaux de neurones profonds entièrement connectés par des matrices de poids aléatoires et des activations arbitraires.
	Nous comparons ces performances à celles obtenues pour des informations à priori séparables parcimonieuses et nous concluons que dans tous les cas analysés les informations à priori génératives présentent un écart statistique-algorithmique bien plus petit que pour des à priori parcimonieux, ce qui confirme théoriquement les observations expérimentales antérieures selon lesquelles les à priori génératifs pourraient être bien plus avantageux en terme de performances algorithmiques.
	En particulier, alors que la parcimonie ne permet pas d'effectuer efficacement une récupération de phase compressive proche de sa limite théorique, nous constatons qu'en utilisant un à priori génératif aléatoire, la récupération de phase devient possible.
	}
	{chap:generative_phase}
	\textbf{Contributions personnelles} : J'ai réalisé l'analyse numérique et l'évaluation des transitions de phase.
\suspend{enumerate}

Enfin, j'ai contribué à un travail supplémentaire, qui n'est pas présenté dans ce manuscrit, qui introduit une implémentation modulaire en python de l'inférence compositionnelle de modèles graphiques, structurés en arbres. Cependant, dans la lignée des travaux précédents, nous présentons dans la \Sec\ref{chap:generative_phase:applications_real} des applications simples de l'algorithme à des problèmes d'estimation avec des à priori génératifs entraînés sur des données réelles.

\resume{enumerate}
	\item \citepublicationnoreffrench{baker2020tramp}{
	Nous introduisons \textbf{tramp}, pour \emph{TRee Approximate Message Passing}, un code python pour l'inférence compositionnelle dans des modèles structurés en arbre et en grande dimension. Le logiciel unifie et fournit un cadre pour étudier plusieurs algorithmes de passage de messages approximatifs et qui s'appliquent à une variété de tâches d'apprentissage automatique, telles que les modèles linéaires généralisés, l'inférence dans les réseaux multicouches, la factorisation matricielle et la reconstruction à l'aide de pénalités non séparables. Pour certains modèles, la performance asymptotique de l'algorithme peut être théoriquement prédite par l'évolution d'état et un formalisme d'entropies libres. L'implémentation est modulaire par construction: chaque module, qui implémente un facteur du modèle graphique, peut être composé à volonté avec d'autres modules pour résoudre des tâches d'inférence complexes. L'utilisateur n'a qu'à déclarer le modèle graphique: l'algorithme d'inférence, l'évolution d'état et l'estimation de l'entropie sont entièrement automatisés. Le code source est accessible au public à \href{https://github.com/sphinxteam/tramp} {\url{https://github.com/sphinxteam/tramp}} et la documentation est accessible à
\href{https://sphinxteam.github.io/tramp.docs} {\url{https://sphinxteam.github.io/tramp.docs}}.\newline
	\textbf{Contributions personnelles}: J'ai contribué à implémenter certaines parties du code source et développé entièrement la documentation en ligne. J'ai principalement étudié les performances de reconstruction du package pour des à priori génératifs avec des poids entraîné avec différentes architectures GAN et VAE sur des données réelles. 
	}
	\item
\end{enumerate}

\clearpage\null\thispagestyle{empty}
\phantomsection
\chapter*{List of publications}

\setstretch{1}
\setlength\bibitemsep{0.8cm}
\addcontentsline{toc}{chapter}{List of publications}
\label{list_publications}
\markboth{\spacedlowsmallcaps{List of publications}}{\spacedlowsmallcaps{List of publications}}

\printbibliography[heading=none, title={Publications}]
\end{refsection}

\thispagestyle{empty}

\ifthenelse{\equal{\format}{oneside}}
	{\clearpage\null\thispagestyle{empty}}
	{\cleardoublepage}

\pagenumbering{arabic} 

\ctparttext{}
\part{An introduction at the crossroads of machine learning and statistical physics} 
\label{part:introduction}
\ifthenelse{\equal{\format}{oneside}}
	{\clearpage\null\thispagestyle{empty}}
	{\cleardoublepage}
\chapter{A short introduction to machine learning}	
\label{chap:review_ml}
	Current \aclink{ML} techniques vastly rely on \aclink{DNN} and pioneered unprecedented advances in various fields of \aclink{AI}. Despite how recently it gained popularity, \aclink{DL} in fact has a long story starting in the 40's. The field of \aclink{AI} was known under different names along its history depending on the most influential research directions and perspectives. Even though the recent progresses might seem very promising, many theoretical challenges on the theoretical foundations of the current \aclink{DNN}-based methods remain unanswered. 	 	
	In this first chapter, we start by describing a few breakthroughs in \aclink{AI} and \aclink{ML} in \Sec\ref{chap:review_ml:ai_history} to provide some context for the recent developments of the field. 
	In \Sec\ref{chap:review_ml:basics}, we provide a short and comprehensive review of modern machine learning basics, so that the unfamiliar reader may correctly follow the rest of the manuscript. The aim is not to provide a fully thorough description, but a qualitative introduction; the interested reader may find more furnished details in reference books such as \cite{murphy2012machine,Mohri12,shalev2014understanding,Goodfellow2016}.
	Finally in \Sec\ref{chap:review_ml:challenges}, in order to fully grasp the scope and the motivations of this work, we take advantage of the opportunity to review the current challenges and fundamental questions which remain unanswered and that statistical physics may contribute to solve.
			
		\section{A brief historical review of artificial intelligence}		
	\label{chap:review_ml:ai_history}
	
	We present a short selection of some key steps in the development of \aclink{AI}\index{Artificial intelligence}, from the early \aclink{ANN}\index{Artificial neural networks} of the 1950's to the modern \aclink{DNN} used  since 2010's. 
	For a more detailed historical overview please refer to \cite{ganascia1993intelligence, hutchins2001machine, schmidhuber2015deep, lazard2016histoire, Goodfellow2016, sejnowski2018deep, skansi2018introduction}.

		\subsection{The first artificial intelligent machines: 1940-1980}
			\paragraph{Symbolists vs Connectionists}
			\aclink{AI} is a wide field whose goal is to design intelligent programs. Inside this field, two main intellectual currents emerged. On one hand \emph{rule} or \emph{knowledge}-based \emph{symbolists}\index{symbolists} whose pioneers are for instance J. McCarthy, M. L Minsky or J. Von Neumann, and in the other hand \emph{learning-based} \emph{connectionists}\index{connectionists}. 
			While symbolists aim to simulate intelligence through  a succession of predefined rules, connectionists investigate instead the possibility that a computer may learn a solution directly from examples and handle complex edge-cases by itself. In other words, \emph{symbolism} refers to feature \emph{engineering} and 
			\emph{connectionism} to feature \emph{learning}.
			Jumping ahead to modern \aclink{AI}, \emph{machine learning} refers to a connectionist kind of \aclink{AI}.
			Notice that in early stages of \aclink{AI}, symbolism and connectionism were two different approaches that many researchers tried to bring together, see \cite{dreyfus1984intelligence}, while nowadays these two approaches have become quite orthogonal. 
			
			Modern \aclink{AI} started with the emergence of computers and the \emph{Turing test} invented in 1950 by A. Turing \cite{turing2009computing} to determine whether a computer may “think” like a human.
			During this same period, the first \aclink{ANN} was developed, initially designed to model the biological learning of the human brain. 
					
			\paragraph{The beginning of neural networks: 1940-1960}
			A significant advance in \aclink{ANN} came with the work of W. McCulloch and W. Pitts (1943) \cite{mcculloch1943logical} who created the first simplified mathematical model of the human brain, an interconnected circuit of binary units, called formal neurons, and demonstrated that it was equivalent to a universal Turing machine. 
			\graffito{A neural-network is a simple supervised model with learnable parameters $\vec{w}$ in which the output $y$ is a linear/non-linear transformation of an input vector $\vec{x}$: $y = \varphi \( \frac{1}{\sqrt{\ndim}} \vec{w} \cdot \vec{x} \) $.\\ \vspace{0.3cm}
			\begin{tikzpicture}
				\node[circle, draw, very thick, minimum size=12pt, inner sep=0pt, fill=none] (Y) at (1.5,1.5) {$y$};
				\foreach \i in {1,2,...,5}{
					\node[draw, very thick, minimum size=12pt, inner sep=0pt, fill=none] (X\i) at (\i/2,0) {$x$\i};
					\path[edge] (X\i) -- (Y);
					\node at (2.5,0.75) {$\vec{w}$};
					}
			\end{tikzpicture}
			}
			A few years later, D. Hebb (1949) \cite{hebb1962organization} reinforced the concept of neurons and pointed out that neural pathways are strengthened each time they are used, introducing for the first time the concept of \emph{plasticity}. 
			Later on in 1955, A. Samuel invented a computer program able to play checkers, combining connectionist and symbolist approaches with a tree search on weights learned with \emph{temporal-difference} adjusted according to the number of errors. Such early experiments, together with the first machine translation results, lead to the Dartmouth conference in 1956 where important figures of the field such as J.McCarthy, M.Minsky and C.Shannon declared the birth of \aclink{AI} and provided a boost to both \aclink{AI} and \aclink{ANN}.
			Finally, the first battle horse of the connectionist empire was introduced in the 1960's by psychologist F. Rosenblatt (1958) \cite{rosenblatt1958perceptron} : the perceptron, an improved version of the McCulloch and Pitts units. Though very simple, such machines are the basic units of what we call today \emph{deep-learning}. 
			The main innovation was to try to simulate the behaviour of biological neurons. In this perspective Rosenblatt added continuously adjustable valued connections, called today weights $\vec{w}$, to enable plasticity of the unit. The weights are then trained in a supervised manner minimizing the number of mistakes with respect to the desired output $y$ for a particular input pattern $\vec{x}$. More significantly, he gave the first convergence proof of the perceptron algorithm, stating that after training the perceptron would perfectly memorize the training set. Later, B. Widrow and M. Hoff \cite{widrow1960adaptive} developed the first \aclink{ANN} to be applied in a real-world problem : (M)ADALINE for (Multiple) ADAptive LINear Elements, for echo suppression on phone lines. 
				
			\paragraph{The first AI winter: the quiet decade 1965-1976}
				The \emph{quiet decade} refers to W. J Hutchins \cite{hutchins2001machine} formulation about the fact that discoveries and progresses in machine translation stalled. 
				At the end of the 1960's there was no more hope in machine learning translation and research fundings in this direction were deeply cut.
				In 1965 \aclink{AI} was soon compared to \emph{alchemy} \cite{dreyfus1965alchemy} because the early successes held only on very simple tasks and led only to disenchantment in complex tasks.
				
				Thus in the 1970's the future of the connectionist \aclink{AI} turned dark. The godfathers of \aclink{AI} themselves, M. Minksy and S. Papert \cite{minsky2017perceptrons}, showed that perceptrons, which are stuck in the realm of linear models, are limited to very simple tasks and moreover are hard to train. They proposed a harsh critique of perceptrons by proving that they could only be trained to solve linear separable problems and fail to learn non-linearly separable rules such as the XOR function, such that $y=1$ for $\vec{x} \in \{ (0,1), (1,0)\}$ and $y=0$ for $\vec{x} \in \{ (0,0), (1,1)\}$. In addition they stated a number of fundamental problems with the neural network research program and they argued that despite being an interesting subject to study, perceptrons were a sterile direction of research.
				This was the first big hit to connectionism. This lead a few years later to the Lighthill report in 1973 \cite{lighthill1973artificial} which came to the conclusion that the early promises of \aclink{AI}, especially in machine translation, were overstated and fundings were accordingly drastically reduced. 
				Study of neural networks  thus fell into a quick decline in the late 1960's due to Minsky and Papert’s campaign and \aclink{AI} research fundings were turned towards other AI projects such as Bobrow's STUDENT program \cite{bobrow1964natural}, Evan's Analogy program \cite{evans1964heuristic} and the Quillian's semantic memory program Teachable Language Comprehender \cite{quillian1969teachable}. Note that in spite of their harsh criticism, M. Minksy and S. Papert continued contributing to neural network research.
				Yet these events put an end to the first phase of connectionist research, see \cite{hecht1989neurocomputer}.
	
		\subsection{From expert systems to machine learning: 1980-2007}
			During the 1970's, most of \aclink{AI} research focused on the symbolist approach. But \aclink{ANN} oriented research continued with a series of works \cite{kohonen1977principle, grossberg1976adaptive} and a deeply philosophical study by Anderson \cite{Anderson393} on the nature of complex systems\footnote{This work was largely influential and is the cornerstone of modern statistical physics.}. Again unfulfilled claims led to a slowdown in funding in \aclink{AI} and \aclink{ANN} research until early 1980's.
		
			\paragraph{The realm of expert systems}
				In the early 1980's, large conferences instigated a rapid increase in interest from industries and governments, showing a renewed interest and hope in \aclink{AI} and expert systems. In particular, the focus was shifted towards commercial products with applications in financial prediction, geological exploration, medical diagnosis or microelectronic circuit design.	
				Instead of being based on neural networks, this \aclink{AI} era was the climax of the symbolist \aclink{AI} approach, during which it was believed that the best approach to perform \aclink{AI} was top-down with handcrafted knowledge-based systems with huge expertise. 
				But as the hype increased, the field started fearing another winter and a corresponding dry up in funding if \aclink{AI} was to disappoint expectancies.
			
			\paragraph{The second AI winter}
			This fear became true. In the following years, the claims of what \aclink{AI} was capable of had to face reality. \emph{Expert systems} were at the heart of the \aclink{AI} revolution and faced many issues. In particular J. McCarthy strongly criticized them as lacking common sense and knowledge about their limitations. Indeed predictions in medicine based on these systems would have killed many patients and many tasks such as vision or speech recognition were still too complicated for engineers to design handcrafted rules that contain all potential edge cases.
				To conclude, the success of expert systems at that time was very limited and failed to reach the broader goal at which these initial \aclink{AI} successes seemed to lead.
				Therefore mid-1990's, again the activity and publications in \aclink{AI} research largely dropped and conferences did not attract that much anymore, leading naturally to another a decrease of fundings.
			
			\paragraph{Machine learning developing in the shadow}
				Fortunately, research continued in the shadow during the second \aclink{AI} winter and surprisingly significant advances were made. After a decade of interruption, connectionist research was back on stage as a significant driving force.
				In 1982, J.  Hopfield \cite{hopfield1982neural} proposed an analysis of the collective behaviors of physical neural networks.
				In 1986, G. Hinton demonstrated that energy-based neural network could be trained efficiently by \emph{back-propagation}\index{Back-propagation} \cite{rumelhart1986learning}. This simple algorithm is still the dominant approach for training deep learning model nowadays.
				 \graffito{The back-propagation technique is based on a simple chain rule computation: $\partial_{\vec{w}} \varphi(\vec{w} \cdot \vec{x}) = \varphi'(\vec{w} \cdot \vec{x})  \times \partial_{\vec{w}} (\vec{w} \cdot \vec{x}) = \varphi'(\vec{w} \cdot \vec{x})  \cdot \vec{x} $} It also provides interesting distribution representations \cite{mcclelland1986appeal,hinton1986learning}, stating that inputs can be represented by many features. This lead to the emergence of a second wave of neural network oriented research. \aclink{AI} started evolving towards a new approach, the so-called \aclink{ML}, based on feature learning, with the first access to datasets. 
				Connectionist advances held strong with Y. LeCun who successfully trained a convolutional \aclink{ANN} to recognize handwritten zip code digits using back-propagation \cite{lecun1989backpropagation}.
				In 1997, \aclink{LSTM} recurrent neural networks were developed to model long sequences such as text \cite{hochreiter1997long}.
				In 1998, a gradient-based learning method was applied to document recognition \cite{lecun1998gradient}.
				In parallel, kernel methods and \aclink{SVM} \cite{cortes1995support,burges1998tutorial,scholkopf1999input} were developed and quickly displayed impressive performances in mainstream tasks. They rapidly took over the \aclink{ML} community and delayed the \aclink{ANN} climax until 2007.
			 	During the early 2000s, the volume of available data was already strongly increasing as well as the range of data sources and types. This marks the beginning of the onslaught of \emph{big data}, but still \aclink{ANN} are not yet democratized because of practical reasons. 
			 
		\subsection{The realm of deep learning: 2007-today}
			\paragraph{Big data age and GPUs}
				Fifty years after the introduction of the perceptron, \aclink{ANN} finally stroke back with a third wave. They were mostly inactive due to practical issues: the computational power was until then insufficient to train large \aclink{DNN} and there was not enough data available to train them. 
				Early 2009 the open-source ImageNet database \cite{imagenet09} was released and set the cat among the pigeons. The dataset contained over 14 million \emph{labeled images} and solved the first technical issue. It was followed by CIFAR-10 \cite{cifar10}. With this first essential ingredient, the \aclink{ML}, and soon \aclink{DL}, revolution was on its way.
				
				Thus \aclink{ML} drastically changed with the confirmation that \emph{big data} helps and started driving the field. Since then gathering data became easier and easier with social and professional networks, and data became a valuable resource. To give an example, there was 5 exa-bytes ($10^{18}$ bytes) created data per year in 2002 against 10 zeta-bytes ($10^{21}$ bytes) in 2019, a factor of 2000 increased ! 
				However the revolution was mainly made possible thanks to another major technological novelty. Computers started becoming faster and faster at processing data with \aclink{CPU} and in parallel another type of processing units called \aclink{GPU} was developed in the late 1990's. While a \aclink{CPU} contains a few large \emph{cores}, a \aclink{GPU}\footnote{GPU are originally used for computer  graphics, image and video processing and gaming.} contains thousands of cores which are particularly suited to perform small tasks in parallel. Therefore \aclink{GPU} are particularly suited to the training of \aclink{ANN} that contain millions of parameters and require millions of simultaneous operations.
				
			\paragraph{Neural networks gaining in popularity}
				As a result \aclink{ANN} started competing with \aclink{SVM} provided better performances even though they were slower. But very interestingly, the performances of \aclink{ANN} continued improving with the number of training data, so that entering the age of the \emph{big data} made it suitable for the climax of the \aclink{DL} and large neural networks.
				\graffito{\aclink{DL} is the field of \aclink{ML} based on deep and wide \aclink{ANN} architectures.}
				This third \aclink{ANN}-oriented research wave started with the breakthrough of \cite{hinton2006fast} that showed that deep belief networks could be trained efficiently using a greedy layer-wise pertaining strategy. G. Hinton had also the idea to mimic the human brain by increasing the network capacity and therefore increasing the number of layers \cite{bengio2007scaling}. Minsky and Papert \cite{minsky2017perceptrons} already knew that multiple layers would be able to solve the perceptron limitations. But at that time there was no practical algorithm to train such large networks.
				Thus it took 17 years for back-propagation to become popularized \cite{rumelhart1986learning}. And \aclink{GPU} power increased by a factor 1000 over ten years, it allowed to finally train large neural networks.
				
				By 2011, the speed of \aclink{GPU} has increased to enable training of large \\ \aclink{CNN} for vision recognition, and marks the beginning of the \aclink{DL} age. 
				The revolution of \aclink{DL} came from the fact that large neural networks were able to be trained with the use of \aclink{GPU}, whose initial graphical goal had been diverted to linear parallel computing, and the access to large datasets with the emergence of internet.
				With the increasing computational power, \aclink{DNN} \cite{LeCun15} such as AlexNet started rising in international pattern recognition competitions. They outperformed the classical feature engineering approach, and the community started believing that the next revolution would be carried by supervised \aclink{DL}. In particular deep \aclink{CNN} succeeded the ImageNet challenge in 2012 \cite{krizhevsky2012imagenet} and the year after the challenge was strikingly dominated by neural networks methods. 
				The power of \aclink{DL} methods compared to symbolist approaches lies in the fact that connectionist were simply asking a computer to minimize an energy-based model to learn automatically the features that symbolists were trying to design by hand.
				
			\paragraph{Explosion of Deep Learning} 					
			\aclink{DL} became very popular in particular thanks to its wide practical successes, even though the early beginning started in the 1950's. However even though \aclink{ANN} were inspired by biological models, the connection between \aclink{DL} and neuroscience is becoming increasingly narrow. Indeed the lack of understanding of the human brain does not drive \aclink{DL} anymore. And the hope of understanding the human brain from shedding light on the learning processes in \aclink{ANN} is still present but weaker and weaker as \aclink{DL} started becoming a standalone discipline. Therefore \aclink{DL} went beyond its biological inspiration and appeals instead to a more general principle of learning hierarchical representations.
				
				Among \aclink{DL} successes, we may cite machine learning translation also called \aclink{NLP} in which great progresses were been made in recent years. The Google translation engine \cite{wu2016google} based on \aclink{LSTM}, which are a special case of \aclink{RNN}, Sequence to Sequence models and Transformers \cite{vaswani2017attention} out-performed state of the art machine learning translation.
				In computer vision, progresses have been made in many applications such as lip reading \cite{chung2017lip}, visual reasoning \cite{santoro2017simple} or face recognition \cite{taigman2014deepface, parkhi2015deep}.
				In  \aclink{GAN} \cite{goodfellow2014generative}, which allow to generate fake images, it is now possible to synthetize them directly from text sentences \cite{reed2016generative} or even to transform images with Image-to-image generation \cite{isola2017image} or image processing \cite{ulyanov2018deep}.
				Works on adversarial attacks \cite{madry2017towards,zhu2017unpaired} opened a new research direction to build \aclink{ML} models more robust to changes in data distribution.
				In reinforcement learning \cite{sutton2000policy}, after the victory of the DeepBlue computer program against chess champion Kasparov in 1996 \cite{campbell2002deep}, it has been generalized to more complicated games such as Atari \cite{mnih2013playing} and more recently Alpha Go beat the world champion Sedol at Go in 2016 \cite{silver2016mastering}.
				Of course this list is not exhaustive and many fields are currently moving to \aclink{DL} methods used in various applications. Among the most recent, we may cite self-driving cars and healthcare. \\
					
			This concludes the non-exhaustive historical overview of \aclink{ML} and \aclink{DL}. 
			As illustrated, \aclink{DL} applies to various domains with complex network architectures and led to considerable successes in \aclink{AI}. Yet despite their wide range of application, high-performances and popularity, many theoretical questions about the efficiency of \aclink{DL} models and algorithms remain unanswered. 
			To fully grasp these burning challenges, in the next section we propose a technical introduction to the \aclink{ML} basics.

		\section{Machine learning basics}
	\label{chap:review_ml:basics}
		As explained in the last section, the great successes of \aclink{ML} --- whether in the \emph{supervised, unsupervised or reinforcement learning} setting --- rely on \aclink{DL} and \aclink{DNN}.
		This section is devoted to accustom the unfamiliar reader to the essential and basic concepts in \aclink{ML} and \aclink{ANN} so that he/she may apprehend correctly the rest of the manuscript and the connection with the statistical physics approach introduced in \Sec\ref{chap:intro:phys_inference}.
		
		For a more detailed introduction to \aclink{ML}, let us mention a few classical references \cite{Bishop2006,murphy2012machine,shalev2014understanding} and a more recent and comprehensive reference on \aclink{DL} \cite{Goodfellow2016} which can be completed by perspectives from different fields \cite{Carleo19,Mehta19}.
				
		\subsection{The machine learning workflow}	
		\label{sec:review_ml:basics:workflow}
			One of the main reasons why \aclink{ML} flew the nest in the recent years lies in 
			\graffito{\say{Data is the new oil} Clive Humby}	
			the ubiquity of internet, which allows to collect large amount of data that naturally became an essential resource.
			In this context \aclink{ML} refers essentially to a branch of applied statistics that makes use of a large amount of data to estimate complex functions. In other words, a \aclink{ML} "algorithm" is nothing but a computer program able to solve a given task from such a set of data as formulated by \cite{Mitchel1997, Goodfellow2016}:
			\say{\emph{A computer program is said to learn from \textbf{experience} $\mE$ with respect to some \textbf{class of tasks} $\mT$  and \textbf{performance measure} $\mP$, if its performance at tasks in $\mT$, as measured by $\mP$, improves with experience $\mE$.}}
			\begin{figure}[htb!]
				\centering
				\begin{tikzpicture}[scale=0.85, transform shape]
				\tikzstyle{sty} = [draw=red!50!black, fill=gray!50, thick, text=black]
				\tikzstyle{myarrow}=[line width=1mm, draw=burntorange, -triangle 60, postaction={draw, line width=2mm, shorten >=4mm, -}]
				\draw[single arrow,fill=burntorange,draw = red!50!black, very thick] (-8.5,0.1) to (-2.1,0.1) to (-2.1,-0.1) to (-8.5,-0.1) to (-8.25,0) to (-8.5,0.1) to (-2.1,0.1);
				\draw[sty](-7,0) circle[radius=25pt] node (t) {Task $\mT$} ;
				\draw[sty](-4,0) circle[radius=25pt] node (d) {} ;
				\node at (-4,0.15) {\scriptsize{Experience} $\mE$};
				\node at (-4,-0.15) {\scriptsize{Dataset} $\bbD$};
				\fill[even odd rule,red!50!black] circle (2.22) circle (1.98);
				\draw[sty](-1.818,-1.05) circle[radius=25pt] node (a) {\footnotesize{Model} $\mM$} ;
				\draw[sty](1.818,-1.05) circle[radius=25pt] node (m) {\scriptsize{Algorithm} $\mA$} ;
				\draw[sty](0,2.1) circle[radius=25pt] node (d) {\tiny{Performances} $\mP$} ;
				\arcarrow{2.0}{2.1}{2.2}{55}{0}{5}{burntorange,draw = red!50!black, very thick}{};
				\arcarrow{2.0}{2.1}{2.2}{175}{120}{5}{burntorange,draw = red!50!black, very thick}{};
				\arcarrow{2.0}{2.1}{2.2}{295}{240}{5}{burntorange,draw = red!50!black, very thick}{};
			\end{tikzpicture}
			\caption{A typical machine learning workflow considers a given task $\mT$ to be solved from the experience $\mE$ of a dataset $\bbD$. The task is eventually solved by a model $\mM$ trained with an algorithm $\mA$, which accuracy is measured by the performance measure $\mP$.}
			\label{fig:main:introduction:ml_workflow}
			\end{figure}
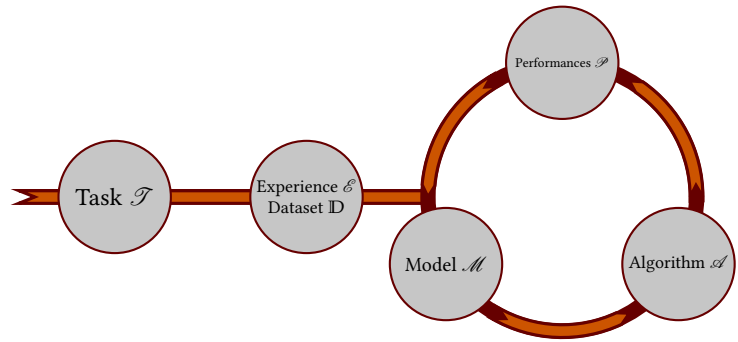
			This formulation can be schematically represented by a \aclink{ML} workflow in \Fig\ref{fig:main:introduction:ml_workflow} and we briefly detail each of its elements in the next sections.
			In more details, a \aclink{ML} program aims to solve a certain task $\mT$, see \Sec\ref{sec:review_ml:machine_learning_task} for a variety of examples, based on observation of a dataset $\bbD$ within a certain experience $\mE$. 
			\aclink{ML} is commonly divided in three main kinds of experiences $\mE$ qualified of \emph{supervised}\index{supervised}, \emph{unsupervised}\index{unsupervised} and \emph{reinforcement learning}\index{reinforcement learning} whose frameworks are briefly presented in \Sec\ref{sec:review_ml:experiences}. 
			To characterize the underlying rule of the task $\mT$, one proposes a probabilistic model $\mM$, very often qualified of \emph{parametric} as assumed to depend on a set of parameters $\btheta$, see \Sec\ref{sec:review_ml:statistical_modeling}. 
			We review the most common kind of models $\mM$ in \Sec\ref{sec:main:introduction:ml:models} and discuss the main difficulties encountered when choosing a model class in \Sec\ref{sec:main:introduction:ml:complexity}-\ref{sec:main:introduction:ml:generalization_bounds}.
			In classical statistics we distinguish two central approaches to estimate the parameters $\btheta$ of the model $\mM$: \emph{frequentist} and \emph{Bayesian} estimators, introduced in \Sec\ref{sec:main:introduction:ml:estimators}.
			Finally the parameters $\btheta$ of this estimator can be computed with an algorithm $\mA$ using the collected dataset $\bbD$. We recall the most common algorithms such as gradient-descent algorithm or sampling methods in \Sec\ref{sec:main:introduction:ml:algorithms}. Finally, the accuracy of the predicted model $\mM$ is measured by a performance measure $\mP$, whose variants are presented in \Sec\ref{sec:review_ml:performances}, and the model is adjusted accordingly.
		
		\subsection{Various machine learning tasks}	
		\label{sec:review_ml:machine_learning_task}
		
			The task $\mT$ denotes the ultimate goal for which the \aclink{ML} algorithm is designed. 
			\graffito{\say{You can have data without information, but you cannot have information without data.} Napoléon Bonaparte}
			For a general task, the \aclink{ML} program aims to recover a hidden underlying structure in a dataset $\bbD$ containing $\nsamples$ \emph{observations}. Each \emph{observation}, also called \emph{example}, represents a collection of \emph{features}, that the program must exploit, either directly if features are meaningful or after processing them to obtain a better features representation, to solve the task $\mT$. 
			Depending on the kind of \emph{task} $\mT$ and \emph{experience} $\mE$, each observation may be either a vector $\vec{x}$, a tensor $\mat{X}$ or an input-output pairs $(y,\vec{x})$. 
			There exists a wide range of specific tasks and we will not present an exhaustive list. Instead, we focus on a series of simple tasks considered later in the contribution part \Part\ref{part:contribution} such as regression, classification and inverse problems, even though current \aclink{ML} enables to handle tasks with increasing difficulty that a human being would not be able to tackle. 
			
			\subsubsection{Regression and classification}
			\label{sec:review_ml:machine_learning_task:reg_class}
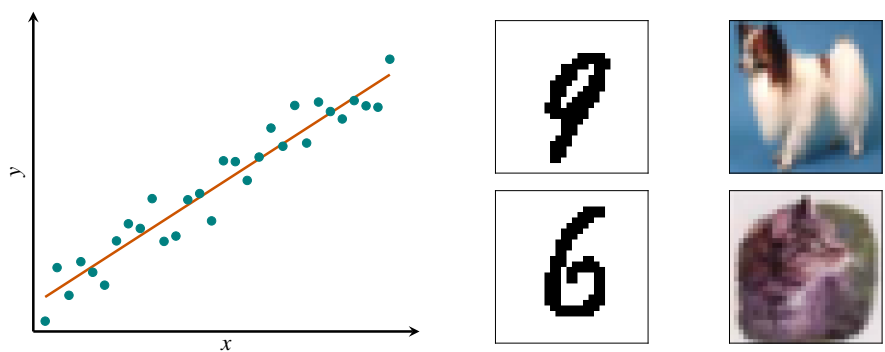
\begin{figure}[htb!]
					\centering
					\begin{minipage}[c]{0.45\linewidth}
						\centering
						\begin{tikzpicture}[scale=0.8, transform shape]
						\pgfplotsset{width=8cm,compat=1.8}
						\pgfmathsetseed{1138}
						\pgfplotstableset{
							create on use/x/.style={create col/expr={42+2*\pgfplotstablerow}},
							create on use/y/.style={create col/expr={(0.6*\thisrow{x}+130)+5*rand}}
						}
					\pgfplotstablenew[columns={x,y}]{30}\loadedtable
					\begin{axis} [
					      xlabel     = $x$, ylabel  = $y$,
					      axis lines = left, axis line style = very thick,
					      clip       = false, xmin = 40,  xmax = 105, ymin = 150, ymax = 200, 					     
					      ticks = none
					    ]
					    \addplot [only marks, teal] table {\loadedtable};
					    \addplot [no markers, very thick, burntorange]
					      table [y={create col/linear regression={y=y}}] {\loadedtable}
					      node [anchor=west] {};
					  \end{axis}
					\end{tikzpicture}
					\end{minipage}	
					\hfill
					\begin{minipage}[c]{0.46\linewidth}
						\centering
						\includegraphics[width=\linewidth]{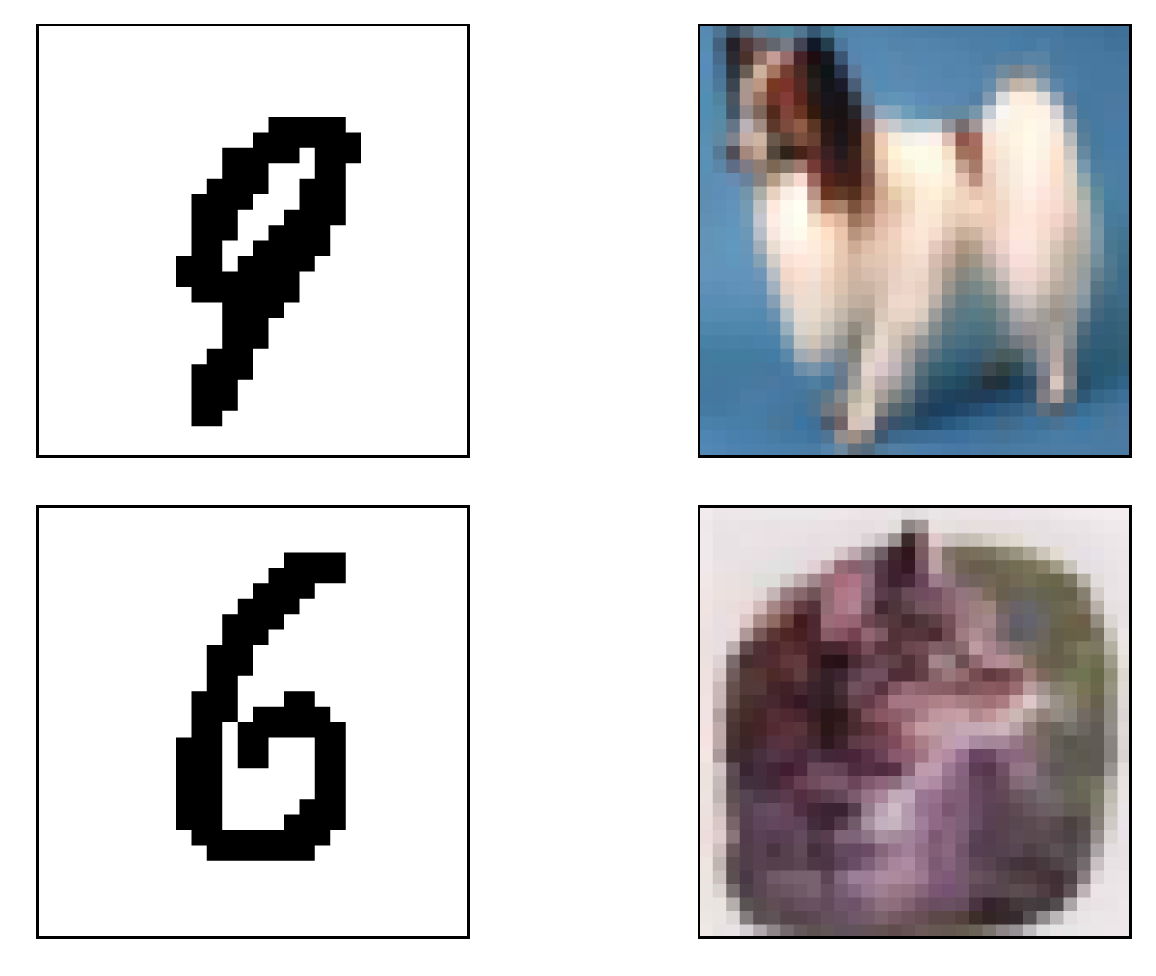}
					\end{minipage}							
					\caption{\Left Illustration of one-dimensional linear regression with $\ndim=1$ and $\nsamples=30$. \Right Images of digits from MNIST and images of a cat and a dog from CIFAR10 to be classified by a machine learning algorithm.}
					\label{fig:main:intro:regression_classification}
					\end{figure}

				The simplest and most common tasks in \aclink{ML} are classification and regression. In these tasks, the goal is to predict a function $f:\bbX \mapsto \bbY$ that maps a given input vector $\vec{x} \in \bbX$ to a numerical value $y \in \bbY$. 	
				The only difference between classification and regression is the output space $\bbY$. In the case of regression, the space $\bbY$ is continuous, while for classification $\bbY$ is discrete and finite so that each output value in $\bbY$ is called a \emph{class}.
				Regression can be applied to various applications such as time series prediction in biology, finance, price prediction, but also to predict the total energy of a molecule. 
				For the sake of the illustration, let us recall the one-dimensional \emph{linear regression} toy example illustrated in \Fig\ref{fig:main:intro:regression_classification} \Left. Observing the input-output pairs $\{x_{\mu}, y_{\mu}\}_{\mu=1}^\nsamples$ (green dots), the simplest \aclink{ML} task consists in finding the best linear fit whether data is linear or not (orange line). While a human is able to easily find a solution to this one-dimensional task, regression becomes harder and harder with increasing problem dimension $\ndim$ while \aclink{ML} algorithms can handle this easily.  
				
				The trendiest example of \emph{classification} is certainly the image recognition task, where one needs to classify pictures of handwritten digits from a dataset such as the MNIST dataset \cite{mnist10}, or classify pictures of cat and dogs from CIFAR-10 \cite{cifar10}, as illustrated in \Fig\ref{fig:main:intro:regression_classification} \Right. Similar tasks consist in recognizing objects from the ImageNet \cite{imagenet09} or Fashion-MNIST \cite{fashionmnist2017} datasets.
				Object recognition is particularly well accomplished with \aclink{CNN} particularly suited to treat images and that allow for instance face recognition, self-driving cars or robots captors, tumor detection, and many other classification tasks.
					
				\subsubsection{Inverse problems}
				\label{sec:review_ml:machine_learning_task:inv_problem}
					In the field of communications and information theory, we are very often interested in a wide class of \emph{inverse problems}
					where one receives a corrupted signal $\vec{y}$ generated from a target signal $\vec{x}$, that we aim to reconstruct, through a noisy channel $\varphi_{\out}$.
					Observing the output of the channel $\vec{y}=\varphi_{\out}(\vec{x})$, the goal of the \aclink{ML} program is to reconstruct the input $\vec{x}$ signal or equivalently to predict the conditional probability $\rP\(\vec{x} \vert \vec{y}\)$. Applications vary according on the form of the noisy channel $\varphi_{\out}$ and the signal dimensions.
					
					\paragraph{Denoising and inpainting}
						The simplest case is when an \emph{additive noise} has been added to the signal $\vec{x}$, equivalent to a channel $\varphi_{\out}(\vec{x})=\vec{x} + \bxi$. The goal of the \emph{denoising} task is therefore to \emph{filter} the noise to reconstruct $\vec{x}$. Note this denoising task may be extended to multiplicative noise.
						
\begin{figure*}[htb!]
							\includegraphics[width=0.47\linewidth]{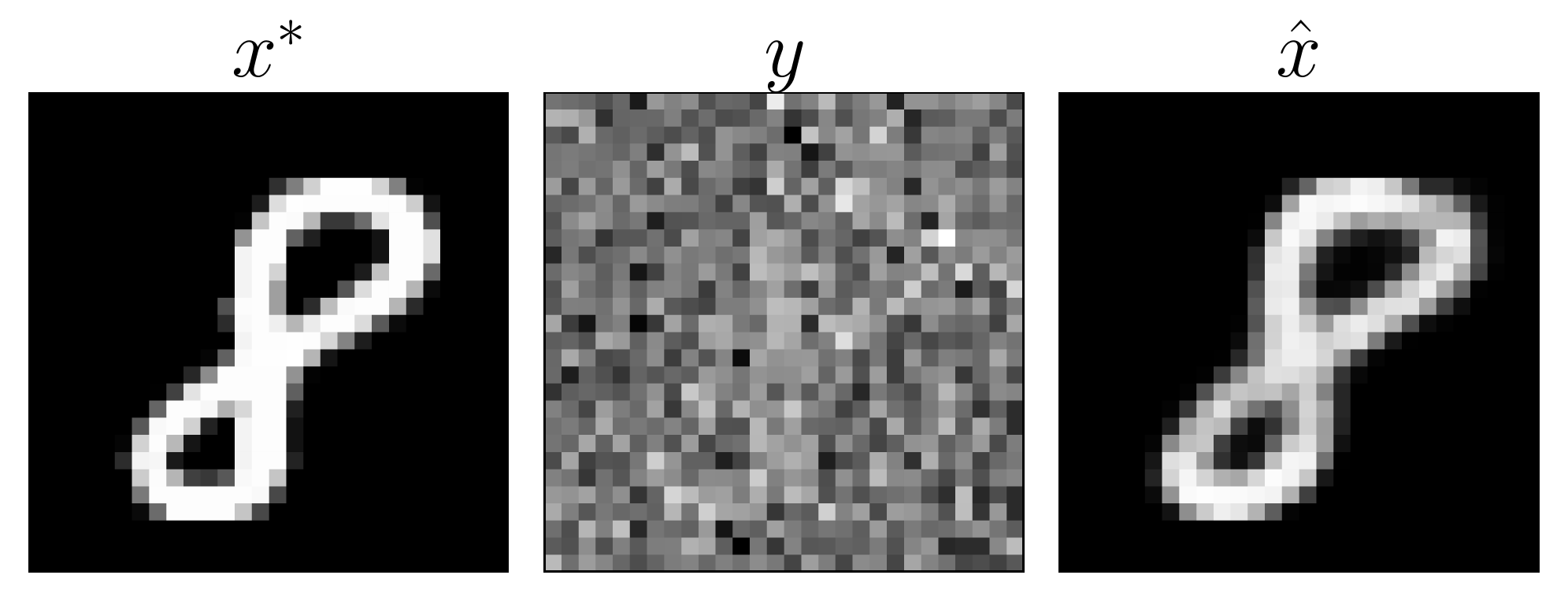}
							\hfill
							\includegraphics[width=0.47\linewidth]{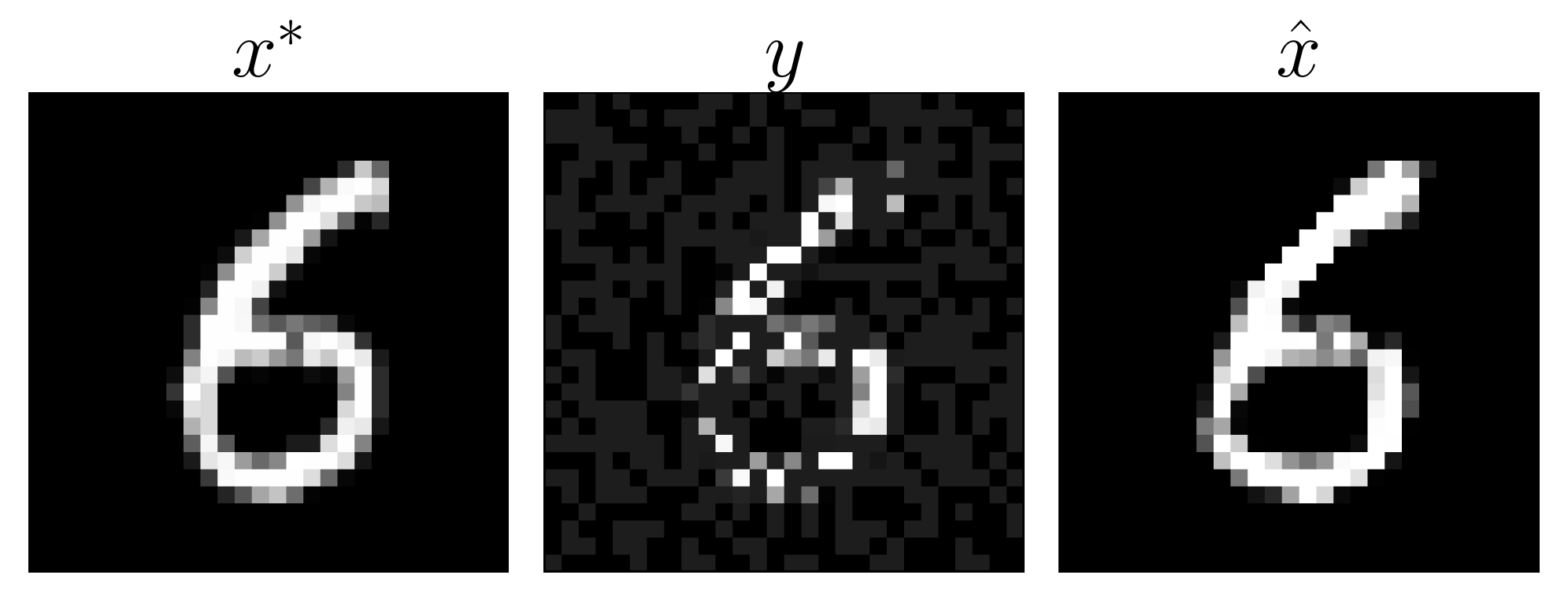}
							\hfill
							\caption{A ground truth image $\vec{x}^\star$ is corrupted and results in an observation $\vec{y}$ for \Left a denoising task and \Right an inpainting task, from \cite{baker2020tramp}. The goal is to reconstruct the ground truth signal $\vec{x}^\star$ from the observation of $\vec{y}$. As an illustration, $\hat{\vec{x}}$ may be the output of a machine learning reconstruction.}
							\label{fig:intro:denoising_inpainting}
						\end{figure*}

						The channel may as well corrupt a few entries of the input vector. The computer program must retrieve the missing entries of the input. For instance, the channel may modify an input image $\vec{x} = (x_1, x_2, \cdots, x_{\ndim-1}, x_\ndim)$ by removing some pixels $x_1,x_{\ndim-1}$ resulting in an observation $\vec{y} = (0, x_2, \cdots, 0, x_\ndim)$. The task to recover the corrupted pixels is known as an \emph{inpainting}. Both tasks are illustrated in \Fig\ref{fig:intro:denoising_inpainting}. 
						
					\paragraph{Compressed sensing and phase retrieval}
						In many applications, the channel involves a multiplication by a known rectangular matrix $\mat{A}$ which applies a linear transformation to the initial signal. This is the case of \emph{compressed sensing} \cite{donoho2006compressed} with $\varphi_{\out}(\vec{x}) = \mat{A} \vec{x}$. We may add and extra difficulty on top of that by adding a non-linearity such as an absolute value $\varphi_{\out}(\vec{x}) = \|\mat{A} \vec{x}\|$. Depending if the matrix and the vector belong to $\bbR$ or $\bbC$, it refers to real or complex \emph{phase retrieval}.
						The phase retrieval task is relevant to many real-life settings in which a detector is only able to capture the amplitude of the signal, for instance in electron microscopy, astronomy, crystallography, optics, etc.
						
					\paragraph{Low-rank matrix factorization}
						Another classical task considered in this work is \emph{low-rank matrix factorization}, used in practice for recommendation systems. The channel is the simple matrix multiplication of rank-$k$ matrices according to $\mat{Y} = \mat{U} \mat{V} + \bxi$, with $\mat{U}\in\bbR^{\nsamples \times k}$ and $\mat{V}\in\bbR^{k \times \ndim}$. Observing the matrix product $\mat{Y}$ that contains a table of users and movies preferences, the aim is to infer separately the latent vectors coding for the users $\mat{U}$ and the movie preferences $\mat{V}$.
						
				\subsubsection{Many others}
					With recent progresses in \aclink{ML}, practical tasks handled in industry are becoming more and more complex than the simple tasks presented above such as the transcription of unstructured representation of some data into discrete textual form such as \emph{optical character recognition} or \emph{speech recognition}. The latter are used by large technological companies to process images, videos or audio recordings, or annotate or describe input data.
					Another useful application is \emph{machine translation} in which the algorithm must translate sentences from a language to another and is referred to \aclink{NLP} \cite{collobert2011natural}. These fields have been the subject of many important advances especially because of the recent use of \aclink{DL} models \cite{sutskever2014sequence, graves2013speech}.
					Let us briefly mention that trying to solve many tasks at the same time is known as \emph{multi-task learning}\index{multi-task learning} \cite{caruana1997multitask}. While learning a given task and trying to apply it to another task, possibly similar enough, refers to \emph{transfer learning}\index{transfer learning} or \emph{domain adaptation} \cite{pan2010survey}. 
			
		\subsection{Supervised, unsupervised and reinforced experiences}	
		\label{sec:review_ml:experiences}
			\aclink{ML} is typically divided in three kinds of paradigms or experiences $\mE$: \emph{supervised}\index{supervised learning}, \emph{unsupervised}\index{unsupervised learning} and \emph{reinforcement}\index{reinforcement learning} learning. Let us present the different frameworks, even though we will focus on the simplest supervised learning case in most of the manuscript. 
			In all these different frameworks, the experience $\mE$ consists in observing a \emph{dataset}\index{dataset} $\bbD$ made of $\nsamples$ \emph{samples}, also called \emph{examples} or \emph{observations}, each being a collection of \emph{features} that the algorithm must process, denoted in full generality by a vector of size $\ndim$, $\vec{x}= \{x_i\}_{i=1}^\ndim$.
			
			\subsubsection{Supervised learning}
			\label{sec:review_ml:experiences:supervised}
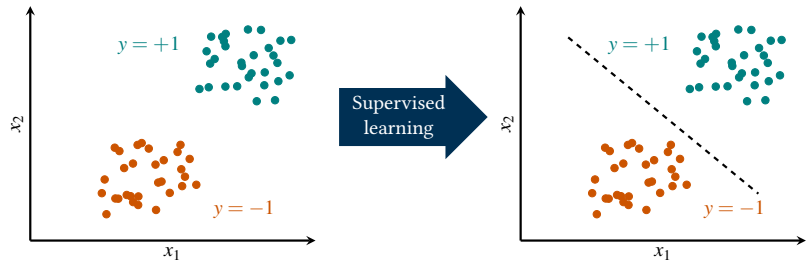
\begin{figure}[htb!]
						\centering
						\begin{tikzpicture}[scale=0.7, transform shape]
							\pgfplotsset{width=7cm,compat=1.8}
							\pgfmathsetseed{1138} 
						\begin{axis} [
						      xlabel     = $x_1$, 
						      ylabel     = $x_2$, 
						      axis lines = left, 
						      axis line style = very thick,
						      clip       = false, 
						      xmin = -3,  xmax = 3, 
						      ymin = -3, ymax = 3, 
						      ticks = none
						    ]
						\pgfplotstableset{ 
    							create on use/x/.style={create col/expr={1.5+rand}},
    							create on use/y/.style={create col/expr={+1.5+rand}}
							}
						\pgfplotstablenew[columns={x,y}]{30}\loadedtable
						\addplot [only marks, teal] table {\loadedtable};
						\pgfplotstableset{ 
    							create on use/x/.style={create col/expr={-.5+rand}},
    							create on use/y/.style={create col/expr={-1.5+rand}}
							}
						\node[text=teal] at (axis cs:-0.5,2) {$y=+1$};
						\node[text=burntorange] at (axis cs:1.5,-2.2) {$y=-1$};
						\pgfplotstablenew[columns={x,y}]{30}\loadedtable
						\addplot [only marks, burntorange] table {\loadedtable};
						\end{axis}
						\node [single arrow, fill=prussianblue, anchor=base, align=center,text width=2cm, text=white] at (7,2.5) {Supervised learning};

						\end{tikzpicture}	
						\begin{tikzpicture}[scale=0.7, transform shape]
							\pgfplotsset{width=7cm,compat=1.8}
							\pgfmathsetseed{1138} 
						\begin{axis} [
						      xlabel     = $x_1$, 
						      ylabel     = $x_2$, 
						      axis lines = left, 
						      axis line style = very thick,
						      clip       = false, 
						      xmin = -3,  xmax = 3, 
						      ymin = -3, ymax = 3, 
						      ticks = none
						    ]
						\pgfplotstableset{ 
    							create on use/x/.style={create col/expr={1.5+rand}},
    							create on use/y/.style={create col/expr={+1.5+rand}}
							}
						\pgfplotstablenew[columns={x,y}]{30}\loadedtable
						\addplot [only marks, teal] table {\loadedtable};
						\pgfplotstableset{ 
    							create on use/x/.style={create col/expr={-.5+rand}},
    							create on use/y/.style={create col/expr={-1.5+rand}}
							}
						\pgfplotstablenew[columns={x,y}]{30}\loadedtable
						\addplot [only marks, burntorange] table {\loadedtable};
						\addplot [ domain=-2:2, samples=3, color=black, dashed, very thick]{-x+0.2};
						\node[text=teal] at (axis cs:-0.5,2) {$y=+1$};
						\node[text=burntorange] at (axis cs:1.5,-2.2) {$y=-1$};
						\end{axis}
						\end{tikzpicture}						
	
						\caption{Illustration of a classification supervised dataset. It contains two clouds of points with different labels $y = \pm 1$ and the algorithm must learn a rule to separate cat images ($y=+1$) from images of dogs ($y=-1$).}
						\label{fig:main:intro:classification_supervised}
						\end{figure}

				The particularity of supervised learning \index{Supervised learning} algorithms lies in the fact that each sample in the dataset is made of a pair of an input features vector $\vec{x}$ and a label or target value $\vec{y}$, so that the dataset $\bbD = \bbX \times \bbY$  contains a collection of input-output pairs $ \{\vec{x}_{\indsamples}, \vec{y}_{\indsamples}\}_{\indsamples=1}^{\nsamples}$ and where each input $\vec{x}_{\indsamples} \in\bbR^{\ndim_x}$ and $\vec{y}_{\indsamples} \in\bbR^{\ndim_y}$. In most of the cases under investigation, we consider scalar outputs, \ie  $\ndim_y=1$ and we use the shorthand $\ndim_x = \ndim$. In this case, as each sample has the same dimension, we may introduce a \emph{design matrix} $\mat{X}\in \bbR^{\nsamples \times \ndim} $ that contains features in columns and different samples in rows.
				We assume that the examples are \aclink{i.i.d} drawn from the joint distribution $\rP(\vec{x}, y)$. Finally, to fix ideas, the input-outputs pairs may represent coordinates $(\vec{x},y)$ in linear regression as illustrated in \Fig\ref{fig:main:intro:regression_classification} or (image of a digit, class of the digit) in an image recognition task.
				Having access to the true labels $\vec{y}$ associated to an input matrix $\mat{X}$, the algorithm must simply estimate a \emph{mapping} function $$f:
						\begin{cases}
							\bbX \mapsto \bbY \\
							\mat{x} \mapsto \vec{y}
						\end{cases}$$ 
				that connects inputs to outputs. Equivalently, this can be understood as estimating the probability distribution $\rP( \vec{y} \vert \mat{x} )$. However, we will see later in \Sec\ref{sec:main:introduction:ml:complexity} that a \emph{good} function $f$ shall not interpolate and memorize every point in the dataset in order to be robust and predict correctly new data-points. 
				More formally, the goal of supervised learning is to predict future outputs $y$ from observations of unseen inputs vector $\vec{x}$, called the \emph{generalization problem}. To fix ideas, in the case of a classification task, we provide examples of images cat and dogs with distinct labels $y =\pm 1$, and the supervised learning algorithm shall separate the feature space according to the labels as illustrated in \Fig\ref{fig:main:intro:classification_supervised}.
				
				This setting is called \emph{supervised} learning in the sense that the labels have been provided by a \emph{teacher} who shows a few examples to an algorithm that aims to understand correctly the underlying rule from them to generalize correctly on unseen cases. 
				Unfortunately this \aclink{ML} setting is very expensive as in a way or another a human intelligence shall assign the labels $y$ to the corresponding input vectors $\vec{x}$. Even though the collection process of data to create \aclink{ML} datasets was incredibly facilitated with the usage of the internet and social networks, yet this reflects the lack of \emph{intelligence} of supervised algorithms. This remains a strong limitation and is the main reason why the community already opened the door to the \emph{unsupervised} learning framework.
				
			\subsubsection{Unsupervised learning}
			\label{sec:review_ml:experiences:unsupervised}				
				In contrast with supervised learning, unsupervised learning \index{Unsupervised learning} involves a collection of a random vectors $\{\vec{x}_{\mu}\}_{\mu=1}^\nsamples$ and consists in learning interesting quantities related to the probability distribution $\rP(\vec{x})$ by observing this dataset. 
				While in supervised learning the algorithm observes both label $y$ and input $\vec{x}$ and estimates the conditional distribution $\rP(\vec{y}|\vec{x})$, in this more involved setting there is no \emph{teacher} to help the algorithm learning a rule: an unsupervised \aclink{ML} algorithm must make sense of the unstructured data and extract structure from data by itself.
				Again for the sake of clarity, this situation is analogous to a baby who still does not understand human language and is able anyway to classify cats and dogs when he/she meets them, even though he/she does not literally know what a dog or a cat means.
					
				As a summary, the specificity of unsupervised learning is that it experiences only features vector without supervision labels: it aims to extract useful informations from a distribution that do not require human labor to annotate examples. The core difficulty is to find a \emph{simple} and \emph{compressed representation} which conserves however as much as information as possible of the distribution $\rP(\vec{x})$. 
				Finding such \emph{dimensionality reduction}\index{Dimensionality reduction} is fundamental in \aclink{ML} as it provides a powerful and meaningful representation to make sense and process the data. It can be reduced to three approaches: attempting to compress the information in \emph{lower-dimensional representation} by \emph{selecting} only a reduced number of the initial features, or embedding the dataset into a higher-dimensional \emph{sparse} representation whose entries contains mostly zeros to \emph{extract} new features from the original ones, or finally finding an \emph{independent representation} to attempt to disentangle underlying features of the data distribution. 
								
				For the sake of conciseness we present the simplest examples: \emph{clustering}, \aclink{PCA} and \emph{density estimation} and we refer the interested reader to \cite{Goodfellow2016} for more details and other applications.
															
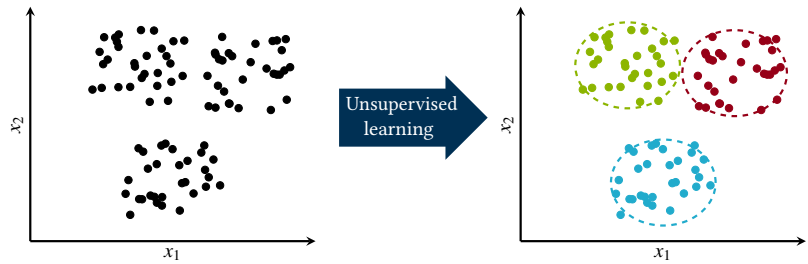
\begin{figure}[htb!]
						\centering
						\begin{tikzpicture}[scale=0.7, transform shape]
							\pgfplotsset{width=7cm,compat=1.8}
							\pgfmathsetseed{1138} 
						\begin{axis} [
						      xlabel     = $x_1$, ylabel     = $x_2$,
						      axis lines = left, axis line style = very thick,
						      clip       = false, 
						      xmin = -3,  xmax = 3, ymin = -3, ymax = 3, 
						      ticks = none
						    ]
						\pgfplotstableset{ 
    							create on use/x/.style={create col/expr={-0.75+rand}},
    							create on use/y/.style={create col/expr={1.5+rand}}
							}
						\pgfplotstablenew[columns={x,y}]{30}\loadedtable
						\addplot [only marks, black] table {\loadedtable};
						\pgfplotstableset{ 
    							create on use/x/.style={create col/expr={0+rand}},
    							create on use/y/.style={create col/expr={-1.5+rand}}
							}
						\pgfplotstablenew[columns={x,y}]{30}\loadedtable
						\addplot [only marks, black] table {\loadedtable};
						\pgfplotstableset{ 
    							create on use/x/.style={create col/expr={1.5+rand}},
    							create on use/y/.style={create col/expr={1.3+rand}}
							}
						\pgfplotstablenew[columns={x,y}]{30}\loadedtable
						\addplot [only marks, black] table {\loadedtable};
						\end{axis}
						\node [single arrow, fill=prussianblue, anchor=base, align=center,text width=2cm, text=white] at (7,2.5) {Unsupervised learning};
						\end{tikzpicture}	
						\begin{tikzpicture}[scale=0.7, transform shape]
							\pgfplotsset{width=7cm,compat=1.8}
							\pgfmathsetseed{1138} 
						\begin{axis} [
						      xlabel     = $x_1$, ylabel     = $x_2$,
						      axis lines = left, axis line style = very thick,
						      clip       = false, 
						      xmin = -3,  xmax = 3, ymin = -3, ymax = 3, 
						      ticks = none
						    ]
						\pgfplotstableset{ 
    							create on use/x/.style={create col/expr={-0.75+rand}},
    							create on use/y/.style={create col/expr={1.5+rand}}
							}
						\pgfplotstablenew[columns={x,y}]{30}\loadedtable
						\addplot [only marks, applegreen] table {\loadedtable};
						\pgfplotstableset{ 
    							create on use/x/.style={create col/expr={0+rand}},
    							create on use/y/.style={create col/expr={-1.5+rand}}
							}
						\pgfplotstablenew[columns={x,y}]{30}\loadedtable
						\addplot [only marks, ballblue] table {\loadedtable};
						\pgfplotstableset{ 
    							create on use/x/.style={create col/expr={1.5+rand}},
    							create on use/y/.style={create col/expr={1.3+rand}}
							}
						\pgfplotstablenew[columns={x,y}]{30}\loadedtable
						\addplot [only marks, carmine] table {\loadedtable};
						\draw[draw=applegreen, radius=1.1, very thick, dashed] (axis cs:-0.75,1.5) circle;
						\draw[draw=carmine, radius=1.1, very thick, dashed] (axis cs:1.5,1.3) circle;
						\draw[draw=ballblue, radius=1.1, very thick, dashed] (axis cs:0,-1.5) circle;
						\end{axis}
						\end{tikzpicture}						
	
						\caption{Illustration of an unsupervised clustering task: the algorithm observes a large cloud of points without labels. The $k$-means algorithm should decide by itself that this large cloud is made of three distinct clusters and assign them different classes.}
						\label{fig:main:intro:clustering_unsupervised}
						\end{figure}

				\paragraph{Clustering}
					The \emph{clustering} approach consists in learning the structure of the dataset by trying to separate the dataset in meaningful unlabelled subgroups whose features are close to each other. This method is in particular used for medical imaging, image segmentation, social network analysis, search result grouping, etc.
					In the absence of labels, the main difficulty is to find a simple representation of the data to appreciate its structure. After being processed, the dataset is split in different \emph{clusters} corresponding to classes defined by the algorithm itself. The battle horse to perform clustering, illustrated in \Fig\ref{fig:main:intro:clustering_unsupervised}, is the $k$-means algorithm that divides the dataset into $k$-clusters, where $k$ is an hyper-parameter that shall be tuned carefully.
					However, the clustering task is inherently ill-posed as there is no single criterion to obtain a good clustering. As a consequence, separating the dataset may be done in several distinct ways and leads to different clusterings. See \cite{kaufman2009finding} for more details on clustering techniques.
					
\begin{figure}[htb!]
				\centering
				\begin{tikzpicture}[scale=0.6]
				\begin{axis} [xlabel = $x_1$, ylabel = $x_2$, axis lines = left, 
						      axis line style = very thick, clip = false, 
						      xmin = -3,  xmax = 3,	ymin = -3, ymax = 3, ticks = none]
				\pgfplotstableread{Plots/data_pca.txt}\bbYdata;
				\addplot [ color=black,only marks, mark=*, mark size=1.5pt, ] table [x expr=\thisrowno{0}, y expr=\thisrowno{1} ] {\bbYdata};
				\end{axis}
				\draw[-latex,burntorange, line width=0.5mm] (0+3, 0+2.5) -- ( -3*0.56194961+3,0.82717147*3+2.5);
				\draw[-latex,teal, line width=0.5mm] (0+3, 0+2.5) -- ( 0.8271714*6+3, 0.56194961*6+2.5);
				\end{tikzpicture}
				\caption{Illustration of Principal Component Analysis of a cloud of random Gaussian matrix $\mat{X}\in\bbR^{\nsamples \times \ndim}$  for $\ndim=2$, $\nsamples = 500$. The orange and green vectors represent the principal components of the observed dataset.}
				\label{fig:main:intro:pca}
				\end{figure}
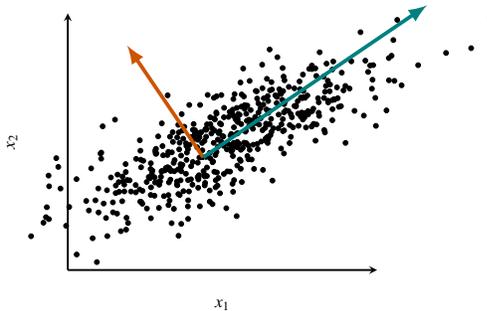

				\paragraph{Dimensionality reduction and PCA}				
				In order to compress data in a meaningful way, we would like to find a \emph{basis} in which the data can be represented in lower dimensionality than the original input, with statistically independent components. This kind of \emph{dimensionality reduction}\index{dimensionality reduction} can be performed for instance with the so-called \aclink{PCA}\index{PCA} method. In the manner of the eigenvalues decomposition of a symmetric positive matrix, \aclink{PCA} is a generalization to any rectangular matrix $\mat{X}\in \bbR^{\nsamples \times \ndim}$. 
				The idea of \aclink{PCA} is to identify patterns in data by linear transformations such as rotating and projecting the matrix in a lower-dimensional subspace whose basis has orthogonal directions, called the \emph{principal components}. Therefore, it builds new independent features that are linear combinations of the initial features.
				In other words it finds the directions of maximum variance in high-dimensional data and projects it in a lower-dimensional sub-space to keep the maximum of essential data in a smaller space. 
				First, the data matrix may be centered by removing its potential mean $\mat{X} \leftarrow \mat{X}- \EE\[\mat{X}\]$, and the principal components are computed as the eigenvectors of the symmetric covariance matrix $\mat{X}^\intercal \mat{X}$. Indeed, the \aclink{SVD} of the data matrix yields $\mat{X} = \mat{U} \bSigma \mat{V}$, with rotationally invariant matrices $\mat{U} \in \bbR^{\nsamples \times k}, \mat{V} \in \bbR^{k \times \ndim} $,  $\mat{U}^\intercal\mat{U}=\rI = \mat{V}\mat{V}^\intercal$. The diagonal matrix $\bSigma$ contains $k$ singular values $\{\bSigma_{i}\}_{i=1}^k$, such that the covariance matrix decomposes as $\mat{X}^\intercal \mat{X} = \mat{V} \bSigma^2 \mat{V}^\intercal$.
				Rotating the data $\mat{X}$ with the rotation matrix $\mat{V}$, the covariance matrix becomes diagonal, so that in this basis the components are mutually uncorrelated as illustrated in \Fig\ref{fig:main:intro:pca}. More details on PCA may be found in \cite{jolliffe1986principal, Goodfellow2016}.
								
				\paragraph{Density estimation and generative modeling}
				Most unsupervised \aclink{ML} adopts a probabilistic approach known as \emph{density estimation}. It consists in approaching the true probability density $\rp(\mat{x})$, from which the dataset $\mat{X}=\{\vec{x}_\mu\}_{\mu=1}^\nsamples$ has been drawn, by an approximate density $\hat{\rp}$. This density may be parametrized according to its hypothesis class as discussed in \Sec\ref{sec:review_ml:statistical_modeling}. We consider therefore a set of parametric densities $\bbK_\btheta = \left\{ \rp_\btheta \(\mat{X}\), \btheta \in \bbR^{\ndim_\theta} \right\}$ such that the set of $\ndim_\theta$ estimated parameters $\hat{\btheta}$ are learned. The corresponding approximate density $\rp_{\hat{\btheta}}$ captures the ground truth distribution and can be used therefore to generate new samples of the distribution, hence the terminology \emph{generative modeling}.
				The training of such density estimation method is performed by maximizing the log-likelihood of the observed dataset $\bbD$, or equivalently the Kullback-Leibler divergence from the approximate distribution $\rP_\btheta$ to the empirical distribution $\rP_\bbD(\vec{x}) = \frac{1}{\nsamples} \sum_{\mu=1}^\nsamples \delta \( \vec{x} - \vec{x}_{\mu}\)$
				\begin{align}
					\hat{\btheta} = \max_{\btheta} \sum_{\mu=1}^\nsamples \log \rP_\btheta \( \vec{x}_{\mu}\)  \Leftrightarrow \hat{\btheta} = \min_{\btheta} \KL\(\rP_\bbD \vert \rP_\btheta \) \,. 
					\label{eq:introduction:ml:max_likelihood}
				\end{align}
				However, expressing and computing in practice the log-likelihood in high-dimensions is very complex and often intractable. Sampling the density in high-dimensions, with for instance a \aclink{MC} method, is very costly and becomes slower and slower with the problem dimension. 
				To circumvent these high-dimensional difficulties, new \aclink{DNN}-based models called \emph{deep generative models} have been recently introduced.
					
				\paragraph{Deep generative models}
				Instead of maximizing the above likelihood \eqref{eq:introduction:ml:max_likelihood}, alternative strategies based on \aclink{DNN} came to light in the recent years such as \aclink{GAN} and \aclink{VAE}, that became very popular thanks to the amazing improvements they brought to the \emph{density estimation} field. Indeed relying on a large amount of data and \aclink{DNN}, they have shown an incredible expressivity and ability to \emph{approximate complex densities} to produce highly realistic images, texts or even sounds. 
				These techniques may be used either to generate new contents or to use as a complex and structured \emph{prior-knowledge} to solve inverse problems, see \Sec\ref{sec:review_ml:machine_learning_task:inv_problem}. 
				The core idea of both \aclink{VAE} or \aclink{GAN} relies on an architecture made of an \emph{encoder} that compresses the data in a low-dimensional representation, and of a  \emph{decoder} that tries to decompresses it. 
				Such systems are trained to minimize the difference between the encoded and decoded signals in an unsupervised manner. This situation is typically referred to an \emph{information bottleneck} \cite{tishby2000information} because the encoder must learn an efficient compression of the data into this lower-dimensional space.
				
				Both \aclink{VAE} and \aclink{GAN} make use of \aclink{DNN} to parametrize the \emph{encoder} and the \emph{decoder}, called the \emph{discriminator}\index{discriminator} and the \emph{generator}\index{generator} in \aclink{GAN} language. Indeed \emph{high-capacity} \aclink{DNN} are of considerable interest in this task in the sense their  wide expressivity allows to approximate any complex density. Also their architecture modularity allows to easily reduce the dimension of the data in a low-dimension latent space. And finally, they have the strong advantage they can be trained and optimized very efficiently using back-propagation algorithm, as specifically presented in \Sec\ref{sec:main:introduction:ml:algorithms:gd}.		
						
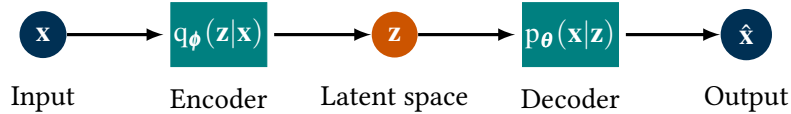
\begin{figure}[htb!]
						\centering
							\begin{tikzpicture}[scale=0.85]
								\node[circle,fill=prussianblue, text=white] (I) at (0,0) {$\vec{x}$};
								\node at (0,-1) {Input};
								\draw[fill=teal,draw=teal, text=white]  (2,-0.5) rectangle (3.5,0.5) node[pos=.5] (E) {$\rq_\bphi (\vec{z} | \vec{x})$};
								\node at (2.75,-1) {Encoder};
								\node[circle,fill=burntorange, text=white] (L) at (5.5,0) {$\vec{z}$};
								\node at (5.5,-1) {Latent space};
								\draw[fill=teal,draw=teal, text=white]  (7.5,-0.5) rectangle (9,0.5) node[pos=.5] (D) {$\rp_\btheta(\vec{x}| \vec{z})$};
								\node at (8.25,-1) {Decoder};
								\node[circle,fill=prussianblue, text=white] (O) at (11,0) {$\hat{\vec{x}}$};
								\node at (11,-1) {Output};
								\draw[-latex,very thick] (I) -- (E);
								\draw[-latex,very thick] (E) -- (L);
								\draw[-latex,very thick] (L) -- (D);
								\draw[-latex,very thick] (D) -- (O);
							\end{tikzpicture}	
							\caption{Illustration of a VAE: An encoder $\rq_\bphi$ maps the input data into a latent space. The decoder $\rp_\btheta$ tries to decode the latent distribution by maximizing the likelihood between the decoded representation $\hat{\vec{x}}$ and the original input $\vec{x}$.}
							\label{fig:main:intro:VAE}
						\end{figure}

					\subparagraph{Variational Auto-Encoders\newline}
					\aclink{VAE}\index{VAE} have been introduced in \cite{kingma2013auto,rezende2014stochastic} and are a \emph{regularized} version of the classical Auto-Encoders \cite{vincent2010stacked}. These generative models are nowadays commonly used to approximate a probability distribution $\rP(\vec{x})$ from a dataset, in the perspective to generate new samples from it. 
					The distribution can be reformulated as the marginalization over some latent variables $\vec{z}$ as follows
					\begin{align}
						\rP(\vec{x})= \int \rp(\vec{x}|\vec{z}) \rp(\vec{z}) \d \vec{z}\,.
						\label{eq:introduction:ml:vae_density}
					\end{align}
					The idea is thus to infer the latent distribution $\rp(\vec{z})$ using the conditional density $\rp(\vec{z}|\vec{x})$, which is however also unknown. Therefore, to use it we should instead approximate this density by using a variational principle.
					Even though we would have access to an approximation $\hat{\rp}(\vec{z}|\vec{x})$, we still need to perform the multidimensional integral \eqref{eq:introduction:ml:vae_density} that is often intractable analytically and hard to evaluate numerically.
					To make this problem tractable, \aclink{VAE} are essentially made up of an \emph{encoder} $\rq_\bphi (\vec{z} | \vec{x})$ parametrized by some parameters $\bphi$ that compresses the input data $\vec{x}$ in a latent representation $\vec{z}$. Yet its particularity lies in the fact that the encoder is regularized during the training in order to ensure that the latent space has \emph{good properties}, that allows to generate appropriate new samples. 
					As illustrated in \Fig\ref{fig:main:intro:VAE}, the encoder is followed by a \emph{decoder} $\rp_\btheta(\vec{x}| \vec{z})$ parametrized by some parameters $\btheta$ that tries to maximize the likelihood with the input data.
			
						The \aclink{VAE} objective can be simply derived as a variational approximation $\rq_\bphi(\vec{z}|\vec{x})$ of the intractable posterior distribution $\rp(\vec{z}|\vec{x})$, see \Sec\ref{main:sec:mean_fields} for more details on variational approximations. The \aclink{KL} divergence defined in \Sec\ref{definition:kullback_leibler_divergence} yields
						\begin{align*}
							&\KL\( \rq_\bphi(\vec{z} | \vec{x})  ~\|~ \rp(\vec{z}|\vec{x})  \)= \EE_{\vec{z} \sim \rq_\bphi(\vec{z}|\vec{x}) } \[ \log  \rq_\bphi(\vec{z} | \vec{x})  - \log  \rp(\vec{z} | \vec{x})  \]\\
							&=\EE_{\vec{z} \sim \rq_\bphi(\vec{z}|\vec{x}) } \[ \log  \rq_\bphi(\vec{z} | \vec{x})  - \log  \rp_\btheta(\vec{x} | \vec{z}) - \log \rp(\vec{z})\] + \log \rp (\vec{x}) \spacecase
							&\Rightarrow \log \rp (\vec{x})  - \KL\( \rq_\bphi(\vec{z} | \vec{x})   ~\|~\rp_\btheta(\vec{z}|\vec{x})  \)  \\
							& \qquad \qquad = \EE_{\vec{z} \sim \rq_\bphi(\vec{z}|\vec{x}) } \[  \log \rp_\btheta(\vec{x} | \vec{z}) - \KL\( \rq_\bphi(\vec{z} | \vec{x})  ~\|~ \rp(\vec{z})\)  \]\,,
						\end{align*}
						 so that the \aclink{VAE} objective $\mL(\bphi, \btheta ; \vec{x})$ is given by maximizing the variational likelihood \emph{lower bound} 
						\begin{align*}
							\mL(\bphi, \btheta ; \vec{x}) &= \EE_{\vec{z} \sim \rq_\bphi(\vec{z}|\vec{x}) } \[  \log \rp_\btheta(\vec{x} | \vec{z}) - \KL\( \rq_\bphi(\vec{z}|\vec{x})   ~\|~ \rp(\vec{z})\)  \]\\
							&\leq \log \rp (\vec{x}) \,.
						\end{align*} 
					 The first term $\log \rp_\btheta(\vec{x} | \vec{z})$ represents the \emph{reconstruction} process of the decoder that should minimize the difference between the decoded signal $\hat{\vec{x}}$ and the initial input data $\vec{x}$ density, or equivalently maximize the likelihood $\EE_{\vec{z} \sim \rq_\bphi(\vec{z}|\vec{x}) } \log  \rp_\btheta(\vec{x} | \vec{z}) $. The second term $\KL\( \rq_\bphi(\vec{z}|\vec{x}) \vert \rp(\vec{z})\)$ should be minimized so that the encoder density closely approaches the latent distribution $\rp(\vec{z})$. 
					In practice this latent distribution is fixed and very often chosen to be Gaussian normal $\mN_{\vec{z}}(\mu(\vec{x}),\bSigma(\vec{x}))$ so that the encoder is trained to return only the two first moments of the Gaussian parametrization. The latent variable is therefore \emph{sampled} and this key step is called the \emph{reparameterization trick}.
					This trick looks like a \emph{regularization} procedure of the latent space, so that \aclink{VAE} can be simply thought as regularized and probabilistic versions of classical Auto-Encoders. 
					Moreover, it makes the computation of the \aclink{KL} divergence explicitly tractable and the optimization possible with for instance classical gradient-descent algorithms. More details may be found in \cite{doersch2016tutorial,kingma2019introduction}.
					Notice that the decoder $\rp_\btheta$ is very often taken as a \aclink{DNN} which is very expressive but also costly to train. Once trained in this \emph{variational} and \emph{unsupervised} fashion, splitting  the encoder from the decoder, the later $\rp_\btheta(\vec{x}| \vec{z})$ allows to generate new samples $\vec{x} \sim \rp_\btheta(\vec{x}| \vec{z})$ of impressive realism, starting from a simple Gaussian noise.
						
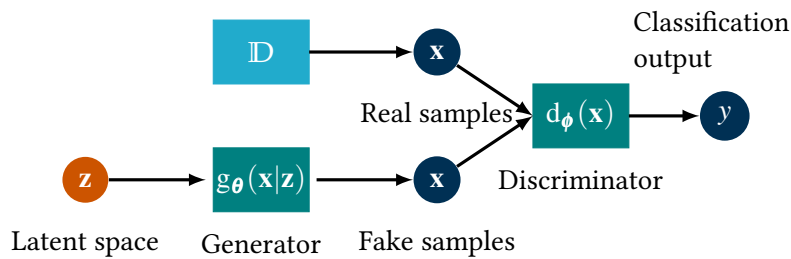
\begin{figure}[htb!]
						\centering
							\begin{tikzpicture}[scale=0.85]
								\node[circle,fill=burntorange, text=white] (I) at (0,0) {$\vec{z}$};
								\node at (0,-1) {Latent space};
								\draw[fill=teal,draw=teal, text=white]  (2,-0.5) rectangle (3.5,0.5) node[pos=.5] (E) {$\rg_\btheta (\vec{x} | \vec{z})$};
								\node at (2.75,-1) {Generator};
								\node[circle,fill=prussianblue, text=white] (FX) at (5.5,0) {$\vec{x}$};
								\node at (5.5,-1) {Fake samples};
								
								\draw[fill=ballblue,draw=ballblue, text=white]  (2,1.5) rectangle (3.5,2.5) node[pos=.5] (R) {$\bbD$};
								\node at (2.75,1) {};
								\node[circle,fill=prussianblue, text=white] (X) at (5.5,2) {$\vec{x}$};
								\node at (5.5,1) {Real samples};
								
								\draw[fill=teal,draw=teal, text=white]  (7,0.5) rectangle (8.5,1.5) node[pos=.5] (D) {$\rd_\bphi(\vec{x})$};
								\node at (7.75,0) {Discriminator};
								\node[circle,fill=prussianblue, text=white] (O) at (10,1) {$y$};
								\node[text width=2cm, above] at (9.75,1.5) {Classification output};
								\draw[-latex,very thick] (I) -- (E);
								\draw[-latex,very thick] (E) -- (FX);
								\draw[-latex,very thick] (3.5,2) -- (X);
								\draw[-latex,very thick] (X) -- (7,1);
								\draw[-latex,very thick] (FX) -- (7,1);
								\draw[-latex,very thick] (8.5,1) -- (O);
							\end{tikzpicture}	
							\caption{Illustration of a GAN: A discriminator $\rd_\bphi$ tries to classify real and fake samples generated from a generator $\rg_\btheta$ that tries to fool the discriminator.}
							\label{fig:main:intro:GAN}
						\end{figure}

					\subparagraph{Generative Adversarial Networks\newline}
					Another kind of common deep generative models are \aclink{GAN} that have enjoyed tremendous success since their introduction \cite{goodfellow2014generative}.
					The idea of \aclink{GAN}\index{GAN} is similar to \aclink{VAE} in the sense it exploits a random latent representation. Its conceptual idea is enlightening by its simplicity and allowed to take a leap forward for generative modeling and density estimation. Again the idea is to compare an approximation of the dataset distribution with the true distribution, which is unknown. 
					The brilliant idea of \aclink{GAN} consists in replacing this direct comparison by two \emph{indirect} ones called \emph{generation} and \emph{discrimination}. The \aclink{GAN} architecture, represented in \Fig\ref{fig:main:intro:GAN}, is therefore made up of a parametric \emph{discriminator} $\rd_\bphi$ that takes samples of some true and fake generated data and tries to classify them as well as possible. On the other hand, a \emph{generator} $\rg_\btheta$ is trained to generate fake samples to fool the discriminator.
						 Therefore, generator and discriminator have \emph{adversarial} missions. The goal of the generator $\rg_\btheta$ is to fool the discriminator $\rd_\bphi$, so the generator computes the probability of samples of belonging to the real dataset $\bbD$ rather than being fake. It is trained to maximize the classification error between real and fake samples. In contrast, the goal of the discriminator is to detect fake generated data, so that it is trained to minimize the final classification error. Therefore, during the training process, the generator promotes the increase of the classification error whereas the discriminator tries to decrease it. This competition can be thought as a mini-max problem and is translated by the \aclink{GAN} adversarial objective
						 \begin{align*}
						 	\mL(\vec{x}) = \min_\bphi \max_\btheta \EE \log \rd_\bphi(\vec{x}) + \log \( 1 -  \rd_\bphi(\rg_\btheta(\vec{z})) \)\,.
						 \end{align*}
						 In practice and as already stressed for \aclink{VAE}, both generator $\rg_\btheta $ and discriminator $d_\bphi$ are commonly chosen as \aclink{DNN} for their wide expressivity and also because they can be easily jointly trained. 
						 \aclink{GAN} are currently used for a variety of tasks such as high quality image or video generation, even though these techniques are not flawless as they can suffer from \emph{mode collapse} issues and raises questions about their ability to really learn the target distribution \cite{arora2018gans}.
						 
			\subsubsection{Reinforcement learning}
			\label{sec:review_ml:experiences:reinforcement}
				\aclink{RL}\index{Reinforcement learning} is the last and more recent class of \aclink{ML} experiences $\mE$. The main specificity of \aclink{RL} is that it interacts with an environment so that there is a feedback loop between the learning system and its actions. 
				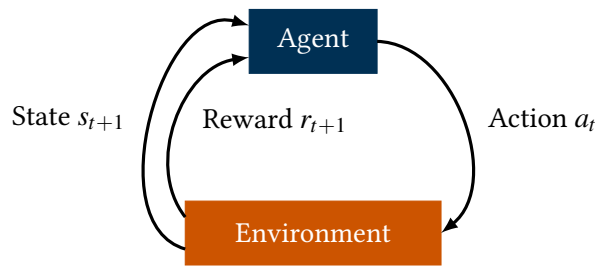
\begin{figure}[htb!]
				\centering
					\begin{tikzpicture}[scale=0.85]
						\draw[fill=prussianblue,draw=prussianblue, text=white] (0,3) rectangle (2,4) node[pos=.5] {Agent};
						\draw[fill=burntorange,draw=burntorange, text=white] (-1,0) rectangle (3,1) node[pos=.5] {Environment};
						\path[-latex, black, very thick] (2,3.5) edge [bend left=70]  node[right=0.2cm] {Action $a_t$} (3,0.5);
						\path[-latex, black, very thick] (-1,0.25) edge [bend left=90]  node[left=0.2cm] {State $s_{t+1}$} (0,3.75);
						\path[-latex, black, very thick] (-1,0.75) edge [bend left=60]  node[right=0.2cm] {Reward $r_{t+1}$} (0,3.25);
					\end{tikzpicture}	
					\caption{Illustration of reinforcement learning.}
					\label{fig:main:intro:reinforcement_learning}
				\end{figure}
				Qualitatively an agent interacts with the environment so that it dynamically learns and decides what actions to take. In more details, the agent takes some actions $a_t$ at time $t$ that lead to a new state of the agent in the environment $s_{t+1}$ and a corresponding reward $r_{t+1}$ whose value depends on the impact of the action on the environment. It is simply illustrated in \Fig\ref{fig:main:intro:reinforcement_learning}. Training such setting to obtain a performant policy $\pi(a, s) = \bbP(a_t = a  \vert s_t = s)$ is largely beyond the scope of this manuscript. Please refer to \cite{sutton1998introduction, sutton2000policy,mnih2013playing} for additional technical details.\\
				
				In the rest of this manuscript, we will principally focus on the simple \emph{supervised learning} type of experience, and only in \Part\ref{part:contribution} we will consider some generative priors generated by deep-generative models such as \aclink{GAN} and \aclink{VAE}, with random weights or trained on real data.
				
		\subsection{Statistical modeling}
		\label{sec:review_ml:statistical_modeling}
			For concreteness, let us summarize the \aclink{ML} workflow in \Fig\ref{fig:main:introduction:ml_workflow}: we have in hand a task $\mT$ that we want to solve, for example the classification of images, within an experience $\mE$, say supervised such that we have access to a dataset $\bbD=\{\mat{X}, \vec{y}\}$ of input images and corresponding labels.
			The next step consists in \emph{modeling} mathematically the underlying rule observed through the dataset and is referred to as \emph{statistical modeling}\index{Statistical modeling}. Statistical modeling and learning from data is the subject of a wide literature and is developed for instance in \cite{cherkassky2007learning}.
			
			\paragraph{Ground truth assumption and dataset}
			\label{sec:review_ml:statistical_modeling:ground_truth}
				In practice this dataset $\bbD$ has been collected without any specification on how samples were generated. 				
				Yet in the perspective of developing an analysis, it is of practical and theoretical interest to assume that some \emph{oracle} or \emph{teacher} knows the generative process of the dataset, even though most of the time it is not available in real industrial applications. In particular, it has the advantage to allow for measuring the model reconstruction performances as explained in \Sec\ref{sec:review_ml:performances}. But of course, for fairness the generative process should be hidden from the algorithm $\mA$ during the learning process, and we introduce it only for a theoretical purpose.
				
				In more details, we assume that there exists either a ground truth function $f^\star$ or equivalently a joint probability $\rP^\star(y, \vec{x})=\rP^\star(y | \vec{x}) \rP^\star(\vec{x})$ accounting for the information contained in the data. The dataset $\bbD = \{\(\vec{x}_{1},y_{1}\),...,\(\vec{x}_{\nsamples} ,y_{\nsamples} \) \} $ is composed of \aclink{i.i.d} samples such that $\forall \mu \in \lb \nsamples \rb, y_{\indsamples} = f^\star\(\vec{x}_{\indsamples} \)$ or equivalently $y_\mu \sim \rP^\star(.|\vec{x}_\mu)$. 
				In the case where the generative process is explicitly known and accessible, the ground truth density $\rp^\star(y | \vec{x})$, which can be can be simply designed by hand in simple theoretical models,
				 is used to generate conveniently new \emph{synthetic datasets}\index{synthetic dataset}. 
				To conclude, as we have a direct access to the ground truth solution, this setup is very close of the \emph{teacher-student}\index{Teacher-student} scenario in planted \emph{spin-glass} models discussed in more details in \Sec\ref{main:intro:phys_ml_together:supervised}, and promotes our statistical physics approach.\\
			
			Under this assumption, \aclink{ML} aims ultimately to select a \emph{model}\index{model} $\mM$ that \emph{estimates} correctly the underlying data distribution $\rP^\star(y | \vec{x})$.
									
			\subsubsection{Hypothesis class}
			\label{sec:review_ml:statistical_modeling:hypothesis}
				To make the estimation problem of the \emph{target function} $f^\star$, or equivalently the \emph{target distribution} $\rP^\star(y | \vec{x})$, tractable we shall consider models $\mM$ in an appropriate \emph{hypothesis class} $\bbH$.
				This is the realm of \emph{statistical modeling} that consists in restricting the whole solution space to a smaller set of hypothesis functions $\bbH = \{ f: \bbX \mapsto \bbY\}$, from the \emph{input space} $\bbX$ to the \emph{target space} $\bbY$. This shall be performed carefully such that the hypothesis class $\bbH$ is rich enough to be contained in the \emph{target class} $\bbH^\star$, to which $f^\star$ belongs.
				In this way a function $f\in \bbH$ may approximate correctly the \emph{target function} $f^\star \in \bbH^\star$.			
				As an illustration to capture the data-points in \Fig\ref{fig:main:intro:regression_classification}, we may consider the set of simple \emph{linear models} parametrized by some weights $\{\vec{w}, w_0\}\in \bbR^{\ndim+1}$: $\bbH_{\textrm{linear}} = \{ f_\vec{w}: \bbX \subseteq \bbR^\ndim \mapsto \bbY : f_\vec{w}(\vec{x}) = \vec{w} \cdot \vec{x} + w_0 \text{ with } \vec{w} \in \bbR^{\ndim}\}$. 
				
				As a remark, notice that finding a good statistical model is at the crossroad of two fields of research: the classical \emph{approximation theory} and the modern \emph{machine learning}. Their discriminating difference lies mainly in the input space dimensionality and the features that are engineered in the first and learned from data in the second. 
											
			\subsubsection{Parametric estimation}	
			\label{sec:review_ml:statistical_modeling:parametric}
				Just as the above linear models class example, we often consider \emph{parametric estimation} by restricting statistical models to parametric hypothesis space $\bbH_\btheta$ that depend on a 
				\graffito{The dimension $n_\theta$ of the parameter $\btheta$ is not specified as it strongly depends on the model.}
				collection of parameters $\btheta\in\bbR^{n_\theta}$. Estimating the model $f_\btheta$ is therefore reduced to computing the parameters $\btheta$. In general, it denotes the set of parameters of the statistical model that could represent either a scalar, a vector or a set of matrices. In particular, in the neural networks language, the parameters are called instead \emph{weights} and will be denoted $\mat{W}$ in the following to represent rectangular matrices.
				As a remark, there exists also \emph{non-parametric} estimation methods such as nearest neighbors regression or decision trees. Being beyond the scope of this work, we do not cover non-parametric estimation in this manuscript. Refer to \cite{tsybakov2008introduction, james2013introduction} for an introduction. 

		\subsection{Measuring the performance}
		\label{sec:review_ml:performances}
			Once the model $\mM$ corresponding to an hypothesis class $\bbH$ has been selected we must introduce a set of tools to measure its validity.
			In a synthetic dataset setting, in which the ground truth is available, the \emph{reconstruction performance} of the parametric model can be directly measured by the \aclink{MSE} between parameters $\btheta$ of the model $f_\btheta$ and $\btheta^\star$ the ones of the target function $f^\star = f_{\btheta^\star}$.
			Otherwise, we need to introduce other statistical tests to measure the model performances.
				
			\subsubsection{Reconstruction measure: the mean squared error} 
				Whenever the ground truth parameters $\btheta^\star$ are available, the performance of the parametric model $\btheta(\bbD)$, estimated on the dataset $\bbD$, can be measured by a direct comparison. 
				The \emph{reconstruction performance} of the estimator is commonly quantified by the \aclink{MSE} between the parameters $\btheta^\star$ and $\btheta(\bbD)$ averaged over all potential dataset and ground truth parameters: 
					\begin{align}
							\MSE(\btheta) = \EE_{\btheta^\star,\bbD} \[ \|\btheta^\star - \btheta\(\bbD\)\|_2^2 \]\,.
					\end{align}
				This is valid only if the parameters $\btheta^\star$ and $\btheta$ have the same dimensions, that is if the target models and statistical models belong to the same hypothesis class $\bbH=\bbH^\star$. 
				In our theoretical analysis of the simple models, we will make use of this reconstruction measure. However, in practice the \aclink{MSE} is rarely used because the ground truth parameters $\btheta^\star$ are not directly available.
				
			\subsubsection{Objective, risks and errors}
			\label{main:introduction:ml:train_val_test}
				As an alternative to this reconstruction measure, which is well suited only in the synthetic setting, 
				most \aclink{ML} tasks are instead formulated as the minimization of a \emph{risk}\index{risk} function, also called \emph{objective}\index{objective} or \emph{error} function. This objective depends on a \emph{criterion}\index{criterion} or \emph{loss function}\index{loss function} $\ell : \bbX \times \bbY \mapsto \bbR$, whose choice specifically depends on the task $\mT$ and experience $\mE$. 
				The validity of the statistical model $f_\btheta \in \bbH$ is thus appreciated from the value of the risk function: achieving a low risk value advocates for a good statistical model.
								
				\paragraph{Population risk and generalization error}
					The learning objective is commonly chosen as 
					the \emph{population risk}\index{population risk} $\mR$ defined as
					\begin{align}
						\mR\(f_\btheta; \ell\) = \EE_{\(\vec{x}, y\) \sim \rP(\vec{x}, y)} \[ \ell\( y, f_\btheta\(\vec{x}\) \) \]\,.
						\label{eq:main:introduction:ml:population_risk}
					\end{align}
					This is also called the \emph{generalization error} in the \aclink{ML} community and it will be equivalently denoted $e_{\gen}\(f_\btheta; \ell \)$. The \emph{loss} function $\ell$ measures pointwise the error between the target value $y$ and the prediction of the model, $f_\btheta(\vec{x})$. The population risk is simply its average over \emph{all possible} examples drawn from the joint distribution $\rP(\vec{x}, y)$. 
					Achieving a low population risk defines a strong criterion of validity of the model and allows for model selection. In fact, the optimal model parameters $\hat{\btheta}$ would be selected by directly minimizing the population risk $\hat{\btheta} = \argmin_{\btheta} \mR\(f_\btheta; \ell\)$.
					Unfortunately, the population risk and the corresponding minimization program are intractable as the average over the high-dimensional joint distribution $\rP(\vec{x}, y)$ 
					is very often complex or unknown. 
					Nonetheless, in this theoretical manuscript, we will be able to compute the generalization error in particular cases with synthetic datasets coming from simple joint distributions $\rP(\vec{x}, y)$.

				\paragraph{Empirical risk, training error and training set}
					In general, we do not have knowledge of the generative process and the distribution $\rP(\vec{x}, y)$. Instead, we only have access to a \emph{finite} \emph{training set} of $\nsamples$ examples $\bbD_\train = \bbX_\train \times \bbY_\train$. 
					Even if it is very large, $\nsamples \gg 1$, this discrete dataset cannot account for the whole continuous and infinite joint distribution $\rP(\vec{x}, y)$.
					As a result the intractable \emph{population average} $\EE_{\(\vec{x}, y\) \sim \rP(\vec{x}, y)}$ is replaced by an \emph{empirical average} over the training set. And consequently the intractable population risk is replaced by the \emph{empirical risk}\index{empirical risk}, also called the \emph{training error}\index{training error} $e_{\train}$, that serves as a proxy of the population risk:
					\begin{align}
						\hat{\mR}\(f_\btheta; \ell, \bbD_\train \) = \frac{1}{\nsamples} \sum_{\indsamples=1}^{\nsamples} \ell\( y_{\mu}, f_\btheta \( \vec{x}_{\mu} \)\) \,.
						\label{eq:main:introduction:ml:empirical_risk}
					\end{align}	
					The population risk gives indications along the training of the model validity. 
					For instance, this criterion gives a practical procedure for many \aclink{ML} algorithms, such as \aclink{ERM}, that minimize the empirical risk $\hat{\btheta} = \argmin_{\btheta} \hat{\mR}\(\btheta; \ell, \bbD_\train \)$, but only as a proxy of the population risk $\mR\(\btheta; \ell\)$. 
					However, minimizing the empirical risk does not guarantee at all a good \emph{generalization} performance of the estimator on unseen data. 
					Indeed, in high-dimensions the empirical and true underlying distributions can be very different, and thus minimizing the population and the empirical risks do not lead to similar results. Addressing this issue and trying to control their difference $| \mR(\btheta) -  \hat{\mR}(\btheta; \ell, \bbD_\train) |$ is at the heart of modern \aclink{ML} and \emph{statistical learning theory}, as illustrated in \Sec\ref{sec:main:introduction:ml:generalization_bounds}.
					
				\paragraph{Test set and error}
					The purpose of \aclink{ML} is essentially to robustly predict the outcomes of unseen data. Thus, it would make little sense to check the validity of the model on data that have been seen and used to estimate the same model.
					Therefore, we must allocate a part of the dataset for testing its validity, so that the dataset $\bbD = \bbD_{\train}  \times \bbD_{\test}$ is split in a \emph{training set}\index{training set} $\bbD_{\train}$ that contains observations the algorithm $\mA$ may use to estimate the model parameters $\btheta(\bbD_{\train})$, and a \emph{testing set}\index{testing set} $\bbD_{\test}$ on which the validity of the model is assessed . 
					Indeed, as suggested in many works such as \cite{zhang2016understanding}, recent \aclink{ML} models can \emph{perfectly} minimize the empirical risk, meaning that the training error is zero and the model has perfectly \emph{memorize} the training set $\bbD_{\train}$.
					As a consequence, reaching zero training error does not ensure the validity of the model, that should be attested instead on the separated \emph{test set} $\bbD_\test$. The error measured on this set is called the \emph{test error}
					\begin{align*}
						e_{\test}\(f_\btheta; \ell, \bbD_\test \) = \hat{\mR}\(f_\btheta; \ell, \bbD_\test \)\,,
					\end{align*} 
					and serves as a finite-size surrogate for the ideal but intractable population risk $\mR\(f_\btheta; \ell\)$ and generalization error.
					
				\paragraph{Hyper-parameters and validation test}
					In addition, as illustrated in \Sec\ref{sec:main:introduction:ml:algorithms}, most of current \aclink{ML} algorithms depend on some \emph{hyper-parameters}. These latter are settings that we can use to control the algorithm and must be fixed in some way. However, the hyper-parameters cannot be learned during the algorithm learning procedure, because it would constantly select high-capacity models that easily fit the training set.
					To circumvent this difficulty, this is often done by introducing a third set, called a \emph{validation set}\index{validation set}, that the algorithm does not observe during the training phase and that is used to select good hyper-parameters. Therefore, the dataset $\bbD = \bbD_\train \times \bbD_{\mathrm{val}} \times \bbD_\test $ is finally decomposed in train/validation/test sets allocated approximately to $70/10/20\%$ of the total size.
					In the case of small datasets, where the statistical significance drastically decreases, an alternative approach, called \emph{cross-validation}\index{cross-validation}, is often used. It consists in a \emph{leave-on-out} strategy of repeating and averaging the training and testing operations on randomized sets. See for instance \cite{goodfellow2014generative} for an extended discussion.
					
			\subsubsection{Choosing a loss function}
				The loss function $\ell$ is strongly task-dependent, and we review the classical choices of loss functions used in the literature and in this work. In general, in order to minimize the empirical risk \eqref{eq:main:introduction:ml:empirical_risk}, with \emph{gradient-based} algorithms, we should prefer smooth loss functions such that the gradients exist, are easy to compute and not too small. 
				The simplest loss is the squared loss $\ell^{\textrm{l2}}(y,\hat{y}) = ( y - \hat{y} )^2 $ particularly suited for real-valued regression tasks, as well as the absolute loss $\ell^{\textrm{l1}}(y,\hat{y}) = |y- \hat{y} |$.
				For classification, where output values are discrete, and very often $\pm 1$, we often use either the hinge loss $\ell^{\textrm{hinge}}(y, \hat{y}) = \max\(0, 1 - y \hat{y} \) $, the logistic loss $\ell^{\textrm{logistic}}(y, \hat{y}) = \log( 1 + \exp(-y \hat{y}) )$, the binary cross entropy loss $\ell^{\textrm{bce}}(y, \hat{y}) = - y \log \hat{y} - (1-y) \log(1-\hat{y})$ or the hard error-counting loss $\ell^{\textrm{hard}}(y,\hat{y}) = \id \[ y \ne \hat{y} \]$, even though it is not differentiable.		
		
		\subsection{Model complexity, limitations and overfitting}
		\label{sec:main:introduction:ml:complexity}
			The ultimate goal of \aclink{ML} is to predict the output of unseen data of the model $f_\btheta$ \emph{trained} from the observation of a training set $\bbD_\train$. To do so, most of \aclink{ML} algorithms $\mA$ minimize the empirical risk so that many \aclink{ML} problems may be reformulated as an \emph{optimization} problem. 
			Yet the main difference between \aclink{ML} and standard optimization fields is that we require instead the algorithm to not find \emph{any} minima, but a minima that \emph{generalizes} correctly on unseen data. In other words, we require that the generalization error (or simply its finite-size estimation, the test error) remains low, as well as the training error optimized during the training. This is called the \emph{generalization} problem.
					
			\subsubsection{Under/over fitting}
			As the test set is drawn before any learning process, the expected test error $e_\test$ will be therefore greater or equal than the training error $e_\train$. 
			Though, in practice a \aclink{ML} algorithm minimizes the training error as a proxy for minimizing the ultimate generalization error, so that we require the \emph{generalization gap} between the test and training errors $e_\test-e_\train$ to be as small as possible. 
			The trade-off between this two conditions may lead to key and burning challenges in the \aclink{ML} community: \emph{underfitting and overfitting}. In one hand \emph{underfitting} refers to a model with large training error and therefore a large test error, while \emph{overfitting} occurs when the generalization gap is too large.
			
			\subsubsection{Model capacity}
			\label{sec:main:introduction:ml:capacity}
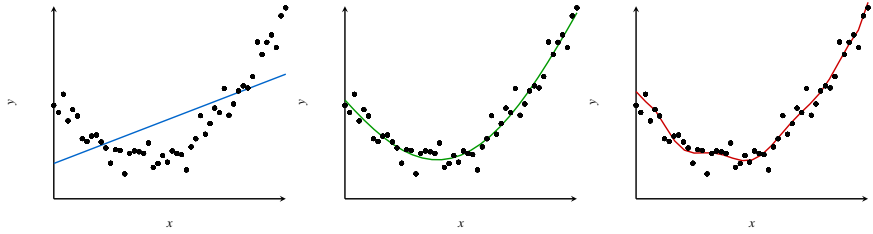
\begin{figure}[htb!]
				\tikzexternalize
				\begin{tikzpicture}[scale=0.45]
				\begin{axis} [xlabel     = $x$, ylabel     = $y$, axis lines = left,axis line style = very thick,
						      clip       = false, xmin = -2,  xmax = 2,ymin = -5, ymax = 15, ticks = none, domain=-2:2,samples=25]
				\pgfplotstableread{Plots/data_ground_truth_x2.txt}\bbYdata;
				\addplot [color=black, only marks, mark=*, mark size=1.5pt] table [x expr=\thisrowno{0}, y expr=\thisrowno{1} ]{\bbYdata};
				\addplot [color=blue, very thick] {2.3187566*x +3.30596572};  
				\end{axis}
				\end{tikzpicture}	
				\begin{tikzpicture}[scale=0.45]
				\begin{axis} [xlabel     = $x$, ylabel     = $y$, axis lines = left,axis line style = very thick,
						      clip       = false, xmin = -2,  xmax = 2, ymin = -5, ymax = 15, ticks = none,domain=-2:2,samples=25]
				\pgfplotstableread{Plots/data_ground_truth_x2.txt}\bbYdata;
				\addplot [color=black, only marks, mark=*, mark size=1.5pt] table [x expr=\thisrowno{0}, y expr=\thisrowno{1} ]{\bbYdata};
				\addplot [color=green, very thick] {-0.1036 * x^4 -0.0357*x^3 + 2.978*x^2 +  2.4079*x -0.4676 }; 
				\end{axis}
				\end{tikzpicture}	
				\begin{tikzpicture}[scale=0.45]
				\begin{axis} [xlabel     = $x$, ylabel     = $y$, axis lines = left,axis line style = very thick,
						      clip       = false, xmin = -2,  xmax = 2,ymin = -5, ymax = 15, ticks = none,domain=-2:2,samples=25]
				\pgfplotstableread{Plots/data_ground_truth_x2.txt}\bbYdata;
				\addplot [color=black, only marks, mark=*, mark size=1.5pt] table [x expr=\thisrowno{0}, y expr=\thisrowno{1} ]{\bbYdata};
				\addplot [color=red, very thick] {  0.174*x^10  -0.011 * x^9 -1.82*x^8 + 0.223*x^7 + 6.74*x^6 -1.0987*x^5 -10.3813*x^4 + 1.667 *x^3 + 8.55518906*x^2 + 1.77304478*x -0.96425398};    
				\end{axis}
				\end{tikzpicture}
				\tikzexternaldisable		
				\caption{Model complexity illustration on a regression task. Input-output example pairs $(x,y)$ of the training set are shown with black points. \Left A linear model cannot fit the training set and leads to a high training error. \Center An intermediate complexity model yields a good estimator with low training error and low test error. \Right A large complexity model interpolates the training points and achieves almost zero training error. But it completely overfits the training set and does not generalize correctly, resulting in a high test error.}
					\label{main:introduction:ml:overfitting}		
				\end{figure}

			Underfitting and overfitting phenomena are closely related to the choice of the hypothesis space $\bbH$ and in particular its \emph{capacity}.
			The capacity of a model refers to its ability to fit a wide range of functions. For example linear models cannot fit non-linearly separable data, while high-degree polynomials can. In general, low capacity models may struggle to fit the dataset, while in contrast high-capacity models can easily \emph{memorize} (and not \emph{learn}!) the dataset, so that they will completely \emph{overfit} at test time. These limiting situations are illustrated in \Fig\ref{main:introduction:ml:overfitting} \Leftn~ and \Rightn.
			A \emph{good} statistical model should strike a balance between high-capacity and small test error: the model capacity should be large enough to solve complex tasks resulting in a low training error, but small enough to not perfectly fit the training set and fail in the test set with high test error as illustrated in \Fig\ref{main:introduction:ml:overfitting} \Center. Therefore the capacity of the model should be adapted to the task $\mT$ difficulty and the size of the dataset $\bbD$.
				
\begin{figure}[htb!]
					\centering
					\begin{tikzpicture}[scale=0.8]
					\begin{axis} [xlabel     = Model capacity, ylabel     = Error, axis lines = left,axis line style = very thick,
							      clip       = false, xmin = 0.1,  xmax = 3,ymin = 0, ymax = 10, ticks = none]
					\pgfplotsset{samples=100}
					\addplot[color=red,domain=0.1:3, very thick]{1/x};
					\addlegendentry{Training error $e_\train$};
					\addplot[color=green,domain=0.1:3, very thick]{1/x + x};
					\addlegendentry{Test error $e_\test$}; 
					\draw[>=latex, <->, black, very thick] (axis cs:2.5,0.5) -- (axis cs:2.5,2.9) node[left, yshift=-0.8cm] {Generalization gap};
					\end{axis}
					\end{tikzpicture}		
						\caption{Illustration of the training and test errors as the function of the model capacity. For small capacity models both the training and test errors are high and fall in the underfitting regime. As the capacity grows, the training error eventually decreases to zero, while the test error reaches a minimum at optimal capacity, before growing again and fall in the overfitting regime. The generalization gap is the difference between the test and training errors $e_\test-e_\train$}
						\label{main:introduction:ml:overfitting_capacity}
					\end{figure}
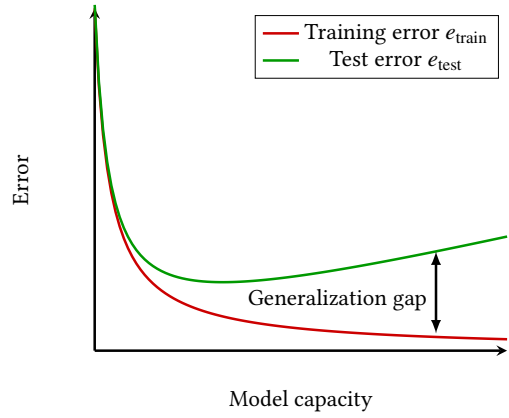

					Typically the generalization gap behavior between the test and training errors is summarized with the U-shaped curve in \Fig\ref{main:introduction:ml:overfitting_capacity}. It can be understood from the \emph{Occam's razor principle} that states that among competing hypotheses that explain a set of observations equally well, we should prefer the hypothesis with the smallest capacity to avoid overfitting.
					Hence by choosing a statistical model, we shall keep in mind that as soon they have small training error,  small capacity functions are more likely to generalize correctly. 
					
				\subsubsection{The bias-variance trade- off}				
					The illustration in \Fig\ref{main:introduction:ml:overfitting} raises the question of how to properly choose the hypothesis class $\bbH$ to not be threaten by overfitting. This is formalized by the evolution of the generalization gap with the model complexity described in  \Fig\ref{main:introduction:ml:overfitting_capacity}. In fact, this non-monotonic behavior is traditionally understood from the bias-variance decomposition. Indeed, bias and variance measure two different sources of errors of a given estimator $\btheta$, as illustrated by the decomposition of the \aclink{MSE} reconstruction error:					
					\begin{align*}
					\begin{aligned}
						&\textrm{MSE}( \btheta) = \EE\[ \(\btheta^\star - \btheta \)^2\] = \EE\[ \(\btheta^\star- \EE[\btheta] + \EE[\btheta] - \btheta  \)^2\]\\
						&=  \EE\[ \(\btheta^\star - \EE[\btheta] \)^2\] + \EE\[ \(\EE[\btheta] - \btheta\)^2\] \\
						& \hspace{1cm} + 2 \EE\[ \btheta^\star - \EE[\btheta] \]  \EE\[ \EE[\btheta] - \btheta\]\\
						&= \EE_{\btheta^\star} \[\btheta^\star - \EE[\btheta] \]^2 + \EE\[ \(\EE[\btheta] - \btheta\)^2\] \equiv \textrm{Bias}\[\btheta\]^2 + \textrm{Var}\[\btheta\]\,,
					\end{aligned}
					\end{align*}
					where the \emph{bias} of the estimator is the expected deviation from the ground truth value $\textrm{Bias}\[\btheta\] \equiv \EE_{\btheta^\star}\[\btheta^\star - \EE[\btheta] \]$ and and the \emph{variance} is the deviation from the expected estimator value $\Var{\btheta} = \EE\[ \(\EE[\btheta] - \btheta\)^2\]$. 
					As the model capacity increases, the prediction accuracy increases so that the bias term decreases whereas the variance term increases. Summing these two terms leads to a U-shaped curve similar to the one in \Fig\ref{main:introduction:ml:overfitting_capacity} and this decomposition is traditionally used to explain underfitting and overfitting behaviors illustrated in \Fig\ref{main:introduction:ml:overfitting}. However, we will describe later that this traditional argument fails explaining the behavior of \aclink{DNN} as suggested in \cite{zhang2016understanding}.
											
				\subsubsection{Regularization}
				\label{sec:main:introduction:ml:regularization}
					As suggested by the above analysis, in order to control the generalization gap, we should act on the model capacity. However in practice, we may prefer to use a fixed model with high-capacity to be able to fit various datasets. Thus in order to avoid overfitting of this high-capacity model, we shall reduce the \emph{effective capacity} of the hypothesis class $\bbH$. This can be done by \emph{promoting} or \emph{biasing} the training algorithm towards particular solutions. In other words, this means that all potential functions $f_\btheta$ in the hypothesis class $\bbH$ are eligible, but a few of them are more likely and have a highest preference. This is commonly done by adding a \emph{regularization term} to the empirical risk \eqref{eq:main:introduction:ml:empirical_risk}:
					\begin{align}
						\hat{\mR}\(f_\btheta; \ell, \bbD_\train \) \leftarrow \hat{\mR}\(f_\btheta; \ell, \bbD_\train \) + \lambda \Omega(\btheta) \,.
						\label{eq:main:introduction:ml:empirical_risk_regulqrized}
					\end{align}
					$\lambda$ is called the regularization strength and we can choose different forms for the regularization term such as the classical $\rL_p$-norm $\Omega(\btheta)=\|\btheta\|_p$, which promotes sparse or small-weights solutions for $p=1,2$. Yet depending on the task $\mT$ and how one wants to restrict the hypothesis class, more complex reguralization terms can be designed. 
					Minimizing the \emph{regularized} empirical risk \eqref{eq:main:introduction:ml:empirical_risk_regulqrized} results in a trade-off between \emph{fitting the training set} and \emph{satisfying the regularization constraint}, \eg
					\emph{keeping small parameters} $\btheta$ in the case of a $\rL_2$ regularization. This avoids the high-capacity model to fully release its expressivity and overfit the training set.
					To summarize, regularization refers to any modification made to the learning problem in order to reduce its generalization gap and is essentially at the heart of \aclink{ML} practical challenged. \\
				
				Since most practical algorithms minimize the (regularized or not) empirical risk, from a theoretical point of view, it would be of great interest to have some \emph{uniform convergence} guarantees that the algorithm simultaneously minimizes the population risk. Bounding the \emph{generalization gap} is a burning challenge widely studied in the statistical learning community and briefly reviewed in the next section.
			
		\subsection{Generalization error bounds}
		\label{sec:main:introduction:ml:generalization_bounds}
			The population risk being out of reach, we are reduced to use the empirical risk as a surrogate. Unfortunately, in high dimensions changing the population average by the empirical average may have strong and damaging consequences. 
			First of all, minimizing the empirical risk $\hat{\mR}$ eq.~\eqref{eq:main:introduction:ml:empirical_risk} is not at all guaranteed to provide the same result than minimizing the true ideal population risk $\mR$ eq.~\eqref{eq:main:introduction:ml:population_risk}.
			This would be correct if we would have a \emph{uniform convergence} theorem that would assert that the \emph{generalization gap} $\| \mR -  \hat{\mR} \|$ decreases quickly with the input dimension $\ndim$ and the number of samples $\nsamples$.
			This question is part of the realm of \emph{statistical learning} theory, pioneered in \cite{vapnik2015uniform, blumer1989learnability, vapnik1994measuring}, and the \aclink{PAC} framework, introduced in \cite{valiant1984theory}, nicely reviewed in \cite{Mohri12, murphy2012machine}.
			Statistical learning theory provides various tools to quantify the model capacity such as the \aclink{VC} dimension $d_\vc$ or the Rademacher complexity $\mathfrak{R}_\nsamples$. Measuring the model capacity allows therefore to bound more finely the generalization gap, \ie the discrepancy between training error and generalization error.
			
			The goal of the next results is to introduce the main quantities that allow to bound the \emph{generalization gap}. First results have been obtained in the case of classification in \cite{vapnik2013nature}.
			The first simple result is that any target function $f^\star$ is learnable using a finite hypothesis set $\bbH$ as soon as $f^\star \in \bbH$. 
				This result is proven with Hoeffding's inequality \cite{hoeffding1994probability} and the union bound argument that states $\forall \delta >0$ with probability $1-\delta$,
				\begin{equation*}
					\forall f \in \bbH, \| \mR(f) -  \hat{\mR}(f, \bbD) \| \leq  \sqrt{\frac{\ln(|\bbH|) + \ln(2/\delta)}{2 \nsamples}}\,.
				\end{equation*}
			The underlying union bound argument is responsible for the presence of the cardinality of the hypothesis class $|\bbH|$ on the right-hand side. Unfortunately, the generalization gap bound becomes vacuous in the case of interest for an infinite class $|\bbH| = \infty$. This issue is circumvented with the use of finer and tighter bounds such as the \aclink{VC} dimension \cite{vapnik2013nature} for classification and more recent distribution-dependent Rademacher complexity \cite{bartlett2002rademacher}.

			\subsubsection{VC dimension}
				 	 The idea of the \aclink{VC} dimension, which is restricted to classification tasks, is to count only hypotheses that provide different labelings of the dataset. This can be formalized with the notion of \emph{dichotomies} that is exactly the number of ways of classifying differently the points of the dataset $\bbD$. To obtain a measure of the richness of the hypothesis class $\bbH$, we introduce the \emph{growth function} $\Delta_{\bbH}(\nsamples)$ which is the maximum number of dichotomies in which the $\nsamples$ points of the dataset can be classified using hypotheses $f \in \bbH$:
				 	\begin{equation*}
					\Delta_{\bbH}(\nsamples) = \max_{\{\vec{x}_1,..,\vec{x}_\nsamples\}\subseteq \bbX } |\{\(f(\vec{x}_1),..,h(\vec{x}_\nsamples)\): f \in \bbH \}  | \leq 2^\nsamples \,,
					\end{equation*} 
					It finally leads to a refinement of the generalization gap bound, called the \aclink{VC} inequality, that states that with probability $1-\delta$,
					\begin{align*}
						\forall f \in \bbH,  \| \mR(f) -  \hat{\mR}(f, \bbD) \| \leq \sqrt{ \frac{8}{\nsamples}\ln(4 \Delta_{\bbH}(2\nsamples)/\delta ) }\,.
					\end{align*}
					To conclude with this generalization bound, we shall compute the growth function $\Delta_{\bbH}(\nsamples)$, that is unfortunately often intractable. Instead we introduce an alternative measure of the hypothesis class complexity: the \aclink{VC} dimension $d_\vc$ which is a combinatorial quantity much easier to compute. It is defined as the size of the largest set that can be fully shattered
					\graffito{A set of $\nsamples$ points is said to be shattered by a hypothesis set $\bbH$ when $\bbH$ realizes all possible dichotomies: $\Delta_{\bbH}(\nsamples) = 2^\nsamples$.}
					\begin{align*}
						 d_\vc \equiv \max \{\nsamples : \Delta_{\bbH}(\nsamples) = 2^\nsamples \} \,.
					\end{align*}
					From Sauer's lemma \cite{sauer1972density,shelah1972combinatorial}, we can show that as soon the \aclink{VC} dimension is finite the growth function verifies $\Delta_{\bbH}(\nsamples) \leq \sum_{i=0}^{d_\vc} {\nsamples \choose i} \leq \Theta \( \(\nsamples e / d_\vc\)^{d_\vc}\)$ so that $\log \Delta_{\bbH}(\nsamples) = \Theta(\log \nsamples)$. Thus, the above generalization bound vanishes with an infinite number of samples. Finally, we obtain the fundamental theorem of statistical learning which states that as soon the \aclink{VC} dimension of hypothesis class $\bbH$ is \emph{finite}, the target function class $\bbH^\star$ is \aclink{PAC} learnable. See \cite{Mohri12} for an extended derivation.
			
				\subsubsection{Rademacher complexity}
				The \aclink{PAC} framework is too restrictive in the sense that it requires the strongest worst-case bound working for any dataset $\bbD$. To relax this strong hypothesis, a more recent generalization bound has been introduced: the Rademacher complexity \cite{bartlett2002rademacher}, which explicitly depends on the data distribution. The Rademacher complexity captures the richness of the family $\bbH$ of functions by measuring the degree to which a hypothesis class can fit random noise. The empirical Rademacher complexity $\hat{\mathfrak{R}}_{\bbD} (\bbH)$ is defined by
				\begin{align*}
					\hat{\mathfrak{R}}_{\bbD} (\bbH) = \EE_{\bsigma} \[ \sup_{f \in \bbH}  \frac{1}{\nsamples} \displaystyle \sum_{\mu=1}^\nsamples \sigma_\mu f\(\vec{x}_{\mu}\))  \] = \EE_{\bsigma} \[\frac{1}{\nsamples}  \sup_{ f \in \bbH} \bsigma \cdot f(\mat{X}) \]\,,
				\end{align*}
				where $\bsigma = \{\pm 1 \}^\nsamples$ is a uniform Rademacher random variable with probability $\frac{1}{2}$. The main classical result states that the empirical Rademacher complexity provides a uniform convergence bound. Informally, for any $\delta>0$, with probability $1-\delta$
				\begin{equation}
						\sup_{f \in \bbH} \| \mR(f) - \hat{\mR}(f, \bbD)  \| \leq 2 \hat{\mathfrak{R}}_{\bbD}(\bbH) + \Theta \( \sqrt{\frac{\ln(2/\delta)}{\nsamples}}\)\,.
				\end{equation}
				Notice that using the Massart's lemma \cite{massart2000some} both the growth function and the Rademacher bound may be reconciled as it follows
				\begin{align*}
					\hat{\mathfrak{R}}_{\bbD}(\bbH) \leq \sqrt{\frac{2 \ln \Delta_{\bbH}(\nsamples)}{\nsamples} } \leq \Theta\( \sqrt{\frac{d_\vc}{\nsamples}} \)\,,
				\end{align*}
				so does the \aclink{VC} dimension.
				To better understand the notion of Rademacher complexity, it is fruitful to notice that it simply measures, on average, the correlation between the prediction of the estimator $f$ and random labels $\bsigma$, which are uncorrelated from the inputs examples $\bbX$.
			To conclude this section, let us mention that the mathematical and statistical learning community largely focussed on such uniform convergence generalization bounds. However, we will discuss that this kind of \emph{worst-case} scenario bounds are believed to be over-pessimistic and fail, therefore, to capture the generalization behavior of practical model classes such as \aclink{DNN}.
		
		\subsection{Statistical estimation}
		\label{sec:main:introduction:ml:estimators}
			Once we have chosen a parametric model $f_\btheta \in \bbH$ within an certain hypothesis class, or equivalently a parametric family of probability distributions $\rP_{\btheta}\( \vec{x}\)$, we shall discuss how to \emph{estimate} in statistics, or equivalently \emph{learn} in \aclink{ML}, the model parameter $\btheta$. 
			\emph{Statistical estimation} of the parameters is divided in two ways of thinking: \emph{frequentist} versus \emph{Bayesian}.
			These approaches undergo long conflicts and the literature is full of debates among statisticians to build proper estimators \cite{aldrich2008ra}. 
			\graffito{\say{Ignorance is preferable to error and he is less remote from the truth who believes nothing than he who believes what is wrong.
				Thomas Jefferson (1781)}}
			In this section, we simply review the two approaches and the most common estimators used in the applications \Part\ref{part:contribution}. We standardly denote $\hat{\btheta}$ the output of the different estimators.
						
		\subsubsection{Frequentist approach}
			The frequentist approach assumes that making use of any \emph{a priori} distribution would be misleading.
			In order to not bias the estimation in the wrong way, frequentists prefer to make no assumption on the a priori probability distributions. The central object of study is the likelihood and follows the work of \cite{Fisher1925}.
			
			\paragraph{Likelihood}
			Let us consider a set of observations $\mat{X} = \left \{\vec{x}_{\mu} \right\}_{\mu=1}^\nsamples$ drawn \aclink{i.i.d} from an underlying data distribution $\rP(\vec{x})$ and a family of distributions parametrized by $\btheta$, $\rP_{\btheta}\( \vec{x}\) \equiv \rP\( \vec{x} | \btheta \) $ that models it. We define the \emph{likelihood} \index{likelihood} function $\mL$ or respectively the \emph{log-likelihood} \index{log-likelihood} $L$ according to the data $\mat{X}$ by 
			\begin{align}
				\mL\(\btheta | \mat{X} \) &: \btheta \mapsto \rP\(\rX=\mat{X} | \btheta \)\,, && L\(\btheta | \mat{X}\): \btheta \mapsto \log \rP\(\rX=\mat{X} | \btheta \)\				
				\label{definition:likelihood}
			\end{align}
			that both measure the probability of obtaining observations $\mat{X}$ for a given value of the model parameters $\btheta$. This likelihood function does not assume any \emph{prior knowledge} on the parameter space and is considered by frequentists to contain all relevant information for statistical inference. 
			
			\paragraph{Maximum Likelihood Estimation}
				Based on the \emph{log-likelihood} $L\(\btheta | \mat{X}\)$, the simplest and most common estimator consists in maximizing the probability of observing the data $\mat{X}$. The \aclink{MLE}\index{maximum likelihood} estimator $\hat{\btheta}_\mle$ is defined as
			\begin{align}
				\hat{\btheta}_\mle \(\mat{X}\) \equiv \argmax_{\btheta} \left\{ L\(\btheta | \mat{X}\) \right \} =  \argmin_{\btheta} \left\{-  L\(\btheta | \mat{X}\) \right \}\,,
			\end{align}
			that can be, equivalently, simply written as a minimization problem. The \aclink{MLE} can be interpreted as a way of matching the empirical distribution of the data and the model distribution:
			\begin{align*}
				\hat{\btheta}_\mle \(\mat{X}\) &\equiv \argmax_{\btheta} \left\{ L\(\btheta | \mat{X}\) \right \} = \argmax_{\btheta} \left\{ \frac{1}{\nsamples} \sum_{\mu=1}^\nsamples \log \rP\(\vec{x}_{\mu}\ | \btheta \) \right \} \\
				&= \argmax_{\btheta} \left\{ \EE_{\vec{x}\sim \hat{\rP} } \log \rP\(\vec{x}\ | \btheta \) \right \}
			\end{align*}
			with the empirical data distribution $\hat{\rP}(\vec{x}) = \frac{1}{\nsamples} \sum_{\mu=1}^\nsamples \delta\( \vec{x} - \vec{x}_{\mu}\)$. Indeed the \aclink{KL} divergence serves as a distance within probability densities (see \Sec\ref{definition:kullback_leibler_divergence}) for more details) and is simply the cross-entropy between the empirical distribution and the model distribution
			\begin{align*}
				\KL\( \hat{\rP} | \rP_{\btheta}  \) = \EE_{\vec{x} \sim \hat{\rP} } \log \hat{\rP}(\vec{x}) - \EE_{\vec{x} \sim \hat{\rP} } \log \rP\(\vec{x} | \btheta \)\,.
			\end{align*}
			As the first term does not depend on the model, the \aclink{MLE} can be thought as minimizing the discrepancy between the empirical data and model distribution, with ideal objective to match the true data-generating distribution $\rP(\vec{x})$.
			
			\paragraph{Conditional likelihood}
			In a supervised learning perspective, where models are trained \emph{end-to-end}, the dataset is in fact made of inputs and outputs $\bbD = \{\mat{X}, \vec{y} \}$ and the likelihood shall be replaced by the conditional likelihood $L\(\btheta | \bbD \) = \btheta \mapsto \log \rP\(\vec{y} | \btheta, \mat{X} \)$. The corresponding maximum likelihood estimator readily generalizes to
			\begin{align*}
				\hat{\btheta}_\mle \(\bbD\) &\equiv \argmax_{\btheta} \left\{ L\(\btheta | \bbD\) \right \} \\
				&= \argmax_{\btheta} \left\{ \EE_{ \( \vec{x}, y \) \sim \hat{\rP}(\vec{x},y)  } \log \rP\( y | \btheta, \vec{x} \) \right \}  \,,
			\end{align*}
			and is a central estimator in most supervised learning settings. It turns out in particular that under the assumption that the ground truth $\rP^\star(\vec{x},y)$ lies within the probability density family $\bbK_\btheta = \left\{ \rP_\btheta \(y \vert \btheta, \vec{x}\), \btheta \in \bbR^{\ndim_\theta} \right\}$, the \aclink{MLE} estimator becomes \emph{optimal} in the asymptotic infinite number of samples $\nsamples \to \infty$. As it converges the fastest towards the true parameters $\btheta^\star$, the estimator is qualified of \emph{consistent} and also \emph{efficient} as moreover its generalization error decreases in this limit. Indeed, the Cramer-Rao bound states that any unbiased estimator has a variance bounded by the inverse of the Fisher information:
				\begin{align*}
					\Var{\hat{\btheta}} \geq \( \EE \[ \(\partial_{\btheta} \log \rP (y | \btheta, \vec{x})\)^2  \]  \)^{-1}\,,
				\end{align*}
				 and this lower bound is attained by the \aclink{MLE} in the large number of sample regime $\nsamples \to \infty$.
				This means that in this regime of large number of data, maximimzing the likelihood should be preferred to any other statistical estimator.
				Unfortunately when the number of data is limited, the \aclink{MLE} is not optimal and leads to overfitting that can be avoided by adding a regularization term. 
				In fact in this regime, more prior-knowledge information is required to perform optimal reconstruction, which can be achieved with the Bayesian approach presented in the next section.
											
		\subsubsection{Bayesian approach}
		\label{main:intro:ml:bayesian_approach}
			In contrast with the frequentist approach, which relies on a \emph{worst-case analysis}, Bayesian statistics makes use of \emph{prior information} or \emph{knowledge beliefs}  and relies on a \emph{typical case analysis}. 
			The Bayesian approach considers all possibles values of the estimators to make a prediction and it follows essentially Bayes \cite{bayes1763lii} and Laplace works.\\
			
			While the frequentist perspective assumes that the ground truth parameter $\btheta^\star$ is unknown but fixed, the Bayesian approach uses probabilities to reflect prior knowledges, so that $\btheta^\star$ is considered as an uncertain random variable with \emph{prior} distribution $\rP(\btheta^\star)$.
			Also while \aclink{MLE} makes predictions using a point-wise estimate, the Bayesian approach makes a predictions using the full distribution over $\btheta$.
			Therefore, observing a supervised learning dataset $\bbD = \{\mat{X}, \vec{y}\}$, we can make use of the observations of the data to model the probability of the parameter $\btheta$ essentially by means of the Bayes formula
			\begin{align}
				\rP\(\btheta|\vec{y}; \mat{X}\) = \frac{\rP\(\vec{y}| \btheta; \mat{X}\) \rP\(\btheta\)}{\rP\(\vec{y} ; \mat{X}\)}
			\end{align}
			where $\rP\(\vec{y}| \btheta; \mat{X}\)$ denotes the \emph{conditional likelihood}, $\rP\(\vec{y} ; \mat{X}\)$ the \emph{evidence}, 
			\graffito{Notice that taking a Gaussian prior $\rP(\btheta) = \mN_\btheta(0,1)$ is equivalent to add a $\rL_2$ regularization term $-\log \rP(\btheta) = \frac{1}{2} \|\btheta\|_2^2$ to the log-likelihood.}
			denoted later in the manuscript $\mZ\(\vec{y} ; \mat{X}\)$ also called the \emph{partition function}, and finally $\rP\(\btheta|\vec{y}; \mat{X}\)$ is called the \emph{a posteriori} or \emph{posterior} distribution.
			With this prior informations, which model the external world, Bayesian methods generalize typically much better when the training set is small and does not contain enough information. 
			However, we can already notice that computing the average over the posterior $\rP\(\vec{y} ; \mat{X}\) = \int_{\bbR^{\ndim}} \d \btheta ~ \rp\(\btheta|\vec{y}; \mat{X}\) $ will strongly suffer in the high-dimensional regime, where $\ndim, \nsamples \to \infty$, 
			and is in fact very often intractable.
			
			\paragraph{How to choose the prior?}
			Bayesian methods make deep use of the prior information $\rP(\btheta)$ which is unknown in general. 
			\graffito{Information theory provides a constructive criterion for setting up probability distributions on the basis of partial knowledge, and leads to a type of statistical inference which is called the maximum entropy estimate. It is least biased estimate possible on the given information; i.e., it is maximally noncommittal with regard to missing information. ET. Jaynes, 1957}
			The prior information is useful in the sense it shifts the probability density towards more probable regions of parameters. 
			In particular, it might be used to promote models that are simpler or more smooth, and can be already understood as a \emph{regularization} factor. 
			As frequentists blame Bayesian to bias estimation by injecting prior information that may be wrong, we should decide how to select \emph{correctly} the prior information $\rP\(\btheta\)$. 
			This question was addressed and answered in \cite{Jaynes57, jaynes03} who advocated that in order to bias as few as possible the estimation, we should select priors according to the \emph{maximum entropy principle}\index{maximum entropy principle} presented in more details in \Sec\ref{sec:main:mean_field:maximum_entropy}. In practice we often start with a Gaussian distribution, which is known to maximize the entropy under certain constraints, with wide variance to reflect the high degree of uncertainty in $\btheta$ and then decrease it along the training.
						
			In practice computing the full posterior distribution in high-dimensions is often intractable. For simplicity it is therefore of practical interest to reduce the problem to simple point-wise estimates such as the mean and the maximum of the posterior distribution $\rP\(\btheta|\vec{y}; \mat{X}\)$ corresponding to the so-called \aclink{MMSE} and \aclink{MAP} estimators.
		
			\paragraph{Minimum Mean Squared Error}
			\label{main:intro:ml:bayesian_approach:mmse}
				
				The \aclink{MMSE} estimator is simply defined as the mean of the posterior distribution
				\begin{align}
					\hat{\btheta}_{\mmse} = \EE_{\rP\(\btheta \vert \vec{y}; \mat{X}\)}\[\btheta\]\,,
				\end{align}
				and will be of central interest in the rest of the manuscript. 
				Indeed, ideally we hope to minimize the reconstruction error with the ground truth parameter $\btheta^\star$, \ie the  Squared Error (SE) 
				\begin{align}
					\textrm{SE}\(\btheta^\star, \hat{\btheta}\) = \frac{1}{\ndim} \|\btheta - \hat{\btheta}\|_2^2\,.
				\end{align}
				However, as very often the ground truth parameter $\btheta^\star$ is not accessible, we would simply require to minimize the error in average, \ie the \aclink{MSE} defined by
				\begin{align}
					\textrm{MSE}\(\hat{\btheta}\) =  \frac{1}{\ndim} \int_{\bbR^\ndim}  \d \btheta ~ \rp\(\btheta \vert \vec{y}; \mat{X} \) \|\btheta - \hat{\btheta}\|_2^2\,.
				\end{align}
				 Taking the derivative with respect to $\btheta$ directly yields the definition of the \aclink{MMSE} estimator $\hat{\btheta}_{\mmse}$ that therefore has the nice property to minimize the \aclink{MSE} reconstruction error. 
				 This estimator is very powerful but unfortunately very rarely tractable in practice as it requires to average over the high-dimensional posterior distribution $\rP\(\btheta \vert \vec{y}; \mat{X}\)$. 
				 An approach to compute this estimator would be to make use of \aclink{MCMC} algorithms to sample the posterior distribution. But in high-dimensions, sampling methods are very inefficient and require a huge number of samples.
				 As a spoiler, a main part of this work is concerned with computing this high-dimensional object with \emph{heuristic methods} from statistical physics. 
							
			\paragraph{Maximum A Posteriori}
			\label{main:intro:ml:bayesian_approach:map}
				The other simple point-wise estimate is taking the maximum of the posterior distribution and not the mean as for the \aclink{MMSE}. The \aclink{MAP} estimator is naturally defined as
				\begin{align*}
					\hat{\btheta}_{\map} &\equiv \argmax_{\btheta}  \log \rP\(\btheta|\vec{y}; \mat{X}\) \\
					&= \argmin_{\btheta} \left\{ - \log  \rP\(\vec{y}| \btheta, \mat{X}\)  - \log \rP\(\btheta\)   \right\}
				\end{align*}
				and can be turned into a minimization problem such as in \aclink{ERM}. Under this formulation, we notice easily that \aclink{MAP} Bayesian estimation with $\rP(\btheta)$ a priori information is strictly equivalent to \aclink{MLE} estimation in the presence of a regularizer $-\log \rP(\btheta)$ and has the advantage to provide a way to design complicated yet interpretable regularization terms.
				In comparison with the \aclink{MLE}, it has the advantage to leverage prior information not contained in the training data at the price to increase the bias.
		
		\subsection{Classical models}
		\label{sec:main:introduction:ml:models}
			In this section, we briefly present the main models and architectures mostly used in modern supervised \aclink{ML}, ranging from linear models to deep neural networks.
			
			\subsubsection{Generalized Linear Models}
			\label{main:introduction:glm_class}
			The simplest and wildest class of models used in many \aclink{ML} applications is \emph{linear models}. 
			To perform classification or regression linear models are very popular because of their simplicity. 
			However, to produce discrete outputs for instance, one often considers a wider hypothesis class known as 
				the \aclink{GLM}\index{generalized linear models} hypothesis class
				\begin{align*}
					\bbH_{\textrm{glm}} = \left\{
					f_{\vec{w}}:
						\begin{cases}
							\bbR^{d} \mapsto \bbR \\
							\vec{x} \mapsto \varphi_\out \( \vec{w}^\intercal \vec{x} + w_0\)\,,
						\end{cases}
					(\vec{w}, w_0) \in \bbR^{\ndim +1}
					\right\}
				\end{align*}
			It contains affine functions parametrized by a vector $\vec{w}\in\bbR^{\ndim}$ applied as a scalar product with the features $\vec{x}$, and a \emph{bias} or \emph{intercept} $w_0$. In addition, $\varphi_\out$ represents a deterministic or stochastic element-wise activation function, potentially non-linear, added on top of the linear operation. In other words, \aclink{GLM} are simple models based on a linear weighted sum of the features components shifted by a bias $w_0$. The parameter $\vec{w}=\{w_i\}_{i=1}^\ndim$ can be thought as the \emph{weights} associated at each sample features $\vec{x}_\mu = \{x_{i\mu}\}_{i=1}^d$.
			This affine operation is called a \emph{formal neuron}. Even though very simple, it is the elementary brick at the origin of more complex modern feed-forward \aclink{DNN}. 
			The decision boundary of linear models is essentially a high-dimensional \emph{hyper-plane} that splits \emph{linearly} the input space. For classification tasks, considering a sign output function $\varphi_\out(z) = \sign(z-K)$ or an Heaviside step function $\varphi_\out(z) = \Theta(K-z)$ refers to the historical \emph{perceptrons}\index{perceptron} with a stability threshold $K$.
			In particular, we will illustrate our statistical physics approach on this simple model class notably in \Sec\ref{main:sec:mean_field:replica_method:example_glm}, \ref{main:intro:mean_field:bp:example_glm}, \ref{main:sec:mean_field:amp:example_glm} and \ref{main:sec:mean_field:se_amp:example_glm}.

			\paragraph{Linear regression: pseudo inverse, ridge \& lasso}
				Consider we want to \emph{predict the output} $y \in \bbR$ of input vector $\vec{x}\in \bbR^{\ndim}$, we first consider a linear predictor that outputs $\hat{y} = \vec{w}^\intercal \vec{x} + w_0$. Taking  the \aclink{MSE} as our performance measure, we would like to minimize the generalization error on the test set $\bbD_\test$
				\begin{align*}
					\MSE_\test(\hat{\vec{w}}) = \EE_{(y,\vec{x}) \sim \bbD_\test} \(y - \hat{y}(\hat{\vec{w}}) \)^2\,.
				\end{align*}
				As as a surrogate, we minimize instead the empirical risk on the training set $\bbD_\train$. The goal is therefore to find an hyperplane that minimizes the sum of the squared errors between the observations $y$ and predictions $\hat{y}$.
				In this simple case, we can derive an explicit expression of the parameters $\hat{\vec{w}}$ that minimize the \aclink{MSE} on the training set. Taking the gradient of the empirical risk to $\vec{0}$, we easily obtain the \emph{pseudo-inverse} estimator, also called the \emph{normal equations}:
				\begin{align*}
					\nabla_\vec{w} \|\vec{y}_\train - \mat{X}_\train \hat{\vec{w}} \|_2^2 = \vec{0} \Rightarrow \hat{\vec{w}}^{\mathrm{pseudo}} = \(\mat{X}_\train^\intercal \mat{X}_\train\)^{-1} \mat{X}_\train^\intercal \vec{y}
				\end{align*}
				However, in practical applications of linear regression, the number of features $\ndim$ is often very large, and even larger than the number of samples $\ndim$, so that in this case the problem has an infinite number of solutions.
				
				To obtain a finite number of solutions, we often try to reduce it by selecting an appropriate set of \emph{features} that describe correctly the underlying distribution. A \emph{feature selection} method consists in projecting the data in a basis where the data are \emph{sparse}, see \cite{hastie2015statistical} for a comprehensive discussion. Nonetheless, the modern \emph{feature selection} approach is to use regularization that slowly pushes the effects of irrelevant features towards zero while keeping only interesting features, see \Sec\ref{sec:main:introduction:ml:regularization}.
				Regularized regression coincides equivalently to penalized models or shrinkage methods. Minimizing the regularized empirical risk \eqref{eq:main:introduction:ml:empirical_risk_regulqrized}, that can be thought as the trade-off between minimizing the squared error and having \emph{small} coefficients, constrains the initial hypothesis class $\bbH$ to particular solutions with small magnitude and fluctuations of the parameters. 
				 
				 In the case of linear regression with the squared loss, three main cases are widely considered: \emph{LASSO} \cite{tibshirani1996regression} with $\Omega(\btheta)=\|\btheta\|_1$, \emph{ridge regression} \cite{hoerl1970ridge} with $\Omega(\btheta)=\|\btheta\|_2^2$ and a combination of them called \emph{elastic net} \cite{zou2005regularization}.
											
\begin{figure}[htb!]
					\centering
					\hfill	
					\begin{tikzpicture}[scale=0.6, transform shape]
						\pgfplotsset{width=7cm,compat=1.8}
						\pgfmathsetseed{1138} 
					\begin{axis} [
					      xlabel     = $x_1$, ylabel     = $x_2$,
					      axis lines = left, axis line style = very thick,
					      clip       = false, xmin = -2.5,  xmax = 2.5, ymin = -2.5, ymax = 2.5, ticks = none
					    ]
						\pgfplotstableset{ 
    							create on use/x/.style={create col/expr={1.5+rand}},
    							create on use/y/.style={create col/expr={+1.5+rand}}
							}
						\pgfplotstablenew[columns={x,y}]{30}\loadedtable
						\addplot [only marks, blue, mark size=2pt] table {\loadedtable};
						\pgfplotstableset{ 
    							create on use/x/.style={create col/expr={-.5+rand}},
    							create on use/y/.style={create col/expr={-1.5+rand}}
							}
						\pgfplotstablenew[columns={x,y}]{30}\loadedtable
						\addplot [only marks, red, mark size=2pt] table {\loadedtable};
						\addplot [ domain=-1.5:2.5, samples=3, color=black, very thick]{-x+0.5};
						\addplot [ domain=-1.5:2.5, samples=3, color=black, dashed, very thick]{-x+1.4};
						\addplot [ domain=-1.5:2, samples=3, color=black, dashed, very thick]{-x-0.3};
						\draw[-latex, very thick] (axis cs:0,0.5) to (axis cs:0.35,1.05);
						\node[text=blue] at (axis cs:0,2) {$+1$};
						\node[text=red] at (axis cs:0.5,-2.2) {$-1$};
						\end{axis}
					\end{tikzpicture}		
					\hfill				
					\begin{tikzpicture}[scale=0.6, transform shape]
					\pgfplotsset{width=7cm,compat=1.8}
					\begin{axis} [
					      xlabel     = $x_1$, ylabel     = $x_2$,
					      axis lines = left, axis line style = very thick,
					      clip       = false, xmin = -4,  xmax = 4, ymin = -4, ymax = 4, ticks = none
					    ]
					\pgfplotstableread{Plots/data_circle_small.txt}\data;
					\addplot [ color=red, only marks, mark=*, mark size=2pt, ] table [x expr=\thisrowno{0}, y expr=\thisrowno{1} ] {\data};
					\pgfplotstableread{Plots/data_circle_large.txt}\data;
					\addplot [ color=blue, only marks, mark=*, mark size=2pt, ] table [x expr=\thisrowno{0}, y expr=\thisrowno{1} ] {\data};
					\node[blue] at (axis cs:3,3) {$+1$};
					\node[red] at (axis cs:1,-1) {$-1$};
					\end{axis}
					\end{tikzpicture}
					\hfill
					\begin{tikzpicture}[scale=0.6, transform shape]
					\pgfplotsset{width=7cm,compat=1.8}
					\begin{axis} [
					      xlabel     = $x_1 x_2$, ylabel     = $x_3$,
					      axis lines = left, axis line style = very thick,
					      clip       = false, xmin = -8,  xmax = 8, ymin = -10, ymax = 3, ticks = none
					    ]
					\pgfplotstableread{Plots/data_circle_small_proj.txt}\data;
					\addplot [ color=red, only marks, mark=*, mark size=2pt, ] table [x expr=\thisrowno{0}, y expr=\thisrowno{1} ] {\data};
					\pgfplotstableread{Plots/data_circle_large_proj.txt}\data;
					\addplot [ color=blue, only marks, mark=*, mark size=2pt, ] table [x expr=\thisrowno{0}, y expr=\thisrowno{1} ] {\data};
					\addplot[color=black, dashed, domain=-7.2:7.2, very thick]{1/4 * x^2 - 10};
					\node[blue] at (axis cs:-5,2) {$+1$};
					\node[red] at (axis cs:-3,-3.5) {$-1$};
					\draw[color=black, dashed, thick] (axis cs:0,-3) ellipse (50pt and 5pt);
					\end{axis}
					\end{tikzpicture}
					\caption{Illustration of a classification task for \Left a linearly separable dataset that can be classified with a large margin SVM, \Center and a non-linearly separable dataset that a generalized linear model cannot fit. \Right Projection of the non-linearly separable dataset into a higher dimensional space $(x_1,x_2) \mapsto (x_1,x_2,x_3)$.}
					\label{fig:linearly_nonlinearly}
					\end{figure}
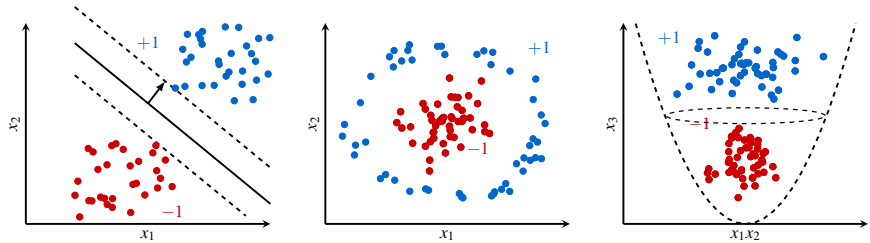

				\paragraph{Binary classification: perceptron \&  logistic}
					For a binary classification task such as represented in \Fig\ref{fig:linearly_nonlinearly} \Left, the decision boundary can be estimated by a linear hyperplane such that on each side of the decision boundary the labels are positive or negative. This setup is known as the classical \emph{perceptron} $y= \sign\( \vec{w}\cdot \vec{x} + w_0 \)$. To train this model and estimate the parameters $\{\vec{w},w_0\}$, the original perceptron algorithm \cite{rosenblatt1958perceptron} and many variant rules have been proposed.			
					The \emph{perceptron} model has been the subject of a rich statistical physics literature, see \cite{engel1993statistical} for a comprehensive review, and it will be discussed in \Sec\ref{chap:phys_ml_together}.
					Modern \aclink{ML} tasks are very often formulated as minimization problems of the empirical risk \eqref{eq:main:introduction:ml:empirical_risk}. Keeping our generalized linear model hypothesis class, we still have the choice of the loss function $\ell$.
					In the case of binary classification, let us mention the widely used \emph{logistic regression}\index{logistic regression} with $\ell^{\textrm{logistic}}(y, \hat{y}) = \log( 1 + \exp(-y \hat{y}) )$, which is equivalent to the binary cross entropy loss $\ell^{\textrm{bce}}(y, \hat{y}) = - y \log \hat{y} - (1-y) \log(1-\hat{y})$ with a sigmoid activation $\hat{y} = \sigma(\vec{w} \cdot \vec{x} + w_0)$. 
										
				\paragraph{Support Vector Machines and hinge loss}				
				In the case where the training examples are \emph{linearly-separable}, the \emph{perceptron}'s solution is ill-defined as there exists an infinite number of hyperplanes that classify correctly the training set. To select a robust solution, the idea of the influential \aclink{SVM} is to select the perceptron with the widest margin \cite{boser1992training, vapnik2013nature}. In the context of a binary classification task, in order to generalize as well as possible to variations of the dataset we should select the hyperplane that maximizes the distance to the nearest examples in the two classes, as illustrated in \Fig\ref{fig:linearly_nonlinearly} \Left.
					In more details, for $(\vec{w}, w_0) \in \bbR^{\ndim+1}$, we require that on the margins $y = \sign \( \vec{w} \cdot \vec{x} + w_0  \)  \Leftrightarrow 1 = y \( \vec{w} \cdot \vec{x} + w_0  \)$, so that the distance of the decision boundary to the margins $\vec{x}_\pm$ is $\vec{w}\cdot \vec{x}_\pm = 1 \mp w_0$ and the width of the margin equals $\gamma = \frac{2}{\|\vec{w}\|_2}$.  As a consequence, to maximize the margin $\gamma$ we may equivalently minimize a $\rL_2$ reguralization term $\frac{1}{2}\|\vec{w}\|_2^2$. To be more precise, the primal form reads 
					\begin{align*}
							\text{Minimize } \frac{1}{2}\|\vec{w}\|_2^2, \text{ under the constraints } y_{\mu} \( \vec{w} \cdot \vec{x}_{\mu} + w_0 \) \geq 1.
					\end{align*}
				The Karush-Kuhn Tucker conditions on the associated dual formulation lead to a well-defined and unique solution and finally reduces the hypothesis class. 
				Indeed while the \aclink{VC} dimension of the \aclink{GLM} hypothesis class $\bbH_{\textrm{glm}}$ is $\ndim+1$, for the \aclink{SVM} the margin constraint $\gamma$ shrinks it to $d_\vc= \min\( \frac{2 R^2}{\gamma}  , \ndim \) +1 $ that can be be much smaller than $\ndim+1$, with $R$ the radius of the smallest sphere comprising the training samples.
				Moreover, the primal problem may be formulated in a practical regularized version 
				\begin{align*}
					\hat{\vec{w}} = \argmin_{\vec{w}} \sum_{\mu=1}^{\nsamples} |1 - y_{\mu} f_\vec{w}(\vec{x}_{\mu}) | + \frac{\lambda}{2} \|\vec{w}\|_2^2\,,
				\end{align*}
				by minimizing the \emph{hinge loss} $\ell^{\textrm{hinge}}(y, \hat{y}) = \max\(0, 1 - y \hat{y} \)$ which is another common choice to perform binary classification.

				\paragraph{Limitations}
					Linear models such as linear regression or binary classification with perceptrons or \aclink{SVM} illustrated in \Fig\ref{fig:main:intro:classification_supervised} are very simple from a practical viewpoint. Unfortunately the low-capacity hypothesis class $\bbH_{\textrm{glm}}$ is extremely limited to simple tasks and dataset and cannot fit correctly more complex tasks. In particular, the XOR function or a more complex donut-like set of points such as in \Fig\ref{fig:linearly_nonlinearly} \Center.
					However, linear models do not allow to classify non-linearly separable points. 
					
				\paragraph{Kernel methods}
					To circumvent this issue, the very elegant idea of \emph{kernels methods}\index{kernels methods} is to change the representation of the input features $\mat{X}=\{\vec{x}_{\mu}\}_{\mu=1}^\nsamples$ by projecting them in a higher-dimensional latent space, in which data become evenetually linearly separable. For more details on kernel methods, see \cite{williams1996gaussian, scholkopf1999input}.
					Kernel methods rely on the \emph{kernel trick}  \cite{aizerman1964theoretical} based on the Mercer's theorem. It follows from the observation that the dot product between the parameters $\vec{w}$ and a feature vector $\vec{x}$ can be written as a linear decomposition with some coefficients $\{\theta_\mu\}_{\mu=1}^\nsamples$ so that
					\begin{align*}
						\vec{w} \cdot \vec{x} + w_0 =\theta_0 + \sum_{\mu=1}^{\nsamples} \theta_\mu \vec{x}^\intercal \vec{x}_{\mu}\,.
					\end{align*}
					Having this trick in mind, it has been extended to more complex \emph{kernels} $k: \bbX \times \bbX \mapsto \bbR$, where $\vec{x}^\intercal \vec{x}_{\mu}$ is replaced by a dot product in a high-dimensional space $k\(\vec{x}, \vec{x}_{\mu} \) = \phi\(\vec{x}\)^\intercal \phi\(\vec{x}_{\mu}\)$. By projecting the features in a new, possibly higher dimensional, space through the mapping $\phi$, we eventually transform the dataset in a linearly separable representation. Indeed, the main interest of kernel methods is that the new estimator parametrized by $\btheta$
					\begin{align*}
						f_{\btheta}(\vec{x}) = \theta_0 + \sum_{\mu=1}^{\nsamples} \theta_i k\(\vec{x}, \vec{x}_{\mu} \)
					\end{align*}
					is non-linear with respect to the examples $\vec{x}_\mu$, yet it is linear in the new features $\phi(\vec{x}_\mu)$. In other words, a kernel method is simply a linear model on pre-processing data in the space $\phi(\mat{X})$. By considering $\phi$ fixed, we only need to optimize over $\btheta$, similarly to linear regression except that the model is now non-linear and more expressive. In particular, \aclink{SVM} may be used in parallel of the kernel trick and are called \emph{kernel}-\aclink{SVM} in this context.
					We need to construct the $\nsamples \times \nsamples$ Gramm matrix $k_{\mu,\nu} = k\(\vec{x}_{\mu}, \vec{x}_{\nu}\)$ from the dot product of $\{\phi(\vec{x}_{\mu})\}_{\mu=1}^\nsamples$. This operation is computationally inefficient as $\Theta(\nsamples^2)$ and certainly hopeless for training sets containing millions of examples. 
					In practice the kernel $k$ is not computed but commonly taken among simple tractable forms such as the Gaussian, also called Radial Basis Function (RBF) kernel $k(\vec{a},\vec{b}) = \mN(\vec{a}-\vec{b},\sigma^2 \rI)$, or even polynomial, Laplace, or sigmoid kernels. 
					Recently, kernel methods started experience a decline in popularity with the advent of \aclink{DL} and \aclink{DNN} and especially when for the first time a neural network outperformed a Gaussian kernel \aclink{SVM} on MNIST \cite{hinton2006fast}.
				
			\subsubsection{Deep feed-forward neural networks}	
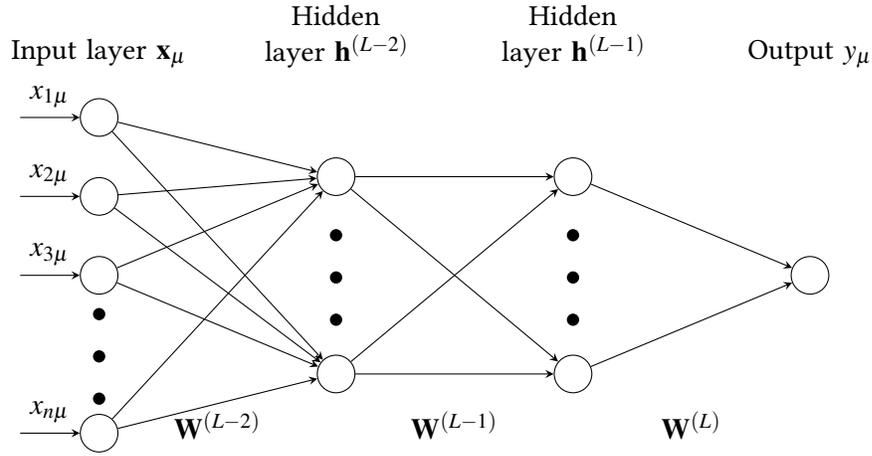
\begin{figure}[htb!]
				\tikzset{
				  every neuron/.style={
				    circle,
				    draw,
				    minimum size=0.5cm
				  },
				  neuron missing/.style={
				    draw=none, 
				    scale=4,
				    text height=0.333cm,
				    execute at begin node=\color{black}$\vdots$
				  },
				}

				\centering
				\begin{tikzpicture}[x=1.5cm, y=1.5cm, >=stealth, scale=0.7]
				
				\foreach \m/\l [count=\y] in {1,2,3,missing,4}
				  \node [every neuron/.try, neuron \m/.try] (input-\m) at (0,2.5-\y) {};
				
				\foreach \m [count=\y] in {1,missing,2}
				  \node [every neuron/.try, neuron \m/.try ] (hidden-\m) at (3,2-\y*1.25) {};

				\foreach \m [count=\y] in {1,missing,2}
				  \node [every neuron/.try, neuron \m/.try ] (hidden2-\m) at (6,2-\y*1.25) {};
				
				\foreach \m [count=\y] in {1}
				  \node [every neuron/.try, neuron \m/.try ] (output-\m) at (9,-0.5) {};
				
				\foreach \l [count=\i] in {1,2,3,n}
				  \draw [<-] (input-\i) -- ++(-1,0)
				    node [above, midway] {$x_{\l\mu}$};
					
				\foreach \i in {1,...,4}
				  \foreach \j in {1,...,2}
				    \draw [->] (input-\i) -- (hidden-\j);

				\foreach \i in {1,...,2}
				  \foreach \j in {1,...,2}
				    \draw [->] (hidden-\i) -- (hidden2-\j);
				    
				\foreach \i in {1,...,2}
				  \foreach \j in {1,...,1}
				    \draw [->] (hidden2-\i) -- (output-\j);
				    
				 \node [above] at (1.5, -2.7) {$\mat{W}^{(L-2)}$};
				 \node [above] at (4.5, -2.7) {$\mat{W}^{(L-1)}$};
				 \node [above] at (7.5, -2.7) {$\mat{W}^{(L)}$};
				  
				\foreach \l [count=\x from 0] in {Input layer $\vec{x}_{\mu}$, Hidden layer $\vec{h}^{(L-2)}$, Hidden layer $\vec{h}^{(L-1)}$, Output $y_{\mu}$}
				  \node [align=center, above, text width=3cm] at (\x*3,2) {\l};
				
				\end{tikzpicture}
				\caption{Representation of a deep feed-forward neural network with depth $L$. Each arrow represents a learnable scalar value and each hidden unit $h_i^{(l)}$ is the result of a formal neuron operation.}
				\label{fig:intro:deep_neural_net}
				\end{figure}

				In the recent years, the wide class of \aclink{DNN} models entered the scene \cite{LeCun15}. Just as the wings of plane are inspired by the wings of birds or many other biomimetics systems, \aclink{DNN} have been inspired by the brain mechanism to simulate \aclink{AI}. Yet the corresponding \aclink{DL} branch of research became far apart of the initial neuroscience field. 
				\aclink{ANN} and \aclink{DNN} are henceforth a class of models made of a cascade connection of simple elementary bricks based on the \emph{perceptron}, as illustrated in \Fig\ref{fig:intro:deep_neural_net}. Connecting several \emph{formal neurons} into complex networks define a richer hypothesis class with higher capacity.
				For instance a \emph{feed-forward}\index{feed-forward} \aclink{DNN} of \emph{depth} $L$ is made of $L$ \emph{hidden layers}\index{hidden layers} $\{\vec{h}^{(l)}\}_{l=1}^L$. Each hidden layer $\vec{h}^{(l)} = \{ h_i^{(l)} \}_{i=1}^{n_l}$ of \emph{width} $n_l$ is composed of $n_l$ \emph{hidden units}. This high-expressivity model is parametrized by a set of weights matrices and bias vectors $\btheta = \left\{ \mat{W}^{(l)}, \vec{b}^{(l)}, \forall l \in \lb L \rb \right\}$. The architecture can be expressed mathematically as the following input to output mapping $\mat{X} \mapsto \vec{y}$:
				 \begin{align*}
				 	\vec{y} = \sigma^{(L)}\left(\mat{W}^{(L)}\sigma^{(L-1)}\left(\cdots\sigma^{(1)}\left(\mat{W}^{(1)}\vec{X} + \vec{b}^{(1)} \right)+\cdots\right) + \vec{b}^{(L)}\right)\,.
				 \end{align*}
				A given layer $\vec{h}^{(l+1)}$ is the result of a linear product of a \emph{matrix of weights} $\mat{W}^{(l)}$ with the result of the previous layer $\vec{h}^{(l)}$ and adding a potential \emph{bias} $\vec{b}^{(l)}$. This is followed by a non-linear operation continuous \emph{activation function} $\sigma$ acting component-wise: $\vec{h}^{(l+1)} = \sigma^{(l)}\( \mat{W} \vec{h}^{(l)} + \vec{b}^{(l)}  \)$ with $\vec{h}^{(L+1)}\equiv y$ and $\vec{h}^{(0)} \equiv \vec{x}$.
				The corresponding very expressive hypothesis class $\bbH$ is largely modular through the architecture of the network, as we can easily tune the depth, width, and activation choice, and it is one of the reasons for its success.
				Especially the \emph{universal approximation theorem} \cite{Cybenko89, Hornik91} showed that a two-layer neural network with $L=2$ can approximate any smooth function. Yet, state-of-the-art \aclink{DNN} used nowadays are not limited to two layers and we observe an explosion of the numbers of layers to apply to various and more complex tasks. To illustrate, famous networks such as AlexNET contains 100 layers \cite{krizhevsky2012imagenet} so does a typical ResNET \cite{he2016deep}.
				Finally, to avoid the \emph{vanishing gradient problem} during the training, typically with a gradient-descent based algorithm, it is preferable to choose smooth activations functions with non-vanishing gradients such as the popular \aclink{ReLU} $\sigma(x) = \max(0,x)$ or, to a lesser extent, the hyperbolic tangent $\sigma(x)=\tanh(x)$.
				
				To conclude, the main advantages of \aclink{DNN} with respect to kernel methods are their expressivity and scalability to be trained on larger and larger datasets.
			
				\paragraph{A wide zoology of networks}
				Nevertheless \aclink{DNN} are not restricted to \emph{feed-forward} neural networks which are particularly suited to regression and classification. Depending on the task $\mT$ and the kind of data, we observed emergence of various kind of networks. 
				Notably, \aclink{CNN} are originally inspired from the biology and the visual cortex. By replacing the matrix product by a convolution product, they are particularly suited to processing arrays of numbers such as images in vision and pattern recognition \cite{lecun1998gradient, lecun1999object, krizhevsky2012imagenet}. In contrast with the classical knowledge-based methods where filters to process images are smartly designed by hand, the power of \aclink{CNN} lies on the fact that these filters are directly learned from data.
				In the context of speech recognition and \aclink{NLP}, to take into account the global meaning of the sentences and correlations between words, \emph{recurrent networks} \cite{rumelhart1986general} such as \aclink{LSTM} are quite popular since their high connectivity allows to simulate \emph{memory}.
		
		\subsection{Practical algorithms}
		\label{sec:main:introduction:ml:algorithms}
			To conclude the global overview of the \aclink{ML} machinery, it remains to address algorithmic questions to perform statistical estimation of the model parameters $\btheta$.
			In this manuscript, we essentially focus on two classes of algorithms depending if the \aclink{ML} estimator is formulated as an \emph{optimization} or an \emph{averaging} problem. 
			In one hand, many problems are formulated as minimizing an objective function or maximizing the likelihood with the data distribution, that can be handled by \emph{gradient-based} algorithms.
			In the other hand, estimators based on the average over certain high-dimensional distributions require either to \emph{sample} or \emph{approximate} it. 
			Notice that there exists other techniques such as \emph{constrained optimization} with Frank–Wolfe algorithms, that we will not discuss in this manuscript.
			
		\subsubsection{Gradient-based algorithms}
		\label{sec:main:introduction:ml:algorithms:gd}
		
		Most of \aclink{ML} algorithms involve the minimization of a certain \emph{smooth} and \emph{differentiable} \emph{objective} function with respect to the model parameters $\btheta$. 
		The most popular objective function is the \emph{negative log-likelihood} with respect to the training set $\bbD_\train$, namely the empirical risk  $\hat{\mR}\(\btheta; \bbD_\train \) = - \EE_{ \( \vec{x}, y \) \sim \hat{\rP}(\vec{x},y)  } \log \rP\( y | \btheta, \vec{x} \)$ so that common estimators such as  \aclink{MLE} and \aclink{MAP} can be formulated as 
			 \begin{align}
			 	\hat{\btheta} =  \argmin_{\btheta} \hat{\mR}\(\btheta; \bbD_\train \) + \lambda \Omega(\btheta) \,,
			 	\label{eq:main:introduction:ml:algorithms:gd:objective}
			 \end{align}
			 where the additional term $\lambda \Omega(\btheta)$ may be added for regularization, see \Sec\ref{sec:main:introduction:ml:regularization}. 
			 The common strategy to train such parametric estimators is to consider simple first-order \emph{gradient-based}\index{gradient-based} algorithms \cite{cauchy1847methode} widely popularized with practical applications in \cite{lecun1998gradient}.
			 Starting with some initial model parameters drawn randomly $\btheta^0 \sim \rP(\btheta^0)$, the underlying idea of a majority of training algorithms consists in performing a \emph{gradient-descent}\index{gradient-descent} on the empirical risk. Following the gradient-descent, the algorithm will certainly end up in a local minima, and eventually in a global one with good \emph{generalization properties}. Recall this is the main difference between \aclink{ML} and optimization: while the later allows any global minima, the first requires at least a local minima that predicts correctly unseen data, in other words: that generalizes.
			 This simple strategy is commonly known as \aclink{GD} defined by an update rule that computes in which way the weights $\btheta$ should be altered so that the proxy objective $\hat{\mR}$ function can reach a minima:
			 	\begin{align}
					\btheta^{t+1} = \btheta^{t} - \gamma^{t} \nabla_{\btheta} \hat{\mR}\(\btheta; \bbD_\train \) = \btheta^{t} - \frac{\gamma^{t}}{\nsamples} \sum_{\mu=1}^\nsamples \nabla_{\btheta} \ell\( y_{\mu}, f_{\btheta^t} \(\vec{x}_{\mu}\) \)  \,,
					\label{eq:main:GD:update}
				\end{align}
			where the hyper-parameter $\gamma^{t}$, called the \emph{learning rate}, controls the size of each decreasing gradient step and is usually fixed by performing line search on a validation set. 
			Notice that this idea may be generalized to second-order methods such as the Newton's method that makes use of the second derivative. However, they are very rarely used in practical applications since computing the Hessian matrix, remains inefficient and costly to compute for a large amount of high-dimensional data.
			\aclink{GD} parameters update rule \eqref{eq:main:GD:update} has the advantage to be easy to implement and to understand, and its trajectory can be analyzed rigorously as soon the objective function is \emph{convex}. Indeed in convex optimization \cite{boyd2004convex}, most of algorithms have convergence guarantees by making strong the assumption that the Hessian of the objective function is always positive semi-definite to ensure there is no saddle points and local minima. 
			
			\paragraph{Convergence}	
			In most practical \aclink{DL} applications, data distribution $\rP(\vec{x},y)$ is very complex and the high-dimensional model may contain millions of parameters. 
					\graffito{In the presence of local minima (red), \aclink{GD} is not guaranteed to converge to the global minima (green), as it depends on the initialization point.
					\begin{tikzpicture}[scale=0.37]
						\begin{axis}[
						    axis lines = left,
						    ylabel = {\Large$\hat{\mR}$},xlabel = {\Large$\btheta$},
    						xtick={},ytick={}, axis line style = very thick,ticks = none,
    						xmin=-1.5,xmax=4,ymin=-2,ymax=4,
						]
						\addplot [
						    domain=-1.5:4, samples=200, color=black,very thick]{1.5* x^2/2 - x^3 + x^4/4};
						\addplot[mark=*,color=red] coordinates {(0,0)};
						\addplot[mark=*,color=green] coordinates {(2.36603,-1.212)};
						\draw[-latex, black, very thick] (axis cs:-1.2,4) to (axis cs:-0.9,2);
						\end{axis}
						\end{tikzpicture}}		
				The corresponding optimization problem \eqref{eq:main:introduction:ml:algorithms:gd:objective} is very often \emph{non-convex} and thus \aclink{GD} algorithm lacks convergence guarantees. In other words, the optimization problem is not guaranteed to converge even to a local minima in a finite time, despite this fact, in practice it often delivers quickly parameters with low values of the objective and good generalization properties.

			 \paragraph{Variants and tricks}			 
			 Even though the \aclink{GD} algorithm is easy to understand and implement, it has the disadvantage to be possibly trapped in local minima and to be computationally inefficient on large datasets, since the full gradient has to be computed.
			 Many variants of this simple gradient algorithms, such as \aclink{SGD}, have been introduced \cite{Robbins07}, where the sum over the gradient of the full training set is replaced by the gradient over a single training example at a time. Thus \emph{stochastic} refers to the randomness in the examples selection at each time step. 
			 Very interestingly, it turned out empirically that this variant was able to find other regions of parameters than simple \aclink{GD}, with low test error and therefore good generalization properties.
			 Hence even though convergence is not guaranteed, these algorithms are strongly used in practice as moreover it solves the computational issue of storing in memory the gradient of the full dataset.
			 In between, \emph{mini-batch} \aclink{GD} is a good compromise and computes the gradient over small \emph{batches}, of size $\nsamples'$ with $ 1 \ll \nsamples' \ll \nsamples$, drawn uniformly from the training set and thus provides a more accurate estimate of the full gradient with some randomness. 
			  In the case where the dataset is redundant, this \emph{mini-batch} version has also the advantage to converge faster than \aclink{GD}, since it does not require to explore the whole dataset to capture the underlying distribution. 
			 In particular, the size $\nsamples'$ of the batch becomes another hyper-parameter we should tune on the validation set. 	
			 In practice the mini-batch size is typically around hundred while the full batch contains millions of examples. Thus \emph{full-batch} \aclink{GD} corresponds to the classical \aclink{GD} while \emph{1-mini-batch} \aclink{GD} refers to \aclink{SGD}. 
			 In particular, these algorithms are widely used because even for infinitely large training set $\nsamples \gg 1$, the \emph{complexity} of mini-batch \aclink{GD} remains $\Theta(1)$.\\
			 
			 As convergence is still not guaranteed, other tricks have been developed to help and accelerate finding minima and avoid oscillations such as adding \emph{momentum} \cite{sutskever2013importance} and \emph{Nesterov accelerated gradient}. See \cite{Goodfellow2016} for a detailed review. 
			 Also as fixing the learning rate may be tricky, new optimization variants with smart update learning-rate rules came to light, such as \emph{Adagrad} \cite{duchi2011adaptive}, \emph{AdaDelta} \cite{zeiler2012adadelta} or \emph{Adam} \cite{kingma2014adam}.
			 To conclude, many tricks and techniques on how training efficiently \aclink{DNN} are comprehensively described in \cite{Bottou10}. In particular, we may briefly mention that the initialization scheme $\rP(\btheta^0)$ seems to play a significant role as well as the batch normalization \cite{ioffe2015batch}, since these tricks suggest to serve as an \emph{inductive bias} and reduce adequately the effective hypothesis class of \aclink{DNN}. 
			 				
			\paragraph{Back-propagation}
			In contrast with kernel methods which training suffers datasets of large size, 
			\aclink{DNN} became very popular because of their scalability made possible thanks to a simple and robust training algorithm. The main difficulty in training a gradient-based algorithm according to the update \eqref{eq:main:GD:update} lies in \emph{computing the gradient} of this loss with respect to the parameters $\btheta$. This has been made possible by the crucial observation that the gradient of the objective \eqref{eq:main:introduction:ml:population_risk} with respect to the parameter $\btheta$ can be computed by the chain rule using simple algorithmic differentiation \cite{griewank1992achieving}. \aclink{DNN} can be trained efficiently, namely in linear time with the size of the network, by applying a simple chain-rule derivative, known as the \emph{back-propagation} algorithm \cite{rumelhart1986learning}. 
			
			In more details, by matrix multiplication, adding biases and applying non-linearities across the different layers, the \emph{forward-propagation} of the input $\vec{x}_{\mu}$ gives access to the predicted output $\hat{y}_\mu^t$ of the model and the loss $l(y_{\mu},\hat{y}_\mu^t)$ at time $t$.
			To fix ideas, consider a two-layer neural network, without bias, of the form
			\begin{align*}
				\vec{z}_1^t &= \mat{W}^{(1)}_t \vec{x}_{\mu}\,, && \vec{z}_2^t =  \mat{W}^{(2)}_t \sigma^{(1)}\(\vec{z}_1^t\)\,, &&\hat{y}_\mu^t = \sigma^{(2)}\( \vec{z}_2^t \)
			\end{align*}
			with parameters at time $t$, $\btheta^t=\{\mat{W}^{(2)}_t,\mat{W}^{(1)}_t\}$. Computing the gradients of the empirical risk \eqref{eq:main:introduction:ml:empirical_risk} with respect to the parameters $\mat{W}^{(2)}$, $\mat{W}^{(1)}$, for the squared loss $\ell(y,\hat{y})=\frac{1}{2}(y-\hat{y})^2$ is simply performed as a succession of linear operations
			\begin{align*}
				\partial_{\mat{W}^{(2)}} \ell(y_{\mu},\hat{y}_\mu^t) &= - (y_{\mu}-\hat{y}_\mu^t) \cdot \partial_{\vec{z}_2}\sigma^{(2)}(\vec{z}_2^t) \cdot \sigma^{(1)}(\vec{z}_1^t) \,,\\
				\partial_{\mat{W}^{(1)}} \ell(y_{\mu},\hat{y}_\mu^t) &=  - (y_{\mu}-\hat{y}_\mu^t) \cdot \partial_{\vec{z}_2}\sigma^{(2)}(\vec{z}_2^t) \cdot \mat{W}^{(2)}_t \partial_{\vec{z}_1}\sigma^{(1)}(\vec{z}_1^t) \cdot \vec{x}_{\mu}\,,
			\end{align*}
			which intermediate results are stored in a computational graph for numerical efficiency.
			\emph{Back-propagating} the derivatives over the whole \aclink{DNN} up to the input layer gives access to all the parameter updates \eqref{eq:main:GD:update} at time $t$. 	
			Linear in the size of the network and the number of data, it allows to scale the training procedure to very large networks.\\
			
			To conclude this section, performing \aclink{GD} at each time step, we often monitor the training error $e_\train$ until convergence. Then we compute the validation error at the end of the training to tune hyper-parameters such as $\gamma$, $n'$. Once the model and hyper-parameters are properly selected, we can finally compute the error on the test set as a surrogate of the generalization performances, see \Sec\ref{main:introduction:ml:train_val_test}.
		
		\subsubsection{Sampling and approximating}
			\label{main:intro:algos:sampling}
			As discussed in \Sec\ref{sec:main:introduction:ml:estimators}, Bayesian estimators such as the \aclink{MMSE} may be formulated instead as an \emph{average} over the posterior distribution $\hat{\btheta} = \EE_{\rP\(\btheta \vert \vec{y}; \mat{X}\)}\[\btheta\]$.
			The average can be done explicitly only in cases where the posterior distribution $\rP\(\btheta \vert \vec{y}; \mat{X}\)$ is explicit and tractable. Unfortunately, in high-dimensions computing it is very often intractable and we shall investigate alternative strategies such as \emph{sampling} or \emph{approximations}. Approximating high-dimensional joint probability distribution such as $\rP\(\btheta \vert \vec{y}; \mat{X}\)$ is the goal of the \emph{mean field methods} presented later on in \Sec\ref{main:sec:mean_fields}.
			Alternatively, among sampling methods, we shall briefly mention \emph{Gibbs sampling} performed with classical \aclink{MC} methods or more performant \aclink{MCMC} variants using notably \emph{importance sampling}. Their simple idea relies on the \aclink{CLT} that insures that the integral over the posterior can be approximated as
			\begin{align*}
				\hat{\btheta} = \int_{\bbR^\ndim} \d \btheta ~ \rp\(\btheta \vert \vec{y}; \mat{X}\) \simeq \frac{1}{\nsamples_{\textrm{mc}}} \sum_{\mu=1}^{\nsamples_{\textrm{mc}}} \btheta_\mu \text{ where }\btheta_\mu \sim \rP\(\btheta \vert \vec{y}; \mat{X}\)
			\end{align*}
			This kind of sampling methods is reviewed in details in \cite{andrieu2003introduction, rupert13, craiu14} and will not be at the heart of this manuscript, especially because they suffer slow convergence rate in very large dimensions and require \emph{acceleration} and \emph{variance reduction} to sample only useful regions of the high-dimensional probability distribution.

		\section{Challenges and open questions in deep learning}	
	\label{chap:review_ml:challenges}				
		\aclink{DL} was designed to overcome the insufficiency of traditional \aclink{ML} to learn complex high-dimensional functions or probability distributions. It impressively brought unprecedented empirical progresses \cite{LeCun15} into various \aclink{ML} applications such as in image, text and speech processing. These recent successes were made possible thanks to the availability of much larger datasets and greater computational resources.
		Yet, as it relies essentially on ingenious engineering tricks, it brought as well many unanswered fundamental questions that still remain open. Strikingly, the early questions raised in \cite{breiman1995reflections} are still of actuality. Because of this lack of theoretical foundations and guarantees, current practical \aclink{DL} is potentially sub-optimal in the sense the current \emph{brut force} approach takes advantage of the computational efficiency of oversized \aclink{DNN} trained on large dataset.
						
			\subsection{Curse of dimensionality and optimization}
				The empirical loss minimization problem \eqref{eq:main:introduction:ml:algorithms:gd:objective} becomes extremely difficult. With the explosion of the features size, this modern optimization problem lies in a high-dimensional space and was shown to be NP-complete \cite{blum1992training}. Indeed, as the problem dimensions increase, the number of configurations --- \ie the number of possible combinations of the different parameters --- is exponential and therefore much larger than the number of training examples in the training set. The \emph{curse of dimensionality}, faced by many computer science tasks in high-dimensions, refers to the statistical challenge to provide accurate predictions on large regions of parameters potentially not explored during the training.\\
				
				On the other hand, our geometric intuition of this high-dimensional space seems to hit a paradox and trivially concludes that a gradient-based algorithm will naturally fall and remain stuck in one of the many existing local minima, if no additional help is given. In most of the practical cases, this task is doable for a large, but finite number of samples. Yet this remains very inefficient since a human baby that would recognize images of dog and cat with a few pictures whereas current state-of-the-art \aclink{ML} models require millions of images. 
				Moreover, while convergence of gradient-based is guaranteed for quasi-convex loss functions, understanding why \aclink{GD} algorithms do not hit poor generalization local minima remains a burning open question.
				Indeed, even though minimizing highly-non convex losses is NP-hard, gradient-based algorithms such as \aclink{GD}, \aclink{SGD} or many other variants \cite{ruder2016overview} strikingly converge to regions of parameters with low generalization error and do not systematically lead to overfitting. Moreover even though many tricks are prescribed to help gradient-based algorithms to converge \cite{Bottou10, bottou2012stochastic}, building theoretical prescriptions is an active line of \aclink{ML} research.					

			\subsection{Generalization problem}
				Classical statistical generalization bounds such as the \aclink{VC} dimensions or the Rademacher complexity are in theory used to justify the learning ability of some \aclink{ML} models. 
				However, nowadays \aclink{DNN} contain millions of parameters and are so large that these classical worst-case statistical bounds became over-pessimistic and fail to predict \aclink{DNN} neural networks behavior. Indeed, as the number of parameters is larger than the number of examples \aclink{PAC} generalization bounds \cite{vapnik2013nature, bartlett2002rademacher} predict they should largely overfit and therefore cannot explain their good generalization behavior observed in practice. 
				Moreover, recent works showed that such traditional \aclink{PAC} bounds do not hold in \aclink{DL}, and should be refined. In particular the experimental work of \cite{zhang2016understanding} showed that \aclink{DNN} were able to simultaneously learn complex rules as well as fitting random labels. 
				Additionally, the traditional bias-variance trade-off to explain generalization performances is therefore obsolete and it is of actuality to understand why heavily parametrized high-capacity neural networks do not overfit the data \cite{neyshabur2017exploring,arora2018stronger}. 
				In fact empirical observations suggest that the optimization procedure induce a bias that reduces the effective dimension of neural networks, that can be captured by only a few order parameters. Highlighting them analytically is of course an intense line of research in the statistical learning community.

			\subsection{Expressive power, universality and architecture}
				Modern \aclink{ML} relying essentially on the \aclink{ANN} and \aclink{DNN}  provide a powerful hypothesis class $\bbH$ with large representation ability as stated by the strong universal approximation theorem \cite{Cybenko89,Hornik91}. However, this result for a two-layer \emph{shallow} network is not \emph{constructive} as it does not prescribe the \emph{width}, \ie the number of hidden units, or the \emph{sample complexity} $\alpha= \frac{\nsamples}{\ndim}$, with $\nsamples$ the number of training examples $\ndim$ the input dimension, to correctly approximate a given target function $f^\star$, neither the estimator or training algorithm $\mA$ to obtain model parameters $\btheta$ with good generalization properties. 
				Also, increasing the depth was known for a long time \cite{minsky2017perceptrons} as a solution to overcome simple perceptron limitations, the intuition that depth provides a natural hierarchal framework to learn different scales and representations across layers was recently advocated \cite{bengio2013representation} as well as the analogy with physics renormalization group \cite{mehta2014exact}. Such intuition as well as theoretical principles on how to choose model-parameters such as the loss, activations, number of layers, sample complexity or hyper-parameters are fragile and the current understanding remains mainly empirical.
				On the unsupervised learning counterpart, even though \aclink{VAE} and \aclink{GAN} showed their impressive ability to produce realistic images, measuring the performance of the generative models by knowing in particular if they provide correct approximations of the true data distribution is an important ongoing line of research \cite{arora2018gans}. Mostly based on \aclink{DNN}, they naturally inherit of the theoretical challenges concerning their architecture, computational cost and training procedure in the supervised setting.

			\subsection{Opening towards statistical physics}
				To conclude, the successes of \aclink{DL} rely essentially on both the type of structured data and substantial biases of gradient-based algorithms, that allow to reduce the hypothesis class and select an estimator with good generalization abilities. 
				As presented in the next sections, statistical physics has a long history with the theory of \aclink{ML} and we believe that powerful statistical physics tools have a role to play in disentangling the joint roles of the data structure, training algorithm and the network architecture. 
				Moreover, since the classical, overly pessimistic, \emph{worst-case} analysis fails to capture high-dimensional generalization behavior of \aclink{DNN}, the \emph{typical analysis} handled by \emph{statistical mechanics} seems to be a fruitfully alternative approach. 
				Indeed, as usual in physics, by dealing with simple architectures and synthetic data, statistical physics tries to highlight universal properties that will potentially hold in general and moreover for a practical usage. In this perspective, we will consider the simplest theoretical case of \emph{supervised learning} with \emph{feed-forward} \emph{shallow networks}. This much simplified set-up for deep learning with a few hidden-layer, without convolutions, pooling, batch-normalization, etc., is already complex to understand and is believed to already capture some of the core difficulties. 
				However, so far the simplicity of the models under consideration by the \emph{statistical physics} community is still far away of being realistic to provide direct and practical guidances of the size, architecture, optimization procedure or sample complexity.\\
		
				In the perspective of handling theoretically simple \aclink{ML} models within this statistical physics framework, we present a general introduction to it in \Chap\ref{chap:statistical_physics} and see how it may help building theoretical foundations of \aclink{DL} in \Chap\ref{chap:phys_ml_together}.

\ifthenelse{\equal{\format}{oneside}}
	{
	\clearpage\null\thispagestyle{empty}
	\clearpage\null\thispagestyle{empty}
	}
	{\clearpage\null\thispagestyle{empty}
	\clearpage\null\thispagestyle{empty}}	
\chapter{An overview of statistical physics and phase transitions}
\label{chap:statistical_physics}
	In this chapter, we introduce the basic tools and concepts of statistical physics that we will use all along this manuscript. In particular, we advocate that statistical physics is a very powerful framework to describe phase transitions appearing in systems composed of a large number of interacting particles.
	In \Chap\ref{sec:main:introduction:stat_phys:motivation}, we introduce the unfamiliar reader to the fundamental concepts of statistical mechanics.
	 \Chap\ref{sec:main:introduction:stat_phys:describing_behavior} and \Chap\ref{sec:main:introduction:stat_phys:disordered} are respectively devoted to describe the set of mathematical tools of statistical mechanics applied to \emph{ordered} and \emph{disordered} systems.

	\section{Why statistical physics matters?}
	\label{sec:main:introduction:stat_phys:motivation}
		This manuscript aims to analyze simple \emph{machine learning models} presented in \Sec\ref{chap:review_ml:basics} through the singular lens of statistical physics. Even though at the first glance it appears unnatural, in this section we advocate that \emph{statistical physics} is a generic framework that applies to various fields outside of pure physics such as computer science or mathematical problems.
			
			Statistical physics is a branch of physics introduced in the 19th century by Maxwell, Boltzmann and Gibbs, whose objective is to understand the collective behavior that emerges from a system built of \emph{many particles} in interaction. Very powerfully, statistical physics directly applies to various fields, starting with phenomenon observed in everyday life. For instance, without it we could not understand the description of the phase transition between solids, liquids and gas, or the difference of behaviors between metals and insulators, nor even we could not understand supraconductivity, or fermionic and bosons quantum systems \cite{balian1986physique,georges2004introduction}.
			But its application range is much wider and goes beyond natural fields. In particular, statistical physics has been successfully applied to social sciences with the Schelling's model \cite{gauvin2009phase},
			information theory with error correcting codes \cite{mezard2009information}, percolation, combinatorial optimization problem \cite{krzakala2007gibbs}, avalanches in
			financial and economy modelization \cite{Mantegna00, bouchaud_potters_2003, voit2013statistical}, 
			as well as simple machine learning models such as perceptrons \cite{Opper1991, engel1993statistical}, and many other systems.
			For a more detailed introduction to statistical physics, please refer to \cite{diu1989elements, sethna2006statistical,kardar2007statistical,ma2018modern}. 
			
		\subsection{From microscopic to macroscopic scales}
			While historically physics strategy focused on describing macroscopic systems by ignoring the precise microscopic details,
			statistical mechanics is a \emph{reductionist} and \emph{statistical} description that deduces the macroscopic properties of a system from the microscopic interactions and laws which govern the behavior of its elementary constituents at smaller scales. 
			At the heart of statistical mechanics, the transition from microscopic to macroscopic scales does not offer a completely new description of the nature, but instead adapt existing tools to describe the macroscopic behavior of systems composed of an extensive number of particles.
			For instance the atmospheric pressure results from collisions between microscopic molecules and statistical mechanics provides a microscopic justification of the laws of thermodynamics that govern macroscopic quantities such as the pressure, temperature and volume.
		
		\subsection{Lagrangian mechanics versus probabilities}
			For the sake of illustration, let us consider a simple glass of water that contains typically $10^{23}$ molecules of water. 
			The large number of particles and degrees of freedom makes the corresponding configuration space so large that tracking over time the positions and speeds of each molecule, which undergoes potentially a huge number of events and interactions, is intractable in practice because of memory usage and precision.
			Thus the classical Lagrangian mechanics cannot be applied directly to properly describe the behavior of such a simple yet large system. 
			%
			Instead of describing in full details the microscopic states of the system at a given time, statistical mechanics takes advantage of a \emph{probabilistic approach} to only quantify the probability of observing the system in a given microscopic configuration during its evolution.
			 		 			
		\subsection{Interactions and collective behavior}
			The emergence of unexpected spectacular \emph{collective behaviors} arises in fact from the interaction of a \emph{very large number} $\ndim$ of particles, that cannot be imagined from the microscopic laws o just a few particles.
			As an illustration increasing the pressure of our glass of water, at a certain \emph{critical} threshold it will undergo a solidification \emph{phase transition} and becomes solid. This common phenomena cannot be explained theoretically without invoking a \emph{sudden change in microscopic interactions} between molecules.
			These observations are well summarized by the famous formulation from P. W. Anderson \cite{Anderson393}
			\say{\emph{More is different}}, that stresses the idea that macroscopic behavior cannot be fully described as the sum of non-interacting agents. In other words, the whole system cannot be thought as the simple sum of its components, and interactions play a fundamental role in the macroscopic behavior. 
			Statistical physics aims to analyze the behavior of such macroscopic system and predict the arising critical \emph{phase transition}\index{phase transition}, such as the classical liquid-gas-solid, para-ferromagnetic or metal-insulator phase transitions. Other spectacular collective behaviors can be observed beyond classical physics systems such as in finance, economy and social sciences in which strongly interacting agents may lead to collective phenomena and rare events such as crashes, reactions of panic and stampedes.

		\subsection{Thermodynamic limit and concentration}
			As in analytical mechanics, analyzing a very large number $\ndim$ of particles is often intractable.  
			Statistical mechanics makes use of this large size system to describe it in the theoretical infinite size limit, the so-called \emph{thermodynamic limit} $\ndim \to \infty$. 
			This \emph{thermodynamic} limit is a favorable and powerful tool as the behavior of the system becomes asymptotically and surprisingly deterministic! 
			Indeed if we assume that particles are \aclink{i.i.d}, the \aclink{CLT} ensures that the \emph{equilibrium probability distribution} of the system concentrates around its most probable value with fluctuations that \emph{decrease} with the size of the system in $\Theta(\ndim^{-1/2})$. 
			Finally even though in practice particles are very often not \aclink{i.i.d} the behavior of real systems will still be typically given by the thermodynamic limit with some finite-size fluctuations, so that describing the behavior in the thermodynamic limit plays a role of the utmost importance in statistical mechanics. 
			
		\section{Describing the system behavior}
		\label{sec:main:introduction:stat_phys:describing_behavior}
			Throughout the manuscript, we consider a set of $\ndim$ interacting particles denoted by a vector $\bsigma = (\sigma_i)_{i=1}^\ndim \in \chi_\ndim$ which lies in the \emph{configuration space} $\chi_\ndim$. $\forall i \in \lb \ndim \rb, \sigma_i$ belongs to an alphabet $\chi$ that represents the degrees of freedom of each spin, and it can be discrete (\eg $\chi_\ndim=\{\pm 1\}^\ndim$) or continuous (\eg $\chi_\ndim= \bbR^\ndim $). The vector $\bsigma$ may represent different physical systems such as particles, magnetic spins, pixels, model parameters, etc., depending on the scope of application among image processing, information theory, computer science, physics, biology, error correcting codes or \aclink{ML}. As physicists we will generally call $\bsigma$ a vector of \emph{spins}\index{spin} for historical reasons and its values a \emph{configuration} that refers to a given realization.
				\begin{figure}[htb!]
				\centering
					\begin{tikzpicture}
						\pgfmathsetseed{2}
						\foreach \i in {1,...,5}
		       				\pgfmathsetmacro\Angle{rand*30}
		    				\draw[-latex,very thick, rotate around={\Angle:(2*\i,0)}] (2*\i,-0.85) -- (2*\i,0.85);
		    				
		    			\foreach \i in {1,...,5}
		    				\node[circle, draw, very thick, minimum size=12pt, inner sep=0pt, fill=burntorange] at (2*\i,0) {$\sigma$\i};
		    			\foreach \i in {1,...,5}
		    				\node at (2*\i,0.5) {\AxisRotator[rotate=90]};
					\end{tikzpicture}
					\caption{System of magnetic moments that can interact with their neighbors.}
					\label{fig:main:magnetic_spins}
				\end{figure}
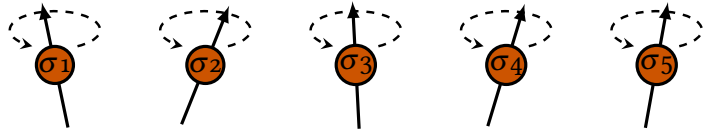
				For the purpose of the illustration, we may imagine $\ndim$ magnetic moments, also called spins, that can precess around a vertical axis so that its extremity evolves on a sphere $\bbS^2 \subseteq \bbR^3$ and interact with the neighboring spins by magnetic interactions. Each spin configuration $\sigma_i \in \chi = \bbS^2$ represents the precession degrees of freedom, as illustrated in \Fig\ref{fig:main:magnetic_spins}.
		
		\subsection{Graphical models and free entropy}
		\label{sec:main:intro:stat_phys:jpd}
				
				\subsubsection{Joint probability distribution}
				As stressed in the introduction \Sec\ref{sec:main:introduction:stat_phys:motivation}, the probabilistic description of the system is inevitable to analyze collective behaviors of systems with many particles. The behavior of the interacting \aclink{RV} $\bsigma \in \chi_\ndim$ is therefore modelled by a \aclink{JPD} 				
						\begin{align}
								\rP_\ndim \(\bsigma\) \equiv \rP_\ndim \(\sigma_1, \cdots, \sigma_\ndim\)\,,
								\label{eq:main:intro:mean_field:jpd}
						\end{align}
					that may hardly be tractable in large dimensions. The ultimate goal is to compute the marginals distributions $\rP(\sigma_i) = \int_{\chi_{d-1}} \d \bsigma_{\setminus i}\rP_\ndim \(\bsigma\) $, \ie the behavior of a single spin variable, by integration over the configurations of the other $\ndim -1$ spins, denoted $\bsigma_{\setminus i}$. Analyzing the \aclink{JPD} \eqref{eq:main:intro:mean_field:jpd} is a complex task at the heart of this manuscript. Yet very often computationally hard, the computation of the marginals rely in most of the cases on mean-field approximations presented in details in \Sec\ref{main:chap:mean_field}.
					However, in the simplest case of \emph{non-interacting} spins, the \aclink{JPD} is degenerated and trivially given by the product of the marginal probabilities $\rP_\ndim \(\bsigma\) \equiv \prod_{i=1}^\ndim \rP\(\sigma_i\)$. 
					Even though the latter factorized decomposition is very useful for approximations, see \Sec\ref{sec:main:intro:variational:naive}, the existence of such non-interacting systems is idealist and essentially pedagogical. Therefore the study of non-interacting variables is instructive but very limited in practice.
					Indeed, interesting and more realistic behaviors mostly appear with the existence of complex interactions between the particles encoded in the \aclink{JPD}.

				\subsubsection{Graphical models}
				\label{sec:main:intro:mean_field:graphical_models}
				However, representing such complex interacting system is difficult, especially in the high-dimensional regime $\ndim \to \infty$, called the \emph{thermodynamic limit}\index{thermodynamic limit}. Indeed in this limit, the main difficulty lies in describing all the interactions between each spin $\sigma_i,\forall i\in\lb \ndim \rb$.
				Therefore, we need a practical way of representing the joint distribution $\rP_\ndim(\bsigma)$ \eq\eqref{eq:main:intro:mean_field:jpd}. 
				To this extent, we introduce \emph{graphical models}\index{graphical model} that give a very generic, intuitive and powerful way to think of probabilistic models for finite or infinite size systems.
				Indeed, interacting spins can be represented conveniently by a \emph{graph}\index{graph} $\mG\(\rV, \rE\)$, either \emph{directed} or \emph{undirected}, composed of 
					\graffito{			
					The graph $\mG\(\rV, \rE\)$ contains 6 vertices  $h, i, j, k, l, m  \in \rV$ and 7 edges $(ij)$, $(ik)$, $(il)$, $(im)$, $(jl)$, $(km)$, $(hm)$ $\in \rE$.
					\tikzstyle{var}=[circle, draw, very thick, minimum size=12pt, inner sep=0pt, fill=burntorange]
					\tikzstyle{edge}=[draw, thick, -]
					\begin{tikzpicture}[scale=0.75, transform shape]
					\centering
					    \foreach \pos/\name/\mathname in { {(0.5,1.5)/a/$ j $}, 
					                                        {(2,1)/b/$ i $}, 
					                                        {(4,1)/c/$ k $},
					                                        {(0,0)/d/$ l $}, 
					                                        {(3,0)/e/$ m $}, 
					                                        {(2,-1)/g/$ h $}}
					        \node[var] (\name) at \pos {\mathname};
					    \foreach \source/ \dest in {b/a,c/b,d/a,d/b,e/b,e/c,g/e}
					        \path[edge] (\source) -- (\dest);
					\end{tikzpicture}
					}	
					\begin{itemize}
						\item a set of nodes $\rV$, also called vertices, that represent the spin configuration $\bsigma = \{\bsigma_i\}_{i=1}^\ndim$, so that $\abs{\rV}=\ndim$ \,,
						\item a set of edges $\rE$ that connects the nodes with $\abs{E}=\nsamples$. The edges represent the statistical dependencies, \ie the interactions, between the random variables $\bsigma$. Directed graphs refers to directed interactions of the form $(i\to j)$, while undirected graphs deal with undirected pairs of vertices $(ij)$.
					\end{itemize}	
					We distinguish \emph{directed} graphical models called \aclink{DAG}\index{DAG} from \emph{undirected} graphical models known as \aclink{MRF}. We will mainly focus on the latter ones in the following and especially in \Sec\ref{sec:main:mean_field:general_markov_fields}.
					To describe more formally the geometry of undirected graphs, we often introduce the \emph{adjacency matrix}\index{adjacency matrix} $\mat{A}$ of size $ \ndim \times \nsamples $ with binary entries such that $ a_{ij} = \id\[(ij) \in \rE \]$. It is in particular useful to compute the \emph{connection degree}\index{connection degree}, \ie the size of set of neighbors of a node $i$, denoted  $\partial_i$
					\begin{equation*}
					    \abs{\partial_i} = \sum_j a_{ij} = \sum_j \id\[(ij) \in \rE \]\,.
					\end{equation*}
					For the sake of the illustration, in the case of independent \aclink{RV}, the \aclink{JPD} factorizes and the set of edges of the corresponding graph $\mG$ reduces to an empty set $\rE = \emptyset$, so that $\forall i \in \rV,  \abs{\partial_i} = 0$.
					
					This simple graphical formulation gives a convenient and geometrical representation to encode the conditional dependencies of a large number of interacting variables $\bsigma$. 
					Interestingly, it will naturally lead to the design of powerful dynamical equations such as the \emph{cavity method}\index{cavity method} and \emph{belief propagation}\index{belief propagation} discussed in \Sec\ref{main:chap:mean_field}. Notice that even though graphical models may be used for finite size systems, their crucial power lies in their ability to represent as well high-dimensional probability distributions.
					The interested reader may find a comprehensive introduction with more details about graphical models in \cite{Yedidia2001, mackay2003information, jordan2004graphical,wainwright2008graphical,koller2009probabilistic}	
						
				\subsubsection{General Markov random fields}
				\label{sec:main:mean_field:general_markov_fields}	
				On non-regular graphs $\mG$ with arbitrary connectivity, counting and describing properly the graph $\mG\(\rV, \rE\)$ associated with the \aclink{JPD} might be tricky. Fortunately, it is often the case that the \aclink{RV} present a certain structure and independence properties.
				For this reason, we introduce the notion of \emph{clique}\index{clique}, defined as a subset $\rC \subseteq \rV$ of fully connected nodes. Indeed the Hammersley and Clifford theorem \cite{hammersley1971markov} insures that if the global independency Markov property
				\begin{align*}
				\forall~\mathrm{V}_1,\mathrm{V}_2,\mathrm{V}_3 \subset \mathrm{V},~~ \bbP\( \bsigma_{\mathrm{V}_1 \cup \mathrm{V}_2 } \vert \bsigma_{\mathrm{V}_3}\) =  \bbP\( \bsigma_{\mathrm{V}_1} \vert \bsigma_{\mathrm{V}_3}\)  \bbP\( \bsigma_{\mathrm{V}_2} \vert \bsigma_{\mathrm{V}_3}\)\,,
				\end{align*}				
				 is verified, the \aclink{JPD} may be decomposed as a general compact \aclink{MRF}\begin{align}
					\rP_\ndim \(\bsigma\) &= \frac{1}{\mZ_\ndim} \prod_{c\in \rC} \Psi_c\(\bsigma_{c}\) = \frac{1}{\mZ_\ndim} \prod_{i=1}^\ndim \phi_i\(\sigma_i\) \prod_{(ij)} \Psi_{ij}\(\sigma_i,\sigma_j\) \prod_{(ijk)} \cdots \,,
					\label{main:intro:mean_field:MRF}
				\end{align}
				where we introduced some \emph{potential functions} corresponding to cliques with different sizes $\{\Psi_c\}_{c \in \rC}$ , see \cite{Yedidia2001,jordan2004graphical}. Partitioning over the sizes reveals successive contributions of many-body interactions. For instance, $\{\phi_i\}_i$ represent the one-spin interactions and $\{\Psi_{ij}\}_{i\ne j}$ the two-spin interactions, etc.
						
				\subsubsection{Factor graph representation}		
				The general \aclink{MRF} formulation \eqref{main:intro:mean_field:MRF} remains quite cumbersome as the size of the cliques may be very large and involve a large number of spins. 
				Therefore, to obtain a more compact representation of the \aclink{JPD} $\rP_\ndim$ that highlights the conditional dependencies between \aclink{RV}, it is helpful to replace them with $\nsamples$ \emph{factors}\index{factor} or \emph{constraints}\index{constraints} $\{\Psi_\mu(\bsigma_{\partial_\mu}): \mu \in \lb \nsamples \rb\}$ that are already factorized, where $\partial_\mu$ denotes the subset of neighboring nodes of the factor $\mu$. 
				\graffito{For clarity, local fields or equivalently one-body interaction are described by leaf factors and drawn in yellow
				\begin{center}
				\begin{tikzpicture}
					\draw[edge] (0,0) -- (1,0); 
					\node[field] at (0,0) {$\phi_i$};
					\node[var] at (1,0) {$\sigma_i$};
				\end{tikzpicture}
				\end{center}
				and many-body interactions are represented with green constraint factors
				\begin{center}
				\begin{tikzpicture}
					\node[var] (s1) at (0,0) {$\sigma_i$};
					\node[var] (s2) at (2,0) {$\sigma_j$};
					\node[var] (s3) at (1,1.5) {$\sigma_k$};
					\node[inter] (f1) at (1,0.5) {$\Psi_{\mu}$};
					\draw[edge] (s1) -- (f1); 
					\draw[edge] (s2) -- (f1); 
					\draw[edge] (s3) -- (f1); 
				\end{tikzpicture}
				\end{center}
				}
				The \aclink{JPD} can therefore be written in full generality as the product over all possible factors
				\begin{align}
					\rP_\ndim \(\bsigma\) = \frac{1}{\mZ_\ndim} 
							\prod_{ i =1 }^\ndim \phi_i\(\sigma_i\) 
							\prod_{ \mu =1 }^\nsamples \Psi_\mu\(\bsigma_{\partial_\mu}\)  \,,
					\label{main:intro:graphical:factor}
				\end{align}
				where $\mZ_\ndim = \sum_{\bsigma} \prod_{ i =1 }^\ndim \phi_i\(\sigma_i\)  \prod_{\mu=1}^\nsamples \Psi_\mu(\bsigma_{\partial_\mu})$ represents a normalizing constant. The above factorization ends up with a simple \emph{bipartite factor graph representation} $\mG = \( \rV , \rF , \rE \)$ of the \aclink{JPD} composed of \emph{variable nodes} $\sigma \in \rV$ represented by circles and \emph{factor nodes} $\phi, \Psi \in \rF$ represented with squares, connected with edges $\rE$, as illustrated in \Fig\ref{fig:main:factor_graph}. In particular, each \emph{non-negative} factor $\Psi_\mu$ is connected to neighboring variables $\bsigma_{\partial_\mu} = \{ \sigma: \sigma \in \partial_\mu \}$.
					
				The factor graph formalism provides a powerful and very convenient representation of the \aclink{JPD} that gained a lot of interest in various fields such as constraint satisfaction and combinatorial optimization, error-correcting codes, bioinformatics, language and speech processing, image processing and spatial statistics. See a review of a wide range of applications in \cite{wainwright2008graphical, koller2009probabilistic}. 
					
				\subsubsection{Connection with physics: Hamiltonian and the Gibbs measure}
				The connection between physics and the factor graph formalism can be made explicit if we exponentiate the above formulation \eqref{main:intro:graphical:factor} 		
				to introduce the fundamental \emph{Hamiltonian}\index{Hamiltonian} energy $\mH_\ndim$ that measures the interaction energies of the local fields $\log \phi_i\(\sigma_i\)$ applied point-wise on each spin $\sigma_i, \forall i \in \lb \ndim \rb$ and the $\nsamples$ constraints $\log \Psi_\mu\(\bsigma_{\partial_\mu}\)$ that the spin configuration $\bsigma$ shall satisfy
				\begin{align}
					\mH_\ndim\(\bsigma\) &= - \sum_{ i =1 }^\ndim \log \phi_i\(\sigma_i\) -
							\sum_{\mu =1 }^\nsamples \log \Psi_\mu\(\bsigma_{\partial_\mu}\)	\,.
					\label{main:intro:stat_phys:hamiltonian}
				\end{align}
				The Hamiltonian of the system describes the microscopic interactions between spin variables $\bsigma = \{\sigma_i\}_{i=1}^\ndim$ so that the corresponding energy $\mH_{\ndim}\( \bsigma \)$ measures the probability of each configuration $\bsigma$ according to the \emph{Gibbs distribution}, also called the \emph{Boltzmann} distribution
				\begin{align}
					\rP_{\ndim} (\bsigma; \beta) \equiv \frac{e^{-\beta \mH_{\ndim}(\bsigma)}}{\mZ_\ndim(\beta)}  \,,
					\label{main:intro:classical_physics:gibbs_distribution}
				\end{align}	
				where we introduced a parameter $\beta$, called the \emph{inverse temperature}, that allows to explore all energy levels above the ground state energy. The inverse temperature will be taken to $\beta=1$ in the most considered cases unless mentioned otherwise. In this case, the Gibbs distribution \eqref{main:intro:classical_physics:gibbs_distribution} is equivalent to the \aclink{JPD} formulation in \eqref{main:intro:graphical:factor}.
				The Gibbs distribution is the central equilibrium measure in statistical physics and its exponential form can be justified by the \emph{maximum entropy principle} detailed in \Sec\ref{sec:main:mean_field:maximum_entropy}. 
				Notice that by construction, the most probable configuration is the one that achieves the smallest Hamiltonian energy \eqref{main:intro:stat_phys:hamiltonian}. It is called the \emph{ground state} configuration and is associated to a \emph{ground state energy}.
				Moreover, notice that we introduced the normalizing constant at inverse temperature $\beta$ of the random measure $\d \rP_{\ndim}$, called the \emph{partition function}\index{partition function}. Indeed imposing the normalization $\int_{\chi_\ndim}  \d \rP_{\ndim} (\bsigma; \beta) = 1 $, the partition function is naturally given by the \emph{sum over all the possible configurations} weighted by their Gibbs weights probability $e^{-\beta \mH_{\ndim}(\bsigma)}$:
				\begin{align}
					\mZ_\ndim\(\beta\) &\equiv \int_{\chi_\ndim}  \d \bsigma ~ e^{-\beta \mH_{\ndim}(\bsigma) } \,.
					\label{main:intro:classical_physics:partition_function}
				\end{align}
				The partition function is a crucial quantity in statistical mechanics as it contains the important informations on the equilibrium distribution of all possible spin configurations of the system. Indeed $\mZ_\ndim\(\beta\)$ is known to be the \emph{moment generating function}, because successive derivatives give access to the moments of the Gibbs measure.
				The \emph{Gibbs average}\index{Gibbs average} over the Gibbs measure \eqref{main:intro:classical_physics:gibbs_distribution} is traditionaly denoted $\langle . \rangle_{\beta}$, and we may also use the notation $\EE_{\bsigma \sim \rP_\ndim}$.
				
			\subsubsection{The free entropy as a cumulant generating function}
		\label{sec:main:introduction:stat_phys:ordered_definitions:generating_function}
				Because the \aclink{JPD} $\rP_\ndim$ of the spin $\bsigma$ becomes exponentially peaked in regions of most probable configurations that dominate the whole distribution, we are only interested in its \emph{large deviation} behavior. Thus, taking the logarithm of the partition function refers and defines the \emph{free entropy} in information theory and statistical physics.
				
				\paragraph{Free entropy and energy}	
				We define respectively the \emph{free entropy} $\Phi_\ndim$
				\graffito{We may choose as well the free energy as historically in statistical physics. Notice that the literature is sometimes confusing and clumsy on the naming of these quantities.}
				and free energy $\varphi_\ndim$ of a system of size $d$ at inverse temperature $\beta$ by:
				\begin{align}
					\Phi_\ndim (\beta) &\equiv \frac{1}{\ndim } \log \mZ_\ndim  (\beta) \,, 
					&& \varphi_\ndim (\beta) \equiv -\frac{1}{\ndim \beta} \log \mZ_\ndim  (\beta) \,.
					\label{main:intro:classical_physics:free_entropy_energy}
				\end{align}
				In order to avoid confusion with sign conventions and temperature pre-factors, we mainly consider the free entropy as our central object of study.
				As stressed in the introduction \Sec\ref{sec:main:introduction:stat_phys:motivation}, in the \emph{thermodynamic limit}\index{thermodynamic limit} $d\to \infty$ the free entropy of many systems \emph{concentrate} around an asymptotic and deterministic value given (when it exists) by
				\begin{align}
					\Phi (\beta) &\equiv \lim_{\ndim \to \infty} \Phi_\ndim (\beta)\,, 
					&& \varphi (\beta)\equiv \lim_{\ndim \to \infty} \varphi_\ndim (\beta)\,.
					\label{main:intro:classical_physics:free_entropy_energy_thermo}
				\end{align}
				As the finite size behavior generally fluctuates around these asymptotic quantities, their computation is of crucial interest to understand the collective behavior of the system. In particular their study reveals potential phase transitions, as illustrated in the analysis of phase transitions in \Sec\ref{main:intro:classical_physics:phase_transitions} and the presentation of simple examples in \Sec\ref{main:intro:classical_physics:examples}.

				\paragraph{Large deviation principle}
				The large deviation theory deals with the exponential decay of the \aclink{JPD} of random systems. In this paragraph, we provide justifications of the fact that computing the free entropy $\Phi$ in the study of equilibrium properties of systems with many-particles in interaction is equivalent to a large deviation theory. See \cite{Yoshitsugu89, varadhan2008, Touchette08} for an extended review.
				Let $\bsigma \in \chi_\ndim$, we say that the \aclink{JPD} $\rP_\ndim \(\bsigma\)$ satisfies a large deviation principle with rate $\mS$ if 
				\begin{align*}
				- \log \rP_\ndim \(\bsigma \) = \ndim \mS + o \(\ndim\) \Rightarrow - \lim_{\ndim\to \infty} \frac{1}{\ndim} \log \rP_\ndim \(\bsigma \) \equiv \mS \,,	
				\end{align*}
				which is equivalent to say that the \emph{dominant} behavior of $\rP_\ndim$ is decaying exponentially with the size of the system and is controlled by the rate function $\mS$, called the \emph{entropy} in physics. 
				Indeed, the Gartner-Ellis theorem  \cite{gartner1977large,ellis1984large} draws an explicit connection between the large deviation principle, the entropy and free entropy. Assuming the latter exists and is differentiable for any temperature $\beta\in \bbR$,
					\begin{align*}
						\Phi(\beta) = \lim_{\ndim \to \infty} \frac{1}{\ndim} \log \int_{\chi_\ndim} \exp\(-\ndim \beta \mH_{\ndim}(\bsigma)\) \d\bsigma \,,
					\end{align*}
					it states that the \aclink{JPD} verifies a large deviation principle
					\begin{align*}
						\lim_{\ndim \to \infty} - \frac{1}{\ndim} \log \rP_\ndim \(\mH_{\ndim}(\bsigma) = \mE\) =  \mS\(\mE\)\,,
					\end{align*}
					where the rate function is given by the \emph{entropy} $\mS(\mE)=\max_{\beta}\( \Phi(\beta) + \beta \mE \)$
					obtained by a Legendre transform detailed in \Sec\ref{main:intro:classical_physics:legendre_transform}.
				Back to statistical physics, proving the existence of a \emph{thermodynamic limit} of the free entropy $\Phi (\beta)$ in 
				\eqref{main:intro:classical_physics:free_entropy_energy_thermo}
				is therefore equivalent to prove a large deviation principle of the Gibbs measure.

				\paragraph{Cumulant generative function}	
				Finally, similarly to the partition function \eqref{main:intro:classical_physics:partition_function}, the free entropy has the advantage to encode for all the useful informations of the system. It can be seen as a \emph{cumulant generative function}. Namely successive cumulants of the Gibbs distribution can be obtained by taking higher order derivatives. 
				In particular, the free entropy gives access to important quantities such as the \emph{magnetization}\index{magnetization}, the corresponding \emph{average energy} or the \emph{ground state energy} associated to ground state configuration. For the sake of illustration, we assume that the local one-body interaction simply reads, as very often in physics, $\forall i \in\lb \ndim \rb,~ \log \phi_i(\sigma_i) = h_i \sigma_i$, where $\vec{h}=\{h_i\}_{i=1}^\ndim$ is called the \emph{external field}.
				
				\subparagraph{Magnetization}
				The \emph{magnetization} $m_\ndim$ at zero external field $\vec{h}=\vec{0}$ is defined as the averaged value of the spin configuration over the Gibbs distribution. It is simply obtained by taking the derivative of the free energy \eqref{main:intro:classical_physics:free_entropy_energy} with respect to the vanishing external field $\vec{h} \to \vec{0}$
					\begin{align}
						m_\ndim &\equiv \left\langle \frac{1}{\ndim} \sum_{i=1}^\ndim \sigma_i \right\rangle_{\beta} = \frac{1}{\ndim \mZ_\ndim(\beta)} \int_{\chi_\ndim} \(\sum_{i=1}^\ndim \sigma_i \) e^{-\beta \mH_{\ndim}(\bsigma)} \d \bsigma 
					\label{main:intro:classical_physics:magnetization}
						\\
						&= - \lim_{\vec{h} \to \vec{0}} \partial_{\vec{h}} \varphi_\ndim (\beta)\,. \nonumber
					\end{align}
					
				\subparagraph{Average energy and variance}
				The average energy at zero external field $\vec{h}=\vec{0}$ is simply the Gibbs average of the Hamiltonian energy given by 
				\begin{align}
					e (\beta) \equiv \left\langle  \frac{\mH_{\ndim}(\bsigma)}{\ndim} \right\rangle_{\beta} = - \lim_{\vec{h} \to \vec{0}} \partial_{\beta} \Phi_\ndim (\beta)\,,
				\end{align}
				while the second cumulant, the variance, is naturally given by the second derivative of the free entropy $\Phi_\ndim$
				\begin{align}
					&\frac{1}{\ndim} \(\langle \mH_{\ndim}(\bsigma)^2 \rangle_\beta -  \langle \mH_{\ndim}(\bsigma) \rangle_\beta^2 \) = \lim_{\vec{h} \to \vec{0}} \partial^2_{\beta^2} \Phi_\ndim \(\beta\)\,.
				\end{align}		

				\subparagraph{Ground state energy}
				The ground state energy $e_{\gs,\ndim}$ is the minimum energy that can be reached by at least one configuration $\bsigma$. It can be computed by taking the \emph{zero-temperature} limit $\beta \to \infty$ of the Gibbs random measure
					\begin{align}
						e_{\gs,\ndim} \equiv \min_{\bsigma \in \chi_\ndim} \left \{\frac{\mH_{\ndim}(\bsigma)}{\ndim} \right\}  = \lim_{\beta \to \infty}\left \langle \frac{\mH_{\ndim}(\bsigma)}{\ndim} \right \rangle_{\beta}\,.
					\end{align}

		\subsubsection{Illustration of simple graphical models}
		For the sake of clarification, in this section we briefly present some simple and common models and their graphical representation such as the \emph{k-SAT} problem, general tree factor graphs and regular pairwise \aclink{MRF}.
		 
\begin{figure}[!htb]
							\centering
							\hfill
							\begin{tikzpicture}[scale=0.75, auto, swap]
							    \foreach \i in {1,...,3}
							        \node[inter] (F\i) at (2*\i,0) {$\Psi$\i};
							    \foreach \i in {1,...,4}
							        \node[var] (X\i) at (-1 + 2*\i,-2) {$\sigma$\i};
							    \foreach \a/\i in {1/4,1/2,1/1, 2/2, 2/3, 2/4, 3/1,3/2, 3/3}
							        \path[edge] (F\a) -- (X\i);
							\end{tikzpicture}
							\hfill
							\begin{tikzpicture}[scale=0.75, auto, swap]
									\foreach \i/\x/\y in {1/1/0, 2/3/0, 4/3/2, 6/5/3, 5/5/1, 3/1/2}
					        			\node[var] (X\i) at (\x, \y) {$\sigma$\i};
					        		\foreach \a/\x/\y in {1/2/1, 2/4/2}
					        			\node[inter] (F\a) at (\x, \y) {$\Psi$\a};
					        		
					        		\foreach \a/\x/\y in {1/0/0, 2/4/0, 4/3/3, 6/6/3, 5/6/1, 3/0/2}
					        			\node[field] (G\a) at (\x, \y) {$\phi$\a};
					        						      
					        	    \foreach \a/\i in {1/1,1/2,1/3,1/4,2/4,2/5,2/6}
					       				\path[edge] (F\a) -- (X\i);
					       				
					       			\foreach \a/\i in {1/1,6/6,2/2,3/3,4/4,5/5}
					       				\path[edge] (G\a) -- (X\i);
					        			
							\end{tikzpicture}
					\caption{Factor graph representations: the red circles represent the spin variables, the green squares represent the factors that account for statistical dependencies between the variables and the yellow squares represent the single variable factors. \Left A $3$-SAT problem in \eqref{eq:sec:main:mean_field:general_markov_fields:3_sat}. \Right A tree factor graph in \eqref{eq:sec:main:mean_field:general_markov_fields:tree}.}
					\label{fig:main:factor_graph}
					\end{figure}
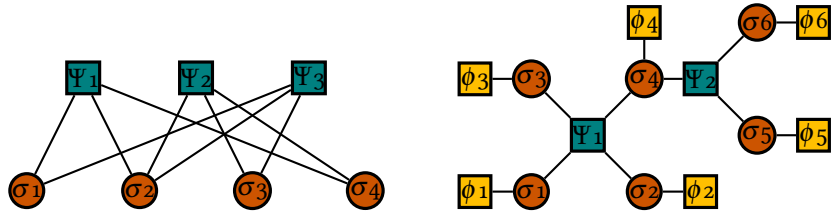

			\paragraph{K-SAT problem}
					The $k$-SAT problem is a \aclink{CSP} at the interface between information theory and error correcting codes. It is specified by $\ndim$ boolean variables $\bsigma \in \chi_\ndim=\{0,1\}^\ndim$ that must verify simultaneously the AND logical operator of $\nsamples$ constraints with $k$-body interactions, also called \emph{clauses}\index{clauses}, that depend on a subset of $k$ boolean variables. Random \aclink{CSP} are a variant in which the clauses are drawn from a random ensemble. 
					As generally in \aclink{CSP}, such as the graph-coloring problem, the traveling salesman problem and many others, the problem can be easily described by a factor graph \eq\eqref{main:intro:graphical:factor} \cite{dechter1988network}. Namely for the $k$-SAT problem, each factor denotes a \emph{hard constraint} represented by the indicator function $\log \Psi_\mu = \id\( \bsigma_{\partial \mu} \)$ so that the Hamiltonian energy counts the number of satisfied clauses
					\begin{align}
						\mH_\ndim(\bsigma) = - \sum_{\mu=1}^\nsamples \id\( \bsigma_{\partial \mu} \)\,.
					\end{align}
					The ground state configuration is reached if the $n$ clauses are verified such that the Hamiltonian energy $\mH_\ndim$ is minimal. As an illustration we give the \aclink{JPD} of a $k$-SAT problem realization at zero temperature for $k=3$, $\ndim = 4$ and $\nsamples=3$
					\begin{align}
						\rP_\ndim \(\bsigma\) \propto \underbrace{\id(\bar{\sigma}_1 \lor \bar{\sigma}_2 \lor \bar{\sigma}_4)}_{\Psi_1(\bsigma_{\partial 1})} \underbrace{\id (\sigma_2 \lor \sigma_3 \lor \sigma_4)}_{\Psi_2(\bsigma_{\partial 2})} \underbrace{\id (\sigma_1 \lor \sigma_2 \lor \sigma_3)}_{\Psi_3(\bsigma_{\partial 3})}\,,
						\label{eq:sec:main:mean_field:general_markov_fields:3_sat}	
					\end{align}
					where $\lor $ denotes the OR operator and $\bar{\sigma}$ the negation of a boolean variable. The corresponding factor graph is represented in
					\Fig\ref{fig:main:factor_graph} \Leftn.
					
					 Notably one important challenge in \aclink{CSP} is to compute the maximum constraints density that can satisfy the spin variables $\bsigma$, called the SAT-\emph{threshold} $\alpha_c = \nsamples_c / \ndim$. The SAT-threshold $\alpha_c$ can be computed from 
					the analysis of the \aclink{JPD} or the free entropy $\Phi_\ndim$ in \eqref{main:intro:classical_physics:free_entropy_energy}, by probing the existence of at least one configuration with strictly positive probability in the thermodynamic limit. See \cite{mezard2009information} and  \Sec\ref{main:intro:phys_ml_together:random_csp} for more details.

				\paragraph{Tree factor graphs}
					Tree-like factor graphs are a class of graphical model with the advantageous property of not presenting any loops, meaning that interactions are local and involve only the nearest neighboring spins. Tree factor graphs play an important practical and theoretical role because the full graph can be scanned with linear time complexity $\Theta\(\ndim\)$ and inference can be performed exactly. As an illustration, we give an example of a tree-like \aclink{JPD} represented in \Fig\ref{fig:main:factor_graph} \Rightn,
						\begin{align}
							\rP_\ndim (\bsigma) = \frac{1}{\mZ_\ndim} \prod_{i=1}^{6} \phi_{i}(\bsigma_i) \Psi_1(\sigma_1, \sigma_2, \sigma_3, \sigma_4 ) \Psi_2( \sigma_4, \sigma_5, \sigma_6 )\,.
							\label{eq:sec:main:mean_field:general_markov_fields:tree}	
						\end{align}

				\paragraph{Pairwise Markov random fields}	
				\label{sec:main:intro:mean_field:pairwise}	
					Among general \aclink{MRF} models, a large class of common models focuses on regular factor graphs with at most \emph{pairwise interactions} known as \emph{pairwise Markov random fields}. Therefore, we consider the \aclink{JPD} in \eq\eqref{main:intro:graphical:factor} by absorbing all potentials $\Psi_\mu$ with strictly more than two-body interactions, such that the \aclink{JPD} simply reads
						\begin{align}
						\rP_\ndim(\bsigma)
						&= \frac{1}{\mZ_\ndim} \prod_{i=1}^{\ndim} \phi_{i}(\sigma_i)  \prod_{(ij)} \Psi_{ij}(\sigma_i, \sigma_j)\,,
						\label{main:intro:graphical:pairwise}
						\end{align}
						where we have introduced pairwise symmetric potentials $\Psi_{ij} = \Psi_{ji}= \Psi_{\mu}$ by re-indexing all interacting pairs $\mu=(ij)=(ji)$, with $\mu \in \lb n \rb$. The corresponding Hamiltonian \eqref{main:intro:stat_phys:hamiltonian} simplifies to
						\begin{align}
							\mH_\ndim \(\bsigma\) &= - \sum_{i=1}^{\ndim} \log \phi_{i}(\sigma_i) - \sum_{(ij)}^{\nsamples} \log \Psi_{ij}(\sigma_i, \sigma_j) \,.
							\label{main:intro:graphical:pairwise:ising}
						\end{align}
						
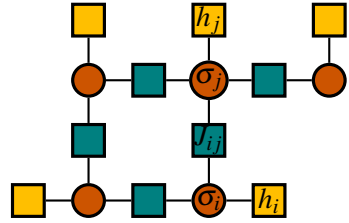
\begin{figure}[htb!]
						\centering
						\begin{tikzpicture}[scale=0.8, auto, swap]
							\tikzstyle{vertex}=[circle, draw, very thick, minimum size=12pt,inner sep=0pt]
							\tikzstyle{var}=[circle, draw, very thick, minimum size=12pt, inner sep=0pt, fill=burntorange]
							\tikzstyle{inter}=[rectangle, draw, very thick, minimum size=12pt, inner sep=0pt, fill=teal]
							\tikzstyle{field}=[rectangle, draw, very thick, minimum size=12pt, inner sep=0pt, fill=amber]
							\tikzstyle{edge} = [draw, thick, -]
						    \node[var] (S1) at (0, 0) {};
						    \node[var] (S2) at (0, 2) {};
						    \node[var] (S3) at (2, 2) {$\sigma_j$};
						    \node[var] (S4) at (2, 0) {$\sigma_i$};
						    \node[var] (S5) at (4, 2) {};
						
						    \node[inter] (J12) at (0, 1) {};
						    \node[inter] (J23) at (1, 2) {};
						    \node[inter] (J34) at (2, 1) {$J_{ij}$};
						    \node[inter] (J14) at (1, 0) {};
						    \node[inter] (J35) at (3, 2) {};
						
						    \foreach \i/\j in {1/2, 2/3, 3/4, 1/4, 3/5} {
						        \path[edge] (J\i\j) -- (S\i);
						        \path[edge] (J\i\j) -- (S\j);
						    }
						    
						    \node[field] (H1) at (-1,0) {};
						    \node[field] (H2) at (0,3) {};
						    \node[field] (H3) at (2,3) {$h_j$};
						    \node[field] (H4) at (3,0) {$h_i$};
						    \node[field] (H5) at (4,3) {};
						
						    \foreach \i in {1,...,5}
						        \path[edge] (H\i) -- (S\i);
						\end{tikzpicture}
						\caption{Factor graph of an Ising-like model on a regular lattice.}
						\label{fig:main:factor_graph_ising}
						\end{figure}

						A large part of the statistical physics literature focuses on such theoretical pairwise \aclink{MRF} on regular lattices, called alternatively \emph{Ising-like models}.
						For the sake of illustration, we consider such a system of magnetic spins, illustrated in \Fig\ref{fig:main:magnetic_spins}, immersed in an uniform external magnetic field $\vec{h}$ and local neighboring interactions. 
						This system can be represented by a spin model $\bsigma \in \chi_\ndim $ associated to a graph $\mG(\rV,\rE)$ with pairwise exchange interactions described by a matrix $ \mat{J} \in \bbR^{\ndim \times \ndim} $ and local external fields $\vec{h} \in \bbR^{\ndim}$, so that	
						\begin{align*}
							\log \Psi_{ij}(\sigma_i, \sigma_j) &= J_{ij} \sigma_i \sigma_j\,, && \log \phi_{i}(\sigma_i) = h \sigma_i \,.
						\end{align*}
						The corresponding factor graph is represented in \Fig\ref{fig:main:factor_graph_ising}. 
						Notice that the energy term $-\log \Psi_{ij}(\sigma_i, \sigma_j) = - J_{ij} \sigma_i \sigma_j$ is a convention that insures that for ferromagnetic interactions $J_{ij}>0$, the energy term decreases the total Hamiltonian energy if the spins $\sigma_i$ and $\sigma_j$ are aligned. The one-body interaction term $-\log \phi_{i}(\sigma_i) = -h \sigma_i $ represents the interaction of each spin with a uniform external field $h$ that tends to align all the spins in its direction.
						The coupling constants $J_{ij}$ represent the strength of the \emph{pairwise interaction}\index{pairwise interaction} between the spin $\sigma_i$ and $\sigma_j$. 
						In particular the interactions may be positive $J_{ij}>0$ or negative $J_{ij}<0$ and the corresponding models are respectively qualified of \emph{ferromagnetic}\index{ferromagnetic} and \emph{antiferromagnetic}\index{anti-ferromagnetic}.		
						The many variants of this general model depend mainly on the geometry and connectivity of the interactions $\rE$, the distribution $\rP(\mat{J})$ of the interaction matrix $ \mat{J} $ and the configuration space $\chi_\ndim$, that lead to a profusion of theoretical models. 
						We briefly recall the different well-know pairwise \aclink{MRF} models with \emph{continuous} or \emph{discrete} variables and pairwise interactions such as the Ising, Potts, XY, Heisenberg and $\Theta(N)$ models. 
												
						\subparagraph{Discrete models: Ising and Potts}
						Discrete models such as the Potts \cite{Potts} model assumes that each spin lies in a discrete alphabet $\chi = \bbZ^q$ with $q$ characters. For instance, it has been notably considered for \emph{hyper-graph coloring} problems, where each alphabet value represent a color, so that positive interaction happen only if interacting spins have the same color $\Psi_{ij}(\sigma_i, \sigma_j)= J_{ij} \delta\(\sigma_i, \sigma_j\)$. In particular, the model reduces to the classical Ising model with two colors for $q=2$ with $\chi_\ndim = \{\pm 1\}^\ndim$.
												
						\subparagraph{Continuous models: continuous Ising, Heisenberg, XY, $\Theta(N)$} 
						The $\Theta(N)$ model \cite{Stanley68,DEGENNES1972, Gaspari86} considers instead $\ndim$ continuous variable in $N$ dimensions so that the interaction term depends on the scalar product between vectorial spins $\Psi_{ij}(\bsigma_i, \bsigma_j)= J_{ij} \bsigma_i \cdot \bsigma_j$. For $N=1$, we recover the continuous Ising model ($\sigma \in \bbR$). The cases $N=2$ and $N=3$ correspond respectively to the XY model ($\bsigma \in \bbR^2$) and the Heisenberg model ($\bsigma \in \bbR^3$).

		\subsection{Phase transitions typology}
		\label{main:intro:classical_physics:phase_transitions}
		
		One of the most spectacular consequences of interactions among particles is the emergence of collective behaviors that would not have been observed in the presence of only a few particles. 
		Indeed in nature, many physical compounds exist under different forms, also called \emph{phases} or \emph{states}. As you change the macroscopic variables of a large system, called \emph{order parameters}\index{order parameters}, sometimes the system will abruptly change and move to another \emph{phase}\index{phase}.
		As these \emph{phase transitions}\index{phase transitions} affect dramatically the macroscopic behavior and properties of the system, they shall correspond to singularities in the free energy. Therefore, studying the free energy, that explicitly describes the interplay between energy and entropy contributions, is crucial to detect phase transitions.
		
			\subsubsection{A first phase transition: the solid-liquid-gas phase transitions}
			For the sake of clarification, let us consider the simplest example observable in everyday life: the phase transitions of water.
			Consider a \emph{solid} ice cube at low temperature $T$ and constant pressure $P$. 			
				\begin{figure}[htb!]
					\centering
					\begin{tikzpicture}[scale=0.85, transform shape]
						\draw[edge, -latex, very thick] (-0.5,0) -- (5,0);
						\draw[edge, -latex, very thick] (0,-0.5) -- (0,5);
						\draw [edge] (1.5,0) to [out=80, in=190] (5,3.5) ;
						\draw [edge] (2.2,1.8) to (1.5,5) ;
						\draw [edge, dashed,-latex, burntorange] (0.5,3) to (5,3) ;
						\node [circle, draw, very thick, minimum size=6pt, inner sep=0pt, fill=burntorange, burntorange] at (0.5, 3) {};
						\node [circle, draw, thick, minimum size=4pt, inner sep=0pt, fill=black, black] at (3.55, 3) {};
						\node [circle, draw, thick, minimum size=4pt, inner sep=0pt, fill=black, black] at (1.95, 3) {};
						\draw [edge, dashed, black] (1.95,3) to (1.95,0);
						\draw [edge, dashed, black] (3.55,3) to (3.55,0);
						\node at (5,-0.3) {$T$};
						\node at (1.95,-0.3) {$T_{\textrm{fus}}$};
						\node at (3.55,-0.3) {$T_{\textrm{vap}}$};
						\node at (-0.2,5) {$P$};
						\node at (1, 1.8) {Solid};
						\node at (3, 1) {Gas};
						\node at (2.5, 4) {Liquid};
					\end{tikzpicture}	
					\caption{Phase diagram $(T, P)$ of water. Increasing the temperature $T$ at constant pressure $P$, the ice will melts at $T_{\textrm{fus}}$ and vaporizes at $T_{\textrm{vap}}$.}
					\label{fig:main:intro:phase_diagram_water}
				\end{figure}
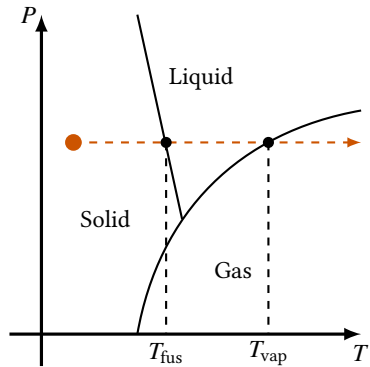
			Increasing the temperature (or decreasing the pressure), we may observe the ice transforming in the \emph{liquid} water at $T_{\textrm{fus}}= 0^{\circ}C$ before vaporizing in the water \emph{vapor} at $T_{\textrm{vap}}= 100^{\circ}C$. The successive transformations are represented in the \emph{phase diagram}\index{Phase diagram} in \Fig\ref{fig:main:intro:phase_diagram_water} and the different states are delimited by some solid lines which represent where the phase transitions occur. Notice that during the phase transitions, the system is still composed of the same number of particles of water. The only difference is the trade-off between the \emph{energetic} and \emph{entropic} contributions in the free energy that are modified so that the system adapts to the most stable collective configuration with the \emph{lowest free energy}.
			We could mention as well the \emph{ferromagnetic–paramagnetic} phase transition of a metal, that will be discussed with the Ising and Currie-Weiss models in \Sec\ref{main:intro:classical_physics:examples:curie_weiss_example} .  
			The analysis of this kind of phase transition in physical systems is particularly suited to statistical physics, whose phenomenology is general and applies to other fields such as information theory, optimization, computer science, biology, social sciences, economy, etc.
			
			\subsubsection{Energy-entropy decomposition: Legendre transform}
			\label{main:intro:classical_physics:legendre_transform}
			We stressed that the Gibbs measure often verifies a large deviation principle and is peaked in the most probable regions. However, we did not discuss \emph{which} or \emph{how many configurations} contribute to this dominant equilibrium configurations. 
			Indeed counting the number of configurations at a certain energy level that participates to the partition sum and the free entropy $\Phi_\ndim$ is very instructive to introduce the energy-entropy decomposition. 
			Let us denote $\Omega$ the number of configurations or equivalently the volume of phase space that achieve a given \emph{energy} $\mE$.
			
			\paragraph{Legendre transform} 
				The partitioning sum $\mZ_\ndim\(\beta\)$ in \eqref{main:intro:classical_physics:partition_function} can be partitioned instead over configurations with a particular level of energy $\mE \equiv \frac{1}{\ndim} \mH_\ndim(\bsigma) $. Introducing $\Omega(\mE) = \int_\bbR \d \bsigma ~ \delta\( \mE - \mH_\ndim(\bsigma) \) $ the number of such configurations, it is linked to the entropy density $\mS(\mE) \equiv \frac{1}{\ndim} \log \Omega(\mE)$. With this decomposition, the partition function writes 
				\begin{align*}
					\mZ_\ndim\(\beta\) & \equiv \exp\(\ndim \Phi_\ndim(\beta)\) \equiv \int_{\chi_\ndim}  e^{-\beta \mH_{\ndim}(\bsigma) } \d \bsigma \\
					&=  \int_{\bbR}  e^{-\beta \ndim \mE } \Omega\(\mE\)  \d \mE  \equiv \int_{\bbR}  e^{\ndim \(\mS(\mE) -\beta \mE \) }  \d \mE
				\end{align*}
				so that using a Laplace method \cite{Rong89} in the thermodynamic limit $\ndim \to \infty$, we obtain 
				\begin{align}
				\begin{aligned}
					\Phi\(\beta\) &= \max_{\mE} \(\mS(\mE) - \beta \mE \) = \mS(\mE^\star) -\beta \mE^\star 
				\end{aligned}
					\label{main:intro:classical_physics:legendre_transform_Phi}
				\end{align}
				where the equilibrium energy $\mE^\star$ verifies $\partial_\mE \mS\vert_{\mE=\mE^\star} = \beta$. This last formulation shows that the \emph{free entropy} $\Phi(\beta)$ is the Legendre transformation \cite{zia2009making} of the \emph{entropy} $\mS(\mE)$. Notice that the free entropy $\Phi(\beta)$ is a function of the inverse temperature $\beta$, which plays the role of a \emph{control parameter}, while the entropy $\mS(\mE)$ is a function of the \emph{response parameter} the energy $\mE$. Indeed the main advantage of the Legendre transformation is to exchange the role of the variables associated with control and response.
				
				\paragraph{Inverse Legendre transform}
				Similarly, with the definitions of $\mZ_\ndim$ and $\Omega$, we can introduce the inverse Laplace transform that leads to the inverse Legendre transform
				\begin{align*}
				\Omega(\mE) &\equiv e^{\ndim \mS(\mE)} =\int_\bbR \mZ_\ndim(\beta) e^{\beta d \mE } \d \beta = \int_{\bbR} \exp\(\ndim \( \Phi_\ndim(\beta) + \beta \mE\) \) \d \beta  \,,
				\end{align*}
				and using again a Laplace method in the thermodynamic limit $d\to \infty$, we obtain that the entropy $\mS(\mE) $ is the Legendre transform of the free entropy $\Phi(\beta)$: 
				\begin{align}
					\mS(\mE) = \max_{\beta} \Phi(\beta) + \beta \mE \Leftrightarrow \mS(\mE) = \Phi(\beta^\star) + \beta^\star \mE \,,
					\label{main:intro:classical_physics:legendre_transform_S}
				\end{align}
				where the critical temperature $\beta^\star$ is such that the slope of the free entropy verifies $\partial_\beta  \Phi \vert_{\beta=\beta^\star} = - \mE$. 
							
			\paragraph{Free entropy decomposition and collective behaviours}
				The Legendre transform \eqref{main:intro:classical_physics:legendre_transform_Phi} reveals that the free energy, or respectively the free entropy, decomposes in two contributions: the energy $\mE^\star$ and the entropy $\mS\(\mE^\star\)$:
				\begin{align*}
					\varphi(\beta) = \mE^\star - \frac{1}{\beta} \mS(\mE^\star)\,.
				\end{align*}
				This decomposition is crucial to understand the emergence of collective behaviors in large systems. Indeed without the entropic term, the free energy would simply be given by the energetic term $\mE^\star$ that measures the \emph{cost} of a typical configuration. It will not change when the inverse temperature $\beta$ or any other control parameter is modified. 
				The appearance of macroscopic collective behavior happens therefore as soon as
				the number of equilibrium configurations $\Omega^\star$ scales exponentially with the size of the system so that the entropic term $\mS(\mE^\star)$ becomes comparable to the energetic term $\mE^\star$. The inverse temperature is a free parameter that plays the role of a tension between the energy and the entropy, and controls the trade-off between being in a disordered phase with a high entropy or in an ordered phase with a low energy.
				 	
			\paragraph{Phase transition: first and second order}
				More generally, the free entropy (or the free energy) is the central object of study to analyze the behavior of large systems because very interestingly it captures the behavior of the different phases according to some carefully chosen \emph{order parameters}\index{order parameters}, denoted $\vec{q}$. 
				Indeed, non trivial behaviors and singularities in the free entropy reveal the \emph{phase transitions}, which are very often abrupt and happen at precise values of the order parameters.
				When it is possible, the free energy of the large size interacting system is computed and mapped to an extremization problem over a set of order parameters:			
				\begin{align}
					\varphi\(\beta\) &= \extr_{\vec{q}} \left\{ \Psi(\vec{q}, \beta) \right \} \,,
				\end{align}
				where we introduced a variational free energy $ \Psi(\vec{q}, \beta)$ that depends on the inverse temperature $\beta$ and the order parameters $\vec{q}$. The equilibrium behavior of the system can be analyzed by solving this simple optimization problem.\\
				
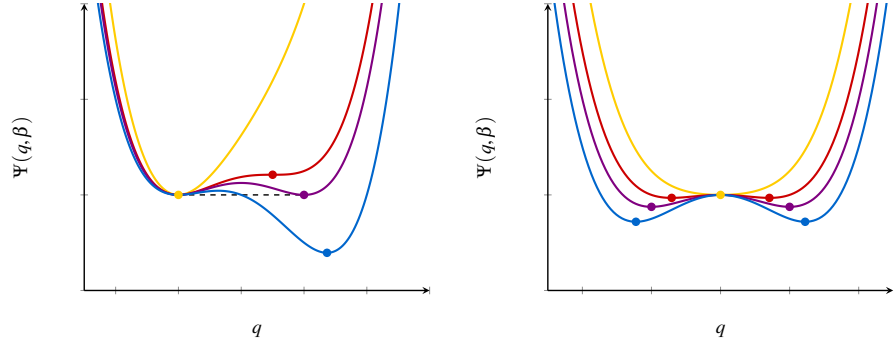
\begin{figure}[htb!]
						\centering
						\begin{tikzpicture}[scale=0.67, transform shape]
						\begin{axis}[
						    axis lines = left,
						    xlabel = $q$,ylabel = {$\Psi(q, \beta)$},
    						xtick={},ytick={},
    						xmin=-1.5,xmax=4,
    						ymin=-2,ymax=4,
    						yticklabels={,,},xticklabels={,,},
    						thick
						]
						\addplot[color=black, dashed] coordinates {(0,0) (2,0)};
						\addplot [
						    domain=-1.5:4, samples=200, color=orange,very thick]{4* x^2/2 - x^3 + x^4/4};
						\addplot [
						    domain=-1.5:4, samples=200, color=red,very thick]{2.25* x^2/2 - x^3 + x^4/4};
						\addplot [
						    domain=-1.5:4, samples=200, color=violet,very thick]{2* x^2/2 - x^3 + x^4/4};
						\addplot [
						    domain=-1.5:4, samples=200, color=blue,very thick]{1.5* x^2/2 - x^3 + x^4/4};
						\addplot[mark=*,color=orange] coordinates {(0,0)};
						\addplot[mark=*,color=red] coordinates {(1.5,0.42187)};
						\addplot[mark=*,color=violet] coordinates {(2,0)};
						\addplot[mark=*,color=blue] coordinates {(2.36603,-1.212)};
						\end{axis}
						\end{tikzpicture}
						\hfill
						\begin{tikzpicture}[scale=0.67, transform shape]
						\begin{axis}[
						    axis lines = left,
						    xlabel = $q$,ylabel = {$\Psi(q, \beta)$},
    						xtick={},ytick={},
    						xmin=-2.5,xmax=2.5,
    						ymin=-2,ymax=4,
    						yticklabels={,,},xticklabels={,,},
    						thick
						]
						\addplot [
						    domain=-4:4, samples=200, color=orange,very thick]{0.5 * x^2/2 + x^4/4};
						\addplot [
						    domain=-4:4, samples=200, color=red,very thick]{-0.5 * x^2/2 + x^4/4};
						\addplot [
						    domain=-4:4, samples=200, color=violet,very thick]{-1 * x^2/2  + x^4/4};
						\addplot [
						    domain=-4:4, samples=200, color=blue,very thick]{-1.5 * x^2/2 + x^4/4};
						\addplot[mark=*,color=orange] coordinates {(0,0)};
						\addplot[mark=*,color=red] coordinates {(-0.7071,-0.0625)};
						\addplot[mark=*,color=red] coordinates {(0.7071,-0.0625)};
						\addplot[mark=*,color=violet] coordinates {(-1,-0.25)};
						\addplot[mark=*,color=violet] coordinates {(1,-0.25)};
						\addplot[mark=*,color=blue] coordinates {(-1.2247,-0.5625)};
						\addplot[mark=*,color=blue] coordinates {(1.2247,-0.5625)};
						\end{axis}
						\end{tikzpicture}
						\caption{\Left First order phase transition for $\Psi(q) = \beta q^2/2 - q^3 + q^4/4$. For $\beta \ll 1$, the free energy has a single minimum $q=0$ (yellow). At $\beta=\beta_{\textrm{sp}}$, a second local minimum with higher free energy appears (red) and the most stable phase remains in $q=0$. At $\beta=\beta^\star$, there exists two global minima (violet). For $\beta > \beta^\star$, the order parameter jumps from $0$ to $q>0$ (blue).
						\Right Second order phase transition for $\Psi(q) = \beta q^2/2 + q^4/4$. For $\beta \ll 1$, the free energy has a single minimum (yellow) in $q=0$. At $\beta=\beta^\star$, this minimum becomes unstable and two global minima appear continuously (red) and becomes more and more stable (violet-blue).}
					\label{fig:main:intro:classical_physics:phase_transition}
						\end{figure}

				Interestingly, depending on the shape of the variational free energy $\Psi(\vec{q})$, we can observe multiple local and global minima that lead to different kind of phase transitions. In general, as well as in this manuscript, phase transitions can be classified into two main classes: \emph{discontinuous} first order and \emph{continuous} second order transitions.				
				This has been formalized by the Landau theory reviewed in \cite{toledano1987landau}. It allows to describe the phase transition phenomenology with a simple formalism by assuming the variational free energy may be written as a polynomial in a scalar order parameter $q$: $\Psi(q, \{\alpha_i\}_i) = \sum_{\{\alpha_i\}} \alpha_i q^i$. Depending on how the shape of the variational free entropy evolves with the control parameter, we may observe continuous or discontinuous phase transitions of the stable phase, which is the one with the lowest free energy.  
				A \emph{second order} phase transition happens when the order parameter of the most stable phase \emph{evolves continuously}. 
				In contrast, a \emph{first order} transition is observed when a local minima becomes at some point the global minima, so that the order parameter \emph{jumps from one phase to the other}. For instance, it is the case of the liquid-solid phase transition. 
				First and second order phase transitions are illustrated in \Fig\ref{fig:main:intro:classical_physics:phase_transition}.
				Interestingly with the Landau approach, 
				distinct systems can be gathered in the same universality class, characterized by the same non-zero coefficients in their variational free energy, such that their phase transition description is identical. 
				Indeed, the \emph{critical exponents}, that describe the behavior of the order and control parameters close to the phase transitions, are believed to be \emph{universal} and can be computed with \emph{renormalization group} \cite{wilson1983renormalization} technics.
			
			\paragraph{Metastable phases and ergodicity breaking}
				The variational free energy landscape can be complex with the presence of various local minimum. As a consequence, \emph{initializing} the system in a configuration close to such a locally stable \emph{state}, if the system is not perturbed, it will remain in this phase. However large fluctuations can destroy this local stability and in this case the system should adapt and move to another phase that corresponds to the global minima of the free energy. This kind of locally stable minima is called a \emph{metastable state}, \ie a state that remains stable if the system is not perturbed too much. 
				Such systems undergo a harmful \emph{ergodicity breaking} of the phase space, which means that the \emph{ensemble average} and the \emph{time average} are no longer equal and breaks the fundamental hypothesis of statistical mechanics. Indeed, by initializing the system in any metastable state, the system should visit all other possible states, eventually after an infinite time. Yet, on finite time scales, we could only observe the system in this state even though it is not the global minima of the free energy. After a finite amount of time, since the \aclink{JPD} remains unchanged, we could conclude that the system reached equilibrium. But in the case of non-ergodic systems, such as structural glasses, they remain stuck in a small portion of the configuration space and do not reach the globally stable equilibrium configuration.\\
				
			To conclude this introduction, we illustrate and apply these notions of metastability with the analysis of  the para/ferro-magnetic phase transition on the celebrated Ising model.
		
		\subsection{A classical example of lattice model}
		\label{main:intro:classical_physics:examples}
						
			\subsubsection{The Ising model}
			\label{main:intro:classical_physics:examples:ising_example}
			
			To describe ferromagnetism observed in metals, the battle-horse model of standard statistical physics is certainly the \emph{Ising model}. Indeed, it is the simplest regular pairwise \aclink{MRF}, that describes the collective behavior of magnetic \emph{spins} $\bsigma \in \chi_\ndim=\{\pm1\}^\ndim$. In general an Ising-like model is defined on a graph $\mG(\rV, \rE)$, so that spins lie on the vertices of the graph $\rV$ and interacts with neighbors, defined by the edges $\rE$, through exchange interactions $\mat{J} \in \bbR^{\ndim \times \ndim}$ and a potential external field $\vec{h} = h \cdot \vec{1}\in \bbR^\ndim$, whose Hamiltonian is given in \eqref{main:intro:graphical:pairwise:ising}. As already stressed, there exists many variants to the Ising model depending on the geometry of the structure of the adjacency matrix. We focus on the simple Ising model defined on a $N$-dimensional regular lattice illustrated in \Fig\ref{fig:main:factor_graph_ising}, such that each spin has $2 N$ interacting nearest-neighbors. It is formalized by the following Hamiltonian 
				\begin{align}
					\mH_\ndim(\bsigma; \mat{J}, \vec{h})&= -\frac{1}{2} \sum_{\langle ij \rangle } J_{ij} \sigma_i \sigma_j - h \sum_i \sigma_i \,,
					\label{main:intro:classical_physics:examples:ising_hamiltonian}
				\end{align}
				where $\langle ij \rangle$ denotes all possible pairs of neighboring spins on the regular lattice $\rV$.
				Solving the Ising model at finite dimensions for $N=1$ is a simple exercise and easy to solve with the transfer matrix technic. Unfortunately the model does not show any phase transition as the \emph{lower critical dimension} of the Ising model is $N_-=1$, under which there does not exist any collective behavior and ordered phase.
				For $N=1$, the fluctuations are so large that they kill the potential ordered phase. The \emph{up-down} symmetry is therefore preserved in a \emph{disordered paramagnetic phase} with zero macroscopic magnetization $m_\ndim=0$.
				Nonetheless, above this lower critical dimension the model exhibits a spontaneous symmetry breaking though and an interesting phase transition. 
				Indeed, for $N=2$, the more cumbersome Ising model has been solved exactly in \cite{onsager1944crystal}, while the much harder case $N=3$ still witnesses important research works.
				In the other hand, above the \emph{upper critical dimension} $N_+=4$, it turns out that the \emph{mean-field approximation} $N \to \infty$ of the Ising model is exact, see \cite{kardar2007statistical} for more details.
				
				As a pedagogical illustration, we present this latter mean-field	 approximation of the Ising model, called in this context the Curie-Weiss model, which is much easier to solve analytically.
				
			\subsubsection{The Curie-Weiss model}			\label{main:intro:classical_physics:examples:curie_weiss_example}
				The Curie-Weiss model \cite{curie1895proprietes} is the mean-field approximation of the Ising model with \emph{fully-connected} interactions in the limit of a high-dimensional lattice.
				As very often, mean-field or fully connected approximation have the advantage to make the model much easier to solve analytically. See \Sec\ref{main:sec:mean_fields} for a different approach to mean-field approximations. 
				In contrast with the Ising model, the interactions of the \emph{mean-field} Curie-Weiss model are fully-connected and long-range such that each spin is connected to all other spins $\bsigma \in \chi_\ndim = \{\pm 1\}^\ndim$ including itself, with a weak homogeneous coupling constant $J_{ij} = \frac{J}{2\ndim}$ scaling with the total number of spins to ensure the existence of the thermodynamic limit. Each spin can also interact with a homogeneous external field $h$ so that the Hamiltonian of the Curie-Weiss model trades the summation over neighboring pairs with all possible long range pairs, such that it reduces to
				\graffito{The factor graph of the Curie-Weiss model is completely symmetric as every spin is connected to every others. For clarity we do not draw the interaction factors $J/(2\ndim)$
				\begin{tikzpicture}[scale=0.6, auto, swap]
					\foreach \a in {1,2,...,9}{
						\draw (\a*360/9: 1.5) node[var,minimum size=1em] (X\a) {\tiny $\sigma$\a};
						\draw (\a*360/9: 2.5) node[field,minimum size=1em] (H\a) {\footnotesize $h$};
						}
					\foreach \b in {1,2,...,9}{
						\foreach \a in {1,2,...,9}{
							\path[edge] (X\b) -- (X\a);
							};
						}
					\foreach \b in {1,2,...,9}{
							\path[edge] (X\b) -- (H\b);
						}
					\end{tikzpicture}
				}
				\begin{align}
					\mH_\ndim(\bsigma; J, h)&= -\frac{J}{2 \ndim} \sum_{i, j = 1 }^{\ndim} \sigma_i \sigma_j - h \sum_{i=1}^\ndim \sigma_i\,.
					\label{main:intro:classical_physics:examples:curie_weiss_hamiltonian}
				\end{align}
				Because of this absence of geometric structure the \aclink{JPD} at inverse temperature $\beta$ is simply given by
				\begin{align*}
					\rP_\ndim(\bsigma; \beta, J, h) = \frac{1}{\mZ(\beta, J, h)}  \prod_{i,j}^\ndim e^{\frac{\beta J}{2\ndim} \sigma_i \sigma_j } \prod_i e^{\beta h \sigma_i}
				\end{align*}
				and is represented by a fully-connected symmetric factor graph.
				Let us introduce a simple order parameter: the \emph{magnetization} in \eq\eqref{main:intro:classical_physics:magnetization}.
				Defined as the macroscopic averaged magnetic moment $m_\ndim \equiv \frac{1}{\ndim} \sum_i \EE_{\bsigma \sim \rP_\ndim} \sigma_i$, it takes $2\ndim+1$ possible values $m_\ndim \in \bbM_\ndim = \{ - 1 + \frac{k}{\ndim}, k \in \lb 0: 2\ndim \rb \}$. At high temperatures, each spin is \emph{free to flip upside down} and is not affected by the interactions with its neighbors. As a consequence, by symmetry the magnetization is, in average, basically zero $m_\ndim=0$ and the system lies in the \emph{paramagnetic phase}.
				The specificity of a ferromagnet is that below a certain \emph{critical temperature} the influence of neighbors increase such that a spontaneous magnetization $m_\ndim \ne 0$ appears in the absence of any external magnetic field $h \to 0$. This transition is the so-called \emph{paramagnetic-ferromagnetic} transition. To describe quantitatively this para-ferro phase transition, let us derive the free entropy  $\Phi_\ndim(\beta, J, h) = \frac{1}{\ndim} \log \mZ_\ndim(\beta, J, h)$ with two different methods: a direct combinatorial one that makes use of the finite set $\bbM_\ndim$ of possible values taken by the magnetization, which is specific to this case, and the general Fourier method that we intensively use in the rest of the manuscript. 
				
				By first introducing the order parameter with a Dirac-delta integral $1 = \int_{\bbM_\ndim} \d m_\ndim \delta\(m_\ndim - \frac{1}{\ndim} \sum_{i=1}^{\ndim} \sigma_i \) $ the partition function can be expressed as
					\begin{align}
						\mZ_\ndim(\beta, J, h) &=  \int_{\chi_\ndim} \d \bsigma \int_{\bbM_\ndim} \d m_\ndim \delta\(m_\ndim - \frac{1}{\ndim} \sum_{i=1}^{\ndim} \sigma_i \)\label{main:intro:classical_physics:examples:curie_weiss_Z} \\
						& \qquad \qquad \qquad \qquad \times  \exp\( \frac{\beta J}{2 \ndim} \(\sum_{i=1}^\ndim \sigma_i\)^2 + \beta h \sum_{i=1}^\ndim \sigma_i \) \nonumber
					\end{align}
					
				\paragraph{Combinatorial method}
				The partition function can be directly computed by a combinatorial argument. Indeed, fixing the total magnetization $m_\ndim$ and denoting $\ndim_+$, $\ndim_-$ the number of positive and negative spins, we therefore have $\ndim \cdot m_\ndim = \ndim_+ + \ndim_-$ and $\ndim = \ndim_+ - \ndim_-$, so that $\ndim_+ = \frac{\ndim(1 + m_\ndim)}{2}$ and $\ndim_- = \ndim - n_+ =  \frac{\ndim (1 - m_\ndim)}{2}$. Defining $\Omega_\ndim(m_\ndim) = \int_{\chi_\ndim} \d\bsigma \delta \( m_\ndim - \frac{1}{\ndim} \sum_{i=1}^{\ndim} \sigma_i \)$ the number of configurations that give the same magnetization $m_\ndim$, it is simply given by the number of possibility to choose $d_+$ positive spins:
				\begin{align*}
					\Omega_\ndim(m_\ndim) &= {\ndim \choose \frac{\ndim (1+m_\ndim)}{2} } = \frac{\ndim !}{\(\ndim\frac{1-m_\ndim}2\)! \(\ndim\frac{1 + m_\ndim}2\)!}\simeq e^{\ndim \rH_{\textrm{binary}}\(\frac{1+m_\ndim}{2}\)} \,.
				\end{align*}
				Up to negligible terms, that do not scale exponentially with the system size, it can be expressed as a function of the Shanon binary entropy $\rH_{\textrm{binary}}(x) = -x \log(x) - (1-x) \log(1-x)$, see \Sec\ref{sec:main:intro:mean_field:entropy}, by using the Stirling approximation of ${\displaystyle d!\sim {\sqrt {2\pi d}}\left({d/e}\right)^{d},}$ in the large size limit $\ndim \to \infty$.
				Finally, the partition function can be transformed as 
				\begin{align*}
					&\mZ_\ndim (\beta, J, h) 
					= \int_{\bbM_\ndim} \d m_\ndim \Omega_\ndim(m_\ndim) \exp\( \ndim\( \frac{\beta J}{2 } m_\ndim^2 + \beta h m_\ndim  \)\)\\
					&\simeq \int_{\bbM_\ndim} \d m_\ndim ~ \exp\( \ndim\( \rH_{\textrm{binary}}\(\frac{1+m_\ndim}{2}\) +  \frac{\beta J}{2 } m_\ndim^2 + \beta h m_\ndim  \) \) \\
					&\equiv \int_{\bbM_\ndim} \d m_\ndim~ \exp\(\ndim \Psi\(m_\ndim; \beta, J, h\)\)  \,,
				\end{align*}
				where we introduced the free entropy potential
				\begin{align}
					\Psi\(m; \beta, J, h\) &= \rH_{\textrm{binary}}\( \frac{1 + m}{2} \) + \frac{\beta J}{2 } m^2 + \beta h m \,.
					\label{main:intro:classical_physics:examples:curie_weiss_Psi}
				\end{align}
				Notice that this mean-field approximation can also be obtained from a more elegant variational principle based on the Gibbs inequality presented in \Sec\ref{main:sec:mean_fields}.
				In the thermodynamic limit $\ndim \to \infty$, since the integral is dominated by its maximum, the partition function can be evaluated with a Laplace method, also called a \emph{saddle point} method \cite{Rong89}, so that the free entropy yields
				\begin{align}
					\Phi(\beta, J, h) = \lim_{\ndim \to \infty} \frac{1}{\ndim} \log \mZ_\ndim \(\beta, J, h\) = \max_{m \in [-1; 1]} \Psi(m; \beta, J, h)
					\label{main:intro:classical_physics:examples:curie_weiss_Phi}
				\end{align}
				where we used the fact that $\bbM_\ndim \underlim{\ndim}{\infty} [-1;1]$. For a pedagogical purpose, we present as well the equivalent computation of the mean-field free entropy with the Fourier transform method that can be generalized 
										
				\paragraph{Fourier transform method}
				The general method consists in introducing the Fourier representation of the Dirac-delta distribution according to
				\begin{align*}
					\delta\(x\) = \frac{1}{2\pi i} \int_{i \bbR} \d\hat{x} e^{\hat{x} x} \,,
				\end{align*}
				in \eqref{main:intro:classical_physics:examples:curie_weiss_Z} so that 
				\begin{align*}
					\mZ_\ndim (\beta, J, h) &\propto  \int_{\bbM_\ndim} \d m_\ndim \int_{i\bbR} \d \hat{m} e^{\ndim\(  \hat{m} m_\ndim  + \frac{\beta J}{2} m_\ndim^2 + \beta h m_\ndim \)}  \int_{\chi_\ndim} \d \bsigma e^{-\hat{m} \sum_{i=1}^{\ndim} \sigma_i } \\
					&\propto  \int_{\bbM_\ndim} \d m_\ndim \int_{i\bbR} \d \hat{m} e^{\ndim\(  \hat{m} m_\ndim  + \frac{\beta J}{2} m_\ndim^2 + \beta h m_\ndim \)}   \( 2 \cosh \hat{m} \)^\ndim\,,
				\end{align*}
				\graffito{The Cauchy theorem states that the integral of a holomorphic function $f:\Gamma\mapsto \bbC $ on a simply connected $\Gamma$ open subset is null $\int_{\Gamma} f(z) dz =0$}
				where we omitted the negligible pre-factors in the thermodynamic limit. Deforming the integration contour with the Cauchy theorem, the free entropy can be formulated as a \emph{saddle point} and evaluated by
				\begin{align*}
					\Phi \equiv \lim_{\ndim \to \infty} \frac{1}{\ndim} \log \mZ_\ndim (\beta, J, h)= \extr_{m,\hat{m}} \td{\Psi}(m, \hat{m}; \beta, J, h)\,,
				\end{align*}
				with $m=\lim_{\ndim \to \infty} m_\ndim \in [-1;1]$ and $\td{\Psi}(m, \hat{m}; \beta, J, h) \equiv \hat{m} m  + \frac{\beta J}{2} m^2 + \beta h m  +  \log \cosh{\hat{m}}$. Taking the saddle point condition over $\hat{m}$ and using the fact that $\rH_{\textrm{binary}}\( \frac{1 + m}{2} \) = - m \atanh(m)   +  \log \cosh{\atanh(m)}$, we finally recover the same free entropy potential \eqref{main:intro:classical_physics:examples:curie_weiss_Psi}
				\begin{align*}
					\td{\Psi}(m, \hat{m}^\star; \beta, J, h)
					&=\frac{\beta J}{2} m^2 + \beta h m  - m \atanh(m)   +  \log \cosh{\atanh(m)}\\
					&=  \frac{\beta J}{2} m^2 + \beta h m + \rH_{\textrm{binary}}\( \frac{1 + m}{2} \) = \Psi(m; \beta, J, h)\,.
				\end{align*}
				\begin{figure}[htb!]
					\centering
					\includegraphics[scale=0.45]{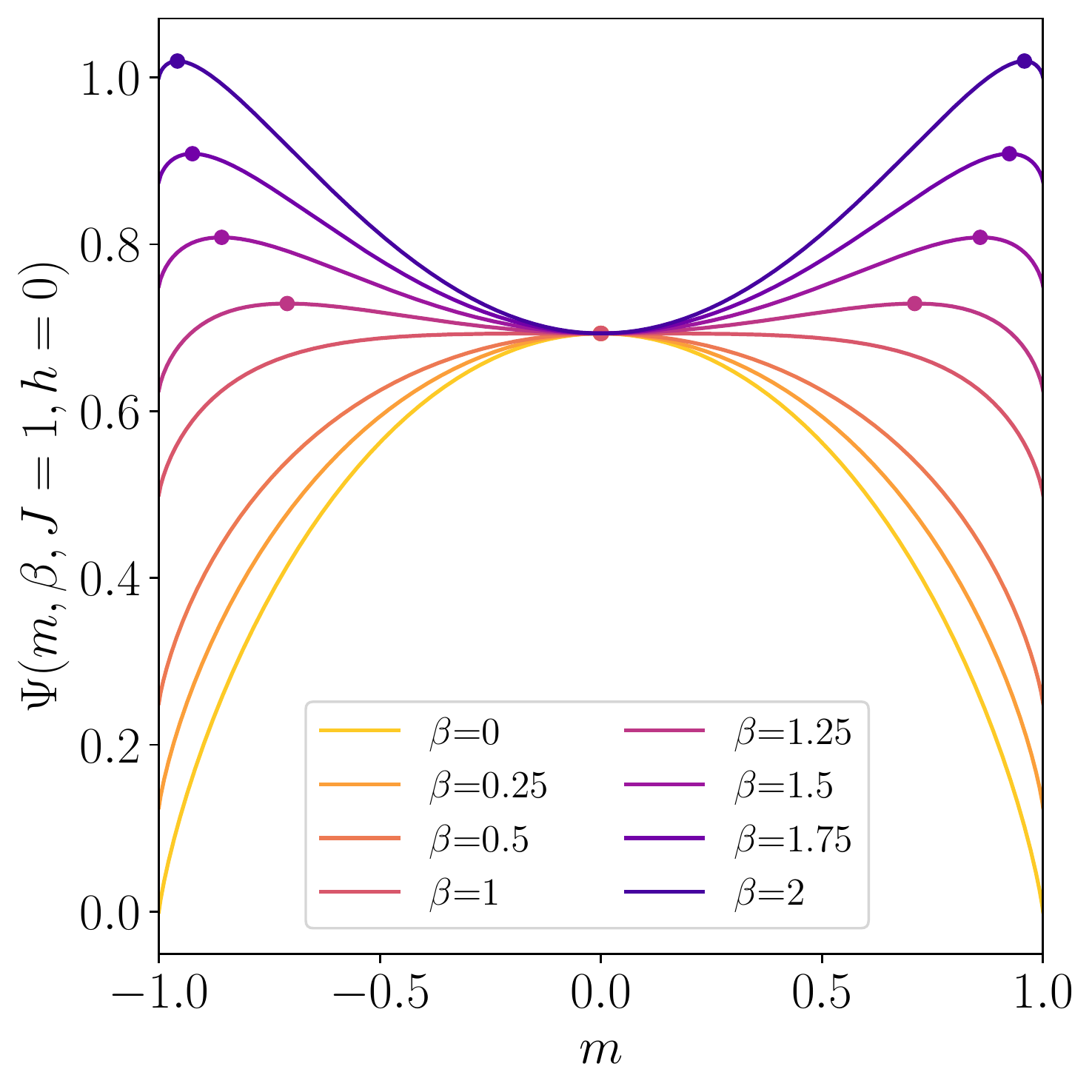}
					\caption{Variational free entropy $\Psi(m; \beta, J=1, h=0)$ of the Curie-Weiss model. Above the critical temperature $\beta < \beta^\star=1$ (yellow-red), the global maxima of the potential is achieved for $m=0$ that corresponds to a paramagnetic phase and no ordered phase exists. For smaller temperature $\beta > \beta^\star$ there exists an ordered phase with strictly positive or strictly negative magnetization that corresponds to ferromagnetic phases (pink-blue).}
				\label{fig:introduction:statistical_physics:curie_weiss}
				\end{figure}

			\paragraph{Paramagnetic-ferromagnetic phase transition}
				To conclude, the computation of the free entropy of $\ndim \to \infty$ interacting spins reduces to a one-dimensional optimization problem over the magnetization order parameter. This extremization \eqref{main:intro:classical_physics:examples:curie_weiss_Phi} can therefore be analyzed easily to finally describe the phase transition occurring in the Currie-Weiss model, which is nothing but the mean-field approximation of the Ising model.
				Taking the extremization over $m \in [-1;1]$, $\partial_m \Psi(m) = 0 $, we obtain that the maximum verifies $m=\tanh\(\beta\( J m + h \) \)$, it can be solved numerically as illustrated in \Fig\ref{fig:introduction:statistical_physics:curie_weiss}.
				In the limit of a vanishing external field $h=0^+$, fixing the coupling constant to $J=1$, we can explore the free entropy behavior as a function of the inverse temperature control parameter $\beta$. For high temperature, \ie $\beta \to 0$, the global maxima of the variational free entropy is given by a paramagnetic phase with $m=0$. At the critical inverse temperature $\beta=1$, we observe the continuous apparition of two global maxima with $m \ne 0$ that correspond to ferromagnetic phases with either a majority of spins \emph{up} $m>0$ or \emph{down} $m<0$, while the paramagnetic local maxima $m=0$ becomes a metastable state.

	\section{Extension to disordered systems and spin glasses}
	\label{sec:main:introduction:stat_phys:disordered}		 
		In the previous section, we focused on systems and models qualified as \emph{ordered} in the sense that all parameters such as the exchange coupling $\mat{J}$ and the external field $\vec{h}$ are deterministic. 
		However in nature, no material is perfectly homogeneous and deterministic. They usually present some sources of randomness like interstitial impurities or random external/local environment.
		For instance, consider a real magnetic material: the description that the local magnetic moments interact in a simple homogeneous way, as illustrated in the Ising and Curie-Weiss models in \Sec\ref{main:intro:classical_physics:examples:ising_example}, is over-idealistic.
		In reality there exists some impurities that modify the interactions and make their behavior more complex.
		In more details, a small fraction of a transition metal may be diluted into a noble metal, to obtain an alloy with magnetic moment randomly localized: this is called a \emph{spin glass}\index{spin glass} \cite{mezard1987spin}.
		Such a system belongs to the large class of \emph{disordered systems} in which some source of randomness emerges in the spin interactions. Statistical physics started to study this new class of models in the 60-70's and was the source of a rich literature since then. 
		Indeed, incorporating randomness in the classical statistical physics tools, presented in \Sec\ref{sec:main:introduction:stat_phys:describing_behavior}, allowed to democratize the approach to various fields and to highlight the existence of new interesting phenomena and phase transitions. 
		In particular, under a \emph{weak disorder} assumption, the description of the ordered phase transitions and critical phenomenon presented in \Sec\ref{main:intro:classical_physics:phase_transitions} may be either conserved or smoothed so that first oder become second order phase transitions. However, in the presence of a \emph{strong disorder} it \emph{strongly affects} and changes the nature of the phase transitions especially at low temperature where we observe the appearance of a singular \emph{glassy phase} with many local metastable states.\\
		
		In this section we discuss several kinds of disorder and models that account for it, and we focus the discussion on spin glasses that are more relevant according to the rest of the manuscript. 
		We present a brief overview of the wide literature of spin glasses mainly based on \cite{mezard1987spin,Castellani2005,dedominicis_2006} to illustrate the basic ideas required to understand the rest of the manuscript. The discssion can be extended with more specific and influent contributions \cite{franz1997phase, bouchaud1998out, biroli2001metastable, cugli2002dynamics, franz2011analytical, Berthier11}.
		
		\subsection{Quenched and annealed disorder}
		\label{main:intro:disordered:quenched_annealed}
			There exists two main types of disordered systems: the ones with \emph{explicit disorder} in the Hamiltonian of the model, and the ones such that the disorder is self-generated. The latter class can be simply illustrated with \emph{structural glasses} in which many interacting particles are moving so that each particle feels a local \emph{random disordered environment}. 
			However, in this manuscript, we consider only systems with explicit disorder and we refer the reader to \cite{kirkpatrick1987stable,mezard2000statistical,lubchenko2007theory,charbonneau2014fractal} for more details on amorphous solids and structural glasses.\\
			
			We therefore consider systems of spins $\bsigma$ with \emph{explicit} disorder in the Hamiltonian \eqref{main:intro:stat_phys:hamiltonian}, for example through the influence of random parameters such as the coupling constant or the external field $(\mat{J},\vec{h})$, that we call for historical reasons \emph{impurities}. We assume that the latter impurities are some \aclink{RV} that evolve at a typical time scale $\tau_q$, while the system of spins $\bsigma$ evolve a time scale $\tau$. Depending on how these time scales compare, we shall distinguish \emph{quenched} and \emph{annealed} disorders.
			\emph{Annealed}\index{annealed} disorder refers to systems such that $\tau_q \simeq \tau$. In other words, the random impurities $(\mat{J},\vec{h})$ and the spins variables $\bsigma$ evolve and fluctuate on a similar time scale \cite{Palmer1982}. Therefore, they play the same role and should be considered on an equal footing. As a consequence, in the presence of an annealed disorder, the impurities are in thermal equilibrium and can simply be included in the statistical description of the system. 
			In contrast, \emph{quenched}\index{quenched} disorder refers to systems such that $\tau_q \gg \tau$: the impurities are \emph{static} and remain fixed while the spin variables $\bsigma$ fluctuate. Each realization of the quenched disorder thus corresponds to a \emph{new experiment} with new sampled parameters. Therefore distinguishing the \emph{slow-evolving} quenched impurities $\mat{J}$ from the thermal spins $\bsigma$ time scales is crucial. In particular, the equilibrium properties and the corresponding thermodynamics cannot be computed in the same way than for systems with annealed disorder.
			In order to take into account the random impurities and not compute properties of the system which depend on a single realization of the randomness, we would like to average over the randomness. However, the specific disorder time scale makes the average over the Gibbs random measure harder. The different time scales and the effect of the randomness in the Hamiltonian require therefore specific analytical treatments that we describe and develop in \Chap\ref{main:chap:mean_field}.
			Notice, nonetheless, that in some specific cases, quenched disordered systems behave as annealed systems and are easier to tackle analytically.\\
			
			Even though other disordered systems such as \aclink{RFIM} \cite{belanger1991random, mezard1992replica} have been considered in the literature \cite{imbrie1984lower,belanger1991random}, in the rest of the manuscript we mainly focus the discussion on \emph{spin glasses} with \emph{quenched disorder}.
			
			\subsection{Spin glasses with quenched disorder}
			\label{main:intro:disordered:quenched_spin_glass}
				\emph{Spin glasses}\index{spin glass} refer historically to metallic alloys in which during the chemical preparation of the sample magnetic impurities substitute to the original atoms in randomly selected positions \cite{binder1986spin, fischer1993spin, mezard1987spin}. 
				In order to theoretically understand the properties of these materials, various models have been proposed based on a spin model with a \emph{quenched} disorder through the random exchange interaction $\mat{J}$ drawn from a distribution $\rP(\mat{J})$. Unlike the simple \aclink{RFIM} case where the randomness only affects the one-body interaction term $\log \phi_i$, the disordered interactions $\mat{J}$ dramatically affect the two-body interactions and the thermodynamics of mean-field models spin glasses.\\
			 
				In this manuscript, we consider essentially models that can be formulated as \emph{spin glass} models with \emph{quenched disorder}. More precisely, we consider a system of $\ndim$ spins with $\bsigma \in \chi_\ndim$ with an Hamiltonian $\mH_\ndim\(\bsigma ; \mat{J}, \vec{h}\)$ that explicitly depends on the quenched \aclink{RV}, \eg the coupling constant $\mat{J}$ completely specified by its probability distribution $\d\rP(\mat{J}) = \rp(\mat{J}) \d \mat{J}$, and a fixed external field 	$\vec{h} \in \bbR^{\ndim} $.
				For the sake of illustration, let us introduce the Ising-like spin glass model, historically considered in \cite{Toulouse1987}, which became the battle horse of the spin glass literature. Consider a graphical model $\mG\(\rV, \rE\)$ with spins at the vertices $\rV$ and pairwise interactions between spins on edges $\rE$:
				\begin{align*}
					\mH_\ndim (\bsigma; \mat{J}, \vec{h}) = -\frac{1}{2} \sum_{(ij) \in \rE} J_{ij} \sigma_i \sigma_j  - \sum_{i\in \rV} h_i \sigma_i  \,,
				\end{align*}
				associated to the Gibbs thermal average and the partition sum
				\begin{align*}
					 \mZ_\ndim \(\beta,\mat{J}, \vec{h}\) &= \int_{\chi_\ndim} \d \bsigma ~ e^{-\beta \mH_\ndim(\bsigma; \mat{J},\vec{h})} \,.
				\end{align*}
				Unlike the case of the Ising ferromagnet \ref{main:intro:classical_physics:examples:ising_example} (respectively anti-ferromagnet) with $\forall (ij) \in \rE, J_{ij}>0$ (respectively $J_{ij}<0$), in spin glass models, the exchange interaction matrix is random so that $J_{ij}$ associated to the edge $(ij)$ has a random sign and can be either positive or negative. The local interaction is called \emph{ferromagnetic}\index{ferromagnetic} (respectively anti-ferromagnet) if $J_{ij}>0$ (respectively $J_{ij}<0$). The coupling being random, in \emph{average}, the model is called ferromagnetic (respectively anti-ferromagnetic) if there exists a \emph{bias} such that $\EE_{\mat{J}} J_{ij} >0$ (respectively $\EE_{\mat{J}} J_{ij}<0$). For conciseness, we leave aside the external field in the following.
								
				\subsection{Frustration}
\begin{figure}[htb!]
					\centering
						\begin{tikzpicture}[scale=0.8, transform shape]
							\pgfmathsetseed{7}
							\foreach \i/\j/\k in {1/0/1, 5/0/2, 9/0/3, 3/1.5/4, 7/1.5/5}
			       				\pgfmathsetmacro\Angle{rand*30}
			       				\pgfmathrandominteger{\dir}{0}{1}
			       				\pgfmathsetmacro\Dir{2*\dir-1}
			    				\draw[-latex,very thick, rotate around={\Angle:(\i/1.5,\j)}] (\i/1.5,-\Dir*0.85+\j) -- (\i/1.5,\Dir*0.85+\j);
			    			\foreach \i/\j/\k/\l/\m/\n in {1/0/1/3/1.5/4, 1/0/1/5/0/2, 9/0/3/7/1.5/5, 3/1.5/4/7/1.5/5}
			    				\draw[-, very thick, dashed, color=red] (\i/1.5,\j) to (\l/1.5,\m);
			    			\foreach \i/\j/\k/\l/\m/\n in {3/1.5/4/5/0/2, 5/0/2/9/0/3,5/0/2/7/1.5/5}
			    				\draw[-, very thick, dashed, color=green] (\i/1.5,\j) to (\l/1.5,\m);
			    			\foreach \i/\j/\k in {1/0/1, 5/0/2, 9/0/3, 3/1.5/4, 7/1.5/5}
			    				\node[circle, draw, very thick, minimum size=12pt, inner sep=0pt, fill=burntorange] (s\k) at (\i/1.5,\j) {$\sigma$\k};
			    			\draw[-,decorate, decoration={snake}, orange, thick] (5/1.5,0.85)  to (7/1.5,0.85+1.5) ;
						\end{tikzpicture}
						\caption{Illustration of frustrated spins on a regular "plaquette": frustration appears when a spin undergoes a positive (green) and a negative (orange) interaction at the same time. Spins $\sigma_2, \sigma_5$ are frustrated as the coupling constant is positive but the corresponding spins are anti-aligned.}
						\label{fig:main:frustration}
					\end{figure}
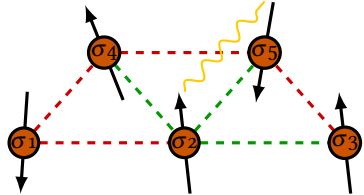

					Understanding spin glasses is more involved that classical ferromagnetic models. Indeed, the quenched disorder may be the source of \emph{frustration} between the spins of the system, so that finding an optimal configuration becomes harder and takes much longer time. 
					In particular, the randomness of the coupling interactions $\mat{J}$ signs breaks down the spatial homogeneity and creates heterogeneity, called \emph{frustration}\index{frustration}. 
					This collective behavior appears when the best possible spins configuration cannot satisfy
					all the local local two-body \emph{constraints} and minimize all interactions terms in the Hamiltonian, as illustrated in \Fig\ref{fig:main:frustration}. 
					As a serious consequence, many distinct configurations may achieve the same energy level, so that one expects the existence of many \emph{local minima} in the free energy landscape leading to a \emph{glassy behavior}.
					Such frustrated systems show non-trivial properties richer than systems without disorder considered in \Sec\ref{sec:main:introduction:stat_phys:describing_behavior} and the quenched random disorder dramatically affects their thermodynamic behavior.	
							
			\subsection{Averaging and self-averaging}
				As discussed in the previous section, the quenched disorder $\mat{J}$ plays a singular role with respect to the thermal fluctuations of the spins $\bsigma$. Since it evolves at a much slower time scale $\tau_q \gg \tau$, for a given experiment the quenched disorder is considered as \emph{static}. It means that at each experiment we draw a new realization of the disorder $\mat{J}$, from the distribution $\rP(\mat{J})$, that is considered to be fixed all along the spin dynamics. As a consequence the free entropy explicitly depends on the realization of the disorder 
				\begin{align}
					\Phi_{\ndim}\(\beta,\mat{J}\)&\equiv  \frac{1}{\ndim} \log \mZ_\ndim\(\beta, \mat{J}\)\,, && \varphi_\ndim (\beta, \mat{J}) \equiv -\frac{1}{\ndim \beta} \log \mZ_\ndim  (\beta, \mat{J})\,.
					\label{main:intro:averaged_free_entropy_free_energy_J}
				\end{align}	
				However, we do not want the description of the system to depend on the realization of the disorder $\mat{J}$. 
				Instead, and specifically to quenched systems, we introduce the \emph{averaged free entropy} and \emph{free energy} by adding the \emph{quenched average} over the disorder, denoted denote $\EE_{\mat{J}}$, on top of \eqref{main:intro:averaged_free_entropy_free_energy_J}, and crucially after the Gibbs thermal average contained in the partition sum
					\begin{align}
						\Phi_\ndim (\beta) &\equiv \EE_{\mat{J}} \Phi_{\ndim}\(\beta,\mat{J}\) \,, && \varphi_\ndim (\beta) \equiv \EE_{\mat{J}} \varphi_\ndim (\beta, \mat{J})\,.
						\label{main:intro:averaged_free_entropy_free_energy}
					\end{align}
					Averaging over all possible disorder \emph{realizations} refers to the so-called \emph{typical scenario}\index{typical scenario}. This is to oppose to the \emph{worst case} scenario, that deals with the worst possible disorder instance to obtain a strong upper-bound of the system description.
					In the perspective of averaging over many experiments, we restrict ourselves to the \emph{typical case} that suits much better our purpose, so that all quantities shall be averaged over the disorder. However, it is important to remark that for the moment there is no reason why the \emph{averaged} free entropy would correctly describe the system for a single realization of the disorder. 
					
					\paragraph{Self-averaging}
					In fact the most remarkable property of spin glasses is that some extensive observables become \emph{self-averaging}\index{self-averaging} in the thermodynamic limit, meaning that they are correctly described by their averaged behavior. In other words, we say that a \aclink{RV} $\rX_\ndim \in \bbR^\ndim$ is self-averaging if it concentrates around its mean, namely if 
					\begin{align}
							\forall \epsilon >0, \lim_{\ndim \to \infty} \bbP\( \|\rX_\ndim - \EE\[ \rX_\ndim \]\| > \epsilon \) = 0\,.
					\end{align}
					Very importantly the free entropy is such a self-averaging quantity. Thus at fixed disorder realization $\mat{J}$, $\Phi_{\ndim}\(\beta,\mat{J}\)$ in \eqref{main:intro:averaged_free_entropy_free_energy_J} \emph{concentrates} in the thermodynamic limit around a deterministic free entropy given by the average over the disorder $\Phi\(\beta\)$ in \eqref{main:intro:averaged_free_entropy_free_energy}. 
					Therefore, in the thermodynamic limit $\ndim \to \infty$ such observables have the same value for each realization of the disorder $\mat{J}$ and this legitimates the fact of considering the \emph{typical} scenario. In other words, the free entropy description no longer depends on the specific realization of the disorder and the sample fluctuations are vanishing in the large system size limit, so that its typical value coincides with the deterministic average value.
					As a final remark, notice that in general observables that involve the sum of an infinite number of particles are expected to self-average. 
					On the contrary, non-self averaging quantities, such as correlation functions, can fluctuate significantly and computing them remains a harsh difficulty.\\
					
					Finally, the self-averageness of the free entropy $\Phi$ can be described for both short and long range systems as explained in the following.
				
				\paragraph{Short-range argument}
				For short-range interactions mode, we recall a simple argument based on the \aclink{CLT} that shows that the free entropy is self-averaging \cite{mezard1987spin, thouless1977solution, Orlandini_2002}. 
				In the case the interactions are short-range, we can split the whole system of total volume $V = L^\ndim$, where $L$ is the typical size of the system, in $N$ macroscopic sub-systems of volume $v = l^\ndim$ with $V=N v$ and $N=\(\frac{L}{l}\)^\ndim$ defined in such way that they weakly interact with each other. 
				If we assume that the interactions have a typical range $\lambda \ll l$, the free entropy decomposes in bulk and surface contributions
				\begin{align*}
					\Phi_\ndim \(\beta,\mat{J}\) &= \frac{1}{\ndim} \log \mZ_\ndim\(\beta, \mat{J}\) = \frac{1}{\ndim} \log \sum_{\bsigma} e^{-\beta \mH_{\textrm{bulk}}(\bsigma; \mat{J}) - \beta \mH_{\textrm{surface}}(\bsigma; \mat{J}) }\\
					& \simeq \frac{1}{\ndim} \log \sum_{\bsigma} e^{-\beta \mH_{\textrm{bulk}}(\bsigma; \mat{J})} + \frac{1}{\ndim} \log \sum_{\bsigma} e^{ - \beta \mH_{\textrm{surface}}(\bsigma; \mat{J}) }\\
					& = \Phi_{\textrm{bulk}}\(\beta,\mat{J}\) + \Phi_{\textrm{surface}}\(\beta,\mat{J}\) \simeq  \Phi_{\textrm{bulk}}\(\beta,\mat{J}\)\\
					&\simeq \sum_{k=1}^N \log \sum_{\bsigma_k} e^{-\beta \mH_{\textrm{bulk}}(\bsigma_k; \mat{J})}\,,
				\end{align*}
				where we first assumed that interactions between the bulk and the surface are negligible, and the last equality generally holds if the surface interaction is negligible with respect to the $N$ blocks of the bulk contribution. In the thermodynamic limit as $l \ll L$, $N \to \infty$ and the \aclink{CLT} applies, the free entropy is therefore the sum of independent variables and becomes a Gaussian variable centered around its average $\Phi_\ndim\(\beta\)\equiv \EE_{\mat{J}} \Phi_{\ndim}\(\beta,\mat{J}\)$ with fluctuations of order
				\begin{align*}
					\frac{\EE_{\mat{J}}\[\Phi_{\ndim}\(\beta,\mat{J}\)^2  \] - \Phi_\ndim\(\beta\)^2 }{\Phi_\ndim\(\beta\)} = \Theta\( \ndim^{-1/2} \) \underlim{\ndim}{\infty} 0\,.
				\end{align*} 
			
				\paragraph{Long-range systems}
				Yet this simple general argument does not apply to long-range interactions. Anyway the self-averaging property is in fact very often assumed and may be proven in some long-range interactions model such as the \aclink{SK} model \cite{guerra2002central,guerra2002thermodynamic} or in more recent works for various models \cite{barbier2016mutual, BarbierM17a, barbier2017phase, Barbier2019a}.
				
			\subsection{Annealed averages}
			\label{main:intro:disordered:quenched_spin_glass:annealed}
				The quenched nature of the disorder imposes to \emph{average the free entropy itself} and \emph{not the partition function}.
				Taking the quenched disorder average of the log-partition function in \eqref{main:intro:averaged_free_entropy_free_energy} is the main challenge and is unfortunately rarely analytically tractable. 
				In order to circumvent this difficulty, the cumbersome replica method presented in \Sec\ref{main:sec:mean_field:replica_method} is a very powerful tool that we use intensively in many applications of this manuscript.
				Another option to avoid computing this average is to exchange instead the order of the expectation and the logarithm. This refers to the \emph{annealed} average \index{annealed average} that leads to the much simpler computation of the \emph{annealed free-entropy} defined by
				\begin{align}
					\Phi_{\ndim}^{\textrm{a}}\(\beta\)&\equiv \frac{1}{\ndim \beta} \log\EE_{\mat{J}}\mZ_\ndim \(\beta,\mat{J}\)\,.
				\end{align}
				As discussed in \Sec\ref{main:intro:disordered:quenched_annealed}, this simplification is justified only for systems with \emph{annealed disorder} so that the spins and the disorder fluctuate with the same time scales $\tau \sim \tau_q$ and appear on the same footing. Consequently, in this case it is necessary to take the thermal Gibbs average $\EE_\bsigma$ and disorder average $\EE_{\mat{J}}$ simultaneously before taking the logarithm, which keeps only the large deviation behavior of the system.	
				\graffito{Jensen inequality states that for a convex function $f$, the secant line of a convex function lies above the graph of the function
					\begin{tikzpicture}[scale=0.75, transform shape]
						\draw[edge, -latex, thick] (-1.5,0) -- (2.5,0);
						\draw[edge, -latex, thick] (0,-1) -- (0, 2);	
						\draw[burntorange, thick]   plot[smooth,domain=-1.5:2.5] (\x, {1/3*\x*\x - 0.25});
						\draw[teal, thick, dashed]  (-1, +1/3 - 0.25) to (1.5, 1/3*1.5*1.5 - 0.25);
					\end{tikzpicture}	
				}
				Even though this simplification is justified for annealed disorder, in the presence of a quenched disorder this abusive annealed simplification for quenched disorder provides in fact an \emph{approximation} of the cumbersome quenched average. 	
				More precisely, because of the concavity of the logarithm and using the Jensen inequality, we observe that the annealed free entropy is an \emph{upper bound} of the quenched free entropy
				\begin{align*}
					\Phi_\ndim\(\beta\) \equiv  \frac{1}{\ndim} \EE_{\mat{J}} \log \mZ_\ndim \(\beta,\mat{J}\) \leq  \frac{1}{\ndim} \log\EE_{\mat{J}} \mZ_\ndim \(\beta,\mat{J}\)  \equiv \Phi_\ndim^{\textrm{a}}\(\beta\)\,.
				\end{align*}
				Finally in the cases where the quenched average is intractable, we can still hope that the simpler annealed average provides a good approximation.
					
			\subsection{On the spin glass phase}
				The thermodynamic behavior of spin glass systems are drastically affected by the appearance of the quenched disorder in the Hamiltonian that is responsible for the emergence of a new collective behavior: the \emph{spin glass} phase.
				Remarkably, as a consequence of the frustration many local constraints may not be satisfied at the same time and thus there eventually exists many distinct ground state configurations with the same, strictly positive, energy level.
				As a result, in contrary with the classical Ising model in \Sec\ref{main:intro:classical_physics:examples:ising_example} where only two phases with positive or negative macroscopic magnetizations emerged, in the spin glass phase we observe a highly non-trivial \emph{ergodicity breaking} of the configuration space such that the Gibbs distribution exhibits many metastable states. In other words, the system is stuck in some sub-regions of the configuration space $\chi_\ndim$ and can take exponential time (in the size of the system $\ndim$) to explore the whole configuration space.\\
				
				To illustrate the remarkable properties observed in experimental spin glasses, we recall the seminal \aclink{EA} model \cite{Edwards1975} that tries to capture the main features of the spin glass phase.
				
				\paragraph{The Edwards-Anderson model}
					Since the interactions in metal alloys are short-range, in order to replace the randomness induced by the impurities positions, it was proposed to consider an Hamiltonian defined on a regular graph $\mG(\rV, \rE)$ with \emph{nearest-neighbor} interactions:
					\begin{align}
				    	\mH_\ndim\(\bsigma; \mat{J}, \vec{h} \) = - \sum_{<ij> \in \rE} J_{ij} \sigma_i \sigma_j - \sum_{i \in \rV} h_i \sigma_i\,.
				    	\label{main:intro:disordered:ea_hamiltonian}
					\end{align} 
					The exchange couplings can be chosen either Gaussian $\rP(J_{ij}) = \mN_{J_{ij}}\(0, \frac{J_0}{\ndim}\)$ or binary $\rP(J_{ij}) = \frac{1}{2} \(\delta(J_{ij}-J_0/\ndim) +  \delta(J_{ij}+J_0/\ndim) \)$ such that $\EE_{\mat{J}} J_{ij}= 0$ and $\EE_{\mat{J}} J_{ij} ^2 = \frac{J_0}{\ndim}$, for some $J_0 >0$. 
					Let us introduce the averaged total magnetization $m_\ndim$ and the celebrated \aclink{EA} order parameter specifically designed to reveal the spin glass phase:
					\begin{align*}
						m_\ndim &= \frac{1}{\ndim} \sum_{i=1}^\ndim \EE_{\mat{J}, \bsigma} \[\sigma_i \] \,, && q_{\textrm{ea}} = \frac{1}{\ndim} \sum_{i=1}^\ndim \EE_{\mat{J}, \bsigma}\[\sigma_i\]^2 \,,
					\end{align*}
					where the average are first taken with respect to the Gibbs distribution before taking the average with respect to the disorder $\mat{J}$. The particularity of the \aclink{EA} model is to present no ferromagnetic nor anti-ferromagnetic phase. 
					In fact, as expected at high-temperature $\beta \to 0$, we observe a paramagnetic phase with a global magnetization $m=0$ and $q_{\textrm{ea}}=0$. At the critical \emph{glass transition} $\beta \geq \beta_g$ the system enters the so-called \emph{spin glass phase} characterized by zero global magnetization $m=0$ but a non-zero \aclink{EA} order parameter $q_{\textrm{ea}} \ne 0$. 
					In other words, even though there is no global ordering of the system as the global magnetization stays zero in average, each individual spin dynamics is still frozen in a preferred orientation.
					Indeed, the time auto-correlation function is non-zero and given by the \aclink{EA} order parameter $C(t) = \frac{1}{\ndim} \sum_{i=1}^\ndim \EE_{\bsigma} \sigma_i(t)\sigma_i(0) \underlim{t}{\infty} q_{\textrm{ea}} \ne 0$. 
					This means that the system at time $t$ $\bsigma(t)$ is strongly correlated to the initial configuration of the system $\bsigma(t=0)$. In other words, the system has a \emph{strong memory} of the \emph{initial preparation} of the system. This \emph{aging} phenomenon \cite{sompolinsky1982relaxational}, measured by the new \aclink{EA} order parameter, has been experimentally observed, for instance, by measuring the magnetic susceptibility with different system initialization \cite{vincent2007ageing}.
					
		\subsection{Spin glass models and computer science}
			After the \aclink{EA} model breakthrough, various disordered models came up to light and boosted the spin glass literature. Let us mention the celebrated \aclink{SK} model \cite{Sherrington1975}, whose dynamics have been studied in \cite{cugliandolo1994out} and rigorously proven in \cite{talagrand1998sherrington, panchenko2013sherrington}, the $p$-spin for $p\geq 3$ interactions with binary or continuous variables for structural glass theory \cite{gardner1985spin, kirkpatrick1987p, kirkpatrick1987dynamics, crisanti1995thouless, Crisanti1992}.
			We shall mention as well the \emph{Random Energy Model} \cite{Derrida1981, mezard2009information}, which is one of the simplest toy model that exhibits a glassy phase; the \emph{KPZ equation} \cite{kardar1986dynamic} that describe the behavior of particles in a rough random landscape whose theory has been recently confirmed numerically by precise simulations \cite{hartman1982ordinary}; or the \emph{Stochastic Block Model} \cite{decelle2011asymptotic} for community detection on random graphs.\\
			
			This kind of glassy dynamics is believed to be present in many systems such as in computer science problems that we will be interested in the main contributions of this manuscript.
			Notably, \emph{combinatorial optimization} and \emph{random constraints satisfaction problems} gained in importance with especially error correcting code such as LDPC in noisy communication channels \cite{shannon1948mathematical, mackay1996near}, \emph{the minimum spanning tree}, \emph{Eulerian circuits}, \emph{Hamiltonian cycles}, the \emph{Travelling salesman problem} or \emph{partitioning} problems \cite{mezard2009information}. 
			These random optimization problems such as random $k$-satisfiability problems \cite{ricci2001simplest, mezard2002random, mezard2005clustering} or graph coloring \cite{jensen2011graph,mulet2002coloring,zdeborova2007phase} can be formulated as generic \aclink{CSP} whose Gibbs distribution was studied in details in \cite{krzakala2007gibbs}.
			Similarly to a physical system being frozen in a sub-region of the configuration space, a similar ergodicity breaking might dramatically impact the algorithmic performances of sampling and optimization algorithms in computer science problems. 
			Indeed it is believed that the existence of exponential metastable states may drastically harm the computational performances and explain the computational hardness in random problems such as \aclink{CSP} \cite{mezard2009information, zdeborova2016statistical} and other optimization problems \cite{moore2011nature}.\\
			
			In the recent years, due to the accession of machine learning and neural networks, we observed a renewed interest of statistical physics in these computer science problems. Interestingly, very often they can be formulated as spin glass models and treated with the corresponding set of powerful tools.
			In the next section, we propose a brief historical review of the exchange of ideas between statistical physics and computer science.

\ifthenelse{\equal{\format}{oneside}}
	{\clearpage\null\thispagestyle{empty}}
	{\cleardoublepage}
\chapter{Statistical physics and machine learning back together}
\label{chap:phys_ml_together}
	Analyzing machine learning problems with statistical physics tools may be unusual to most of the computer science community. Yet, there exists a rich literature with influential connections between these two fields, that we briefly review in this chapter. On one hand, statistical physics aims to understand collective behaviors of matter and phase transitions as illustrated in \Chap\ref{chap:statistical_physics}. 
	However, its powerful formalism readily applies to various fields such as \emph{statistical inference}, whose goal is to \emph{detect} and \emph{recover} a \emph{hidden signal} from observations. 
	In particular, the \emph{high-dimensional statistical regime} in which the number of data and parameters diverge fits perfectly the underlying fundamental large-size hypothesis of the statistical physics framework. 
	Thus, approaching high-dimensional inference and other machine learning problems with statistical physics has a long tradition and an intimate connection which is widely depicted in the literature of \emph{statistical physics of machine learning} \cite{nishimori2001statistical,mackay2003information,mezard2009information,grassberger2012statistical,zdeborova2016statistical, Advani2016b,zdeborova2017machine,biehl2019statistical, NatureZdeborova}.\\
			
	In this section, we first recall the main interactions between these two fields by presenting a short historical overview in \Sec\ref{chap:phys_ml_together:physics_history}. 
	Then we focus on the main contributions that influenced the current statistical physics approach in \Sec\ref{chap:phys_ml_together:physics_history_current}, before presenting our global approach in \Sec\ref{chap:intro:phys_inference} that we deeply use in the main contributions \Part\ref{part:contribution}. 
	In particular, we depict how mean-field methods such as the replica method or message passing algorithms, presented in details in \Sec\ref{main:sec:mean_field:replica_method}-\ref{main:intro:mean_field:bp}, became central to analyze the phase transitions of inference problems and can lead to the design of new algorithms. 
	All along this work, we try to especially highlight and compare the algorithmic phase transitions to optimal statistical thresholds in light of glassiness behaviors and computational hardness.

	\section{A common history of machine learning and statistical physics}
	\label{chap:phys_ml_together:physics_history}
		The intimate connection between machine learning and statistical physics basically started in the 80's and was recently renewed with the democratization and accession of \aclink{ANN}. After the recent successes of \aclink{DL} in many applications, the scientific communities from various fields try to address many of the theoretical challenges raised by their empirical successes. In particular, statistical physics experienced a renewed interest in \aclink{ANN} research with in particular the emergence of rigorous justifications of former heuristic statistical physics methods. 
		The goal of this short section is to briefly recount the main influential works of the statistical physics approach on open \aclink{ML} questions as well as the intricate story between physics and neural networks.
		
		\subsection{From spin glass theory to rigorous machine learning}
			The connection of \aclink{IT} with physics dates probably back to the end of the 1900's early 2000's with Maxwell, Boltzmann, Szilard that study the entropy in thermodynamical systems. It opened a breach for \aclink{IT} whose Shannon became the pioneer followed later on in the 60's by the Gibbs-Bogoluibov-Feynman variational principle that became a central tool in approximate statistical inference, as detailed in \Sec\ref{main:sec:mean_fields}. 
						
			\subsubsection{Emergence of the spin glass community}
				Later on, during the second \aclink{AI} winter, the physicists Hopfield \cite{hopfield1982neural} revived the \aclink{ANN}-oriented research by proposing the celebrated eponym \emph{Hopfield model}\index{Hopfield model} to explain \emph{associative memory} as a variant of the Ising model with pairwise interactions generated from a set of $\nsamples$ patterns $\{\bxi_\mu\}_{\mu=1}^\nsamples$ such that $J_{ij} = \frac{1}{\nsamples} \sum_{\mu=1}^\nsamples \xi_{i \mu} \xi_{j \mu}$. This energy based model crystallized particularly the interest of physicists and is certainly responsible of the emergence of an entire branch of the statistical physics dedicated to \aclink{ML} models.
				By taking advantage of the heuristic tools of the spin glass community, mainly developed with the previous \aclink{EA} and \aclink{SK} models \cite{Edwards1975, Sherrington1975, thouless1977solution}, such that the \aclink{TAP} approach, the Hopfield model is analyzed heuristically \cite{amit1985storing, amit1987statistical} and opened the door to more complex \aclink{ANN} models. 
				The same heuristic mean-field methods are then used to analyze \aclink{ANN} and started being popularized in \cite{amit1985spin}. For instance Boltzmann machines, which are nothing more than a stochastic version of the \aclink{SK} spin glass model, have been brought to light in the computer science community \cite{ackley1985learning}, making the connection of statistical physics and machine learning even closer. 
				In parallel, computer scientists developed the \aclink{PAC} theory to analyze the generalization property of neural networks \cite{valiant1984theory}, which turns out to be completely orthogonal to the physicist approach used to analyze the Hopfield model. Hence, even though the physics community deeply contributed in the early analysis of \aclink{ANN} models, the computer science community largely ignored the corresponding approach based on heuristic techniques. 

				In the late 90's, E. Gardner introduced the \emph{replica method} to analyze the \emph{maximum storage capacity} \cite{gardner1987maximum, gardner1988space} that is known to be closely related to the \aclink{VC} dimension mostly considered in the \aclink{ML} community. These very influential works introduced a powerful technic used to compute the typical configurations space volumes in order to count how many networks achieve a certain level of error. Many heuristic papers followed \cite{derrida1987exactly, gardner1988optimal,krauth1989storage} and readily apply the \emph{Gardner approach} to supervised learning with \emph{randomly-quenched disorder} for which the random labels are not correlated with the inputs. 
				In parallel, the training method of \aclink{SVM} \cite{boser1992training} was inspired by a physics intuition \cite{krauth1987learning}. During the same years, a deeply influential review \cite{mezard1987spin} gathered the main mean-field treatments from the fruitful research on spin glasses whose publication accelerated and democratized their use.
	
			\subsubsection{From random labels to learning a rule}
				After having widely studied the storage capacity problem with random quenched disorder and random labels, the research shifts towards the \emph{statistical inference} of a hidden \emph{signal}, that a supervised model shall recover from observations.
				The idea that the training set contains a hidden \emph{planted configuration}, representing a \emph{crystal} configuration in the physicist language, also called a \emph{rule}, refers to the so-called \aclink{T-S} scenario. Both these \emph{random} and \emph{structured} settings were in fact already introduced in the seminal work \cite{gardner1989three}.
				The replica method and the \aclink{TAP} approach started being applied to more general inference problems such as this \aclink{T-S} for the simplest \aclink{ANN}, the perceptron. The first learning curves and physics-like phase transitions \cite{gyorgyi1990first} are derived and exhibit interesting physics: first and second order phase transitions with the existence of metastable states are observed \cite{Sompolinsky1990, Opper1990, hansel1990learning, Kinzel96} keeping the interest of physicists at the highest level.
				This simple model architecture is then pushed forward with a second untrained layer: the \emph{committee machine} \cite{schwarze1992generalization,schwarze1993learning,schwarze1993generalization}. In another direction, the usage of gradient-descent-like algorithms is studied in an online setting \cite{saad1995line}, where a single example, from an unrealistic infinite reservoir of examples, is observed at each time step.\\
				
				Unfortunately, physics contributions had almost no impact in the \aclink{ML} community that largely frustrated the physics community. Even though the approach was very elegant for the physicist oriented mind, it was not taken fully seriously mainly because of its lack of rigor.
				Moreover with the decline of \aclink{AI} attraction, in the late 90's the physics research globally stopped in this direction. 
				 
			\subsubsection{A renewed interest of physicists and rigorous justifications}
				With the recent flourishing numerical successes of \aclink{ML} and \aclink{DL}, the theoretical research activity around these disciplines grew up again in the recent years.  
				Especially because traditional \aclink{ML} theory, based on data-independent \aclink{PAC} generalization bounds \cite{vapnik1994measuring}, predicted that models such as \aclink{DNN} with a number of parameters similar to a number of data should overfit. Thus it failed explaining the empirical and striking \emph{generalization problem} of \aclink{DNN} that does not experienced overfitting.
				Therefore, statistical physics community stroke back and started to work in this direction as they believe that their singular typical case approach, yet on unreasonable simple models, may contribute to understand this challenge and answer fundamental \aclink{ML} questions.
				In order to finally bring impact of the physicists heuristic methods to the \aclink{ML} community, the mathematical-physics research started to prove rigorously results previously derived in the spin glass literature with the so-called \emph{replica method}, see \Sec\ref{main:sec:mean_field:replica_method}. 
				This stage starts with a first rigorous tentative \cite{haussler1996rigorous} where they rigorously showed the existence of phase transitions in the learning curve behavior. The analogy between spin glasses of dynamical systems and machine learning seem very interesting but again failed to fully break through.				
				Finally, with the works of \cite{guerra2002thermodynamic,talagrand2003spin,panchenko2013sherrington}, which rigorously proved heuristic results of the 80's in the context of the \aclink{SK} model, \aclink{IT} started slowly to consider the statistical mechanics approach. This renewed approach of statistical mechanics techniques are currently gaining in popularity as well in \aclink{ML} because of their recent rigorous justifications for instance in the case of the \aclink{GLM} \cite{barbier2016mutual,Reeves2016} and the committee machine \cite{Aubin2018}, whose models have been both studied heuristically in the 90's, or deeper architectures \cite{Gabrie2017}.
								
			\subsubsection{Algorithms and computational complexity}
				The influence of statistical physics is even more obvious in combinatorial optimization and \aclink{CSP}. Indeed, early 2000 many graphical model algorithms such as \aclink{BP} \cite{pearl1982reverend, yedidia2001generalized, Yedidia2001} are popularized. These algorithms derived in different fields under different names such as the Viterbi algorithm, Pearl's \aclink{BP}, Gallager codes, Kalman filter, transfer-matrix approach \cite{Yedidia2001} are closely related to the physics \emph{cavity method} \cite{mezard2009information}. The simplification of the \aclink{BP} equations under a set of assumptions, see \Sec\ref{main:sec:mean_field:amp}, leads to \aclink{AMP} algorithms, introduced in the context 
				of \aclink{CS} in \cite{donoho2006compressed, maleki2011approximate} and popularized in \cite{montanari2012graphical,rangan2011generalized}. 
				These physics-inspired algorithms are applied to various \aclink{CSP} whose general Gibbs measure description was studied in \cite{krzakala2007gibbs}. Very importantly, this renewed line of \aclink{ANN} research made a clear connection with the algorithmic computational complexity \cite{moore2011nature} that was never considered in the early statistical physics literature.
				 											
		\subsection{Recent and current line of research}
		\label{chap:phys_ml_together:physics_history_current}
			Statistical physics is currently pursuing actively this line of research and attempting to answer fundamental questions raised by the increasing use of \aclink{DNN}. We present below a short and,  inevitably, biased selection of important research directions from a physicist point of view.
						
			\subsubsection{From AMP to the analysis of GD algorithms}
				In many models the \aclink{BP} algorithm and variants such as \aclink{AMP} are of theoretical interest since they have been shown and believed to achieve the optimal statistical performances in large regions of parameters. Easily derived and implemented for finite sizes, their performances are nonetheless not guaranteed.
				But powerfully, the statistical physics approach turns out very useful as it allows to derive and prove the high-dimensional asymptotic performances of the corresponding \aclink{AMP} algorithms, called in this context the \emph{state evolution}. While this requires in principle to compute a high-dimensional \aclink{JPD}, the physics mean-field methods reduce it to a simple optimization problem over a small set of order parameters. 
				Yet the prevalence of gradient-based algorithms in \aclink{DL} recently shifted the current research towards understanding \aclink{GD} dynamics. Indeed, dynamics of \aclink{GD} is believed to be very important as it induces a bias that reduces the wide hypothesis class along training by diffusion and is responsible for good generalization. 
				Even though the analysis of \aclink{GD} dynamics has been performed for linear models in \cite{baldi1991temporal,baldi1995learning,dunmur1993learning,krogh1992simple,advani2017high}, generalizing it to non-linear models remain challenging.
				Yet, first steps in this direction have been recently performed. 
				Following the early works of \cite{Saad1995a}, the dynamics of \emph{online} \aclink{SGD} was studied in more details and generalized to more complex architectures \cite{goldt2019dynamics}.
				In the other hand, following the dynamical approach studied in the early works \cite{cugliandolo1993analytical, arous2006cugliandolo} in the context of the $p$-spin model, it was recently extended to the perceptron \cite{agoritsas2018out}, the spiked matrix model \cite{mannelli2020marvels} and a Gaussian mixture classification task \cite{mignacco2020dynamical}. Generalizing this dynamical approach to more complex architectures and data structures is certainly a fruitful direction of research.		
				
			\subsubsection{The role of data: from iid to a manifold}
				Another essential ingredient in understanding \aclink{DL} performances is definitely the essential role of data. Most theoretical statistics works commonly assume that data come from a \aclink{i.i.d} factorized probability distribution, without explicitly modelling the training dataset. As a consequence, these approaches lack capturing the deep correlations of real datasets and their fundamental impact on the training of \aclink{DNN}.
				A first step to overtake this \aclink{i.i.d} limitation was performed in \cite{kabashima2008inference} by generalizing it to rotationally invariant inputs in perceptrons and later on to the weights in \aclink{DNN} \cite{gabrie2018entropy}.
				Moreover, the original \aclink{T-S} scenario fed with \aclink{i.i.d} samples is also gradually challenged as the dynamics of \aclink{DNN} on real-life tasks such as MNIST classification do not reveal the same dynamics than for a \aclink{T-S} synthetic dataset. 
				To capture this particular learning dynamics on MNIST, \cite{goldt2019modelling} introduced the \emph{Hidden manifold model} to represent the input data by a low-dimensional structure, that was studied later on in \cite{gerace2020generalisation} in the context of random features. This rich data modelling idea is another promising step to take into account the importance of real-data distributions in the learning dynamics. 	
					
			\subsubsection{From a few hidden units to deep/wide layers}
				\paragraph{Multi-layer and over-parametrization}
				Finally, the last ingredient responsible for the \aclink{DNN} success is certainly the wide and deep architectures of networks that form a large hypothesis class with a great expressivity.
				While early rigorous works in statistical physics focused on simple single-layer perceptron \cite{barbier2016mutual}, the current trend consists in analyzing models with increasing sizes, starting with a simple two-layers extension \cite{Aubin2018}.
				In parallel, another mean-field scaling limit was recently proposed, where the number of hidden units $K$ is much larger than the input size $\ndim = o(K)$. 
				In a recent line of research, \cite{jacot2018neural, du2018gradient, allen2019convergence, arora2019exact, lee2019wide} observed that with a scaling of the weights as $\Theta(K^{-1/2})$, the dynamics enters a \emph{lazy regime} governed by the Neural Tangent Kernel (NTK). As a consequence, it remains stuck close to the initialization and can be therefore trivially analysed. 
				In contrast, in the same infinitely wide limit, but under a different scaling $\Theta(K^{-1})$, \cite{chizat2018global,mei2018mean,rotskoff2018neural} observed another, yet more interesting, \emph{feature learning} regime in which the NTK really learns. 
				Closely related, \emph{random features} was the subject of various works notably to understand the learning curve behavior and double descent generalization phenomena \cite{belkin2019reconciling, mei2019generalization,d2020double}.
				It was also studied with \emph{Random Matrix Theory} (RMT) applied to 
				single-layer random neural networks by analyzing the Gram matrix of the hidden units \cite{louart2018random, couillet2011random}.
				In another direction, in order to evaluate the \emph{information bottleneck} theory in \aclink{DNN} \cite{tishby2015deep} suggested a connection with representation compression whereas \cite{gabrie2018entropy} developed a rigorous scalable formula for mutual information between layers of multi-layer neural networks. Finally, another approach \cite{mehta2014exact} consists in applying ideas of the physics renormalization group to \aclink{DNN} whose idea is to learn hierarchal representations across layers. 
				
				\paragraph{Beyond separable priors}
				Recently, we observed a practical and intense use of deep neural-network-based generative priors for estimation problems \cite{bora2017compressed, tramel2016inferring}. Whereas in classical statistics, we often assume that the hidden ground truth signal is drawn from a separable prior, this overly simple hypothesis may be replaced by a generative prior to model a more complex, non-separable \aclink{JPD} of the signal. 
				Therefore the practical use of generative priors angled research towards understanding them in simple estimation problems such as compressed sensing, phase retrieval of spike matrix models \cite{aubin2019spiked,aubin2019exact} and to design multi-layer approximate message passing algorithms \cite{manoel2017multi, fletcher2018inference}.

	\section{Statistical inference and CSP as a statistical physics problem}
	\label{chap:intro:phys_inference}
		In this section, we present the general approach used in the main contributions of this manuscript to analyze various models. 
		Among them, we will consider two large classes of problems already mentioned: \aclink{SI} and \aclink{CSP}. Both kind of problems can be formulated as a statistical physics model, and classical tools of disordered systems, presented in \Sec\ref{main:chap:mean_field}, readily apply in certain scaling limits.
		The connection between statistical physics, \aclink{SI} and \aclink{CSP} is not relatively new and has in fact a long history as sketched in \Sec\ref{chap:phys_ml_together:physics_history}. Especially, influential and seminal works \cite{shannon1948mathematical,Jaynes57} brought to light the link between \aclink{IT}, Bayesian inference, thermodynamics and statistical physics. The connection was more recently renewed during the second \aclink{AI} winter \cite{grassberger2012statistical} and was celebrated during a recent summer school \emph{Statistical Physics and Machine Learning back together}.
		All along this manuscript, we stress and make an intense use of the deep connection between Bayesian inference and spin glass techniques \cite{mezard1987spin} that were early applied to error correcting codes \cite{sourlas1989spin}, perceptrons \cite{seung1992statistical, watkin1993statistical} and sparse random graphs \cite{mezard2001bethe, mezard2003cavity}. See \cite{nishimori2001statistical, mezard2009information} for an extended review.
		Moreover, as early heuristic works were not focussing on algorithmic considerations, in this manuscript we will put the accent on rigorous results and algorithmic thresholds.
		Before closing this chapter and presenting in details the mean-field methods, we provide a high-level perspective of the general approach of this work. The same approach will be used on different problems and it is therefore useful to summarize it once for all. Finally we present the generic phase diagram descriptions of \aclink{CSP} and \aclink{SI} models.
		
		\subsection{Bayesian inference in the high-dimensional regime}
		\label{chap:intro:phys_inference:high_dim}
		\label{chap:intro:phys_inference:bayesian}
		
		\subsubsection{High dimensional regime}
		Classical statistics traditionally considers models with a finite number of parameters $\ndim$. Yet, the recent progresses of \aclink{DNN} drove the usage of modern \aclink{ML} models with increasing number of parameters. Additionally with the widely increasing availability of data, the classical statistical regime must be rethought.
		In contrast with classical statistics, the sizes of the dataset $\nsamples$ and the number of parameter $\ndim$ are assumed to very large and even \emph{infinite}. These limits are particularly suitable to the statistical physics approach that requires the \emph{thermodynamic limit} $\ndim \to \infty$ to proceed. Therefore, we mostly consider that both the number of parameters $\ndim \to \infty$ as well as the number of observed data $\nsamples \to \infty$. To be able to tackle analytically these ill-defined behaviors we assume that they both go to infinity with a fixed and finite ratio $\alpha \equiv \frac{\nsamples}{\ndim} = \Theta(1)$. This simplifying assumption is however not arbitrary and reflects quite correctly the practical dimensions. For instance the MNIST \cite{mnist10} dataset contains $60000$ images and can be learned correctly by a two-layers network with biases with $\ndim=(784+1) \times 32 + (32+1) \times 10 = 25450$ parameters, so that the ratio $\alpha$ is indeed of order one.

		\subsubsection{Statistical inference}
			Let us consider a set of interacting variables $\bsigma$ defined on a graph $\mG(\rV, \rE)$. The overall goal in \aclink{CSP} and \aclink{SI} problems is to compute the \emph{marginal distributions} $\rP_\ndim(\sigma_i) = \int_{\chi_{\ndim-1}} \d \bsigma_{\setminus i} \rP_\ndim(\bsigma) $ accessible from the knowledge of the \aclink{JPD} $\rP_\ndim(\bsigma)$. However in the high-dimensional regime this \aclink{JPD} is a high-dimensional object that is very often not tractable analytically. 
			That is where statistical physics comes into play. Indeed, statistical physics with its long-history provides a suitable and powerful set of tools, see \Sec\ref{chap:statistical_physics}, to analyze and characterize the corresponding high-dimensional \aclink{JPD} $\rP_\ndim(\bsigma)$. 
			Moreover its extension to disordered systems and spin glasses, see \Sec\ref{sec:main:introduction:stat_phys:disordered}, makes it singular and a very powerful approach to analyze a \aclink{JPD} of the form $\rP(\bsigma \vert \vec{y})$ in presence of a \emph{quenched disorder} $\vec{y}$, such as the randomness in the \emph{observed data}.
			In the perspective that we will apply these tools to supervised \aclink{ML} applications, let us draw the correspondence between physics spin models and \aclink{SI}. 
			While in physics spin models, the randomness steps in through the exchange interactions $\mat{J}$ that follow a particular distribution $\rP(\mat{J})$, see \Sec\ref{main:intro:disordered:quenched_annealed}, in most of \aclink{ML} applications, the randomness intervenes in the distribution of the input data $\mat{X}$ and the corresponding labels $\vec{y}$ in a supervised setting. 
			Thus, the spin configuration $\bsigma$ will naturally denote the value of the \emph{parameters} of the \aclink{ML} model. More details on the connection between statistical physics and Bayesian-inference can be found in \cite{engel1993statistical,nishimori2001statistical, mezard2009information, grassberger2012statistical, zdeborova2016statistical, Advani2016b}.		
			
			\subsubsection{Bayesian inference as a statistical physics model}
				In order to compute the \aclink{JPD}, we use a particularly suitable Bayesian approach based on the \emph{Bayes-formula} decomposition, see \Sec\ref{main:intro:ml:bayesian_approach},		
				\begin{align}
					\rP_\ndim \(\bsigma \vert \vec{y} \) = \frac{\rP\(\vec{y} \vert \bsigma\) \rP\(\bsigma\) }{\rP\(\vec{y}\)} = \frac{1}{\rP\(\vec{y}\)} \prod_{\mu=1}^\nsamples \rP \(y_\mu \vert \bsigma_{\partial_\mu} \) \prod_{i=1}^\ndim \rP(\sigma_i)\,,
					\label{main:eq:inference:jpd}
				\end{align}
				where we used the fact that, in many examples, the joint \emph{channel} and \emph{prior} distributions $\rP\(\vec{y} \vert \bsigma\)$ and $\rP\(\bsigma\)$ respectively factorize over the $\nsamples$ observations and $\ndim$ input dimensions. This decomposition is very interesting in the sense it explicitly shows the distributions used to model the observations: the prior distribution $\rP\(\bsigma\)$ describes the prior knowledge we have on the variables $\bsigma$, \eg discrete binary, Laplace, Gaussian, etc. whereas the distribution $\rP\(\vec{y} \vert \bsigma\)$ models how the observations are related to the variables, for instance through a noisy Gaussian channel, a linear matrix multiplication, etc.
				To properly cast this problem into a statistical physics formalism, we shall introduce the Hamiltonian
				\begin{align}
					\mH_\ndim\(\bsigma; \vec{y} \) \equiv - \sum_{\mu=1}^\nsamples \log   \rP\(y_\mu \vert \bsigma_{\partial_\mu} \)  - \sum_{i=1}^\ndim \log \rP(\sigma_i)
				\end{align}
				so that the \aclink{JPD} in \eq\eqref{main:eq:inference:jpd} can be formulated as a Gibbs distribution
				\begin{align}
					\rP_\ndim\(\bsigma \vert \vec{y} \) = \frac{ e^{-\beta \mH_\ndim\(\bsigma; \vec{y} \) }}{\mZ_\ndim(\vec{y})}\,, \text{ with } \mZ_\ndim(\vec{y}) \equiv \int_{\chi_\ndim} \d \bsigma ~ \rp\(\vec{y} \vert \bsigma\) \rp\(\bsigma\) \,,
					\label{main:eq:inference:Gibbs}
				\end{align}
				as soon as the inverse temperature is set to $\beta=1$. Yet, the temperature parameter may be freely chosen depending on the statistical estimator that we will consider. For instance to obtain the \aclink{MAP} behavior, we should take the zero temperature limit $\beta \to \infty$.
				Moreover, in a physics language, $\log \rP(\sigma_i)$ is exactly analogous to the local external field interaction $h_i \sigma_i$ in spin systems, while the term $\log \rP(y_\mu \vert \bsigma_{\partial_\mu} )$ represents the interaction term between $|\partial_\mu|$ variables. In particular, this analogy allows to compute easily the marginal probability $\rP(\sigma_i) = \int_{\chi_{\ndim-1}} \d \bsigma_{\setminus i} \rP(\bsigma) $ as a simple local magnetization.
				Written under this general formulation with generic prior distributions, it has the deep advantage to encompass a large class of models: Ising, \aclink{SK}, $p$-spin models, \aclink{GLM}, committee machine,... with a various choice of prior distributions ranging from discrete to continuous variables.

			\subsubsection{Free entropy and replica computation}
				As stressed in \Sec\ref{sec:main:introduction:stat_phys:ordered_definitions:generating_function}, the averaged free entropy $\Phi_\ndim =  \frac{1}{\ndim} \EE_{\vec{y}}  \log \mZ_\ndim  (\vec{y}) $
				being effectively the cumulant generative function of many useful quantities is therefore a central object in statistical physics.
				In the high-dimensional regime, $\ndim\to \infty, \nsamples \to \infty, \alpha=\Theta (1)$, we focus instead in the asymptotic averaged free entropy $\Phi = \lim_{\ndim \to \infty} \Phi_\ndim$ that can be computed with the \emph{replica method} that plays a crucial role in this work and detailed in \Sec\ref{main:sec:mean_field:replica_method}.
				To quickly summarize, while computing the high-dimensional \aclink{JPD} is often intractable, the replica method allows to reduce the high-dimensional inference problem to a simple \emph{optimization problem} of a free entropy potential $\Psi$ over a set of a few \emph{order parameters}, \eg $q,\hat{q}$,
				\begin{align}
					\Phi = \extr_{q, \hat{q}} \{ \Psi(q,\hat{q}) \}\,.
					\label{eq:intro:phys_inference:free_entropy}
				\end{align}
				The free entropy behavior allows to detect statistical thresholds and exhibit potential information theoretical phase transitions.
				Indeed, these \emph{order parameters} $q,\hat{q}$, called \emph{overlaps}, have a deep interpretation as they directly provide knowledge on the solution space: either the correlation with the ground truth solution in the case of \aclink{SI}, or the typical distance between solutions in the context of \aclink{CSP}. 
				
			\subsubsection{Towards rigor}
				Yet, the replica method, that we use intensively, is unfortunately not a rigorous method in the mathematical sense: a few important steps are not justified and may even seem absurd. But it turns out that in many cases the result was either proven or believed to be \emph{correct}. 
				As stressed in the historical part in \Sec\ref{chap:phys_ml_together:physics_history}, progresses and results in statistical physics have very often not been taken seriously in the mathematics and computer science community because of this lack of rigor.
				That is one of the reasons why researchers at the interface between mathematics and physics recently undertook to rigorously prove these results, which may have been obtained 20-30 years earlier by physicists.
				Even though proving the heuristic results does not provide new essential understanding of the system behavior, it nonetheless has the profound benefit to bring greater impact and visibility outside of physics. 
				To give a flavor, most of the proofs of this manuscript will be based on \emph{Guerra-interpolation} \cite{guerra2002central, Guerra2003, Talagrand06} that require fundamentally to previously derive the heuristic replica result. Other proofs are simply based on moment bounds \cite{friedgut1999sharp, achlioptas2002asymptotic,bremaud2017discrete} and Gordon's Convex Gaussian Min-max Theorem (CGMT) \cite{gordon1985some}.
					
		\subsection{Algorithmic perspectives}
			The free entropy computation gives access to the information theoretical phase transitions of the system, also called the \emph{statistical thresholds}.
			However, for practical purposes, we are interested in algorithms that are able to reach these theoretical performances. 
			Another interesting feature of statistical physics approach is that, very often, computations can be turned in very powerful polynomial algorithms, whose behaviors show new phase transitions. 
			Among them, in this dissertation we will focus on \emph{message passing} algorithms which have a long history with physics. On top of that, they have the advantageous property of being proven optimal in many applications \cite{maleki2011approximate,donoho2013information, barbier2016mutual} or believed so.
			To derive such algorithms, the first step is to represent the high-dimensional \aclink{JPD} with a factor graph, as illustrated in \Fig\ref{fig:main:factor_graph_1}.
			\begin{figure}[htb!]
			\centering
			\begin{tikzpicture}[scale=0.7, auto, swap]
			    \foreach \i in {1,...,6}
			        \node[fun] (F\i) at (1.5*\i,0) {};
			    \foreach \i in {1,...,5}{
			        \node[var] (X\i) at (0.75+1.5*\i,-1.5) {};
			        \node[field] (H\i) at (0.75+1.5*\i,-2.5) {};}
			    \foreach \a/\i in {1/1, 1/2, 1/3, 2/2, 2/3, 3/2, 3/4, 4/2, 4/3, 4/5, 5/4, 6/4, 6/5}
			        \path[edge] (F\a) -- (X\i);
			    \foreach \i in {1,...,5}
			    	\path[edge] (X\i) -- (H\i);
			    \node at (4.5,0) {$\Psi_\mu$};
			    \node at (6,0) {$\Psi_\nu$};
			    \node at (3.75 ,-1.5) {$\sigma_i$};
			    \node at (5.25 ,-1.5) {$\sigma_j$};
			    \node at (6.75 ,-1.5) {$\sigma_k$};
			    \node at (3.75 ,-2.5) {$h_i$};
			    \node at (5.25 ,-2.5) {$h_j$};
			    \node at (6.75 ,-2.5) {$h_k$};
			\end{tikzpicture}
			\hfill
			\begin{tikzpicture}[scale=0.7, auto, swap]
					    \foreach \i in {1,...,3}
					        \node[fun] (F\i) at (2*\i,0) {$\Psi$\i};
					    \foreach \i in {1,...,4}
					        \node[var] (X\i) at (-1 + 2*\i,-1.5) {$\sigma$\i};
					        
					    \foreach \a in {1, ..., 3}
        					\foreach \i in {1, ..., 4} 
        					{
           					\path[edge] (F\a) -- (X\i);
        					}
			\end{tikzpicture}
			\caption{Factor graph representation of \Left the joint probability distribution $\rP_\ndim(\bsigma \vert \vec{y})$ \eq\eqref{main:eq:inference:jpd}, \Right the linear system \eq\eqref{eq:intro:phys_inference:linear_system}.}
			\label{fig:main:factor_graph_1}
			\end{figure}
			From this factor graph, we may apply the \emph{cavity method} \cite{mezard2001bethe, mezard1987spin} or equivalently the \aclink{BP} equations, which can be simplified in an \aclink{AMP} algorithm under a Gaussian assumption in the thermodynamic regime. Notice that for more complex models, a more general approach was developed known as \emph{survey propagation} \cite{braunstein_Survey_propagation}.
			This set of iterative equations of the form $\forall i \in \lb \ndim \rb,~ \hat{\sigma}_i^{t+1} = f_i(\hat{\bsigma}^{t})$ can be iterated and gives an estimate of the marginal probabilities. 
			In addition of being very often optimal, \aclink{AMP} algorithms have the exceptional particularity that in certain situations their average infinite behavior, called \emph{state evolution}, is exactly characterize by the replica free entropy potential, that allows to compare their performances to the information theoretical statistical thresholds.
			Notably, if we define for instance the self-overlap parameter
			\begin{align*}
				q^{t}  =\lim_{\ndim \to \infty} \frac{1}{\ndim} \EE_{\vec{y}} \[ \bsigma^t \cdot \bsigma^t \] \,,
			\end{align*}
			at convergence and under certain conditions, the performance measures such as the generalization error or the \aclink{MSE} are characterised by the asymptotic overlap $q^{t=\infty}$, which is, strikingly,
			\graffito{Polynomial refers to the space complexity, meaning that at each time iteration, the algorithm requires a polynomial (in the size of system) number of operations.}
			equivalently the solution of the replica free entropy extremization problem in \eq\eqref{eq:intro:phys_inference:free_entropy}. In particular, this means that at each iteration \aclink{AMP} follows the gradient of the replica free entropy, until convergence to a maxima.
			However, whereas the information theoretical performances are characterized by the global maxima of the replica free entropy, since \aclink{AMP} iterations start with non-informative initializations, the algorithm may converge to some local maximum and achieve sub-optimal performances. This key observation reveals in particular the existence of \emph{hard} algorithmic regions as soon as the free entropy potential presents metastable states. 
			To conclude, we already see that the replica free entropy and the \aclink{AMP} algorithm are two sides of the same coin, and this observation will follow in extended discussions in all the considered applications. \\		
					
		Hereafter, we present the two types of problems we will mainly describe in the application part of this manuscript: \aclink{CSP} and \aclink{SI}, which are very similar but do not show the same phase transitions typology because the quenched disorder is, crucially, not of the same nature. In fact historically physics was first interested in \aclink{CSP} \cite{mezard1986replica,mezard1986mean, gardner1988optimal, krauth1989storage, mezard2002random, krzakala2007gibbs, zdeborova2007phase} before recently shifting towards phase transitions in \aclink{SI} \cite{decelle2011asymptotic,krzakala2012probabilistic, lesieur2017constrained, barbier2017phase}. Even though the phase transitions are slightly different, the above general Bayesian approach readily apply to their analysis. 
			
		\subsection{Random constraint satisfaction problems}
		\label{main:intro:phys_ml_together:random_csp}						
			Let us introduce general combinatorial optimization problems, generally called \aclink{CSP} \cite{apt2003principles,mezard2009information, tsang2014foundations}. This is a general definition that applies to various problems such as the $k$-SAT, sphere packing, Eulerian and Hamiltonian paths, travelling salesman, the q-coloring and the vertex-covering problems and many others. 
			A \aclink{CSP} is specified by $\ndim$ variables $\bsigma = \{ \sigma_i\}_{i=1}^\ndim \in \chi_\ndim$, where $\chi_\ndim$ denotes an alphabet, that must satisfy a set of $\nsamples$ constraints $\{ \Psi_\mu (\bsigma_{\partial_\mu}) \}_{\mu=1}^\nsamples$ within a given collection. 
			We say that a constraint is satisfied by the variables (resp. non-satisfied ) if $\Psi_\mu (\bsigma_{\partial_\mu})=1$ (resp. $0$). The problem is satisfiable (SAT) if there exist at least one configuration that satisfies all the constraints, and UNSAT otherwise.
			As a generalization, \aclink{rCSP} are a particular case when the constraints are drawn randomly among the collection \cite{franco1983probabilistic}. The randomness may be based either on the graph geometry by a random connectivity or any other random source in the constraints for fully connected variables. Such models can be seen as spin glass models with \emph{random quenched disorder} as the constraints are completely random.
			In order to maximize the number of satisfied constraints, we introduce an Hamiltonian defined as the number of violated constraints, that measures the energy of a configuration $\bsigma$
			\begin{align*}
				\mH_\ndim(\bsigma) \equiv \sum_{\mu=1}^\nsamples  \( 1-\Psi_\mu\(\bsigma_{\partial_\mu}\) \)\,,
			\end{align*}
			and the Gibbs distribution at finite temperature $1/\beta$, that represents the level of exigence we require on the satisfiability of the constraints,
			\begin{align*}
					\rP_\ndim \(\bsigma\) = \frac{1}{\mZ_\ndim(\beta)} e^{-\beta \mH_\ndim(\bsigma) } \underlim{\beta}{\infty} \frac{1}{\mZ_\ndim} \prod_{ \mu =1 }^\nsamples \Psi_\mu\(\bsigma_{\partial_\mu}\) \,,
			\end{align*}
			converges to a product of indicator functions at zero temperature $\beta \to \infty$. We introduce the rescaled number of constraints by the number of variables of the problem $\alpha=\frac{\nsamples}{\ndim}$. To fix ideas, let us provide a simple example: solving an \emph{affine system}. The interested reader may find various other examples in the literature starting with the reference book \cite{mezard2009information}. 
			
			\paragraph{A toy example: affine system of equations} 
			Consider you have a set of $\nsamples$ \emph{linear} equations, represented by the matrix $\mat{A} \in \bbR^{\nsamples \times \ndim}$ and a vector $\vec{b}\in \bbR^{\nsamples}$ with real coefficients, depending on $\ndim$ variables $\vec{x}\in \bbR^{\ndim}$
			\begin{align}
				\mat{A} \vec{x} - \vec{b} = \vec{0} \hhspace \Leftrightarrow \hhspace \prod_{\mu=1}^\nsamples \Psi_\mu\(\vec{x}\) = 1\,.
				\label{eq:intro:phys_inference:linear_system}
			\end{align}
			This linear system can be rewritten as a fully-connected \aclink{CSP} involving the product of constraints $\{\Psi_\mu\}_{\mu=1}^\nsamples$ with $\Psi_\mu(\vec{x}) = \id\[ \vec{a}_\mu \cdot \vec{x} - b_\mu \]$, where $\vec{a}_\mu$ represents the $\mu$-th row vector of the matrix $\mat{A}$. The problem may be represented by a factor graph illustrated in \Fig\ref{fig:main:factor_graph_1} \Rightn.
			For a random or deterministic matrix $\mat{A}$, we would like to know when the linear system has at least one solution. We shall remember that if the number of constraints is smaller than the number of variables  $\nsamples < \ndim$, the linear system is undetermined and there exists many degenerated solutions. While if $\nsamples > \ndim$ the system is over-constrained and there does not exist any solution, so that there exists an intermediate critical value $\alpha_{\textrm{sat}}=1$ such that solutions no longer exist, called the SAT-UNSAT phase transition.
			
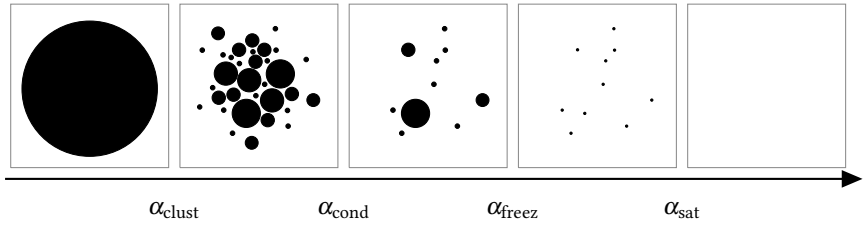
\begin{figure}[htb!]
				\centering
					\begin{tikzpicture}[scale=0.15]
				    \tikzstyle{dot}=[circle,minimum size=10pt, scale=0.4]
				    \tikzstyle{green}=[fill=green!70!black]
				    \tikzstyle{orange}=[fill=orange!90!black]
				    \tikzstyle{red}=[fill=red!90!black]
				    \tikzstyle{yellow}=[fill=yellow!95!black]
				    \tikzstyle{red}=[fill=red!90!black]
				    \tikzstyle{annot}=[text width=3cm, text centered, font=\footnotesize]
				    \tikzstyle{line}=[-,thick, black]
					\draw [->,thick, black] (0,-1) to (76,-1);
					\foreach \i in {0,1,2,3,4}
						\draw[draw=gray, fill=none] (15*\i + 0.5,0) to (15*\i + 0.5,14.5) to (15*\i + 14.5,14.5) to (15*\i +  14.5, 0) to (15*\i + 0.5, 0);
					\draw[fill=black](7.5,7) circle[radius=6cm];
					\foreach \x/\y/\r in 
						{6.36/2.20/0.57, 
						11.82/5.99/0.57,
						3.38/11.91/0.57,
						1.75/5.38/0.195,
						5.88/4.81/1.25,
						8.90/8.31/1.25,
						4.07/8.34/1.03,
						6.14/7.75/1.03,
						8.165/5.96/1.03,
						7.78/4.21/0.59,
						3.44/6.2/0.59,
						9.92/6.53/0.59,
						4.75/6.47/0.59,
						6.70/9.39/0.59,
						5.24/10.45/0.59,
						7.48/10.45/0.59,
						6.41/11.28/0.59,
						4.66/3.05/0.2,
						9.59/3.68/0.2,
						9.53/5.07/0.2,
						3.88/5.10/0.2,
						6.73/6.38/0.2,
						7.51/7.39/0.2,
						2.90/7.09/0.2,
						1.98/10.42/0.2,
						3.82/10/0.2,
						4.54/9.77/0.2,
						5.27/9.23/0.2,
						7.74/9.47/0.2,
						6.47/10.27/0.2,
						8.52/10.42/0.2,
						11.19/9.59/0.2,
						8.46/12.32/0.2
						}
						{
						\draw[fill=black](15.5+\x,\y) circle[radius=\r cm];
						}
									
					\foreach \x/\y/\r in 
						{
						11.82/5.99/0.57,
						5.88/4.81/1.25,
						5.24/10.45/0.59,
						4.66/3.05/0.2,
						9.59/3.68/0.2,
						3.88/5.10/0.2,
						7.51/7.39/0.2,
						7.74/9.47/0.2,
						8.52/10.42/0.2,
						8.46/12.32/0.2
						}
						{
						\draw[fill=black](30.5+\x,\y) circle[radius=\r cm];
						\draw[fill=black](45.5+\x,\y) circle[radius=0.1 cm];
						}

					\node[annot, below = 0.25cm] at (15, -0.5) {$\alpha_{\textrm{clust}}$}  ;
					\node[annot, below = 0.25cm] at (30, -0.5) {$\alpha_{\textrm{cond}}$}  ;
					\node[annot, below = 0.25cm] at (45, -0.5) {$\alpha_{\textrm{freez}}$}  ;
					\node[annot, below = 0.25cm] at (60, -0.5) {$\alpha_{\textrm{sat}}$}  ;
					
					\end{tikzpicture}
				\caption{Illustration of the solution configuration space of a random CSP crossing clustering, condensation, freezing and SAT-UNSAT phase transitions as a function of the constraint density $\alpha$, inspired from \cite{krzakala2007gibbs}.}
				\label{fig:intro:csp_phases}
				\end{figure}

			\paragraph{On phase transitions of rCSP}
			As illustrated with the above simple example, we understand intuitively that \aclink{CSP} and \aclink{rCSP} may undergo phase transitions as the constraints density $\alpha$ varies. In particular above a large number of constraints, if the system is heavily over-constrained it is intuitive that no configuration can be solution. In contrast, if the system is largely under-constrained, there will eventually exist many solutions.
			The SAT-UNSAT phase transition appears at the constraint density above which no solution exists. For instance in the case of the linear system, the SAT-UNSAT threshold is simply given by $\nsamples = \ndim \Leftrightarrow \alpha_{\textrm{sat}}=1$.
			Yet, the phase diagram of \aclink{rCSP} is not limited to this SAT-UNSAT phase transition and reveals a richer description. We invite the reader to read more about it in \cite{krzakala2007gibbs, mezard2009information} where the whole phase transition phenomenology of \aclink{rCSP} is described with the \emph{cavity} formalism. 
			In more details, for a small constraint density $\alpha \ll 1$, we expect that many solutions exist in a large connected sub-region of parameters. 
			As the density of constraints increases, a first remarkable \emph{clustering} phase transition, also known as the dynamical phase transition in the context of structural glasses \cite{parisi2010mean, charbonneau2017glass}, is encountered at $\alpha_{\textrm{clust}}$ and is characterized by the decomposition of the large set of solutions in an exponential number of disconnected sets, called clusters.
			Further increasing the density $\alpha$, we observe that this exponential number of clusters reduces to a sub-exponential number of clusters at the \emph{condensation} phase transition $\alpha_{\textrm{cond}}$.
			If the alphabet $\chi_\ndim$ is discrete, the system may as well undergo another phase transition: a \emph{rigidity} or \emph{freezing} phase transition \cite{semerjian2008freezing} at $\alpha_{\textrm{freez}}$ such that each cluster shrinks and contains only a finite number of solutions.
			Finally, above a certain constraint density $\alpha_{\textrm{sat}}$, the problem becomes UNSAT, meaning that no configuration can satisfy all the hard constraints simultaneously such that at least one constraint is violated and the ground state energy is strictly positive. 
			The description of the different phase transitions by increasing $\alpha$ is illustrated in \Fig\ref{fig:intro:csp_phases}, that can be completed by quantitative definitions \cite{krzakala2007gibbs, Gabrie2017}.
				
		\subsection{Statistical inference and supervised learning}
		\label{main:intro:phys_ml_together:supervised}

			\aclink{SI} denotes the process of extracting useful informations and properties of an underlying high-dimensional joint probability distribution $\rP_\ndim(\bsigma)$ from the observations of data. 
			With the large amount of data nowadays available, statistical inference apply to many applications from computer science with \aclink{ML}, \aclink{DL}, signal processing and \aclink{IT}, to natural sciences with medicine, neuroscience, biology, social sciences or economy, etc. \aclink{SI} can be thought as the action of extracting informations from a large set of data or in other words recovering a hidden signal from a set of observations. The literature about \aclink{IT} and \aclink{SI} is very rich and more details can be found in \cite{barber2012bayesian,mackay2003information}.  
			
			\paragraph{Ground truth representation}
			The main difference between \aclink{SI} and \aclink{CSP}, 
			\graffito{
			We may imagine that the data projected in a particular space, potentially in higher dimensions, contains two principal components representing the images of cats and dogs
			\begin{tikzpicture}[scale=0.3]
			\pgfplotsset{compat=1.7}
			\draw[rotate=45, thick] (0, 0) ellipse (4.5cm and 1.0cm);
			\draw[thick, dashed, -latex] (-5, 0) -- (5, 0) node[right]{};
			\draw[thick, dashed, -latex] (0, -5) -- (0, 5) node[above]{};
			\draw[thick, -latex, red] (-4, -4) -- (4, 4) node[above right]{$\btheta_c^\star$};
			\draw[thick, -latex, blue] (3, -3) -- (-4, 4) node[above left]{$\btheta_d^\star$};
			\end{tikzpicture}
			}
			depicted in the previous section, is that we assume there always exists a hidden solution to the problem, called the \emph{ground truth} $\btheta^\star$ or \emph{planted solution},
			that we aim to recover. De facto, the SAT-UNSAT transition does not exist in inference problems and we expect the phase transition phenomenology to be different, yet even richer. 
			The existence of the ground truth $\btheta^\star$ is ensured in many applications such as noisy communication channels, compressed sensing, phase retrieval, matrix factorization, and many others. In the case of supervised learning of real datasets, the ground truth is not explicit but we shall still assume it exists for our theoretical purposes. Indeed, even though the generative process of the data is hidden, we shall assume that the collected dataset contains a common hidden representation. For instance in a dataset containing images of cats and dogs, it seems natural to assume there exists some ground truth representations, yet unaccessible, $\btheta_c^\star, \btheta_d^\star$ that commonly characterize the images of cats and dogs.
				
		\paragraph{The teacher-student scenario and the planted ensemble}
			In practice, the ground truth representation $\btheta^\star$ is not available for a direct comparison.
			Yet, for theoretical purposes, in order to measure the reconstruction performances of the hidden signal and depict the corresponding phase transitions, we naturally need to have access to the ground truth representation.
			To circumvent this difficulty, it gems from this idea the notion of hidden \emph{rule} based on a signal $\btheta^\star$ that a \emph{teacher} uses to generate a training set $\bbX_{\textrm{train}}$ \cite{Patarnello_1987,gardner1989three,tishby89,Sompolinsky1990,seung1992statistical,watkin1993statistical,Gyorgyi2001}. This is called the \aclink{T-S} scenario: \emph{student} aims to recover the hidden rule for the observations in the corresponding \emph{synthetic} training set $\bbX_{\textrm{train}}$ generated by the \emph{teacher}. 
			Statistical inference and statistical physics show a narrow connection through the lens of this \aclink{T-S}\index{teacher-student scenario} and the \emph{planted spin glass ensemble}. 		
			Indeed, in this context inferring the ground truth vector in statistical inference is similar to recovering a \emph{crystal configuration} in planted spin glasses. 
			One of the main advantage of this setting is that the Bayesian approach easily suits this framework and gives an optimal strategy that can be furthermore analyzed by statistical physics tools in the high-dimensional regime. 
			Even though the \aclink{T-S} scenario and the randomly-quenched disorder in \aclink{rCSP} can be analyzed in a similar Bayesian approach, we shall keep in mind their striking difference that lead to very different phase diagrams typology. 
			In particular, in the case of randomly-quenched disorder in \aclink{rCSP}, the observations correspond to independent random constraints, whereas in the case of the \aclink{T-S} scenario, the observations are not independent as they all depend on the ground truth hidden representation.
			
			
		\paragraph{Statistical inference and estimators}				
			Let us define general inference problems we will focus on. 
			One considers a $\ndim-$dimensional hidden ground truth variable $\btheta^\star = \{ \theta_i^\star \}_{i=1}^\ndim \in \chi_\ndim$, drawn from a probability distribution $\rP_{\btheta^\star}$. The goal of \aclink{SI} is to infer it from $\nsamples$ observations $\{\mat{X},\vec{y}\} \in \bbX_\train$ generated according to a \emph{generative process}
			\begin{align}
				\vec{y} = \varphi_{\out}^\star\(\mat{X}, \btheta^\star \) \qquad \Leftrightarrow \qquad \vec{y} \sim \rP_{\out}^\star \(.\) \,,
			\label{appendix:teacher_model}	
			\end{align}
			where $\varphi_{\out}^\star$ represents a deterministic or stochastic function equivalently associated to a distribution $\rP_{\out}^\star$. 
			Again, we introduce the parameter $\alpha$ as the ratio of the number of observations over the dimension of the problem, namely here $\alpha=\frac{\nsamples}{\ndim}$.
			Inferring the above statistical model from observations $\{\mat{X},\vec{y}\}$ can be tackled in several ways and consists in building an estimator $\hat{\btheta}$ that approaches the ground truth planted solution $ \btheta^\star$. For instance we often use in this case the \aclink{MSE} $\ell (\btheta^\star, \hat{\btheta} ) =  \|  \btheta^\star - \hat{\btheta} \|_2^2$, to measure the distance between the estimator $\hat{\btheta}$ and the hidden parameter $\btheta^\star$.
			Our Bayesian framework \eqref{main:eq:inference:jpd} is particularly suited to the analysis of two common estimators based on the high-dimensional, often intractable, posterior distribution \eqref{main:eq:inference:Gibbs}. 
			In one hand, the \aclink{MMSE} estimator for $\beta=1$, consists in computing the mean of the of the posterior $\rP\(\btheta | \mat{X}, \vec{y}\)$ according to 
			$$\hat{\btheta}_{\mmse} = \EE_{\rP\(\btheta | \mat{X}, \vec{y}\)}\[\btheta\]\,,$$ 
			which is well-known to minimize the \aclink{MSE} reconstruction error. 
			In the other hand the \aclink{MAP} estimator consists in computing the maximum of the posterior distribution, that can be performed in the limit $\beta=\infty$. It can be formulated as a minimization problem according to
			$$\hat{\btheta}_{\map} = \argmax_\btheta \rP\(\btheta | \mat{X}, \vec{y}\) = \argmin_{\btheta} \[ \sum_{\mu=1}^\nsamples \ell \(\btheta ; y_\mu, \vec{x}_\mu \) +   r\(\btheta \) \] \,,$$
			where the loss is simply mapped to $ \ell \(\btheta; \vec{y}, \mat{X} \) =  - \log \rP\(\vec{y} | \btheta, \mat{X} \)$ and the regularizer $r\(\btheta\) = -\log \rP\(\btheta\)$, so that \aclink{ERM} can be analyzed in this framework. Thus both the study of \aclink{MAP} and \aclink{MSE} estimations can be casted in this general Bayesian approach and are simply reduced to the analysis of the posterior.
			Moreover, while the \aclink{MMSE} estimator is exactly the one performed by classical \aclink{AMP} algorithms, the \aclink{MAP} estimator can be thought as the ground state of the physical system and is closely related to \aclink{ERM} estimation performed by practical \aclink{GD} whose asymptotic behavior can be analyzed within this framework.

		\paragraph{Bayes-optimal estimation and the Nishimori conditions}
			In the idealistic case where the \emph{student} knows all the correct prior distributions $\rP_{\btheta} = \rP_{\btheta^\star}$, \newline $\rP_\out\(\vec{y} | \btheta ; \mat{X} \)=\rP_{\out^\star}\(\vec{y} | \btheta^\star ; \mat{X} \)$, this scenario is called the \emph{Bayes-optimal} setting. In the context of \aclink{MSE} reconstruction loss, performing the \aclink{MMSE} estimation in the Bayes-optimal case, yet unrealistic in practice, will be an important theoretical optimal baseline all along this manuscript.
			In this very specific and idealistic Bayes-optimal case, \aclink{SI} turns out to deeply simplify thanks to the Nishimori conditions \cite{opper1991calculation, iba1999nishimori, nishimori2001statistical}, presented in \App\ref{appendix:replica_computation:nishimori}, and that will be intensely used in the following.
			These Nishimori conditions simply state that in average there is no statistical difference between the ground truth configuration and a configuration sampled uniformly at random from the posterior distribution, so that \emph{overlaps} between the ground truth and the estimator is essentially the self-overlap of the estimator. As a consequence, under the Bayes-optimal assumption, the free entropy turns out to be exactly given by the \emph{replica symmetric} ansatz. However, these powerful identities do not hold in the practical \emph{mismatched setting} where the correct ground truth prior distributions are hidden during estimation.
						
		\paragraph{Information theoretical phase transitions}
				For the moment, without any algorithmic consideration, assuming we have access to exponential time and resources, we can already depict various \emph{phase transitions} in the above \aclink{SI} problem, based on \emph{information theoretical} predictions. 
				They can be formalized from a quantitative analysis of the free entropy potential, but for conciseness we propose to only describe qualitatively the different phase transitions of the \emph{optimal} estimator:
				\begin{itemize}
					\item With very few observations $\alpha \ll 1$, any algorithm is unlikely to infer correctly the hidden signal $\btheta^\star$. The estimator cannot \emph{extract} any information correlated with the ground truth solution and the loss reaches its maximal value $\rho$: $\ell(\hat{\btheta}, \btheta^\star) = \rho$. This region is called the \emph{undetectable phase} for $\alpha < \alpha_{\textrm{weak}}$.
					\item From a certain number of samples $\alpha_{\textrm{weak}}$, the estimator can \emph{partially} reconstruct the signal such that the loss decreases but does not reach is minimal value $0<\ell(\hat{\btheta}, \btheta^\star) < \rho$, that corresponds to the \emph{weak-recovery phase} $\alpha_{\textrm{weak}} \leq \alpha < \alpha_{\textrm{IT}}$.
					\item Above a critical observations density $\alpha_{\textrm{IT}}$, it becomes theoretically possible to \emph{perfectly} reconstruct the signal such that the loss $\ell(\hat{\btheta}, \btheta^\star) = 0$. This regime is called the \emph{easy phase} for $\alpha \geq \alpha_{\textrm{IT}}$.
				\end{itemize} 
				
		\paragraph{Algorithmic phase transitions and computational efficiency}				
				With the recent success of \aclink{ML} applications, while statistics was often not concerned with algorithmic performances, the increasing number of parameters in the models raises the question of the computational efficiency. 
				Hence, for practical purposes, we are interested in knowing if a particular algorithm can achieve the above \emph{information theoretical} performances.
				Optimality of most algorithms is far to be theoretically guaranteed. 
				Yet, in the case of \aclink{MMSE} estimation, the \aclink{AMP} algorithms under consideration are proven (or believed) to achieve \emph{information theoretical} performances. 
				However, very often, there exists some regions of parameters in which the optimal algorithmic reconstruction is not possible, while, theoretically, it is should be the case. Therefore, the \emph{easy phase} shall be revised under this algorithmic perspective with \emph{finite} resources. This region is called a \emph{hard phase} that slots into the \emph{weak recovery} phase and the \emph{easy} phase: $\alpha_{\textrm{IT}} < \alpha < \alpha_{\textrm{alg}}$. It is related to the notions of computational complexity and notably to the distinctions between P, NP, and NP-complete classes. A more accurate discussion may be found in \cite{monasson1999determining,percus2006computational, arora2009computational,moore2011nature}.	
				\begin{figure}[htb!]
				\centering
				\begin{tikzpicture}[scale=0.16]
			    \tikzstyle{dot}=[circle,minimum size=10pt, scale=0.4]
			    \tikzstyle{green}=[fill=green!70!black]
			    \tikzstyle{orange}=[fill=orange!90!black]
			    \tikzstyle{red}=[fill=red!90!black]
			    \tikzstyle{yellow}=[fill=yellow!95!black]
			    \tikzstyle{red}=[fill=red!90!black]
			    \tikzstyle{annot}=[text width=3cm, text centered, font=\footnotesize]
			    \tikzstyle{line}=[-,thick, black]
				\draw[draw=none, red] (0,-6) to (0,-6.5) to (15,-6.5) to (15, -6) to (0, -6);
				\draw[draw=none, orange] (15,-6) to (15,-6.5) to (35,-6.5) to (35, -6) to (15, -6);
				\draw[draw=none, yellow] (35,-6) to (35,-6.5) to (55,-6.5) to (55, -6) to (35, -6);
				\draw[draw=none, green] (55,-6) to (55,-6.5) to (70,-6.5) to (70, -6) to (55, -6);
				\draw [->,thick, black] (0,-6) to (71,-6);
				\node[text centered, font=\footnotesize, below = 0.25cm] at (7, -6.2) {\textbf{Undetectable}}  ;
				\node[text centered, font=\footnotesize, below = 0.25cm] at (26, -6.2) {\textbf{Weak recovery}}  ;
				\node[annot, below = 0.25cm] at (45, -6.2) {\textbf{Hard}}  ;
				\node[annot, below = 0.25cm] at (62.5, -6.2) {\textbf{Easy}}  ;
				\node[annot, below = 0.25cm] at (71, 0) {$\alpha$}  ;
				\node[annot, below = 0.25cm] at (16, 0) {$\alpha_{\textrm{weak}}$}  ;
				\draw [-, black] (15,-6.5) to (15,-5.5);
				\draw [-, black] (35,-6.5) to (35,-5.5);
				\draw [-, black] (55,-6.5) to (55,-5.5);
				\node[annot, below = 0.25cm] at (35, 0) {$\alpha_{\textrm{IT}}$}  ;
				\node[annot, below = 0.25cm] at (55, 0) {$\alpha_{\textrm{alg}}$}  ;
				\end{tikzpicture}
				\caption{Illustration of the phase transitions happening in inference problems.}
				\label{fig:intro:inference_phases}
				\end{figure}
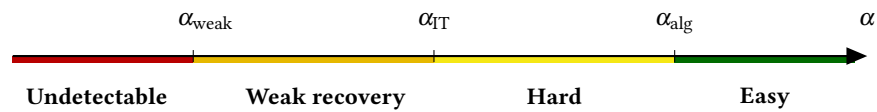
				As a conclusion, the schematic phase diagram of \aclink{SI} is represented in \Fig\ref{fig:intro:inference_phases}. In the next sections, we will provide more details on these phase transitions and especially stress they have a clear and deep interpretation in the physics formalism.\\
				
				In the next chapter, we finally introduce the \emph{mean-field} methods to analyze quantitatively the free entropy potential and depict quantitatively the phase diagrams of simple \aclink{CSP} and \aclink{SI} models.

\ifthenelse{\equal{\format}{oneside}}
	{\clearpage\null\thispagestyle{empty}}
	{\cleardoublepage}
\chapter{From mean-field methods to algorithms}
\label{main:chap:mean_field}
	In this chapter, we propose a short review of the main mean-field methods used to study analytically spin glass models along this manuscript. Namely we start by presenting the \emph{replica} method in  \Sec\ref{main:sec:mean_field:replica_method}, which is at the heart of this dissertation since it provides a powerful technique to compute the averaged quenched free entropy. It naturally gives access to the free entropy potential from which phase diagrams can be directly described. 
	Next in \Sec\ref{main:sec:mean_fields}, we discuss variational principles to derive various general mean-field methods. Finally, we present in \Sec\ref{main:intro:mean_field:bp} the \aclink{BP} equations that are a set of iterative equations, closely related to the cavity method \cite{mezard1987spin}, and leading to a perfect inference algorithm on tree-like graphs. Under a Gaussian projection, the set of \aclink{BP} equations can be simplified to the \aclink{AMP} algorithm, highlighted in \Sec\ref{main:sec:mean_field:amp}. The literature is quite extensive on the subject and the interested reader may find more details in \cite{mezard1987spin, mezard2009information, zdeborova2016statistical, advani2017high, gabrie2020mean}.

	\section{The replica method: a powerful heuristic mean-field method}
	\label{main:sec:mean_field:replica_method}
		This section is devoted to present the powerful \emph{replica method}\index{replica method} introduced in \cite{Kac1968,Edwards1975} and reviewed in \cite{mezard1987spin,dedominicis_2006,parisi_urbani_zamponi_2020}. This method allows to tackle the logarithmic difficulty in the computation of the average over the quenched disorder $\mat{J}$ in the free entropy \eqref{main:intro:averaged_free_entropy_free_energy} 
		\begin{align}
			\Phi_{\ndim}\(\beta,\mat{J}\)&\equiv  \frac{1}{\ndim} \EE_{\mat{J}} \log \mZ_\ndim\(\beta, \mat{J}\)\,.
		\end{align}
		The method fundamentally relies on the so-called \emph{replica trick}, which is a simple mathematical identity, that carries nonetheless profound physical consequences.

		\subsection{Replica trick}
		\label{main:intro:replicas:replica_trick}
		The replica trick is a simple identity that allows to exchange the expectation over the disorder and the logarithm,  in exchange of computing the $r \in \bbN$ moments of the partition function $\mZ_\ndim^r$ according to 
			\begin{align}
				\EE_{\mat{J}} \[\log \mZ_\ndim \] = \lim_{r \to 0} \frac{\partial \log \EE_{\mat{J}}\[\mZ_\ndim^r\]}{\partial r} \,.
			\end{align}
			\begin{proof}
			Suppose $ r \in \bbR $ close to zero, then
			\begin{align*}
			    \mZ_\ndim^r = e^{r \log \mZ_\ndim} = 1 + r \log \mZ_\ndim + \mathrm{o}(r)
			    \quad \Rightarrow \quad
			    \log \mZ_\ndim = \lim_{r \to 0} \frac{\mZ_\ndim^r - 1}{r}\,.
			\end{align*}
			By exchanging the limit $r \to 0$ and the expectation, and assuming that $r\in\bbN$, we obtain
			\begin{align*}
				\EE_{\mat{J}} \[ \log \mZ_\ndim \] &= \EE_{\mat{J}} \[ \lim_{r \to 0} \frac{\mZ_\ndim^r - 1}{r}  \] = \lim_{r\to 0} \partial_r \log \( \EE_{\mat{J}} \[ \mZ_\ndim^r  \]\)\,.
			\end{align*}
			\end{proof}
			As a result, the replica trick reduces the quenched average of the logarithm to the average of the moments of the partition function $\mZ_\ndim^r$, that are more tractable. Moreover, as soon as $r\in \bbN$, the moment $\mZ_\ndim^r$ represents in fact the product of $r$ identical partition functions, namely the partition function of a system containing $r$ non-interacting copies, called \emph{replicas}, of the original system
			\begin{align}
					\mZ_\ndim^{r}\(\beta,\mat{J}\) = \prod_{a=1}^r \mZ_\ndim^a\(\beta,\mat{J}\) = \prod_{a=1}^r \int_{\chi_\ndim} \d \bsigma^a ~ e^{-\beta \mH_\ndim\(\bsigma^a; \mat{J}\) }\,,
				\end{align}
			where $a \in \lb r \rb$ denotes the replica indices. However, under the disorder average, the initial $r$ non-interacting replicas are transformed in a highly non-trivial interacting particles problem charaterized by a matrix order parameter $\mat{Q} \in \bbR^{r\times r}$
			\begin{align}
				\EE_{\mat{J}} \mZ_\ndim^{r}\(\beta,\mat{J}\) = \int_{\bbR^{r\times r}} \d \mat{Q} ~ e^{ \Phi^{(r)}(\mat{Q}) }\,,
				\label{eq:main:intro:lagrangian_replicas}
			\end{align}
			where $\Phi^{(r)}$ denotes the replica potential. This simple mathematical trick has profound consequences as non-trivial properties can emerge from the interactions between these \emph{coupled} \emph{replicas}\index{replicas}. 
			Additionally, notice that the average of the replicated partition function has substituted the initial exponentially large summation $\EE_\bsigma$ by an analytical formula involving a new order parameter.
			In return, the difficulty is from now on to analyze the complex structure of the order parameter $\mat{Q}$. In particular, the initial invariance of the replicas can be conserved in certain situations. This solution is called the \aclink{RS} Ansatz, in contrast to \aclink{RSB} Ansätze in which the mean-field solution breaks the initial invariance of the replicas permutation. As soon as the symmetry is broken, choosing the correct structure for the matrix $\mat{Q}$ in the replica space is highly non-trivial. 
			As a conclusion, the replica trick and the average of the replicated partition function substituted the complex analysis of interacting disordered models to finding the values of a matrix order parameter of finite size. In general, they can be found as the solution of a closed set of non-linear equations that require only a polynomial number of operations.

		\subsection{Pure states and overlap distribution}
			Analyzing the overlap matrix distribution becomes essential to understand the behavior of this new interacting problem. In this end, we introduce the probability distribution averaged over the quenched disorder $\mat{J}$			\begin{align*}
				\rP(q) = \EE_{\mat{J}} \int_{\chi_\ndim} \d \bsigma ~ e^{-\beta \mH_\ndim(\bsigma, \mat{J}) } \int_{\chi_\ndim} \d \bsigma' ~ e^{-\beta \mH_\ndim(\bsigma', \mat{J}) } \frac{\delta\( q  - \frac{1}{\ndim} \bsigma \cdot \bsigma'  \)}{\mZ_\ndim(\beta)^2} \,.
			\end{align*}
			that two configurations $\bsigma,\bsigma'$ have a mutual overlap $q$ at equilibrium.
			The overlap distribution $\rP(q)$ reveals important knowledges about the thermodynamics of the model and especially the distance between typical equilibrium configurations. In particular, the Gibbs measure at equilibrium is carried by a few \emph{pure states} \cite{mezard1987spin} that respectively describe distinct ergodic connected components of the configuration space. 
			Indeed, denoting $\alpha$ these pure states, the Gibbs average can be decomposed as
			\begin{align*}
				\langle \cdots \rangle_\beta &=  \sum_{\alpha} \underbrace{\frac{\mZ_\alpha(\beta)}{\mZ(\beta)}}_{w_\alpha(\beta)}  \underbrace{\int \d \bsigma_\alpha \cdots \frac{e^{—\beta \mH_\ndim(\bsigma,\mat{J})}}{\mZ_\alpha(\beta)}}_{\langle \cdots \rangle_\alpha} = \sum_{\alpha} w_\alpha(\beta) \langle \cdots \rangle_\alpha\,,
			\end{align*}			
			with $w_\alpha(\beta)$ the thermodynamic weight of the state $\alpha$ that contributes to the non-trivial structure of overlap distribution $\rP(q)$.
			\graffito{
			Overlap distribution for a single paramagnetic pure state and two ferromagnetic states
			\begin{tikzpicture}[scale=0.3]
				    \tikzstyle{annot}=[text width=3cm, text centered, font=\footnotesize]
				    \tikzstyle{line}=[-,thick, black]
					\draw [->,thick, black] (-5,0) to (5,0);
					\draw [-,thick, black] (0,-0.3,0) to (0,0.3);
					\draw [-,very thick, teal] (0,0) to (0,4);
					\node[annot, below = 0.1cm] at (5,0) {$q$}  ;
					\node[annot, below = 0.1cm] at (0,0) {$0$}  ;
			\end{tikzpicture}
			\begin{tikzpicture}[scale=0.3]
			\hspace{-0.75cm}
			\tikzstyle{annot}=[text width=3cm, text centered, font=\footnotesize]
				    \tikzstyle{line}=[-,thick, black]
					\draw [->,thick, black] (10,0) to (20,0);
					\draw [-,thick, black] (15,-0.3,0) to (15,0.3);
					\draw [-,very thick, burntorange] (12.5,0) to (12.5,2);
					\draw [-,very thick, burntorange] (17.5,0) to (17.5,2);
					\node[annot, below = 0.1cm] at (20,0) {$q$}  ;
					\node[annot, below = 0.1cm] at (12.5,0) {$-m^2$}  ;
					\node[annot, below = 0.1cm] at (17.5,0) {$m^2$}  ;   
			\end{tikzpicture}
			}
			In the presence of different pure states $\alpha, \beta$, we define the overlap between them $q_{\alpha \beta} = \frac{1}{\ndim} \sum_{i=1}^{\ndim} \langle \sigma_i\rangle_\alpha \langle \sigma_i\rangle_\beta$
			so that the averaged overlap distribution reads $\rP(q) = \sum_{\alpha, \beta} w_\alpha w_\beta \delta\(q - q^{\alpha \beta}\)$ because of the \emph{clustering property} $\langle \sigma_i \sigma_j \rangle_\alpha = \langle \sigma_i \rangle_\alpha \langle \sigma_j \rangle_\alpha$. 
			As an illustration, in the Curie-Weiss model presented in \Sec\ref{main:intro:classical_physics:examples:curie_weiss_example}, we observed that below the critical inverse temperature $\beta^\star=1$ there exists a single pure paramagnetic state $q=0$, such that the distribution contains a single ergodic component $\rP(q)=\delta(q)$. Above the critical temperature, we observed the emergence of two ferromagnetic states $q=-m^2$ and $q=m^2$ with $m>0$, such that the distribution splits into two connected components $\rP(q)=\frac{1}{2} \delta(q-m^2) + \frac{1}{2} \delta(q + m^2)$.
					
			\subsection{Replica Ansatz}
			\label{main:sec:mean_field:replica_method:replica_ansatz}
				In general, the full replica computation boils down to a Lagrangian similar to \eqref{eq:main:intro:lagrangian_replicas} expressed in terms of a symmetric matrix order parameter $\mat{Q}$. 
				The computation for unconstrained symmetric matrices is unfortunately intractable, and \cite{Parisi1983} proposed an iterative scheme to approximate the corresponding overlap distribution $\rP(q)$. 
				We present the \aclink{RS} and \aclink{RSB} simple Ansätze that turn out to be stable in various models.
										
 			\subsubsection{Replica Symmetric}
\begin{figure}[htb!]
					\centering
					\begin{tikzpicture}[scale=0.2]
				    \tikzstyle{annot}=[text width=3cm, text centered, font=\footnotesize]
				    \tikzstyle{line}=[-,thick, black]
				    \tikzstyle{Q}=[fill=teal]
				    \tikzstyle{q0}=[fill=burntorange]
				    
				    \draw[line, dashed, q0] (0,0) rectangle (16,16);
				    \draw[line, Q, thick] (0,16) rectangle (4,12);
				    \draw[line, Q, thick, dashed] (4,12) rectangle (8,8);
				    \draw[line, Q, thick, dashed] (8,8) rectangle (12,4);
				    \draw[line, Q, thick] (12,4) rectangle (16,0);
					\node[annot] at (2,14) {$Q$} ;
					\node[annot] at (6,10) {$Q$} ;
					\node[annot] at (10,6) {$Q$} ;
					\node[annot] at (14,2) {$Q$} ;
					\node[annot] at (12,12) {$q_0$} ;
					\node[annot] at (3,3) {$q_0$} ;
					\draw[<->,thick, black] (12,-0.5) to (16,-0.5);
					\draw[<->,thick, black] (0,-3) to (16,-3);
					\node[annot] at (8,-4) {$r$} ;
					\node[annot] at (14,-2) {$x_0=1$} ;
					\end{tikzpicture}					
					\caption{Illustration of the replica symmetric overlap matrix $\mat{Q}^{(\rs)}$.}
					\label{fig:introduction:rs_ansatz}
				\end{figure}
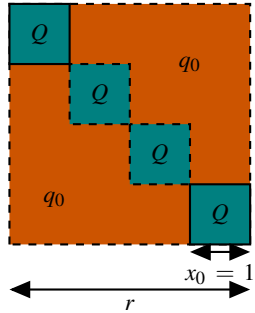

				The simplest \aclink{RS} Ansatz is particular as it assumes that the initial permutation invariance of the fictive replicas is conserved so that the overlap between two arbitrary replicas is identical and fixed to  $q_0 = \frac{1}{\ndim} \bsigma^{(a)} \cdot \bsigma^{(b)},~\forall (a,b) \in \lb r \rb^2$.
				The overlap distribution is therefore given by $\rP^{(\rs)}(q) =\delta(q-q_0)$ such that the overlap matrix $\mat{Q}^{(\rs)}= (Q-q_0) \mat{I}_r + q_0 \mat{J}_r \in \bbR^{r \times r}$, illustrated in \Fig\ref{fig:introduction:rs_ansatz}, where $Q=\frac{1}{\ndim} \|\bsigma\|_2^2$ denotes the self-overlap.
				This Ansatz turned out to be stable in many situations such as on the Nishimori line in the context of the \aclink{SK} model \cite{nishimori1980exact, nishimori1981internal, georges1985exact} or in the \emph{Bayes-optimal} setting in \aclink{SI}, where the Nishimori conditions detailed in \App\ref{appendix:replica_computation:nishimori} allow to rigorously prove the validity of the \aclink{RS} Ansatz.
				
			\subsubsection{Ansatz stability: de Almeida-Thouless transition}
				Otherwise, the correctness of a given Ansatz can be highlighted by estimating its \emph{stability} \cite{Almeida1978, thouless1977solution}. For instance, in the case of the \aclink{RS} Ansatz, its stability is evaluated by expanding the replica potential $\Phi^{(r)}(\mat{Q})$ in \eq\eqref{eq:main:intro:lagrangian_replicas} around the \aclink{RS} fixed point
 			\begin{align}
 					\Phi^{(r)}(\mat{Q}) &= \Phi^{(r)}(\mat{Q}^{(\textrm{rs})}) - \frac{1}{2} \sum_{a<b,c<d} \delta \mat{Q}^{ab} \mM^{ab,cd} \delta \mat{Q}^{cd}\,.
 			\end{align}
 			By studying the fluctuations and eigenvalues of the Hessian matrix $\mM^{ab,cd} = - \frac{\partial^2 \Phi^{(r)}}{\partial \mat{Q}^{ab} \partial \mat{Q}^{cd}}\vert_{\mat{Q}=\mat{Q}^{(\textrm{rs})}}$, we can detect the so-called \aclink{dAT} transition that occurs when the \aclink{RS} Ansatz becomes unstable, \ie when the first negative eigenvalue appears, called in this context the replicon eigenvalue. Notice that the technique is not limited to this latter Ansatz and can be applied to more complex ones.		
		
			\subsubsection{Replica symmetry breaking and ergodicity breaking}
			In case the \aclink{RS} Ansatz is unstable and leads sometimes to unphysical results such as negative entropies \cite{gardner1988optimal}, more complex Ansätze should be investigated above the \aclink{dAT} line. 
			Constructing such an Ansatz is not easy as it should respect physical constraints such as the positivity of the entropy and the overlap distribution $\rP(\mat{Q})$, and it should be stable with respect to Gaussian fluctuations.
			The first step forward was introduced in \cite{blandin1978theories,blandin1980mean} where the idea of breaking the replicas symmetry into blocks emerged and it was further developed in \cite{sommers1978, sommers1979sherrington, dedominicis1979, Bray1980}. 
			Yet the permutation symmetry may be broken in many ways such that finding the correct Ansatz was the main focus of most theoretical research works in the spin glass literature \cite{Sherrington1975,Derrida1981,gardner1988optimal,Crisanti1992}.
			Finally, the general solution was delivered by Parisi in a series of works \cite{parisi1979infinite, Parisi1980_magnetic, Parisi1980_sequence,Parisi1983}, in which he proposed a general scheme, which respect all the physical constraints, for progressively breaking the replica symmetry, called \aclink{RSB}, that eventually leads to the correct solution			
			This scheme predicts that the stable Ansatz should perform a \emph{infinite} and \emph{continuous} hierarchy of symmetry breaking, the so-called \aclink{FRSB} Ansatz. However, very often the \aclink{1RSB},\aclink{2RSB} Ansätze give very accurate approximations that avoid to solve numerically the cumbersome \aclink{FRSB} equations.
			In the context of the \aclink{SK} model \cite{Sherrington1975}, this \aclink{FRSB} Ansatz turned out to be exact and was rigorously proven later on in \cite{Guerra2003, talagrand2006parisi}.
			
			\paragraph{RSB Parisi's scheme}
			For the sake of illustration, let us illustrate the Parisi's scheme for breaking the replicas symmetry. The overlap matrices and distributions in the \aclink{1RSB} and \aclink{2RSB} Ansätze can be written as follows
			\begin{align*}
				\mat{Q}^{(\textrm{1rsb})} &= \( Q - q_1 \) \mat{I}_r  + \( q_1 - q_0 \) \mat{I}_{\frac{r}{x_0}} \otimes \mat{J}_{x_0} +  q_0 \mat{J}_r \\
				\rP^{(\textrm{1rsb})}(q) 
				&\underlim{r}{0} (1-x_0)\delta(q-q_1) + x_0 \delta(q-q_0) 
			\end{align*}
			and
			\begin{align*}
				\mat{Q}^{(\textrm{2rsb})} &=  \( Q - q_2 \) \mat{I}_r + \( q_2 - q_1 \) \mat{I}_{\frac{r}{x_1}} \otimes \mat{J}_{x_1} \\
				& \qquad \qquad \qquad \qquad + \( q_1 - q_0 \) \mat{I}_{\frac{r}{x_0}} \otimes \mat{J}_{x_0} + q_0 \mat{J}_r\\
				 \rP^{(\textrm{2rsb})}(q) & \underlim{r}{0} (1-x_1)\delta(q-q_2) + (x_1-x_0) \delta(q-q_1) + x_0 \delta(q-q_0)
 			\end{align*}
			and the corresponding overlap matrices are depicted in \Fig\ref{fig:introduction:rsb_ansatz}. 
			Therefore, the \aclink{RSB} leads to consider that replicas play different roles and are clustered in different states with different inner and outer correlations, respectively $q_1$ and $q_0$ in the context of the \aclink{1RSB} Ansatz with $q_1>q_0$. In contrast the \aclink{RS} Ansatz, in the context of the \aclink{RSB} the ergodicity is broken in a nontrivial way and the phase space is organized into a hierarchical structure of pure states. 
			In particular, given the multiplicity of ergodic components in the \aclink{RSB} Ansatz, that mainly appear at low temperature, the thermodynamic averages performed with the Gibbs measure are not equivalent to the average inside one state but it rather takes into account the presence of all the states. 
			The overlap matrix $\mat{Q}$ is therefore hierarchically constant by \emph{blocks} for a finite number $k$ of \aclink{RSB} steps. Nonetheless, the Parisi scheme can be repeated for an infinite number of steps $k=\infty$, reaching a continuous limit and the so-called \aclink{FRSB} solution scheme as illustrated in \Fig\ref{fig:introduction:overlap_distribution}. As an illustration, by iterating the Parisi's scheme, the \aclink{2RSB} Ansatz can be obtained by simply imposing a similar fractal structure within the smallest blocks of the \aclink{1RSB} Ansatz.
				\begin{figure}[htb!]
					\centering
					\begin{tikzpicture}[scale=0.2]
				    \tikzstyle{annot}=[text width=3cm, text centered, font=\footnotesize]
				    \tikzstyle{line}=[-,thick, black]
				    \tikzstyle{Q}=[fill=teal]
				    \tikzstyle{q0}=[fill=burntorange]
				    \tikzstyle{q1}=[fill=carnelian]
				    \tikzstyle{q2}=[fill=carmine]
				   
				    \draw[line, dashed, q0] (0,0) rectangle (16,16);
				  	\foreach \x/\y in {0/16, 4/12, 8/8, 12/4}
						{
						\draw[line, q1, dashed, thick] (\x,\y) rectangle (\x+4,\y-4);
						}
					\foreach \x/\y in {0/16, 1/15, 2/14, 3/13, 4/12, 5/11, 6/10, 7/9, 8/8, 9/7, 10/6, 11/5, 12/4, 13/3, 14/2, 15/1}
						{
						\draw[line, Q, thick] (\x,\y) rectangle (\x+1,\y-1);
						\node[annot] at (\x+0.5,\y-0.5) {\tiny $Q$} ;
						}
					
					\draw[line, dashed, q1] (20,2) rectangle (36,18);
					\foreach \x/\y in {0/16, 4/12, 8/8, 12/4}
						{
						\draw[line, q2, dashed, thick] (20+\x,2+\y) rectangle (20+\x+4,2+\y-4);
						}
					\foreach \x/\y in {0/16, 1/15, 2/14, 3/13, 4/12, 5/11, 6/10, 7/9, 8/8, 9/7, 10/6, 11/5, 12/4, 13/3, 14/2, 15/1}
						{
						\draw[line, Q, thick] (20+\x,2+\y) rectangle (20+\x+1,2+\y-1);
						\node[annot] at (20+\x+0.5,2+\y-0.5) {\tiny $Q$} ;
						}
					
					\draw[-,thick, black] (12,0) to (20,2);
					\draw[-,thick, black] (12,4) to (20,18);
					
					\node[annot] at (1,13) {$q_1$} ;
					\node[annot] at (3,15) {$q_1$} ;
					\node[annot] at (5,9) {$q_1$} ;
					\node[annot] at (7,11) {$q_1$} ;
					\node[annot] at (9,5) {$q_1$} ;
					\node[annot] at (11,7) {$q_1$} ;
					\node[annot] at (13,1) {$q_1$} ;
					\node[annot] at (15,3) {$q_1$} ;
					\node[annot] at (12,12) {$q_0$} ;
					\node[annot] at (3,3) {$q_0$} ;
					\draw[<->,thick, black] (12,-0.5) to (16,-0.5);
					\draw[<->,thick, black] (0,-2) to (16,-2);
					\node[annot] at (8,-3) {$r$} ;
					\node[annot] at (14,-1.5) {$x_0$} ;
					
					\node[annot] at (21,15) {$q_2$} ;
					\node[annot] at (23,17) {$q_2$} ;
					\node[annot] at (25,11) {$q_2$} ;
					\node[annot] at (27,13) {$q_2$} ;
					\node[annot] at (29,7) {$q_2$} ;
					\node[annot] at (31,9) {$q_2$} ;
					\node[annot] at (33,3) {$q_2$} ;
					\node[annot] at (35,5) {$q_2$} ;
					\node[annot] at (32,14) {$q_1$} ;
					\node[annot] at (24,5) {$q_1$} ;
					
					\draw[<->,thick, black] (32,1.5) to (36,1.5);
					\draw[<->,thick, black] (20,0) to (36,0);
					\node[annot] at (28,-1) {$x_0$} ;
					\node[annot] at (34,0.5) {$x_1$} ;
					\end{tikzpicture}
					\caption{Illustration of the Parisi scheme: the 2-step RSB Ansatz $\mat{Q}^{(\textrm{2rsb})}$ is obtained by repeating the hierarchal structure inside the red block of the 1-step RSB Ansatz $\mat{Q}^{(\textrm{1rsb})}$ \Leftn.}
					\label{fig:introduction:rsb_ansatz}
				\end{figure}
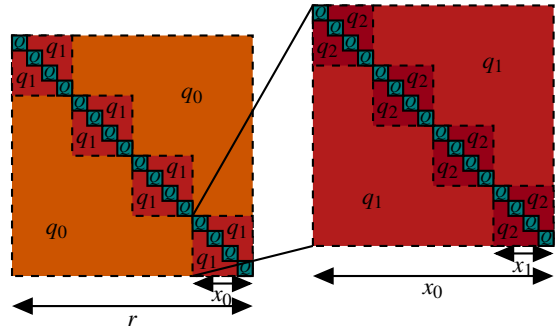  
				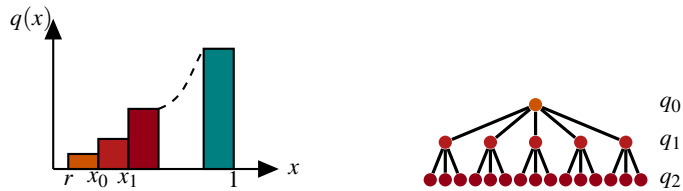
\begin{figure}
					\centering
					\begin{tikzpicture}[scale=0.2]
						\tikzstyle{annot}=[text width=3cm, text centered, font=\footnotesize]
					    \tikzstyle{line}=[-,thick, black]
					    \tikzstyle{Q}=[fill=teal]
					    \tikzstyle{q0}=[fill=burntorange]
					    \tikzstyle{q1}=[fill=carnelian]
					    \tikzstyle{q2}=[fill=carmine]
					    
						\draw[->,thick, black] (0,0) to (0,10);
						\draw[->,thick, black] (0,0) to (15,0);
						\node[annot] at (-1.5,10) {$q(x)$} ;
						\node[annot] at (16,0) {$x$} ;
						
						\draw[line, q0] (1,0) rectangle (3,1);
						\draw[line, q1] (3,0) rectangle (5,2);
						\draw[line, q2] (5,0) rectangle (7,4);
						\draw[line, Q] (10,0) rectangle (12,8);
						\draw[dashed, thick, black] (7,4) to[out=20,in=-120] (10,8) ;
						\node[annot] at (1,-0.7) {$r$} ;
						\node[annot] at (3,-0.7) {$x_0$} ;
						\node[annot] at (5,-0.7) {$x_1$} ;
						\node[annot] at (12,-0.7) {$1$} ;
					\end{tikzpicture}
					\begin{tikzpicture}[scale=0.2]
						\tikzstyle{annot}=[text width=3cm, text centered, font=\footnotesize]
					    \tikzstyle{line}=[-,thick, black]
					    \tikzstyle{Q}=[fill=teal]
					    \tikzstyle{q0}=[fill=burntorange]
					    \tikzstyle{q1}=[fill=carnelian]
					    \tikzstyle{q2}=[fill=carmine]
					    \node[circle,inner sep=0pt,minimum size=5pt, q0] (RS) at (0,10) {};
					    \foreach \x/\y in {-6/1,-3/2,0/3, 3/4,6/5}
							{
							\node[circle,inner sep=0pt,minimum size=5pt, q1] (1RSB\y) at (\x,7.5) {};
							}
					    \foreach \x/\y in {-7/1,-6/2,-5/3,-4/4,-3/5, -2/6,-1/7, 0/8, 1/9, 2/10, 3/11, 4/12, 5/13, 6/14, 7/15}
							{
							\node[circle,inner sep=0pt,minimum size=5pt, q2] (2RSB\y)  at (\x,5) {};
							}
						\foreach \x in {1,2,3,4,5}
							{
							\draw[-, very thick, black] (RS) to (1RSB\x);
							}
						\foreach \x in {1,2,3,4,5}
							{
							\foreach \y in {1,2,3}{
								\pgfmathtruncatemacro{\yn}{3*(\x -1) + \y }%
								\draw[-, very thick, black] (1RSB\x) to (2RSB\yn);
								}
							}
						\node[annot] at (9, 10) {$q_0$};
						\node[annot] at (9, 7.5) {$q_1$};
						\node[annot] at (9, 5) {$q_2$};
						
					\end{tikzpicture}
					\caption{Illustration of the Parisi iterative scheme of the overlap distribution which reflects the  multiplicity of ergodic components in RSB solutions.
					\Left Evolution of the distribution of overlaps from a constant by parts to a continuous function $q(x)$ for an infinite number of RSB steps. \Right Illustration of the hierarchical structure of the overlaps. 
					 }
					\label{fig:introduction:overlap_distribution}
				\end{figure}
					
		\subsection{Complexity and metastable states}
			The replica method and the hierarchical of the \aclink{FRSB} scheme naturally reveal the ergodicity breaking and the existence of \emph{metastable states} in spin glasses. 
			A metastable state represents a region of configuration space separated from the rest of the space by a free energy barrier that diverge with the size of the system. 
			Therefore, to escape this locally attracting valley, we shall go across higher free energy barriers. Equivalently, from the dynamic point of view, a metastable state is a region where the system will remain confined for finite times and could escape it only in a time scaling with the size of the system.
			The analysis of the \emph{p-spin} model in particular \cite{thouless1977solution,Rieger1992,Crisanti1992} revealed that this spin glass model had a very large number of metastable states, \ie locally stable solution with free energy higher than the ground state free energy. 
			Moreover, this number of states $\mN_\Phi$ turned out to scale exponentially with the size $\ndim$ of the system. As a result, to take into account the multiplicity of the metastable states, we define a new entropy measure, called the \emph{complexity}\index{complexity} or the \emph{configurational entropy} in the glass community, defined as the logarithm of $\Omega(\Phi)$ the number of states at a given free entropy, \ie $\Sigma(\Phi) = \frac{1}{\ndim} \log \Omega(\Phi)$.		
			The existence of such metastable states has potentially harmful consequences on dynamic systems such as structural glasses or optimization algorithms since they could eventually get stuck in metastable local minimum for an exponential time.
			
			\paragraph{Complexity computation}
				In order to quantify the existence of metastable states, \cite{monasson1995weight} proposed a general method, comprehensively reviewed in \cite{Zamponi2010}, to compute the complexity $\Sigma(\Phi)$ as a function of the free entropy of the states. The idea consists in considering $m$ \emph{real replicas} of the original system that are coupled by a small interacting term that will push all copies in the same pure state.
				The total replicated free entropy of the $m$ replicas can be well approximated by the sum of the contributions over all the states	
				\begin{align*}
					\mZ_m &\equiv e^{d\Phi_m(m, \beta)} = \sum_{\alpha} \exp\(\ndim m \Phi_\alpha\)  = 
					 \int \d\Phi ~ \sum_{\alpha} \delta(\Phi - \Phi_\alpha) e^{ \ndim m \Phi }\\
					 &=\int \d\Phi \Omega(\Phi) e^{ \ndim m \Phi} = \int \d\Phi ~ \exp\(\ndim\( m \Phi + \Sigma\(\Phi\) \)  \)\,.
				\end{align*}
				In the thermodynamic limit $d\to \infty$, a Laplace method \cite{Rong89} allows to write the replicated free entropy as the Legendre transform of the complexity:
				\begin{align}
					\Phi_m(m,\beta) = \max_{\Phi}\{m\Phi + \Sigma(\Phi)\} = m \Phi^\star(\beta) + \Sigma\(\Phi^\star\)\,,
				\end{align}
				where the equilibrium free entropy $\Phi^\star$ can be computed as				
				\begin{align*}
						\Phi^\star(\beta) = \frac{\d \Phi_m(m,\beta)}{\d m} \qquad \text{and} \qquad 
						\Sigma\(m,\beta\)  =  \Phi_m(m,\beta) - m \Phi^\star(\beta)\,.
				\end{align*}
				Varying the Legendre parameter $m$ at fixed temperature $\beta$, we can reconstruct the full complexity function $\Sigma(m,\beta)$ from the knowledge of the replicated partition function, which turns out to be closely related to the \aclink{1RSB} free entropy $\Phi_m(m,\beta) = m \Phi^{(\textrm{1rsb})}(\beta)$.
										
	\subsection{\textbf{Application} - Replica computation of the GLM}
	\label{main:sec:mean_field:replica_method:example_glm}
		In this section, we finally illustrate how the replica method developed in the context of the spin glass theory can be readily applied to simple supervised \aclink{ML} models such as the \aclink{GLM}, defined in \Sec\ref{main:introduction:glm_class}.\\
		
		The generalized linear estimation problem consists to fit $\nsamples$ observations $\bbX_\train = \{\mat{X}, \vec{y}\}$ with a linear parametric model with weights $\vec{w}\in \bbR^{\ndim}$ according to
		\begin{align*}
			\vec{y} = \varphi_\out\(\frac{1}{\sqrt{\ndim}} \mat{X}\vec{w}\)\,.
		\end{align*}
		In other words, we try to fit the observation $\vec{y}\in \bbR^{\nsamples}$, which can be either discrete \emph{labels} or continuous \emph{outputs}, with a linear transformation of the \emph{input} data matrix $\mat{X} \in \bbR^{\nsamples \times \ndim}$, up to a component-wise non-linear activation function $ \varphi_{\out^\star}: \bbR \mapsto \bbR$ which can be deterministic or stochastic.
		 Moreover, we assume that the matrix of data inputs $\mat{X} \in \bbR^{\nsamples \times \ndim }$ is drawn \aclink{i.i.d} with density $\rp_{\X}$. Specifically we will consider them to be Gaussian with zero mean and unit variance, namely $\forall \mu \in \lb \nsamples \rb,~ \vec{x}_\mu \sim \mN_{\vec{x}}\(\vec{0}, \mat{I}_{\ndim}\)$.
		 
		\paragraph{On the data generative process}
			As stressed in \Chap\ref{chap:intro:phys_inference}, different settings have been considered in the physics literature on how the ground truth observations $\vec{y}$ relate to the inputs $\mat{X}$. In particular \cite{gardner1989three} in their influential paper introduced the two main generative processes constantly studied in the subsequent literature:
			\begin{itemize}
				\item The \emph{random labels} setting: the labels $\vec{y}$ are uncorrelated from the input data $\mat{X}$. Namely, 
				\begin{align}
					\forall \mu \in \lb\nsamples\rb, ~ y_\mu \sim \rP_{\y}(.) \andcase ~ \vec{x}_\mu \sim \rP_{\textrm{x}}(.) \text{ with } y_\mu \perp \vec{x}_\mu\,.
					\label{main:eq:replicas:random_labels}
				\end{align}
				This setting has been studied in particular for perceptrons in \cite{gardner1988optimal, krauth1989storage} in the context of \aclink{rCSP}, see \Sec\ref{main:intro:phys_ml_together:random_csp}. Indeed the \emph{randomly quenched disorder} over the input $\mat{X}$ and $\vec{y}$ are not correlated such that trying to fit this dataset can be equivalently seen as trying to satisfy random constraints. 
				\item The \emph{teacher-student} scenario or equivalently the \emph{planted spin glass} model: the labels $\vec{y}$ are generated from a synthetic model designed by a \emph{teacher}, from which, in the context of \aclink{SI}, the \emph{student} aims to recover the teacher's parameters. Here we consider the ground truth as a linear model with weights $\vec{w}^\star$ according to the channel $\vec{y} = \varphi_{\out^\star}\(\frac{1}{\sqrt{\ndim}}\mat{X}\vec{w}^\star\)$ or equivalently
				\begin{align}
					\vec{y} &\sim \rP_{\y}(\vec{y}|\mat{X}) = \int_{\bbR^\nsamples} \d \vec{z}^\star ~ \rp_{\out^\star}\(\vec{y} | \vec{z}^\star \) \label{main:eq:replicas:synthetic_data_set} \\
					&  \qquad \qquad \qquad \qquad \times \int_{\bbR^\ndim} \d\vec{w}^\star ~ \rp_{\w^\star}(\vec{w}^\star) \delta\(\vec{z}-\frac{1}{\sqrt{\ndim}}\mat{X}\vec{w}^\star\) \,,
					\nonumber
				\end{align}
				with generic teacher densities $\rp_{\w^\star}, \rp_{\out^\star}$. 
				This \aclink{T-S} scenario perfectly fits in a supervised learning setting mentioned in \Sec\ref{main:intro:phys_ml_together:supervised}. 
				In this section, we assume that the student must infer the rule designed by the teacher, where both \emph{teacher} and \emph{student} belong to the same hypothesis class.
				\end{itemize}
			The full computation in the case of \emph{random labels}, used in particular in \Chap\ref{chap:binary_perceptron}-\ref{chap:rademacher}, detailed in \App\ref{appendix:replica_computation:random_labels:iid} is very similar and even simpler.
			In the \aclink{T-S} setting, the replica computation in the \aclink{GLM} for \aclink{i.i.d} data has been performed in many works such as \cite{schulke2016statistical} and has been generalized to rotationally invariant matrices in \cite{kabashima2008inference}. 
			For the sake of illustration, in this section we show only the main steps of the replica computation for the \aclink{GLM} and we leave the cumbersome details in \App\ref{appendix:replica_computation:committee}, presented in the context of the more general \emph{committee machines} hypothesis class. Committee machines, that we investigate in \Chap\ref{chap:committee_machine}, use instead $K$ \aclink{GLM} estimators simultaneously, so that its parameters is a matrix $\mat{W}\in \bbR^{\ndim \times K}$, to fit the training set according to
		\begin{align*}
			\vec{y} = \varphi_\out \(\frac{1}{\sqrt{\ndim}} \mat{X} \mat{W} \) = \varphi_\out \( \left\{ \frac{1}{\sqrt{\ndim}} \mat{X} \vec{w}_k \right \}_{k=1}^K \)\,,
		\end{align*}
			where $\varphi_\out : \bbR^K \mapsto \bbR$.
			As a consequence, the classical \aclink{GLM}, we present in this section, is a particular case of committee machines for $K=1$. Nonetheless, \aclink{GLM} are a wide class of linear models with various applications such as
			\begin{itemize}
				\item Compressed sensing: $\varphi_{\out^\star}(y|z) = z + \sqrt{\Delta} \xi$\,,
				\item Phase retrieval: $\varphi_{\out^\star}(y|z) = |z| + \sqrt{\Delta} \xi$\,,
				\item Perceptron: $\varphi_{\out^\star}(y|z) = \sign(z) + \sqrt{\Delta} \xi$\,,
			\end{itemize}
			where $\xi \sim \mN(0,1)$ represents a potential Gaussian noise scaled by a variance $\Delta \geq 0$. Moreover the ground truth vector $\vec{w}^\star$ can be drawn according to common prior distributions such as
			\begin{itemize}
				\item Gaussian weights: $\rP_{\w^\star}(\vec{w}^\star) = \mN_{\vec{w} ^\star}\(\vec{0}, \rho_{\w^\star} \rI_\ndim \)$\,,
				\item Spherical weights: $\rP_{\w^\star}(\vec{w}^\star) = \delta\( \|\vec{w}^\star\|_2^2 - \ndim \)$\,,
				\item Binary weights: $\rP_{\w^\star}(\vec{w}^\star) = \prod_{i=1}^\ndim \frac{1}{2} \(\delta(w_i^\star - 1\) + \(\delta(w_i^\star + 1\) $\,.
			\end{itemize}
			
		\paragraph{On statistical estimation}
		As stressed in \Sec\ref{main:intro:ml:bayesian_approach}, \aclink{MMSE} and \aclink{MAP} estimations boil down to the analysis of the joint distribution $\rP_\ndim\(\vec{y}, \mat{X}\)$ involved in the high-dimensional posterior \aclink{JPD} according to the Bayes formula
			\begin{align}
				\rP_\ndim\(\vec{w} | \vec{y}, \mat{X}\)  &= \frac{\rP\(\vec{y} | \vec{w}, \mat{X} \) \rP\(\vec{w}\) }{\rP_\ndim\(\vec{y},\mat{X}\)} = \frac{\rP_{\out} \(\vec{y} | \vec{w}, \mat{X} \) \rP_{\w}\(\vec{w}\)}{\mZ_\ndim\(\{\vec{y}, \mat{X}\} \)}  \,.
				\label{appendix:bayes_formula}
			\end{align}
			To explicitly connect with the spin glass approach, the distribution $\rP_\ndim\(\vec{y}, \mat{X}\)=\mZ_\ndim\(\{\vec{y}, \mat{X}\} \)$ is also called the \emph{partition function} and we define the corresponding Hamiltonian, for separable prior distributions $\rP_{\out}, \rP_{\w}$ as
		\begin{align*}
			\mH_\ndim\(\vec{w},\{\vec{y},\mat{X}\}\) &=- \log \rP_{\out} \(\vec{y} | \vec{w}, \mat{X} \) -  \log \rP_{\w}\(\vec{w}\)\,,\\
			&= - \sum_{\mu=1}^\nsamples \log \rP_{\out} \(y_\mu | \vec{w}, \vec{x}_\mu \) - \sum_{i=1}^\ndim \rP_{\w}\(w_i\)\,.
		\end{align*}
		The spin variables denote the linear model weights $\vec{w} \in \bbR^{\ndim}$ that interact through the quenched  dataset $\{\vec{y}, \mat{X}\}$, which plays the role of the exchange interaction. However, here the interactions are not \emph{pairwise}, but instead \emph{fully connected}, meaning that each variable $w_i$ is connected to $\{w_j\}_{j \in \partial i \setminus i }$ through the factors $\rP_{\out} \(y_\mu | \vec{w}, \vec{x}_\mu \)$.
			\begin{figure}[htb!]
			\centering
			\begin{tikzpicture}[scale=0.8, auto, swap]
			    \foreach \i in {1,...,6}
			        \node[var] (X\i) at (1.5*\i,0) {};
			    \node at (11, 0) {$ w_i $};
			
			    \foreach \mu in {1,...,4}
			        \node[inter] (Y\mu) at (1.5+1.5*\mu,-2) {};
			    \foreach \i in {1,...,6}
			        \foreach \mu in {1,...,4}
			            \path[edge] (X\i) -- (Y\mu);
			    \node at (10, -2) {};
			    \node (F) at (11, -2) {$ \rP_{\out}\(y_\mu | \vec{w}, \vec{x}_\mu \) $};			
			    \foreach \i in {1,...,6} {
			        \node[field] (P\i) at (1.5*\i,1) {};
			        \path[edge] (X\i) -- (P\i);
			    }
			    \node at (11, 1) {$ \rP_\w(w_i) $};
			\end{tikzpicture}
			\caption{Factor graph representing the GLM class. The variables $w_i$ are fully connected through the factor $\rP_{\out} \(y_\mu | \vec{w}, \vec{x}_\mu \)$ that represent the constraint imposed by the $\mu$-th example in the dataset. Each variable is connected to a one-body interaction with a separable prior distribution $\rP_\w(w_i)$.}
			\label{fig:main:factor_graph_glm}
			\end{figure}
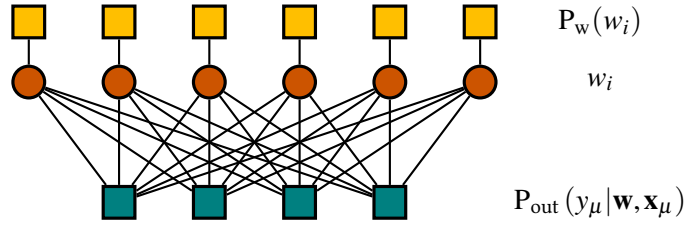
			The corresponding factor graph is represented in \Fig\ref{fig:main:factor_graph_glm} and the partition function at temperature $\beta$ is defined by 
			\begin{align}
			&\mZ_\ndim \(\{\vec{y}, \mat{X}\}; \beta\) \equiv \rP_\ndim \(\vec{y},\mat{X}\) = \int_{\bbR^\ndim} \d\vec{w} ~ e^{-\beta \mH_\ndim \(\vec{w},\{\vec{y},\mat{X}\}\)}  \\
			&= \int_{\bbR^\ndim} \d\vec{w} ~ e^{\beta\( \log \rp_{\out} \(\vec{y} | \vec{w}, \mat{X} \) + \log \rp_{\w}\(\vec{w}\) \)} = \int_{\bbR^\ndim} \d\vec{w} ~ \rp_{\out} \(\vec{y} | \vec{w}, \mat{X} \) \rp_{\w}\(\vec{w}\)\,, \nonumber
			\end{align}
			and can be mapped to Bayesian estimation for $\beta=1$.
			In the considered modern high-dimensional regime with $\ndim \to \infty$, $\nsamples \to\infty$ with $\alpha = \nsamples/\ndim = \Theta(1)$, we are interested to compute the \emph{free entropy} $\Phi$ \emph{averaged} over the input data $\mat{X}$ and teacher weights $\vec{w}^\star$ or equivalently over the output labels $\vec{y}$, 
			defined as
			\begin{align}
				\Phi(\alpha) \equiv \lim_{\ndim \to \infty}  \frac{1}{\ndim}  \EE_{\vec{y},\mat{X}} \[\log  \mZ_\ndim\(\vec{y}, \mat{X}\) \]\,.
				\label{main:glm:free_entropy}
			\end{align}
			The replica method described in \Sec\ref{main:sec:mean_field:replica_method} allows to compute the above average over the dataset $\{\mat{X},\vec{y}\}$, that plays the role of the planted quenched disorder in usual spin glasses.
			The details of the computation can be found in \App\ref{appendix:replica_computation:committee} for committee machines in the case of a synthetic dataset  $\rP_\y\(\vec{y}|\mat{X}\)$ in \eqref{main:eq:replicas:synthetic_data_set}, whereas the similar computation for random labels \eqref{main:eq:replicas:random_labels} is derived in \App\ref{appendix:replica_computation:random_labels:iid}.
									
			\paragraph{Replica computation}
			We present here the replica computation of the averaged free entropy $\Phi(\alpha)$ in eq.~\eqref{main:glm:free_entropy} for general \emph{student} prior and channel distributions $\rP_{\w}$ and $\rP_{\out}$. The average in eq.~\eqref{main:glm:free_entropy} is intractable in general, and the computation relies on the so called \emph{replica trick}, see \Sec\ref{main:intro:replicas:replica_trick}, that consists in applying the identity  
			\begin{align}
				\EE_{\vec{y},\mat{X}} \[ \lim_{\ndim \to \infty} \frac{1}{\ndim} \log  \mZ_\ndim\(\vec{y}, \mat{X}\) \] =  \lim_{r \to 0} \[ \lim_{\ndim \to \infty}  \frac{1}{\ndim}  \frac{\partial \log \EE_{\vec{y},\mat{X}} \[  \mZ_\ndim\(\vec{y}, \mat{X}\)^r\] }{\partial r} \]\,.
				\label{main:glm:replicas:replica_trick}
			\end{align}
			This is interesting in the sense that it reduces the intractable average to the computation of the moments of the averaged partition function, which are easier quantities to compute. Note that for $r \in \bbN$, $\mZ_\ndim\(\vec{y}, \mat{X}\)^r = \prod_{a=1}^r \mZ_\ndim\(\vec{y}, \mat{X}\)$ represents the partition function of $r \in \bbR{\ndim}$ identical non-interacting copies of the initial system, called \emph{replicas}. Taking the quenched average over the disorder will correlate the replicas, before taking the number of replicas $r\to 0$.
			Therefore, we assume there exists an analytical continuation so that $r\in \bbR$ and the limit is well defined. Finally, notice we exchanged the order of the limits $r \to 0$ and $\ndim \to \infty$. These technicalities are crucial points but are not rigorously justified and we will ignore them in the rest of the computation.
			Next, in order to decouple the contributions of the channel $\rP_\out$ and the prior $\rP_\w$, we introduce the variable $\vec{z} = \frac{1}{\sqrt{\ndim}} \mat{X} \vec{w}$ with a Dirac-delta integral
			\begin{align*}
				\mZ_\ndim\(\{\vec{y}, \mat{X}\}\) &= \int_{\bbR^\nsamples} \d\vec{z} ~ \rp_{\out}\(\vec{y} | \vec{z} \) \int_{\bbR^\ndim} \d\vec{w} ~ \rp_{\w}\(\vec{w}\) \delta\(\vec{z} - \frac{1}{\sqrt{\ndim}} \mat{X} \vec{w}\)\,,
			\end{align*} 
			so that the replicated partition function in eq.~\eqref{main:glm:replicas:replica_trick} can be  written as
			\begin{align}
			\begin{aligned}
			\EE_{\vec{y},\mat{X}} \[  \mZ_\ndim\(\vec{y}, \mat{X}\)^r \] &= \EE_{\mat{X}} \int_{\bbR^\nsamples} \d \vec{y}  \prod_{a=0}^r \int_{\bbR^\nsamples} \d\vec{z}^a ~ \rp_{\out^a}\(\vec{y} | \vec{z}^a \) \\
			& \times \int_{\bbR^\ndim} \d\vec{w}^a ~ \rp_{\w^a}\(\vec{w}^a\) \delta\(\vec{z}^a - \frac{1}{\sqrt{\ndim}}\mat{X} \vec{w}^a\)\,,
			\label{main:glm:average_Zn}
			\end{aligned}
			\end{align}
			with the decoupled channel $\rp_{\out}\(\vec{y} | \vec{z} \) = \prod_{\mu=1}^\nsamples \rp_{\out}\(y_{\mu} | z_\mu \)$ and prior $\rp_\w\(\vec{w}\) = \prod_{i=1}^\ndim \rp_\w\(w_i\) $ densities.
			Interestingly the average over $\vec{y}$ is equivalent to the one over the ground truth vector $\vec{w}^\star$ in the \aclink{T-S} scenario. Making use of the analogous formulation in \eqref{main:eq:replicas:synthetic_data_set}, the average can simply be considered as a new replica $\vec{w}^0$ with index $a=0$ leading to a total of $r+1$ replicas. 
			In the case of \emph{random labels} \eqref{main:eq:replicas:random_labels}, $\rP_\y$ is independent of $\mat{X}$ and therefore the computation is similar with only $r$ replicas and an additional average over $\rP_\y$, see \App\ref{appendix:replica_computation:random_labels:iid}.
			
			\paragraph{Average over the iid input data $\mat{X}$}
			We suppose that inputs are drawn from an \aclink{i.i.d} distribution, for example a Gaussian $\rP_{\textrm{x}}(\vec{x}) = \mN_{\vec{x}}\(\vec{0},\mat{I}_\ndim\)$. More precisely, for $(i,j)\in \lb \ndim\rb^2$, $(\mu,\nu) \in \lb \nsamples \rb^2$, $\EE_\mat{X} \[ x_{\mu i} x_{\nu j} \] =  \delta_{\mu\nu} \delta_{ij}$. 
			Hence $z_{\mu}^a =\frac{1}{\sqrt{\ndim}} \sum_{i=1}^\ndim x_{\mu i} w_i^a$ is the sum of \aclink{i.i.d} random variables. The \aclink{CLT} insures that in the thermodynamic limit $z_{\mu}^a \sim \mN\(\EE_{\mat{X}}[z_\mu^a]  ,\EE_{\mat{X}}[z_\mu^a z_\mu^b] \)$, with the two first moments given by:
			\begin{align*}
					\EE_{\mat{X}}[z_\mu^a] &= \frac{1}{\sqrt{\ndim}} \sum_{i=1}^\ndim \EE_{\mat{X}}\[x_{\mu i}\] w_i^a =0\,, \spacecase
					\EE_{\mat{X}}[z_\mu^a z_\mu^b] &= \frac{1}{\ndim} \sum_{ij} \EE_{\mat{X}}\[x_{\mu i} x_{\mu j}\] w_i^a w_j^b  = \frac{1}{\ndim} \sum_{ij}  \delta_{ij} w_i^a w_j^b = \frac{1}{\ndim} \vec{w}^a \cdot \vec{w}^b   \,.
			\end{align*}
			Note that averaging over the quenched disorder induces correlations between replicas, which were initially independent.
			In the following we introduce the symmetric \emph{overlap} matrix that measures the correlations between the replicated vector $\vec{w}^a$: $\mat{Q}(\{\vec{w}^a\})\equiv\(\frac{1}{\ndim} \vec{w}^a \cdot \vec{w}^b\)_{a,b=0..r}$. 
			Let us define $\td{\vec{z}}_{\mu} \equiv (z^a_{\mu})_{a=0..r}$ and $\td{\vec{w}}_i \equiv (w_i^a)_{a=0..r}$ the replicated vectors. 
			The vector $\td{\vec{z}}_{\mu}$ follows a multivariate Gaussian distribution $\td{\vec{z}}_{\mu} \sim \rP_{\tilde{\z}}(\tilde{\vec{z}};\mat{Q}) = \mN_{\tilde{\vec{z}}}( \tbf{0}_{r+1}, \mat{Q})$ and as the \aclink{i.i.d} prior and channel distributions factorize $
			        \rp_{\w}(\vec{w}) = \prod_{i=1}^\ndim \rp_{\w}(w_i)$ and $\rp_{\out}(\vec{y} | \vec{z}) = \prod_{\mu=1}^\nsamples \rp_{\out}(y^{(\mu)}~|~ z^{(\mu)})$, it follows
			 \begin{align*}       
			  &\EE_{\vec{y},\mat{X}} \[  \mZ_\ndim\(\vec{y}, \mat{X}\)^r \] \\
			  &= \EE_{\mat{X}}  \int_{\bbR^\nsamples} \d \vec{y} ~\prod_{a=0}^r \int_{\bbR^\nsamples} \d\vec{z}^a ~ \rp_{\out^a}\(\vec{y} | \vec{z}^a \) \\
			  & \qquad \qquad \qquad \qquad \times \int_{\bbR^\ndim} \d\vec{w}^a ~ \rp_{\w^a}\(\vec{w}^a\) \delta\(\vec{z}^a -\frac{1}{\sqrt{\ndim}} \mat{X} \vec{w}^a\) \\
			  &= \[ \int_{\bbR} \d y \int_{\bbR^{r+1}} \d\td{\vec{z}} ~ \rp_{\out}\(y | \td{\vec{z}} \) \rp_{\tilde{\z}}(\tilde{\vec{z}};Q(\td{\vec{w}}))    \]^{\nsamples} \[\int_{\bbR^{r+1}} \d\td{\vec{w}} ~ \rp_{\td{\w}}\(\td{\vec{w}}\)\]^{\ndim}\,,
			\end{align*}     
			where we introduced $\rP_{\td{\w}}\(\td{\vec{w}}\) = \prod_{a=0}^r \rP_{\w}\(w^a\)$ the distribution of the replicated vector of weights. To finish decoupling the channels, we use the Fourier representation of a Dirac-delta function of a variable $x \in \bbR$ as a function of a purely imaginary parameter $\hat{x}$:
			\begin{align*}
				\delta\(x\) = \frac{1}{2 i \pi} \int_{i\bbR} \d \hat{x} ~ e^{-\hat{x} x }\,.
			\end{align*}
			 Applying the above identity to the following change of variable
			 \begin{align*}
			 	&1 = \int_{\bbR^{r+1 \times r+1}} \d \mat{Q} \prod_{0 \leq a \leq b \leq r} \delta \(\ndim Q_{ab}-\sum_{i=1}^\ndim w_i^a w_i^b \)\\
				&\propto \int \int_{\bbR^{r+1 \times r+1}} \d \mat{Q} \d\hat{\mat{Q}} ~ \exp \(- \ndim \tr{\mat{Q} \hat{\mat{Q}}} \)    e^{\frac{1}{2} \sum_{i=1}^\ndim  \td{\vec{w}}_i^{\intercal} \hat{\mat{Q}} \td{\vec{w}}_i + \td{\vec{w}}_i^{\intercal} \diag{\hat{\mat{Q}}} \td{\vec{w}}_i } \,,
			\end{align*} 
			that involves a new ad-hoc purely imaginary matrix parameter $\hat{\mat{Q}}$.
			Hence, multiplying the replicated partition function by $1$, it becomes an integral over the symmetric matrices $\mat{Q} \in \bbR^{r+1 \times r+1}$ and $\hat{\mat{Q}} \in \bbR^{r+1 \times r+1}$, that can be evaluated using a Laplace method \cite{Rong89} in the $\ndim \to \infty$ limit,
			\begin{align}
			\begin{aligned}
				\EE_{\vec{y},\mat{X}} \[  \mZ_\ndim\(\vec{y}, \mat{X}\)^r \] &= \int_{\bbR^{r+1 \times r+1}} ~ \d \mat{Q} ~ \int_{\bbR^{r+1 \times r+1}} \d\hat{\mat{Q}} e^{\ndim \Phi^{(r)} (\mat{Q},\hat{\mat{Q}} ) }\\
				&\underset{\ndim \to \infty}{\simeq} \exp\(\ndim \cdot  \extr_{\mat{Q}, \hat{\mat{Q}}} \left\{ \Phi^{(r)}\(\mat{Q},\hat{\mat{Q}}\) \right\}\), 
			\end{aligned}
			\end{align}
			where we omitted the sub-leading factors and defined the free entropy potential
			\begin{align}
			     \begin{aligned}
			     \Phi^{(r)} (\mat{Q},\hat{\mat{Q}}) &= -\tr{\mat{Q}\hat{\mat{Q}}} +\log \Psi_{\w}^{(r)} (\hat{\mat{Q}})+\alpha\log \Psi_{\out}^{(r)}(\mat{Q})\,,\spacecase
			      \Psi_{\w}^{(r)} (\hat{\mat{Q}}) &= \displaystyle \int_{\mathbb{R}^{r+1}} \d \td{\vec{w}}~ \rp_{\tilde{\w}}(\td{\vec{w}})  e^{ \frac{1}{2}\td{\vec{w}}^{\intercal} \hat{\mat{Q}} \td{\vec{w}} + \frac{1}{2} \td{\vec{w}}^{\intercal} \diag{\hat{\mat{Q}}} \td{\vec{w}}  }   \spacecase
			      \Psi_{\out}^{(r)}(\mat{Q}) &=  \displaystyle \int \d_\bbR y \int_{\mathbb{R}^{r+1}}  \d\td{\vec{z}}~ \rp_{\tilde{\z}}(\td{\vec{z}};\mat{Q}) \rp_{\out}(y| \td{\vec{z}})\,,
			     \end{aligned}
			    \label{main:intro:replicas:Phi_r}
			\end{align}
			and $\rP_{\td{z}} (\td{\vec{z}};\mat{Q}) = \displaystyle e^{-\frac{1}{2}\td{\vec{z}}^\intercal \mat{Q}^{-1} \td{\vec{z}}}/\det{2\pi \mat{Q}}^{1/2}$. Recall that the average over the teacher vector has been merged as a new replica so that $\rp_{\out^0}=\rp_{\out^\star}$, $\rp_{\w^0}=\rp_{\w^\star}$.
			Finally switching the two limits $r\to 0$ and $\ndim \to \infty$, the quenched free entropy $\Phi$ simplifies to a saddle point equation
			\begin{equation}
				\Phi (\alpha) = \extr_{ \mat{Q}, \hat{\mat{Q}} } \left\{\lim_{r\rightarrow 0} \frac{\partial \Phi^{(r)}(\mat{Q},\hat{\mat{Q}})}{\partial  r} \right\} ,
				\label{main:intro:replicas:Phi_r_fonctional}
			\end{equation}
			over symmetric matrices $\mat{Q}\in \bbR^{r+1 \times r+1}$ and $\hat{\mat{Q}} \in \bbR^{r+1 \times r+1}$. In the following we will assume a simple Ansatz for these matrices in order to first obtain an analytic expression in $r$ before taking the derivative with respect to $r$. Note that the partition function of this fully connected model can be expressed as a saddle point only because distributions $\rP_\out$ and $\rP_\w$ factorize so that a pre-factor scaling with the system size dominates the exponential distribution.
						
						
			\paragraph{Replica Symmetric free entropy}
			Let's compute the functional $\Phi^{(r)}(\mat{Q},\hat{\mat{Q}})$ in \eq\eqref{main:intro:replicas:Phi_r_fonctional} in the simplest Ansatz: the \aclink{RS} Ansatz. This latter assumes that all the replicas remain equivalent with a common overlap $q = \frac{1}{\ndim} \vec{w}^a \cdot \vec{w}^b$ for $a \ne b$, a norm $Q= \frac{1}{\ndim} \|\vec{w}^a\|_2^2$, and an overlap with the ground truth $m =\frac{1}{\ndim} \vec{w}^a \cdot \vec{w}^\star$, leading to the following expressions of the replica symmetric matrices $\mat{Q}^{(\textrm{rs})} \in \mathbb{R}^{r+1\times r+1}$ and $\hat{\mat{Q}}^{(\textrm{rs})} \in \mathbb{R}^{r+1\times r+1}$:
			\begin{equation}
			\begin{aligned}[c]
			\mat{Q}^{(\textrm{rs})} =\scalemath{0.85}{\begin{pmatrix} 
			Q^\star & m & \cdots & m \\
			 m & Q & q & q  \\
			 \vdots & q & \ddots & q  \\
			 m & q & q & Q    \\
			\end{pmatrix}}
			\end{aligned}
			\hhspace
			\textrm{and} 
			\hhspace
			\begin{aligned}[c]
			\hat{\mat{Q}}^{(\textrm{rs})}=\scalemath{0.85}{\begin{pmatrix} 
			 -\frac{1}{2} \hat{Q}^\star & \hat{m} & \cdots & \hat{m}\\
			\hat{m} &-\frac{1}{2}\hat{Q} & \hat{q} & \hat{q}  \\
			\vdots & \hat{q} & \ddots & \hat{q}  \\
			\hat{m} & \hat{q} & \hat{q} & -\frac{1}{2}\hat{Q}\\  
			\end{pmatrix}}
			\end{aligned}
			\end{equation}
			with $Q^\star=\frac{1}{\ndim} \|\vec{w}^\star\|_2^2$. The factor $-\frac{1}{2}$ is not necessary but useful to recover commonly used formulations. The functional $\Phi^{(r)}(\mat{Q},\hat{\mat{Q}})$can be computed with this Ansatz: the first is a trace term, the second term $\Psi_{\w}^{(r)}$ depends on the prior distributions $\rP_\w$, $\rP_{\w^\star}$ and finally the third term $\Psi_{\out}^{(r)}$ depends on the channel distributions $\rP_{\out^\star}$, $\rP_\out$.
			
			\paragraph{Replica trick $r \to 0$ limit}		
				The last step of the computation is to take properly the limit $r\to 0$. We obtain that
				\begin{align}
				-\lim_{r \to 0} \partial_r \left. \tr{\mat{Q}\hat{\mat{Q}}}  \right|_{\textrm{rs}} = - m \hat{m} + \frac{1}{2} Q \hat{Q} + \frac{1}{2} q \hat{q}\,.
				\label{main:intro:replicas:Phi_r}
				\end{align}
				and 
			\begin{align}
			&\lim_{r \to 0} \partial_r \left. \log \Psi_{\w}^{(r)} (\hat{\mat{Q}})\right|_{\textrm{rs}} = \nonumber \\ 
			& \hhspace\hhspace\hhspace \EE_{\xi, w^\star} \mZ_{\w^\star}\(\hat{m}\hat{q}^{-1/2} \xi, \hat{m}^2\hat{q}^{-1} \) \log \mZ_\w\(\hat{q}^{1/2} \xi , \hat{Q} + \hat{q}\)\,,\nonumber \\
			&\lim_{r \to 0} \partial_r \left. \log  \Psi_{\out}^{(r)} (\mat{Q})\right|_{\textrm{rs}} = \label{main:intro:replicas:log_Psi_w_out}\\ 
			&\hhspace\hhspace\hhspace \int \d y \EE_{\xi} \mZ_{\out^\star}\( mq^{-1/2}\xi, Q^\star - m^2 q^{-1} \)  \log \mZ_{\out}\(q^{1/2} \xi, Q-q  \)\,, \nonumber
			\end{align}
			with denoising functions $\mZ_{\out^\star},\mZ_{\out}, \mZ_{\w^\star}, \mZ_{\w}$ defined in \App\ref{appendix:update_functions}.
			
			\subsubsection{Summary}
			Gathering eq.~(\ref{main:intro:replicas:Phi_r}, 
				\ref{main:intro:replicas:log_Psi_w_out}), we finally obtain the \aclink{RS} free entropy $\Phi_{\textrm{rs}}$.
			\begin{align}
				&\Phi_{\textrm{rs}}(\alpha) \equiv \extr_{Q,\hat{Q},q,\hat{q},m,\hat{m} } \left\{\lim_{r\rightarrow 0} \partial_r \Phi^{(r)}(\mat{Q}^{(\textrm{rs})},\hat{\mat{Q}}^{(\textrm{rs})}) \right\} \nonumber \\
				&=  \extr_{Q,\hat{Q},q,\hat{q},m,\hat{m}} \left \{  - m \hat{m} + \frac{1}{2} Q \hat{Q} + \frac{1}{2} q \hat{q} \right. \label{main:intro:replicas:free_entropy_non_bayes} \\
				&  \qquad \qquad \qquad \qquad \qquad \left. + \Psi_\w\(\hat{Q},\hat{m},\hat{q}  \) + \alpha \Psi_\out\(Q,m,q ;  \rho_{\w^\star}\) \right\} \,, \nonumber
			\end{align}
			where $\rho_{\w^\star}= \lim_{\ndim \to \infty} \EE_{\vec{w}^\star} \frac{1}{\ndim} \|\vec{w}^\star\|_2^2$ and the channel and prior integrals are defined by
			\begin{align}
				&\Psi_\w\(\hat{Q},\hat{m},\hat{q}  \) \equiv \EE_{\xi} \[ \mZ_{\w^\star}\( \hat{m}  \hat{q}^{-1/2}  \xi , \hat{m}^2 \hat{q}^{-1}  \) \log \mZ_{\w} \(  \hat{q}^{1/2}\xi  , \hat{Q} + \hat{q} \)   \]\,, \nonumber \spacecase
				&\Psi_\out\(Q,m,q; \rho_{\w^\star}\) \equiv \EE_{y, \xi } \[ \mZ_{\out^\star} \( y,  m q^{-1/2}\xi, \rho_{\w^\star} - m q^{-1}m  \)\right. \\
				& \qquad \qquad \qquad \qquad \qquad \qquad \left. \times \log \mZ_{\out} \( y,  q^{1/2}\xi, Q - q \)  \]\,, \nonumber
			\end{align}
			 for generic $\rP_{\out^\star}, \rP_{\out}$ and $\rP_{\w^\star}, \rP_{\w}$ distributions and corresponding update functions $\mZ_{\out^\star},\mZ_{\out}, \mZ_{\w^\star}, \mZ_{\w}$ are defined in \App\ref{appendix:update_functions}.
			As a conclusion, we notice remarkably that the behavior of the initial complex high-dimensional inference problem is charaterized by an optimization problem over only six scalar order parameters, and is therefore controlled by a set of six fixed point equations. Finally, let us mention that \aclink{MMSE} estimation can be performed in the Bayes-optimal setting for $\beta=1$, while \aclink{MAP} estimation requires to take properly $\beta=\infty$ as detailed later on in \Chap\ref{chap:erm}.
			
			\paragraph{Bayes-optimal free entropy}
				In the Bayes-optimal setting, we perform inference using the knowledge of the \emph{ground truth} distributions so that the student denoising functions are exactly the ones used to generate the dataset, namely $\rP_{\out} = \rP_{\out^\star}$ and $\rP_\w = \rP_{\w^\star}$
				so that $\mZ_{\out}=\mZ_{\out^\star}$, $\mZ_{\w}=\mZ_{\w^\star}$. 
				The Nishimori's conditions in the Bayes-optimal case, derived in \App\ref{appendix:replica_computation:nishimori}, imply that $Q=Q^\star\equiv\rho_{\w^\star}$, $m=q \equiv q_\bayes$, $\hat{Q}=\hat{Q}^\star=0$, $\hat{m}=\hat{q}\equiv\hat{q}_\bayes$.
				Therefore, in the Bayes-optimal setting, the free entropy of the high-dimensional inference problem eq.~\eqref{main:intro:replicas:free_entropy_non_bayes} simplifies as an optimization problem over scalar \emph{overlaps} parameters $q_{\textrm{b}}, \hat{q}_{\textrm{b}}$:
				\begin{align}
					\Phi_{\textrm{rs}}^\bayes(\alpha) &=  \extr_{q_{\textrm{b}},\hat{q}_{\textrm{b}}} \left \{ - \frac{1}{2} q_\bayes \hat{q}_{\textrm{b}}  + \Psi_{\w}^{\textrm{b}}\(\hat{q}_{\textrm{b}}  \) + \alpha \Psi_{\out}^{\textrm{b}}\(q_{\textrm{b}}; \rho_{\w^\star}\) \right\} \,,
					\label{appendix:free_entropy_bayes}
				\end{align}
				with free entropy terms $\Psi_{\w}^\bayes$ and $\Psi_{\out}^\bayes$ given by
				\begin{align*}
						\Psi_{\w}^\bayes\(\hat{q}_{\textrm{b}}\) &= \EE_{\xi} \[ \mZ_{\w^\star} \(  \hat{q}_{\textrm{b}}^{1/2}\xi,   \hat{q}_{\textrm{b}} \) \log \mZ_{\w^\star} \(  \hat{q}_{\textrm{b}}^{1/2}\xi,   \hat{q}_{\textrm{b}} \)   \] \,, \spacecase 
						\Psi_{\out}^\bayes \(q_{\textrm{b}}; \rho_{\w^\star}\) &= \EE_{y, \xi } \[ \mZ_{\out^\star} \( y,  q_{\textrm{b}}^{1/2}\xi, \rho_{\w^\star} - q_{\textrm{b}} \)\right. \\
						& \qquad \qquad \qquad \qquad \qquad  \left. \log \mZ_{\out^\star} \( y,  q_{\textrm{b}}^{1/2}\xi, \rho_{\w^\star} - q_{\textrm{b}} \)  \]\,.
				\end{align*} 
				Notice that the above Bayes-optimal replica symmetric free entropy for the \aclink{GLM} class has been rigorously proven in \cite{barbier2017phase}. Taking the derivatives with respect to $q_{\textrm{b}}, \hat{q}_{\textrm{b}}$, we obtain the stationary conditions verified by the overlap parameters
				\begin{align}
				\begin{aligned}	
					q_{\textrm{b}} &= \alpha \EE_{y, \xi} \mZ_{\out^\star} \( y,  q_{\textrm{b}}^{1/2}\xi, \rho_{\w^\star} - q_{\textrm{b}} \)  f_{\out^\star} \( y,  q_{\textrm{b}}^{1/2}\xi, \rho_{\w^\star} - q_{\textrm{b}} \)^2 \\
					\hat{q}_{\textrm{b}} &= \EE_{\xi} \mZ_{\w^\star}\(\hat{q}_{\textrm{b}}^{1/2}\xi,   \hat{q}_{\textrm{b}} \) f_{\w^\star}\(\hat{q}_{\textrm{b}}^{1/2}\xi,   \hat{q}_{\textrm{b}} \)^2 \,,
					\label{main:intro:replicas:glm:fixed_point}
				\end{aligned}	
				\end{align}
				that will turn out to be strongly connected to the infinite-size behavior of the \aclink{AMP} algorithm, the so-called \emph{state evolution} equations. \\
				
				In this section, we have presented the heuristic replica method which provides a powerful technique to directly compute the free entropy, associated to a complex \aclink{JPD}, and to describe the statistical thresholds of the corresponding phase diagram. 
				Next, we present other mean-field methods to perform approximate inference of this same \aclink{JPD}. 
				Interestingly, even though these techniques do not directly yield the result like the replica method, however, they have the profound advantage of leading to interesting algorithmic perspectives and insights to complete the phase diagram.
									
	\newpage
	\section{On variational mean-field methods}	
		\label{main:sec:mean_fields}
		Assume we consider a statistical model associated to a \aclink{JPD} $\rP_{\ndim} (\bsigma; \beta)$ and an Hamiltonian energy function $\mH_\ndim(\bsigma)$. The main challenge is to compute its marginal probabilities, moments or even more complex observable of the \aclink{JPD}. 
								
		\paragraph{Intractability of exact inference}
			However, computing analytically the posterior, with or without the replica method, is very rarely possible.
			In general, even though the replica method provides a quick and strong tool to calculate it in some particular cases,
			computing the marginal probabilities of a high-dimensional \aclink{JPD} $\rP_{\ndim} (\bsigma; \beta)$ according to $\rP(\sigma_i) = \int_{\chi_{\ndim-1}} \d \bsigma_{\setminus i} \rP_{\ndim} (\bsigma; \beta)$,
			for some $i\in \lb \ndim \rb$, remains a hard task. Indeed computing the corresponding continuous or discrete sum requires very often a number of operations that scales exponentially with the size of the system and becomes critical in the high-dimensional regime that we consider $\ndim\to \infty$. 
			Of course in the case where the spins are restricted to one-body interactions and do not interact, the \aclink{JPD} distribution factorizes and the sum over $\bbR^{\ndim}$ reduces to a sum over $\bbR$ and deeply simplify the computation.
			Yet, this kind of simplification remains very limited and, moreover,complex and interesting behaviors arise very often only when \emph{interactions} come on stage. 
					
		\paragraph{On Tree factor graphs}
			Let us first draw attention on very simple factor graphs and corresponding \aclink{JPD}.
			In the case where the factor graph under consideration is a \emph{tree}\index{tree}, as an illustration see for instance \Fig\ref{fig:main:factor_graph} \Right, the computation of the \aclink{JPD} can be performed in linear time complexity, in contrast with the exponential complexity mentioned above. 
			Indeed using the Markov property and conditional expectation \cite{pearl1982reverend,pearl1986fusion}, it is possible to compute the whole \aclink{JPD} as a product of $\Theta(\ndim)$ terms.
			Moreover, this procedure may be turned into a dynamical algorithm called the \emph{sum-product algorithm} or \aclink{BP} equations that, as we just stressed, is \emph{exact on tree} factor graphs. The corresponding algorithm reaches the fixed point of a well designed free energy approximation, the \emph{Bethe free energy}, detailed in \Sec\ref{main:intro:mean_field:bp}.
			More interestingly, it can approximate correctly the target \aclink{JPD} on loopy \emph{factor graphs} as well, even though it is not guaranteed to converge. 
			
		\paragraph{Approximate variational methods}
		On general factor graphs, we therefore have to resort to \emph{approximate inference}, to circumvent this difficulty and compute \emph{approximately} and \emph{efficiently} the marginal probabilities $\rP(\sigma_i)$.
		Sampling methods relying on \aclink{MCMC} algorithms, see \Sec\ref{main:intro:algos:sampling}, are widely used in practice. Yet, they are not very performant especially in the high-dimensional inference regime of interest. 
		To address this issue, instead trying to sample a huge number of examples, other \emph{variational mean-field method} have been designed, in particular in physics, to compute a good approximation of the posterior distribution \cite{opper2001advanced}.\\
		
		The design of such mean-field approximations requires, first, to recall and introduce some useful \aclink{IT} quantities in \Sec\ref{sec:main:introduction:stat_phys:info_graphs:info_theory}, that naturally lead to the \emph{Gibbs free energy} and its variational formulation presented in \Sec\ref{sec:main:intro:variational:gibbs_free_energy}. 
		Finally, in \Sec\ref{sec:main:intro:variational:naive_tap}, we recall the \emph{naive} mean-field approach and its extension to more complex approximations, such as the \aclink{TAP} approach. 
		For an extended introduction on variational mean-field methods, let us mention the comprehensive review \cite{blei2017variational}.
		
	\subsection{Information theory quantities}
	\label{sec:main:introduction:stat_phys:info_graphs:info_theory}
		In the perspective of comparing and constructing approximations of the complex \aclink{JPD} associated to interacting systems, we introduce the classical tools from \aclink{IT} to compare distribution families such as the Shanon entropy, the \aclink{KL} divergence and the mutual information. More details can be found in \cite{mackay2003information, koller2009probabilistic}. 
						
		\subsubsection{Shanon Entropy} 
		\label{sec:main:intro:mean_field:entropy}
			Let $\rX$ be a \aclink{RV} with probability distribution $\rP$ and density $\rp(s) \equiv \d \rP / \d x$ on a set $\bbX$, the Shannon entropy $\rH(\rX)$ measures the quantity of information carried by the \aclink{RV} $\rX$ and is defined by
				\begin{align}
					\rH(\rX) &= - \EE_{\rX \sim \rP} \[ \log \rP(x) \] = - \int_{\bbX} \d x ~ \rp(x) \log \rp(x) \,.
					\label{main:intro:info_theory:entropy}	
				\end{align}	
							
		\subsubsection{The Kullback-Leibler divergence}
		\label{definition:kullback_leibler_divergence}
			Consider two probability distributions $\rQ$ and $\rP$, with densities $q,p$ on a set $\bbX$.
			The \aclink{KL} divergence \index{Kullback-Leibler divergence} is used to compare two arbitrary distributions $\rP$ and $\rQ$, defined as 
				\begin{align}
				\begin{aligned}
					\mD_{\textrm{KL}}\( \rQ \parallel \rP \) &= \EE_{\rX \sim  \rQ } \[\log  \rQ(x) - \log \rP(x) \] \\
					&= \int_{\bbX} \d x ~ \rq(x) \log\(\frac{\rq(x)}{\rp(x)}\) \,, 
				\end{aligned}
				\end{align}
				with densities $\rp, \rq$ defined by $\d \rP \equiv \rp(x) \d x$, $ \d \rQ  \equiv \rq(x) \d x$.
				Because it is not symmetric under the exchange of $\rQ$ and $\rP$, $\mD_{\textrm{KL}}\( \rQ \parallel \rP \)\ne \mD_{\textrm{KL}}\( \rP \parallel \rQ \)$ and does not verify the triangle inequality, the \aclink{KL} divergence is not formally a distance in the rigorous mathematical sense. 
				However, it plays exactly the role of a distance in the space of probability densities as it is always positive as stated by the \emph{Gibb’s inequality}:
					\begin{proposition}[From \cite{cover2012elements}] Consider two distributions $\rP, \rQ$ with densities $\rp, \rq$, then
						{$\mD_{\textrm{KL}}\( \rQ \parallel \rP \) \geq 0$} and $ \mD_{\textrm{KL}}\( \rQ \parallel \rP \) = 0 \Leftrightarrow \rQ = \rP$.
					\end{proposition}
					\newpage
					\begin{proof}
					\graffito{Recall that a function $f$ is concave if for $x_1 \leq x\leq x_2 $ the point $(x,f(x))$ is above the line joining the points $(x_1, f(x_1))$, $(x_2, f(x_2))$.
					\begin{tikzpicture}
						\draw[->, very thick] (-1,0)--(1.2,0) node[right]{$x$};
						\draw[->, very thick] (0,-1)--(0,1) node[above]{$f(x)$};
						\draw [burntorange, very thick] plot [smooth, tension=1] coordinates {(-1,-1) (0,0.8) (1,0.5)};
						\draw [black, dashed, thick] plot coordinates {(-1,-1) (1,0.5)};	
						\draw[black,fill=black] (-1,-1) circle[radius=0.5pt];
						\draw[black,fill=black] (1,0.5) circle[radius=0.5pt];
					\end{tikzpicture}
					}
						As the \emph{logarithm} is concave, from the Jensen inequality we obtain
						\begin{align*}
							-\mD_{\textrm{KL}}\( \rQ \parallel \rP \) = \int_{\bbX} \d x ~ \rq(x) \log\(\frac{\rp(x)}{\rq(x)}\)  \leq  \int_{\bbX} \d x ~ \rq(x) - \rp(x)  = 0 \,.
						\end{align*}
					\end{proof}
	
		\subsubsection{The mutual information}
			Consider two random variables $\rX$ and $\rY$ jointly distributed according to $\rP_{\rX,\rY}$, the mutual information specifically measures the \aclink{KL} divergence from the product $\rP_{\rX} \rP_{\rY}$ to the joint distribution $\rP_{\rX,\rY}$:
				\begin{align}
					\mI\(\rX; \rY\) &=  \mD_{\textrm{KL}}\( \rP_{\rX,\rY} \parallel \rP_{\rX} \rP_{\rY} \) = \int_{\bbX^2} \d x \d y  ~ \rp(x, y) \log\(\frac{\rp(x, y)}{\rp(x) \rp(y))}\)  \notag \\
					&= \rH \(\rX\) - \rH\( \rX \vert \rY \) = \rH \(\rY\) - \rH\( \rY \vert \rX \) \\
					&= \rH \(\rX\) + \rH \(\rY\) - \rH\( \rX , \rY \)\,, \notag
				\end{align}
			where the marginal densities write $\rp(x) = \int_\bbX \d y ~ \rp(x,y)$ and $\rp(y) = \int_\bbX \d x ~ \rp(x,y)$.	
				
	\subsection{Gibbs free energy and variational principle}
	\label{sec:main:intro:variational:gibbs_free_energy}
			Let us consider a \aclink{JPD} that we aim to approximate, for instance $\rP_{\ndim}(\bsigma; \beta) \equiv e^{-\beta \mH_\ndim(\bsigma)} / \mZ_\ndim(\beta)$, associated to the Hamiltonian $\mH_\ndim(\bsigma)$ for some spins $\bsigma \in \chi_\ndim$.
			For any arbitrary probability distribution $\rQ$, we define the Gibbs free energy as the trade-off between the variational energy $\rU[\rQ] \equiv \EE_{\bsigma \sim \rQ } \[\mH_\ndim(\bsigma)\] $ and the entropy of the distribution $\rH[\rQ]$ according to
			\begin{align}
			   \varphi^{\mathrm{gibbs}}_{\ndim}[\rQ]& \equiv \rU[\rQ] - \frac{1}{\beta} \rH[\rQ]\,,
			    \label{main:intro:variational:gibbs_free_energy}
			\end{align}
			where $\beta$ is a free inverse temperature parameter. In order to find a good mean-field approximation, we introduce the Gibbs variational principle that states that the Gibbs free energy is minimal when the mean-field approximation equals the target \aclink{JPD} distribution $\rP_{\ndim}$. 
						
		\subsubsection{Gibbs variational principle}
			The Gibbs variational principle follows from the fact that for any arbitrary distribution $\rQ$, the Gibbs free energy may be rewritten as
			\begin{align}
			    &\varphi^{\mathrm{gibbs}}_\ndim[\rQ]
			    = \EE_{\bsigma \sim \rQ} \sqbrs{ \mH_\ndim(\bsigma) } \nonumber \\
			    & \qquad \qquad \qquad \qquad + \frac{1}{\beta} \int_{\chi_\ndim} \d \bsigma ~ \rq(\bsigma)\( \log \frac{\rq(\bsigma)}{\rp(\bsigma)} +  \rq(\bsigma) \log \rp(\bsigma) \)  \nonumber \\
			    &= \frac{1}{\beta} \mD_{\textrm{KL}}\( \rQ || \rP_\ndim \) + \varphi_\ndim(\beta)
			    \ge \varphi_\ndim(\beta)\,,
			\label{main:intro:variational:gibbs_variational_principle}
			\end{align}
			where we first used the definition of the target free energy $\varphi_\ndim(\beta) =- \frac{1}{\beta} \log$\newline $\mZ_\ndim(\beta)$ and the positivity of the \aclink{KL} divergence. 
			The last inequality is known as the \emph{Gibbs variational principle} also called the \emph{Gibbs-Bogoliubov-Feynman inequality}.
			As a consequence the Gibbs free energy of any approximate distribution $\rQ$ is larger than the true free energy $\varphi_\ndim(\beta)$ associated to the $\rP_\ndim$, namely $\varphi^{\mathrm{gibbs}}_\ndim[\rQ] \geq \varphi_\ndim(\beta)$. Moreover, the inequality is saturated if the approximation exactly equals the Gibbs distribution $\rQ = \rP_{\ndim}$. 
			 This variational principle allows to measure the correctness of a given approximation. 
			 However, this variational principle cannot be solved in full generality, and we therefore need to restrict the set of possible probability distributions to a practical set.
			 Instead of choosing arbitrarily a potential set, we present in the next section the \emph{maximum entropy principle} which allows to restrict approximations to simple distributions families.
			
			\subsubsection{Maximum entropy principle}
			\label{sec:main:mean_field:maximum_entropy}
			To restrict the space of probability densities, first we consider only the ones that verify the \emph{moments matching conditions} of a set of moments $\{\phi_k\}_{k\in \bbK}$, such that $\EE_{\bsigma \sim \rQ}[\phi_k(\bsigma)] = \mu_k$. 
			\graffito{The maximum entropy principle is similar to the middle age philosophical principle known as the Ockham Razor. It has been slightly modified and popularized by \emph{the Shadoks} from Jacques Rouvel.}
			\graffito{\includegraphics[scale=0.25]{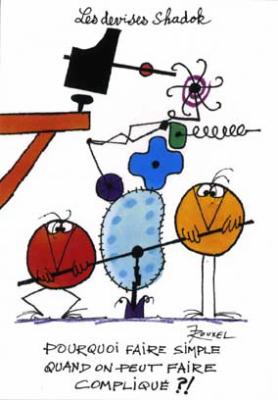}}
			In other words, in expectation the $k$-th moments in the index set $\bbK$ of the approximated distribution $\rQ$ should match the true moments $\{\mu_k\}_{k \in \bbK}$ of the target distribution $\rP_\ndim$. 
			Yet, imposing the moments matching constraints does not determine uniquely the distribution, since in general it exists an infinite number of solutions verifying them. 
			The  \emph{maximum entropy principle}, introduced by \cite{Jaynes57,jaynes03, wainwright2008graphical} and very close to the \emph{least action principle} in analytical mechanics, allows to prescribe
			\emph{good choices} for the approximation $\rQ$ of the \aclink{JPD} $\rP_\ndim$.
			Assume we have access to $\nsamples$ observations $\{\bsigma^{(1)},\cdots, \bsigma^{(\nsamples)} \}$ drawn from the target Gibbs distribution, or any other distribution we try to approximate, the maximum entropy principle simply states that the probability distribution which best represents the \emph{current state of knowledge} is the one with the \emph{largest entropy}.	
			In more details, imposing the normalization and the moments matching, the least action principle can be formulated mathematically as a Lagrangian problem over the distribution $\rQ$:
				\begin{align*}
					\mL\[\rQ\] &= \rH[\rQ] - \sum_{k\in \bbK}  \lambda_k  \(\int_{\chi_\ndim} \d \bsigma ~ \rq (\bsigma) \phi_k(\bsigma)  - \mu_k \) \\
					& \qquad \qquad \qquad \qquad \qquad  \qquad  - \lambda_0 \(\int_{\chi_\ndim} \d \bsigma ~ \rq(\bsigma)  - 1 \)
				\end{align*}
				\emph{Minimizing the action}, \ie the Lagrangian $\mL\[\rQ\]$, yields
				\begin{align*}
					0 = \frac{\partial \mL}{\partial \rQ} = \log \rQ(\bsigma) + 1 - \sum_{k\in \bbK} \lambda_k \phi_k(\bsigma)  -\lambda_0 \,.
				\end{align*}
				Imposing the normalization $\int_{\chi_\ndim} \d \bsigma ~ \rq (\bsigma) =1 $  we finally recover the so-called \emph{exponential family} \cite{jordan1999introduction}
				\begin{align}
					\rQ(\bsigma) = \frac{1}{\mZ} \exp\(-\sum_{k\in \bbK} \lambda_k \phi_k(\bsigma)\) \,, 
					\label{eq:main:intro:mean_field:exponential_family}
				\end{align}	
				where $\mZ = e^{1-\lambda_0} = \int_{\chi_\ndim} \d \bsigma ~ \e^{-\sum_{k\in\bbK} \lambda_k \phi_k(\bsigma)} $ with potential additional cut-offs to avoid the normalizing constant to diverge, especially for $k=1$.
				To summarize, the least biased distribution to consider to approximate the Gibbs distribution are the ones belonging to the exponential family. 
								
		\subsection{Naive and TAP mean-field approximation}
		\label{sec:main:intro:variational:naive_tap}
		We present two simple approximations considered in the physics literature \cite{opper2001advanced}, starting with the naive mean-field approximation and the \aclink{TAP} approach, which can be derived from the Gibbs variational principle.
				
		\subsubsection{Naive mean-field approximation}
		\label{sec:main:intro:variational:naive}
		The \emph{naive} mean-field approximation consists in a simple factorized density approximation $\rQ^{\textrm{naive}}(\bsigma) = \prod_{i=1}^\ndim \rQ_i(\sigma_i)$ of $\ndim$ independent spins. 
		It has been introduced in classical physics long time ago in the celebrated Curie-Weiss model \cite{curie1895proprietes,weiss1907hypothese} to study magnetic properties of materials. Hundreds years later, the naive mean-field has been largely democratized and used 
		in various communities \cite{jordan1999introduction,jaakkola2000bayesian,wainwright2008graphical}.
		Computing its Gibbs free energy by injecting the naive mean-field approximation in \eqref{main:intro:variational:gibbs_free_energy} yields
		\begin{align*}
		    &\varphi^{\mathrm{gibbs}}_\ndim[\rQ^{\textrm{naive}}]
		    = \EE_{\bsigma \sim \rQ^{\textrm{naive}}} \sqbrs{ \mH_\ndim(\bsigma) } + \frac{1}{\beta} \sum_{j=1}^\ndim \int_\chi \dd \sigma_j ~ \rq_j(\sigma_j) \log \rdbrs{ \rq_j(\sigma_j) } \\
		    &= \frac{1}{\beta} \int_\chi \dd \sigma_i ~ \rq_i(\sigma_i) \log \rdbrs{ \rq_i(\sigma_i) } + \frac{1}{\beta} \sum_{j \ne i} \int_\chi \dd \sigma_j ~ \rq_j(\sigma_j) \log \rdbrs{ \rq_j(\sigma_j) }\\
		    & + \int_\chi \dd \sigma_i ~ \rq_i(\sigma_i)  \underbrace{ \sqbrs{ \int_\chi \prod_{j \ne i} \dd \sigma_j ~ \rq_j(\sigma_j)  \mH_\ndim(\bsigma) } }_{ \equiv \EE_{\bsigma_{\setminus i}} \[ \mH_\ndim(\bsigma) \] }
		\end{align*}
		where we denote $ \bsigma_{\setminus i} $ the vector formed by deleting the $i$-th component of the spin configuration $\bsigma$ and $ \EE_{\bsigma_{\setminus i}} \sqbrs{ \mH_\ndim(\bsigma) } $ the conditional expectation of $ \mH_\ndim(\bsigma) $ when we fix $\sigma_i$. 
		Defining $ \rQ_{\setminus i}(\sigma_i) \equiv \frac{ e^{ -\beta \EE_{\bsigma_{\setminus i}} \sqbrs{ \mH_\ndim(\bsigma) } } }{ \mZ_{\setminus i}(\beta) }$ with $\mZ_{\setminus i}(\beta) \equiv \int_\chi \dd \sigma_i ~ e^{ -\beta \EE_{\bsigma_{\setminus i}} \sqbrs{ \mH_\ndim(\bsigma) } }$, 
		the Gibbs free energy $ \varphi^{\mathrm{gibbs}}_\ndim[\rQ^{\textrm{naive}}] $ can be ingeniously decomposed as
		\begin{align*}
		    &\varphi^{\mathrm{gibbs}}_\ndim[\rQ^{\textrm{naive}}]
		    = \frac{1}{\beta} \sum_{j \ne i} \int \dd \sigma_j ~ \rq_j(\sigma_j) \log \rdbrs{ \rq_j(\sigma_j) } \\
		    & \qquad \qquad \qquad + \frac{1}{\beta} \int \dd \sigma_i ~ \rq_i(\sigma_i) \sqbrs{ \log \rdbrs{ \rq_i(\sigma_i) } - \log \rdbrs{ e^{ -\beta \, \EE_{\bsigma_{\setminus i} } \sqbrs{ \mH_\ndim(\bsigma) } } } } \\
		    &= \frac{1}{\beta} \sum_{j \ne i} \int \dd \sigma_j ~ \rq_j(\sigma_j) \log \rdbrs{ \rq_j(\sigma_j) } \\
		    & \qquad \qquad \qquad + \frac{1}{\beta} \int \dd \sigma_i ~ \rq_i(\sigma_i) \sqbrs{ \log \rdbrs{ \frac{ \rq_i(\sigma_i) }{ \rq_{\setminus i}(\sigma_i) } } - \log \rdbrs{ \mZ_{\setminus i}(\beta) } } \\
		    &= \underbrace{ \frac{1}{\beta} \sum_{j \ne i} \int \dd \sigma_j ~ \rq_j(\sigma_j) \log \rdbrs{ \rq_j(\sigma_j) } - \frac{1}{\beta} \log \rdbrs{ \mZ_{\setminus i}(\beta) } }_{\varphi_{\setminus i}} \\
		    & \qquad \qquad \qquad \qquad \qquad \qquad \qquad \qquad  \qquad + \frac{1}{\beta}  \mD_{\textrm{KL}}\( \rQ_i || \rQ_{\setminus i} \)\,,
		\end{align*}
		where the first term $\varphi_{\setminus i}$ is independent of the marginal density $\rq_i$. Therefore, minimizing the Gibbs free energy $\varphi^{\mathrm{gibbs}}_\ndim[\rQ^{\textrm{naive}}]$, the Gibbs variational principle \eqref{main:intro:variational:gibbs_variational_principle} prescribes the marginal densities to
		\begin{align}
		    \rQ_i(\sigma_i) \equiv \frac{1}{ \mZ_{\setminus i}(\beta)} e^{ -\beta \EE_{\bsigma_{\setminus i}} \sqbrs{ \mH_\ndim(\bsigma) } }\,.
		    \label{eq:main:intro:variational:naive_mf}
		\end{align}
		Applied to the Curie-Weiss model, which is only the mean-field Ising model, see \Sec\ref{main:intro:classical_physics:examples:curie_weiss_example}, the naive mean-field approximation \eqref{eq:main:intro:variational:naive_mf} allows in particular to recover the well-known set of implicit equations verified by the magnetizations		
		\begin{align}
			m_i \equiv \EE_{\rQ_i}[\sigma_i] = \tanh\( \beta\(h_i + \sum_{j=1}^{\ndim} J_{ij} m_j \)\)\,.
			 \label{eq:main:intro:variational:naive_mf_ising}
		\end{align}
		
		As a conclusion, the naive mean-field approximation has the advantage to treat the surrounding interactions of each spin $\sigma_i$ as an average interaction of all the other spins, but at the cost of discarding, eventually, important statistical correlations. In the case where interactions between spins are weak enough, this naive mean-field approach might be exact, as for instance in the case of the Curie-Weiss model they vanish in the thermodynamic limit $\ndim \to \infty$.
		Consequently, the naive mean-field approximation can only poorly describe the behavior of finite-size systems or strongly interacting systems and it is more of pedagogical interest than of real practical utility.
					
		\subsubsection{TAP, Plefka, Georges-Yedidia high-temperature expansion}
			In fact, it turns out that the naive mean-field approximation can be recovered from the truncation of more complex approximations \cite{opper2001advanced}. 
			Especially in the context of disordered systems with densely connected spin glass, namely the \aclink{SK} model \cite{Sherrington1975} with Gaussian random $J_{ij} \sim \mN(0,J_0/\ndim)$ couplings, the \aclink{TAP} equations \cite{thouless1977solution} provide a more accurate approximation than the naive mean-field equations.
			Their derivation is closely related to the cavity method \cite{mezard1987spin} or equivalently the Bethe approximation presented in \Sec\ref{sec:main:intro:variational:bethe_approximation}. 
			Similarly to the cavity method, the idea is to approximate the marginal probability $\rQ_i$ by considering a reduced set of $\ndim-1$ spins $\bsigma_{\setminus i}$ where the spin $\sigma_i$ has been removed. It is finally possible to write a consistent set of non-linear equations of the form
			\begin{align}
				m_i^{t+1} = \tanh\( \beta\(h + \sum_{j=1}^{\ndim} J_{ij} m_j^{t}  \)  - \beta^2 m_i^{t-1}\sum_{j=1}^{\ndim} J_{ij}^2 (1-(m_j^{t})^2)  \)\,,
				\label{eq:main:intro:mean_field:TAP_indices}
			\end{align}
			called, without the time indices, the \aclink{TAP} equations. We immediately observe that these equations are very similar, yet, more complex than the naive ones in \eqref{eq:main:intro:variational:naive_mf_ising}. They simply include a correction term to the effective local field, known as the \emph{Onsager reaction term}, to take into account the absence of the spin variable $\sigma_i$, which was not present in the oversimplified naive approximation. 
			Indeed, later on, the corresponding \aclink{TAP} free energy has been derived with the so-called Plefka expansion \cite{plefka1982convergence}. It turned out that it was simply the second term of a high-temperature expansion proposed in a more general setting \cite{Georges1991}. In the context of the \aclink{SK} model, keeping only the first term leads therefore to the naive mean-field approximation, whereas truncating at the second order, by incorporating the Onsager term, turned out to be exact since other contributions are sub-leading and vanish in the thermodynamic limit.
			Moreover, the latter derivations insured that the fixed points of the \aclink{TAP} equations are the stationary points of the \aclink{TAP} free energy. However, finding a stationary solution is often achieved by turning them into an iterative procedure until convergence towards fixed points. Unfortunately neither the \aclink{TAP} equations, the Plefka expansion nor the Georges-Yedidia high-temperature expansion include the time indices to iteratively solve them correctly and gives free rein to interpretation. 
			By simply and naturally assuming times $t+1$ on the left hand side of \eqref{eq:main:intro:mean_field:TAP_indices} and $t$ on the other magnetizations on the right hand side led to convergence issues of the \aclink{TAP} equations first observed in \cite{kabashima2003cdma}. 
			It turns out that this simple arbitrary prescription of the time indices was wrong and responsible for the convergence issues. The time indices were corrected in \cite{bolthausen2014iterative} that finally leads to \eqref{eq:main:intro:mean_field:TAP_indices}.\\
	
			 In the following, we present an alternative mean-field method based on the \aclink{BP} equations, that provides by construction the correct time indices of the iterative procedure and leads especially to performant algorithms.

	\newpage
	\section{Belief propagation and the Bethe free energy}
	\label{main:intro:mean_field:bp}

	In this dissertation, we make deeply use of \emph{message passing algorithms} such as \aclink{AMP} that can be simply derived from the more general set of \aclink{BP} iterative equations.
	The \aclink{BP} equations have a long history and have started in physics with the Bethe-Peierls approximation \cite{bethe1935statistical,peierls1936statistical}. Very interestingly it can be seen as an extended version of the the \aclink{TAP} \cite{thouless1977solution} and Plefka approach \cite{plefka1982convergence}, Georges-Yedidia expansions \cite{Georges1991} presented in the previous section.
	Moreover as inference problems arose in many various fields, local message passing algorithms have been discovered simultaneously under different names. The \aclink{BP} approach was first introduced in information theory \cite{gallager1962low} and in Bayesian inference \cite{pearl1982reverend}, whereas it was known under the name of \emph{cavity method} in statistical physics of disordered systems \cite{mezard1987spin, mezard2009information}. 
	The different approaches are reviewed in \cite{Aji2000,Yedidia2001} that connects especially the \aclink{BP} to variational mean-field approach.\\
	
	 In this section, we review the main results of \cite{Yedidia2001, Yedidia2002, Yedidia2005, wainwright2008graphical, mezard2009information} starting by presenting in \Sec\ref{sec:main:intro:variational:bethe_approximation} the Bethe approximation and the Bethe free energy. Theses latter naturally give rise to the set of \aclink{BP} iterative equations presented in \Sec\ref{sec:main:intro:variational:bp}. Finally, in the perspective to derive the \aclink{AMP} algorithm for the \aclink{GLM} class, we present the \aclink{BP} equations for this model class in \Sec\ref{main:intro:mean_field:bp:example_glm}.

		\subsection{The Bethe approximation}
		\label{sec:main:intro:variational:bethe_approximation}
			
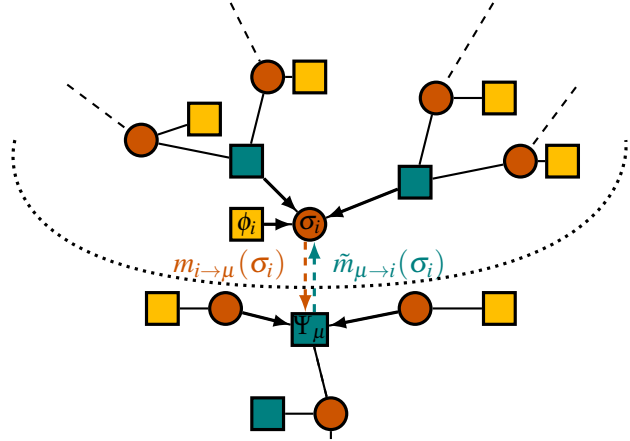
\begin{figure}[htb!]
		\centering
		\begin{tikzpicture}[scale=0.8, auto, swap]
			\def\mylen{0.35};
		    \foreach \pos/\name/\mathname in {  {(13*\mylen,2*\mylen)/X1/$ $},
		                                        {(8*\mylen,7*\mylen)/X2/$  $},
		                                        {(17*\mylen,7*\mylen)/X3/$  $},
		                                        {(12*\mylen,11*\mylen)/X4/$ \sigma_i $},
		                                        {(4*\mylen,15*\mylen)/X5/$  $},
		                                        {(10*\mylen,18*\mylen)/X6/$  $},
		                                        {(18*\mylen,17*\mylen)/X7/$  $},
		                                        {(22*\mylen,14*\mylen)/X8/$  $}}
		        \node[var] (\name) at \pos {\mathname};
		    \foreach \pos/\name/\mathname in {  {(12*\mylen,6*\mylen)/F1/$ \Psi_\mu $},
		                                        {(9*\mylen,14*\mylen)/F2/$  $},
		                                        {(17*\mylen,13*\mylen)/F3/$ $},
		                                        {(10*\mylen,2*\mylen)/F4/$  $}}
		        \node[inter] (\name) at \pos {\mathname};
		    \foreach \pos/\name/\mathname in {  
		                                        {(5*\mylen,7*\mylen)/F5/$ $},
		                                        {(21*\mylen,7*\mylen)/F6/$  $},
		                                        {(24*\mylen,14*\mylen)/F7/$  $},
		                                        {(21*\mylen,17*\mylen)/F8/$  $},
		                                        {(12*\mylen,18*\mylen)/F9/$  $},
		                                        {(7*\mylen,16*\mylen)/F10/$  $},
		                                        {(9*\mylen,11*\mylen)/F15/$ \phi_i $}}
		        \node[field] (\name) at \pos {\mathname};
		    \foreach \pos/\name/\mathname in {  {(0*\mylen,18*\mylen)/F11/$  $},
		                                        {(8*\mylen,22*\mylen)/F12/$  $},
		                                        {(21*\mylen,22*\mylen)/F13/$  $},
		                                        {(25*\mylen,18*\mylen)/F14/$  $},
		                                        {(13*\mylen,0*\mylen)/F16/$  $}}
		        \node (\name) at \pos {\mathname};
		    \foreach \source/ \dest in {    X6/F9,X5/F10,X7/F8,X8/F7,X5/F2,F2/X6,F2/X4,X7/F3,F3/X8,
		                                    F3/X4,X4/F15,F5/X2,X2/F1,F1/X3,X3/F6,F1/X1, F4/X1}
		        \path[edge] (\source) -- (\dest);
		    \foreach \source/ \dest in {    F11/X5,
		                                    F12/X6,
		                                    F13/X7,
		                                    F14/X8,
		                                    X1/F16,
		                                    X1/F1}
		        \path[edge,dashed] (\source) -- (\dest);
		    \path[-latex, burntorange, very thick, dashed] ([xshift=-0.075cm]X4.south) edge node[left=0.1cm, yshift=0.15cm]{$m_{i \to \mu}(\sigma_i)$} ([xshift=-0.075cm]F1.north);
			\path[-latex, teal, very thick, dashed] ([xshift=0.075cm]F1.north) edge node[right=0.1cm,yshift=0.15cm]{$\td{m}_{\mu \to i}(\sigma_i)$} ([xshift=0.075cm]X4.south);
			 \path[-latex, black, very thick] (F2) edge (X4);
			 \path[-latex, black, very thick] (F3) edge (X4);
			 \path[-latex, black, very thick] (F15) edge (X4);
			 \path[-latex, black, very thick] (X2) edge (F1);
			 \path[-latex, black, very thick] (X3) edge (F1);
		
		    \draw [black, very thick, dotted] plot [smooth, tension=2] coordinates { (-2*\mylen,15*\mylen) (12*\mylen,8*\mylen) (27*\mylen,15*\mylen)};
		\end{tikzpicture}
		\caption{Tree-like factor graph on which the Belief Propagation iterative equations can be decomposed.}
		\label{main:fig:factor_graph_bp}
		\end{figure}

		The Bethe approximation\index{Bethe approximation} plays a central role among approximations that take into account interactions between spins. In particular, it allows to incorporate correlations between the variables to describe more complex models.
		Consider the \aclink{JPD} described by a factor graph $\mG\(\rV, \rF, \rE\)$ represented in \Fig\ref{main:fig:factor_graph_bp}, the Bethe approximation assumes that the \aclink{JPD} can be written as
			\begin{align}
				\rQ^{\textrm{bethe}}(\bsigma) = \frac{ \prod_{\mu=1}^\nsamples \td{m}_\mu (\bsigma_{\partial{_\mu}} ) }{\prod_{i=1}^\ndim m_i(\sigma_i)^{\abs{\partial_i}-1}}\,,
			\label{main:intro:variational:bethe_approximation}	
			\end{align}
			where $m_i(\sigma_i)$ denotes the marginals of the variable $\sigma_i$, and $\td{m}_\mu$ the marginals of the cliques $\bsigma_{\partial{_\mu}}$ around the factors $\mu$.  
			The Bethe approximation can be easily derived on tree-like factor graphs, such as the one in \Fig\ref{main:fig:factor_graph_bp}, by simply taking the product of the marginals of all the cliques $\td{m}_\mu (\bsigma_{\partial{_\mu}} )$ and dividing by the variables marginals $m_i(\sigma_i)$ to remove the marginals already taken into account. Therefore, the number of neighbouring factors of the variable $\sigma_i$, $\abs{\partial_i}$, is present in the denominator to avoid counting repetitions.
			Consequently, the formulation \eqref{main:intro:variational:bethe_approximation} has the main advantage to be rigorously exact on tree-like connected factor graphs with no loops, and to be strongly connected to the \aclink{BP} algorithmic procedure \cite{kabashima1998belief} and the \aclink{TAP} equations.
			Moreover, this latter formulation can  also be applied to more general factor graphs, providing a powerful approximation but loosing in return its exactness. 
						
		\subsection{The Bethe free energy}
		\label{sec:main:intro:variational:bethe_free_energy}

			Plugging the Bethe approximation \eqref{main:intro:variational:bethe_approximation} in the Gibbs free energy at $\beta=1$ \eqref{main:intro:variational:gibbs_free_energy} leads to the corresponding Bethe free energy \cite{Yedidia2001, mezard2009information} which can be written as a functional over the marginals $\{\td{m}_\mu \}_{\mu=1}^\nsamples \cup\{ m_i \}_{i=1}^\ndim$
			\begin{align}
				\varphi^{\textrm{bethe}}_\ndim \[\{\td{m}_\mu\}_\mu, \{m_i\}_i\] &=  \rU^{\textrm{bethe}}\[\{\td{m}_\mu\}_\mu, \{m_i\}_i\] \label{eq:main:intro:variational:bethe_functional} \\ 
				& \qquad\qquad\qquad\qquad - \rH^{\textrm{bethe}}\[\{\td{m}_\mu\}_\mu, \{m_i\}_i\]\,, \nonumber
			\end{align}
			where $\rU^{\textrm{bethe}}$, $\rH^{\textrm{bethe}}$ denote the variational energy and entropy			
			\begin{align*}
				\rU^{\textrm{bethe}}\[\{\td{m}_\mu\}_\mu, \{m_i\}_i\]&\equiv \sum_{\mu} \int \d \bsigma_{\partial_\mu} \td{m}_\mu (\bsigma_{\partial{_\mu}} ) \log \Psi_\mu(\bsigma_{\partial_\mu}) \\
				 & \qquad \qquad \qquad  + \sum_{i} \int \d \sigma_i m_i (\bsigma_{i} ) \log \phi_i(\sigma_i)\,, \\ 
				\rH^{\textrm{bethe}}\[\{\td{m}_\mu\}_\mu, \{m_i\}_i\]&\equiv \sum_{\mu} \rH\[\td{m}_\mu\] + \sum_{i}(\abs
				{\partial_i}-1) \rH\[m_i\]\,,
			\end{align*}
			and $\rH[\rp]$ the entropy of the probability density $\rp$ defined in \eqref{main:intro:info_theory:entropy}.
			Enforcing the self-consistency marginalization and normalization constraints $m_i(\sigma_i) = \int \d \bsigma_{\partial_\mu \setminus i} \td{m}_\mu(\bsigma_{\partial_\mu}) $, $\int \d \sigma_i m_i(\sigma_i) = 1 = \int \d \bsigma_{\partial_\mu } \td{m}_\mu(\bsigma_{\partial_\mu}) $, with some Lagrange multipliers \cite{Yedidia2001,Yedidia2005,  wainwright2008graphical}, the extremization of the Lagrangian leads to the following expressions of the marginals estimate
			\begin{align*}
				\td{m}_\mu(\bsigma_{\partial_\mu}) &\propto \Psi_\mu(\bsigma_{\partial_\mu}) \prod_{i \in \partial \mu} m_{i \to \mu}(\sigma_i)\,, && m_i(\sigma_i) \propto \prod_{\mu \in \partial_i } \td{m}_{\mu \to i}(\sigma_i)\,,
			\end{align*}			
			that involve approximate beliefs $\{m_{i \to \mu}, \td{m}_{\mu \to i}\}$ over the variable $\sigma_i$ if we respectively cut the edge $(i\mu)\in\rE$ of the factor graph between the variable $\sigma_i$ and the factor $\Psi_\mu$, as illustrated in \Fig\ref{main:fig:factor_graph_bp}.
			In the context of pairwise \aclink{MRF}, the above conditions are crucial to understand the link between \aclink{BP}, introduced in \Sec\ref{main:intro:mean_field:bp}, and the Bethe approximation as stressed in \cite{kabashima1998belief, Yedidia2001}. 
			Indeed, since the \aclink{BP} marginal densities are obtained by extremizing the Bethe free energy, the fixed point of the \aclink{BP} algorithm are by construction the stationary points of the Bethe free energy. 	 
			Finally, under the Bethe approximation \eqref{main:intro:variational:bethe_approximation}, the Bethe free energy  can be written as a function of the one and two-body interactions $\{\phi_i, \Psi_\mu\}$ and the beliefs $\{m_{i \to \mu },\td{m}_{\mu \to i}\}$:
			\begin{align}
				\varphi^{\textrm{bethe}}_\ndim = -\sum_{i \in \rV} \log \mZ_i -\sum_{ \mu \in \rF} \log \mZ_\mu  + \sum_{(i\mu)\in\rE} \log \mZ_{i \mu}\,, 
				\label{main:intro:variational:bethe_free_energy_bp}
			\end{align}
			with 
			\begin{align*}
				\mZ_i &= \int \d \sigma_i \phi_i(\sigma_i)  \prod_{i\in \partial_\mu} \td{m}_{\mu \to i}(\sigma_i)\,, \hhspace \mZ_{i\mu} = \int \d \sigma_i  \td{m}_{\mu \to i}(\sigma_i) m_{i \to \mu }(\sigma_i)\,,\\
				\mZ_\mu &= \Psi_{\mu}(\bsigma_{\partial_\mu}) \int \prod_{i \in \partial_\mu} \d \sigma_i  \prod_{i \in \partial_\mu} m_{i \to \mu }(\sigma_i)\,.
			\end{align*}

		\subsection{Belief propagation equations}
		\label{sec:main:intro:variational:bp}
		The \aclink{BP}\index{Belief propagation} algorithm is an inference algorithm that computes an approximation of the marginal densities of a complex \aclink{JPD}. In particular, it makes use of the fact that many \aclink{JPD} are locally factorizable to reduce the estimation of the full complex problem into tractable sub-problems on each factor of the factor graph. 
		The set of \aclink{BP} iterative equations can be obtained directly from the variational principle and the Bethe free energy as presented in the previous section. Yet, for a more intuitive and practical perspective they can be directly obtained from the factor graph as we detail in the following.
		Depending on the problem under consideration, the \aclink{BP} approach can be expressed in two variants: the \emph{sum-product} or the \emph{max-sum} equations. The sum-product approach estimates the marginal densities and directly leads to \aclink{MMSE} estimation. In contrast, the max-sum approach is more suitable to \aclink{MAP} estimation and the corresponding equations can be found in \cite{mezard2009information, Advani2016b}.
		
		\subsubsection{Sum-product equations}
		Let us present the sum-product version of the \aclink{BP} procedure for a general \aclink{MRF} illustrated in \Fig\ref{main:fig:factor_graph_bp}.
		Importantly, we first attach two kinds of auxiliary functions $\{m_{i\to \mu},\td{m}_{\mu \to i}\}_{}$ to the edges of the factor graph, called \emph{messages}. These messages are interpreted \cite{mezard2009information, Yedidia2005} as the estimates of the marginal $\rQ(\sigma_i)$ if we remove the edges $(i\mu)=\{i\to \mu, \mu \to i\}$. 
		In other words, $m_{i\to \mu}(\sigma_i)$ denotes the message from the variable $\sigma_i$ to the factor node $\Psi_\mu$ delivering the estimation of the marginal density $\rQ(\sigma_i)$ in the partial visited graph represented by the top part of the graph in \Fig\ref{main:fig:factor_graph_bp} and delimited by the dotted line. Similarly, $\td{m}_{\mu \to i}(\sigma_i)$ is the message from the factor node $\Psi_\mu$ to the variable $\sigma_i$ that transmits an estimation of the marginal density $\rQ(\sigma_i)$ in the bottom part of the graph, called the \emph{cavity graph}.
		Essentially, the \aclink{BP} algorithm consists in letting the variables and factors communicate their \emph{beliefs} to their neighbours based on the informations captured from the nodes and factors already visited along the tree. Iterating the procedure, we expect qualitatively that the beliefs converge to an average value of their neighbouring beliefs. 
		\begin{figure*}[htb!]
		\centering
			\begin{tikzpicture}[scale=0.85, auto, swap]
			\centering
				\def\mylen{0.75};
				\node[var] (X) at (0,0) {$ \sigma_i $};
				\node[inter] (F1) at (2*\mylen,2*\mylen) {$ \Psi_{\nu_1} $};
				\node[inter] (F2) at (-2*\mylen,2*\mylen) {$ \Psi_{\nu_2} $};
				\node[inter] (F3) at (0,-2*\mylen) {$ \Psi_{\mu} $};
				\node[field] (H1) at (-2*\mylen,0) {$ \phi_{i} $};
 				\path[-latex, burntorange, very thick, dashed] (X) edge node[right]{$m_{i \to \mu}(\sigma_i)$} (F3);
			 	\path[-latex, black, very thick] (F1) edge node[right]{\footnotesize $\td{m}_{\nu_1 \to i}(\sigma_i)$} (X);
			 	\path[-latex, black, very thick] (F2) edge node[left]{ \footnotesize$\td{m}_{\nu_2 \to i}(\sigma_i)$} (X);
			 	\path[-latex, black, very thick] (H1) edge (X);
			\end{tikzpicture}
			\hspace{1cm}
			\begin{tikzpicture}[scale=0.8, auto, swap]
				\def\mylen{0.75};
				\node[inter] (F) at (0,0) {$ \Psi_{\mu} $};
				\node[var] (X1) at (0,3*\mylen) {$ \sigma_{k_1} $};
				\node[var] (X2) at (-2*\mylen,1*\mylen) {$ \sigma_{k_2} $};
				\node[var] (X3) at (2*\mylen,1*\mylen) {$ \sigma_{k_3} $};
				\node[var] (X4) at (0.5*\mylen,-2*\mylen) {$ \sigma_{i} $};
 				\path[-latex, teal, very thick, dashed] (F) edge node[right]{$\td{m}_{\mu \to i}(\sigma_i)$} (X4);
			 	\path[-latex, black, very thick] (X1) edge node[above right]{\footnotesize $m_{k_1 \to \mu}(\sigma_{k_1})$} (F);
			 	\path[-latex, black, very thick] (X2) edge node[below left]{ \footnotesize$m_{k_2 \to \mu}(\sigma_{k_2})$} (F);
			 	\path[-latex, black, very thick] (X3) edge node[below right]{ \footnotesize$m_{k_3 \to \mu}(\sigma_{k_3})$} (F);
			\end{tikzpicture}
			\caption{Local representation of the factor graph around the variable $\sigma_i$ and the factor $\Psi_\mu$.}
		\label{main:fig:factor_graph_bp_decomposition}
		\end{figure*}
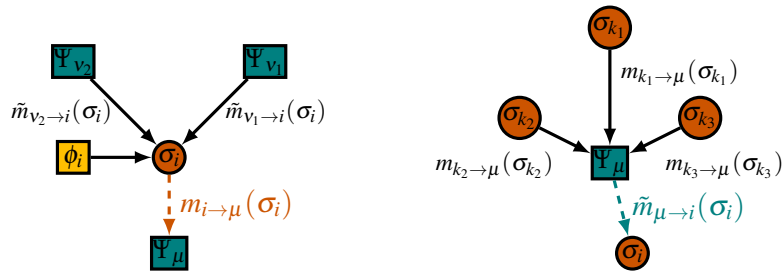
		To formalize this procedure, the sum-product equations \cite{gallager1962low,pearl1982reverend,wainwright2008graphical,mezard2009information} make use of the crucial tree-like assumption, originating from the Bethe approximation, that guarantees that the incoming messages to the variable $\sigma_i$ are independent. Thereby, the messages are given by the self-consistency rule for the messages $m_{i \to \mu}^{t+1}$ and $\td{m}_{\mu \to i}^t$,  $\forall i\in\lb \ndim \rb$, $\mu\in\lb\nsamples\rb$ according to
			\begin{align}
			\begin{aligned}
				m_{i \to \mu}^{t+1}(\sigma_i)  &=  \frac{1}{\mZ_{i\to\mu}} \phi_i(\sigma_i) \prod_{\nu \in \partial_i \setminus \mu} \td{m}_{\nu \to i}^{t}(\sigma_i) \\
				\td{m}_{\mu \to i}^{t}(\sigma_i) &= \frac{1}{\mZ_{\mu\to i}} \sum_{\bsigma_{\partial_{\mu} \setminus i }} \Psi_\mu\( \bsigma_{\partial_{\mu} \setminus i} \) \prod_{k \in \partial_\mu \setminus i } m_{k \to \mu}^{t}(\sigma_k)\,.
				\label{main:intro:bp_equations}
			\end{aligned}
			\end{align}
			These update rules can be easily understood by looking at the local decomposition of the factor graph \Fig\ref{main:fig:factor_graph_bp_decomposition}. The message $m_{i \to \mu}$ in dashed orange is built from the incoming messages of the neighbouring factors of the spin $\sigma_i$ if we remove the edge $i \to \mu \in \rE$. Similarly the message $\td{m}_{\mu\to j}$ in dashed green is obtained by summing over all the possible values of the messages coming from the neighbouring variables if we remove the edge $\mu \to i \in \rE$.
			
		\subsubsection{BP algorithm and properties}
			The \aclink{BP} algorithm is the procedure that consists in iterating the set of dynamical equations \eqref{main:intro:bp_equations} over time. 
			Eventually if it converges, it provides at convergence an estimation of the Bethe free energy \eqref{main:intro:variational:bethe_free_energy_bp} and especially of the marginal probabilities given by $\forall i,~ \rQ(\sigma_i) \propto \phi_i(\sigma_i) \prod_{\mu \in \partial_i} \td{m}_{\mu \to i}(\sigma_i)$ where $\partial_i = \{\mu : (\mu i) \in \rE\}$ represents all the neighbouring factors of the variable $i$.
			However the \emph{messages independence} is crucial for writing the \aclink{BP} equations \eqref{main:intro:bp_equations}, such that the obtained marginal estimation and the Bethe free energy are exact only in the case of \aclink{DAG} factor graphs for which there is no correlation between the incoming messages. In other words, by construction the convergence of \aclink{BP} to the fixed points of the Bethe free energy is guaranteed only for tree-like factor graphs.
			\paragraph{Loopy BP}
			Nevertheless, the powerful \aclink{BP} algorithm can be used as an approximation in more complex \aclink{MRF} that do not factorize as \aclink{DAG} and thus contain some loops. 
			The influence of \emph{loops}\index{loops} in the graph can induce strong correlations and harm the convergence of \aclink{BP}. Violating the messages independence hypothesis breaks the convergence guarantees, but provides anyway an \emph{approximate} algorithmic procedure that, hoping for the best, may still converge. 
			Notice that there exists some cases for which the presence of loops may not be that harmful. 
			In particular, in the case of \emph{long enough loops} and if the \emph{correlations decrease fast enough} with the Hamming distance, since the factor graph remains locally tree-like, we expect the message independence to still hold. 
			In this context, under the name of \emph{loppy}-\aclink{BP}, the algorithm may sometimes succeed converging and provide good approximations of the marginals, loosing in return convergence guarantees. As an alternative, since the \aclink{BP} algorithm is not guaranteed to converge, we could instead directly find the minimum of the Bethe free energy \cite{yuille2001double}, even though it is much slower and the Bethe free energy does not provide anymore a variational Gibbs free energy upper-bound.
			\paragraph{State Evolution}
			In addition of being a general procedure adaptable to any \aclink{MRF}, the main interest of \aclink{BP} lies in the possibility to predict its asymptotic performances. Indeed, in the thermodynamic limit $\ndim \to \infty$, it is possible to fully characterize the dynamics of the \aclink{BP} fixed point equations, known as the \aclink{SE} equations \index{state evolution}. They have been introduced in \cite{bayati2011dynamics} and their interest considerably increased with the regain of activity in the high-dimensional regime.
			In the next section, we will show in the context of the \aclink{GLM} class that the \aclink{SE} equations of \aclink{AMP}, which is nothing more than a Gaussian simplification of the \aclink{BP} algorithm, can be equivalently obtained from the replica computation, which therefore provides an efficient way for obtaining the asymptotic behaviour of such message passing algorithms. 

		\subsection{\textbf{Application} - BP equations for the GLM}
		\label{main:intro:mean_field:bp:example_glm}
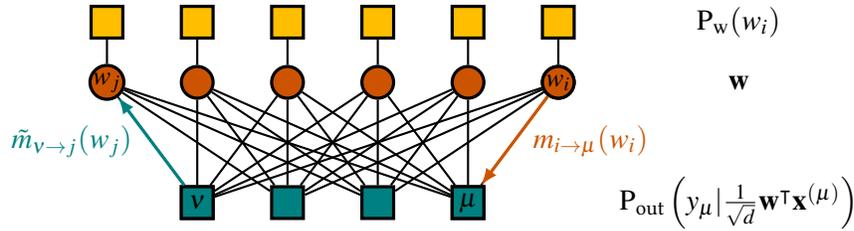
\begin{figure}[htb!]
				\centering
			\begin{tikzpicture}[scale=0.8, auto, swap]
			    \foreach \i in {1,...,6}
			        \node[var] (X\i) at (1.5*\i,0) {};
			    \node at (12, 0) {$ \vec{w} $};
			    \node[var] (X1) at (1.5,0) {$w_j$};
			    \node[var] (X6) at (9,0) {$w_i$};
			
			    \foreach \mu in {1,...,4}
			        \node[inter] (Y\mu) at (1.5+1.5*\mu,-2) {};
			    \node[inter] (Y1) at (1.5+1.5,-2) {$\nu$};
			    \node[inter] (Y4) at (7.5,-2) {$\mu$};
			        
			    \foreach \i in {1,...,6}
			        \foreach \mu in {1,...,4}
			            \path[edge] (X\i) -- (Y\mu);
			    \node at (10, -2) {};
			    \node (F) at (12, -2) {$ \rP_{\out}\(y_\mu | \frac{1}{\sqrt{\ndim}} \vec{w}^\intercal\vec{x}^{(\mu)} \) $};			
			    \foreach \i in {1,...,6} {
			        \node[field] (P\i) at (1.5*\i,1) {};
			        \path[edge] (X\i) -- (P\i);
			    }
			    \node at (12, 1) {$ \rP_\w(w_i) $};
			    \path[-latex, teal, very thick] (Y1) edge node[left=0.1cm]{$\td{m}_{\nu \to j}(w_j)$} (X1);
			    \path[-latex, burntorange, very thick] (X6) edge node[right=0.1cm]{$m_{i \to \mu}(w_i)$} (Y4);
			\end{tikzpicture}
			\caption{Factor graph corresponding to the posterior distribution \eqref{main:intro:bp:posterior_glm} associated to the GLM hypothesis class. The variable $w_i$ send a message $m_{i \to \mu}(w_i)$ to the factor $\mu$, and reciprocally it sends back a message $\td{m}_{\mu \to i}(w_i)$ to the variable based on the corresponding cavity graph.}
			\label{main:fig:factor_graph_glm_bp}
			\end{figure}

			As a central illustration, we present the instructive and systematic derivation of the \aclink{rBP} equations starting with the \aclink{BP} equations, before deriving the corresponding \aclink{AMP} algorithm in the next section, for the \aclink{GLM} class. The corresponding \aclink{JPD} can be written as
			\begin{align}
				\rP_\ndim\(\vec{w} | \vec{y}, \mat{X}\)  &=  \frac{\prod_{\mu=1}^\nsamples \rP_\out(y_\mu | z_\mu) \prod_{i=1}^\ndim \rP_{\w}(w_i)  }{\mZ_\ndim(\vec{y},\mat{X})} \,,
				\label{main:intro:bp:posterior_glm}
			\end{align}
			already considered in \Sec\ref{main:sec:mean_field:replica_method:example_glm}, and where we assumed that the channel and prior distributions factorize over factors $\Psi_\mu=\rP_{\out,\mu}$ and spin variables $\phi_i=\rP_{\w,i}$. The posterior distribution can be naturally represented by the factor graph \Fig\ref{main:fig:factor_graph_glm_bp}.
			We define the auxiliary variable $z_\indsamples = \frac{1}{\sqrt{\ndim}}\vec{w}^\intercal \vec{x}_\indsamples=\Theta(1)$ which is of order one thanks to the crucial rescaling pre-factor $1/\sqrt{d}$. Indeed, even though the factor graph is fully-connected and contains short loops, this weak coupling insures that the messages remain slightly correlated and the \aclink{BP} equations hold true.
			 The details of the computations for the more general committee machine hypothesis class can be found in \App\ref{appendix:amp:committee}.
			 
			\paragraph{BP equations for the GLM}
			Let us consider a set of messages $\{m_{i\to \mu},\tilde{m}_{\mu \to i}\}$ on the edges of the bipartite factor graph \Fig\ref{main:fig:factor_graph_glm_bp}. These messages correspond to the marginal probabilities of $w_i$ if we remove the edges $(i \to \mu)$ and $(\mu \to i)$. The sum-product \aclink{BP} equations \eqref{main:intro:bp_equations} are simply follow as
			\begin{align}
					m_{i\to \mu}^{t+1} (w_i) &= \displaystyle \frac{1}{\mZ_{i\to \mu}} \rp_\w (w_i) \prod\limits_{\nu \neq \mu}^\nsamples \tilde{m}_{\nu \to i}^t (w_i)\,, \label{main:intro:bp_equations_glm}
					\\
					\tilde{m}_{\mu \to i}^t (w_i) &=  \displaystyle \frac{1}{\mZ_{\mu \to i}} \int \prod\limits_{j\neq i}^\ndim \d w_j ~ \rp_\out \(y_{\mu} |  \frac{1}{\sqrt{\ndim}} \sum_{j=1}^\ndim  x_{\mu j}w_{j} \)  m_{j \to \mu}^t (w_j )\,. \notag
			\end{align}
			
		\paragraph{Towards relaxed-Belief Propagation equations}	
			The idea is to expand, in the limit $\ndim \to \infty$, the set of $\Theta(\ndim^2)$ messages $\{\td{m}_{\mu \to i}\}_{i,\mu}$ before plugging them back in the messages $\{m_{i\to \mu}\}_{i,\mu}$. Truncating the expansion and keeping only terms of order $\Theta\(1/\ndim\)$, messages become \textit{Gaussian} and therefore messages can be parametrized only by the mean $\hat{w}_{i\to \mu}^t$ and the variance $\hat{c}_{i\to \mu}^t$ of the marginal distribution estimate $m_{i\to \mu}^t$ at time $t$:
			\begin{align}
			\begin{aligned}
				\hat{w}_{i\to \mu}^t &\equiv \displaystyle \int_{\bbR} \d w_i ~
					 m_{i\to \mu}^t (w_i) w_i\,,  \spacecase
					 \hat{c}_{j\to \mu}^t &\equiv \displaystyle \int_{\bbR} \d w_i ~
					 m_{i\to \mu}^t (w_i) w_i^2 - (\hat{w}_{i\to \mu}^t)^2\,.
			\end{aligned}
			\label{main:intro:amp_glm:what_chat}
			\end{align}
			Using a Fourier representation of $\rP_\out$ in $\tilde{m}_{\mu \to i}^t$ in \eqref{main:intro:bp_equations_glm} to decouple its fully-connected argument, and expanding it in the large size limit $\ndim \to \infty$, we obtain that the set of \aclink{BP} equations finally closes over Gaussian beliefs $\{m_{i\to \mu}\}_{i,\mu}$ 
			\begin{equation}
				 m_{i\to \mu}^{t+1} (w_i) = \frac{1}{\mZ_{i\to \mu}} \rp_\w (w_i) \prod\limits_{\nu \neq \mu}^\nsamples \sqrt{\frac{A_{\nu \to i}^t}{(2\pi)}} e^{-\frac{A_{\nu \to i}^t}{2}\(w_{i} - (A_{\nu \to i}^t)^{-1}b_{\nu \to i}^t \)^2   }\,,
				 \label{main:intro:amp_glm:m_i_mu}
			\end{equation} 
			with natural parameters $b_{\mu \to i}^t$ and the precision $A_{\mu \to i}^t$ defined as
			\begin{align}
					b_{\mu \to i}^t &\equiv  \frac{x_{\mu i}}{\sqrt{\ndim}} f_\out (y_{\mu}, \omega_{i\mu}^t, V_{i\mu}^t) \,,
					&& A_{\mu \to i}^t \equiv - \frac{x_{\mu i}^2}{\ndim}  \partial_\omega f_\out(y_{\mu}, \omega_{i\mu}^t, V_{i\mu}^t)
				\label{main:intro:amp_glm:A_B_mu_to_i}
			\end{align}
			with the \emph{channel denoising functions} $f_\out, \partial_\omega f_\out$, defined in \App\ref{appendix:update_functions},
			which depend on the mean and variance of the \emph{channel belief}
			\begin{align}
				\omega_{\mu \to i}^t &\equiv  \frac{1}{\sqrt{\ndim}} \sum\limits_{j\neq i}^\ndim x_{\mu j}  \hat{w}_{j\to \mu}^t \,, &&
				V_{\mu \to i}^t \equiv  \frac{1}{\ndim} \sum\limits_{j\neq i}^\ndim x_{\mu j}^2  \hat{c}_{j \to \mu}^t\,.
				\label{main:intro:amp_glm:omega_V_mu_to_i}
			\end{align}
			Finally the mean and variance \eqref{main:intro:amp_glm:what_chat} of the message $m_{i\to \mu}$ are updated by
			\begin{align}
				\hat{w}_{i\to \mu}^{t+1} &= f_\w( \gamma_{\mu \to i}^t, \Lambda_{\mu \to i}^t  )\,,
				&& \hat{c}_{i \to \mu}^{t+1}=\partial_\gamma f_\w( \gamma_{\mu \to i}^t, \Lambda_{\mu \to i}^t)\,,
				\label{main:intro:amp_glm:w_cw}
			\end{align}
			with the \emph{prior denoising functions} $f_\w, \partial_\gamma f_\w$, defined in \App\ref{appendix:update_functions},
			where the mean $\gamma_{\mu \to i}^t$ and variance $\Lambda_{\mu \to i}^t$ of the \emph{prior belief} are defined by
			\begin{align}
				\gamma_{\mu \to i}^t &=  \sum\limits_{\nu \ne \mu}^\nsamples  b_{\nu \to i}^t \,, 
				&&\Lambda_{\mu \to i}^t =  \sum\limits_{\nu \ne \mu}^\nsamples  A_{\nu \to i}^t\,.
				\label{main:intro:amp_glm:gamma_Lambda_mu_to_i}
			\end{align}	
			The set of equations (\ref{main:intro:amp_glm:A_B_mu_to_i}, \ref{main:intro:amp_glm:omega_V_mu_to_i},
				\ref{main:intro:amp_glm:w_cw},
				\ref{main:intro:amp_glm:gamma_Lambda_mu_to_i}) form the set of $\Theta(\ndim^2)$ \aclink{rBP} equations, which are simply the projection of the \aclink{BP} equations over any parametrized family, namely the Gaussian family in the presented case. 
			
	\section{Approximate message passing}
	\label{main:sec:mean_field:amp}			
		
		\aclink{AMP} algorithms start to emerge \cite{boutros2002iterative,montanari2006analysis} and being popular when applied to dense models such as 
		 \aclink{CS} \cite{donoho2009message, bayati2011dynamics} and later to \aclink{GLM} with the \aclink{GAMP} algorithm \cite{rangan2011generalized}. 
		 These algorithms are closely related to the so-called \aclink{TAP} equations in the context of the spin glass theory and the \aclink{SK} model \cite{thouless1977solution,Sherrington1975} as the latter mean-field equations and \aclink{TAP} free energy can be recovered from the Bethe free energy. 
		 However, \aclink{AMP} algorithms largely overtook these previous mean-field methods presented in \Sec\ref{sec:main:intro:variational:naive_tap}
		 as they naturally provide the correct time indices to iterate the self-consistent set of fixed point equations. In contrast, the \aclink{TAP} approach struggled to solve them as no explicit iteration scheme is prescribed by the method. 
		 This connection with statistical mechanics was recently renewed with notably \cite{tanaka2002statistical, guo2005randomly, rangan2009asymptotic, krzakala2012probabilistic} and this manuscript falls within this same approach by applying \aclink{AMP} to the theoretical understanding of \aclink{ANN}. 
		 Moreover, being popularized to various applications, \aclink{AMP} algorithms underwent various extensions such that BiGAMP for bilinear estimation \cite{parker2014bilinear} or ML-AMP for multi-layer estimation \cite{manoel2017multi}.
		Informally the \aclink{AMP} algorithms can be seen as a Taylor expansion of the loopy-\aclink{BP} approach \cite{mezard1987spin,mezard2009information,wainwright2008graphical}.
		The general procedure starts with the set of loopy-\aclink{BP} equations \eqref{main:intro:bp_equations} associated to the corresponding \aclink{JPD} and factor graph. 
		After performing the asymptotic expansion and parametrize the beliefs with Gaussians that allows to track only two parameters, the mean and variance, per message, we finally end up with a set of $\Theta(\ndim^2)$ messages, the so-called set of \aclink{rBP} equations.
		Their latter computational cost can be reduced with additional expansions around the \emph{full messages} to remove the target-node dependency at the cost of making appear \emph{Onsager terms} at time previous steps. Finally, in the large size limit $\ndim \to \infty$, keeping only the leading terms, the set of equations can be reduced to a set of only $\Theta(\ndim)$ messages.
		Notice that the discrepancy between the \aclink{BP} algorithm and the obtained \aclink{AMP} algorithm are not quantified rigorously as anyway, for general factor graphs with loops, the loopy-\aclink{BP} provides only an approximate estimation, so does \aclink{AMP}. 
		However, the resulting \aclink{AMP} has the strong advantage to be rigorously provable in a roundabout way, from the so-called \aclink{SE} equations that can be obtained from the replica computation, proven with a Guerra-like interpolation \cite{Guerra2003}.\\
		
		For the sake of clarity, we present a pedagogical and instructive derivation of the \aclink{GAMP} algorithm, following closely the one of \cite{zdeborova2016statistical}. 
		
		\subsection{\textbf{Application} - AMP for the GLM}
		\label{main:sec:mean_field:amp:example_glm}
	
			The \aclink{rBP} set of equations for the \aclink{GLM} (\ref{main:intro:amp_glm:A_B_mu_to_i}, \ref{main:intro:amp_glm:omega_V_mu_to_i},
				\ref{main:intro:amp_glm:w_cw}, \ref{main:intro:amp_glm:gamma_Lambda_mu_to_i}) contain $\Theta(\ndim^2)$ messages of the form $x_{i\to\mu}$. 
			However it is worth observing that the messages \emph{depend weakly} on the \emph{target node} $\mu$, as the missing message in the sum vanishes in the limit $\ndim \to \infty$. This crucial observation allows to expand the previous \aclink{rBP} equations around the \emph{full} messages by completing the sum that do not show anymore the target-node dependence:
			\begin{align}
			\begin{aligned}
				\omega_{\mu}^t &\equiv \sum\limits_{j = 1}^\ndim\frac{x_{\mu j}}{\sqrt{\ndim}}   \hat{w}_{j\to \mu}^t\,, 
				&& V_{\mu}^t \equiv  \sum\limits_{j=1}^\ndim  \frac{x_{\mu j}^2}{\ndim}  \hat{c}_{j\to \mu}^t\,, \\
				\gamma_{i }^t & \equiv   \sum\limits_{\nu =1}^\nsamples  b_{\nu \to i}^t \,,
				&& \Lambda_{i  }^t \equiv \sum\limits_{\nu =1}^\nsamples  A_{\nu \to i}^t \,.
			\end{aligned}		
			\end{align}
			Performing the expansion of the \aclink{rBP} (\ref{main:intro:amp_glm:A_B_mu_to_i},\ref{main:intro:amp_glm:omega_V_mu_to_i},
				\ref{main:intro:amp_glm:w_cw},
				\ref{main:intro:amp_glm:gamma_Lambda_mu_to_i}), the set can be reduced to $\Theta(\ndim)$ iterative equations, at the cost of introducing \emph{memory terms} at previous time steps, the Onsager terms. The lengthy, yet straightforward, computation is shown for the committee machines hypothesis class in \App\ref{appendix:amp:derivation:amp_eqs}. We finally end up with the \aclink{GAMP} algorithm \cite{rangan2011generalized} as a set of $\Theta(\ndim)$ messages presented in \Alg\ref{main:alg:AMP_glm}. 
			\begin{algorithm} 
			\begin{algorithmic}
			    \STATE {\bfseries Input:} vector $\vec{y} \in \bbR^\nsamples$ and matrix $\mat{X}\in \bbR^{\nsamples \times \ndim}$:
			    \STATE \emph{Initialize}: $\hat{w}_i$, $f_{\out,\mu} \in \bbR$ and $\hat{c}_i$, $\partial_{\omega} f_{\out , \mu} \in \bbR$ for $ 1 \leq i \leq \ndim $ and $ 1 \leq \mu \leq \nsamples $ at $t=0$.
			    \REPEAT   
			    \STATE Channel: Update the mean $\omega_{\mu} \in \bbR$ and variance $V_{\mu}\in \bbR^+$: \spacecase
			    $\omega_{\mu}^t = \sum\limits_{i = 1}^\ndim \frac{x_{\mu i}}{\sqrt{\ndim}}\hat{w}_{i}^t  -  V_{\mu}^t f_{\out,\mu}^{t-1} $\,, 
			    $V_{\mu}^t = \sum\limits_{i=1}^{\ndim}\frac{x_{\mu i}^2}{\ndim} \hat{c}_{i}^t $\spacecase
			    \STATE Update $f_{\out, \mu} \in \bbR$ and $\partial_{\omega} f_{\out , \mu} \in \bbR^+$: \spacecase
			    $f_{\out, \mu}^t = f_\out (y_{\mu}, \omega_{\mu}^t, V_{\mu}^t) $\,, $ \partial_{\omega} f_{\out, \mu}^t = \partial_{\omega}  f_\out (y_{\mu}, \omega_{\mu}^t, V_{\mu}^t) $ \spacecase
			    \STATE Prior: Update the mean $\gamma_i \in \bbR$ and variance $\Lambda_i \in \bbR^+$:\spacecase
			    $\gamma_i^t =  \sum\limits_{\mu =1}^\nsamples
			      \frac{x_{\mu i}}{\sqrt{\ndim}} f_{\out,\mu}^t  + \Lambda_{i}^t \hat{w}_{i}^t $\,, $ \Lambda_{i}^t = -\sum\limits_{\mu =1}^\nsamples \frac{x_{\mu
			      i}^2}{\ndim}  \partial_\omega f_{\out,\mu}^t $\spacecase
			    \STATE Update the estimated marginals $\hat{w}_i \in \bbR$ and $\hat{c}_i \in \bbR^+$: \spacecase
			   $\hat{w}_i^{t+1} = f_\w( \gamma_i^t,  \Lambda_i^t  ) $\,, $   \hat{c}_i^{t+1} = \partial_\gamma f_\w(\gamma_i^t,  \Lambda_i^t  )$\spacecase
			    \STATE ${t} = {t} + 1$ 
			    \UNTIL{Convergence on
			    $\hat{\vec{w}}$, $\hat{\vec{c}}$.} 
			    \STATE {\bfseries Output:}
			    $\hat{\vec{w}}$ and $\hat{\vec{c}}$.
			\end{algorithmic}
			\caption{Approximate Message Passing algorithm for Generalized Linear Models.}
  			\label{main:alg:AMP_glm}
			\end{algorithm}
			The \aclink{GAMP} algorithm can be interpreted in a series of iterative steps starting by the estimation of the mean $\bomega$ and variance $\mat{V}$ of the variable $\vec{z} = \frac{1}{\sqrt{\ndim}} \mat{X} \vec{w}$. The estimate of $\vec{z}$ provides a potential output that is compared to the true output vector $\vec{y}$ through the denoising functions $f_\out$, $\partial_\omega f_\out$. This comparison gives a feedback to update the mean $\bgamma$ and variance $\bLambda$ of the variable $\vec{w}$ which is updated to provide a new estimation with the denoising functions $f_\w$ and $\partial_\gamma f_\w$. Moreover, \aclink{AMP} provides a general inference algorithm valid on single instance of finite size for generic prior and channel distributions. As a consequence, it is valid in the Bayes-optimal case for \aclink{MMSE} estimation when $\rP_\out = \rP_{\out^\star}$ and $\rP_\w = \rP_{\w^\star}$ or the mismatched setting for arbitrary distribution such as for \aclink{MAP} estimation and \aclink{ERM}.
			Finally looking beyond the cumbersome appearances of \Alg\ref{main:alg:AMP_glm}, in contrast with most of state-of-the-art gradient-based algorithms that suffer theoretical understanding, \aclink{AMP} algorithms have the main advantage that their asymptotic behavior and convergence performances can be rigorously tracked for large \aclink{i.i.d} input matrices through their \aclink{SE} equations. 
			At the heart of this manuscript, we should stress that these \aclink{SE} equations connect surprisingly to the results obtained by the replica computation.
			
		\subsection{State evolution equations - Connection with replicas}
		\label{main:sec:mean_field:se_amp}
		
			One of the main interests of the \aclink{AMP} algorithm is certainly that we can analyze its average behavior in the thermodynamic limit.
			Indeed, taking the average over the quenched disorder and introducing proper order parameters, the so-called \emph{overlaps}, we can obtain an asymptotic closed set of equations, called the \aclink{SE} equations. They characterize the performances of the \aclink{AMP} algorithm \cite{donoho2009message, bayati2011dynamics,javanmard2013state} in the large size limit $\ndim \to \infty$.
			The derivation usually starts with the set of \aclink{rBP} equations. By assuming the fundamental message independence, using the \aclink{CLT}, 
			and defining a correctly chosen set of order parameters, the statistical analysis ends up to the set of \aclink{SE} equations.
			Very importantly, under the strong \aclink{i.i.d} assumption and in the Bayes-optimal case, these \aclink{SE} equations converge to the stationary points of the \aclink{RS} replica free entropy. 
			We illustrate the derivation again on the \aclink{GLM} hypothesis class and draw the intimate connection with the replica computation.  The full computation for the committee hypothesis class can be found in \App\ref{appendix:amp:derivation:se_eqs}. 	

		\subsection{\textbf{Application} - SE for the GLM}	
		\label{main:sec:mean_field:se_amp:example_glm}		
			The derivation of the \aclink{SE} equations starts by defining a series of order parameters, called \emph{overlaps}			
			\begin{align}
				m^t &\equiv \lim_{\ndim \to \infty} \EE_{\vec{w}^\star, \mat{X}} \[ \frac{1}{\ndim}  \hat{\vec{w}} \cdot \vec{w}^\star \]\,, && q^t \equiv \lim_{\ndim \to \infty} \EE_{\vec{w}^\star, \mat{X}} \[ \frac{1}{\ndim}  \hat{\vec{w}}^t \cdot \hat{\vec{w}}^t \]\,, \nonumber \\
				\rho_{\w^\star} &\equiv \lim_{\ndim \to \infty} \EE_{\vec{w}^\star} \[ \frac{1}{\ndim}  \vec{w}^\star \cdot \vec{w}^\star \]\,, && \Sigma^t \equiv \lim_{\ndim \to \infty} \EE_{\vec{w}^\star, \mat{X}} \[ \frac{1}{\ndim}  \hat{\vec{c}}^t \cdot  \vec{1} \]\,,
				\label{main:intro:amp:overlap_definitions}
 			\end{align}
 			that measure the correlations between the ground truth vector $\vec{w}^\star$ and the estimator $\hat{\vec{w}}^t$ at time $t$ of the \aclink{AMP} algorithm in \Alg\ref{main:alg:AMP_glm}. In a \aclink{T-S} scenario, they allow to quantify properly the reconstruction performance of the \aclink{AMP} algorithm. For the purpose of the derivation, we need to define other ad-hoc overlaps which have less direct physical meaning
			\begin{align}
			\begin{aligned}
					\hat{q}^t & \equiv \alpha \EE_{\omega,z} \[ f_{\out} (\varphi_{\out^\star}(z), \omega^t, \Sigma^t )^2 \] \,,\\
					\hat{m}^t &\equiv \alpha \EE_{\omega, z} \[ \partial_z f_{\out}(\varphi_{\out^\star}(z), \omega^t, \Sigma^t ) \]  \,, \\
					\hat{\chi}^t &\equiv \alpha \EE_{\omega, z} \[ - \partial_\omega f_{\out}(\varphi_{\out^\star}(z), \omega^t, \Sigma^t ) \] \,. \\
			\end{aligned}
			\end{align}
			Under the \aclink{BP} independent messages assumption and using the \aclink{CLT}, we obtain in \App\ref{appendix:amp:derivation:se_eqs} the message statistics of the \aclink{rBP} messages. Computing the average of the overlaps defined above and making use of the latter statistics, we finally obtain a set of six \aclink{SE} equations, that can be reduced using the Nishimori conditions in the Bayes-optimal case \cite{opper1991calculation, iba1999nishimori}
			\begin{align*}
					m^t &= q^t\,, && \hat{q}^t  = \hat{m}^t = \hat{\chi}^t\,, && \Sigma^{t} = \rho_{\w^\star} - q^t\,,
			\end{align*} 
			to only two \aclink{SE} equations 
				\begin{align}
					&q^{t+1}  \displaystyle = \EE_{\xi} \[ \mZ_{\w^\star}\((\hat{q}^t)^{1/2}\xi, \hat{q}^t \) f_{\w^\star}\((\hat{q}^t)^{1/2}\xi, \hat{q}^t \)^2 \]\,, \nonumber\\
					&\hat{q}^{t} = \alpha \int_\bbR \d y ~ \EE_{\xi} \mZ_{\out^\star}\(y, (q^t)^{1/2}\xi, \rho_{\w^\star} - q^t \)^2 \label{main:intro:amp:se} \\
					& \qquad \qquad \qquad \qquad \qquad \qquad \times f_{\out^\star}\(y, (q^t)^{1/2}\xi, \rho_{\w^\star} - q^t \)^2 \,. \nonumber
				\end{align}
			As a crucial conclusion, we finally observe that the set of \aclink{SE} equations \eqref{main:intro:amp:se}, which characterize the asymptotic behavior of the \aclink{AMP} algorithm in the Bayes-optimal setting, are connected to the ones obtained by the \aclink{i.i.d} replica computation in \eqref{main:intro:replicas:glm:fixed_point}.			
			Indeed, similarly to the \aclink{TAP} approach, while the replica result does not provide the time indices to solve the fixed point equation, the \aclink{SE} \eqref{main:intro:amp:se} fully determine the dynamics of the \aclink{AMP} algorithm at any time $t$. As as consequence, it turns out that, under a \aclink{T-S} scenario the \aclink{SE} of the \aclink{AMP} algorithm follows exactly the gradient of the \aclink{RS} free entropy \eqref{appendix:free_entropy_bayes} in the Bayes-optimal setting, that intrinsically grasp the importance of the overlaps in the considered \aclink{JPD}.
			Importantly, the connection between \aclink{AMP}, the replica formalism and the possibility to prove them rigorously with Guerra-like interpolation breaks down in the \emph{mismatched setting}. In this case, the prediction of the replica method fails delivering the correct behavior of the \aclink{AMP} algorithm under the simple \aclink{RS} assumption and reveal a more complex \aclink{RSB} structure of the phase space. Yet, it is not the case for convex optimization as shown in \Chap\ref{chap:erm}, where the \aclink{RS} turned out to hold rigorously correct even in the mismatched \aclink{MAP} estimation setting.
			
		\subsection{Beyond i.i.d matrices and AMP}
			Even though the derivation of \aclink{AMP} in the case of the \aclink{GLM} does not assume any hypothesis on the input matrix $\mat{X}$, it has been observed that \aclink{AMP} may experience divergences for non-\aclink{i.i.d} inputs, even for non-pathological matrices \cite{rangan2019convergence}. 
			In practice, to circumvent this issue, we can try to improve the stability by using mean removal, \emph{damping}\index{damping} \cite{vila2015adaptive, rangan2019convergence}, sequential updating or other tricks. These stability techniques are partially successful but convergence may still fail and often needs specific tuning. 
			Moreover even the convergence of \aclink{AMP} is proven \cite{bayati2011dynamics, bolthausen2014iterative} only under particular restrictive conditions, such as the \aclink{i.i.d} hypothesis. As a consequence it is not surprising that correlated statistics can therefore breaks down the message independence assumption in \aclink{AMP} and leads to divergences for more complex input matrices.
			To overcome this fundamental limitation, many efforts have been made to generalize the mean-field approaches to more complex matrix statistics such as the high-temperature expansion for orthogonal matrices \cite{parisi1995mean} or the ADA-\aclink{TAP} approach for dense graphical models with generic weight statistics \cite{opper2001adaptive,opper2001tractable}. The corresponding approaches were understood later as a particular case of the \aclink{EP} approximate inference algorithm \cite{minka2001family, Minka2013, heskes2005approximate, heskes2012expectation}. Similarly to \aclink{BP} with the Bethe free energy, the \aclink{EP} procedure  is associated to an approximate free energy called the \aclink{EC} \cite{opper2005expectation} and solution of the Gibbs variational principle by enforcing the moments matching constraints.
			Analogously to \aclink{AMP} with \aclink{BP}, the \aclink{EP} procedure was applied by projecting the messages to a Gaussian parametrization leading to the \aclink{VAMP} algorithm \cite{rangan2019vector}. 
			For the sake of clarity, the main difference with \aclink{AMP} lies on the fact that the factor graph is \emph{vector valued}, meaning that a vector variable with a separable prior is represented by a single variable and factor nodes, which insight comes from \cite{cakmak2014s}. It reduces the global inference problems to sub-vectorial-problems by imposing connecting Dirac-delta constraints.
			\aclink{AMP} and \aclink{VAMP} are conjectured to be equivalent for \aclink{i.i.d} matrices, and asymptotically are proved to be rigorously identical as the corresponding \aclink{RS} free energies are equivalent. 
			However, \aclink{VAMP} experienced less convergence issues than the classical \aclink{AMP}. \aclink{VAMP} converges for orthogonally invariant matrices and is more stable and robust to ill-conditioned matrices at the cost of computing matrix inversion or \aclink{SVD}. 
			Very importantly, similarly to \aclink{AMP} algorithms, the \aclink{VAMP} algorithm asymptotic performances are remarkably characterized by a set of \aclink{SE} equations for orthogonally invariant matrices. 
			The \aclink{SE} are again related to the stationary point of the replica free energy computed for this matrix statistics as observed in \cite{tulino2013support} and shown rigorously in \cite{barbier2018mutual} in linear estimation and \cite{gabrie2018entropy,reeves2017additivity} in \aclink{DNN}.

\ifthenelse{\equal{\format}{oneside}}
	{\clearpage\null\thispagestyle{empty}\newpage
	\clearpage\null\thispagestyle{empty}\newpage
	}
	{\clearpage\null\thispagestyle{empty}\newpage
	\clearpage\null\thispagestyle{empty}\newpage
	}
	
\ctparttext{}
\part{Main contributions}
\label{part:contribution}
\ifthenelse{\equal{\format}{oneside}}
	{\clearpage\null\thispagestyle{empty}\newpage
	}
	{\cleardoublepage}

\subpartpage{II~ A}{Bayes-optimal, empirical risk minimization and worst-case analysis in simple feed-forward neural networks}
\ifthenelse{\equal{\format}{oneside}}
	{\clearpage\null\thispagestyle{empty}\newpage}
	{\newpage\clearpage}
	
		\chapter*{Outline and motivations}
	\chaptermark{Outline and motivations}
		The mean-field methods originating from statistical physics presented in \Chap\ref{main:chap:mean_field} have been extensively used in the past to analyze the equilibrium behavior of common estimators for simple model classes such as single-layer neural networks. 
		This statistical physics approach, seeking to understand the \emph{typical} behavior of such systems, focused essentially on a simple data generative process: the \emph{teacher-student} scenario glimpsed in \Chap\ref{chap:phys_ml_together}.
		
		In this part, we essentially revisit this setting in light of the modern challenge of algorithmic complexity, and second, we develop and extend rigorous proofs of earlier heuristic analysis.
		With these contributions, we propose an overview of the complementary approaches used in different communities to understand simple classes of feed-forward neural networks.
		Especially, in the context of the classical \emph{perceptron} and its multi-layer generalization, the \emph{committee machine}, we try to reconcile the \emph{Bayes-optimal} setting, the \emph{wort-case} analysis and \emph{empirical risk minimization methods} in a unified statistical physics framework.
		
			\paragraph{Bayes-optimal analysis and computational complexity}
			In \Chap\ref{chap:committee_machine}, we revisit the \aclink{T-S} scenario to a more sophisticated class of two-layers neural networks, the \emph{committee machines}. It naturally extends and encompasses the classical \aclink{GLM} class \cite{barbier2016mutual}, which was restricted to linearly separable data, to higher complexity models.			
			We analyze the Bayes-optimal setting, in the case where the dataset has been generated by a ground truth \emph{teacher} committee machine with weights $\mat{W}^\star \in \bbR^{\ndim \times K}$ and prior distributions $\rP_{\out^\star}$ and $\rP_{w^\star}$.
			Under the Bayes-optimal hypothesis, the \emph{student} seeks to fit the dataset based on the exact model architecture and by having access to the ground truth prior distributions $\rP_\out = \rP_{\out^\star}$ and $\rP_\w = \rP_{\w^\star}$. 
			This idealistic setting provides nonetheless a crucial information theoretical lower-bound of optimal statistical estimation. It naturally provides answers to the questions of knowing \emph{under what conditions}, and without any algorithmic consideration, if \emph{it is possible to recover the structure in the data}? And \emph{how many examples} are needed in that case?
			In the asymptotic regime where the size of the input vector $\ndim$ and the number of training examples $\nsamples$ diverge, using Guerra interpolation we first provide a rigorous justification that the \aclink{RS} free entropy, initially derived with the heuristic replica method \cite{schwarze1992generalization,schwarze1993generalization}, is \emph{exact} for \aclink{i.i.d} input data $\mat{X}\in \bbR^{\nsamples \times \ndim}$. Moreover we provide expressions for the corresponding optimal generalization error learning curves. 
			Secondly, we develop an extension of the polynomial time \aclink{GAMP} algorithm \cite{rangan2011generalized} for the committee machine hypothesis class. 
			This algorithmic perspective allows us to answer the burning questions of computational complexity that focuses on knowing \emph{if the algorithm is efficient with respect to the information-theoretical baseline} and \emph{how many examples it requires to achieve it}. By locating the phase transitions, we highlight hard regions where the best algorithm fails delivering the optimal predictions.
			
			To summarize, our approach capitalizes on the trade-off between information theoretical statistical inference and the computational efficiency of the conjectured best polynomial \aclink{AMP}  algorithm for \aclink{i.i.d} data. 
			By making an intense use of the description of metastability, first and second order phase transitions phenomenology, borrowed to statistical physics and depicted in \Chap\ref{chap:statistical_physics}, we can study in details the phase statistical and algorithmic phase transitions. Especially, we unveil the existence of large computational gaps, even in the Bayes-optimal case, for small and extensive hidden-layer sizes $K$.
	
			\paragraph{Worst case analysis: from the storage capacity and ground state energies to the VC dimension and the Rademacher complexity}		
			In practice, the previous Bayes-optimal approach is harshly criticized for its lack of fairness: the analysis requires a strong prior knowledge with the access to the prior distributions involved in the ground truth generative process.
			An alternative approach from the statistical learning theory literature consists instead in evaluating the \emph{worst-case} performances of statistical inference, by quantifying generalization error upper-bounds. This is classically done with the \aclink{VC} dimension and the Rademacher complexity \cite{vapnik1994measuring, bartlett2002rademacher}.
			Alternatively, the physics approach deeply focused on the \emph{Gardner capacity} \cite{gardner1988optimal, krauth1989storage, engel1993statistical} that provides essentially a lower-bound of its twin from statistical learning theory, the \aclink{VC} dimension. The Gardner storage capacity is essentially the maximum number of examples that a model, namely the perceptron, is able to \emph{memorize}. Indeed, under a randomly quenched disorder, input vectors $\vec{x}$ and output labels $y$ are uncorrelated. 
			Therefore, in a \aclink{rCSP} language the storage capacity is equivalently the maximum number of random input-output constraints the model parameters $\vec{w}$ can satisfy simultaneously.  
			As a consequence, above this critical SAT-UNSAT threshold, the perceptron can no longer satisfy all the random constraints without making a prediction error.
			
			In \Chap\ref{chap:binary_perceptron}, we present the Gardner-like computation of the storage capacity for the binary perceptron with various activation functions. We show that, unlike for the usual step-function-binary-perceptron \cite{gardner1988optimal}, the critical capacity in simple symmetric variants is rigorously given by the annealed computation. Moreover by studying the structure of the configuration space, we unveil a \aclink{f1RSB} structure using simple first and second moment methods.
			
			By definition, above the SAT-UNSAT threshold the best configuration of the model parameters cannot satisfy all the random constraints and inevitably makes classification errors. 
			 Counting this minimal number of mistakes for a given constraint density is equivalent of computing the ground state energy of the system: below the storage capacity the energy vanishes whereas it becomes strictly positive above it. 
			 This \aclink{rCSP} approach can be naturally extended above the SAT-UNSAT transition to compute the corresponding ground state energies within the same framework.
			 
			 In \Chap\ref{chap:rademacher}, we reveal the deep connection between the Rademacher worst-case generalization bound, which measures if a function can fit random noise, and the ground state energies from statistical physics. Finally, while statistical learning theory computes the generalization bounds up to asymptotic scalings, we are able to explicitly compute the Rademacher complexity for the spherical and binary perceptrons.			

			\paragraph{Empirical risk minimization in Generalized Linear Models for synthetic i.i.d data}
			The two previous approaches provide the optimal and worst-case predictions that define the operating range of any statistical estimator. As a consequence, it turns out that \emph{practical} machine learning estimators and algorithms are not described correctly neither by the pessimistic worst case analysis nor the idealistic Bayes-optimal. 
			
			In \Chap\ref{chap:erm}, we present how to analyze rigorously the behavior of practical \aclink{ERM} for regularized linear models, such as ridge, logistic or hinge regression.
			We focus on a common supervised classification task of a synthetic dataset, whose labels are generated by feeding a one-layer neural network with random \aclink{i.i.d} inputs. 
			In this convex optimization task, the replica computation, under the \aclink{RS} ansatz, turns out to be correct and matches exactly the results of the Gordon convex Gaussian min-max theorem.
			After observing that, unlike ridge regression, logistic and hinge regressions surprisingly approach closely the Bayes-optimal generalization error, we design an optimal loss and regularizer that provably lead to Bayes-optimal generalization error performances.
			
			As a conclusion, we summarize and reconcile the different approaches in a global picture. We conclude that, unlike the generalization error bounds, the Bayes-optimal analysis, even though its requires strong prior knowledges, captures the good scaling behaviors of the practical algorithms.

	\ifthenelse{\equal{\format}{oneside}}
	{\clearpage\null\thispagestyle{empty}\newpage}
	{\cleardoublepage}
	
	\chapter{The committee machine: Computational to statistical gaps in learning a two-layers neural network}
	\chaptermark{The committee machine: computational to statistical gaps}
	\label{chap:committee_machine}
	While the traditional approach to learning and generalization follows the \aclink{VC} \cite{vapnik1998statistical} and Rademacher \cite{bartlett2002rademacher} worst-case type bounds, there has been a considerable body of theoretical work on calculating the generalization ability of neural networks for data arising from a probabilistic model within the framework of statistical mechanics \cite{seung1992statistical,watkin1993statistical,monasson1995learning,monasson1995weight,engel2001statistical}. In the wake of the need to understand the effectiveness of neural networks and also the limitations of the classical approaches \cite{zhang2016understanding}, it is of interest to revisit the results that have emerged thanks to the physics perspective. This direction is currently experiencing a strong revival, see e.g. \cite{Chaudhari2016,martin2017rethinking,barbier2017phase,baity2018comparing}.

Of particular interest is the so-called \aclink{T-S} approach, where labels are generated by feeding \aclink{i.i.d} random samples to a neural network architecture (the \emph{teacher}) and are then presented to another neural network (the \emph{student}) that is trained using these data. Early studies computed the information theoretic limitations of the supervised learning abilities of the teacher weights by a student who is given $\nsamples$ independent $\ndim$-dimensional examples with $\alpha\!\equiv\!\nsamples / \ndim \!=\!\Theta(1)$ and $ \ndim \to \infty$ \cite{seung1992statistical,watkin1993statistical,engel2001statistical}. These works relied on non-rigorous heuristic approaches, such as the replica and cavity methods \cite{mezard1987spin,mezard2009information}. Additionally no provably efficient algorithm was provided to achieve the predicted learning abilities, and it was thus difficult to test those predictions, or to assess the computational difficulty. 

Recent developments in statistical estimation and information theory ---in particular of \aclink{AMP} \cite{donoho2009message,rangan2011generalized,bayati2011dynamics,javanmard2013state}, and a rigorous proof of the replica formula for the optimal generalization error \cite{barbier2017phase}--- allowed to settle these two missing points for single-layer neural networks (i.e. without any hidden variables). In the present chapter, we leverage on these works, and provide rigorous asymptotic predictions and corresponding message passing algorithm for a class of two-layers networks.

\section{Main contributions and related works}
\begin{figure}
\centering
\begin{tikzpicture}[scale=1.2, every node/.style={transform shape}]
    \tikzstyle{factor}=[rectangle,minimum size=4pt,draw=black, fill opacity=1.]
    \tikzstyle{latent}=[circle,minimum size=19pt,draw=black, fill opacity=1.,fill=white]
    \tikzstyle{output}=[circle,minimum size=19pt,draw=black, fill opacity=1.,fill=white]
        \tikzstyle{noise}=[circle,minimum size=18pt,draw=black, fill opacity=1.,fill=white]
    \tikzstyle{output_y}=[rectangle,minimum size=15pt,draw=black, fill opacity=1.,fill=white]
    \tikzstyle{annot} = [text width=3cm, text centered]
    \tikzstyle{annot} = [text width=3cm, text centered]
    \tikzstyle{annotLarge} = [text width=5cm, text centered]
    \def\NX{15}
    \def\NK{2}
    \def\NY{1}
    \def\middle{0}
    \foreach \i in {1,...,\NX}
     	  \pgfmathparse{0.9*rnd+0.3}
          \definecolor{MyColor}{rgb}{\pgfmathresult,\pgfmathresult,\pgfmathresult}
    	\node[factor, fill=MyColor] (X-\i) at (\middle 0, 0.24*0.5*\NX+0.12 - 0.24*\i ) {}; 
    \foreach \k in {1,...,\NK}
    	\node[latent] (K-\k) at (2,\middle  \k - 1.5) {}; 
    \foreach \y in {1,...,\NY}
    	\node[output] (Y-\y) at (3.5,0) {}; 	
    \node[output_y] (Y-2) at (4.5,0) {}; 
    
    \foreach \i in {1,...,\NX}
    	\foreach \k in {1,...,\NK}
    		\path[-] (X-\i) edge (K-\k);
   	\foreach \k in {1,...,\NK}
    	\foreach \y in {1,...,\NY}
    		\path[-] (K-\k) edge (Y-\y);
    \path[-] (Y-1) edge (Y-2);
    \node[annotLarge] at (-1.5,0) {$(\vec{x}_{\mu})_{\mu=1}^{\nsamples}$ \\ samples};
    \node[annot] at (1.5,-1.6) {$\mat{W}^\star \in \bbR^{\ndim \times K}$}  ;
    \node[annot] at (4.525,0) {${y_{\mu}}$}  ;
    \node[annot] at (3.2,-0.8) {$\vec{w}^{(2)}\in\bbR^{K}$}  ;
    \node[annot] at (2,-0.45) {\small $ f^{(1)}$}  ;
    \node[annot] at (2,0.55) {\small $ f^{(1)}$}  ;
    \node[annot] at (3.5,0.05) {\small $ f^{(2)}$}  ;
    \node[annot] at (0,2.2) {\small $\ndim$ features}  ;
    \node[annot] at (2,1.3) {\small $K$ hidden\\ units}  ;
    \node[annot] at (4.5,0.7) {\small output}  ;
	\end{tikzpicture}	
	\caption{Illustration of the \emph{committee machine}: it is one of the simplest models belonging to the considered model class \eqref{model}, and on which we focus to illustrate our results. It is a two-layers neural network with sign activation functions $f^{(1)},f^{(2)}=\sign$ and weights $\vec{w}^{(2)}$ fixed to unity. It is represented for $K=2$.}
	\label{fig:committee}
\end{figure}
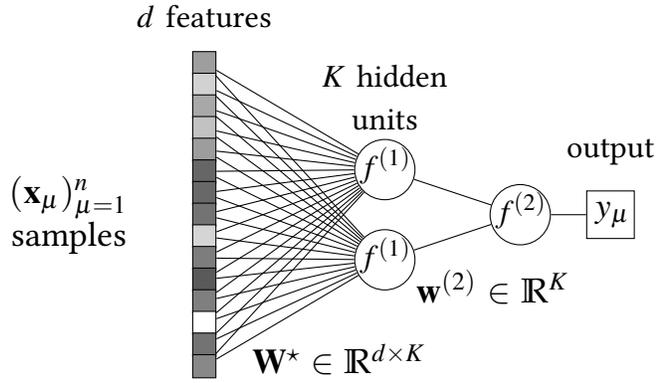

While our results hold for a rather large class of non-linear activation functions, we illustrate our findings on a case considered most commonly in the early literature: the \emph{committee machine}.  This is possibly the simplest version of a two-layers neural network where all the weights in the second layer are fixed to unity, and we illustrate it in Fig.~\ref{fig:committee}. Denoting $\forall \mu \in \lb \nsamples \rb, ~ y_\mu$ the label associated with a $\ndim$-dimensional sample $\vec{x}_\mu$, and $w_{il}^\star$ the weight connecting the $i$-th coordinate of the input to the $k$-th node of the hidden layer, it is defined by:
\begin{equation}
y_\mu = \sign \[\sum_{k=1}^K \sign \( \sum_{i=1}^\ndim x_{\mu i}  w_{i k}^\star \) \] = \sign \[\sum_{k=1}^K \sign \( \vec{x}_\mu^\intercal \mat{W}^\star \) \]  \,, \label{model:com}
\end{equation}
where $\mat{W}^\star \in \bbR^{\ndim \times K}$.
We concentrate here on the \aclink{T-S} scenario: The teacher generates \aclink{i.i.d} data samples with \aclink{i.i.d} standard Gaussian coordinates $x_{\mu i} \sim\mathcal{N}(0,1)$, then she/he generates the associated labels $y_\mu$ using a committee machine as in \eqref{model:com}, with \aclink{i.i.d} weights $w_{il}^\star$ unknown to the student. In the proof though, we will consider the more general case of a distribution for the weights of the form $\prod_{i=1}^n \rP_\w(\{w_{il}^\star\}_{k=1}^K)$, but in practice we consider the fully separable case. The student is then given the $\nsamples$ input-output pairs $(\vec{x}_{\mu}, y_\mu)_{\mu=1}^\nsamples$ and knows the distribution $\rP_\w$ used to generate $w_{il}^\star$. The goal of the student is to learn the weights $w_{il}^\star$ from the available examples  $(\vec{x}_{\mu}, y_\mu)_{\mu=1}^\nsamples$ in order to reach the smallest possible generalization error, \ie to be able to predict the label the teacher would generate for a new sample not present in the training set.  

There have been several studies of this model within the non-rigorous statistical physics approach in the limit where $\alpha\equiv \nsamples / \ndim =\Theta(1)$, $K=\Theta(1)$ and $\ndim \to \infty$~\cite{schwarze1993learning,schwarze1992generalization,schwarze1993generalization,MatoParga92,monasson1995weight,engel2001statistical}. A particularly interesting result in the \aclink{T-S} setting is the \emph{specialization of hidden neurons} (see sec. 12.6 of \cite{engel2001statistical}, or \cite{saad1995line} in the context of online learning): For $\alpha<\alpha_{\textrm{spec}}$, where $\alpha_{\textrm{spec}}$ is a certain critical value of the sample complexity, the permutational symmetry between hidden neurons remains conserved even after an optimal learning, and the learned weights of each of the hidden neurons are identical. For $\alpha>\alpha_{\textrm{spec}}$, however, this symmetry gets broken as each of the hidden units correlates strongly with one of the hidden units of the teacher. Another remarkable result is the calculation of the optimal generalization error as a function of $\alpha$.

Our first contribution consists in a proof of the replica formula conjectured in the statistical physics literature, using the adaptive interpolation method of \cite{BarbierM17a,barbier2017phase}, that allows to put several of these results on a rigorous basis.
However, this proof uses a technical unproven assumption. 
Our second contribution is the design of an \aclink{AMP}-type of algorithm that is able to achieve the optimal generalization error in the above limit of large dimensions for a wide range of parameters. 
The study of \aclink{AMP} ---that is widely believed to be optimal between all polynomial algorithms in the above setting \cite{donoho2013accurate,zdeborova2016statistical,deshpande2015finding,bandeira2018notes}--- unveils,
in the case of the committee machine with a large number of hidden neurons $K \to \infty$ with $K = o(\ndim)$, the existence a large \emph{hard phase} in which learning is information-theoretically possible, leading to a good generalization error decaying asymptotically as $1.25 K/\alpha$ (in the $\alpha = \Theta(K)$ regime), 
but where \aclink{AMP} fails and provides only a poor generalization that does not go to zero when increasing $\alpha$. This strongly suggests that no efficient algorithm exists in this hard region and therefore there is a computational gap in learning such neural networks. 
In other problems where a hard phase was identified its study boosted the development of algorithms that are able to match the predicted thresholds and we anticipate this will translate to the present model.

We also want to comment on a related line of work that studies the loss-function landscape of neural networks. While a range of works show under various assumptions that spurious local minima are absent in neural networks, others show under different conditions that they do exist, see e.g.~\cite{safran2017spurious}. 
The regime of parameters that is hard for \aclink{AMP} must have spurious local minima, but the converse is not true in general. It might be that there are spurious local minima, yet 
the \aclink{AMP} approach succeeds. Moreover, in all previously studied models in the Bayes-optimal setting the generalization error obtained with the \aclink{AMP} is the best 
known and other approaches, \eg noisy gradient-based, spectral algorithms or semidefinite programming, are not better 
in generalizing even in cases where the \emph{student} models are over-parametrized. Of course in order to be in the Bayes-optimal setting one needs to know the model used by the teacher which is not the case in practice.

\section{Technical results}

\subsection{A general model}
While in the illustration of our results we shall focus on the model (\ref{model:com}), all our formulas are valid for a broader class of models: Given $\nsamples$ input samples $\mat{X} = \{x_{\mu i}\}_{\mu,i=1}^{\nsamples, \ndim}$, we denote $\mat{W}^\star = \{w_{ik}^\star\}_{i=1..\ndim}^{k=1..K}$ the teacher-weight connecting for all $(i,k) \in\lb \ndim \rb\times \lb K \rb$, the $i$-th input, \ie the visible unit, to the $k$-th node of the hidden layer, represented in \Fig\ref{fig:committee}. For a generic function $\varphi_\out: \bbR^K \times \bbR \to \bbR$ one can formally write the output as
\begin{equation}\label{model}
y_\mu = \varphi_\out \Big(\Big\{\frac{1}{\sqrt \ndim }\sum_{i=1}^\ndim x_{\mu i} w_{i k}^\star \Big\}_{k=1}^K, a_\mu \Big)~~~\text{or}~~~y_\mu \sim \rP_\out\Big( \cdot\Big| \Big\{\frac{1}{\sqrt \ndim }\sum_{i=1}^\ndim x_{\mu i} w_{i k}^\star \Big\}_{k=1}^K \Big)\, ,
\end{equation}
where $(a_\mu)_{\mu=1}^\nsamples$ are \aclink{i.i.d} real valued random variables with
known distribution $\rP_a$, that form the probabilistic part of the model, generally accounting for noise. 
For deterministic models the second argument is simply absent and the distribution $\rP_\out$ is a Dirac mass. We can view alternatively \eqref{model} as a channel where the transition kernel $\rP_\out$ is directly related to $\varphi_\out$. As discussed above, we focus on the \aclink{T-S} scenario where the teacher generates Gaussian \aclink{i.i.d} data $x_{\mu i}\sim{\cal N}(0,1)$, and \aclink{i.i.d} weights $w_{ik}^\star \sim \rP_\w$. The student then learns $\mat{W}^\star\in\bbR^{\ndim \times K}$ from the data $(\vec{x}_{\mu},y_\mu)_{\mu=1}^\nsamples$ by computing marginal means of the posterior probability distribution (\ref{posterior-measure}).

Different scenarii fit into this general framework. Among those, the committee machine is obtained when choosing $\varphi_\out({h})=\sign(\sum_{k=1}^K {\sign} ( h_k) )$ while another model considered previously is given by the parity machine, when $\varphi_\out({h})=\prod_{k=1}^K {\sign} ( h_k)$, see e.g.  \cite{engel2001statistical} and \Sec \ref{sec:parity} for the numerical results in the case $K=2$. A number of layers beyond two has also been considered, see \cite{MatoParga92}. Other activation functions can be used, and many more problems can be described, e.g. compressed pooling \cite{alaoui2016decoding,el2017decoding} or multi-vector compressed sensing \cite{zhu2017performance}. 

\subsection{Two auxiliary inference problems} 
Denote $\mS_K$ the finite dimensional vector space of $K\times K$ matrices,
$\mS_K^+$ the convex set of
semi-definite positive $K\times K$ matrices, $\mS_K^{++}$ for
positive definite $K\times K$ matrices, and
$\forall \,\mat{N} \in \mS_K^+$ we set
$\mS_K^{+}(\mat{N}) \equiv \{\mat{M} \in \mS_K^+ \text{ s.t. } \mat{N}-\mat{M} \in \mS_K^+\}$. Note that ${\cal S}_K^+(\mat{N})$ is convex and compact. Exceptionally in this section, parameters denoted with lowercase letters such as $\mat{q}, \hat{\mat{q}}, \brho^\star$ represent matrices of size $K \times K$.
Stating our results requires introducing two simpler auxiliary $K$-dimensional estimation problems: \\

$\bullet$ The first one consists in retrieving a $K$-dimensional input vector $\vec{w} \sim \rP_\w$ from the output of a Gaussian vector channel with $K$-dimensional observations 
$$\vec{y}_0= \hat{\mat{q}}^{1/2} \vec{w} + \vec{z}_0\,,$$ 
$\vec{z}_0 \sim \mN(\vec{0}, \mat{I}_K)$ and the \emph{channel gain}
matrix $\hat{\mat{q}} \in \mS_K^+ \subseteq \bbR^{K \times K}$. The posterior distribution on $\vec{w} = (w_k)_{k=1}^K \in \bbR^K$ is 
\begin{align}
 \rP(\vec{w} \vert \vec{y}_0)=\frac{1}{{\mZ}_{\w}}~\rP_\w(\vec{w})~ \exp\(\vec{y}_0^\intercal \hat{\mat{q}}^{1/2} \vec{w} - \frac{1}{2} \vec{w}^\intercal \hat{\mat{q}} \vec{w}\)\, ,
\label{aux-model-1}
\end{align}
and the associated \emph{free entropy} (or minus \emph{free energy}) is given by the expectation over $\vec{y}_0$ of the log-partition function $\Psi_{\w}(\hat{\mat{q}}) \equiv \EE \log {\mZ}_{\w}$ and involves $K$ dimensional integrals. \\

$\bullet$ The second problem considers $K$-dimensional \aclink{i.i.d} vectors $\vec{v}, \vec{u}^\star \sim \mN(\vec{0}, \mat{I}_K )$ where $\vec{v}$ is considered to be known and one has to retrieve $\vec{u}^\star $ from a scalar observation obtained as 
$$ \td{y}_0 \sim \rP_{\out}(\,\cdot\,| \mat{q}^{1/2} \vec{v} + (\brho^\star-\mat{q})^{1/2}\vec{u}^\star )$$ 
where the second moment 
matrix $\brho^\star \equiv \mathbb{E}[\vec{w} \vec{w}^\intercal] \in \mS_K^+$, where $\vec{w}\in \bbR^{K} \sim \rP_\w$, and the so-called \emph{overlap matrix}
$\mat{q}$ is in $S_K^{+}(\brho^\star)$. The associated posterior is 
\begin{align}
 \rP(\vec{u} \vert \td{y}_0, \vec{v}) = \frac{1}{{\mZ}_{\out}}\frac{e^{-\frac{1}{2}\vec{u}^\intercal \vec{u}}}{{(2\pi)^{K/2}}}~\rP_{\out}\big( \td{y}_0 | \mat{q}^{1/2} \vec{v} + (\brho^\star - \mat{q})^{1/2}\vec{u} \big)\, ,
\label{aux-model-2}
\end{align}
and the free entropy reads this time $\Psi_{\out}(\mat{q};\brho^\star) \equiv \EE \ln {\mZ}_{\out}$, with the expectation over $\td{y}_0$ and $\vec{v}$, and also involves $K$ dimensional integrals.

\subsection{The free entropy}
 The central object of study leading to the optimal learning and generalization errors in the present setting is the posterior distribution of the weights:
\begin{align}
 \rP( \{\vec{w}_k\}_{k=1}^K  \mid  \{\vec{x}, y_\mu\}_{\mu}^{\nsamples})  &=   \frac{1}{{\mZ}_\ndim}\prod_{i=1}^\ndim  \rP_\w(\{w_{ik}\}_{k=1}^K) \label{posterior-measure} \\
 & \times \prod_{\mu=1}^\nsamples  \rP_{\out}\Big(y_\mu\Big|\Big\{\frac{1}{\sqrt \ndim}\sum_{i=1}^\ndim x_{\mu i} w_{i k}\Big\}_{k=1}^K\Big)\,,\nonumber
\end{align}
where the normalization factor is nothing else than a \emph{partition function}, \ie the integral of the numerator over $\{w_{ik}\}_{i,l=1}^{\ndim, K}$. The symbol $\mathbb{E}$ will generally denote an expectation over all random variables in the ensuing expression (here $\{\mat{X}, \vec{y}\}$). Subscripts will be used 
only when we take partial expectations or if there is an ambiguity. The expected free entropy is by definition
\begin{align}
	 \Phi_\ndim \equiv  \frac1\ndim \mathbb{E}\ln {\mZ}_\ndim \,. \label{freeent}
\end{align}
The replica formula gives an explicit (conjectural) expression of
$\Phi_\ndim$ in the high-dimensional limit $\ndim,\nsamples\to \infty$ with $\alpha = \nsamples /\ndim $
fixed. We show in \App\newline
\ref{appendix:replica_computation:committee} how the heuristic replica
method \cite{mezard1987spin,mezard2009information} yields the
formula. This computation was first performed, to the best of our knowledge, by \cite{schwarze1993learning} in the case of the committee machine. 
Our first contribution is a rigorous proof of the corresponding free entropy formula using an interpolation method \cite{Guerra2003,talagrand2003spin,BarbierM17a}, under a technical assumption, see \Sec~5.3 of \cite{Aubin2018}. 

In order to formulate our results, we add an arbitrarily small Gaussian regularization noise $z_\mu\sqrt{\Delta}$ to the first expression of the model \eqref{model}, where $\Delta>0$, $z_\mu\sim\mN(0,1)$, which thus becomes
\begin{align}
\label{modelnoise}
y_\mu = \varphi_\out \(\Big\{\frac{1}{\sqrt \ndim }\sum_{i=1}^\ndim x_{\mu i} w_{i k}^\star \Big\}_{k=1}^K, a_\mu \)+ z_\mu\sqrt{\Delta}\,,	
\end{align}
so that the channel kernel is for $\vec{u} \in \mathbb{R}^K$,
\begin{align}\label{new-kernel}
 \rP_{\out}(y | \vec{u}) = \frac{1}{\sqrt{2\pi\Delta}} \int_{\mathbb{R}} \d  \rP_a(a) e^{-\frac{1}{2\Delta}(y -\varphi_\out(\vec{u}, a))^2}\,.
\end{align}
Let us define the \aclink{RS} \emph{potential} as 
\begin{align}\label{RSpot}
\Phi^{(\rs)}(\mat{q}, \hat{\mat{q}}; \brho^\star) \equiv -\frac{1}{2} \tr{\hat{\mat{q}} \mat{q}} + \Psi_{\w}(\hat{\mat{q}}) +\alpha \Psi_{\out}(\mat{q};\brho^\star) \,,
\end{align}
where $\alpha\equiv \nsamples / \ndim $, and $\Psi_{\out}(\mat{q} ;\brho^\star)$ and $\Psi_{\w}(\hat{\mat{q}})$
are the free entropies of the two simpler $K$-dimensional estimation problems \eqref{aux-model-1} and \eqref{aux-model-2}. Notice that the expression is obtained from the replica computation in \eqref{appendix:free_entropy_bayes}.\\

All along this chapter, we assume the following hypotheses for our rigorous statements:
\begin{enumerate}
	\item \label{hyp:1} The prior $\rP_\w$ has bounded support in $\mathbb{R}^K$.
	\item \label{hyp:2} The activation $\varphi_\out: \mathbb{R}^K \times\mathbb{R} \to \mathbb{R}$ is a bounded $\cC^2$ function with bounded first and second derivatives with respect to its first argument, in $\mathbb{R}^K$-space.
	\item \label{hyp:3} For all $\mu \in \lb \nsamples \rb$ and $i\in\lb\ndim\rb$ we have \aclink{i.i.d} $x_{\mu i} \sim {\cal N}(0,1)$. 
\end{enumerate}

We finally rely on a technical hypothesis, stated as Assumption~1 in \Sec~5.3 of \cite{Aubin2018}.
\begin{thm}[Replica formula]\label{main-thm}
 Suppose \ref{hyp:1}, \ref{hyp:2} and \ref{hyp:3}, and Assumption~1. Then for the model
 \eqref{modelnoise} with kernel \eqref{new-kernel} the limit of the free entropy is:
\begin{align}\label{repl-1}
\Phi_\rs \equiv \lim_{\ndim\to \infty}\Phi_\ndim \equiv \lim_{\ndim \to \infty}\frac1\ndim \mathbb{E}\log {\mZ}_\ndim = {\adjustlimits \sup_{\hat{\mat{q}} \in \mS^{+}_K} \inf_{\mat{q} \in\mS_K^+(\rho)}} \Phi^{(\rs)}(\mat{q}, \hat{\mat{q}}; \brho^\star )\,.
\end{align}
\end{thm}
This theorem extends the recent progress for generalized linear models of \cite{barbier2017phase}, which includes the case $K=1$ of the present contribution, to the phenomenologically richer case of two-layers problems such as the committee machine. The proof sketch based on an \emph{adaptive interpolation method} recently developed in \cite{BarbierM17a} is outlined in \Sec 5 of \cite{Aubin2018} and the details can be found in the corresponding \Sec A. 
\begin{remark}[Relaxing the hypotheses]
Note that, following similar approximation arguments as in \cite{barbier2017phase}, the hypothesis \ref{hyp:1} can be relaxed to the existence of the second moment of the prior; thus covering the Gaussian case, \ref{hyp:2} can be dropped and thus include model \eqref{model:com} and its $\sign(\cdot)$ activation and \ref{hyp:3} extended to data matrices $\mat{X}$ with \aclink{i.i.d} entries of zero mean, unit variance and finite third moment. Moreover, the case $\Delta=0$ can be considered when the outputs are discrete, as in the committee machine \eqref{model:com}, see \cite{barbier2017phase}. The channel kernel becomes in this case $\rP_{\mathrm{out}}(y |\vec{u}) =  \int \d \rP_a(a) \id\[ y -\varphi_{\mathrm{out}}(\vec{u}, a) \]$ and the replica formula is the limit $\Delta\to0$ of the one provided in Theorem~\ref{main-thm}. In general this regularizing noise is needed for the free entropy limit to exist.	
\end{remark}

\subsection{Learning the teacher weights and optimal generalization error} 
A classical result in Bayesian estimation is that the estimator $\hat{\mat{W}}$ that minimizes the mean-square error with the ground-truth $\mat{W}^\star$ is given by the expected mean of the posterior distribution. Denoting $\mat{q}^\star $ the extremizer in the replica formula (\ref{repl-1}), we expect from the replica method that in the limit $\ndim \to\infty,~ \nsamples / \ndim =\alpha$, and with high probability, $\hat{\mat{W}}^\intercal  \mat{W}^\star /\ndim  \to  \mat{q}^\star $. We refer to proposition~5.3 and to the proof in \Sec A of \cite{Aubin2018} for the precise statement, that remains rigorously valid \emph{only} in the presence of an additional (possibly infinitesimal) side-information. This condition is similar to the \emph{small magnetic field} used to select a given Gibbs state in the Ising model in statistical physics. 
From the overlap matrix $\mat{q}^\star$, one can compute the Bayes-optimal generalization error when the student tries to classify a new, yet unseen, sample $\vec{x}_{\mathrm{new}}\in\bbR^{1 \times \ndim}$. The estimator of the new label $\hat{y}_{\mathrm{new}}$ that minimizes the mean-square error with the true label is given by computing the posterior mean of $\varphi_\out(\vec{x}_{\mathrm{new}}  \mat{W})$ ($\vec{x}_{\mathrm{new}}$ is a row vector). Given the new sample, the optimal generalization error is then 
\begin{align}
 \frac{1}{2} \EE_{\mat{X},\mat{W}^\star} \[\(\EE_{\mat{w}|\mat{X},\vec{y}} \big[ \varphi_\out(\vec{x}_{\mathrm{new}}  \mat{W} )\big] - \varphi_\out(\vec{x}_{\mathrm{new}} \mat{W}^\star )\)^2\]\xrightarrow[\ndim \to \infty]{}\epsilon_g(\mat{q}^\star),
\end{align}
where $\mat{W}$ is distributed according to the posterior measure \eqref{posterior-measure}. Note that this Bayes-optimal computation differs from the so-called Gibbs estimator by a factor $2$.
Indeed, one can naturally define the \emph{Gibbs generalization error} as:
\begin{align}\label{eq:Gibbs_gen_error}
\epsilon_g^{\gibbs} &\equiv \frac{1}{2} \EE_{\mat{W}^\star,\mat{X}} \big\langle\[\varphi_{\out}\(\vec{x} \mat{W}\) - \varphi_{\out}\(\vec{x} \mat{W}^\star \) \]^2\big\rangle,
\end{align}
and define the \emph{Bayes-optimal generalization error 
} as:
\begin{align}\label{eq:def_eg_Bayes}
\epsilon_g^{\textrm{bayes}} &\equiv \frac{1}{2} \EE_{\mat{W}^\star, \mat{X}} \big[\big(\braket{\varphi_{\out}\(\vec{x} \mat{W}\)} - \varphi_{\out}\(\vec{x} \mat{W}^\star \)\big)^2 \big].
\end{align}
Using the Nishimori identity \ref{prop:nishimori}, one obtains:
\begin{align*}
\epsilon_g^{\textrm{bayes}} &= \frac{1}{2} \EE_{\mat{X},\mat{W}^\star} \[ \varphi_{\out}\(\vec{x} \mat{W}^\star \)^2\] + \frac{1}{2} \EE_{\mat{X},\mat{W}^\star} \[ \braket{ \varphi_{\out}\(\vec{x} \mat{W} \)}^2\] \nonumber\\
&\hspace{1cm} - \EE_{\mat{X},\mat{W}^\star} \braket{ \varphi_{\out}\(\vec{x} \mat{W}^\star \) \varphi_{\out}\(\vec{x} \mat{W} \)} , \\
&= \frac{1}{2} \EE_{\mat{X},\mat{W}^\star} \[ \varphi_{\out}\(\vec{x} \mat{W}^\star \)^2\]  \\
&\hspace{1cm} - \frac{1}{2} \EE_{\mat{X},\mat{W}^\star} \braket{ \varphi_{\out}\(\vec{x} \mat{W}^\star \) \varphi_{\out}\(\vec{x} \mat{W} \)}\,.
\end{align*}
Using again the Nishimori identity one can write:
\begin{align*}
\epsilon_g^{\gibbs} &= \EE_{\mat{X},\mat{W}^\star} \[ \varphi_{\out}\(\vec{x} \mat{W}^\star \)^2\]  \\
&\hspace{1cm} -  \EE_{\mat{X},\mat{W}^\star}\braket{ \varphi_{\out}\(\vec{x} \mat{W}^\star \) \varphi_{\out}\(\vec{x} \mat{W} \)},
\end{align*}
which shows that $\epsilon_g^{\gibbs} = 2 \epsilon_g^{\textrm{bayes}}$. Note finally that since the distribution of $\mat{X}$ is rotationally invariant, the quantity $\EE_{\mat{X}} \[\varphi_{\out}\(\vec{x} \mat{W}^\star \) \varphi_{\out}\(\vec{x} \mat{W} \)\]$ only depends on the \emph{overlap} $\mat{q} \equiv \mat{W}^\intercal \mat{W}^\star$. As the overlap is shown to concentrate under the Gibbs measure, and as we expect that the value it concentrates on is the optimum $\mat{q}^\star$ of the replica formula (such fact is proven, \eg, for random linear estimation problems in \cite{barbier_ieee_replicaCS}), the generalization error can itself be evaluated as a function of $\mat{q}^\star$. Examples where it is done include  \cite{Kinzel96,seung1992statistical,schwarze1993learning,barbier2017phase}.\\

In particular, when the data $\mat{X}$ is drawn from the standard Gaussian distribution on $\bbR^{\nsamples \times \ndim}$, and is thus rotationally invariant, it follows that this error only depends on $ \mat{W}^\intercal  \mat{W}^\star/ \ndim $, which converges to $\mat{q}^\star $. Then a direct algebraic computation gives a lengthy but explicit formula for $\epsilon_g(\mat{q}^\star)$ presented below.

\subsubsection{The generalization error at $K = 2$}
From the definition of the generalization error, one can directly give an explicit expression of this error in the $K = 2$ case. Recall our committee-symmetric assumption on the overlap matrix, which here reads 
\begin{align*}
\mat{q} &= \begin{pmatrix}
q_d + \frac{q_a}{2} & \frac{q_a}{2} \\
 \frac{q_a}{2} & q_d + \frac{q_a}{2} 
\end{pmatrix}.
\end{align*} 
For concision, we denote here $\sign(x) = \sigma(x)$. One obtains from \eqref{eq:def_eg_Bayes}:
\begin{align}
	&\frac{1}{2} -2 \epsilon_g^{\textrm{bayes},K=2} \nonumber \\
	&\equiv \EE \int \D \vec{x} ~ \sigma \( \sigma\( \frac{1}{\sqrt{\ndim}} \vec{x} \cdot \vec{w}_1^\star\) +  \sigma\( \frac{1}{\sqrt{\ndim}} \vec{x}  \cdot \vec{w}_2^\star \) \) \nonumber \\
	& \qquad \times \sigma\( \sigma\(\frac{1}{\sqrt{\ndim}}\vec{x} \cdot \vec{w}_1 \) +  \sigma\( \frac{1}{\sqrt{\ndim}}\vec{x} \cdot \vec{w}_2\)\)  \nonumber \\
	&= \EE \frac{1}{(2\pi)^4} \int_{\bbR^4} \d \vec{x}  \sigma \( \sigma(x_1) + \sigma(x_2) \) \sigma \( \sigma(x_3) + \sigma(x_4) \) \nonumber \\
	& \qquad \times \int_{\bbR^4} \d \hat{\vec{x}} ~ e^{i \hat{\vec{x}}^\intercal \vec{x}}  \int \D \vec{x} ~ e^{-\frac{i}{\sqrt{\ndim}} \hat{\vec{x}}^\intercal \td{\mat{W}}^\intercal \vec{x} } \\
	&= \EE \frac{1}{(2\pi)^4} \int_{\bbR^4} \d \vec{x} ~ \sigma \( \sigma(x_1) + \sigma(x_2) \) \sigma \( \sigma(x_3) + \sigma(x_4) \) \nonumber \\  
	& \qquad \times \int_{\bbR^4} \d \hat{\vec{x}} ~  e^{i \hat{\vec{x}}^\intercal \vec{x}}  e^{-\frac{1}{2} \hat{\vec{x}}^\intercal \bSigma \hat{\vec{x}}} \nonumber \\
	&= \int_{\bbR^4} \d \vec{x} ~ \sigma \( \sigma(x_1) + \sigma(x_2) \) \sigma \( \sigma(x_3) + \sigma(x_4) \) \mN_{\vec{x}} ( \vec{0} , \bSigma ) \nonumber
\end{align}
where $\td{\mat{W}} = \( \vec{w}_1^\star, \vec{w}_2^\star, \vec{w}_1, \vec{w}_2 \) \in\bbR^{\ndim \times K}$ with $ \bSigma = \frac{1}{\ndim} \EE \td{\mat{W}}^\intercal \td{\mat{W}} \underlim{\ndim}{\infty} \scalemath{0.6}{\begin{bmatrix} \mat{I}_2 & \mat{q}\\
	\mat{q} & \mat{I}_2
\end{bmatrix}}$.
This expression can be reformulated also as 
\begin{align}
\frac{1}{2}& -2 \epsilon_g^{\textrm{bayes},K=2} = \int_{\bbR^4} \D \vec{x} \, \sigma\[\sigma(x_1) + \sigma(x_2) \] \nonumber \\
& \times \sigma\left\{ \sigma\[(\frac{q_a}{2} + q_d)x_1  +  \frac{q_a}{2} x_2 + x_3 \sqrt{1-\frac{q_a^2}{2} - q_a q_d - q_d^2}\] \right. \nonumber \\
& \left. \quad + \sigma\[\frac{q_a}{2} x_1 +(\frac{q_a}{2} + q_d)x_2  - x_3 \frac{q_a (q_d + \frac{q_a}{2})}{\sqrt{1-\frac{q_a^2}{2} - q_a q_d - q_d^2}} \right. \right.  \nonumber \\
& \left. \left. \quad \quad  +x_4 \sqrt{\frac{(1-q_d^2)(1-(q_a+q_d)^2)}{1-\frac{q_a^2}{2} - q_a q_d - q_d^2}}\] \right\}\,.
\end{align}
Note that one could possibly simplify this expression by using an appropriate orthogonal transformation on $\vec{x}$. These integrals were then computed using \aclink{MC} methods to obtain the generalization error in the left and middle plots of Fig.~\ref{fig:phaseDiagramK2}.

\subsubsection{The generalization error at large $K$}
Recall the definition of the generalization error in \eqref{eq:def_eg_Bayes}, one can compute it
using \eqref{eq:Gibbs_gen_error} after a tedious, yet straightforward, calculation:
\begin{align}
\epsilon_g^{\textrm{bayes}} = \frac{1}{2}\epsilon_g^{\gibbs}  = \frac{1}{\pi} \arccos \[\frac{2}{\pi} \(q_a + \arcsin q_d\)\] + \Theta(K^{-1}).
\end{align}
This expression is the one used in the computation of the generalization error in the left panel of Fig.~\ref{fig:phaseDiagramKlarge}.

\subsection{Approximate message passing and its state evolution} 
Our next result is based on an adaptation of a popular algorithm to solve random instances of generalized linear models, the \aclink{AMP} algorithm \cite{donoho2009message,rangan2011generalized}, for the case of the committee machine and models described by \eqref{model}. 
The \aclink{AMP} algorithm can be obtained as a Taylor expansion of loopy belief-propagation, see \App\ref{appendix:amp:committee} for the derivation, and also originates in earlier statistical physics works \cite{thouless1977solution,mezard1989space,opper1996mean,kabashima2008inference,Baldassi26062007,zdeborova2016statistical}. It is conjectured to perform the best among all polynomial algorithms in the framework of these models. It thus gives us a tool to evaluate both the intrinsic algorithmic hardness of the learning and the performance of existing algorithms with respect to the optimal one in this model. 
			\begin{algorithm} 
			\begin{algorithmic}
			    \STATE {\bfseries Input:} vector $\vec{y} \in \bbR^\nsamples$ and matrix $\mat{X}\in \bbR^{\nsamples \times \ndim}$:
			    \STATE \emph{Initialize}: $\hat{\vec{w}}_i$, $\vec{f}_{\out,\mu} \in \bbR^K$ and $\hat{\mat{C}}_i, \hat{\mat{V}}_i$, $\partial_{\bomega} \vec{f}_{\out, \mu} \in \bbR^{K\times K}$ for $ 1 \leq i \leq \ndim $ and $ 1 \leq \mu \leq \nsamples $ at $t=0$.
			    \REPEAT   
			    \STATE \noindent Channel: Update the mean $\omega_{\mu} \in \bbR^K$ and variance $V_{\mu}\in \bbR^{K\times K}$: \spacecase
			    \indent $\mat{V}_{\mu}^t = \sum\limits_{i=1}^\ndim  \frac{x_{\mu j}^2}{\ndim}  \hat{\mat{C}}_{i}^t $\\ 
			    \indent $\bomega_{\mu}^t = \sum\limits_{i = 1}^{\ndim} \frac{x_{\mu i}}{\sqrt{\ndim}} \hat{\vec{w}}_{i}^t -   \mat{V}_{\mu}^t \vec{f}_{\out,\mu}^{t-1}$\,, \\
			    \STATE \noindent Update $\vec{f}_{\out, \mu}$ and $\partial_\bomega \vec{f}_{\out,\mu}$: \spacecase
			    $\vec{f}_{\out,\mu}^t = \vec{f}_\out \( y_{\mu}, \bomega_{\mu}^t, \mat{V}_{\mu}^t \)$\,, $ \partial_\bomega \vec{f}_{\out^,\mu}^t = \partial_\bomega\vec{f}_\out \( y_{\mu}, \bomega_{\mu}^t, \mat{V}_{\mu}^t \)$ \spacecase
			    \STATE \noindent Prior: Update the mean $\bgamma_i \in \bbR^K$ and variance $\bLambda_i \in \bbR^{K\times K}$:\spacecase
			    $ \bLambda_{i}^t =  - \sum\limits_{\mu =1}^\nsamples \frac{x_{\mu i}^2}{\ndim}  \partial_\bomega \vec{f}_{\out, \mu}^t $\spacecase
			    $\bgamma_i^t = \sum\limits_{\mu =1}^\nsamples    \frac{x_{\mu i}}{\sqrt{\ndim}} \vec{f}_{\out,\mu}^t + \bLambda_{i}^t \hat{\vec{w}}_{i}^t $\,,
			    \STATE Update the estimated marginals $\hat{w}_i \in \bbR$ and $\hat{c}_i \in \bbR^+$: \spacecase
			   $\hat{\vec{w}}_{i}^{t+1}= \vec{f}_\w\( \bgamma_{i}^t, \bLambda_{i}^t \)$\,, $\hat{\mat{C}}_{i}^{t+1}= \partial_\bgamma  \vec{f}_\w\(\bgamma_{i}^t, \bLambda_{i}^t \)$\spacecase
			    \STATE ${t} \leftarrow {t} + 1$ 
			    \UNTIL{Convergence on
			    $\hat{\vec{w}}_i$, $\hat{\mat{C}}_i$.} 
			    \STATE {\bfseries Output:}
			    $\{\hat{\vec{w}}\}_{i=1}^\ndim$ and $\{\hat{\mat{C}}_i\}_{i=1}^\ndim$.
			\end{algorithmic}
			\caption{Approximate Message Passing for the committee machine.\label{alg:AMP}}  
			\end{algorithm}
			
The \aclink{AMP} algorithm is summarized by its pseudo-code in \Alg\ref{alg:AMP}, where the update
functions $\vec{f}_{\out}$, $\partial_{\bomega} \vec{f}_{\out}$, $\vec{f}_\w$ and
$\partial_\bgamma \vec{f}_\w$ are related, again, to the two auxiliary problems
(\ref{aux-model-1}) and (\ref{aux-model-2}) and defined in \App\ref{appendix:update_functions}.
The functions $\vec{f}_\w(\bgamma, \bLambda)$ and $\partial_\bgamma \vec{f}_\w(\bgamma, \bLambda)$ are respectively the mean and variance under the posterior distribution \eqref{aux-model-1} when $\hat{\mat{q}} \to \bLambda$ and $\vec{y} \to  \bLambda^{-1/2} \bgamma
$, while $\vec{f}_{\out}(y_\mu, \bomega_{\mu}, \mat{V}_{\mu})$ is given by the product of $\mat{V}_{\mu}^{-1/2}$ and the mean of $\vec{u}$ under the posterior \eqref{aux-model-2} using
$\td{y}_0 \to y_{\mu} $, $\brho^\star-\mat{q} \to \mat{V}_{\mu}$ and $\mat{q}^{1/2}\vec{v} \to \bomega_{\mu}$. 

After convergence, $\hat{\mat{W}}$ estimates the weights of the teacher-neural
network. The label of a sample $\vec{x}_{\textrm{new}}$ not seen in the training
set is estimated by the \aclink{AMP} algorithm as 
\begin{align}
      y^t_{\textrm{new}} = \int_\bbR \d y ~ \int_{\bbR^K} \d \vec{z} ~ y ~
  \rP_{\out}(y | \vec{z}) \mN_{\vec{z}}(\bomega_{\textrm{new}}^t ,
  \mat{V}_{\textrm{new}}^t)\, ,    \label{AMP_gen}
\end{align}
where $\bomega_{\textrm{new}}^t = \sum_{i=1}^\ndim x_{{\textrm{new}},  i} \hat{\vec{w}}_i^t$ is
the mean of the normally distributed variable $\vec{z}\in {\mathbb R}^K$,
and $\mat{V}_{\textrm{new}}^t=\brho^\star-\mat{q}_{\textrm{amp}}^t$ is the $K\times K$ covariance
matrix (see below for the definition of $\mat{q}_{\textrm{amp}}^t$). We provide a demonstration code of the algorithm on \href{https://github.com/benjaminaubin/TheCommitteeMachine}{GitHub}  \cite{GitHub_AMP_Aubin}.

\aclink{AMP} is particularly interesting because its performance can be tracked rigorously, again in the asymptotic limit when $\ndim \to
\infty$, via a procedure known as \aclink{SE}, which is a rigorous version
of the cavity method in physics \cite{mezard2009information}, see \cite{javanmard2013state}. \aclink{SE} tracks the value of the overlap between
the hidden ground truth $\mat{W}^*$ and the \aclink{AMP} estimate $\hat{\mat{W}}^t$, defined as $\mat{q}_{\textrm{amp}}^t\equiv\lim_{\ndim\to\infty}({\hat{\mat{W}}^t})^\intercal   \mat{W}^\star/\ndim $, via the iteration of the following equations:
\begin{equation}
\label{main:StateEvolution}
	\mat{q}_{\textrm{amp}}^{t+1} = 2 \nabla\Psi_{\w}(\hat{\mat{q}}_{\textrm{amp}}^t) \, ,	 \hspace{1cm}
	\hat{\mat{q}}_{\textrm{amp}}^{t+1} = 2 \alpha \nabla \Psi_{\out}(\mat{q}_{\textrm{amp}}^t;\brho^\star)\, .
\end{equation}
See sec.~G of \cite{Aubin2018} for more details and note that the fixed points of these equations correspond surprisingly to the critical points of the replica free entropy (\ref{repl-1}).
Let us comment further on the convergence of the
  algorithm. In the large $\ndim$ limit, and if the integrals are
  performed without errors, then the algorithm is guaranteed to converge. This is a consequence of the \aclink{SE} combined with
the Bayes-optimal setting. In practice, of course, $\ndim$ is finite and integrals are
approximated. In that case convergence is not guaranteed, but is
robustly achieved in all the cases presented in this paper.  
We also expect, by experience with the single layer case, that if the
input-data matrix is not random, which is beyond our assumptions, then we
will encounter convergence issues, which could be fixed by moving to
some variant of the algorithm such as \aclink{VAMP} \cite{rangan2019vector}.

\section{From two to more hidden neurons and the specialization phase transition}
\sectionmark{The specialization phase transition}
\subsection{Two neurons committee machine $K=2$} 

\begin{figure}[t]
\centering
\includegraphics[width=1.0\linewidth]{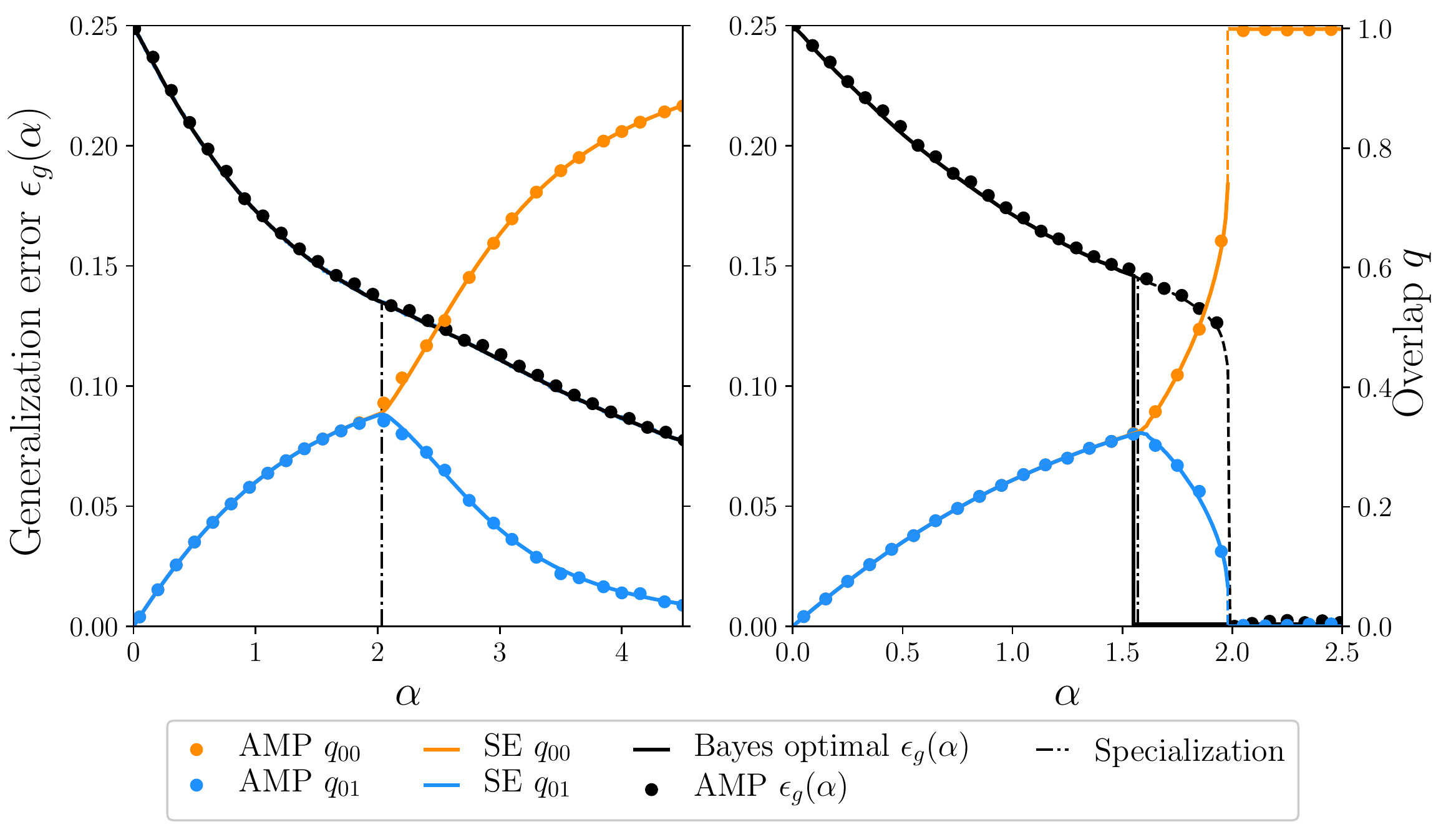}	
 \caption{Generalization error and order parameter for a committee
   machine with two hidden neurons ($K=2$) with \Left Gaussian weights, \Right
   binary/Rademacher weights. These are shown as a function
   of the ratio $\alpha=\nsamples /\ndim$ between the number of samples $\nsamples$ and the
   dimensionality $\ndim$. Lines are obtained from the state evolution (SE)
   equations (dominating solution is shown in full line), data-points from the AMP algorithm averaged over 10
   instances of the problem of size $\ndim =10^4$.
   $q_{00}$ and $q_{01}$ denote diagonal and off-diagonal overlaps of the matrices $\mat{q}^\star$ and $\mat{q}_{\textrm{amp}}$,
   and their values are given by the labels on the far-right of the figure.
   }
\label{fig:phaseDiagramK2}
\vspace{-0.5cm}
\end{figure}

\begin{figure}[t]
\centering
\includegraphics[width=1.0\linewidth]{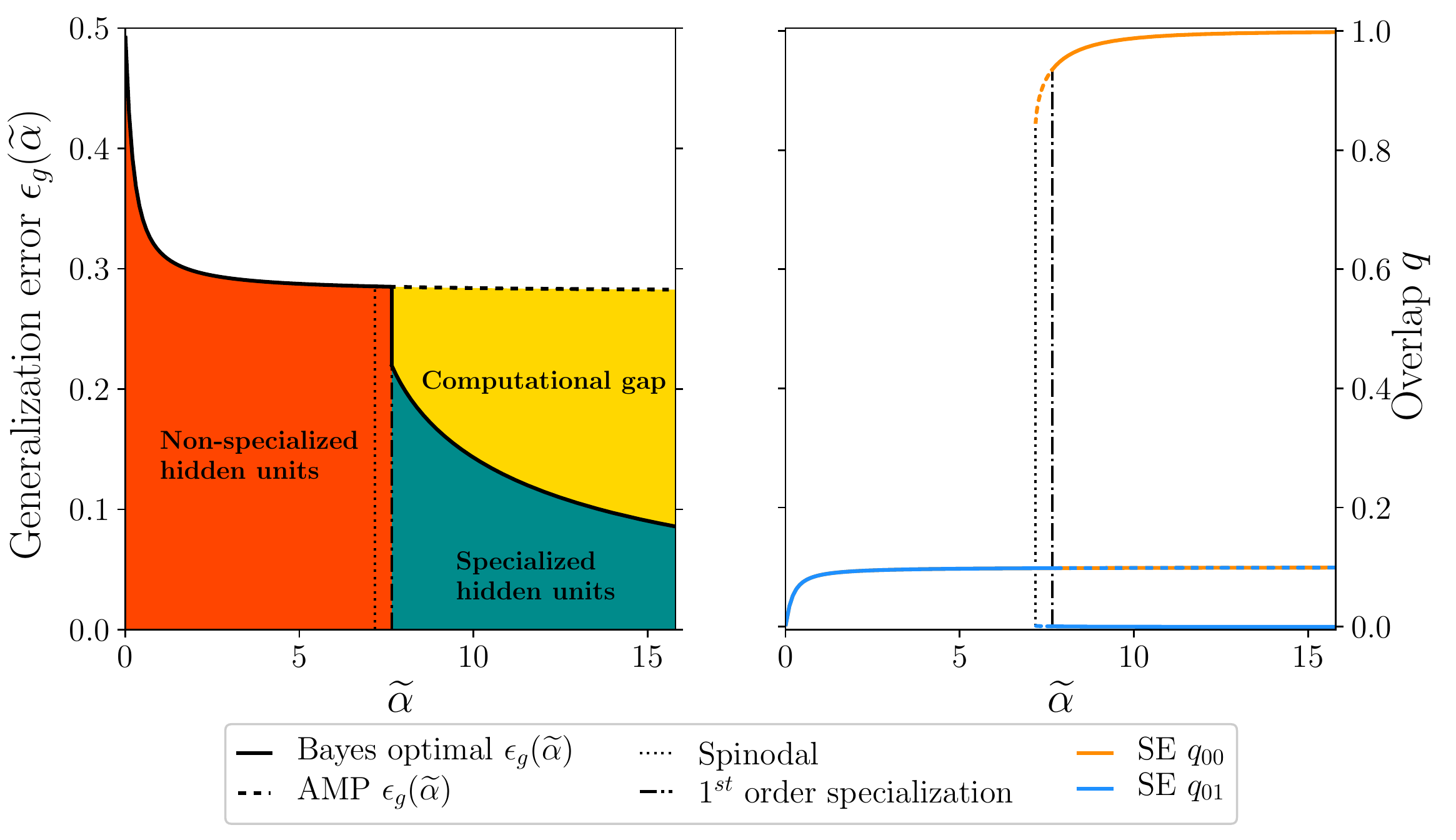}
 \caption{\Left Bayes optimal and AMP generalization errors and \Right diagonal and off-diagonal overlaps $q_{00}$, $q_{01}$ for a committee
   machine with a large number of hidden neurons $K$ and Gaussian weights, as a function
   of the rescaled parameter $\tilde{\alpha}=\alpha/K$. Solutions corresponding to global and local minima of the replica free entropy are respectively represented with full and dashed lines. The dotted line marks the spinodal at $\widetilde \alpha^G_{\textrm{spinodal}}\simeq 7.17$, \ie the apparition of a local minimum in the replica free entropy, associated to a solution with specialized hidden units. The dotted-dashed line shows the first order specialization transition at $\widetilde \alpha^G_{\textrm{spec}} \simeq 7.65$, at which the specialized fixed point becomes the global minimum. For $\widetilde \alpha < \widetilde \alpha^G_{\textrm{spec}}$, AMP reaches the Bayes optimal generalization error and overlaps, corresponding to a non-specialized solution (red area). However, for $\widetilde \alpha > \widetilde \alpha^G_{\textrm{spec}}$, the AMP algorithm does not follow the optimal specialized solution (green area) and is stuck in the non-specialized solution plateau, represented with dashed lines. Hence it unveils a large computational gap (yellow area).}
\label{fig:phaseDiagramKlarge}
\vspace{-0.5cm}
\end{figure}
Let us now discuss how the above results can be used to
study the optimal learning in the simplest non-trivial case of a
two-layers neural network with two hidden neurons, that is when model (\ref{model:com}) is simply
$$y_{\mu}={\sign}\Big[{\sign} \Big( \sum_{i=1}^\ndim x_{\mu i}
w^\star_{i1}\Big)+{\sign} \Big( \sum_{i=1}^\ndim x_{\mu i} w^\star_{i2}\Big)\Big]\,,$$ and is represented in Fig.~\ref{fig:committee}, with the
convention that ${\sign}(0)=0$. We remind that the
input-data matrix $\mat{X}$ has \aclink{i.i.d} $\mN(0,1)$ entries, and the
teacher-weights $\mat{W}^\star$ used to generate the labels $\vec{y}$ are taken \aclink{i.i.d} from $\rP_\w$.
In Fig.~\ref{fig:phaseDiagramK2} we plot the optimal generalization error as a
function of the sample complexity $\alpha=\nsamples / \ndim$. In the left panel the weights are Gaussian
(for both the teacher and the student), while in the right panel they are
binary/Rademacher. The full line is obtained from the fixed point of
the \aclink{SE} of the
\aclink{AMP} algorithm \eqref{main:StateEvolution}, corresponding to the extremizer of the replica free entropy \eqref{repl-1}. The points are results of the \aclink{AMP} algorithm run till convergence averaged
over 10 instances of size $\ndim=10^4$. 
{\color{black} In this case and with random initial conditions the \aclink{AMP} algorithm did converge in all our trials.}
As expected we observe excellent agreement
between the \aclink{SE} and \aclink{AMP}.

In both left and right panels of Fig.~\ref{fig:phaseDiagramK2} we
observe the so-called \emph{specialization} phase transition. 
Indeed \eqref{main:StateEvolution} has two types of fixed points: a {\emph
  non-specialized} fixed point where every matrix element of the $K\times K$
order parameter $\mat{q}$ is the same (so that both hidden neurons learn the
same function) and a \emph{specialized} fixed point where the diagonal
elements of the order parameter are different from the non-diagonal
ones. We checked for other types of fixed points for $K=2$ (one where
the two diagonal elements are not the same), but have not found
any. In terms of weight-learning, this means for the non-specialized fixed point
that the estimators for both $\vec{w}_{1}$ and $\vec{w}_{2}$, with $\hat{\mat{W}}=(\vec{w}_{1},\vec{w}_{2})$ are the same,
whereas in the specialized fixed point the estimators of the weights
corresponding to the two hidden neurons are different, and that the
network ``figured out'' that the data are better described by a model that 
is not linearly
separable. The specialized fixed point is associated
with lower error than the non-specialized one (as one can see in
Fig.~\ref{fig:phaseDiagramK2}). The existence of this phase transition
was discussed in statistical physics literature on the
committee machine, see
e.g. \cite{schwarze1992generalization,saad1995line}.

For Gaussian weights (Fig.~\ref{fig:phaseDiagramK2} left), the
specialization phase transition arises continuously at
$\alpha^G_{\textrm{spec}}(K=2)\simeq 2.04$. This means that for
$\alpha<\alpha^G_{\textrm{spec}}(K=2)$ the number of samples is too small,
and the student-neural network is not able to learn that two different
teacher-vectors $\vec{w}_1^\star$ and $\vec{w}_2^\star$ were used to generate the observed
labels. For $\alpha>\alpha^G_{\textrm{spec}}(K=2)$, however, it  is able to
distinguish the two different weight-vectors and the generalization
error decreases fast to low values (see
Fig.~\ref{fig:phaseDiagramK2}). For completeness we remind that in the
case of $K=1$ corresponding to single-layer neural network no such
specialization transition exists. We show in sec.~E of \cite{Aubin2018} that it is absent also in multi-layer neural networks as long as the activations remain linear. The
non-linearity of the activation function is therefore an essential
ingredient in order to observe a specialization phase transition.

The right part of Fig.~\ref{fig:phaseDiagramK2} depicts the fixed
point reached by the state evolution of \aclink{AMP} for the case of binary
weights. We observe two phase transitions in the performance of \aclink{AMP} in
this case: (a) the specialization phase transition at
$\alpha_{\textrm{spec}}^{B}(K=2) \simeq 1.58$, and for slightly larger
sample complexity a transition towards \emph{perfect generalization}
(beyond which the generalization error is asymptotically zero) at
$\alpha^B_{\textrm{perf}}(K=2)\simeq 1.99$. The binary case with $K=2$
differs from the Gaussian one in the fact that perfect generalization
is achievable at finite $\alpha$. While the specialization transition
is continuous here, the error has a discontinuity at the 
transition of perfect generalization. This discontinuity is associated
with the 1st order phase transition, in the physics nomenclature, leading
to a gap between algorithmic (\aclink{AMP} in our case) performance and
information-theoretically optimal performance reachable by exponential
algorithms. To quantify the optimal performance we need to evaluate
the global extremum of the replica free entropy (not the local one
reached by the state evolution). In doing so that we get that information
theoretically there is a single discontinuous phase transition towards
perfect generalization at $\alpha^B_{\textrm{IT}}(K=2)\simeq 1.54$.

While the information-theoretic and specialization
phase transitions were identified in the physics literature
on the committee machine
\cite{schwarze1992generalization,schwarze1993generalization,seung1992statistical,watkin1993statistical},
the gap between the information-theoretic performance and the
performance of \aclink{AMP} ---that is conjectured to be optimal
among polynomial algorithms--- was not yet discussed in the context of
this model. Indeed, even its understanding in simpler models than those
discussed here, such as the single layer case, is more recent
\cite{donoho2009message,zdeborova2016statistical,donoho2013accurate}.

\subsection{Two neurons parity machine $K=2$} 
\label{sec:parity}
Although we mainly focused on the committee machine, another classical two-layers neural network is the parity machine \cite{engel2001statistical} and our proof applies to this case as well.  While learning is known to be computationally hard for general $K$, the case $K=2$ is special, and in fact can be reformulated as a committee machine, where the sign activation function has been replaced by $f_2 (z) = \id (z\ne 0) - \id (z = 0)$: 
\begin{equation} y_\mu = {\sign}\Big[\prod_{k=1}^K {\sign} \Big( \sum_{i=1}^\ndim x_{\mu i} w_{i k}^\star \Big) \Big]\,= f_2\Big[\sum_{k=1}^K {\sign} \Big( \sum_{i=1}^\ndim x_{\mu i} w_{i k}^\star \Big) \Big]\, . 
\end{equation}

\begin{figure}[t]
\center
\includegraphics[width=1.0\linewidth]{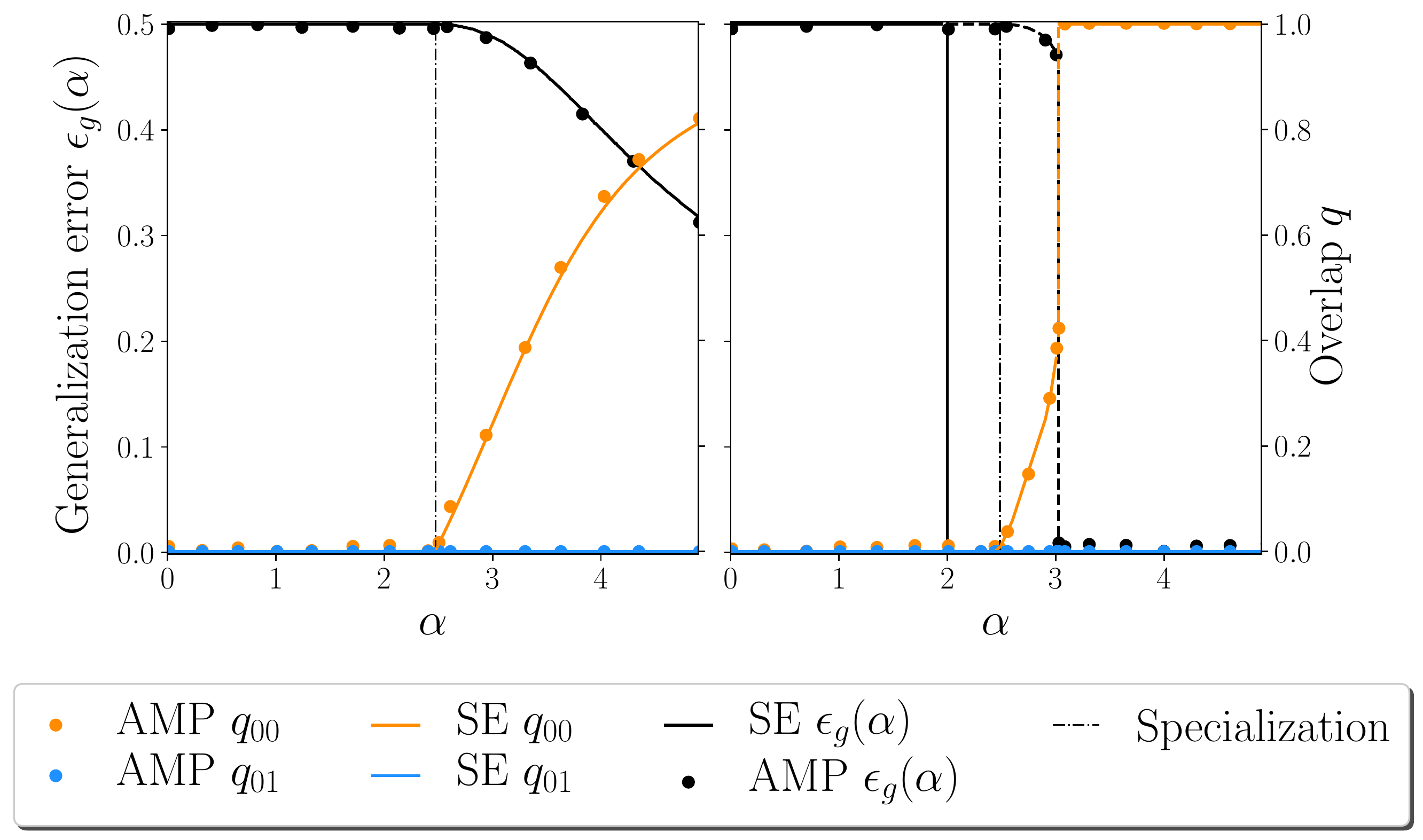}
 \caption{Similar plot as in Fig. \ref{fig:phaseDiagramK2} but for the parity machine with two hidden neurons. Value of the order parameter and the optimal generalization error for a parity
   machine with two hidden neurons with \Left Gaussian weights and \Right
   binary/Rademacher weights. SE and AMP overlaps are respectively represented in full line and points.}
\label{fig:phaseDiagramK2_parity}
\vspace{-0.5cm}
\end{figure}

We have repeated our analysis for the $K=2$ parity machine and the phase diagram is summarized in Fig.~\ref{fig:phaseDiagramK2_parity} where we show the generalization error and the elements of the overlap matrix for Gaussian \Left and binary weights \Rightn, with the results of the \aclink{AMP} algorithm (points). 
Below the specialization phase transition $\alpha < \alpha_{\textrm{spec}}$, the symmetry of the output imposes the non-specialized fixed point $q_{00}=q_{01}=0$ to be the only solution, with $\alpha_{\textrm{spec}}^G(K=2) \simeq 2.48 $ and $\alpha_{\textrm{spec}}^B(K=2)\simeq 2.49$. Above the specialization transition $\alpha_{\textrm{spec}}$, the overlap becomes specialized with a non-trivial diagonal term. 
Additionally, in the binary case, an information theoretical transition towards a perfect learning occurs at $\alpha_{\textrm{IT}}^B(K=2)\simeq 2.00$, meaning that the perfect generalization fixed point ($q_{00}=1,q_{01}=0$) becomes the global optimizer of the free entropy. It leads to a first order phase transition of the \aclink{AMP} algorithm which retrieves the perfect generalization phase only at $\alpha_{\textrm{perf}}^B(K=2)\simeq 3.03$. This is similar to what happens in single layer neural networks for the symmetric door activation function, see \cite{barbier2017phase}. Again, these results for the parity machine emphasize a gap between information-theoretical and computational performance.

\subsection{More is different $K \to \infty$} It becomes more difficult to study the
replica formula for larger values of $K$ as it involves (at least)
$K$-dimensional integrals. Quite interestingly, it is possible to
work out the solution of the replica formula in the large $K$ limit (thus taken {\emph after} the large $\ndim$ limit, so that $K/\ndim$ vanishes).
It is indeed natural to look for solutions of the replica formula, as
suggested in \cite{schwarze1993learning}, of the form $\mat{q} = q_d
\mat{I}_K + ({q_a}/{K}) \vec{1}_K \vec{1}_K^\intercal$, with the
unit vector $\vec{1}_K = (1)_{l=1}^K$. Since both $\mat{q}$ and $\brho^\star$
are assumed to be positive, this scaling implies that $0\leq q_d \leq 1$ and $0 \leq q_a +
q_d \leq 1$, as it should, see sec.~D of \cite{Aubin2018}. We also detail in this same section the corresponding large $K$ expansion of the free entropy
for the teacher-student scenario with Gaussian weights. Only the
information-theoretically reachable generalization error was computed
\cite{schwarze1993learning}, thus we concentrated on the analysis of
performance of \aclink{AMP} by tracking the \aclink{SE} equations. In doing
so, we unveil a large computational gap.

In the right panel of Fig.~\ref{fig:phaseDiagramKlarge} we show the fixed
point values of the two overlaps $q_{00} = q_d + q_a/K$ and $q_{01} = q_a/K$ and the resulting generalization error, plotted in the left panel. As discussed in \cite{schwarze1993learning} it
can be written in a closed form as $\epsilon_g= \arccos
\left[2\left(q_a + \arcsin q_d\right)/\pi\right]/\pi$, represented in the left panel of Fig.~\ref{fig:phaseDiagramKlarge}. The specialization
transition arises for $\alpha=\Theta(K)$ so we define $\widetilde
\alpha\equiv \alpha/K$. The specialization is now a first order phase
transition, meaning that the specialization fixed point first appears
at $\widetilde \alpha^G_{\textrm{spinodal}}\simeq 7.17$ but the
free entropy global extremizer remains the one of the non-specialized fixed point until
$\widetilde  \alpha^G_{\textrm{spec}}\simeq 7.65$. This has
interesting implications for the optimal generalization
error that gets towards a plateau of value $\varepsilon_{\textrm{plateau}}
\simeq 0.28$ for $\widetilde \alpha < \widetilde \alpha^G_{\textrm{spec}} $ and then
jumps discontinuously down to reach a decay aymptotically as $1.25/ \widetilde \alpha$. See left panel of Fig.~\ref{fig:phaseDiagramKlarge}.

\aclink{AMP} is conjectured to be optimal among all polynomial algorithms (in the considered limit) and thus analyzing its \aclink{SE} sheds light on possible computational-to-statistical gaps
that come hand in hand with first order phase transitions. In the
regime of $\alpha = \Theta(K)$ for large $K$ the non-specialized fixed
point is always stable implying that \aclink{AMP} will not be
able to give a lower generalization error than $\varepsilon_{\textrm{plateau}}$. Analyzing the replica formula for large $K$ in more details, see sec.~D of \cite{Aubin2018}, we concluded that \aclink{AMP} will not reach the optimal generalization for any
$\alpha < \Theta(K^2)$. This implies a rather sizable gap between the
performance that can be reached information-theoretically and the one
reachable tractably (see yellow area in Fig.~\ref{fig:phaseDiagramKlarge}). Such large computational gaps have been
previously identified in a range of inference problems ---most
famously in the planted clique problem \cite{deshpande2015finding}---
but the committee machine is the first model of a multi-layer neural
network with realistic non-linearities (the parity machine is another
example but use a very peculiar non-linearity) that presents such large gap. 

\section*{Conclusion}
In this chapter, we revisited a model for two-layer
neural network known as the committee machine in the \aclink{T-S}
scenario that allows for explicit evaluation of Bayes-optimal learning
errors. This model has been solved in early statistical physics
literature using the non-rigorous replica method. We built on recent
progress in proving the replica formulas rigorous in the Bayes-optimal
setting and extend these proof to the case of the committee machine. 

One of our contributions is the design of an \aclink{AMP}-type 
algorithm that is able to achieve the Bayes-optimal learning error in
the limit of large dimensions for a range of parameters out of the
so-called hard phase. The hard phase is associated with first order phase
transitions appearing in the solution of the model. In the case of the
committee machine with a large number of hidden neurons we identify
a large hard phase in which learning is possible
information-theoretically but not efficiently. In other problems where
such a hard phase was identified, its study boosted the development
of algorithms that are able to match the predicted threshold. We anticipate
this will also be the same for the present model. We should, however,
note that for larger $K>2$ the present \aclink{AMP} algorithm includes higher-dimensional integrals
that hamper the speed of the algorithm. Our current strategy to
tackle this is to combine the large-$K$ expansion and use it in the
algorithm. Detailed account of the corresponding results are left for
future work. 

We studied the Bayes-optimal setting where the student-network is the same as the teacher-network, for which the replica method can be
readily applied. 
The method still applies when the number of hidden units in the
student and teacher are different, while our proof does not generalize easily to this case. It is an interesting subject for future work to see how the hard phase evolves
under over-parametrization and what is the interplay between the
simplicity of the loss-landscape and the achievable generalization
error. We conjecture that in the present model over-parametrization
will not improve the generalization error achieved by \aclink{AMP} in the
Bayes-optimal case.\\  

Even though we focused on a two-layers neural network,
the analysis and algorithm can be readily extended to a multi-layer
setting, see \cite{MatoParga92}, as long as the number of layers as well as the number of
hidden neurons in each layer is held constant, and as long as one
learns only weights of the first layer, for which  the proof already applies. The numerical evaluation of the phase diagram would be more
challenging than the cases presented in this paper as multiple
integrals would appear in the corresponding formulas. 
In future works, we also plan to analyze the case where the weights of
the second and subsequent layers (including the biases of the
activation functions) are also
learned. This could be done for instance with a combination of Expectation Maximization and
\aclink{AMP} along the lines of \cite{krzakala2012probabilistic, arXiv:1207.3859} where this is done for the simpler single layer case. 

Concerning extensions of the present work, an important open case is
the one where the number of samples per dimension $\alpha = \Theta(1)$ and
also the size of the hidden layer per dimension $K/\ndim = \Theta(1)$ as $\ndim\to
\infty$, while in this paper we treated the case $K = \Theta(1)$ and $\ndim \to
\infty$. This other scaling where $K/\ndim = \Theta(1)$ is challenging
even for the non-rigorous replica method.


	\ifthenelse{\equal{\format}{oneside}}
	{
	\clearpage\null\thispagestyle{empty}\newpage
	\clearpage\null\thispagestyle{empty}\newpage
	}
	{
	\clearpage\null\thispagestyle{empty}\newpage
	\clearpage\null\thispagestyle{empty}\newpage
	\cleardoublepage}
	
	\chapter{Storage capacity in symmetric binary perceptrons}
	\chaptermark{Storage capacity in symmetric binary perceptrons}
	\label{chap:binary_perceptron}
	In this chapter, we revisit the problem of computing 
the capacity of the binary perceptron 
\cite{gardner1988optimal,krauth1989storage} for storing random
patterns. This problem lies at the core of early statistical physics 
studies of neural networks and their learning and generalization 
properties, for reviews see \eg 
\cite{watkin1993statistical,seung1992statistical,engel2001statistical,nishimori2001statistical}.
While the perceptron problem is motivated by studies of simple
artificial neural
networks as discussed in detail in the above literature, in this
paper we view it as a \aclink{rCSP} where the
vector of binary weights $\vec{w} \in \{ \pm 1\}^\ndim$ (a 
\emph{solution}) must satisfy $\nsamples$
\emph{step} constraints of the type 
\begin{equation}
           \sum_{i=1}^\ndim  x_{\mu i} w_i \ge K\, ,  \label{eq_constraints}
\end{equation}
where $\mu\in\lb \nsamples \rb$, $K\in \bbR$ is the 
\emph{threshold}, the random variables $x_{\mu i}$ are \aclink{i.i.d} 
Gaussian variables with zero mean and variance $1/\ndim$, and the rows 
of the matrix $\mat{X}\in \mathbb{R}^{\nsamples \times \ndim}$ are called patterns. We 
define an indicator function associated to the perceptron with a step 
constraint as $\varphi^s(z) = \id\[\displaystyle z \geq K\]$.

We say that a given vector $\vec{w}$ is a solution of the
perceptron instance if all $\nsamples$ constraints given by eq.~\eqref{eq_constraints}
are satisfied. The \emph{storage capacity} is then defined similarly to
the satisfiability threshold in random constraint satisfaction
problems: we denote the constraint density as $\alpha \equiv \nsamples/
\ndim $ and define the storage capacity $\alpha_c(K)$ as the 
infimum of densities $\alpha$ such that in the limit $\ndim \to \infty$, 
with high probability over the choice of the matrix $\mat{X}$ there are 
no solutions. It is natural to conjecture that the converse also holds, 
\ie the storage capacity $\alpha_c(K)$ equals the supremum of $\alpha$ 
such  that in the limit $\ndim\to \infty$ solutions exist with high 
probability. In this case we would say the storage capacity is a \emph{sharp threshold} according to the definition:
\begin{equation}
	\exists \epsilon > 0 \textrm{ / }  \forall \alpha > \alpha_c + \epsilon, \lim_{\nsamples, \ndim \to \infty} \bbP\[\nexists \vec{w} ~/~ \forall \mu\in\lb \nsamples \rb ,~  \varphi\(\vec{x}_{\mu} \cdot \vec{w} \) \] = 1 \,.
\end{equation} 

Gardner and Derrida in their paper \cite{gardner1988optimal} assume the
storage capacity $\alpha_c(K)$ is a sharp threshold and they apply the replica calculation to
compute it, but reach a result inconsistent with a simple upper bound
obtained by the first moment method. M\'ezard and Krauth \cite{krauth1989storage} found
a way to obtain a consistent prediction from the replica calculation
and concluded that the storage capacity $\alpha^s_c(K)$ for the step binary perceptron
(SBP), i.e. associated to the constraint $\varphi^s$, is given by the largest $\alpha$ for which the following quantity, the  \emph{entropy} in physics, is positive:
\begin{align}
\Phi^{(\rs)}_{s}(\alpha,K) &= \extr_{q_0,\hat{q}_0}\left\{ \frac{1}{2}\left(q_0 - 1\right)\hat{q}_0 +  \int \D \xi ~ \log \left[ 2 \cosh \left( \sqrt{\hat{q}_0 \xi}\right)   \right]\right. \nonumber \\ 
 & \left. \qquad \qquad \qquad  \hhspace  + \alpha \int \D \xi ~ \log\left[  \int^{\infty}_{\frac{K-\xi\sqrt{q_0}}{\sqrt{1-q_0}}} \D z      \right] \right\} \, , 
 \label{RS_capacity}
\end{align}
where $\D \xi = \frac{e^{-\xi^2/2}}{\sqrt{2\pi}} \d \xi$ is a normal Gaussian
measure, and $\extr$ means that the expression is evaluated where the derivatives on the curl-bracket, with
respect to $q_0\ge 0$ and $\hat q_0 \ge 0$, are zero.

Several decades of subsequent research in the statistical physics of disordered systems are consistent with the conjectured M\'ezard-Krauth formula for the storage capacity of the binary perceptron. 
Despite the simplicity of the above conjecture and decades of impressive progress in the mathematics of spin glasses and related problems, (see e.g. \cite{talagrand2006parisi,talagrand2003spin, mezard2009information,achlioptas2011solution,panchenko2014parisi,ding2015proof} and many others), the storage capacity of the binary perceptron remains an open mathematical problem. 
In fact, even the very existence of a sharp threshold, \ie the fact that in the limit $\ndim \to \infty$ the probability that patterns can be stored drops sharply from one to zero at the capacity, is an open problem. 
Up to very recently only widely non-matching upper bounds and lower
bounds for the storage capacity of the binary perceptron were available
\cite{kim1998covering,stojnic2013discrete}. As the present work was
being finalized Ding and Sun \cite{Sun2018} proved in a remarkable paper a lower
bound on the capacity that matches the
Krauth and Mezard conjecture (note that much like Theorem~\ref{thmMain} below, the main theorem in~\cite{Sun2018} depends on a numerical hypothesis).  
A matching upper bound remains an open challenge in
mathematical physics and probability theory. \\

In this chapter, we introduce two simple \emph{symmetric} variants of the binary
perceptron problem.  Let $z_\mu (\vec{w}) = \sum_{i=1}^\ndim  x_{\mu i} w_i = \vec{x}_\mu \cdot \vec{w}$. For a threshold $K\in \bbR^+$, we consider two different types of symmetric constraints: 
\begin{itemize}
	\item The rectangle binary perceptron (RBP) requires $|z_\mu|\le K, 
          \forall \mu\in\lb \nsamples \rb$. Its associated indicator function is $\varphi^r(z) = \id\[\displaystyle |z| \leq K\]$.
	\item The $u$-function binary perceptron (UBP)
          requires $|z_\mu|\ge K, \forall \mu\in\lb \nsamples \rb $. Its associated indicator function is $\varphi^u(z) = \id\[\displaystyle |z| \geq K\]$.
\end{itemize}
These constraints are symmetric in the sense that if $\vec{w}$ is a
solution then $-\vec{w}$ is a solution as well. Our motivation
behind these symmetric variants of the perceptron is that this
symmetry simplifies greatly the mathematical treatment of the problem,
while keeping the relevant physical properties intact. Thus, results
that remain open questions for the canonical perceptron can be established
rigorously for these symmetric versions. Symmetric perceptron models are also directly related to the  problem of determining the
discrepancy of a random matrix or set system 
\cite{BansalSpencer19}, a problem of interest in combinatorics.\\

Our main result, presented in \Sec\ref{section:proof}, is a proof, subject to a numerical hypothesis, of a formula 
for the storage capacity, defined in the same way as for the step-function binary
perceptron above. In particular, we show that in these
symmetric variants the first
moment upper bound, corresponding to the annealed capacity in physics,
on the storage capacity is tight (except for $K > K^* \simeq 0.817$
for the UBP case). We prove this statement using the
second moment method.
We note that the existing physics literature on perceptron-like
problem contains other cases of models where the first moment upper
bound on the storage capacity was observed to be tight, in particular the parity
machine \cite{opper1995statistical}, and the reversed-wedge binary
perceptron \cite{bex1995storage,hosaka2002statistical}. Those works,
however, rely on the comparison of the first moment bound on the capacity with the
result of the replica method, rather than providing a rigorous justification.
To formally state our main result,
let $z \sim  \mN(0,1)$, and for $K \in \bbR^{+}$ let 
$p_{r,K}= \bbP[|z| \le K]$ and $p_{u,K} = \bbP[|z| \ge K]$. 
\begin{itemize}
	\item The storage capacity for the rectangle binary perceptron is:
	\begin{equation}
 \alpha_c^{r}(K) = \frac{-\log(2)}{\log(p_{r,K})} \hspace{0.5cm}
 \forall K \in \mathbb{R}^+  \, . \label{cap_rec}
\end{equation}
 \item The storage capacity for the $u-$function binary perceptron is: 
 \begin{equation}
      \alpha_c^{u}(K)= \frac{-\log(2)}{\log(p_{u,K})} \hspace{0.5cm}  \text{for }\hspace{0.2cm} 0<K < K^* \simeq 0.817\, . \label{cap_ss}
\end{equation}
\end{itemize}
The constant $K^* \simeq 0.817 $ stems from the properties of the
second moment entropy eq.~\eqref{main:AT_second_moment}. In the physics terms it is defined as the point of intersection between the annealed
capacity $\alpha_a^{u}(K)$ and the local stability of the \aclink{RS} solution $\alpha_\textrm{AT}^{u}(K)$ eq.~\eqref{main:AT_crossover_RS}. That is,  $K^*$ is the solution of the following implicit equation:
\begin{equation}
        \pi ~ p_{u,K}^2 ~ \exp\(K^2\) ~ \log(p_{u,K}) = -2 ~ \log (2) ~ K^2\, .
         \label{AT_crossover_RS}
\end{equation}

The two symmetric variants of the perceptron problem considered here share many
of the intriguing geometric properties of the original step-function
binary perceptron problem. Most significant is the conjectured \aclink{f1RSB} \cite{krauth1989storage} nature of the
space of solutions that splits into well separated clusters of vanishing entropy at
any $\alpha>0$. Remarkably, this \aclink{f1RSB} property can  be deduced
from the form of the second moment entropy as we explain in section
\ref{section:frozen}. Our justification of the \aclink{f1RSB} property does not rely on
the replica method and is hence of independent interest. 

For the UBP and $K > K^*$, the second-moment proof technique fails, and
this failure marks tightly the onset of the replica symmetry breaking
region. In that region, we evaluate the \aclink{1RSB} approximation for the storage capacity, but conclude
that \aclink{FRSB} would be needed to
obtain the exact result. While the \aclink{FRSB} equations can be written along
the lines of \cite{20}, they are more involved than the ones for the
Sherrington-Kirkpatrick model \cite{parisi1979infinite,parisi1980sequence,parisi1980order}, and solving them numerically or
getting additional insight from them is a challenging task left for
future work. We present the replica analysis in section~\ref{section:replicas}. Table \ref{tab_summary} contains the summary of our main results along with the
predictions for the step-function perceptron. 

\setlength{\tabcolsep}{10pt}
\renewcommand{\arraystretch}{0.32}
\begin{table}
\begin{adjustbox}{width=\columnwidth,center}
\centering
\begin{tabular}{|c||c|c|c|c|}
\hline
Binary perceptron & Constraint & Constraint function & Range of $K$ & Storage capacity \\
\hline
Step-function   & $z \geq K$      & $\varphi^s(z)=\id\[\displaystyle z \geq K \] $ \vspace{0.01cm}     & $\forall K \in {\mathbb R}$  & \aclink{RS} eq.~\eqref{RS_capacity}     \\
\hline
Rectangle   & $|z| \leq K$     & $\varphi^r(z) =\id\[\displaystyle |z| \leq K \] $ \vspace{0.01cm}      & $\forall K  \in {\mathbb R}^+$ & Annealed eq.~\eqref{cap_rec}       \\
\hline
$U$-function   & $|z|\geq K$   &  $\varphi^u(z)=\id\[\displaystyle |z| \geq K \] $ \vspace{0.01cm}     & $ 0< K<K^*=0.817$ & Annealed eq.~\eqref{cap_ss}
                                              \\
\hline
$U$-function &  $|z|\geq K$   &  $\varphi^u(z)=\id\[\displaystyle |z| \geq K \] $ \vspace{0.01cm}    & $\forall K>K^*=0.817$ & \aclink{FRSB} ?           \\
\hline
\end{tabular}
\end{adjustbox}
\caption{This table summarizes results for storage capacity in binary
  perceptrons with different types of constraints. The result for canonical step-function is from \cite{krauth1989storage}. The results for the rectangle and $u$-function
  are obtained in this paper.}
\label{tab_summary}
\end{table}

Finally let us comment on the simpler and more commonly considered case of spherical perceptron where the binary constraint on the vector $\vec{w}$ is replaced by the spherical constraint $\|\vec{w}\|_2^2 = \sum_{i=1}^\ndim w^2_i = \ndim$. For $K=0$ the spherical perceptron reduces to the famous problem of intersection of half-spaces with capacity $\alpha_c=2$ as solved by Wendell~\cite{wendel1962problem} and Cover \cite{cover1965geometrical}. For $K>0$ the Gardner-Derrida solution \cite{gardner1988optimal}
 is correct as proven in \cite{shcherbina2003rigorous,stojnic2013another}. For $K<0$ the situation is more challenging and \aclink{FRSB} is needed to compute the storage capacity; for recent progress in physics see \cite{franz2016simplest,20}, while mathematical considerations about this case were presented in \cite{stojnic2013negative}.
\section{Proof of correctness of the annealed capacity}
\label{section:proof}

To precisely state the main results, we introduce some definitions. Let $\mat{X}({\ndim,\nsamples})$ be the random $\nsamples \times \ndim $ pattern matrix. Define the partition functions
\begin{align}
\begin{aligned}
	\mathcal Z_r(\mat{X}) &= \displaystyle \sum_{ \textbf{w} \in \{\pm 1\}^\ndim} \prod_{\mu = 1}^\nsamples  \varphi^r ( \displaystyle z_\mu (\textbf{w}) )\\
	\mathcal Z_u(\mat{X}) &= \displaystyle \sum_{ \textbf{w} \in \{\pm 1\}^\ndim} \prod_{\mu = 1}^\nsamples \varphi^u ( \displaystyle z_\mu (\textbf{w}) ) \,,
\end{aligned}
\label{main:binary_perceptron:partition_functions}
\end{align}
which count respectively the number of solutions for the rectangle and $u-$function constraints.
Let  $\cE^r(\ndim, \nsamples)$ and  $\cE^u(\ndim, \nsamples)$ be the events that $\mathcal Z_r(\mat{X})\ge1$  and $\mathcal Z_u(\mat{X})\ge1$, we formally define the storage capacity as follows.
\begin{definition}
The storage capacity $\alpha_c^r(K)$ is 
\begin{align*}
\alpha_c^r(K) &= \inf \left\{ \alpha: \lim_{\ndim \to \infty} \bbP[ \cE^r(\ndim,\lfloor \alpha \ndim \rfloor ) ] = 0 \right\} \,,
\end{align*}
and likewise for $\alpha_c^u(K)$.
\end{definition}
It is believed that there is a sharp threshold for the existence of solutions.  
\begin{conjecture}
\label{conjSharp}
The storage capacity is a sharp threshold:
\begin{align*}
\alpha^r_c(K) &= \sup \left\{ \alpha: \lim_{\ndim \to \infty} \bbP[ \cE^r(\ndim ,\lfloor \alpha \ndim \rfloor ) ] = 1 \right\} \,,
\end{align*}
and likewise for $\alpha_c^u(K)$.
\end{conjecture}

The corresponding conjecture for the random k-SAT model is the celebrated \emph{satisfiability threshold conjecture} proved for $k$ large by Ding, Sly, and Sun~\cite{ding2015proof}.
Next, couple two standard Gaussians $z_1, z_\beta$ by letting $z$ and $z'$ be independent standard Gaussians and setting $z_1 = \sqrt{\beta} z + \sqrt{1-\beta} z'$ and $z_\beta = \sqrt{\beta} z - \sqrt{1-\beta} z'$. Let 
\begin{align}
\begin{aligned}
q_{r,K}(\beta) &\equiv \bbP[ |z_1 | \le K \wedge |z_\beta| \le K ] = q_K
(\beta)\,,  \spacecase 
q_{u,K}(\beta) &\equiv  \bbP[ |z_1 | \ge K \wedge |z_\beta| \ge K ] = 1 -2 p_{r,K} + q_K (\beta)  \label{qK}\,, 
\end{aligned}
\end{align}
with $q_K(\beta)$ the probability that two standard Gaussians with correlation $2\beta-1$ are both at most $K$ in absolute value, explicitly given by
\begin{align*}
		q_K (\beta) &=  \frac{1}{2\pi} \int_{-K}^{K} \d y ~ \int_{\frac{-K+(1-2\beta)y}{2\sqrt{\beta(1-\beta)}}}^{\frac{K+(1-2\beta)y}{2\sqrt{\beta(1-\beta)}}} \d x ~ \exp\(-\frac{x^2+y^2}{2}\)  \, .      
\end{align*}

Note that $q_{t,K}(1) = p_{t,K}$ and $q_{t,K}(1/2) = p_{t,K}^2$ for $t \in \{r,u\}$.
We now  introduce the functions that dictate the effectiveness of the second moment bound.   Let
\begin{align}
\begin{aligned}
F_{r,K, \alpha}(\beta) &=  \rH(\beta) +  \alpha \log q_{r,K}(\beta)\,, \spacecase 
F_{u,K,\alpha}(\beta) &=  \rH(\beta) +  \alpha \log q_{u,K}(\beta)   \label{FK}
\end{aligned}
\end{align}
where $\rH(\beta) = -\beta \log \beta -(1-\beta) \log (1-\beta)$ is the Shannon entropy function. We state a numerical hypothesis in terms of the derivatives of these two functions. 
\begin{hypothesis}
\label{hypo}
For all choices of $K>0$ and $\alpha>0$ so that $F''_{r,K,\alpha}(1/2) <0$, there is exactly one $\beta \in (1/2,1)$ so that $F'_{r,K,\alpha}(\beta) =0$.  The same holds for $F_{u,K,\alpha}$.
\end{hypothesis}

Our main theorem is a proof, under Hypothesis~\ref{hypo}, that the storage capacity is given by the annealed computation.
\begin{theorem}
\label{thmMain}
Under the assumption of Hypothesis~\ref{hypo}, the following hold. 
\begin{enumerate}
\item For all $K>0$, we have $\alpha_c^r(K) = -\log(2) / \log (p_{r,K})$. 
\item For all $K \in (0, K^*)$, we have $\alpha_c^u(K) = -\log(2) / \log (p_{u,K})$.
\end{enumerate}
\end{theorem}

Under our definition of $\alpha_c^r(K)$ and $\alpha_c^u(K)$, we must prove two statements to show that $\alpha_c^r(K) = -\log(2) / \log (p_{r,K})$ (and similarly for $\alpha_c^u(K)$).  We use the first moment method to show that for $\alpha> -\log(2) / \log (p_{r,K})$, \\$\lim_{\ndim \to \infty}\bbP[\cE^r(\ndim, \nsamples)] =0$; then we use the second moment method to show that for $\alpha < -\log(2) / \log (p_{r,K})$, $\liminf_{\ndim \to \infty}\bbP[\cE^r(\ndim, \nsamples)] >0$ (a result analogous to what Ding and Sun prove for the more challenging step binary perceptron~\cite{Sun2018}).  Conjecture~\ref{conjSharp} asserts the stronger statement that for $\alpha < -\log(2) / \log (p_{r,K})$, $\lim_{\ndim \to \infty}\bbP[\cE^r(\ndim, \nsamples)] =1$.  

\subsection{First moment upper bound}
\label{proof:first_moment}

\begin{prop}[First moment upper bound]
$ $
\begin{enumerate}
	\item If $\alpha > \alpha_a^{r}(K) =\frac{-\log(2)}{\log(p_{r,K}) } $, then with high probability there is no satisfying assignment to the binary perceptron with the rectangle activation function.
	\item If $\alpha > \alpha_a^{u}(K) =\frac{-\log(2)}{\log(p_{u,K}) } $, then with high probability there is no satisfying assignment to the binary perceptron with the $u$-function activation function.
	\end{enumerate}  
\end{prop}

\begin{proof}
We give the proof for the rectangle function as the proof for the $u$-function is identical. Let $\epsilon = \alpha - \alpha_a^{r}(K)  >0$. Let  $\vec{1}_\ndim$ denote the vector of dimension $\ndim$ with all $1$ entries.
\begin{align*}
\bbP[ \cE^r (\ndim, \alpha \ndim) ] &\le \EE\[ \mZ_{r}\( \mat{X}(\ndim,\alpha \ndim) \) \] = 2^\ndim ~ \EE\left [ \prod_{\mu =1}^{\alpha \ndim} \id \[ \left |z_{\mu}(\vec{1}) \right| \le K \] \right] \\
&= 2^\ndim ~ p_{r,K}^{\alpha \ndim} = \exp(\ndim (\log(2)+\alpha \log( p_{r,K})  )) \\
&= \exp (\ndim \epsilon \log (p_{r,K}))   \to 0  \text{ as } \ndim \to \infty \,.
\end{align*}
\end{proof}

\subsection{Second moment lower bound}
\label{proof:second_moment}

\begin{prop}[Second moment lower bound]
$ $
\label{prop2ndMoment}
\begin{enumerate}
\item If $\alpha < \frac{-\log(2)}{\log(p_{r,K}) }$, then  
$ \liminf_{\ndim \to \infty} \bbP[  \cE^r (\ndim, \alpha \ndim)  ] > 0.$
\item If $K < K^*$ and $\alpha <\frac{-\log(2)}{\log(p_{u,K}) }$, then 
$ \liminf_{\ndim \to \infty} \bbP[  \cE^u (\ndim, \alpha \ndim)   ]  >0.$ 
\end{enumerate}
\end{prop}

To prove Proposition~\ref{prop2ndMoment}  we will apply the second-moment method in a similar fashion to Achlioptas and Moore~\cite{achlioptas2002asymptotic} who determined the satisfiability threshold of random $k$-SAT to within a factor $2$ by considering not-all-equal satisfying assignments (not-all-equal satisfiability (NAE-SAT) constraints are symmetric in the same way the rectangle and $u$-function constraints are symmetric). 
Recall the Paley-Zygmund inequality.
\begin{lemma}
Let $\rX$ be a non-negative random variable.  Then
\begin{align*}
\bbP[\rX >0 ] &\ge \frac{\E[\rX]^2 }{\E[\rX^2]} \,.
\end{align*}
\end{lemma}

We will also use the following application of Laplace's method from Achlioptas and Moore~\cite{achlioptas2002asymptotic}.
\begin{lemma}
\label{lemLaplace}
Let $g(\beta)$ be a real analytic function on $[0,1]$ and let 
\begin{align*}
G(\beta) &=  \frac{g(\beta)}{\beta^{\beta} (1-\beta)^{1-\beta} } \, .
\end{align*}
If $G(1/2) > G(\beta)$ for all $\beta \ne 1/2$ and $G''(1/2) <0$, then there exists constants $c_1, c_2$ so that for all sufficiently large $\ndim$
\begin{align*}
c_1 G(1/2)^\ndim \le \sum_{l = 0}^\ndim  \binom{\ndim}{l} g(l/\ndim )^\ndim  \le c_2 G(1/2)^\ndim \,.
\end{align*}
\end{lemma}

\subsubsection{Rectangle binary perceptron}
We calculate 
\begin{align*}
&\EE [\mathcal Z_r (\mat{X})^2   ] =  \sum_{\vec{w}_1,\vec{w}_2 \in \{ \pm 1\}^\ndim} \bbP[\vec{w}_1,\vec{w}_2  \textrm{ satisfying}] \\
&= 2^\ndim \sum_{\vec{w} \in \{ \pm 1\}^\ndim} \bbP[\vec{1}, \vec{w}  \textrm{ satisfying}] = 2^\ndim \sum_{l=0}^\ndim ~ \binom{\ndim}{l} q_{r,K}(l/\ndim)^{\alpha \ndim} \\ 
&= \exp\left( \ndim (\log(2) +  F_{r,K,\alpha}(\beta) )   \right) \, .
\end{align*}
where we recall $q_{r,K}$ from eq.~\eqref{qK}. Define 
\begin{align}
G_{r,K,\alpha}(\beta) &\equiv \exp( F_{r,K,\alpha}(\beta)) =  \frac{ q_{r,K}(\beta)^{\alpha} }{ \beta^{\beta} (1-\beta)^{1-\beta}  } \,.
\label{eqGtoF}
\end{align}
If we can show that $G_{r,K,\alpha}(1/2) > G_{r,K,\alpha}(\beta)$ for all
$\beta \ne 1/2$ and $G_{r,K,\alpha}''(1/2)$ $<0$, then by Lemma~\ref{lemLaplace}, we have  
\begin{align*}
\EE [\mathcal Z_r (\mat{X})^2   ] & \le c_2 ~ 4^\ndim ~ q_{r,K}(1/2)^{\alpha \ndim} =c_2 ~ 4^\ndim ~  p_{r,K}^{2 \alpha \ndim} \,.
\end{align*}
Then since $\mZ_r (\mat{X})$ is integer valued, we have 
\begin{align*}
\bbP[ \mathcal Z_r (\mat{X}) \ge 1] &\ge \frac { \EE[ \mathcal Z_r (\mat{X})]^2 }{ \EE[\mathcal Z_r (\mat{X})^2   ]   }  =  \frac { ( 2^\ndim  p_{r,K}^{\alpha \ndim} )^2 }{ \EE[\mathcal Z_r (\mat{X})^2   ]   }   \ge  \frac{ ( 2^\ndim  p_{r,K}^{\alpha \ndim} )^2 }{c_2 4^\ndim  p_{r,K}^{2 \alpha \ndim}}  = \frac{1}{c_2} >0 \,.
\end{align*}
It remains to show that when $\alpha < \frac{-\log(2)}{\log(p_{r,K}) }$, then $G_{r,K,\alpha}(1/2) > G_{r,K,\alpha}(\beta)$ for all $\beta \ne 1/2$ and $G_{r,K,\alpha}''(1/2) <0$.  By eq.~\eqref{eqGtoF} and the fact that $G_{r,K,\alpha}'(1/2) = 0$, it is enough to show the same for $F_{r,K,\alpha}$. 
Certainly one necessary condition is that  $F_{r,K,\alpha}(1/2) > F_{r,K,\alpha}(1)$.  This reduces to the condition $2 p_{r,K}^{2\alpha} > p_{r,K}^{\alpha}$ or $\alpha < \frac{-\log (2)}{\log( p_{r,K})}$  which is exactly the condition of Proposition~\ref{prop2ndMoment}. Next consider $F_{r,K,\alpha}''(1/2)$. A straightforward calculation shows that
\begin{align*}
F_{r,K,\alpha}''(1/2) &= 4 \left( -1 + \frac{2}{\pi} \frac{\alpha K^2 e^{-K^2} }{ p_{r,K}^2}  \right ) \,.
\end{align*}
In particular, $F_{r,K,\alpha}''(1/2) <0$ if and only if $\alpha < \frac{\pi}{2} \frac{ p_{r,K}^2}{ K^2 ~ \exp(-K^2) }$.
But another calculation also shows that 
\begin{align*}
- \frac{\log(2) }{\log (p_{r,K})} <  \frac{\pi}{2} \frac{ p_{r,K}^2}{ K^2 e^{-K^2} }  
\end{align*}
for all $K>0$ and so the condition of Proposition~\ref{prop2ndMoment} implies that $F_{r,K,\alpha}''(1/2) <0$.
Moreover, since $F_{r,K,\alpha}(\beta)$ is symmetric around $\beta=1/2$ and it has a local maximum at $\beta=1/2$, Hypothesis~\ref{hypo} implies that the global maximum of $F_{r,K,\alpha}(\beta)$ occurs at either $1/2$ or $1$, and since $F_{r,K,\alpha}(1/2) > F_{r,K,\alpha}(1)$, we have that  $F_{r,K,\alpha}(1/2) > F_{r,K,\alpha}(\beta)$ for all $\beta \ne 1/2$, completing the proof of Proposition~\ref{prop2ndMoment} for the rectangle binary perceptron.  

\subsubsection{$u$-function binary perceptron}
\label{proof:second_moment_u}
The proof for the $u$-function is similar.  We can calculate
\begin{align*}
\EE[\mZ_u (\mat{X})^2   ] &= 2^\ndim \sum_{l=0}^\ndim \binom{\ndim }{l} q_{u,K}(l/\ndim )^{\alpha \ndim } \\
&= \exp\left( \ndim (\log(2) +  F_{u,K,\alpha}(\beta) )  \right) \, ,
\end{align*}
where we recall $q_{u,K}$ from eq.~\eqref{qK}. Using Lemma~\ref{lemLaplace} and Hypothesis~\ref{hypo} again, it suffices to show that for $0 < K< K^*$ and $\alpha< \frac{-\log(2)}{\log ( p_{u,K})}$ we have $F_{u,K,\alpha}(1/2) > F_{u,K,\alpha}(1)$ and $F_{u,K,\alpha}''(1/2) <0$. 
The first follows immediately from the fact that $\alpha< \frac{-\log(2)}{\log ( p_{u,K})}$.  For the second, we have
\begin{align*}
F_{u,K,\alpha}''(1/2) &= 4 \left (-1 +  \frac{2}{\pi} \frac{\alpha K^2 e^{-K^2} }{ p_{u,K}^2}   \right) 
\end{align*}
and so $F_{u,K,\alpha}''(1/2) < 0$ if and only if $\alpha < \frac{\pi}{2} \frac{ p_{u,K}^2}{ K^2 e^{-K^2} }$.
Unlike with the rectangle function it is not true that 
\begin{align}
- \frac{\log(2)}{\log (p_{u,K})} < \frac{\pi}{2} \frac{ p_{u,K}^2}{ K^2 e^{-K^2} }	
\label{main:AT_second_moment}
\end{align}
 for all $K$: the left and right sides of the inequality cross at $K= K^*$, which  implicitly defines $K^*$.
 Thus for $K<K^*$ and $\alpha < - \frac{\log(2)}{\log (p_{u,K})}$ we have $F_{u,K,\alpha}''(1/2)$ $< 0$, which completes the proof of Proposition~\ref{prop2ndMoment} for the $u$-function binary perceptron. 

\begin{figure}[htb!]
\centering
\includegraphics[width=0.49\linewidth]{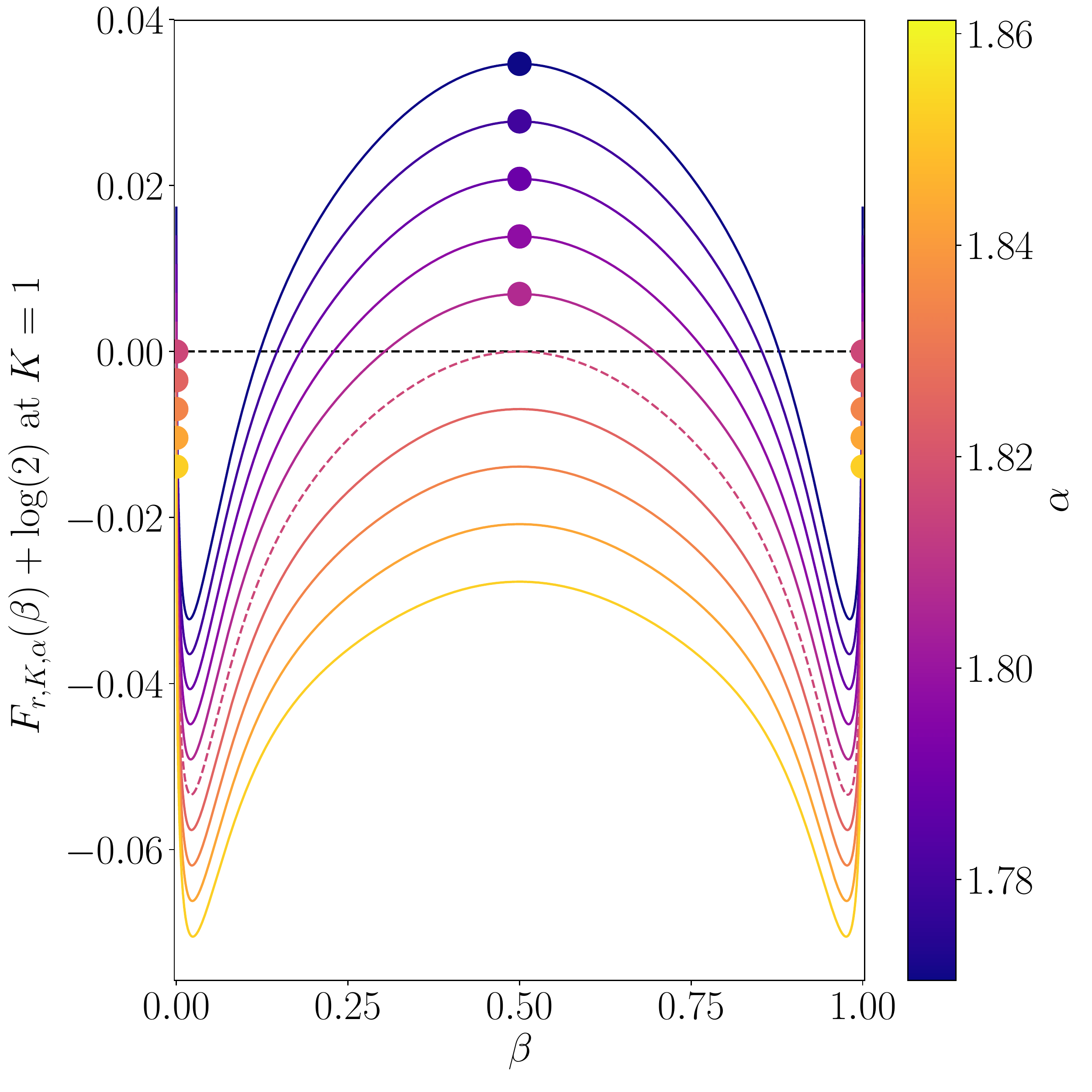}
\includegraphics[width=0.49\linewidth]{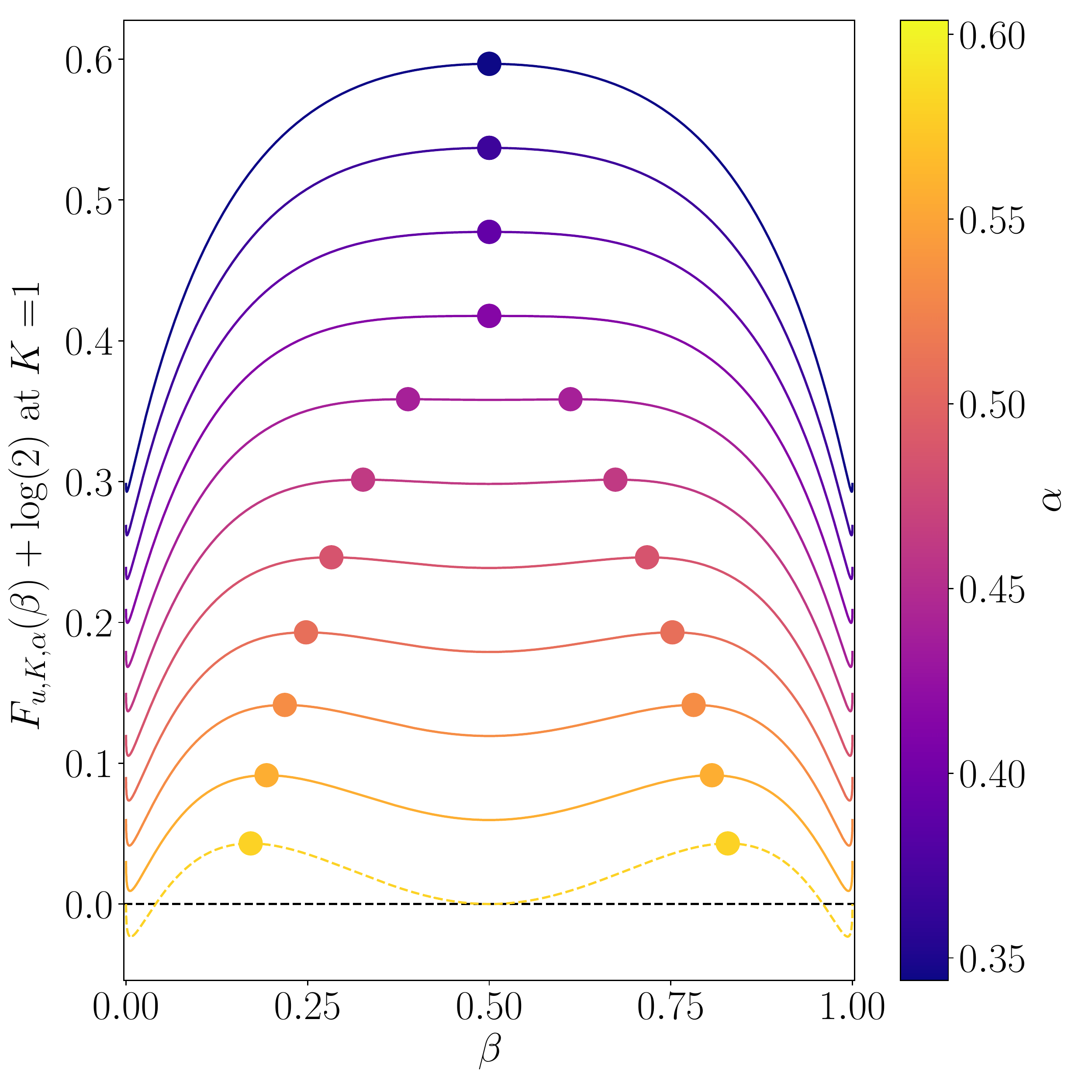}
\caption{Second moment entropy densities. \Left the rectangle binary perceptron for $\alpha \leq \alpha_a^r
  =1.816$ (dashed pink), $\beta=\frac{1}{2}$ is the global
  maximizer. For $\alpha \geq \alpha_a^r$, $\beta=0$ and $\beta=1$ are
  the maximizers. \Right the $u$-function binary perceptron for $\alpha \leq \alpha^*=0.430$, $\beta=\frac{1}{2}$ is the maximizer while for $\alpha^*\leq \alpha \leq \alpha_a^u = 0.604$ (dashed yellow), the maximizer is non-trivial $\beta \ne 0$.}
\label{fig:plot_second_moment_rectangle_symstep}
\end{figure}

\subsubsection{Illustration}
As an illustration, we plot the second moment entropy density $\lim_{\ndim \to \infty} \frac{1}{\ndim} \log$ $\EE [ \mZ_t^2 ] = \log(2) + F_{t,K,\alpha} $ for $t\in\{r,u\}$ at $K=1>K^*$ in \Fig\ref{fig:plot_second_moment_rectangle_symstep}. For the rectangle function \Leftn, the second moment is tight: the maximum is reached for $\beta=1/2$ for all $\alpha$ smaller than the first moment $\alpha_a^r$ (dashed pink). Exactly the same happens for the $u-$function with $K<K^*$. However for $K>K^*$, the second moment method fails \Rightn: $\beta=1/2$ becomes a minimum and the maximum is obtained for non trivial values $\beta \ne 1/2$ for constraint density smaller than the first moment $\alpha_a^u$  (dashed yellow).

\section{Frozen-1RSB structure of solutions in binary perceptrons}
\label{section:frozen}

One of the most striking properties of the canonical step-function perceptron is
the predicted \aclink{f1RSB} \cite{krauth1989storage} nature of the space of solutions. This means
that the dominant, \ie with measure tending to one, part of the space of solutions splits
into well separated clusters each of which has vanishing entropy
density at any $\alpha>0$.
This \aclink{f1RSB} scenario and quantitative properties of the solution space were studied in detail recently \cite{16,huang2014origin}. Following up on conjectures
that such a frozen structure of solutions implies computational hardness in
diluted constraint satisfaction problems
\cite{zdeborova2008constraint}, it was argued that finding a satisfying assignment in the binary
perceptron should also be algorithmically hard since its solution space is dominated by
clusters of vanishing entropy density \cite{huang2014origin}. Yet this
conjecture contradicted empirical results of
\cite{braunstein2006learning}. This paradox was resolved in \cite{baldassi2015subdominant} where the authors identified that there
are subdominant parts (\ie parts of measure converging to zero as
the system size diverges) of the solution space that form extended
clusters with large local entropy and all the algorithms that work well
always find a solution belonging to one of those  large-local-entropy
clusters. These sub-dominant clusters are not frozen and somewhat strangely are not captured in the canonical \aclink{1RSB} calculation
\cite{baldassi2015subdominant}.   It was argued that existence of these
large-local-entropy clusters bears more general consequences on the
dynamics of learning algorithms in neural networks,
see e.g. \cite{baldassi2016unreasonable}. 

While \aclink{f1RSB} structure has also been identified in \aclink{CSP} on sparse graphs \cite{zdeborova2008locked,zdeborova2011quiet}, we want to note
that its nature in the binary perceptron is of a rather different
nature. In sparse systems a simple argument using expansion
properties of the underlying graph and properties of the constraints show that each cluster with high
probability contains only one solution. In the perceptron model, which has a
fully connected bipartite interaction graph, this argument from sparse
models does not apply.

In the present work, we deduce from the second moment
calculation of the previous section that the space of solutions in the symmetric binary
perceptrons is also of the \aclink{f1RSB} type and this property moreover extends
to any finite temperature (with energy being defined as the number of
unsatisfied constraints). This is different from the locked \aclink{CSP} of \cite{zdeborova2008constraint,zdeborova2011quiet} living on
diluted hyper-graphs, where the solution-clusters have extensive
entropy at any non-zero temperature. Another difference is that
whereas in the locked \aclink{CSP} the size of
each cluster is one with high probability, in the binary perceptron
there are still many solutions in the clusters, it is only their entropy
density, \ie the logarithm of their number per variable, that vanishes
as $\ndim \to \infty$. 
Investigation of the large local
entropy clusters and their implications for learning in the symmetric perceptrons is also
of great interest, but left for future work. Clearly since mathematically
the symmetric perceptrons are simpler than the step-function one, they
should also be the proper playground to deepen our understanding of
the large local entropy clusters and their relation to learning and generalization. 

We present the \aclink{f1RSB} scenario as a conjecture and then below indicate how the second moment calculation gives evidence for this conjecture.  Given an instance $\mat{X}$ and a solution $\vec{w}$, let $\Gamma(\vec{w},\td{d})$ denote the set of solutions $\vec{w}'$ with Hamming distance at most $\td{d}$ from $\vec{w}$.
\begin{conjecture}
\label{conjFrozen}
For every $K > 0$ and every $\alpha \in (0,\alpha_c^r(K))$ there exists a Hamming distance $\td{d}_{\text{min}}>0$ so that with high probability over the choice of the random instance $\mat{X}$ from the RBP, the following property holds: for almost every solution $\vec{w}$, 
\begin{align*}
\frac{1}{\ndim} \log | \Gamma(\vec{w},\td{d}_{\text{min}})| \underlim{\ndim}{\infty} 0
\end{align*}
The same holds for the UBP for all $K \leq K^*$.  
\end{conjecture}

\subsection{The link between the second-moment entropy and size of
  clusters} 
In this section we use $t \in\{r,u\}$ and note that the form of the second moment entropy density
$\frac{1}{\ndim} \log \EE [ \mathcal{Z}_t^2 ]$ has very direct implications on the structure of
solutions in the corresponding models. 
As we defined it above, the second moment entropy is the normalized logarithm of
the expected number of pairs of solutions of overlap $\beta$. 

For problems such as the symmetric binary perceptrons where the quenched
and annealed entropies are equal in leading order, there is a
striking relation between the planted and the random ensemble of the model \cite{achlioptas2008algorithmic,krzakala2009hiding}. The \emph{random
ensemble} is the problem we have considered so far, while the \emph{planted ensemble} is
defined by starting with a configuration of the weights (a solution) and then including only constraints that are 
satisfied by this \emph{planted} configuration. As long as the quenched and 
annealed entropies of the random ensemble are equal in leading order the planted and random ensembles should be 
contiguous, meaning that high-probability properties that hold in one ensemble also hold in the other. Moreover the 
planted configuration in the planted ensemble has all the properties of a configuration sampled uniformly at random 
in the random ensemble. These properties follow on the heuristic
level from the cavity method reasoning \cite{krzakala2009hiding}. They
were established fully rigorously in a range of models, see
e.g. \cite{achlioptas2008algorithmic,mossel2015reconstruction,coja2018information}. 
In the present case of symmetric binary perceptrons we have not yet managed
to prove contiguity between the random and the planted
ensemble, and so we leave a rigorous mathematical result for future work.  (In fact the missing ingredient is a version of Friedgut's sharp threshold result~\cite{friedgut1999sharp} suitable for perceptrons; such a result combined with Theorem~\ref{thmMain} would also prove Conjecture~\ref{conjSharp}).   We hence rely on the above heuristic argument and assume it
holds in what follows.

Given a planted solution $\vec{w}$ and a configuration $\vec{w}_\beta$  that agrees with $\vec{w}$  on $\beta \ndim$ coordinates, the probability that $\vec{w}_\beta$ is a solution in the planted model is $( q_{t,K}(\beta)/ p_{t,K})^\nsamples$, and thus the expected number of solutions $\mZ_\beta$ at Hamming distance $\beta \ndim$ from the planted solution in the planted ensemble is 
\begin{align*}
 \EE[\mathcal Z_\beta ]= \binom{\ndim}{\beta \ndim} ( q_{t,K}(\beta)/ p_{t,K})^\nsamples \,,
\end{align*}
and its entropy density is 
\begin{align}
	\omega_t(\beta) \equiv \lim_{\ndim \to \infty} \frac{1}{\ndim} \log \EE[\mathcal{Z}_\beta]  = F_{t,K,\alpha}(\beta)- \alpha \log{p_{t,K}} \textrm{ for } t \in\{r,u\} \,.
\label{omega}
\end{align}
Recalling that \emph{contiguity} implies that the planted
solution has the properties of a uniformly chosen solution in the random ensemble then this entropy gives
us direct access to properties of the solution space in the random ensemble at equilibrium. Most
notably we notice (see derivation in section~\ref{frozen_2nd} below) that the derivative of
$\omega_t(\beta)$ at $\beta=1$ is $+\infty$ thus implying that $\forall
\epsilon>0$ with high probability there are no solutions at overlap $\beta \in [\td{d}_\textrm{min}(\alpha,K),
(1-\epsilon)]$. In turn, this means that the dominant (measure
converging to one as $\ndim\to \infty$) part
of the solution space splits into clusters each of which has
vanishing entropy density (i.e. logarithm of the number of solutions
in the cluster divided by $\ndim$ goes to zero as $\ndim\to \infty$).   The missing ingredient in a full proof of Conjecture~\ref{conjFrozen} is a proof of the contiguity statement. 

\subsection{Form of the 2nd moment entropy implying frozen-1RSB}
\label{frozen_2nd}

\begin{figure}[htb!]
		    \centering
	   		\includegraphics[width=0.45\linewidth]{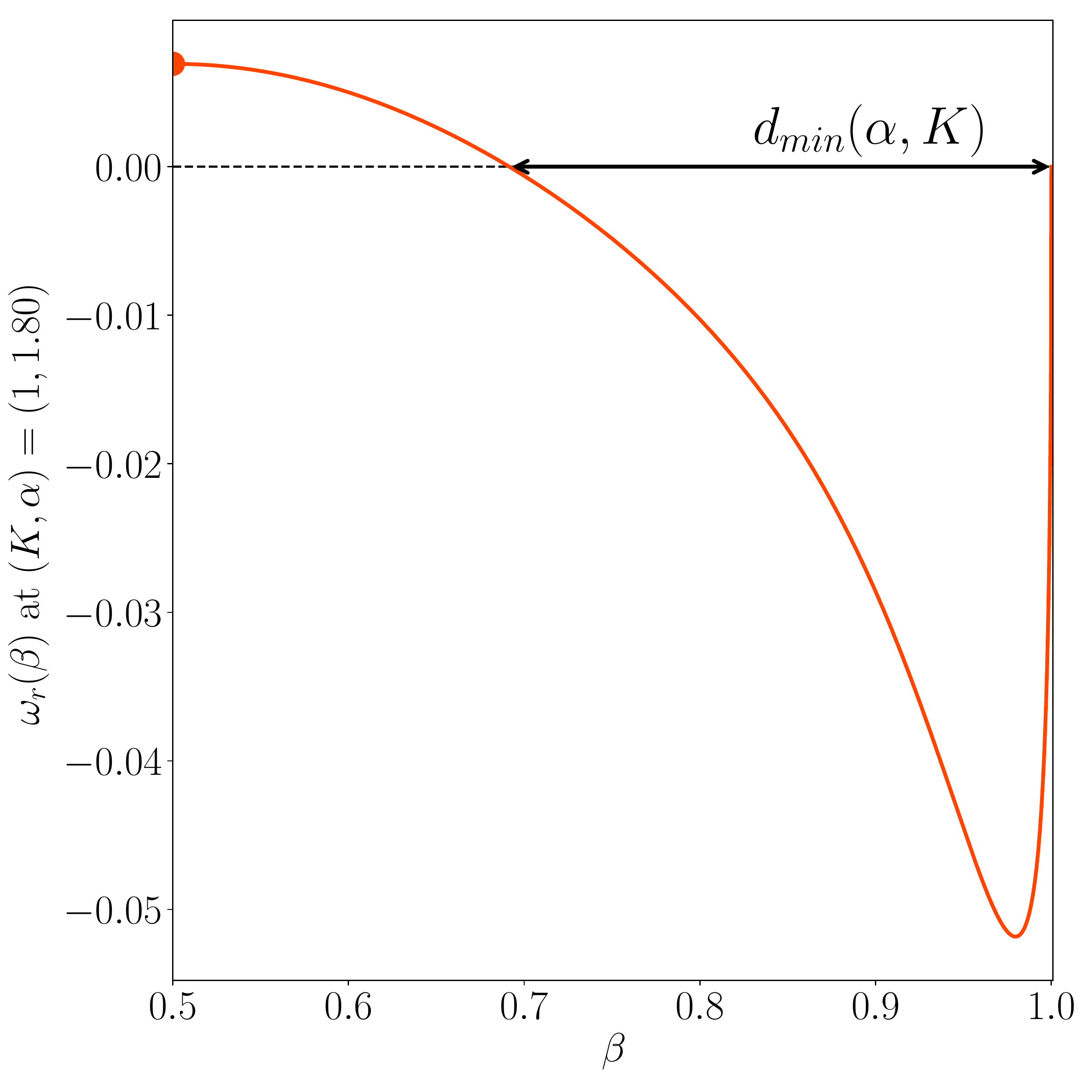}
	   		\hspace{0.2cm}
	   		\includegraphics[width=0.45\linewidth]{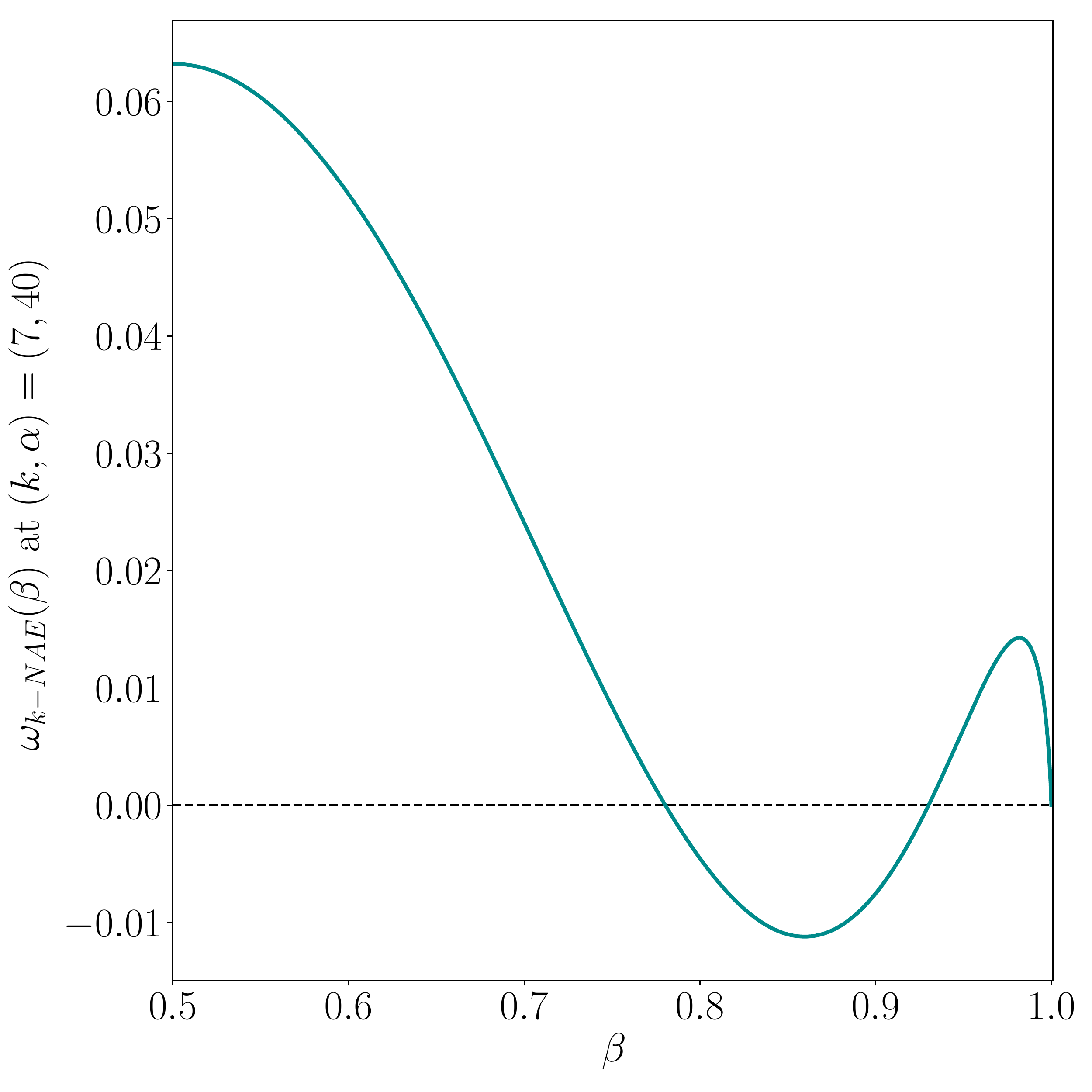}
\caption{
\textbf{(Left)} Density of the annealed entropy of solutions at overlap $\beta$ from a random solution in the rectangle binary perceptron at $K=1$, $\alpha = 1.80 \le \alpha_c^{r}(K=1)$. We see there are no solution in an interval of overlaps $(1-\td{d}_\textrm{min}, 1-\epsilon)$. This curve is obtained from the second moment entropy and contiguity between the random and planted ensembles. It implies the frozen-1RSB nature of the space of solutions. The same holds for the $u-$function.
\textbf{(Right)} To compare we plot the density of the annealed entropy of solutions at overlap $\beta$ from a random solution in the $k$-NAE SAT model \cite{achlioptas2002asymptotic} at $k=7$, $\alpha = 40$. We see the density is positive in a large region close to $\beta=1$, showing the absence of frozen-1RSB structure in this problem.
} 
\label{fig_frozen}
\end{figure}

In \Fig\ref{fig_frozen} \Leftn, we plot $\omega_r(\beta)$ for the rectangle
binary perceptron, at $K=1$,
$\alpha=1.80\le \alpha_c^{r}(K=1)$. Thanks to the contiguity between the planted and random ensembles that holds as long as the second moment entropy
density is twice the first moment entropy density, this curve
represents also the annealed entropy of solutions at overlap $\beta$
with a random reference solution. We see notably that there is an
interval of distances in which no solutions are present. Analytically
we can see from the properties of the functions $F_{t,K,\alpha} (\beta)$ and
$\log{p_{t,K}}$ that $F_{t,K,\alpha}(1) = \alpha \log{p_{t,K}}$ and the derivative of
$F_{t,K,\alpha}(\beta) \to \infty$. This is in contrast with, for instance, the
satisfiability problems studied in \cite{achlioptas2002asymptotic},
where the function corresponding to $F_{t,K,\alpha}(\beta)$ would have a negative
derivative in $\beta=1$, see \Fig\ref{fig_frozen} \Rightn. There could still be an interval of \emph{forbidden} distance, but the bump in entropy for $\beta \approx 1$
corresponds to the size of the clusters to which typical solutions
belong and those would be extensive.  

\subsubsection{Frozen 1RSB in rectangle binary perceptron}
In the rectangle binary perceptron, the random and planted ensembles are conjectured to be contiguous for
all $K >0$ and $\alpha \in (0, \alpha^r_c(K))$. Using eq.~\eqref{FK}, the first derivative of $\omega_r(\beta)$, eq.~\eqref{omega}, is given by
\begin{align*}
	 \frac{\partial \omega_r }{\partial \beta} &= \frac{\partial F_{r,K,\alpha}}{\partial \beta} =  \log\left (\frac{1-\beta}{\beta} \right) \\
	 &+  \frac{\alpha}{ q_{r,K,T}(\beta)} \frac{1}{\pi \sqrt{\beta(1-\beta)}} \left( e^{-\frac{K^2}{2(1-\beta)}} \left( e^{\frac{(2\beta-1)K^2}{2(1-\beta)\beta}} -1  \right)  \right) \xrightarrow[\beta \to 1]{} +\infty \,,
\end{align*}
where the computation is detailed in \Sec E of \cite{Aubin2019_storage}. It diverges for all $K \in \mathbb{R}^+$, $\alpha>0$ in the limit $\beta \to 1$, that implies vanishing entropy density of clusters to which typical
solutions belong. 

\subsubsection{Frozen 1RSB in the $u$-function binary perceptron}

In the $u$-function binary perceptron, the random and planted ensembles are conjectured to be contiguous for
all $0 < K \le K^*$ and $\alpha \in (0, \alpha^u_c(K))$. Using eq.~\eqref{FK}, the first derivative of $\omega_u(\beta)$ eq.~\eqref{omega}, is given by
\begin{align*}
	& \frac{\partial \omega_u }{\partial \beta} = \frac{\partial F_{u,K,\alpha}}{\partial \beta} =  \log\left (\frac{1-\beta}{\beta} \right)\\ 
	 &+  \frac{\alpha}{ q_{u,K,T}(\beta)} \frac{1}{\pi \sqrt{\beta(1-\beta)}} \left( e^{-\frac{K^2}{2(1-\beta)}} \left( e^{\frac{(2\beta-1)K^2}{2(1-\beta)\beta}} -1  \right)  \right) \underset{\beta \to 1}{\longrightarrow} +\infty \, ,
\end{align*}
thus reaching the same conclusion on presence of \aclink{f1RSB}.\\

In \Sec E of \cite{Aubin2019_storage}, we extend the second moment calculation to finite
temperature (for both the rectangle and $u-$function case). This means that we define the energy of a configuration
${\cal E}(\tbf{w})$ as
the number of constraints that are violated by this
configurations. Then the corresponding partition function is defined
${\cal Z}(T) = \sum_{\tbf{w}} e^{-{\cal E}(\tbf{w})/T}$. There is a
one-to-one mapping between the temperature $T$ and energy density
$e={\cal E}/\ndim$, consequently 
the corresponding finite-temperature second moment entropy density
counts the number of pairs of solutions at overlap $\beta$ and energy
density $e$. In \Sec E of \cite{Aubin2019_storage}, we apply the same
argument as here connecting the random and planted ensemble, and deduce that  the
finite-temperature solution space of the models is of also of the \aclink{f1RSB} type
for any $T<\infty$. 

\subsection{Frozen-1RSB as derived from the replica analysis}
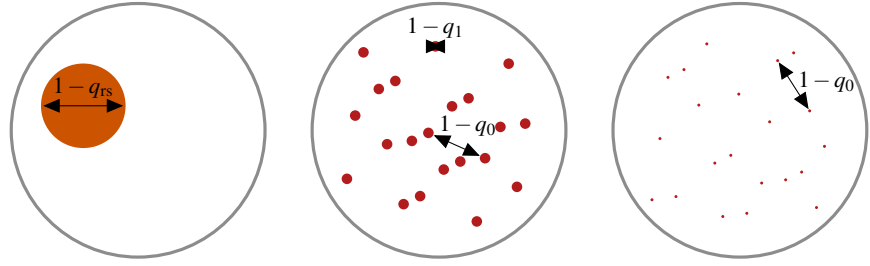
\begin{figure}[htb!]
	\centering
	\begin{tikzpicture}[scale=0.8, every node/.style={transform shape}]
    \tikzstyle{factor}=[rectangle,fill=black,minimum size=7pt,inner sep=1pt]
    \tikzstyle{annot} = [text width=2.5cm, text centered]
    \tikzstyle{RSCluster} = [circle,fill=burntorange,minimum size=40pt,inner sep=1pt,]]
    \tikzstyle{RSBCluster} = [circle,fill=carnelian,minimum size=5pt,inner sep=1pt]]
    \tikzstyle{fRSBCluster} = [circle,fill=carnelian,minimum size=0.5pt,inner sep=0.5pt]
    \tikzstyle{configSpace} = [circle,minimum size=120pt,inner sep=3pt,draw=gray, fill opacity=1.,fill=white, very thick]
    
    \node[configSpace] (C-1) at (-5,0) {};
    \node[configSpace] (C-1) at (0,0) {};
    \node[configSpace] (C-1) at (5,0) {};
    
    \pgfmathsetseed{3}
    \coordinate (mycenterpoint) at ($ (-5,0)+{rand}*(1,0)+ {rand}*(0,1)$);
    \node[RSCluster] (RS-1) at (mycenterpoint) {};
    \node (C) at ([shift=({-0.8cm,0cm})]RS-1) {};
	\node (D) at ([shift=({0.8cm,0cm})]RS-1) {};
	\draw [<->,black] (C) -- (D);
	\path [<->] (C) -- node [above] {$1-q_{\rs}$} (D) ;
    
    \foreach \l in {10,...,30}
      {	\pgfmathsetseed{\l}
          \coordinate (mycenterpoint) at (${rand}*(1.55,0)+ {rand}*(0,1.55)$);
          \node[RSBCluster] (RSB-\l) at (mycenterpoint) {};
      }
    
    \node (A) at ([shift=({-0.25cm,0cm})]RSB-10) {};
	\node (B) at ([shift=({0.25cm,0cm})]RSB-10) {};
	\draw [<->,black] (A) -- (B);
    \path [<->] (A) -- node [above] {$1-q_{1}$} (B) ;

	\draw [<->,black] (RSB-28) -- (RSB-30);
    \path [<->] (RSB-28) -- node [above] { $1-q_{0}$} ++(-0.5,0.4) (RSB-30) ;
    
    \foreach \l in {40,...,60}
      {	\pgfmathsetseed{\l}
          \coordinate (mycenterpoint) at ($(5,0)+ {rand}*(1.5,0)+ {rand}*(0,1.5)$);
          \node[fRSBCluster] (fRSB-\l) at (mycenterpoint) {};
      }
      
    \draw [<->,black] (fRSB-48) -- (fRSB-60);
    \path [<->] (fRSB-48) -- node [above] { $1-q_{0}$} ++(1.6,-1.3) (fRSB-60) ;
    \end{tikzpicture}
	\caption{Illustration of the configuration space for the different phases \Left RS: solutions are concentrated in a single cluster of typical size $1-q_{\rs}$. \Center 1RSB: solutions form clusters of size $1-q_1$ at a distance $1-q_0$ from each other. \Right f1RSB: clusters are point-like ($1-q_1\simeq 0$) at a distance $1-q_0 = 1-q_{\rs}$ from each other.}
	\label{configurationSpace}
\end{figure}

We stress that we derived the \aclink{f1RSB} nature of the space of
solutions without the use of replicas. For completeness we summarize
here how this translates to the properties of the
one-step-replica-symmetry breaking solution. This is the way this
phenomena was originally discovered and described in
\cite{martin2004frozen,16}. For readers not familiar with the
replica method this section should be read after reading section~\ref{section:replicas}.
In general, three kinds of fixed points of the \aclink{1RSB} equations are possible: 
\begin{itemize}
	\item The replica symmetric \aclink{RS} solution $q_0= q_1 = q_{\rs}<1$\,,
	\item The frozen-1RSB solution \aclink{f1RSB} $(q_0,q_1)=(q_{\rs},1)$\,,
	\item The \aclink{1RSB} solution  $(q_0,q_1)$ with $q_1\ne 1$\,.
\end{itemize}
The \aclink{f1RSB} is characterized by
an inner-cluster overlap $q_1=1$ and an inter-cluster overlap
$q_0=q_{\rs}$, which means that clusters have vanishing entropy density and
remain far from each other.
Mathematically \aclink{RS} and \aclink{f1RSB} solutions are equivalent in the sense that
these solutions have the same free entropy $\Phi^{(\textrm{1rsb})}\{q_0 = q_{\rs}, q_1 = q_{\rs}\}=\Phi^{(\textrm{1rsb})}\{q_0 = q_{\rs},q_1 = 1\}$, and the complexity
of the \aclink{f1RSB} solution equals the \aclink{RS} entropy $\Sigma(\Phi=0) =
\Phi^{(\rs)}$ \eq(\ref{main:complexity}, \ref{main:phi_RS}). However, \aclink{RS} and \aclink{f1RSB} do
not share the same configuration space. The \aclink{RS}
phase is associated to a single cluster of solution with typical size
$1-q_{\rs}$, while the \aclink{f1RSB} configuration space is composed of many
point-like solutions of size $q_1\simeq 1$ and at distance $1-q_0 =
1-q_{\rs}$ of each other, see \Fig\ref{configurationSpace}. From this point of view \aclink{f1RSB} is the correct
description of the phase space. 
\section{Replica calculation of the storage capacity}
\label{section:replicas}

In this section we provide the replica free entropies leading to the expression of the storage capacity in the step-function binary perceptron \eqref{RS_capacity}. We show that
in the symmetric binary perceptrons the annealed calculation is
reproduced by the replica symmetric result. For the $u-$function binary
perceptron we show that $K^*$ coincides with the onset of replica
symmetry breaking and we evaluate the \aclink{1RSB} capacity for $K>K^*$.  
The details of the computation is presented in \App\ref{appendix:replica_computation:random_labels:iid} for the constraint function at zero temperature
\begin{equation}
    \mathcal{C}(\vec{z}) \equiv \prod_{\mu=1}^\nsamples
        \varphi(z_{\mu})  \vspace{0.2cm} \textrm{ with }\vspace{0.2cm} z_{\mu}= \vec{x}_{\mu} \cdot \vec{w}\,,
\end{equation}
and $\rP_\y(\vec{y}) = \delta(\vec{y} - \vec{1})$ if we use the Gauge transformation $\vec{x} \to y \vec{x}$, $y \to 1$ by symmetry of the labels and the data. The replica computation of the quenched average of the partition functions  \eqref{main:binary_perceptron:partition_functions} $\mathbb{E}_{\vec{y},\mat{X}}[\log(\mZ_\ndim(\mat{X}))]$
\begin{equation*}
	\mZ_\ndim (\vec{y}, \mat{X}) =  \int_\bbR \d y ~ \P_\y(y) ~
        \int_{\bbR^\ndim} \d \vec{w} ~ \P_\w(\vec{w}) \int \d \vec{z} ~ \,
        \mathcal{C}(\vec{z})\delta(\vec{z}-\mat{X}\vec{w}) \,,
\end{equation*}
and boils down to a free entropy formulation
\begin{equation}
\Phi (\alpha) = \extr_{\mat{Q}, \hat{\mat{Q}}} \left\{\lim_{r\rightarrow 0} \frac{\partial \Phi^{(r)} (\mat{Q},\hat{\mat{Q}}, \alpha)}{\partial  r} \right\} ,
\label{appendix:free_entropy2}
\end{equation}
as a function of symmetric overlap matrices $\mat{Q}\in\bbR^{r \times r}$ and $\td{\mat{Q}}\in\bbR^{r \times r}$ in the limit $r\to 0$:
\begin{align}
\begin{aligned}
     \Phi^{(r)}\(\mat{Q}, \hat{\mat{Q}}, \alpha\) &\equiv  -\frac{1}{2}\tr{\mat{Q}\hat{\mat{Q}}} +\log \Psi_{\w}^{(r)} (\hat{\mat{Q}})+\alpha\log \Psi_{\out}^{(r)}(\mat{Q})\,,
      \spacecase
      \Psi_{\w}^{(r)} (\hat{\mat{Q}}) &= \displaystyle \int_{\mathbb{R}^r} \d \rP_{\w}(\td{\vec{w}})  e^{ \frac{1}{2}\td{\vec{w}}^{\intercal} \hat{\mat{Q}} \td{\vec{w}} }  \,, \spacecase
     \Psi_{\out}^{(r)}(\mat{Q}) &=  \displaystyle \int_{\mathbb{R}^r}  \d \rP_{\z}(\td{\vec{z}},\mat{Q}) \mC(\td{\vec{z}}).
\end{aligned}
\end{align}
To obtain a tractable expression of the free entropy, in the following we perform the so-called \aclink{RS} and \aclink{1RSB} ansatz.

\subsection{RS calculation and stability}

\subsubsection{RS entropy}
The simplest ansatz is to assume that the overlap matrix $\mat{Q}$ is \aclink{RS}, which means that all replicas play the same role: the correlation between two arbitrary, but different, replicas is denoted $q_0$, and therefore the \aclink{RS} ansatz reads: 
\begin{equation*}
\forall (a,b) \in \lb r \rb^2, \qquad \frac{1}{\ndim} (\vec{w}^a\cdot\vec{w}^b) = 
	\begin{cases}
		q_0  \textrm{ if } a\ne b\, ,\\
		Q = 1 \textrm{ if } a = b \, .
	\end{cases}	
\end{equation*}
It enforces the matrix $\td{\mat{Q}}$ to present the same symmetry, respectively with parameters $\hat{q}_0$ and $\hat{Q}=1$. Using this Ansatz and the $r\to 0$ limit, the \aclink{RS} entropy can be expressed as a set of saddle point equations over scalar parameters $q_0$ and $\hat{q}_0$, evaluated at the saddle point, see \App\ref{appendix:replicas_iid:rs}, 
\begin{align}
\begin{aligned}
    \Phi^{(\rs)}(\alpha) &= \underset{ q_0, \hat{q}_0}{\textbf{extr}}     \left\{  \frac{1}{2}(q_0\hat{q}_0-1) +  \Psi_{\w}^{(\rs)}(\hat{q}_0)   +\alpha  \Psi_{\out}^{(\rs)}(q_0)    \right\},
\end{aligned}
\label{main:phi_RS}
\end{align} 
with 
\begin{align}
	\Psi_{\w}^{(\rs)}(\hat{q}_0) &\equiv \EE_{\xi_0}  \log ~ g_0^w (\xi_0,\hat{q}_0)\,, && \Psi_{\out}^{(\rs)}(q_0) \equiv \EE_{\xi_0}  \log ~ f_0^z(\xi_0, q_0) \,, 
\end{align}
\begin{align}
\begin{aligned}
		g_i^w (\xi_0,\hat{q}_0) &\equiv   \EE_{w} \[ w^i ~ \exp \( {\frac{(1 - \hat{q}_0  )}{2}  w^2}+ \xi_0 \sqrt{\hat{q}_0} w \)  \]\,, \\
		f_i^z(\xi_0,q_0) &\equiv \EE_{z} \[ z^i ~ \varphi(\sqrt{q_0} \xi_0 +
                \sqrt{1-q_0} z) \]\,,
           \end{aligned}
\label{main:f_z_g_w_rs}
\end{align}
for $i \in \bbN$ and where $\xi_0, z \sim \mN(0,1)$, $w\sim \rP_\w(.)$. In
  the binary perceptron case, the function $\rP_\w$ is defined as
  $\rP_\w(w)= [ \delta(w-1) + \delta(w+1) ]$ (note that this is not a
  probability distribution because of the normalization), and recall
  $\varphi(z)$ is the indicator function, checking that a constraint
  on the argument is satisfied (e.g in the step case, $\varphi^s(z) = 1$ if $z>K$).

While in the step binary perceptron (SBP) the fixed point solution
$(q_0,\hat{q}_0)$ is non-trivial, the symmetry of the activation
function in the RBP and UBP cases enforces the configuration space to
be symmetric and the fixed point $(q_0,\hat{q}_0)= (0,0)$ to exist. If
this symmetric fixed point is stable and has the lowest free energy,
the \aclink{RS} free entropy matches the annealed entropy
$\Phi^a_t (\alpha) = \log(2) + \alpha \log(p_{t,K}) = \lim_{\ndim \to \infty} \frac{1}{\ndim} \log \EE_{\mat{X}} [ \mZ_t ( \mat{X} ) ]$ from section~\ref{proof:first_moment} and \App\ref{appendix:annealed_calculation} with $t \in \{r,u\}$. 

\paragraph{Rectangle}
Solving numerically the corresponding saddle point equations leads to the single symmetric fixed point $(q_0,\hat{q}_0)=~(0,0)$. Hence the \aclink{RS} entropy saturates the first moment bound: 
\begin{align*}
	\Phi^{(\rs)}_r (\alpha) =  \log(2) + \alpha \log\( p_{r,K} \)  = \Phi^a_{r}(\alpha)\, ,
\end{align*}
and the \aclink{RS} capacity equals the annealed capacity eq.~\eqref{proof:first_moment}:
\begin{align*}
	\alpha_{\textrm{rs}}^{r}(K) = \alpha_a^{r}(K) = \frac{-\log(2)}{\log\( p_{r,K} \)} \, .
\end{align*}

\paragraph{$U$-function}
\begin{itemize}
		\item For $K  \leq K^*$, only the symmetric fixed point $(q_0,\hat{q}_0)= (0,0)$ exists, which leads again to the annealed free entropy:
		\begin{align*}
			\Phi^{(\rs)}_{u} (\alpha) =   \log(2)
                                + \alpha \log\( p_{u,K}
                               \)  =  \Phi^a_{u}(\alpha) \, ,
		\end{align*}
		and annealed capacity eq.~\eqref{proof:first_moment}:
		\begin{align*}
			\alpha_{\rs}^{u}(K) = \alpha_a^{u}(K) = \frac{-\log(2)}{\log\( p_{u,K} \)} \, .
		\end{align*}
		\item For $K >  K^*$, the \aclink{RS} entropy does not match
                  the annealed entropy because the fixed point
                  $(q_0,\hat{q}_0)\neq(0,0)$ corresponds to a lower
                  free energy than the symmetric fixed point
                  $(0,0)$. The symmetric fixed point becomes unstable for
                  $K>K^*$, where $K^*$ is remarkably given by the same value as in the independent section~\ref{proof:second_moment_u}. Hence it naturally verifies eq.~\eqref{AT_crossover_RS} even though its definition derives from the stability of the \aclink{RS} solution, that we study in the next section.
\end{itemize}

\subsubsection{RS stability}
\begin{figure}[htb!]
\centering
\includegraphics[scale=0.23]{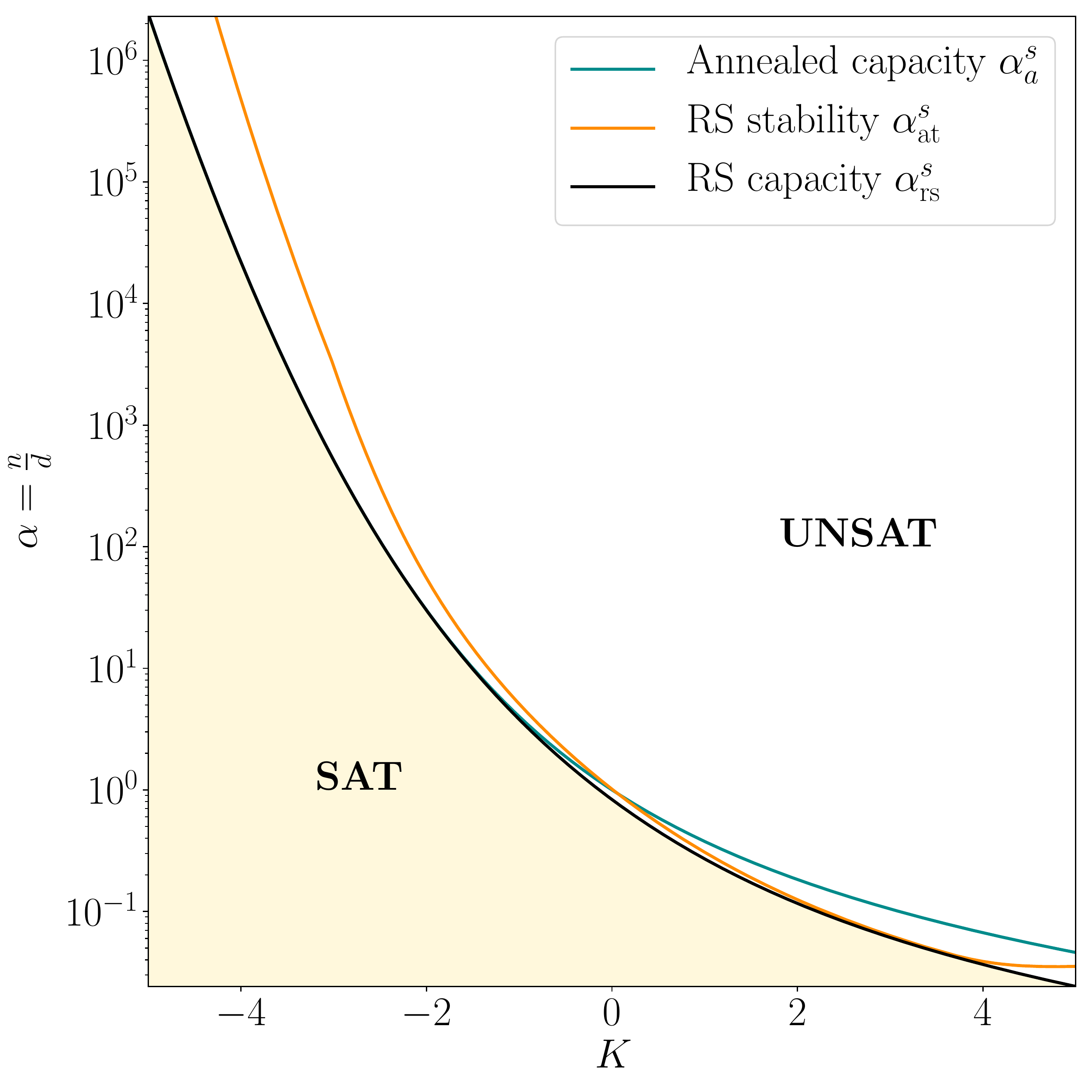}
\hfill
\includegraphics[scale=0.23]{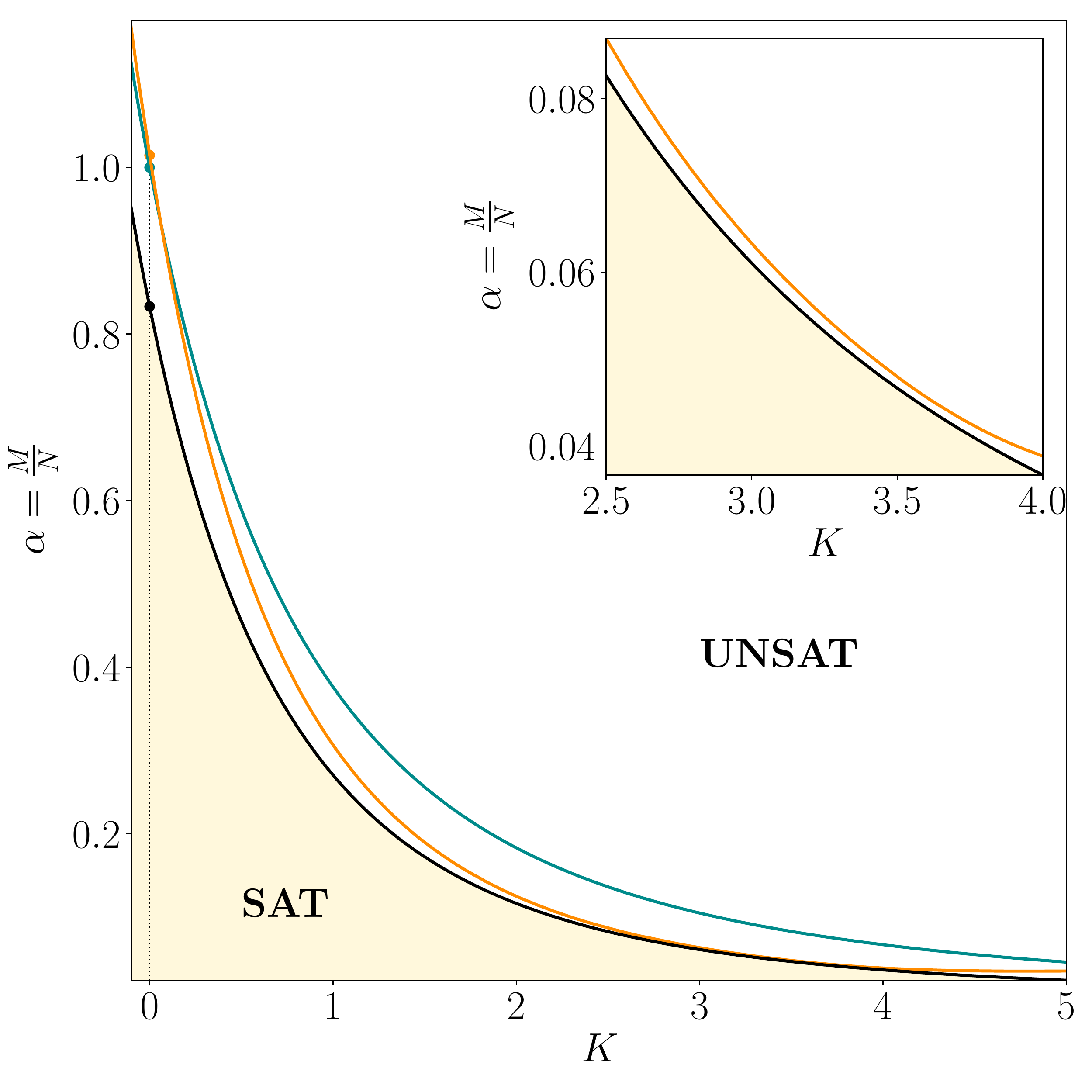}
\caption{Step binary perceptron (SBP): the RS capacity $\alpha_{\textrm{rs}}^s$
  (black) does not match the annealed capacity $\alpha_a^s$ (blue) and
  is always below the dAT-line $\alpha_{\textrm{at}}^s$ (orange). 
  The dAT-line is
  closest to the annealed capacity for $K_{\textrm{min}} \simeq 3.62$ where the
  difference $\alpha_{\textrm{at}}^s - \alpha_a^s \simeq 0.0012$. For $K=0$, we retrieve well known results \cite{krauth1989storage}:
  $\alpha_{\textrm{rs}}^r \simeq 0.833$, $\alpha_{\textrm{at}}^r \simeq 1.015$ and
  $\alpha_a^r = 1$. The left and right hand sides, and the inset, represent the same
  data on different scales. The satisfiable (SAT) phase is represented by the beige shaded area and is located below the RS capacity, while the unsatisfiable (UNSAT) starts at the capacity (black line) and extends for a larger number of constraints.}
	\label{main:plot_RS_capacity_step}
\end{figure}

\begin{figure}[htb!]
\centering
\includegraphics[scale=0.23]{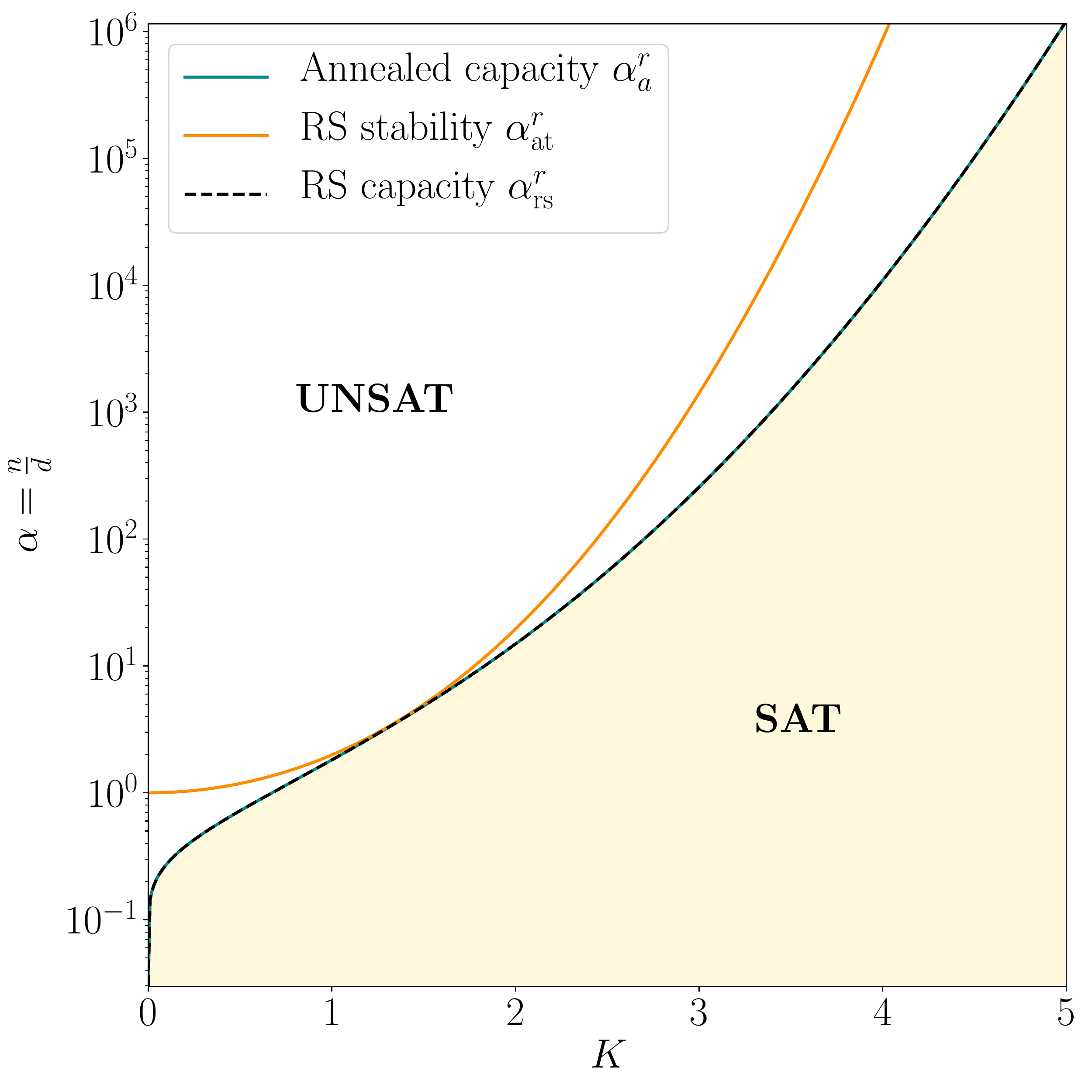}
\hfill
\includegraphics[scale=0.23]{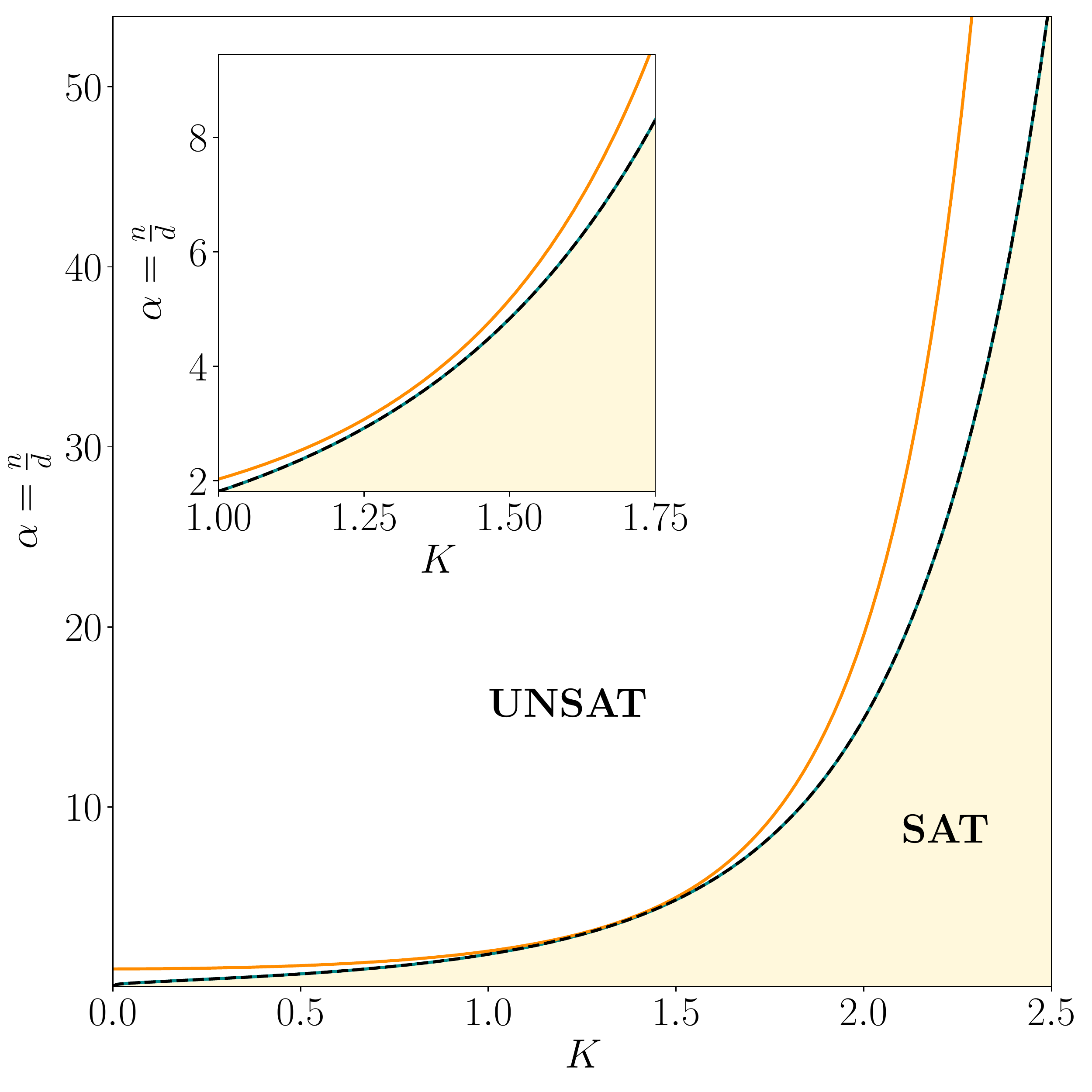}
\caption{Rectangle binary perceptron (RBP): the RS capacity
  $\alpha_{\textrm{rs}}^{r}$ (black) matches the annealed bound $\alpha_{a}^{r}$ (blue), and
  the RS solution is locally stable for all $K$:
  $\alpha_{\textrm{rs}}^{r}<\alpha_{\textrm{at}}^{r}$. The dAT-line (orange) is
  closest to the annealed capacity for $K_{\textrm{min}} \simeq 1.24$ where the
  difference $\alpha_{\textrm{at}}^s - \alpha_a^s \simeq 0.15$. The left and right hand sides, and the inset, represent the same
  data on different scales. The satisfiable (SAT) phase is represented by the beige shaded area and is located below the RS capacity, while the unsatisfiable (UNSAT) starts at the capacity (black line) and extends for a larger number of constraints.}
	\label{main:plot_RS_capacity_rectangle}
\end{figure}

\begin{figure}[htb!]
\centering
\includegraphics[scale=0.23]{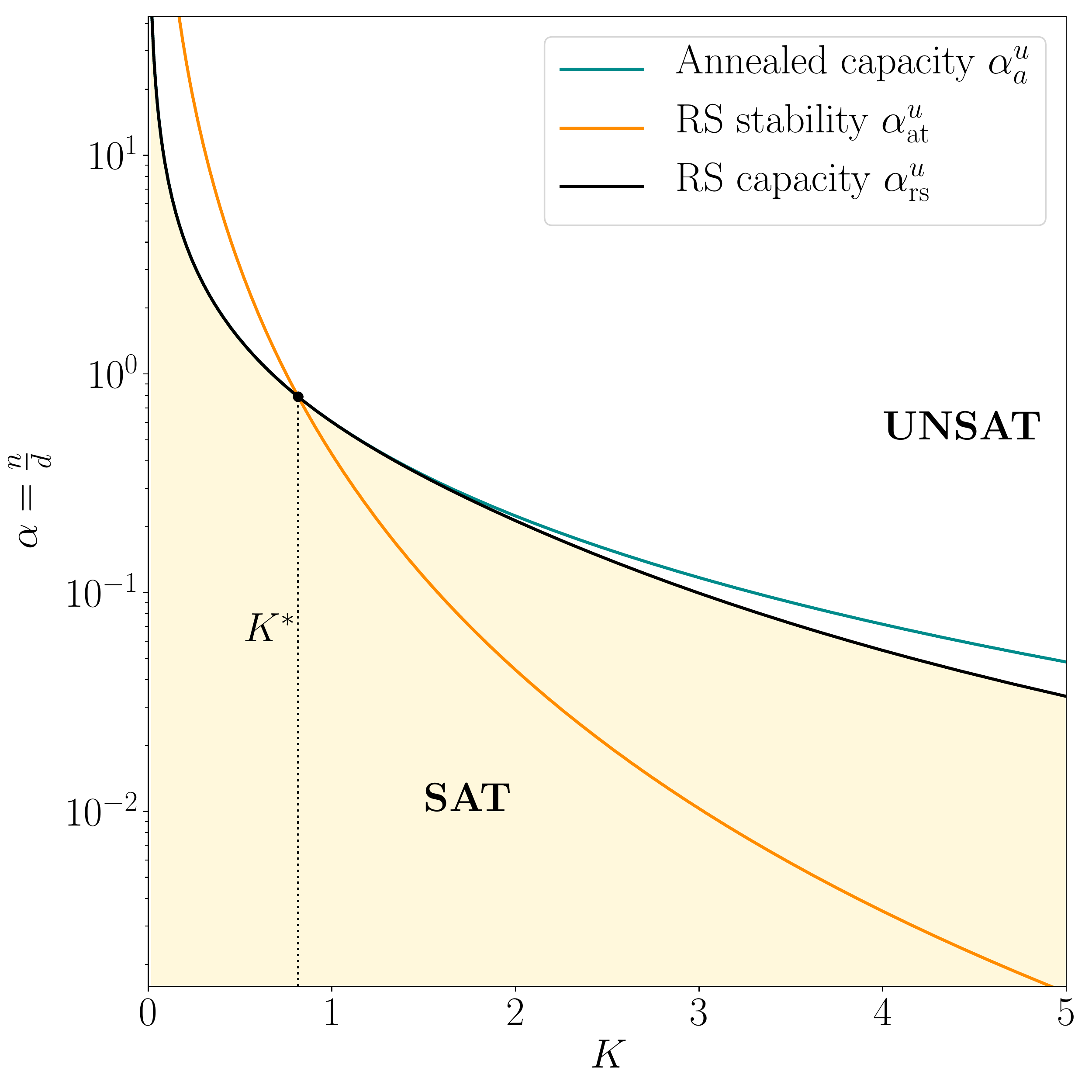}
\hfill
\includegraphics[scale=0.23]{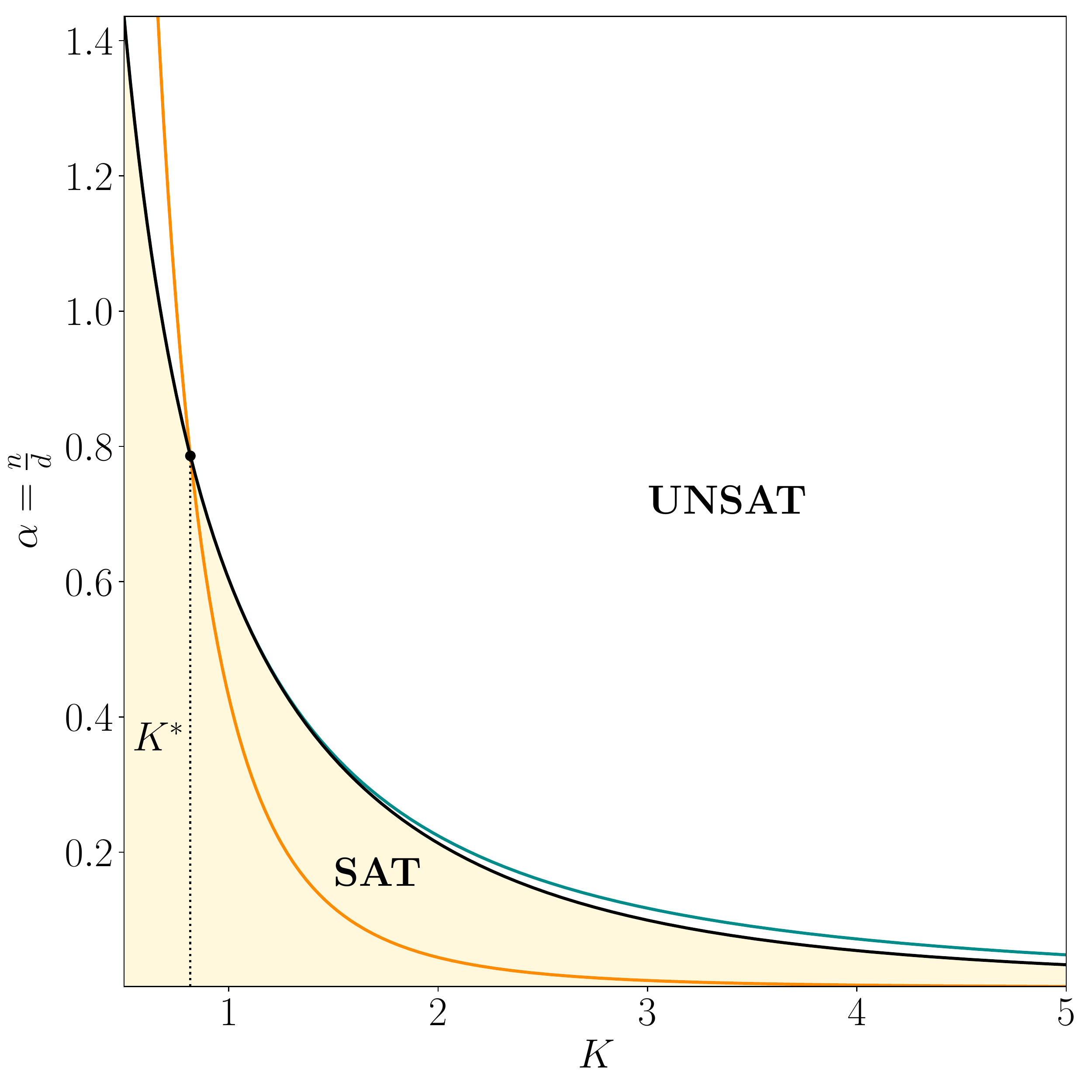}
\caption{$U-$function binary perceptron (UBP): the RS capacity Â§black) matches the
  annealed bound (blue) for $K<K^*$. At $K=K^*$, the RS capacity crosses the
  dAT-line (orange). For $K>K^*$, the RS solution is unstable and the RS
  capacity deviates from the annealed capacity. The left and right hand sides, and the inset, represent the same
  data on different scales. The satisfiable (SAT) phase is represented by the beige shaded area and is located below the RS capacity, while the unsatisfiable (UNSAT) starts at the capacity (black line) and extends for a larger number of constraints.}
	\label{main:plot_RS_capacity_symstep}
\end{figure}

The local stability of the \aclink{RS} solution can be studied using \aclink{dAT} method \cite{Almeida1978}, based on the positivity of the Hessian of $-\Phi^{(r)}(\mat{Q},\td{\mat{Q}} )$. The replica symmetric \aclink{dAT}-line $\alpha_{\textrm{at}}$ is given by the solution of the following implicit equation, derived in \App\ref{appendix:AT_stability}: 
\begin{align*}
\frac{1}{\alpha} &= \frac{1}{(1-q_0(\alpha))^2}
\EE_{\xi_0}\[\frac{\(f_0^{z}(f_0^{z}-f_2^{z}) + (f_1^{z})^2
  \)^2}{(f_0^{z})^4}( \xi_0 ,q_0(\alpha)) \]\\
  & \qquad \qquad \times \EE_{\xi_0} \[ \frac{
  \(g_0^{w}g_2^{w} -(g_1^{w})^2  \)^2
}{(g_0^{w})^4}(\xi_0,\hat{q}_0(\alpha)) \]\,.
 \end{align*}
As illustrated above, for the rectangle and $u-$function, the symmetry of the weights $\rP_\w$ and the constraint $\varphi$ imposes the existence of the symmetric fixed point $(q_0,\hat{q}_0)=(0,0)$. This simplifies the previous condition and becomes equivalent to the linear stability condition of the symmetric fixed point $(q_0,\hat{q}_0)=(0,0)$, see \App\ref{appendix:AT_stability}, 
\begin{equation*}
	\frac{1}{\alpha_{\textrm{at}}}=
        \(\frac{\td{f}_2^{z}-\td{f}_0^{z}}{\td{f}_0^{z}}\)^2
        \(\frac{\td{g}_2^{w} }{\td{g}_0^{w}}\)^2\,,
\end{equation*}
where for $i\in \bbN$
\begin{align*}
	\td{g}_i^{w} &= \EE_{w \sim\rP_\w} \[w^i ~ \exp \(w^2/2\)\] \,, &&
	\td{f}_i^{z} = \EE_{z \sim \mN(0,1) } \[ z^i ~ \varphi(z) \] \,.
\end{align*}
We plot the annealed capacity, the \aclink{RS} capacity and
the \aclink{dAT}-line for the step, rectangle and $u$-function binary
perceptrons as functions of $K$ in
\Fig\ref{main:plot_RS_capacity_step}, \ref{main:plot_RS_capacity_rectangle},
\ref{main:plot_RS_capacity_symstep}.
 
\paragraph{Step binary perceptron}
We note that for the step binary perceptron the \aclink{RS} solution is always stable towards \aclink{1RSB}, even for negative threshold $K<0$. This is interesting in the view of recent work on the spherical perceptron with negative threshold where the replica symmetry breaks for all $K<0$, and \aclink{FRSB} is needed to evaluate the storage capacity \cite{20}.

\paragraph{Rectangle}
As the \aclink{RS} capacity $\alpha_{\textrm{rs}}^{r}$ is always below the \aclink{dAT}-line $\alpha_{\textrm{at}}^{r}$, the \aclink{RS} solution is always locally stable.

\paragraph{$U$-function }
There is a crossing between the values of the  \aclink{RS} capacity $\alpha_{\textrm{rs}}^{u}$ and the  \aclink{dAT}-line $\alpha_{\textrm{at}}^{u}$, which defines implicitly the value $K^*\simeq 0.817$, and matches the equality in eq.~\eqref{main:AT_second_moment}: 
\vspace{-0.3cm}
\begin{equation}
	 \frac{-\log\(2 \)}{ \log\(p_{u,K^{*}}  \)} =
         \frac{\pi}{2}\frac{\( p_{u,K^{*}} \)^2}{ \exp\(-(K^*)^2\) (K^*)^2}\, .
	 \label{main:AT_crossover_RS}
\end{equation}
For $K\leq K^*$, the \aclink{RS} solution is locally stable, while for $K>K^*$ the \aclink{RS} solution becomes unstable, and a symmetry breaking solution appears.

\subsection{1RSB calculation and stability}

\subsubsection{1RSB entropy}
In the previous section we concluded that the replica symmetric
solution is unstable in the $u-$function binary perceptron for $K>K^*$,
we analyze therefore the first-step of replica symmetry breaking \aclink{1RSB}
Ansatz in this section. This ansatz and calculations is due to seminal
works of G. Parisi and is classic in the field of disordered systems and well 
presented in the literature
\cite{mezard1987spin,parisi1979infinite,parisi1980sequence,parisi1980order}, we
thus mainly give the key formulas and defer the
details into \App\ref{appendix:replica_1RSB}. 
\newpage
The \aclink{1RSB} Ansatz assumes that the space of configurations
splits into states. Consequently replicas are not symmetric anymore and instead $r$ replicas are organized in $\frac{r}{m}$ groups containing $m$ replicas each:
\begin{equation}
\forall (a,b) \in \lb r \rb^2, \qquad \frac{1}{\ndim}(\vec{w}^a\cdot\vec{w}^b) = 
	\begin{cases}
		q_1  \textrm{ if $a, b$ belong to the same state,}\\
		q_0 \textrm{ if $a, b$ do not belong to the same state,}\\
		Q = 1 \textrm{ if } a = b \, .
	\end{cases}	
	\label{main:1RSB_ansatz}
\end{equation}
Following \cite{monasson1995learning}, the replicated partition function $\mZ_m$
associated to $m$ replicas falling in the same state is expressed as a sum over all possible
states $\psi$ weighted by their corresponding free entropy $\Phi$:
\begin{align}
	\mZ_m &= \sum_{\{\psi\}} \exp(\ndim m \Phi(\psi)) =\sum_{\{\Phi\}} {\cal N}_{\Phi} \exp(\ndim m \Phi) \spacecase
	&= \sum_{\{\Phi\}}  \exp(\ndim \Sigma(\Phi)) \exp(\ndim m \Phi) \sim \int \d \Phi
	\exp(\ndim(m\Phi+\Sigma(\Phi))) \nonumber
\end{align}
where we introduced the number of states at a given free
entropy $\Phi$: $\mN_{\Phi}\equiv\exp(\ndim \Sigma(\Phi))$ and the
complexity $ \Sigma(\Phi)$, also called the configurational entropy. 
Using the saddle point method in the $\ndim \to \infty$ limit, the \aclink{1RSB}
replicated free entropy $\Phi_m^{(\textrm{1rsb})}$ is written as a function of the Parisi parameter $m$, the free entropy $\Phi$ and the complexity $\Sigma(\Phi)$:
\begin{equation}
	\Phi_m^{(\textrm{1rsb})}(m,\alpha) \equiv  \lim_{\ndim \to \infty} \frac{1}{\ndim}
        \mathbb{E}_{\mat{X}}\left[\log( \mZ_m(\mat{X}) )\right]  = m
        \Phi + \Sigma(\Phi) \, .
        \label{main:phi_1RSB_phi_m_Sigma}
\end{equation}
Injecting the \aclink{1RSB} ansatz eq.~\eqref{main:1RSB_ansatz}, the \aclink{1RSB} replicated free
entropy $\Phi_m^{(\textrm{1rsb})} = m \Phi^{(\textrm{1rsb})} $ is written as a saddle point equation over
$\vec{q}=(q_0,q_1)$ and $\td{\vec{q}}=(\hat{q}_0,\hat{q}_1)$, see \App\ref{appendix:replica_1RSB}:
\begin{multline}
	\Phi_m^{(\textrm{1rsb})}(m, \alpha) = m \cdot \underset{ \vec{q}, \hat{\vec{q}}}{\textbf{extr}} \left\{  \frac{1}{2} \(  q_1\hat{q}_1 - Q\hat{Q} \) + \frac{m}{2} \(q_0\hat{q}_0 - q_1\hat{q}_1 \)   \right. \\
	\left. + \td{\Psi}_{\w}^{(\textrm{1rsb})}(\hat{\vec{q}}, m)   +\alpha \td{\Psi}_{\out}^{(\textrm{1rsb})}(\vec{q}, m)    \right\} \,,
\end{multline}
with
\begin{align}
\begin{aligned}
	\td{\Psi}_{\w}^{(\textrm{1rsb})}(\hat{\vec{q}}, m) &\equiv \frac{1}{m}
                \EE_{\xi_0} \log\( \EE_{\xi_1} \(g_0^w\)^{m}\) \,,  \\
    \td{\Psi}_{\out}^{(\textrm{1rsb})}(\vec{q}, m) &\equiv \frac{1}{m}
                 \EE_{\xi_0}  \log\(  \EE_{\xi_1} \(f_0^z\)^{m} \)\,,       
\end{aligned}
\end{align}
and
\begin{align}
\begin{aligned}
	g_i^w (\bxi,\vec{q})  &= \EE_w \[w^i ~ \exp\(
                  \frac{(1 - \hat{q}_1  )}{2} w^2 +
                  \(\sqrt{\hat{q}_0}\xi_0+\sqrt{\hat{q}_1-\hat{q}_0}\xi_1
                  \)w \)  \] \,,\\
	f_i^z (\bxi,\vec{q})  &= \EE_{z \sim \mN(0,1) } \[ z^i ~ \varphi\( \sqrt{1-q_1} z + \sqrt{q_0} \xi_0 +
                \sqrt{q_1-q_0} \xi_1 \) \]\,,
\label{main:f_i_g_i}
\end{aligned}
\end{align}
for  $\bxi=(\xi_0, \xi_1)$ and for $i\in \bbN$. Taking the derivative of $\Phi_m^{(\textrm{1rsb})}$ with respect to $m$, the \aclink{1RSB} free entropy $\Phi^{(\textrm{1rsb})}$ and complexity $\Sigma$ can be expressed as: 
\begin{align}
		&\Phi^{(\textrm{1rsb})}(\alpha) = \frac{\partial \Phi_m^{(\textrm{1rsb})}(m,\alpha)  }{\partial m }  \\
		&= \underset{\vec{q} , \td{\vec{q}}, m}{\extr} \left\{ \frac{1}{2} (  q_1\hat{q}_1 - 1 ) +   m \(q_0\hat{q}_0 - q_1\hat{q}_1 \) \right. \nonumber \\
	& \left.  \qquad \qquad \qquad \qquad \qquad +\Psi_{\w}^{(\textrm{1rsb})}(\td{\vec{q}}, m) + \alpha
        \Psi_{\out}^{(\textrm{1rsb})}(\vec{q}, m) \right\} \,, \nonumber \\
	&\Sigma (\Phi^{(\textrm{1rsb})}) =  \Phi_m^{(\textrm{1rsb})}- m \Phi^{(\textrm{1rsb})} \label{main:complexity} \\
	&= \underset{\vec{q} ,
          \td{\vec{q}}, m}{\extr} \left\{
          \frac{m^2}{2}(q_1\hat{q}_1 - q_0\hat{q}_0) +
          m\(\td{\Psi}_{\w}^{(\textrm{1rsb})} - \Psi_{\w}^{(\textrm{1rsb})}\)(\td{\vec{q}}, m) \right. \nonumber	\\     
    & \left. \quad \quad \qquad \qquad \qquad \qquad  + m \alpha \(\td{\Psi}_{\out}^{(\textrm{1rsb})}
         -\Psi_{\out}^{(\textrm{1rsb})}\)(\vec{q}, m)\right\}\,, \nonumber	
\end{align}
with
\begin{align*}
		\Psi_{\w}^{(\textrm{1rsb})}(\td{\vec{q}}, m) &=\partial_m \(m \td{\Psi}_{\w}^{(\textrm{1rsb})} \) = \EE_{\xi_0} \[ \frac{\EE_{\xi_1}\[  \log\(g_0^w(\bxi, \vec{q})\) g_0^w(\bxi, \vec{q})^m \]}{\EE_{\xi_1}\[ g_0^w(\bxi, \vec{q})^m \]} \]\,, \spacecase 
		\Psi_{\out}^{(\textrm{1rsb})}(\vec{q}, m) &= \partial_m
                  \(m \td{\Psi}_{\out}^{(\textrm{1rsb})} \) = \EE_{\xi_0} \[ \frac{\EE_{\xi_1}\[  \log\(f_0^z(\bxi, \vec{q})\) f_0^z(\bxi, \vec{q})^m \]}{\EE_{\xi_1}\[ f_0^z(\bxi, \vec{q})^m \]} \]\,.
\end{align*}

\subsubsection{1RSB results for UBP}
\begin{figure}
\centering
\includegraphics[width=0.45\linewidth]{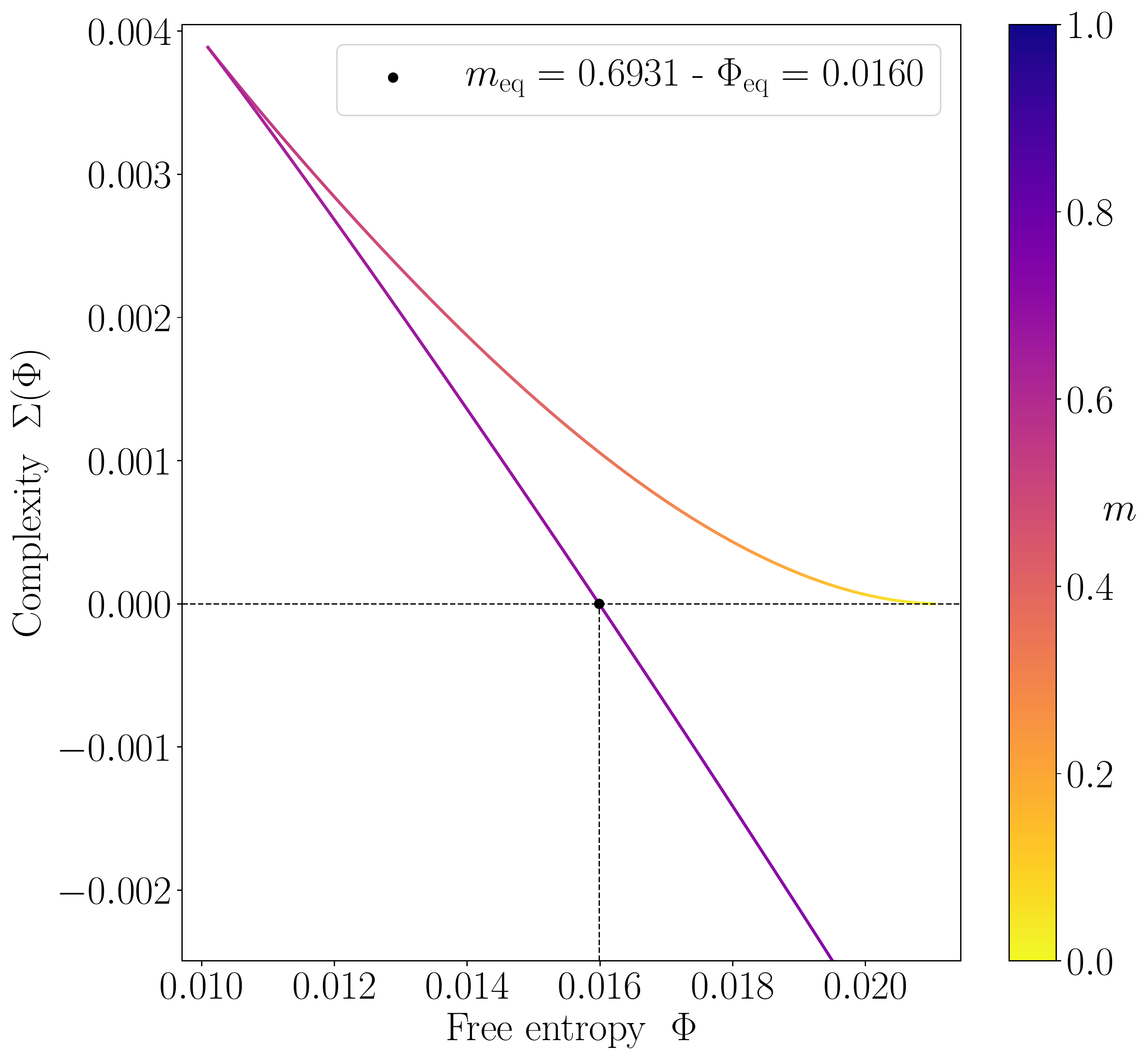}
	\hspace{0.2cm}
	\includegraphics[width=0.45\linewidth]{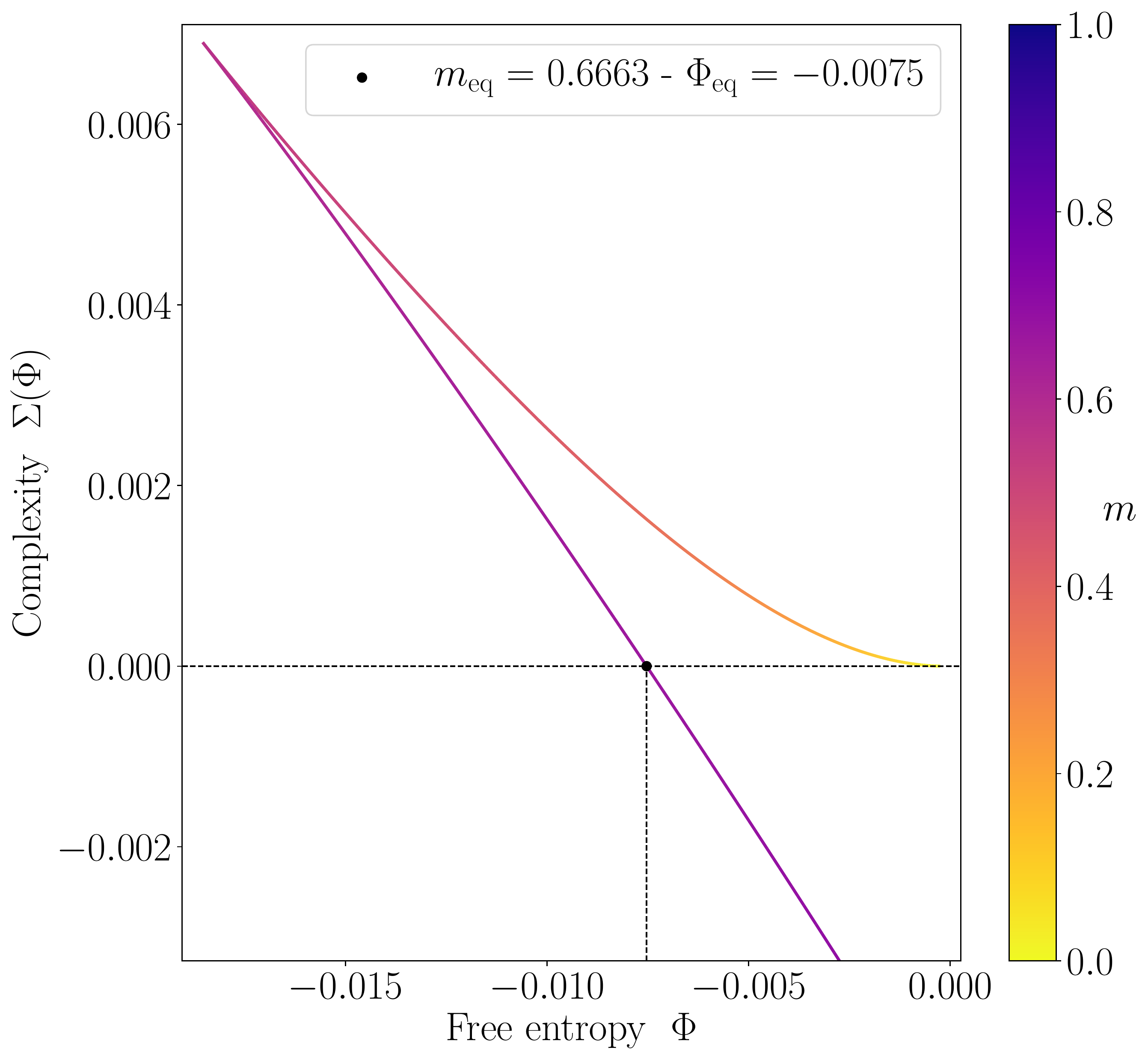}
	\caption{Complexity $\Sigma(\Phi)$ as a function of the free
          entropy $\Phi$ for the $u-$function binary perceptron  at
          $K=1.5>K^*$. Complexity reaches $\Sigma=0$ (black dot) at
          $m_{\textrm{eq}}$. For $K=1.5$ and $\alpha=0.33$ \Left the free-entropy
          corresponding to $m_{\textrm{eq}}$ is positive $\Phi_{\textrm{eq}} >0$, whereas for
          $\alpha=0.34$, \Right the free entropy at $m_{\textrm{eq}}$ is negative $\Phi_{\textrm{eq}} < 0$ and therefore there is no part of the curve where both
          complexity and free entropy are positive: thus this value
          of $\alpha$ is beyond the 1RSB storage capacity, and the capacity is in the interval $[0.33;0.34]$.}
	\label{main:complexity_curves}
	\end{figure}

From now on, we only consider the $u-$function binary perceptron, whose \aclink{RS} solution is unstable for $K>K^*$.
To describe the equilibrium of the system in the SAT phase, we need to find the value of the Parisi parameter at equilibrium $m_{\textrm{eq}}$. The complexity $\Sigma(\Phi)$ is the entropy of clusters having internal entropy $\Phi$. In order to capture clusters that carry almost all configurations, we need to maximize the total entropy $\Phi_{\textrm{tot}}  = \Sigma(\Phi) + \Phi$ under the constraint that the free entropy and complexity are both positive $\Phi \geq 0$ and $\Sigma(\Phi)\geq 0$. Hence from eq.~\eqref{main:phi_1RSB_phi_m_Sigma}, the equilibrium Parisi parameter $m_{\textrm{eq}}$ verifies 
\begin{equation}
	\Phi_{\textrm{eq}}=\underset{\Phi \geq 0,\Sigma \geq
          0}{\textrm{argmax }}{ \{ \Phi + \Sigma(\Phi) \} } \hspace{1cm} \textrm{and} \hspace{1cm}
    m_{\textrm{eq}} =    \left. - \frac{d \Sigma}{d \Phi} \right
    |_{\Phi_{\textrm{eq}}}\, .  \label{eq:meq}
\end{equation}
As a side remark, we note that in the Parisi's replica theory the more commonly known condition for
$m_{\textrm{eq}}$ is obtained by extremizing the (rescaled) replicated free entropy
$\Phi_m^{(\textrm{1rsb})}(m,\alpha) /m$ which leads using
(\ref{main:phi_1RSB_phi_m_Sigma}) to the condition $-\frac{\Sigma(\Phi_\textrm{eq})}{m^2}=0$. This extrema is in fact a minima as $\Sigma(\Phi)$ is concave and $m = -\frac{d \Sigma}{d \Phi}$. This is, however, only valid when $m_{\textrm{eq}}<1$, and is
moreover highly counter-intuitive as physical systems maximize entropy
whereas here one minimizes it. We hence prefer to use the formulation
of eq.~(\ref{eq:meq}) which we find physically better justified.
Using the expressions eq.~\eqref{main:complexity} and varying the Parisi
parameter $m\in[0;1]$, we obtain the curve of the complexity
$\Sigma(\Phi)$ as shown in \Fig\ref{main:complexity_curves}. At
$m=1$, the complexity is negative. Decreasing $m$, the complexity
increases and becomes positive at the value $m_{\textrm{eq}}$. Besides for
small values of $m$, an unphysical (convex) branch appears, as commonly
observed in other systems solved by the replica method.

We note that as $\alpha$ increases both the equilibrium complexity and
free entropy decrease. In \aclink{CSP} such as
k-SAT or random graph coloring the mechanism in which the
satisfiability threshold appears is that the maximum of the complexity
becomes negative. In the present UBP problem it is actually both the
free entropy and the complexity that vanish together, as illustrated
in \Fig\ref{main:complexity_curves}.\\

\begin{figure}[htb!]
\centering
\includegraphics[width=0.55\linewidth]{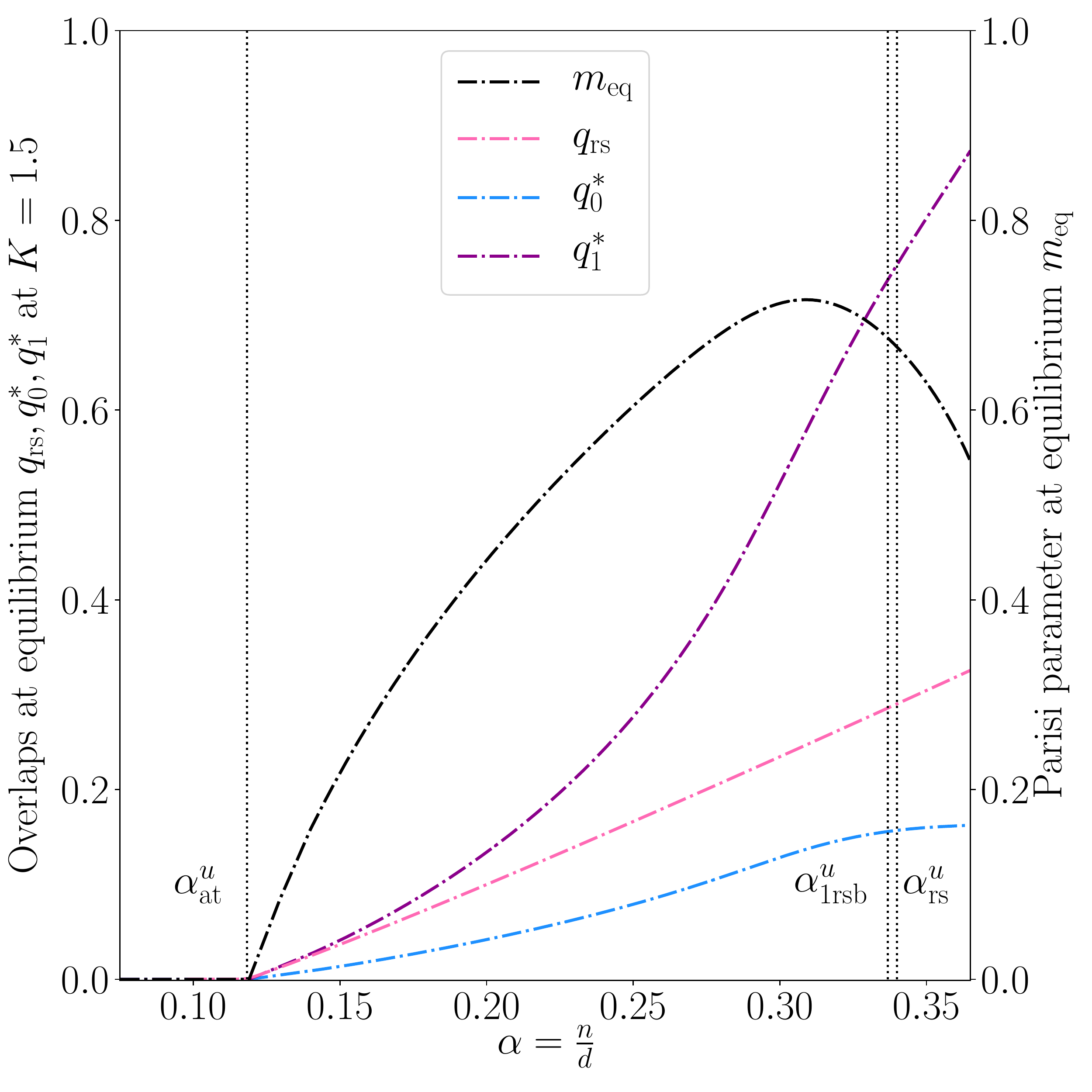}
	\caption{Equilibrium values of the overlap
          $q_0^*\ne q_{\textrm{rs}}$, $q_1^*$ and the Parisi parameter $m_{\textrm{eq}}$ for the UBP at
          $K=1.5$. For $K<K^*$, the RS solution is stable and the only fixed point is $q_0^*=q_1^*=q_{\rs}=0$. }
	\label{main:plot_1RSBmeq}
	\label{main:plot_entropy_1RSB}
\end{figure}

\begin{figure}[htb!]
		    \centering
		    \includegraphics[width=0.49\linewidth]{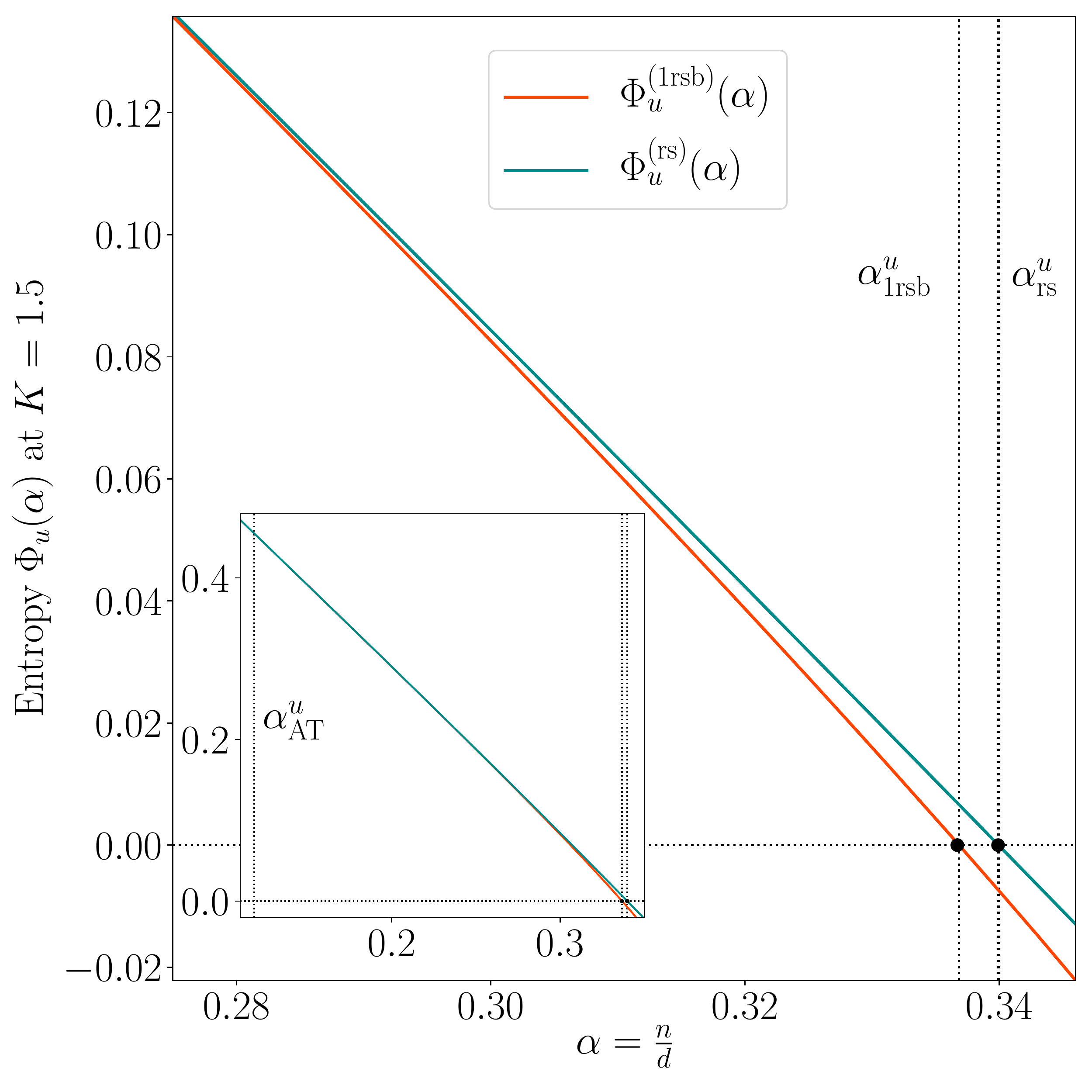}
		    \hfill
	   		\includegraphics[width=0.49\linewidth]{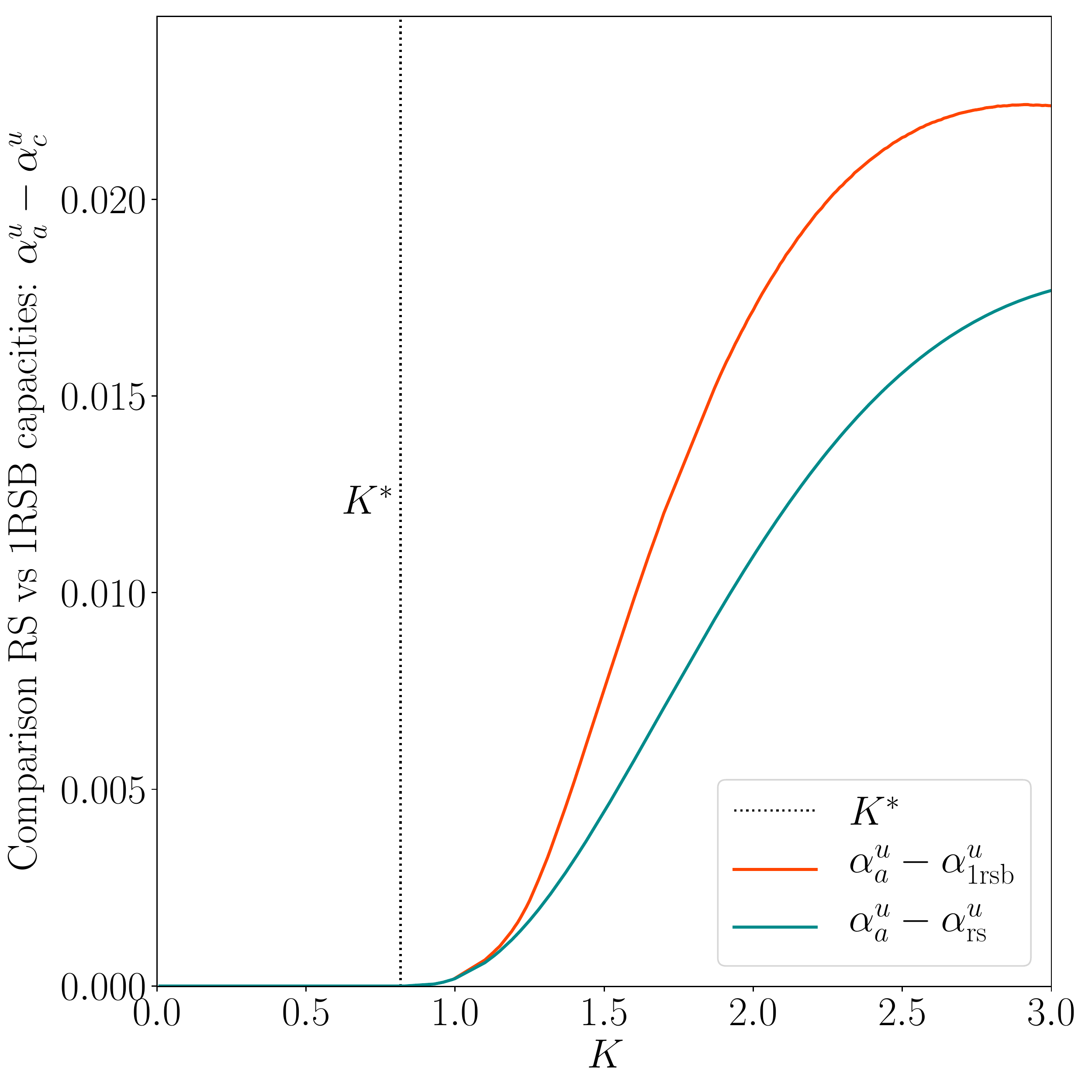}
	   		\caption{\Left Comparison of the RS (blue) and 1RSB (orange) entropy
          for the UBP at $K=1.5$. For $\alpha<\alpha_{\textrm{at}}\simeq 0.118 $, RS and
          1RSB entropies are equalled. For $\alpha>\alpha_{\textrm{at}}$,
          1RSB entropy deviates slightly of the RS entropy before vanishing respectively at $\alpha_{\textrm{1rsb}}^{u}\simeq 0.337 $ and $\alpha_{\textrm{rs}}^{u}\simeq 0.334$. The inset represents the same data on a different scale. \Right Difference between the annealed upper
                          bound and the 1RSB capacity
                          $\alpha_a^{u}-\alpha_{\textrm{1rsb}}^{u}$ (orange) and the RS capacity
                          $\alpha_a^{u}-\alpha_{\textrm{rs}}^{u}$ (blue). Below $K^*$ the RS solution is stable: RS and 1RSB entropies match exactly. Above $K^*$, the RS solution is unstable: the 1RSB entropy deviates slightly from the RS solution.}
	   		\label{main:plot_annealed_minus_replicas}
\end{figure}

Computing the equilibrium value $m_{\textrm{eq}}(\alpha)$, we have access to
the corresponding equilibrium overlaps $q_0^*$ and $q_1^*$, that we
may compare with the \aclink{RS} solution $q_{\textrm{rs}}$. All these are depicted
in \Fig\ref{main:plot_1RSBmeq}. 
The function $m_{\textrm{eq}} ( \alpha )$ shows a non monotonic behaviour as it has been previously observed, e.g. in the \aclink{SK} model as a function of temperature \cite{mezard1987spin}.
 We also compute the \aclink{1RSB} entropy that verifies $\Phi^{(\textrm{1rsb})}_{u} \leq \Phi^{(\rs)}_{u}$ and which vanishes at the \aclink{1RSB} capacity $\alpha_{\textrm{1rsb}}^{u}$ as
depicted in \Fig\ref{main:plot_annealed_minus_replicas}~\Leftn. We note that the above inequality is as predicted by Parisi's replica theory \cite{mezard1987spin}, taking into account that we are working at strictly zero energy, where the entropy becomes minus the free energy. 
The \aclink{1RSB} solution provides a small correction to the \aclink{RS} result for storage capacity, as illustrated in
\Fig\ref{main:plot_annealed_minus_replicas}~\Rightn, where we plotted the
difference between the annealed upper bound and the capacity for the \aclink{RS}
and \aclink{1RSB} solutions: $\alpha_a^{u}-\alpha_{\textrm{rs}}^{u}$ and
$\alpha_a^{u}-\alpha_{\textrm{1rsb}}^u$. 

\subsubsection{1RSB Stability}
In the previous section we evaluated the \aclink{1RSB} storage capacity of the
$u-$function binary perceptron for $K>K^*$. In this section we will argue that this cannot be an exact solution to the problem. 
We could investigate the stability of \aclink{1RSB} towards further levels of
replica symmetry breaking along the same lines we did for the \aclink{RS}
solution. However, in the present case we do not need to do that to
see that the obtained solution cannot be correct. The explanations
lies in the breaking of the up-down symmetry in the problem. This
symmetry must either be broken explicitly as in the ferromagnet,
where the system would acquire an overall magnetization, but we have
not observed any trace of this in the present problem. Or this up-down 
symmetry must be conserved in the final correct solution. The
conservation of the up-down symmetry is manifested in the value
$q_0=0$ in the replica symmetric phase. The fact that in the \aclink{1RSB}
solution evaluated above we do not observe $q_0=0$, but instead
$q_0>0$ is a sign of the fact that we are evaluating a wrong
solution. The only possible way to obtain an exact solution we foresee
is to evaluate the full-step replica symmetry breaking with a
continuity of overlaps $q(x)$, the smallest one of them should be $0$
in order to restore the up-down symmetry.  We let the evaluation of
the \aclink{FRSB} for future work. 
 
Finally let us note that the \aclink{1RSB} solution obtained in the previous
section can be interpreted as frozen-\aclink{2RSB}. In \aclink{2RSB} we would have 3
kinds of overlaps, $q_0$, $q_1$ and $q_2$. In frozen \aclink{2RSB} we would have
$q_2=1$, $q_1=q_1^{\textrm{1rsb}}$, $q_0=q_0^{\textrm{1rsb}}$. 
\section*{Conclusion}
In this chapter we analyzed a class of symmetric binary perceptron
problems that are simple variants of the canonical step-function
binary perceptron. 
The step-function binary perceptron has thus far eluded a
rigorous establishment of the conjectured storage capacity, eq.~\eqref{RS_capacity}. This prediction
is expected to be exact because of the \aclink{f1RSB} nature of the
problem \cite{krauth1989storage}. At the same time the work of
\cite{baldassi2015subdominant} sheds light on the fact that the
structure of the space of solutions is not fully described by the
\aclink{f1RSB} picture, and that rare dense and unfrozen regions exist and
in fact are amenable to dynamical procedures searching for solutions. It
remains to be understood how is it possible that the \aclink{1RSB} calculation
does not capture these dense unfrozen regions of solutions
\cite{baldassi2015subdominant}. 
They do not dominate the equilibrium, but the \aclink{RSB} calculation is
expected to describe rare events via their large deviations, which in
this case it does not. 

We focus on two cases of the binary perceptron with
symmetric constraints, the rectangle binary perceptron and the
$u-$function binary perceptron. We prove (up to a numerical assumption) using the second moment method that the
storage capacity agrees in those cases with the annealed upper bound,
except for the $u-$function binary perceptron for $K>K^*$ eq.~\eqref{AT_crossover_RS}. 
We analyze the \aclink{1RSB} solution in that case and indeed obtain a lower
prediction for the storage capacity. However, we do not expect the
\aclink{1RSB} to provide the exact solution because it does not respect the
up-down symmetry of the problem. Though the precise nature of the satisfiable phase for the $u-$function binary
perceptron for $K>K^*$ remains illusive, we can conjecture it is
\aclink{FRSB} \cite{parisi1979infinite,parisi1980sequence,parisi1980order}. Establishing this rigorously would provide much deeper understanding and remains a challenging subject for future work. 
	
	\ifthenelse{\equal{\format}{oneside}}
	{
	\clearpage\null\thispagestyle{empty}\newpage
	\clearpage\null\thispagestyle{empty}\newpage
	}
	{\clearpage\null\thispagestyle{empty}\newpage
	\clearpage\null\thispagestyle{empty}\newpage
	\cleardoublepage}
	
	\chapter{Rademacher complexity and spin glasses: A link between the replica and statistical theories of learning}
	\chaptermark{Rademacher complexity and ground state energies of perceptrons}
	\label{chap:rademacher}
	\aclink{ERM} is the workhorse of most of modern supervised machine learning successes. Consider for instance a data-set $\{y_{\mu} ,\vec{x}_\mu \}_{\mu=1}^\nsamples$ of $\nsamples$ examples ${\bx}_\mu \in \mathbb{R}^d$ assumed to be drawn from a distribution $\rP_x(.)$, with labels $y_\mu \in \{ -1, +1 \}$ used for a binary classification task. We consider an estimator $f_{\vec{w}}(.)$ that belongs to a \emph{hypothesis class} $\mF$, for instance a neural network or a linear function, with respective weights or parameters $\vec w$. The latter are typically computed by minimizing the empirical risk
\begin{equation}
    {\mR}_{\textrm{empirical}}^\nsamples ( f_{\vec{w}} ) = \frac{1}{\nsamples} \sum_{\mu=1}^\nsamples {\mL}\left( y_\mu ,f_{\vec w} \left( \vec{x}_\mu\right) \right)  \nonumber \,
\end{equation}
over $\vec{w}$, where $\mL$ denotes a loss function, e.g. the mean-squared-loss ${\cal L}(a,b)=(a-b)^2$. The main theoretical issue of statistical learning theory concerns the performance of the estimator $f_{\vec{w}}(.)$ obtained by such a minimization on yet unseen data, namely the \emph{generalization problem}. In fact, what we really hope to minimize is the population risk, defined as
\begin{equation}
    {\mR}_{\textrm{population}} (f_{\vec{w}}) = 
    \mathbb{E}_{y,{\textbf x}} \left[{\cal L}(y ,f_{\vec{w}}({\vec{x})})\right]  \nonumber \,.
\end{equation}
Since we are optimizing the empirical risk instead, the difference between the two might be arbitrarily large. Bounding this difference between \emph{empirical} and \emph{population} risks is therefore a major problem of statistical learning theories.

In a large part of the literature, statistical learning analysis, see e.g.~\cite{bartlett2002rademacher,vapnik2013nature,shalev2014understanding} relies on the \aclink{VC} analysis and on the so-called \emph{Rademacher complexity}. The latter is a measure of the complexity of $\mF$,  the hypothesis class spanned by $f_{\vec{w}}(.)$, to bound ${{\mR}_{\textrm{population}} - {\cal R}^\nsamples_{\textrm{empirical}}}$, the \emph{generalization gap}. A gem within the literature is the Uniform Convergence result which states the following: if the Rademacher complexity or the \aclink{VC} dimension is finite, then for a large enough number of samples the generalization gap will vanish uniformly over all possible values of parameters ${\vec w}$. Informally, uniform convergence tells us that with high probability, for any weights value $\vec{w}$, the generalization gap satisfies
\vspace{-0.25cm}
\begin{equation}
  {\mR}_{\textrm{population}}(f_{\vec{w}}) - {\cal R}_{\textrm{empirical}}^\nsamples(f_{\vec{w}}) = \Theta\left(\sqrt{\frac{d_{\vc}(\mF)}{\nsamples}}\right)\,,
  \label{eq:VC-bound}
  \vspace{-0.2cm}
\end{equation}
where $d_{\vc}(\mF)$ denotes the \aclink{VC} dimension of the hypothesis class $\mF$. Tighter bounds can be obtained using the Rademacher complexity. These  bounds, although useful, do not seem to fully explain the success of current deep-learning architectures \cite{zhang2016understanding}.

Over the last four decades, a different vision of generalization  --- based on the analysis of \emph{typical case} problems with synthetic data created from simple generative models --- was developed to a large extent in the statistical physics literature, see e.g. \cite{seung1992statistical,watkin1993statistical,opper1995statistical,engel2001statistical} for a review. The link with the \aclink{VC} dimension was discussed in many of these works, notably via its connection with its  twin from statistical physics, the \emph{Gardner capacity} \cite{gardner1988optimal}. In particular, one can show that the \aclink{VC} capacity is always larger than half of the Gardner one \cite{engel2001statistical}. We shall review this discussion later on in this paper. 
However, to the best of our knowledge the Rademacher complexity was absent from these considerations. This omission is unfortunate: not only does the Rademacher complexity give tighter bounds than the \aclink{VC} dimension, it also intrinsically connects with a quantity that physicists are familiar with and have been computing from the very beginning of their studies, namely the average \emph{ground-state energy}. 

The goal of the present chapter is to bridge this gap and unveil the deep link between ground-state energy and Rademacher complexity, and how this connection is valuable to both parties. The chapter is organized as follows: After giving proper definitions of common generalization bounds in sec.~\ref{sec:rademacher}, we detail calculations of Rademacher complexities for simple function classes in sec.~\ref{sec:iid}. These sections serve as an introduction to the readers not familiar with these notions. The subsequent sections~\ref{sec:statistical physics} and \ref{sec:applications_rademacher} provide the original content of this work.
\paragraph{Here we summarize the main contributions of this work:}
\begin{itemize}
 \setlength\itemsep{0.1em}
    \item  We point out the one-to-one connections between the Rademacher complexity in statistical learning, and the ground-state energies and Gardner capacity from statistical physics.
    \item We show how the heuristic replica method from statistical physics can be used to compute the Rademacher complexity in the high-dimensional statistics limit and reinterpret classical results of the statistical physics literature as Rademacher bounds in the case of perceptron and committee machines models with \aclink{i.i.d} data.
    \item  We contrast these results with the generalization in the teacher-student scenario, illustrating the worst-case nature of the Rademacher bound that fails to capture the typical-case behavior.
    \item We finally show \emph{en passant}, that learning theory also bears consequences for the spin glass physics and the related replica symmetry breaking scheme by showing it implies  strong constraint on the ground-state energy of some spin glass models.
\end{itemize}

\section{A primer on Rademacher complexity \label{sec:rademacher}}

The bound of the generalization gap involving the \aclink{VC} dimension is specific to binary classification, and does not depend on the data distribution. While this is a strong property, the Rademacher approach does depend on data distribution and allows for tighter bounds. Moreover, it generalizes to multi-class classification and regression problems. We recall the definition of the Rademacher complexity:
\begin{definition}
	Let $f_{\vec{w}}$ be any function in the hypothesis class $\mF$, and let $\bepsilon \in \{ \pm 1 \}^\nsamples$ be drawn uniformly at random. The \textbf{empirical Rademacher complexity} is defined as
	\begin{align}
		\hat{\mathfrak{R}}_\nsamples \( \mF, \mat{X} \) \equiv \EE_{\bepsilon} \[ \sup_{f_{\vec{w}} \in \mF} \frac{1}{\nsamples} \sum_{\mu =1}^\nsamples  \epsilon_\mu  f_{\vec{w}}\(\vec{x}_\mu\) \] \,,
		\label{main:def_empirical_rademacher}
	\end{align} 
	and depends on the sample examples $\mat{X} = \{\vec{x}_{1}, \dots \vec{x}_{\nsamples} \} \in \bbR^{d \times \nsamples}$. The \textbf{Rademacher} \textbf{complexity} is defined as the population average
	\begin{align}
		\mathfrak{R}_\nsamples \( \mF \) \equiv \EE_{\mat{X}} \[\hat{\mathfrak{R}}_\nsamples \( \mF, \mat{X} \) \] \,.
		\label{main:def_rademacher}
	\end{align} 
\end{definition}

In this chapter, we focus on binary classification and consider the corresponding loss function $\mL(a,b) = \id \[ a \ne b \]$ that counts the number of misclassified samples. We will be therefore interested in a hypothesis class $\mF =\left \{ f_{\vec{w}}: \bbR^d\longrightarrow \{\pm 1 \} \right\}$. Defining the training $\epsilon_{\train}^\nsamples(.)$ and generalization errors $\epsilon_{\textrm{gen}}(.)$ for any function $f_{\vec{w}}\in \mF$ by
\begin{align}
\begin{aligned}
    \epsilon_{\train}^\nsamples (f_{\vec{w}}) & \equiv \frac{1}{\nsamples} \sum_{\mu=1}^{\nsamples} \id \[ y_\mu \ne f_{\vec{w}}\(\vec{x}_\mu\) \]\,, \\
    \epsilon_{\textrm{gen}} (f_{\vec{w}}) &\equiv \EE_{y,\vec{x}} \[ \id \[ y \ne f_{\vec{w}}\(\vec{x}\) \] \]\,,
\end{aligned}
\end{align}
the Rademacher complexity  provides a generalization error bound as expressed by the following theorem, and many of its variants, see e.g. \cite{bartlett2002rademacher,vapnik2013nature,shalev2014understanding,Mohri12}:
\begin{theorem}{Uniform convergence bound - Binary classification\\}
Fix a distribution $\rP_x$ and let $\delta > 0$. Let $\mat{X}= \{\vec{x}_1, \dots \vec{x}_\nsamples \} \in \bbR^{\ndim \times \nsamples}$ be drawn \aclink{i.i.d} from $\rP_x$. Then
with probability at least $1 - \delta$ (over the draw of $\mat{X}$),
\begin{align}
      \forall f_{\vec{w}} \in \mF,  \epsilon_{\textrm{gen}} (f_{\vec{w}}) - \epsilon_{\train}^\nsamples (f_{\vec{w}})   \leq  \mathfrak{R}_\nsamples(\mF) + \sqrt{ \frac{\log(1/\delta)}{\nsamples}}  \,.
      \label{eq:main-bound}
    \end{align}
\end{theorem}
Thus, the Rademacher complexity is a uniform bound of the generalization gap. In the high-dimensional limit when both $\nsamples$ and $\ndim$ goes to infinity that we will consider in the remaining of the paper, we shall see that we can discard the $\delta-$dependent term and that only the first term will remains finite. 
Note that this theorem can be used to recover the classical result \eqref{eq:VC-bound}. Indeed it can be shown \cite{massart2000some,ledoux2013probability,dudley1967sizes} that the Rademacher complexity can be bounded by the \aclink{VC} dimension so that for some constant value $C$,
\begin{equation}
\mathfrak{R}_\nsamples(\mF) \le  C{\sqrt {\frac {d_{\vc}(\mF)}{\nsamples}}}  \,.
\label{boundVC}
\end{equation}

We remind the reader that the \aclink{VC} dimension is the size of the set that can be fully shattered by the hypothesis class $\mF$. Informally, if $\nsamples > d_{\vc}$ then for all set of $\nsamples$ data points, there exists an assignment of labels that cannot be fully fitted by the function class \cite{vapnik2013nature}.
\begin{proof}
	\label{proof_rademacher_vc}
	    Applying Massart's lemma \cite{massart2000some} for $\mF_{\mat{X}} = \{ f(\vec{x}_{1}), .., f(\vec{x}_{\nsamples} )  \}$ $\subset \bbR^{\nsamples} $ with $f: \bbR^{\ndim} \longrightarrow \{\pm 1\}$. Hence $\sup_{\vec{x} \in \mF_{\mat{X}}} \| \vec{x} \|_2 = \sqrt{\nsamples}$ and it follows
	    \begin{align*}
	        &\mathfrak{R}_\nsamples \(\mF\) \equiv \EE_{\mat{X}} \[ \hat{\mathfrak{R}}_\nsamples \(\mF, \mat{X}\) \] \leq \EE_{\mat{X}} \[  \sup_{\vec{x} \in \mF_{\mat{X}}} \| \vec{x} \|_2 \frac{ \sqrt{2 \log |\mF_{\mat{X}}| }}{\nsamples} \]  \\
	        &\leq
	        \EE_{\mat{X}} \[  \sqrt{ \frac{ 2 \log \max_{\{\vec{x}_{1},..\vec{x}_{\nsamples}\}}|\mF_{\mat{X}}| }{\nsamples}} \]  \\
	        &=\sqrt{ \frac{ 2 \log \Pi_{\mF}(\nsamples)}{\nsamples}}  \leq  \Theta\( \sqrt{ \frac{ d_{\vc}(\mF) }{\nsamples}}  \)\,,
	    \end{align*}
	    where $\Pi_{\mF}(\nsamples) \equiv  \max_{\{\vec{x}_{1},..\vec{x}_{\nsamples}\}} \{ f(\vec{x}_{1}), .., f(\vec{x}_{\nsamples}) \} \leq 2^\nsamples$ is the growth function of the hypothesis class $\mF$. The last inequality comes from the fact that the VC dimension of the hypothesis class $\mF$ is defined as the maximum sample size dataset that can be shattered $d_{\vc}(\mF)= \max_\nsamples\{\nsamples : \Pi_\mF(\nsamples) = 2^\nsamples\} $.
 \end{proof}

\section{Synthetic models in the high dimensional statistics limit}
\label{sec:iid}
In this section, we consider data generated by a simple generative model. We suppose that each vector of input data points $\mat{X}=\{\vec{x}_{1}, \cdots, \vec{x}_{\nsamples}\} \in \bbR^{ \ndim \times \nsamples}$ has been generated \aclink{i.i.d} from a factorized, e.g. Gaussian, distribution, that is
$\forall \mu \in \lb \nsamples \rb, {\rP_x\left(\vec{x}_{\mu}\right) = \prod_{i=1}^{\ndim} \rP_x(x_{i\mu})}$.
In the following, we will focus on this simple data distribution, but sec.~\ref{sec:rot_inv_mat} presents a generalization to rotationally invariant data matrices $\mat{X}$ with arbitrary spectrum. The main interest of such settings is to use the analysis of \emph{typical case} problems with synthetic data created from simple generative models as means of getting additional insight on real world applications where data are not worst case \cite{seung1992statistical,watkin1993statistical,opper1995statistical,engel2001statistical,zdeborova2016statistical}. In particular, we shall be interested in the high-dimensional statistics limit when $\nsamples, \ndim \longrightarrow \infty$, with $\alpha=\frac{\nsamples}{\ndim} = \Theta(1)$. In the following, the aim is to compute exactly, rather than merely bounding, and asymptotically the Rademacher complexity for such problems.

\subsection{Linear model}
As the simplest example, we first tackle the computation of the Rademacher complexity for a simple function class containing all linear models with weights $\vec{w} \in \bbR^{\ndim}$,
 \begin{align}
    \mF_{\textrm{linear}} = \left\lbrace f_{\vec{w}}: 
\begin{cases}
\bbR^{\ndim} \longrightarrow \bbR\\
\vec{x} \longrightarrow  \frac{1}{\sqrt{\ndim}} \vec{w}^\intercal \vec{x}
\end{cases}  , \vec{w} \in \bbR^{\ndim}~~/~~\|\vec{w}\|_2 = \Gamma \sqrt{\ndim} \right\rbrace \,. 
\label{main:linear_model}
\end{align}
From eq.~\eqref{main:def_rademacher}, computing the empirical Rademacher complexity amounts to finding the vector $\vec{w}^\star$ that maximizes the scalar product between $\vec{y}$ (that replaces the variable $\bepsilon$) and $\mat{X}^\intercal \vec{w}$. It is thus sufficient to take $\vec{w}^\star = \frac{\mat{X} \vec{y}}{\|\mat{X} \vec{y}\|_2} \|\vec{w}\|_2$ and the empirical Rademacher complexity \eqref{main:def_rademacher} thus reads
\begin{align}
    &\mathfrak{R}_\nsamples\left(\mF_{\textrm{linear}}\right) = \EE_{\vec{y}, \mat{X}} \[ \sup_{f \in \mF_{\textrm{linear}}}  \frac{1}{\nsamples} \sum_{\mu =1}^\nsamples  y_{\mu}  f\(\vec{x}_{\mu}\) \]\\
    &= \EE_{\vec{y}, \mat{X}} \[ \frac{1}{\nsamples} \vec{y}^\intercal \( \frac{1}{\sqrt{\ndim}} \mat{X}^\intercal\vec{w}^\star \) \]
    = \EE_{\vec{y}, \mat{X}} \[ \frac{1}{\nsamples} \vec{y}^\intercal \( \frac{1}{\sqrt{\ndim}} \mat{X}^\intercal  \frac{\mat{X} \vec{y}}{\|\mat{X} \vec{y}\|_2} \|\vec{w}\|_2  \) \] \nonumber \\
    &= \EE_{\vec{y}, \mat{X}} \left[  \frac{1}{\nsamples} \frac{1}{\sqrt{\ndim}} \|\mat{X} \vec{y}\|_2 \|\vec{w}\|_2 \right]\,. \nonumber
\end{align}
$\mat{X}$ having \aclink{i.i.d} entries, we can apply the \aclink{CLT}, which enforces $\forall i \in \llbracket d \rrbracket$, $\left(\mat{X} \vec{y}\right)_i = \sum_{\mu=1}^\nsamples x_{i\mu} y_\mu \sim \mN\left( 0 , \nsamples \right)$ hence ${\EE_{\vec{y}, \mat{X}} \|\mat{X} \vec{y}\|_2 = \sqrt{ \ndim \nsamples}}$. Assuming that weights are restricted to lie on the sphere of radius $\Gamma$ in $\bbR^{\ndim}$, we set $\|\vec{w}\|_2 = \Gamma \sqrt{\ndim}$ and finally obtain
\begin{align}
     \mathfrak{R}_\nsamples \left(\mF_{\textrm{linear}}\right) &= \frac{\Gamma}{\sqrt{\alpha}}\,\,,
\end{align}
where recall $\alpha = \frac{\nsamples}{\ndim}$. The above result for the simple linear function hypothesis class allows to grasp the meaning of the Rademacher complexity: At fixed input dimension $\ndim$, it decreases with the number of samples as $\alpha^{-1/2}$, closing the generalization gap in the infinite $\alpha$ limit. Illustrating the bias-variance trade-off, we also see that increasing the radius of the weights expands the function complexity (and might help for fitting the data-set), but unfortunately leads to a looser generalization bound.

Note also that the fact that the Rademacher complexity is $\Theta(\alpha^{-1/2})$ shows that it remains finite in the high-dimensional statistics limit. In this case, we see indeed that we can disregard the term $\sqrt{\log(1/\delta)/\nsamples}$ that goes to zero as $\nsamples \!\to\!\infty$ in eq.~(\ref{eq:main-bound}).

\subsection{Perceptron model}
The scaling of Rademacher complexity inverse as $\sqrt{\alpha}$ in the high-dimensional statistics limit is actually \emph{not} restricted to the linear model but appears to be a universal property, at least at large enough $\alpha$. To see this we now focus on a different hypothesis class: the perceptron, denoted $\mF_{\sign}$. This class contains linear classifiers which output binary variables, and will fit much better labels in the binary classification task. The class writes
 \begin{align}
    \mF_{\sign} = \left\lbrace f_{\vec{w}}: 
\begin{cases}
\bbR^{\ndim} \longrightarrow \{\pm1\}\\
\vec{x} \longrightarrow  \sign\( \frac{1}{\sqrt{\ndim}} \vec{w}^\intercal \vec{x} \)
\end{cases}  , \vec{w} \in \bbR^{\ndim} \right\rbrace \,. 
\label{main:sign_model}
\end{align}
Let us consider a sample \aclink{i.i.d} matrix $\mat{X} \in \bbR^{d \times m}$ with $\vec{x}_{\mu}\sim \mN(\vec{0},\mat{I}_{\ndim})$.
\begin{theorem} For the perceptron model class  eq.~\eqref{main:sign_model} with random i.i.d. input data in the high-dimensional limit, $\mathfrak{R}_\nsamples \left(\mF_{\sign}\right) = \Theta\left(\frac{1}{\sqrt{\alpha}} \right) $\,.
\end{theorem}
\begin{proof}
In a nutshell, the proof uses the fact that the Rademacher complexity is upper-bounded by the \aclink{VC} dimension divided by $\alpha^{1/2}$, and lower-bounded by one particular example of its function class, when the weights are chosen according to Hebb's rule, which also gives a behavior scaling as $\alpha^{-1/2}$.

{\textbf{Upper bound}\\}
For a linear classifier with binary outputs such as the perceptron, the \aclink{VC} dimension is easy to compute and $d_{\vc} = d$. Hence we know from Massart theorem's \cite{massart2000some} that
\begin{align*}
	\mathfrak{R}_\nsamples (\mF_{\sign}) \leq \Theta \(\sqrt{\frac{d_{\vc}(\mF_{\sign})}{\nsamples}}\) = \Theta\(\sqrt{\frac{\ndim}{\nsamples}}\) = \Theta\(\alpha^{-1/2}\)  \,.
\end{align*}

{\textbf{Lower bound}\\}
Let us consider the following estimator, known as the Hebb's rule \cite{hebb1962organization}: $\displaystyle \vec{w}^\star = \frac{1}{\sqrt{\ndim}} \sum_{\nu=1}^\nsamples y_\nu \vec{x}_\nu$. Hence for a given sample $\vec{x}_{\mu}$ the above estimator outputs
$$f_{\vec{w}^\star} \(\vec{x}_{\mu}\) = \sign\(  \frac{1}{\sqrt{\ndim}} \vec{w}^{\star \intercal} \vec{x}_{\mu} \) = \sign \( \(\displaystyle  \frac{1}{\ndim} \sum_{\nu=1}^\nsamples y_\nu \vec{x}_\nu \)^\intercal \vec{x}_{\mu} \)\,.$$ 
Injecting its expression in the definition the Rademacher complexity eq.~\eqref{main:def_rademacher} one obtains:
\begin{align*}
	&\mathfrak{R}_\nsamples (\mF_{\sign}) \equiv  \EE_{\vec{y}, \mat{X}} \[ \sup_{\vec{w}} \frac{1}{\nsamples} \sum_{\mu =1}^\nsamples  y_{\mu}  f_{\vec{w}} \(\vec{x}_{\mu}\) \] \\
	& \geq   \EE_{\vec{y}, \mat{X}} \[ \frac{1}{\nsamples} \sum_{\mu=1}^\nsamples y_{\mu} f_{\vec{w}^\star}\(\vec{x}_{\mu} \) \] \\
	&=   \EE_{\vec{y}, \mat{X}} \[ \frac{1}{\nsamples} \sum_{\mu=1}^\nsamples \sign \( y_{\mu}  \frac{1}{\ndim} \(\sum_{\nu=1}^\nsamples y_\nu \vec{x}_\nu \)^\intercal \vec{x}_{\mu}\) \] \\
	&=  \EE_{\vec{y}, \mat{X}} \[ \frac{1}{\nsamples} \sum_{\mu=1}^\nsamples \sign \(1 + \frac{1}{\ndim}\sum_{\nu \ne \mu}^\nsamples y_{\mu}  y_\nu \vec{x}_\nu^{\intercal} \vec{x}_{\mu} \)\] \,.
\end{align*}
As $\vec{x}_{\mu} \sim \mN\(\vec{0},\mat{I}_d\)$ and the labels are drawn uniformly $y_{\mu} \sim \mathcal{U}(\pm 1)$, $\vec{z}_\mu \equiv y_{\mu} \vec{x}_{\mu} \sim \mN\(\vec{0},\mat{I}_d\)$.
Hence let us define the Gaussian random variable
\begin{align*}
    \theta_\mu \equiv \frac{1}{\ndim}\sum_{\nu \ne \mu}^\nsamples y_{\mu}  y_\nu \vec{x}_\nu^{\intercal} \vec{x}_{\mu} = \frac{1}{\ndim}\sum_{\nu \ne \mu}^\nsamples \vec{z}_\nu^{\intercal} \vec{z}_\mu\,,
\end{align*}
and compute its two first moments
\begin{align*}
	\EE \[\theta_{\mu}\] &= \EE_{\vec{z}}\[ \frac{1}{\ndim}\sum_{\nu \ne \mu}^\nsamples \vec{z}_\nu^{\intercal} \vec{z}_\mu \] = \EE_{\vec{z}}\[ \frac{1}{\ndim}\sum_{\nu \ne \mu}^\nsamples  \sum_{i=1}^d z_{i \nu}z_{i\mu} \] = 0 \,,\spacecase
	\EE \[\theta_{\mu}^2\] &= \EE\[ \frac{1}{d^2} \(\sum_{\nu \ne \mu}^\nsamples \vec{z}_\nu^{\intercal} \vec{z}_\mu\)^2 \] = \frac{(\nsamples -1)}{\ndim} \underset{\nsamples \to \infty}{\longrightarrow}  \alpha \,.
\end{align*}
Hence because of the \aclink{CLT}, in the high-dimensional limit $\theta_\mu \sim \mN(0, \alpha)$, 
\begin{align*}
	\mathfrak{R}_\nsamples (\mF_{\sign}) &\geq  \EE_{\btheta} \[ \frac{1}{\nsamples} \sum_{\mu=1}^\nsamples \sign\(1 + \theta_\mu \)\] =\EE_{\theta} \[ \sign\( 1 + \theta\)  \] \\
	&=   \bbP\[ \theta \geq -1 \] -  \bbP\[\theta \leq -1 \] = 2 \bbP\[ \theta \geq -1 \] - 1 \,.
\end{align*}
Finally, noting that 
\begin{align*}
	\bbP\[ \theta \geq -1 \] &= \int_{-\frac{1}{\sqrt{\alpha}}}^\infty \Diff_\theta = \frac{1}{2} {\textrm{erfc}}\(- \frac{1}{\sqrt{2\alpha}} \) \underset{\alpha \to \infty}{\simeq} \frac{1}{2} - \frac{1}{\sqrt{2\pi \alpha}}\,,
\end{align*}
we obtain a lower bound for the Rademacher complexity
\begin{align*}
	\mathfrak{R}_\nsamples \(\mF_{\sign}\) \geq \sqrt{\frac{2}{\pi}} \frac{1}{\sqrt{\alpha}} = \Theta\(\frac{1}{\sqrt{\alpha}}\)\,.
\end{align*}
\end{proof}

Heuristically, this result generalizes as well to a two-layer neural network with $K$ hidden neurons. Indeed, the two-layer function class contains, as a particular case, the single layer one, so the lower bounds goes through. The upper bound is however harder to control rigorously. Since neural networks have a finite \aclink{VC} dimension, the Rademacher complexity is again lower-bounded by $\Theta(1/\sqrt{\nsamples})$; However, we do not know of any theorem that would ensure that the \aclink{VC} dimension is bounded by  $\Theta(d)$ \cite{bartlett2003vapnik}. Nevertheless, anticipating on the statistical physics approach, we indeed expect from the concentration (self-averaging) properties of the ground-state energy \cite{talagrand2003spin} in the high-dimensional limit that it will yield a Rademacher complexity that is a function of $\alpha=\nsamples / \ndim $ \emph{only} at fixed $K$. From this argument, we expect  that the $\Theta\left(\frac{1}{\sqrt{\alpha}} \right)$ dependence of the Rademacher complexity to be very generic in the high-dimensional limit.

\section{The statistical physics approach\label{sec:statistical physics}}

\subsection{Average case problems: Statistical physics of learning}
As anticipated in the previous chapter, the approach inspired by statistical physics to understand neural networks considers a set of data points coming from known distributions. Again, for the purpose of this presentation we focus on a simple example, where $\vec x \sim \rP_{x}(.)$ with $\rP_x(\vec{x})= {\mN}_\vec{x}(\vec{0},\mat{I}_d)$. \Sec\ref{sec:rot_inv_mat} is devoted to a generalization to random input data corresponding to random matrices with arbitrary singular value density.

Consider a function class, for instance we can again use the \emph{perceptron} one $\mF_{\sign}$: $\{f_{\vec{w}}: \vec{x} \to \sign{\(\frac{1}{\sqrt{\ndim}} \vec{w}^\intercal\vec{x}\)}\}$; a typical question in the literature was to compute how many misclassified examples can be obtained for a given rule used to generate the labels \cite{engel2001statistical}. Given $\nsamples$ samples $\{y_\mu, \vec{x}_\mu \}_{\mu=1}^\nsamples$, in order to count the number of wrongly classified training samples, we define the Hamiltonian, or \emph{energy} function \cite{Mezard1986}:
\begin{align}
\begin{aligned}
	\mH_\ndim\(\{\vec{y},\mat{X}\}, \vec{w}\) &\equiv \sum_{\mu=1}^\nsamples \id \[ y_\mu \ne 
    f_{\vec{w}}\( \vec{x}_\mu\) \] \\
    &= \frac{1}{2}\( \nsamples- \sum_{\mu=1}^\nsamples   y_\mu f_{\vec{w}}\( \vec{x}_\mu\) \) \,.
\label{main:hamiltonian}
\end{aligned}
\end{align}
A classical problem in statistical physics is to compute the random capacity also called \emph{Gardner capacity} $\alpha_c$ \cite{gardner1989three} studied in \Chap\ref{chap:binary_perceptron}: given $\nsamples$ examples $\{\vec{x}_\mu\}_{\mu=1}^{\nsamples}$ and labels  $\{y_\mu\}_{\mu=1}^{\nsamples}$ randomly chosen between $\pm 1$, it consists in finding how many samples $\nsamples_c$ can be correctly classified.

It turns out there exists a deep connection between the Gardner capacity and the \aclink{VC} dimension, as their common aim is to measure the maximum number of points $\nsamples_c$ such that there exists a function in the hypothesis class being able to fit the data set. In particular, using Sauer's lemma \cite{sauer1972density} in the large size limit $\nsamples , \ndim \longrightarrow \infty$, keeping $\alpha_{c}=\frac{\nsamples_c}{\ndim} = \Theta(1)$ and $\alpha_{\vc} = \frac{d_{\vc}}{\ndim} = \Theta(1)$, it is possible to show that the Gardner  capacity $\alpha_c$ provides a lower-bound of the VC dimension \cite{engel2001statistical}:
\begin{align}
    \alpha_c \leq 2  \alpha_{\vc} \,.
\end{align}{}
To illustrate this inequality, let us consider again the perceptron classifier hypothesis class $\mF_{\sign}$ for which the above inequality is saturated. In fact, the \aclink{VC} dimension is in this case (linear classification with binary outputs) simply $\ndim_{\vc} = \ndim$. Hence on one hand $\alpha_{\vc} = 1$, on the other hand the Gardner capacity amounts to $\alpha_c=2$ \cite{cover1965geometrical,gardner1989three}.

It is fair to say that a large part of the statistical physics literature focused mainly on the Gardner capacity, in particular in a series of works in the 90's \cite{gardner1989three, krauth1989storage} that led to more recent rigorous works \cite{talagrand2003spin, talagrand2006parisi, Sun2018, Aubin2019_storage}. 

\subsection{The Rademacher complexity and the ground state energy}
As we shall see now, computing the Rademacher complexity for random input data can be directly reduced to a more natural object in the physics literature: the \emph{ground state energy}. Defining the Gibbs measure at inverse temperature $\beta$, that weighs configurations with their respective cost, as
\begin{align}
 \langle \dots \rangle_\beta  \equiv  \frac {\int_{\bbR^\ndim} \d \vec{w} ~ \dots ~ e^{-\beta \mH_\ndim(\{\vec{y},\mat{X}\}, \vec{w}) }}{\int_{\bbR^\ndim} \d\vec w ~ e^{-\beta \mH_\ndim(\{\vec{y},\mat{X}\}, \vec{w})}}\, ,
\end{align}
we observe that averaging the Hamiltonian in eq.~\eqref{main:hamiltonian} over $\{\vec{y},\mat{X}\}$ and the Gibbs measure for any function $f_{\vec{w}} \in \mF$ provides
\begin{align}
   	\EE_{\vec{y}, \mat{X}} \left \langle  \frac{ 	\mH_\ndim\(\{\vec{y},\mat{X}\}, \vec{w}\) }{\ndim} \right\rangle_\beta &=  \frac{\alpha}{2} \[ 1 - \EE_{\vec{y}, \mat{X}} \left\langle \frac{1}{\nsamples} \sum_{\mu=1}^\nsamples y_\mu  f_{\vec{w}}\(\vec{x}_\mu\) \right\rangle_\beta   \] \,,
\end{align}
where $\alpha = \frac{\nsamples}{\ndim} = \Theta(1)$.
Taking the zero temperature limit, \ie $\beta \to \infty$, in the above equation, we finally obtain the ground state energy $e_{\textrm{gs}}$, a quantity commonly used in physics. Interestingly, we recognize the definition of the Rademacher complexity $\mathfrak{R}_\nsamples(\mF)$
\begin{align}
\begin{aligned}
   	e_{\textrm{gs}} &\equiv \lim_{\beta \to \infty} \lim_{d \to \infty} \EE_{\vec{y}, \mat{X}} \left \langle  \frac{ 	\mH_\ndim\(\{\vec{y},\mat{X}\}, \vec{w}\) }{\ndim} \right\rangle_\beta\\ 
   	&= \frac{\alpha}{2} \[ 1 - \EE_{\vec{y}, \mat{X}} \sup_{f_{\vec{w}}\in \mF}  \frac{1}{\nsamples} \sum_{\mu=1}^\nsamples y_\mu f_{\vec{w}}\(\vec{x}_\mu\)  \] \\
   	&=  \frac{\alpha}{2} \[ 1 -  \mathfrak{R}_\nsamples\(\mF\) \]\,,
   	\label{main:link_gs_rademacher}
\end{aligned}
\end{align}
where random labels $\vec{y}$ play the role of the Rademacher variable $\bepsilon$ in \eqref{main:def_rademacher}. The above equation shows a simple correspondence between the ground state energy on the perceptron model with randomly quenched disorder and the Rademacher complexity of the corresponding hypothesis class, and shall bring insights from both the machine learning and statistical physics communities. Consequently, as we shall see, this connection means that the Rademacher complexity can be computed, rather than bounded, for many models using the replica method from statistical physics. As far as we are aware, this basic connection between the ground state energy and Rademacher complexity was not previously stated in literature. 

\subsection{An intuitive understanding on the Rademacher bounds on generalization}
At this point, the Rademacher complexity becomes a more familiar object to the physics-minded reader. However, could we understand more intuitively why the Rademacher complexity, or equivalently the ground state energy, is involved in the generalization gap bound? Let us present an intuitive hand-waving explanation. Consider the fraction of mistakes performed by a classifier $f_{\vec{w}}$ on unknown samples, namely the generalization error $\epsilon_{\gen}(f_{\vec{w}})$, and on the training set the training error $\epsilon_{\textrm{train}}^\nsamples(f_{\vec{w}})$. The worst case scenario that could occur is trying to fit while there exists no underlying rule, meaning that labels are purely random uncorrelated from input. The estimator will purely overfit and its generalization error will remain constant to $1/2$ in any case. This leads to the following heuristic generalization bound:
\begin{align}
\begin{aligned}
   &\epsilon_{\gen}(f_{\vec{w}}) - \epsilon_{\textrm{train}}^\nsamples(f_{\vec{w}})  \leq \epsilon^{\textrm{random~labels}}_{\gen}(f_{\vec{w}}) - \epsilon^{{\textrm{random~labels}}, \nsamples}_{\textrm{train}}(f_{\vec{w}}) \\
   &= \frac{1}{2} -  \epsilon^{{\textrm{random~labels}}, \nsamples}_{\textrm{train}}(f_{\vec{w}}) = \frac{1}{2} \(1 - 2\epsilon^{{\textrm{random~labels}}, \nsamples}_{\textrm{train}}(f_{\vec{w}}) \)\\
   &= \frac{1}{2} \hat{\mathfrak{R}}_\nsamples \(\mF\)\,.
\end{aligned}
\end{align}
Note that this heuristic reasoning does not give the \emph{exact} Rademacher generalization bound. In fact, the actual stronger and uniform over all possible $\vec{w}\in\bbR^{\ndim}$ bound does not have a factor $1/2$, and surely cannot be fully captured by the simple above argument. Nevertheless, this argument reflects the crux of the Rademacher bound: it provides a very pessimistic bound by assuming the worst possible scenario: \ie fitting data and trying to make predictions while the labels are random. Of course, in real data problems the rule is not random; it is then no surprise that the Rademacher bound is not tight \cite{zhang2016understanding}. Indeed, real problems labels are \emph{not} randomly correlated with the inputs.

\section{Consequences and bounds for simple models}
\label{sec:applications_rademacher}
In this section, we illustrate our previous arguments and the connection between the spin glass approach and the Rademacher complexity still for the case of Gaussian \aclink{i.i.d} input data matrix $\mat{X}$ in the high-dimensional limit when $\nsamples,\ndim \to \infty$.

\subsection{Ground state energies of the perceptron}
\label{magic}
For a number of samples smaller than the Gardner capacity $\alpha_c$, also called the SAT-UNSAT threshold, it is by definition possible to fit all random labels $\vec{y}$. Accordingly, the number of misclassified examples is zero and the ground state energy $e_{\gs }=0$. This means that the Rademacher complexity is asymptotically equal to $1$ for $\alpha<\alpha_{c}$. However above the Gardner capacity $\alpha > \alpha_c$, the estimator $f_{\vec{w}}$ cannot perfectly fit the random labels and will misclassify some of them, equivalently $e_{\gs}>0$. From the arguments given in sec.~\ref{sec:iid}, we thus expect 
\begin{equation}
\begin{aligned}
 	\mathfrak{R}_\nsamples \( \mF \) &= 1 \text{~for $\alpha<\alpha_c$}\,, \\
 	\mathfrak{R}_\nsamples \( \mF \) &\approx \Theta \left(\sqrt{\frac{\alpha_c}{\alpha}}\right) \text{for $\alpha \gg \alpha_c$}\,. 
 \end{aligned}
\end{equation}
This relation is already non-trivial, as it yields a link between the Gardner capacity and the Rademacher complexity. 
Using the replica method from spin glass analysis, and the mapping with ground state energies \eqref{main:link_gs_rademacher}, we shall now see how one can go beyond these simple arguments, and compute the actual precise asymptotic value of the Rademacher complexity.

\subsection{Computing the ground-state energy with the replica method}
Knowing that statistical physics literature focused mainly on the Gardner capacity, the connection between the ground-state energy and the Rademacher complexity suggests that it would be worth looking at these old results in a new light. In fact, the replica method allows for an \emph{exact} computation of the Rademacher complexity for random input data in the large size limit. In the following, we handle computations by focusing on a simple generalization of the linear functions hypothesis class. Fix any activation function ${\varphi:\bbR\longrightarrow\{\pm 1\}}$, we define the following hypothesis class
\begin{align}
    \mF_\varphi \equiv \left\lbrace f_{\vec{w}}: 
\begin{cases}
\bbR^{\ndim} \longrightarrow \{\pm 1\}\\
\vec{x} \longrightarrow \varphi \( \frac{1}{\sqrt{\ndim}} \vec{w}^\intercal \vec{x}\)
\end{cases}  , \vec{w} \in \bbR^{\ndim} \right\rbrace \,. 
\label{main:glm_hypothesis_class}
\end{align}
Starting with the posterior distribution
\begin{align}
	\bbP ( \vec{w} | \vec{y}, \mat{X} ) = \frac{\bbP(\vec{y} | \vec{w}, \mat{X}) \bbP(\vec{w})  }{ \bbP ( \vec{y}, \mat{X} )  } = \frac{ e^{-\beta \mH_\ndim( \{\vec{y}, \mat{X}\}, \vec{w})} \rP_\w\( \vec{w} \)}{\mZ_\ndim( \{\vec{y}, \mat{X}\}, \alpha, \beta)} \,,
\end{align}
we introduced the partition function associated to the Hamiltonian eq.~\eqref{main:hamiltonian} at inverse temperature $\beta$
\begin{align}
	\mZ_\ndim( \{\vec{y}, \mat{X}\}, \alpha, \beta) = \int_{\bbR^d} \d \rP_\w\( \vec{w} \) ~ e^{-\beta \mH_\ndim( \{\vec{y}, \mat{X}\}, \vec{w})} \,.
	\label{main:partition_function}
\end{align}
In the large size limit $\ndim \to \infty$, the posterior distribution becomes highly peaked in particular regions of parameters. In physics we are interested in these dominant regions and focus on the free energy at inverse temperature $\beta$ defined as
\begin{align}
    \varphi_{\vec{y}, \mat{X}} (\{\vec{y}, \mat{X}\}, \alpha, \beta) \equiv - \lim_{d \to \infty} \frac{1}{d \beta} \log \mZ_\ndim( \{\vec{y}, \mat{X}\}, \alpha, \beta) \,.
    \label{main:free_energy_not_avg}	
\end{align}
The free energy is closely related to the free entropy that can be equivalently considered according to $\varphi_{\vec{y}, \mat{X}} = - \Phi_{\vec{y}, \mat{X}}$.
However, as we are interested in computing quantities in the \emph{typical case}, we want to average over all potential training sets $\{\vec{y}, \mat{X}\}$ and compute instead the averaged free energy
\begin{align}
\varphi(\alpha, \beta) \equiv \EE_{\vec{y}, \mat{X}} \[ \varphi_{\vec{y}, \mat{X}} \( \{\vec{y}, \mat{X}\}, \alpha, \beta\) \].
\end{align}
Computing directly this average rigorously is difficult, hence we will carry out the computation using the so-called \emph{replica method}, starting by writing the \emph{replica trick}
\begin{align}
    - \frac{1}{d \beta} \EE_{\vec{y}, \mat{X}} \[ \log \mZ_\ndim \] = - \frac{1}{d \beta}  \lim_{r \to 0}  \frac{\partial \log \EE_{\vec{y}, \mat{X}} \[ \mZ_\ndim(\{\vec{y}, \mat{X}\}, \alpha, \beta)^r\] }{\partial r} \,,
    \label{main:replica_trick}	
\end{align}
which replaces the expectation of $\log \mZ_\ndim$ by the moments of $\mZ_\ndim$, which are easier to compute. Taking the limit $\ndim \rightarrow \infty$, and assuming that we can revert it with the limit $r \rightarrow 0$, we finally obtain
\begin{align}
    \varphi(\alpha, \beta) = \lim_{r \to 0} \[ \lim_{d \to \infty}  - \frac{1}{d \beta}  \frac{\partial  \log \EE_{\vec{y}, \mat{X}} \[ \mZ_\ndim(\{\vec{y}, \mat{X}\}, \alpha, \beta)^r\] }{\partial r} \] \,.
    \label{main:free_energy_trick}
\end{align}

We give some details on the replica computation in the context of the \aclink{GLM} with randomly quenched disorder in \App\ref{appendix:replica_computation:random_labels:iid}, and we also refer the reader to the relevant literature in physics \cite{Mezard1986,Hertz1993,engel2001statistical,mezard2009information,zdeborova2016statistical} and in mathematics \cite{talagrand2003spin,talagrand2006parisi,bolthausen2007spin,panchenko2004bounds,panchenko2018free}. 
Notice that in this randomly quenched setting where the labels are uncorrelated from the input vector, the replica computation is exactly the same than the one used for the storage capacity problem in \Chap\ref{chap:binary_perceptron}.
The computation of the free energy by the replica method is done by deriving a hierarchy of approximate ansatz, named \aclink{RS}, \aclink{1RSB}, \aclink{2RSB} \ldots While in some problems the \aclink{RS} or the \aclink{1RSB} ansatz is sufficient, in others only the infinite step solution \aclink{FRSB} gives the exact ansatz \cite{mezard1989space,talagrand2003spin,talagrand2006parisi}, although the \aclink{1RSB} approach is usually an accurate approximation. 

Computing the ground state energy consists in taking the zero temperature limit $\beta \to \infty$ above the capacity $\alpha>\alpha_c$ in the replica free energy $\varphi(\alpha,\beta) = e(\alpha,\beta) - \beta^{-1} s(\alpha,\beta)$; where $e, s$ denote respectively the densities of the energy and entropy contributions. The simplest form of the replica computation is known as \aclink{RS} and the next simplest is \aclink{1RSB} which plugged in eq.~\eqref{main:free_energy_trick} leads to expressions \cite{Majer1993a, Erichsen1992, Whyte1996}
\begin{align}
\label{main:free_energy_iid}
\varphi^{(\textrm{rs})}_{\textrm{iid}} (\alpha, \beta) &= -\frac{1}{\beta} \underset{ q_0, \hat{q}_0}{\textbf{extr}}     \left\{\frac{1}{2}\( q_0\hat{q}_0 -1 \)+  \Psi_{\w}^{(\textrm{rs})}(\hat{q}_0)   +\alpha  \Psi_{\out}^{(\textrm{rs})}(q_0, \beta)    \right\} \,, \nonumber \\ 
\varphi^{(\textrm{1rsb})}_{\textrm{iid}}(\alpha, \beta) &=  - \frac{1}{\beta}\underset{ q_0, q_1, \hat{q}_0, \hat{q}_1, x}{\textbf{extr}} \left\{  \frac{1}{2} \left(  q_1\hat{q}_1 - 1 \right) + \frac{x}{2} \left(q_0\hat{q}_0 - q_1\hat{q}_1 \right)   \right. \\
 &	\left. \hspace{1.5cm} + \Psi_{\w}^{(\textrm{1rsb})}(\hat{q}_0, \hat{q}_1, x)   +\alpha \Psi_{\out}^{(\textrm{1rsb})}(q_0, q_1,\beta, x)    \right\} \nonumber \,,
\end{align}
with auxiliary functions
\begin{align}
\begin{aligned}
    &\Psi_{\w}^{(\textrm{rs})}(\hat{q}_0) \equiv \EE_{\xi_0}  \log   \EE_{w} \[ \exp \(             {\frac{(1- \hat{q}_0  )}{2}  w^2}+ \xi_0 \sqrt{\hat{q}_0} w \)  \]  \,, \\
	&\Psi_{\out}^{(\textrm{rs})}(q_0, \beta) \equiv  \EE_y \EE_{\xi_0}  \log  \EE_{z} \[\mC \(y \big | \sqrt{1 - q_0} z + \sqrt{q_0}\xi_0, \beta \)  \]\,, \\
	&\Psi_{\w}^{(\textrm{1rsb})}(\hat{q}_0,\hat{q}_1, x) \equiv \frac{1}{x}
                \EE_{\xi_0} \log\( \right. \\
    & \hspace{1cm} \left. \EE_{\xi_1} \EE_w \[\exp\left(
                  \frac{(1- \hat{q}_1  )}{2} w^2 +
                  \left(\sqrt{\hat{q}_0}\xi_0+\sqrt{\hat{q}_1-\hat{q}_0}\xi_1
                  \right)w \right)  \]^{x}\)  \,,\\
    &\Psi_{\out}^{(\textrm{1rsb})}(q_0, q_1,\beta, x) \equiv \frac{1}{x}
                \EE_y \EE_{\xi_0}  \log\left(
                \right. \\
    & \hspace{1cm} \left.  \EE_{\xi_1} \EE_{z} \[\mC( y \big | \sqrt{q_0} \xi_0 +
                \sqrt{q_1-q_0} \xi_1  + \sqrt{1-q_1} z,\beta) \]^x \right)\,.    
\end{aligned}
\end{align} 
We introduced a temperature-dependent constraint function $\mC(y|z) = $ \newline $\exp ( - \beta V(y|z))$ where the generic cost function $V$ reads in our case $V(y|z) = \id\[ y  \ne \varphi (z) \]$ and $y \sim \rP_y(.)$ the distribution of the random labels. Above expressions are valid for any generic weight distribution $\rP_\w(.)$ and non-linearity $\varphi$. The detailed computation is left in \App\ref{appendix:replica_computation:random_labels:iid}, in particular eq.~\eqref{appendix:replicas:free_energy_rs_out_w} and eq.~\eqref{appendix:free_energy_1rsb}. Then the general method to find the ground state energy it to take the zero temperature limit $\beta \to \infty$
\begin{align}
\label{main:def_gs}
    e_{\gs, \iid}(\alpha) \equiv \lim_{\beta \to \infty} \varphi_{\iid} (\alpha, \beta) \,,
\end{align}
while handling carefully the scaling of the optimized order parameters in this limit.
\begin{figure}[t] 
\centering
    \includegraphics[width=\linewidth]{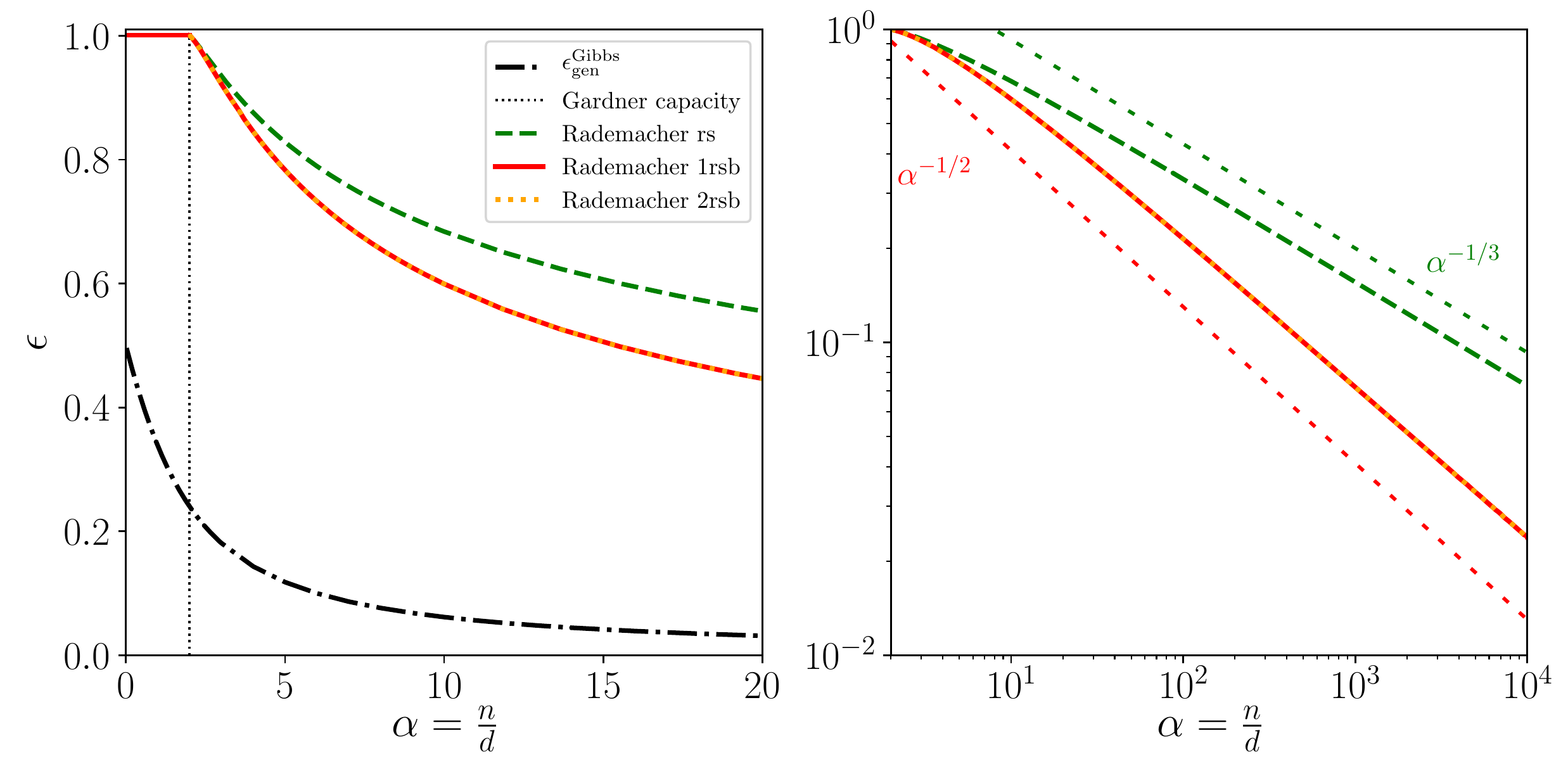}
    \caption{
    Explicit Rademacher complexity for the spherical perceptron ($\alpha_c=2$). \Left For $\alpha<\alpha_c$ the problem is satisfiable so the number of error is zero and the Rademacher complexity is constant to unity. For $\alpha> \alpha_c$, the problem becomes unsatisfiable and $e_{\gs}>0$. \Right In the case of the spherical perceptron, RS (dashed green) and 1RSB (red) ansatz provide really different results that scale respectively with $\alpha^{-1/3}$ and $\alpha^{-1/2}$ (scaling are represented with colored dashed lines). Performing 2RSB ansatz (dashed orange) does not change the scaling and difference with respect to 1RSB is visually imperceptible.
    The black dotted-dashed curve is the generalization error in the teacher-student scenario \cite{barbier2017phase}. Note the large gap between the worst case Rademacher bound and the actual teacher-student generalization error.}
    \label{fig:rademacher_spherical}
\end{figure}
\subsubsection{Spherical perceptron}
The most commonly studied model \cite{gardner1988optimal,gardner1989three,gardner1988optimal} with continuous weights is the spherical model with $\vec{w}\in\bbR^{\ndim}$ such that $\|\vec{w}\|_2^2 = \ndim$. The spherical constraint allows to have a well-defined model which excludes diverging or vanishing weights. In this case, the Gardner capacity is rigorously known to be equal to $\alpha_c=2$ \cite{cover1965geometrical}. 

We computed both the \aclink{RS} and \aclink{1RSB} free energies \cite{Majer1993a, Erichsen1992, Whyte1996}, see also \App\ref{appendix:replicas_ground_state_spherical}. Taking the zero temperature limits $\beta \to \infty,q_0 \to 1$ and $q_1 \to 1, x\to 0$ in the \aclink{1RSB} case, while keeping $\chi \equiv \beta (1-q_0)$ and $\Omega_0 \equiv \frac{\beta x}{\chi}$ finite leads to the following expressions of the ground states energies:
\begin{align}
\label{main:ground_states_spherical}
    &e_{\textrm{rs}, \iid}^{(\textrm{rs})} =  \extr_{\chi} \left\{ -\frac{1}{2 \chi} + \alpha  \EE_{y, \xi_0} \min_{z} \[ V(y|z) + \frac{\(z - \xi_0\)^2}{2 \chi}  \]   \right\} \spacecase
    &e_{\textrm{rs}, \iid}^{(\textrm{1rsb})} = \extr_{\chi,\Omega_0,q_0} \left\{  \frac{1}{2\Omega_0 \chi} \log\( 1 + \Omega_0 (1-q_0) \) \right. \nonumber \\
    & \left. \qquad + \frac{q_0}{2\chi \( 1 + \Omega_0 (1-q_0) \) }   \right.  \\
	& \left. \qquad  + \frac{\alpha}{\chi \Omega_0}  \EE_{\xi_0}  \log \EE_{\xi_1} e^{-\Omega_0 \chi \min_z \[V(y | z) + \frac{1}{2 \chi } \(z - \sqrt{q_0} \xi_0 -
                \sqrt{1-q_0} \xi_1 \)^2 \]}   \right\}\,, \nonumber
\end{align}
where the cost function $V(y|z) = \id\[ y  \ne \varphi (z) \]$. The details of the  derivation via the replica methods are given in Appendix~\ref{appendix:replicas_ground_state_spherical}. The results for Rademacher variable $y$ with $\varphi(z) = \sign(z)$ are depicted in Fig.~\ref{fig:rademacher_spherical}. 

Interestingly, the bounds on the Rademacher complexity also imply consequences for the ground state energy. Indeed the Rademacher complexity scales as $\alpha^{-1/2}$ for large values of $\alpha$ --- namely there exists a constant $\mC$ such that $\mathfrak{R}_\nsamples \( \mF \) \underset{\alpha \to \infty}{\approx} \frac{C}{\sqrt{\alpha}}$ --- therefore the ground state energy behaves for large $\alpha$ as
\begin{equation}
e_{\gs} (\alpha) = \frac{\alpha}{2} \(1-\mathfrak{R}_\nsamples \( \mF \)\) \underset{\alpha \to \infty}{ \longrightarrow} \frac{\alpha}2 \left(1-\frac{C}{\sqrt{\alpha}}\right)\,.
\label{scaling-rsb-from-VC}
\end{equation}
We first notice that the replica symmetric \aclink{RS} solution complexity fails to deliver the correct scaling as sketched in Fig.~\ref{fig:rademacher_spherical}, so the scaling in eq.~(\ref{scaling-rsb-from-VC}) must not be entirely trivial. On the other hand, the \aclink{1RSB} solution we used, which is expected to be numerically very close to the harder to evaluate \aclink{FRSB} one, seems to yield the correct scaling, see Fig.~\ref{fig:rademacher_spherical}. 
It is rather striking that the statistical learning connection allows to predict, through eq.~(\ref{scaling-rsb-from-VC}), the scaling of the energy in the large $\alpha$ regime, that is only satisfied with replica symmetry breaking ansatz. This yields an open question for replica theory: in practice, can one compute exactly the value of the constant $C$? Given the \aclink{FRSB} solution is notoriously hard to evaluate, this might be an issue worth investigating in mathematical physics.

\subsubsection{Binary perceptron}
\begin{figure}[t]
\centering
    \includegraphics[width=\linewidth]{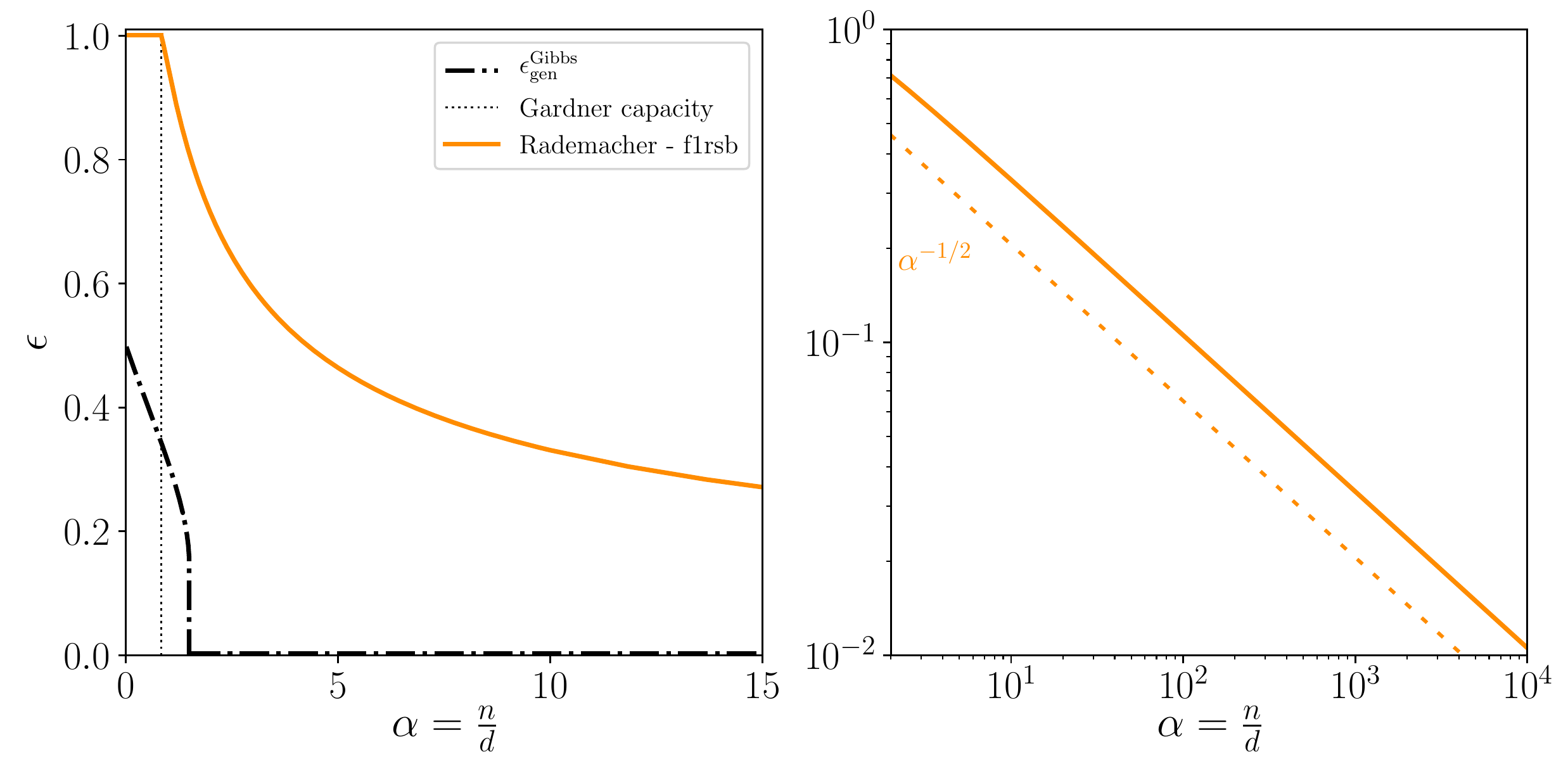}
    \caption{
    Explicit Rademacher complexity for \Left the binary perceptron ($\alpha_c=0.83\ldots$). The replica solution (orange) leads again \Right to a $\alpha^{-1/2}$ scaling (dashed orange) of the Rademacher complexity at large $\alpha$. The dotted-dashed black curve is the generalization error in the teacher-student scenario. Note the gap between the worst case bound (Rademacher) and the teacher-student generalization error.
    }
    \label{fig:rademacher_binary}
\end{figure}

Another common choice for the weights distribution is the binary prior ${\rP_\w(w)= \delta(w-1) + \delta(w+1)}$ studied e.g. in \cite{krauth1989storage}. In this case, the Gardner capacity is predicted to be $\alpha_c \approx 0.83\ldots$, a prediction which, remarkably, is still not entirely rigorously proven, but see \cite{Sun2018, Aubin2019_storage}.
To see this, we use eq.~\eqref{main:free_energy_iid}. In the binary perceptron, the landscape of the model is said to be \aclink{f1RSB}, \ie clustered in point-like dominant solutions as discussed in \Chap\ref{chap:binary_perceptron}, and the \aclink{RS} and \aclink{1RSB} free energies are the same, even though their entropies are different $\varphi(\alpha,\beta) = e(\alpha,\beta) - \beta^{-1} s(\alpha,\beta)$. In this case computing the ground state can be tackled via finding the effective temperature $\beta^\star$ such that the $s(\alpha, \beta^\star) = 0$, that can be plugged back to find the ground state energy $e_{\gs}(\alpha) =\varphi(\alpha,\beta^\star)$. Again, we note that even though the \aclink{1RSB} ansatz is unstable and should be replaced by a more complex (and ultimately \aclink{FRSB}) solution, it already gives the good scaling $\mathfrak{R}_\nsamples(\mF) \sim \alpha^{-1/2}$, and satisfies the scaling eq.~(\ref{scaling-rsb-from-VC}) for large $\alpha$, as in the case of the spherical model, see Fig.~\ref{fig:rademacher_binary}.

\subsection{Teacher-student scenario versus worst case Rademacher}

The Rademacher bounds are really interesting as they depend only on the data distribution, and are valid for \emph{any rule} used to generate the labels, no matter how complicated. In this sense, it is a worst-case scenario on the rule that prescribes labels to data. A different approach, again pioneered in statistical physics \cite{gardner1989three}, is to focus on the behavior for a given rule, called the \emph{teacher} rule. Given the Rademacher bounds tackle the worst case with respect to that rule, it is interesting to consider the generalization error one actually gets for the \emph{best case}, i.e. fitting the labels according to the same teacher rule..  This is the  so-called \aclink{T-S} approach.
In the wake of the need to understand the effectiveness of neural networks, and the limitations of the classical approaches, it is of interest to revisit the results that have emerged thanks to the physics perspective. 

We shall thus assume that the \emph{actual labels} are given by the rule
\begin{equation}
    y = {\sign}{\left( \frac{1}{\sqrt{\ndim}} \vec{w}^{\star \intercal} \vec{x}\right)}\,,
\end{equation}
with $\vec{w}^\star$, the \emph{teacher weights} that can be taken as Rademacher $\pm1$ variables, or Gaussian ones. Now that labels are generated by feeding \aclink{i.i.d} random samples to a neural network architecture (the teacher) and are then presented to another neural network (the student) that is trained using this data, it is interesting to compare the worst case Rademacher bound with the actual generalization error of this student on such synthetic data.

We now  consider the error of a \emph{typical} solution $\vec w$ from the posterior distribution (this is often called the Gibbs rule) for the student. Given the rule is outputting $\pm 1$ variables, this yields
\begin{align}
    \epsilon_{\textrm{gen}}^{\gibbs} = 1 - \EE_{\vec{x},\vec{w}^\star} \[ \langle  f_{\vec{w}^\star}(\vec{x}) \times f_{\vec{w}}(\vec{x}) \rangle \] = 1- q^\star
\end{align}
where $q^\star=\EE_{\vec{x},\vec{w}^\star} \[ \langle  f_{\vec{w}^\star}(\vec{x}) \times f_{\vec{w}}(\vec{x}) \rangle \] $. Computing $q^\star$ can be done within the statistical mechanics approach \cite{seung1992statistical,watkin1993statistical,opper1995statistical,engel2001statistical} and can be rigorously done as well \cite{barbier2017phase}. Notice that this error is equal to the Bayes-optimal error for the quadratic loss, see as well \cite{barbier2017phase}.

The two \emph{optimistic} (teacher-student) and \emph{pessimistic} (Rademacher) errors can be seen in Fig.~\ref{fig:rademacher_spherical} for spherical and in Fig.~\ref{fig:rademacher_binary} for binary weights. In this case, since a perfect fit is always possible, the training error is zero and the Rademacher complexity is itself the bound on the generalization error. These two figures show how different the worst and teacher-student case can be in practice, and demonstrate that one should perhaps not be surprised by the fact that the empirical Rademacher complexity does not always give the correct answer \cite{zhang2016understanding}, as after all it deals only with worst case scenarios.
\subsection{Committee machine with Gaussian weights}
Given the large gap between the Rademacher bound and the teacher-student setting, we can ask whether we can find a case where the Rademacher bound is void in the sense that the Rademacher complexity is $1$ yet generalization is good for the teacher-student setting? This can be done by moving to two-layer networks. Consider a simple version of this function class, namely the committee machine \cite{engel2001statistical}. It is a two-layer network where the second layer has been fixed, such that only weights of the first layer $\mat{W}= \{\vec{w}_1, \cdots,\vec{w}_K \}\in\bbR^{\ndim \times K}$ are learnt. The function class for a committee machine with $K$ hidden units, already considered in \Chap\ref{chap:committee_machine}, is defined by
\begin{align}
    \mF_{\textrm{com}} \equiv \left\lbrace f_{\mat{W}}: 
\begin{cases}
\bbR^{\ndim} \longrightarrow \{-1,1\}\\
\vec{x} \longrightarrow {\sign}{\left( \sum_{k=1}^K   {\sign}{\left( \frac{1}{\sqrt{\ndim}} \vec{w}_k^\intercal \vec{x}\right)}\right)}
\end{cases}\,,   \mat{W}\in\bbR^{\ndim \times K} \right\rbrace \,. 
\label{eq:comm}
\end{align}
Instead of computing the Rademacher complexity with the replica method, it is sufficient for the purpose of this section to understand its rough behavior. As discussed in sec.~\ref{magic}, this requires knowing the Gardner capacity. A generic bound by \cite{Mitchison1989} states that it is upper bounded by $\Theta(K \log(K))$. Additionally, the Gardner capacity has been computed by the replica method in \cite{monasson1995weight,urbanczik1997storage,xiong1998storage} who obtained  that $\alpha_c = \Theta(K \sqrt{\log(K)})$. We thus expect that 
\begin{align}
\begin{aligned}
 	\mathfrak{R}_\nsamples \( \mF_{\textrm{com}} \) &= 1 \text{~for $\alpha<\Theta\(K \sqrt{\log(K)}\)$}\,,\\
 	\mathfrak{R}_\nsamples \( \mF_{\textrm{com}} \) &\approx \Theta  \left(\sqrt{\frac{K\sqrt{\log{K}}}{\alpha}}\right) \text{for $\alpha \gg \Theta \(K \sqrt{\log{K}}\)$}\,.
 \end{aligned}
\end{align}

\begin{figure}[t]
\centering
    \includegraphics[width=\linewidth]{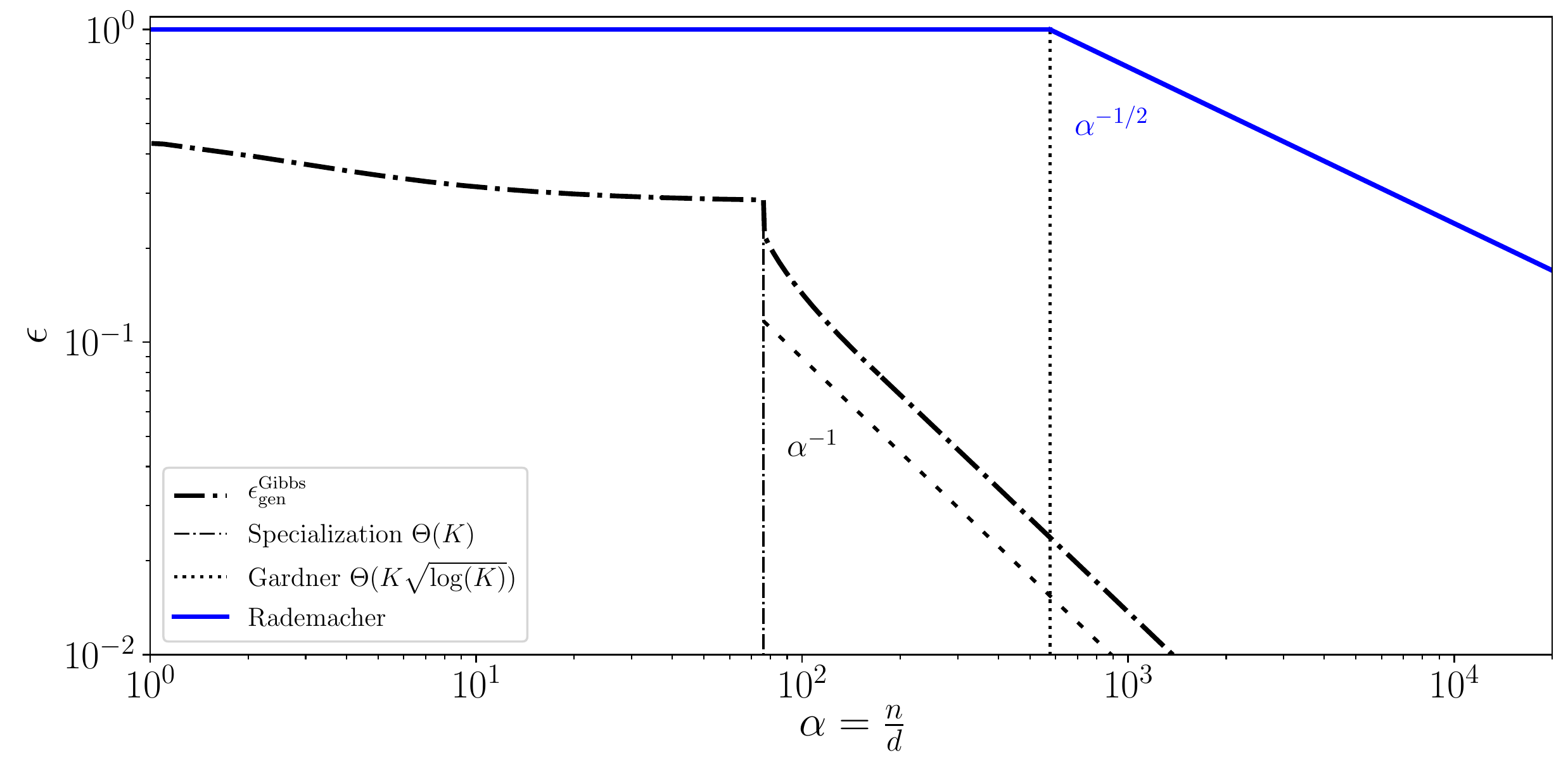}
    \caption{Illustration of the scaling of the Rademacher complexity (blue) for the fully connected committee machine, drawn together with the exact generalization error in the teacher-student scenario (dotted-dashed black), scaling as $\alpha^{-1}$ at large $\alpha$. Notice the large gap between the worst case bound (Rademacher) and the teacher-student result.}
    \label{fig:rademacher_committee}
\end{figure}

To compare with the teacher-student case, when the labels are  produced by a teacher committee machine as
\begin{equation}
      y = {\sign}{\left( \sum_{k=1}^K {\sign}{\left( \frac{1}{\sqrt{\ndim}} \vec{w}^{\star \intercal}_k \vec{x}\right)}\right)}\,,
\end{equation}
the error of the Gibbs algorithm reads
\begin{align}
    \epsilon_{\textrm{gen}}^{\gibbs} = 1 - \EE_{\vec{x},\vec{w}^\star} \[ \langle  f_{\vec{w}^\star}(\vec{x}) \times f_{\vec{w}}(\vec{x}) \rangle \] = 1- q^\star
\end{align}
where, again $q^\star=\EE_{\vec{x},\vec{w}^\star} \[ \langle  f_{\vec{w}^\star}(\vec{x}) \times f_{\vec{w}}(\vec{x}) \rangle \] $, has been computed in a series of papers in statistical physics \cite{Hertz1993,schwarze1993learning}, and using the Guerra interpolation method in \Chap\ref{chap:committee_machine} and \cite{Aubin2018}. Interestingly, in this case, one can get an error that decays as $1/\alpha$ as soon as $\alpha \gg \Theta(K)$. One thus observes a huge gap between the Rademacher bound that scales as $ \mathfrak{R}_\nsamples\(\mF_{\textrm{com}}\) =\Theta\(\sqrt{K\sqrt{\log(K)}/\alpha}\)$ and the actual generalization error $\epsilon_\gen = \Theta(K/\alpha)$ for large sample size. This large gap further illustrates the considerable difference in behavior one can get between the worst case and teacher-student case analysis, see Fig.~\ref{fig:rademacher_committee}.

\subsection{Extension to rotationally invariant matrices}
\label{sec:rot_inv_mat}
The previous free entropy and ground state energy computation for \aclink{i.i.d} data matrix $\mat{X}$ can be generalized to \aclink{RI} random matrices $\mat{X} = \mat{U}\mat{S}\mat{V}$ with rotation matrices $\mat{U} \in \mat{O}\(d\)$, $\mat{V} \in \mat{O}\(\nsamples\)$ independently sampled from the Haar measure, and $\mat{S} \in \bbR^{\ndim \times \nsamples}$ a diagonal matrix of singular values. Computation for this kind of matrices can be handled again using the replica method \cite{kabashima2008inference,barbier2018mutual,gabrie2018entropy} and leads to \aclink{RS} and \aclink{1RSB} free energies
\begin{align}
\begin{aligned}
&\varphi^{(\textrm{rs})}_{\textrm{RI}}(\alpha, \beta)= - \dfrac{1}{\beta} \extr_{\chi_w, \chi_u, q_w, q_u} \left\lbrace  \mathcal{A}_0^{(\textrm{rs})} (\chi_w, \chi_u, q_w, q_u)  \right.   \\
    & \left. \qquad \qquad + \mathcal{A}_w^{(\textrm{rs})} (\chi_w, q_w) + \alpha \mathcal{A}_u^{(\textrm{rs})} (\chi_u, q_u, \beta)\right\rbrace\,, \\
&\varphi^{(\textrm{1rsb})}_{\textrm{RI}} (\alpha, \beta)=- \dfrac{1}{\beta} \extr_{\chi_w, \chi_u, v_w, v_u, q_w, q_u, x} \left\lbrace \right. \\
& \qquad \qquad \left.  + \mathcal{A}_0^{(\textrm{1rsb})} (\chi_w, \chi_u, v_w, v_u, q_w, q_u, x) \right.  \\ 
    & \left. \qquad \qquad + \mathcal{A}_w^{(\textrm{1rsb})}(\chi_w, v_w, q_w, x)  + \alpha \mathcal{A}_u^{(\textrm{1rsb})}(\chi_u, v_u, q_u, x, \beta)\right\rbrace\,,
    \label{main:free_energy_RI}
\end{aligned}
\end{align}
where each term is properly defined in Appendix~B.2 of \cite{abbara2020rademacher}. Note that taking $\mat{X}$ a random Gaussian \aclink{i.i.d} matrix, the eigenvalue density $\rho(\lambda)$ follows the Marchenko-Pastur distribution and~\eqref{main:free_energy_RI} matches free energies eq.~\eqref{main:free_energy_iid}, and ground states energies eq.~\eqref{main:ground_states_spherical} in the spherical case.  The ground state energy (and therefore the Rademacher complexity) can be again computed as in the \aclink{i.i.d} case, taking the zero temperature limit $\beta \to \infty$
\begin{align}
    e_{\gs, \textrm{RI}}(\alpha) = \lim_{\beta \to \infty}   \varphi_{\textrm{RI}}(\alpha, \beta)\,,
\end{align}
keeping in particular $\beta \chi_w$ and $\omega = x \beta$ finite in the limits $\beta \to \infty, x\to 0, \chi_w \to 0$.

\newpage
\section*{Conclusion}
\label{sec:discussion}
In this chapter, we discussed the deep connection between the Rademacher complexity and some of the classical quantities studied in the statistical physics literature on neural networks, namely the Gardner capacity, the ground state energy of the random perceptron model \Chap\ref{chap:binary_perceptron}, and the generalization error in the \aclink{T-S} model discussed in \Chap\ref{chap:committee_machine}.  We believe it is rather interesting to draw the link with approaches inspired by statistical physics, and compare its findings with the worst-case results. In the wake of the need to understand the effectiveness of neural
networks and also the limitations of the classical approaches, it is of interest to revisit the results that have emerged thanks to the physics perspective. This direction is currently experiencing a strong revival, see e.g. \cite{Chaudhari2016,martin2017rethinking, advani2017high, baity2018comparing}. The connection discussed in the paper opens the way to a unified presentation of these often contrasted approaches, and we hope this paper will help bridging the gap between researchers in traditional statistics and in statistical physics. There are many possible follow-ups, the more natural one being the computation of Rademacher complexities from statistical physics methods for more complicated and realistic models of data, starting for instance with correlated matrices discussed in \Sec\ref{sec:rot_inv_mat}.

	\ifthenelse{\equal{\format}{oneside}}
	{
	\clearpage\null\thispagestyle{empty}\newpage
	\clearpage\null\thispagestyle{empty}\newpage
	}
	{
	\clearpage\null\thispagestyle{empty}\newpage
	\clearpage\null\thispagestyle{empty}\newpage
	\cleardoublepage}
	
	\chapter{Generalization error in high-dimensional perceptrons: Approaching Bayes error with convex optimization}
	\chaptermark{Analyzing convex empirical risk minimization}
	\label{chap:erm}
	High-dimensional statistics, where the ratio $\alpha={\nsamples}/{\ndim}$ is kept finite while the dimensionality $\ndim$ and the number of samples $\nsamples$ grow, often display interesting non-intuitive features. Asymptotic generalization performances for such problems in the so-called \aclink{T-S} setting, with synthetic data, have been the subject of intense investigations spanning many decades \cite{seung1992statistical,watkin1993statistical,engel2001statistical,bayati2011lasso,el2013robust,Donoho2016}.
To understand the effectiveness of modern machine learning techniques, and also the limitations of the classical statistical learning approaches \cite{zhang2016understanding,belkin2019reconciling}, it is of interest to revisit this line of research. Indeed, this direction is currently the subject to a renewal of interests, as testified by some very recent, yet already rather influential papers \cite{Candes18,barbier2017phase,Hastie19,belkin2019two,mei2019generalization}. The present work subscribes to this line of work and studies high-dimensional classification within one of the simplest models considered in statistics and machine learning: convex linear estimation with data generated by a teacher \emph{perceptron} \cite{gardner1989three}. We will focus on the generalization abilities in this problem, and compare the performances of Bayes-optimal estimation to the more standard \aclink{ERM}. We then compare the results with the prediction of standard generalization bounds that illustrate in particular their limitation even in this simple, yet non-trivial, setting.

\paragraph{Synthetic data model ---} We consider a supervised machine learning task, whose dataset is generated by a single layer neural network, often named a \emph{teacher} \cite{seung1992statistical,watkin1993statistical,engel2001statistical}, that belongs to the \aclink{GLM} class. Therefore we assume the $\nsamples$ samples are drawn according to
\begin{align}
	\vec{y} = \varphi_{\out}^\star\(\frac{1}{\sqrt{\ndim}} \mat{X} \vec{w}^\star \) \Leftrightarrow \vec{y} \sim \rP_{\out}^\star \(.\) \,,
\label{main:teacher_model}	
\end{align}
where $\vec{w}^\star \in \bbR^\ndim$ denotes the ground truth vector drawn from a probability distribution $\rP_{\w^\star}$ with second moment $\rho_{\w^\star}\equiv \lim_{\ndim \to \infty} \frac{1}{\ndim} \EE\[ \|\vec{w}^\star\|_2^2 \]$ and $\varphi_{\out}^\star$ represents a deterministic or stochastic activation function equivalently associated to a distribution $\rP_{\out}^\star$. The input data matrix $\mat{X}=\(\vec{x}_\mu\)_{\mu=1}^\nsamples \in \bbR^{\nsamples \times \ndim}$ contains \aclink{i.i.d} Gaussian vectors, \ie $\forall \mu \in \lb \nsamples \rb,~\vec{x}_\mu \sim \mN\(\vec{0},\mat{I}_\ndim\)$. Even though the framework we use and the theorems and results we derived are valid for a rather generic channel in eq.~\eqref{main:teacher_model} ---including regression problems---
we will mainly focus the presentation on the commonly considered perceptron case: a binary classification task with data given by a $\sign$ activation function $\varphi_{\out}^\star\(\vec{z}\) = \sign\(\vec{z}\)$, with a Gaussian weight distribution $\rP_{\w^\star}(\vec{w}^\star)= \mN_{\vec{w}^\star}\(\vec{0}, \rho_{\w^\star} \rI_{\ndim}\)$. The $\pm 1$ labels are thus generated as
\begin{align}
\vec{y} &= \sign\(\frac{1}{\sqrt{\ndim}} \mat{X}\vec{w}^\star \)\,, \hhspace \text{with} \hhspace \vec{w}^\star \sim \mN_{\vec{w}^\star}\(\vec{0}, \rho_{\w^\star} \rI_{\ndim}\) .
\label{main:teacher_sign}	
\end{align}

\paragraph{Empirical Risk Minimization ---}
The workhorse of machine learning is \aclink{ERM}, where one minimizes a \emph{loss function} in the corresponding high-dimensional parameter space
$\bbR^{\ndim}$. To avoid overfitting of the training set one often adds a \emph{regularization term} $r(\vec{w})$. 
\aclink{ERM} then corresponds to estimating $\hat{\vec{w}}_{\textrm{erm}} =  \argmin_{\vec{w}} \[ \mL\(\vec{w}; \vec{y}, \mat{X}\) \]$ where the regularized training loss $\mL$ is defined by, using the notation $z_\mu \(\vec{w}, \vec{x}_\mu\) \equiv \frac{1}{\sqrt{\ndim}} \vec{x}_\mu^\intercal\vec{w}$,
\begin{align}
	 \mL\(\vec{w}; \vec{y}, \mat{X}\) = \sum_{\mu=1}^\nsamples l\(y_\mu, z_\mu\(\vec{w}, \vec{x}_\mu\) \) +   r\(\vec{w}\) \,.
	\label{main:training_loss}
\end{align}
The goal of the present chapter is to discuss the generalization performance of these estimators for the classification task (\ref{main:teacher_sign}) in the high-dimensional limit.
We focus our analysis on commonly used loss functions $l$, namely the square $l^{\textrm{square}}(y,z)=\frac{1}{2}(y-z)^2$, logistic $l^{\textrm{logistic}}(y, z)=\log(1+\exp(-y z))$ and hinge losses $l^{\textrm{hinge}}(y, z)=\max\(0,1-yz\)$.
We will mainly illustrate our results for the $\rL_2$ regularization $r\(\vec{w}\) = {\lambda} \|\vec{w}\|_2^2/2$, where we introduced a regularization strength hyper-parameter $\lambda$. The same analysis can be performed for any other convex-separable regularization.

\paragraph{Related works ---} The above learning problem has been extensively studied in the statistical physics community using the heuristic replica method \cite{gardner1989three,seung1992statistical,watkin1993statistical,Kinzel96,engel2001statistical}. Due to the interest in high-dimensional statistics, they have experienced a resurgence in popularity in recent years. In particular, rigorous works on related problems are much more recent. 
The authors of \cite{barbier2017phase} established rigorously the replica-theory predictions for the Bayes-optimal generalization error. Here we focus on standard \aclink{ERM} estimation and compare it to the results obtained in \cite{barbier2017phase}. 
Authors of \cite{Thrampoulidis16} analyzed rigorously M-estimators for the regression case where data are generated by a linear-activation teacher. Here we analyze classification with a more general and non-linear teacher, focusing in particular on the sign-teacher.  
The case of max-margin loss was studied in \cite{Montanari19} with a technically closely related proof, but with a focus on the over-parametrized regime, thus not addressing the questions that we focus on. A range of unregularized losses was also analyzed for a sigmoid teacher (that is very similar to a sign-teacher) again in the context of the double-descent behavior in \cite{deng2019model,kini2020analytic}. Here we focus instead on the regularized case as it drastically improves generalization performances of the \aclink{ERM} and that allows us to compare with the Bayes-optimal estimation as well as to standard generalization bounds.  
Our proof, as in the above mentioned works and \cite{Mignacco2020}, is based on Gordon's minimax formalism, including in particular the effect of the regularization. 
\paragraph{Main contributions ---} 

Our first main contribution is to provide rigorously, in Sec.~\ref{sec:fixed_point}, the classification generalization performances of empirical risk minimization with the loss given by (\ref{main:training_loss}) in the high-dimensional limit, for any convex loss and an $\ell_2$ regularization. Note that the proof is easily extended to any convex separable regularization.
Additionally, we provide a proof of the equivalence between the results of our paper and the ones initially obtained by the replica method, which is of additional interest given the wide range of  application of these heuristics statistical-physics technics in machine learning and computer science \cite{mezard2009information,NatureZdeborova}. In particular, the replica predictions in \cite{Opper1990, Opper1991, Kinzel96} follow from our results.
Another approach that originated in physics are the so-called \aclink{TAP} equations \cite{mezard1989space,kabashima2003cdma,kabashima2004bp} that lead to the \aclink{AMP} algorithm for solving linear and generalized linear problems with Gaussian matrices \cite{donoho2009message,rangan2011generalized}. This algorithm can be analyzed with the so-called \aclink{SE} method \cite{bayati2011dynamics}, and it is widely believed, and in fact proven for linear problems \cite{bayati2011lasso,gerbelot2020asymptotic} that the fixed-point of the \aclink{SE} gives the optimal error in high-dimensional convex optimization problems. 
The \aclink{SE} equations are in fact equivalent to the one given by the replica theory and therefore our results vindicate this approach as well. We also demonstrate numerically that these asymptotic results are very accurate even for moderate system sizes, and they have been performed with the \textsf{scikit-learn} library \cite{scikit-learn}.

Secondly, and more importantly, we provide in Sec.~\ref{sec:applications} a detailed analysis of the generalization error for standard losses such as square, hinge (or equivalently support vector machine) and logistic, as a function of the regularization strength $\lambda$ and the number of samples per dimension $\alpha$. We observe, in particular, that while the ridge regression never closely approaches the Bayes-optimal performance, the logistic regression with optimized $\rL_2$ regularization gets extremely close to optimal. And so does, to a lesser extent, the hinge regression and the max-margin estimator to which the unregularized logistic and hinge converge \cite{Rosset04}. It is quite remarkable that these canonical losses are able to approach the error of the Bayes-optimal estimator for which, in principle, the marginals of a high-dimensional probability distribution need to be evaluated. Notably, all the later losses give ---for a \emph{good choice} of the regularization strength $\lambda$--- generalization errors scaling as $\Theta\(\alpha^{-1}\)$ for large $\alpha$, just as the Bayes-optimal generalization error~\cite{barbier2017phase}. This is found to be at variance with the prediction of Rademacher and max-margin-based bounds that predict instead a $\Theta\(\alpha^{-1/2}\)$ rate \cite{vapnik2006estimation,shalev2014understanding}, which therefore appear to be vacuous in the high-dimensional regime.

Third, in Sec.~\ref{sec:optimality}, we design a custom, non-convex, loss and regularizer that provably gives a plug-in estimator that efficiently achieves Bayes-optimal performances, including the optimal $\Theta\(\alpha^{-1}\)$ rate for the generalization error. Our construction is related to the one discussed in \cite{gribonval2011should,NIPS2013_4868,Advani2016}, but is not restricted to convex losses.

\section{Main technical results}
\label{sec:fixed_point}

In the formulas that arise for this statistical estimation problem, the correlations between
the estimator $\hat{\vec{w}}$ and the ground truth vector
$\vec{w}^\star$ play a fundamental role and we thus define two scalar overlap parameters to measure the statistical reconstruction:
\begin{align}
	m &\equiv \frac{1}{\ndim}\EE_{\vec{y},\mat{X}} \[ \hat{\vec{w}}^\intercal\vec{w}^\star \]\,, 
	&& q \equiv \frac{1}{\ndim}\EE_{\vec{y},\mat{X}}\[\|\hat{\vec{w}}\|_2\]^2\,.
\end{align}
In particular, the generalization error of the estimator $\hat{\vec{w}}(\alpha)
\in \bbR^{\ndim}$, obtained by performing \aclink{ERM} on the training loss $\mL$ in eq.~\eqref{main:training_loss}
with $\nsamples = \alpha \ndim$ samples,
\begin{align}
e_{\textrm{g}}^{\textrm{erm}}(\alpha) \equiv \EE_{y, \vec{x}}  \id\[ y \ne \hat{y}\(\hat{\vec{w}}(\alpha); \vec{x} \) \]\,,
\end{align}
where $\hat{y}\(\hat{\vec{w}}(\alpha); \vec{x}\)$ denotes the predicted label, has both at finite size $\ndim$ and in the asymptotic limit an explicit expression depending only on the above overlaps $m$ and $q$:
\begin{proposition}[Generalization error of classification]
\label{main:thm:generalization_errors}
In our synthetic binary classification task, the 
generalization error of \aclink{ERM} (or equivalently the test error) is given by
\begin{align}
	e_{\textrm{g}}^{\textrm{erm}}\(\alpha\) &= \frac{1}{\pi} \textrm{acos}\( \sqrt{\eta} \)\,, \label{main:generalization_errors}	
\end{align}
with 
\begin{align*} 
	\eta & \equiv \frac{m^2}{\rho_{\ndim} ~ q}\,, && \rho_{\ndim} \equiv \frac{1}{\ndim} \EE\[ \|\vec{w}^\star\|_2^2 \]\,.
	\end{align*}
\end{proposition}
\begin{proof}
The proof is a simple computation based on Gaussian integration.
The generalization error $e_{\textrm{g}}$ is the prediction error of the estimator $\hat{\vec{w}}$ on new samples $\{\vec{y}, \mat{X}\}$, where $\mat{X}$ is a Gaussian matrix with \aclink{i.i.d} entries and $\vec{y}$ are $\pm 1$ labels generated according to \eq\eqref{main:teacher_model} $\vec{y} = \varphi_{\out^\star}\(\vec{z}\)$ with $\vec{z} = \frac{1}{\sqrt{\ndim}} \mat{X} \vec{w}^\star$.
As the model fitted by \aclink{ERM} may not lead to binary outputs, we add a non-linearity $\varphi:\bbR \mapsto \{\pm 1\}$ (for example a sign or a soft-sign) on top of it to ensure to obtain binary outputs $\hat{\vec{y}}\pm 1$ according to $\hat{\vec{y}} = \varphi_\out\(\hat{\vec{z}}\)$ with $\hat{\vec{z}} = \frac{1}{\sqrt{\ndim}} \mat{X} \hat{\vec{w}}$. 
The classification generalization error is given by the probability that the predicted labels $\hat{\vec{y}}$ and the true labels $\vec{y}$ do not match. To compute it, first note that the vectors $(\vec{z},\hat{\vec{z}})$
averaged over all possible ground truth vectors $\vec{w}^\star$ (or equivalently labels $y$) and input matrix $\mat{X}$ follow in the large size limit a joint Gaussian distribution with zero mean and covariance matrix 
\begin{align}
	\bsigma =  \frac{1}{\ndim} \EE_{\vec{w}^\star, \mat{X}} 
\begin{bmatrix}
	\|\vec{w}^{\star}\|_2^2 & \vec{w}^{\star \intercal} \hat{\vec{w}} \\
	\vec{w}^{\star \intercal} \hat{\vec{w}} & \| \hat{\vec{w}}\|_2^2
\end{bmatrix}
\equiv
\begin{bmatrix}
	\rho_{\ndim} & \sigma_{\w^\star \hat{\w}} \\
	\sigma_{\w^\star \hat{\w}} & \sigma_{\hat{\w}}
\end{bmatrix}\,.
\end{align} 
The asymptotic generalization error depends only on the covariance matrix $\bsigma$ and as the samples are \aclink{i.i.d} it reads
\begin{align}
	&e_{\textrm{g}}(\alpha) =  \id\[ y \ne \hat{y}\(\hat{\vec{w}}(\alpha); \vec{x} \) \] = 1 - \bbP[y = \hat{y}\(\hat{\vec{w}}(\alpha); \vec{x} \)] \nonumber \\
	&=  1 - 2  \int_{\(\bbR^+\)^{2}}  \d \vec{x} ~ \mN_{\vec{x}} \( \vec{0}, \bsigma\) \label{main:generalization_error:general} \\ 
	&=  1 - \( \frac{1}{2}  + \frac{1}{\pi}  \atan \(\sqrt{\frac{\sigma_{\w^\star\hat{\w}}^2}{\rho_{\ndim} \sigma_{\hat{\w}} - \sigma_{\w^\star \hat{\w}}^2}}\) \)  = \frac{1}{\pi} \acos\(\eta\) \nonumber \,,
\end{align}
where we used the fact that $\atan(x) = \frac{1}{2}\(\pi - \acos\(\frac{x^2-1}{1+x^2}\)\)$, $\frac{1}{2} \acos(2x^2-1) = \acos(x)$ and defined $\eta =\frac{\sigma_{\w^\star\hat{\w}}}{\sqrt{\rho_{\ndim} \sigma_{\hat{\w}}}}$.
For the \aclink{ERM} estimator, the parameters $\sigma_{\hat{\w}} = \frac{1}{\ndim} \EE_{\vec{w}^\star,\mat{X}} \| \hat{\vec{w}}\|_2^2 = q$ and $\sigma_{\w^\star\hat{\w}} =  \frac{1}{\ndim}  \EE_{\vec{w}^\star,\mat{X}}\(\hat{\vec{w}}\)^\intercal \vec{w}^\star = m$, 
 such that the generalization error for classification is given by \eqref{main:generalization_error:general} with $\eta \equiv \frac{m^2}{\rho_{\ndim}q}$.
\end{proof}
To obtain the generalization performances of \aclink{ERM}, it remains to obtain the asymptotic values of $m$, $q$ (and thus of $\eta$), in the limit $d\to \infty$. 
For any $\tau > 0$, let us first recall the definitions of the Moreau-Yosida regularization $\mM_\tau$ and the proximal operator $\mP_\tau$ of a convex loss function $(y,z) \mapsto \ell(y \cdot z)$:
\begin{align}
\begin{aligned}
\label{eq:MY_reg}
	\mM_\tau(z) &=  \min_x \Big\{\ell(x) + \frac{(x-z)^2}{2\tau} \Big\}\,,\\
	\mP_\tau(z) &=  \argmin_x \Big\{\ell(x) + \frac{(x-z)^2}{2\tau} \Big\}\,.
\end{aligned}
\end{align}
 With the $\ell_2$ regularization, the asymptotic overlaps are characterized by a set of fixed point equations and follow from the Gordon's Convex Gaussian Min-max Theorem (CGMT) states in the next theorems.
\begin{theorem}[Gordon's min-max fixed point - Regression/Classification with $\rL_2$ regularization]
	\label{apppendix:thm:gordon_fixed_points:classification}
As $\nsamples, \ndim \to \infty$ with $\nsamples/ \ndim = \alpha = \Theta(1)$, the overlap parameters $m, q$ concentrate to
\begin{align}
m & \underlim{\ndim}{\infty} \sqrt{\rho_{\w^\star}} \nu^\ast\,, && q \underlim{\ndim}{\infty} (\nu^\ast)^2 + (\delta^\ast)^2 && \rho_{\ndim}\underlim{\ndim}{\infty} \rho_{\w^\star}\,.
\end{align}
For \textbf{regression} the parameters $\nu^\ast, \delta^\ast$ are the solutions of
\begin{align}\label{eq:pot_func_general}
(\nu^\ast, \delta^\ast) &= \underset{\nu, \delta \ge 0}{\arg\min} \ \sup_{\tau > 0} \left\{\frac{\lambda(\nu^2 + \delta^2)}{2} - \frac{\delta^2}{2\tau} \right.\\ 
 	& \hspace{2cm} \left. + \alpha \EE_{g, s} \mM_{\tau}[l(\varphi_{\out^\star}(\sqrt{\rho_{\w^\star}} s),.)](\nu s + \delta g)\right\}, \nonumber
\end{align}
while for \textbf{classification},  $\nu^\ast, \delta^\ast$ are the solutions of
\begin{align}
\label{eq:pot_func}
(\nu^\ast, \delta^\ast) &= \underset{\nu, \delta \ge 0}{\arg\min} \ \sup_{\tau > 0} \left\{\frac{\lambda(\nu^2 + \delta^2)}{2} - \frac{\delta^2}{2\tau} \right.\\ 
 	& \hspace{2cm} \left.  + \alpha \EE_{g, s} \mM_\tau[\delta g + \nu s \varphi_{\out^\star}(\sqrt{\rho_{\w^\star}} s)]\right\}\,. \nonumber
\end{align}
Here, $g, s$ are two \aclink{i.i.d} standard Gaussian normal random variables. The solutions $(\nu^\ast, \delta^\ast)$ of \eqref{eq:pot_func} for classification can be reformulated as a set of fixed point equations 
\begin{align}
	\nu^\ast &= \frac{\alpha}{\lambda \tau^\ast + \alpha } \EE_{g,s} [s \times \varphi_{\out^\star}(\sqrt{\rho_{\w^\star}} s) \nonumber\\ 
	& \qquad \qquad \qquad \qquad  \times \mP_{\tau^\ast}(\delta^\ast g+ \nu^\ast s \varphi_{\out^\star}(\sqrt{\rho_{\w^\star}} s))]\,, \nonumber \spacecase
	\delta^\ast &= \frac{\alpha}{\lambda \tau^\ast + \alpha -1} \EE_{g,s} [g \times \mP_{\tau^\ast}(\delta^\ast g+ \nu^\ast s \varphi_{\out^\star}(\sqrt{\rho_{\w^\star}} s))]\,, \nonumber \spacecase
	(\delta^\ast)^2 &= \alpha \EE_{g,s} [\((\delta^\ast g + \nu^\ast s \varphi_{\out^\star}(\sqrt{\rho_{\w^\star}} s)) \right.  \label{main:fixed_point_equations_gordon} \\ 
	&  \qquad \qquad \qquad \qquad \left. - \mP_{\tau^\ast}(\delta^\ast g + \nu^\ast s \varphi_{\out^\star}(\sqrt{\rho_{\w^\star}} s)) \)^2] \,. \nonumber
\end{align}
\end{theorem}
\begin{proof}
Since the teacher weight vector $\vec{w}^\star$ is independent of the input data matrix $\mat{X}$, we can assume without loss of generality that $\vec{w}^\star = \sqrt{d ~ \rho_d} \vec{e}_1$ where $\vec{e}_1$ is the first natural basis vector of $\bbR^d$, and $\rho_d = \norm{\vec{w}^\star}_2^2/ d$. As $d \to \infty$, $\rho_d \to \rho_{\w^\star}$. 
Accordingly, it will be convenient to split the data matrix into two parts $\mat{X} = \[\vec{s}, \mat{B}  \]$, 
where $\vec{s} \in \bbR^{n \times 1}$ and $\mat{B} \in \bbR^{n \times (d-1)}$ are two sub-matrices of \aclink{i.i.d} standard normal entries. The weight vector $\vec{w}$ can also be written as $\vec{w} = [\sqrt{d} \nu, \vec{v}^\intercal]^\intercal$, where $\nu \in \bbR$ denotes the projection of $\vec{w}$ onto the direction spanned by the teacher weight vector $\vec{w}^\star$, and $\vec{v} \in \bbR^{d-1}$ is the projection of $\vec{w}$ onto the complement subspace. These representations serve to simplify the notations in our subsequent derivations. For example, we can now write the output as $y_\mu = \varphi_{\out^\star}(\sqrt{\rho_d} s_\mu)$ where $s_\mu$ is the $\mu$-th entry of the Gaussian vector $\vec{s}$.
Let $\Phi_d$ denote the cost of the \aclink{ERM} according to the loss \eqref{main:training_loss}, normalized by $\ndim$. Using our new representations introduced above, we have
\begin{equation}
\label{eq:Phi}
	\Phi_d = \min_{\nu, \vec{v}} \frac{1}{d}\sum_{\mu=1}^\nsamples l\(y_\mu, \nu s_\mu + \tfrac{1}{\sqrt{d}}\vec{b}_\mu^\intercal \vec{v} \) +   \frac{\lambda (d\nu^2 + \norm{\vec{v}}^2)}{2d},
	\end{equation}
where $\vec{b}_\mu^\intercal$ denotes the $i$-th row of $\mat{B}$. Since the loss function $l(y_\mu, z)$ is convex with respect to $z$, we can rewrite it as $l(y_\mu, z) = \sup_q \{q z - l^\ast(y_\mu, q)\}$, 
where $l^\ast(y_\mu, q) = \sup_z \{qz - l(y_\mu, z)\}$ is its convex conjugate. Substituting $l$ into \eqref{eq:Phi}, we obtain
\begin{align*}
\Phi_d &= \min_{\nu, \vec{v}} \, \sup_{\vec{q}}  \left\{\frac{\nu \vec{q}^\intercal \vec{s}}{d} + \frac{1}{d^{3/2}} \vec{q}^\intercal \mat{B} \vec{v} \right.\\
&\qquad \qquad \qquad \qquad \qquad \left. - \frac{1}{d} \sum_{\mu=1}^n l^\ast(y_\mu, q_\mu) + \frac{\lambda \(d\nu^2 + \norm{\vec{v}}^2\)}{2d}\right\}. 
\end{align*}
Now consider a new optimization problem
\begin{align*}
\label{eq:Phi_s}
	\td{\Phi}_d &= \min_{\nu, \vec{v}} \, \sup_{\vec{q}}  \left\{\frac{\nu \vec{q}^\intercal \vec{s}}{d} + \frac{\norm{\vec{q}}}{\sqrt{d}} \frac{\vec{h}^\intercal \vec{v}}{d} + \frac{\norm{\vec{v}}}{\sqrt{d}} \frac{\vec{g}^\intercal \vec{q}}{d} \right.\\
	&\qquad \qquad \qquad \qquad \qquad  \left.- \frac{1}{d} \sum_{\mu=1}^n l^\ast(y_\mu, q_\mu) + \frac{\lambda \(d\nu^2 + \norm{\vec{v}}^2\)}{2d}\right\},
\end{align*}
where $\vec{h} \sim \mN\(\vec{0},\rI_{\ndim-1}\)$ and $\vec{g} \sim  \mN\(\vec{0},\rI_{n}\)$ are two independent standard normal vectors. It follows from Gordon's minimax comparison inequality, see \eg \cite{gordon1985some, pmlr-v40-Thrampoulidis15}, that
\begin{equation}\label{eq:cgmt}
\mathbb{P}(\abs{\Phi_d - c} \ge \epsilon) \le 2 \mathbb{P}\(\abs{\td{\Phi}_d-c} \ge \epsilon\)\,,
\end{equation}
for any constants $c$ and $\epsilon > 0$. This implies that $\td{\Phi}_d$ serves as a surrogate of $\Phi_d$. Specifically, if $\td{\Phi}_d$ concentrates around some deterministic limit $c$ as $d \to \infty$, so does $\Phi_d$. In what follows, we proceed to solve the surrogate problem for $\td{\Phi}_d$. First, let $\delta = \norm{\vec{v}}/\sqrt{d}$. It is easy to see that $\td{\Phi}_d$ can be simplified as
\begin{align*}
	&\td{\Phi}_d = \min_{\nu, \delta \ge 0} \, \sup_{\vec{q}}  \left\{\frac{\vec{q}^\intercal (\nu\vec{s} + \delta \vec{g})}{d} - \delta \frac{\norm{\vec{q}}}{\sqrt{d}} \frac{\norm{\vec{h}}}{\sqrt{d}} \right. \\  
	& \qquad \qquad \qquad \qquad \qquad \qquad  \left. - \frac{1}{d} \sum_{\mu=1}^n l^\ast(y_\mu, q_\mu) + \frac{\lambda (\nu^2 + \delta^2)}{2}\right\}\\
	&\overset{(a)}= \min_{\nu, \delta \ge 0} \, \sup_{\tau > 0} \, \sup_{\vec{q}} \left\{-\frac{\tau \norm{\vec{q}}^2}{2d} - \frac{\delta^2 \norm{\vec{h}}^2}{2\tau d} +\frac{\vec{q}^\intercal (\nu\vec{s} + \delta \vec{g})}{d} \right. \\  
	&\qquad \qquad \qquad \qquad \qquad \qquad  \left. - \frac{1}{d} \sum_{\mu=1}^n l^\ast(y_\mu, q_\mu) + \frac{\lambda (\nu^2 + \delta^2)}{2}\right\}\\
	&=\min_{\nu, \delta \ge 0} \, \sup_{\tau > 0}\left\{ \frac{\lambda (\nu^2 + \delta^2)}{2}-\frac{\delta^2 \norm{\vec{h}}^2}{2\tau d}  \right. \\  
	& \qquad \qquad \qquad \qquad  \left. - \frac{\alpha}{n}\inf_{\vec{q}} \[\frac{\tau \norm{\vec{q}}^2}{2} - \vec{q}^\intercal (\nu\vec{s} + \delta \vec{g})+ \sum_{\mu=1}^n l^\ast(y_\mu, q_\mu)\]\right\}\\
	&\overset{(b)}=\min_{\nu, \delta \ge 0} \, \sup_{\tau > 0}\left\{ \frac{\lambda (\nu^2 + \delta^2)}{2}-\frac{\delta^2 \norm{\vec{h}}^2}{2\tau d} \right. \\  
	& \qquad \qquad \qquad \qquad \qquad  \left. - \frac{\alpha}{n}\sum_{\mu=1}^n \mM_{\tau}[l(y_\mu,.)](\nu s_\mu + \delta g_\mu)\right\}.
\end{align*}
In $(a)$, we have introduced an auxiliary variable $\tau$ to rewrite $- \delta \frac{\norm{\vec{q}}}{\sqrt{d}} \frac{\norm{\vec{h}}}{\sqrt{d}}$ as
\begin{align*}
- \delta \frac{\norm{\vec{q}}}{\sqrt{d}} \frac{\norm{\vec{h}}}{\sqrt{d}} = \sup_{\tau > 0} \left\{-\frac{\tau \norm{\vec{q}}^2}{2d} - \frac{\delta^2 \norm{\vec{h}}^2}{2\tau d}\right\}\,,
\end{align*}
and to get $(b)$, we use the identity 
\begin{align*}
\inf_q \left\{\frac{\tau}{2} q^2 - q z + \ell^\ast(q)\right\} = - \inf_x \left\{\frac{(z-x)^2}{2\tau} + \ell(x)\right\} \,,
\end{align*}
that holds for any $z$ and for any convex function $\ell(x)$ and its conjugate $\ell^\ast(q)$. As $d \to \infty$, standard concentration arguments give us $\frac{\norm{\vec{h}}^2}{d} \to 1$ and $\frac{1}{n}\sum_{\mu=1}^n \mM_{\tau}[l(y_\mu,.)](\nu s_\mu + \delta g_\mu) \to \mathbb{E}_{g, s} \mM_{\tau}[l(y,.)](\nu s + \delta g)$  uniformly over $\tau, \nu$ and $\delta$. Using \eqref{eq:cgmt}, we can conclude that the normalized cost of the \aclink{ERM} $\Phi_d$ converges to the optimal value of the deterministic optimization problem in \eqref{eq:pot_func_general}. Finally, since $\lambda > 0$, one can show that the cost function of \eqref{eq:pot_func_general} has a unique global minima at $\nu^\ast$ and $\delta^\ast$. It follows that the empirical values of $(\nu, \delta)$ also converge to their corresponding deterministic limits $(\nu^\ast, \delta^\ast)$.

To obtain the result for \emph{classification}, we note that
\begin{align*}
\mM_{\tau}[l(y,.)](z) &= \min_x \left\{l(y; x) + \frac{(x-z)^2}{2\tau}\right\}= \min_x \left\{\ell(yx) + \frac{(x-z)^2}{2\tau}\right\}\\
&=\min_x \left\{\ell(x) + \frac{(x-yz)^2}{2\tau}\right\} = \mM_\tau(yz),
\end{align*}
where to reach the last equality we have used the fact that $y \in \{\pm 1\}$. Substituting this special form into \eqref{eq:pot_func_general} and recalling $y_\mu = \varphi_{\out^\star}(\sqrt{\rho_d} s_\mu)$, we obtain the result.
Finally, to obtain the fixed point equations, we simply take the partial derivatives of the cost function with respect to $\nu, \delta, \tau$, and use the following well-known calculus rules for the Moreau-Yosida regularization \cite{hiriartUrruty1993}:
\begin{align*}
\frac{\partial \mM_\tau(z)}{\partial z} &= \frac{z - \mP_\tau(z)}{\tau}\,, && \frac{\partial \mM_\tau(z)}{\partial \tau} = -\frac{(z - \mP_\tau(z))^2}{2\tau^2}.
\end{align*}
\end{proof}

Interestingly, this set of fixed point equations \eqref{main:fixed_point_equations_gordon} can be finally mapped to the ones
obtained by the heuristic \emph{replica} method from statistical
physics, whose heuristic derivation is shown in
SM.~III.3 of \cite{aubin2020generalization}, as well as the \aclink{SE} of the \aclink{AMP} \cite{kabashima2003cdma,rangan2011generalized,zdeborova2016statistical}. Thus their validity for this convex estimation problem is rigorously established by the following theorem:
\begin{corollary}[Equivalence Gordon-replicas]
\label{main:corollary:equivalence_gordon_replicas_formulation_l2}
As $\nsamples, \ndim \to \infty$ with $\nsamples/ \ndim = \alpha =
\Theta(1)$, the overlap parameters $m, q$ concentrate to the
fixed point of the following set of equations:
\begin{align}
	m &= \alpha \Sigma \rho_{\w^\star} \times \EE_{y, \xi } \[ \mZ_{\out^\star} \(y,  \sqrt{\rho_{\w^\star} \eta} \xi, \rho_{\w^\star}\(1 - \eta\)\)  \right. \nonumber \\  
	&\qquad \qquad  \left. \times f_{\out^\star} \(y,  \sqrt{\rho_{\w^\star} \eta} \xi, \rho_{\w^\star}\(1 - \eta\)\) \times f_{\out} \( y,  q^{1/2}\xi, \Sigma \)  \]  \nonumber\\
	q &= m^2/\rho_{\w^\star} + \alpha \Sigma^2 \times \EE_{y, \xi } \[ \mZ_{\out^\star} \( y,  \sqrt{\rho_{\w^\star}\eta} \xi, \rho_{\w^\star}\(1 - \eta\)  \)  \right. \label{main:fixed_point_equations_replicas} \\  
	&\qquad \qquad \qquad \qquad \qquad \qquad \qquad \left. \times f_{\out} \( y,  q^{1/2}\xi, \Sigma \)^2  \] \nonumber \\
	\Sigma &=   \( \lambda - \alpha \times \EE_{y, \xi } \[ \mZ_{\out^\star} \( y,  \sqrt{\rho_{\w^\star} \eta} \xi, \rho_{\w^\star}\(1 - \eta\)  \) \right.\right. \nonumber\\  
	& \qquad \qquad \qquad \qquad \qquad \qquad \qquad \left. \left. \times  \partial_\omega f_{\out} \( y,  q^{1/2}\xi, \Sigma \)    \] \)^{-1} \nonumber
\end{align}
with $\eta \equiv \frac{m^2}{\rho_{\w^\star}q}$ and where $\xi, z$ denote two \aclink{i.i.d} standard normal random variables, and $\EE_y$ the continuous or discrete sum over all possible values $y$ according to $\rP_{\out^\star}$. The corresponding functions $\mZ_{\out^\star}$, $f_{\out^\star}$ and $f_{\out}$, $\partial_\omega f_{\out}$ are defined in \App\ref{appendix:definitions:distributions:committee}-\ref{appendix:definitions:updates:committee}.
\end{corollary}
For clarity, the proof is left in SM.~III.3 of \cite{aubin2020generalization}. Moreover, an equivalent set of six equations for the whole \aclink{GLM} class (classification and regression) with any separable and convex regularizer different than $\rL_2$ are shown in \App\ref{appendix:fixed_point_erm} and in SM.~III.2 of \cite{aubin2020generalization}.

\paragraph{Bayes optimal baseline ---} Finally, we shall 
compare the \aclink{ERM} performances to the Bayes-optimal generalization
error. Being the information-theoretically best possible estimator, we will use it as a reference baseline for comparison.
The expression of the Bayes-optimal generalization was derived in
\cite{Opper1991} and  proven in \cite{barbier2017phase} and we recall here the result:
\begin{theorem}[Bayes asymptotic performance, from \cite{barbier2017phase}]
\label{main:thm:fixed_point_equations_bayes}
For the model \eqref{main:teacher_model} with $\rP_{\w^\star}(\vec{w}^\star)= \mN_{\vec{w}^\star}\(\vec{0}, \rho_{\ndim} \rI_{\ndim}\)$, such that $\rho_{\ndim} \underlim{\ndim}{\infty} \rho_{\w^\star}$, the Bayes-optimal generalization error is quantified by two scalar parameters $q_\bayes$ and $\hat{q}_\bayes$ that verify asymptotically the set of fixed point equations
	\begin{align}
		q_\bayes &= \frac{\hat{q}_\bayes}{1+\hat{q}_\bayes}
		\,, \label{main:fixed_point_equations_bayes} \\
		\hat{q}_\bayes &= \alpha \EE_{y, \xi } \[ \mZ_{\out^\star } \( y,  q_\bayes^{1/2}\xi, \rho_{\w^\star} - q_\bayes \) \cdot  f_{\out^\star } \( y,  q_\bayes^{1/2}\xi, \rho_{\w^\star} - q_\bayes \)^2    \]  \,, \nonumber
	\end{align}
and is expressed by
	\begin{align}
        e_{\textrm{g}}^{\textrm{bayes}}\(\alpha\) = \frac{1}{\pi}
          \textrm{acos}\(\sqrt{\eta_\bayes}\) &\text{~~with~~~} \eta_\bayes = \frac{q_\bayes}{\rho_{\w^\star}}\,.
          \label{main:generalization_error_bayes}
	\end{align}
\end{theorem}
\begin{proof}
The Bayes estimator $\hat{\vec{w}}$ is the average over the posterior distribution, denoted $\langle . \rangle$, knowing the teacher prior $\rP_{\w^\star}$ and channel $\rP_{\out^\star}$ distributions so that $\hat{\vec{w}}=\langle \vec{w} \rangle$. Hence we obtain $m=q=q_{\bayes}$ and the parameters $\lim_{\ndim \to \infty}  \sigma_{\hat{\w}} = \lim_{\ndim \to \infty} \frac{1}{\ndim}  \EE_{\vec{w}^\star,\mat{X}} \|\langle \vec{w} \rangle\|_2^2  \equiv q_{\bayes}$ and $\lim_{\ndim \to \infty} \sigma_{\w^\star\hat{\w}} =  \lim_{\ndim \to \infty}  \frac{1}{\ndim}\EE_{\vec{w}^\star,\mat{X}}  \langle \vec{w} \rangle^\intercal \vec{w}^\star \equiv m_{\bayes}$. Using the Nishimori identity, see \App\ref{appendix:replica_computation:nishimori}, the generalization error \eqref{main:generalization_errors} simplifies in the Bayes-optimal setting to \eqref{main:generalization_error:general} with $\eta_\bayes = \frac{q_\bayes}{\rho_{\w^\star}}$.
\end{proof}
\section{Generalization errors}
\label{sec:applications}
We now move to the core of the contribution and analyze the set of fixed point equations \eqref{main:fixed_point_equations_gordon}, or equivalently \eqref{main:fixed_point_equations_replicas}, leading to the generalization performances given by \eqref{main:generalization_errors},
 for common classifiers on our synthetic binary classification task. As already stressed, even though the results are valid for a wide range of regularizers, we focus on estimators based on \aclink{ERM} with $\rL_2$ regularization $r(\vec{w}) = \lambda \|\vec{w}\|_2^2/2$, and with square loss (ridge regression) $l^{\textrm{square}}(y,z)=\frac{1}{2}(y-z)^2$, logistic loss (logistic regression) $l^{\textrm{logistic}}(y, z) = \log(1+\exp(-y z))$ or hinge loss (\aclink{SVM}) $l^{\textrm{hinge}}(y, z) = \max\(0,1-yz\)$. In particular, we study the influence of the hyper-parameter $\lambda$ on the generalization performances and the different large $\alpha$ behavior generalization rates in the high-dimensional regime, and compare with the Bayes results.
 We show the solutions of the set of fixed point equations eqs.~\eqref{main:fixed_point_equations_replicas} in Figs.~\ref{fig:gen_error_ridge},~\ref{fig:gen_error_hinge},~\ref{fig:gen_error_logistic} respectively for ridge, hinge and logistic $\rL_2$ regressions.   Ridge regression is a special case, for which its quadratic loss allows to derive and fully solve the equations, see SM.~V.3 of \cite{aubin2020generalization}. However in general the set of equations has no analytical closed form and needs therefore to be solved numerically. It is in particular the case for logistic and hinge, whose Moreau-Yosida regularization is, however, analytical.  

 First, to highlight the accuracy of the theoretical predictions, we compare in Figs.~\ref{fig:gen_error_ridge}-\ref{fig:gen_error_hinge}-\ref{fig:gen_error_logistic} the \aclink{ERM} asymptotic ($\ndim\to \infty$) generalization error with the performances of numerical simulations ($\ndim = 10^{3}$, averaged over $n_s=20$ samples) of \aclink{ERM} of the training loss eq.~\eqref{main:training_loss}. Presented for a wide range of number of samples $\alpha$ and of regularization strength $\lambda$, we observe a perfect match between theoretical predictions and numerical simulations so that the error bars are barely visible and have been therefore removed. This shows that the asymptotic predictions are valid even with very moderate sizes. As an information theoretical baseline, we also show the Bayes-optimal performances (black) given by the solution of eq.~\eqref{main:fixed_point_equations_bayes}.

\subsection{Ridge estimation}
As we might expect the square loss gives the worst performances. For low values of the generalization, it leads to an interpolation-peak at $\alpha=1$. The limit of vanishing regularization $\lambda \to 0$ leads to the  \emph{least-norm} or \emph{pseudo-inverse} estimator $\hat{\vec{w}}_{\textrm{pseudo}} = \(\mat{X}^\intercal\mat{X}\)^{-1}\mat{X}^\intercal \vec{y}$. The corresponding generalization error presents the largest interpolation-peak and achieves a maximal generalization error $e_{\textrm{g}}=0.5$. These are well known observations, discussed  as early as in \cite{Kinzel96, Opper1990}, that are object of a renewal of interest under the name \emph{double descent}, following a recent series of papers \cite{Hastie19,Geiger19,geiger2020scaling,belkin2019reconciling,Mitra2019,mei2019generalization, gerace2020generalisation,d2020double}. This double descent behavior for the pseudo-inverse is shown in Fig.~\ref{fig:gen_error_ridge} with a yellow line. On the contrary, larger regularization strengths do not suffer this peak at $\alpha=1$, but their generalization error performance is significantly worse than the Bayes-optimal baseline for larger values of $\alpha$. Indeed, as we might expect, for a large number of samples, a large regularization biases wrongly the training. However, even with optimized regularizations, performances of the ridge estimator remains far away from the Bayes-optimal performance.

\subsection{Hinge and logistic estimation} 
Both these losses, which are the classical ones used in classification problems, improve drastically the generalization error. First of all, let us notice that they do not display a double-descent behavior. This is due to the fact that our results are illustrated in the noiseless case and that our synthetic dataset is always linearly separable. Optimizing the regularization, our results in Fig.~\ref{fig:gen_error_ridge}-\ref{fig:gen_error_hinge}-\ref{fig:gen_error_logistic} show both hinge and logistic \aclink{ERM}-based classification approach very closely the Bayes error. 
 To offset these results, note that performances of logistic regression on non-linearly separable data are however very poor, as illustrated by our analysis of a \emph{rectangle door} teacher, see SM.~V.6 of \cite{aubin2020generalization}.
  
 \subsection{Max-margin estimation} As discussed in \cite{Rosset04}, both the logistic and hinge  estimator converge, for vanishing regularization $\lambda\to0$, to the \emph{max-margin} solution.  Taking the $\lambda \to 0$ limit in our equations, we thus obtain the \emph{max-margin} estimator performances. While this is not what gives the best generalization error (as can be seen in Fig.\ref{fig:gen_error_logistic} the logistic with an optimized $\lambda$ has a lower error), the max-margin estimator gives very good results, and gets very close to the Bayes-error.

 \subsection{Optimal regularization} Defining the regularization value that optimizes the generalization as
\begin{align}
	\lambda^{\textrm{opt}}\(\alpha\) &= \argmin_{\lambda} ~ e_{\textrm{g}}^{\textrm{erm}}\(\alpha, \lambda\)\,,
\end{align} 
we show in Figs.~\ref{fig:gen_error_ridge}-\ref{fig:gen_error_hinge}-\ref{fig:gen_error_logistic} that both optimal values $\lambda^{\textrm{opt}}\(\alpha\)$ (dashed-dotted orange) for logistic and hinge regression decrease to $0$ as $\alpha$ grows and more data are given. Somehow surprisingly, we observe in particular that  the generalization performances of logistic regression with optimal regularization are \emph{extremely close} to the Bayes performances. The difference with the optimized logistic generalization error is barely visible by eye, so that we explicitly plotted the difference, which is roughly of order $10^{-5}$. 
Ridge regression Fig.~\ref{fig:gen_error_ridge} shows a singular behavior: there exists an  optimal value (purple) which is moreover independent of $\alpha$ achieved for $\lambda^{\textrm{opt}} \simeq 0.5708$. This value was first found numerically and confirmed afterwards semi-analytically in SM.~V.3 in \cite{aubin2020generalization}.

\begin{figure}
    \centering
    \begin{subfigure}[b]{\textwidth}
		\centering
		\includegraphics[width=0.49\linewidth]{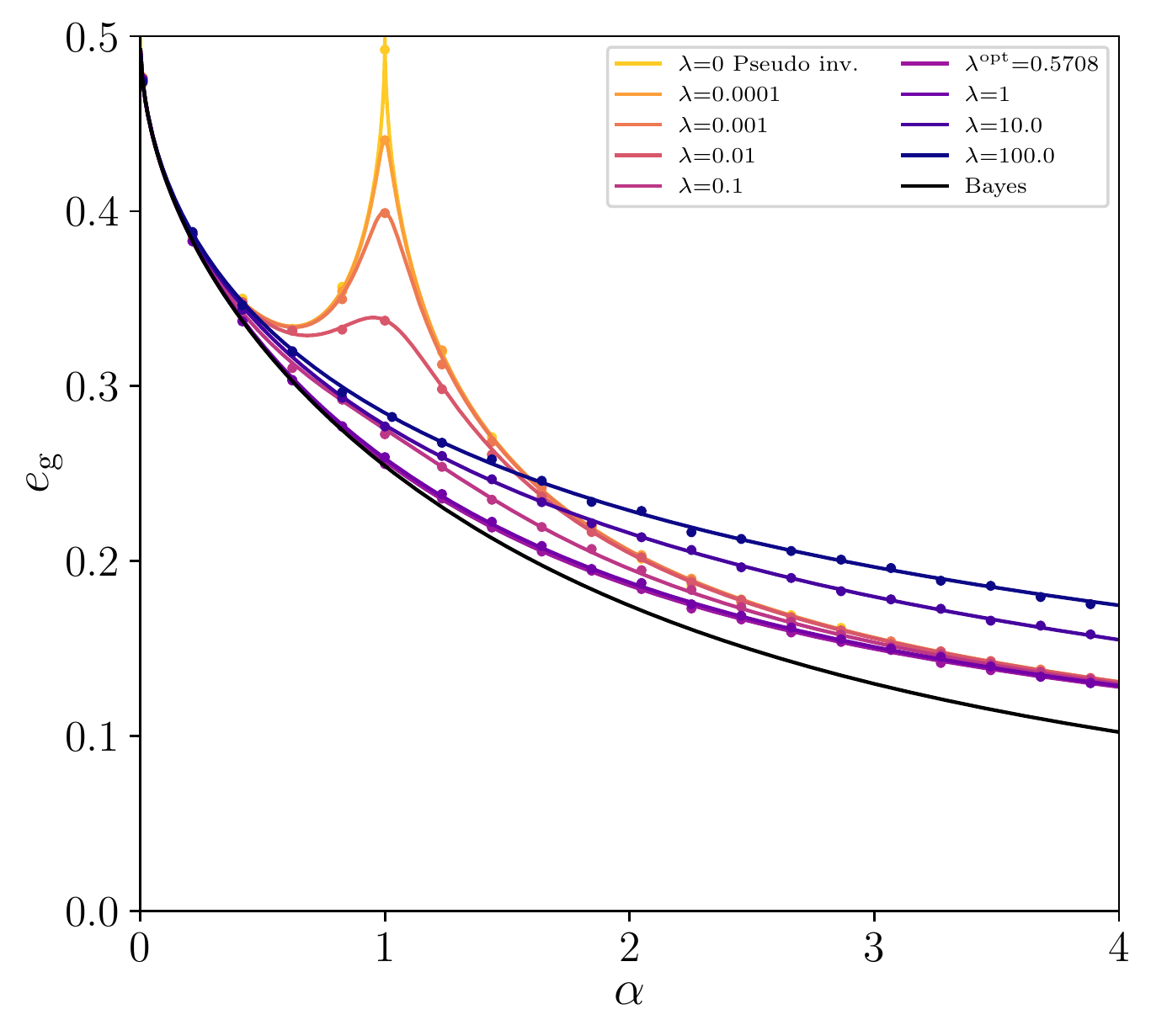}
		\hfill
		\includegraphics[width=0.49\linewidth]{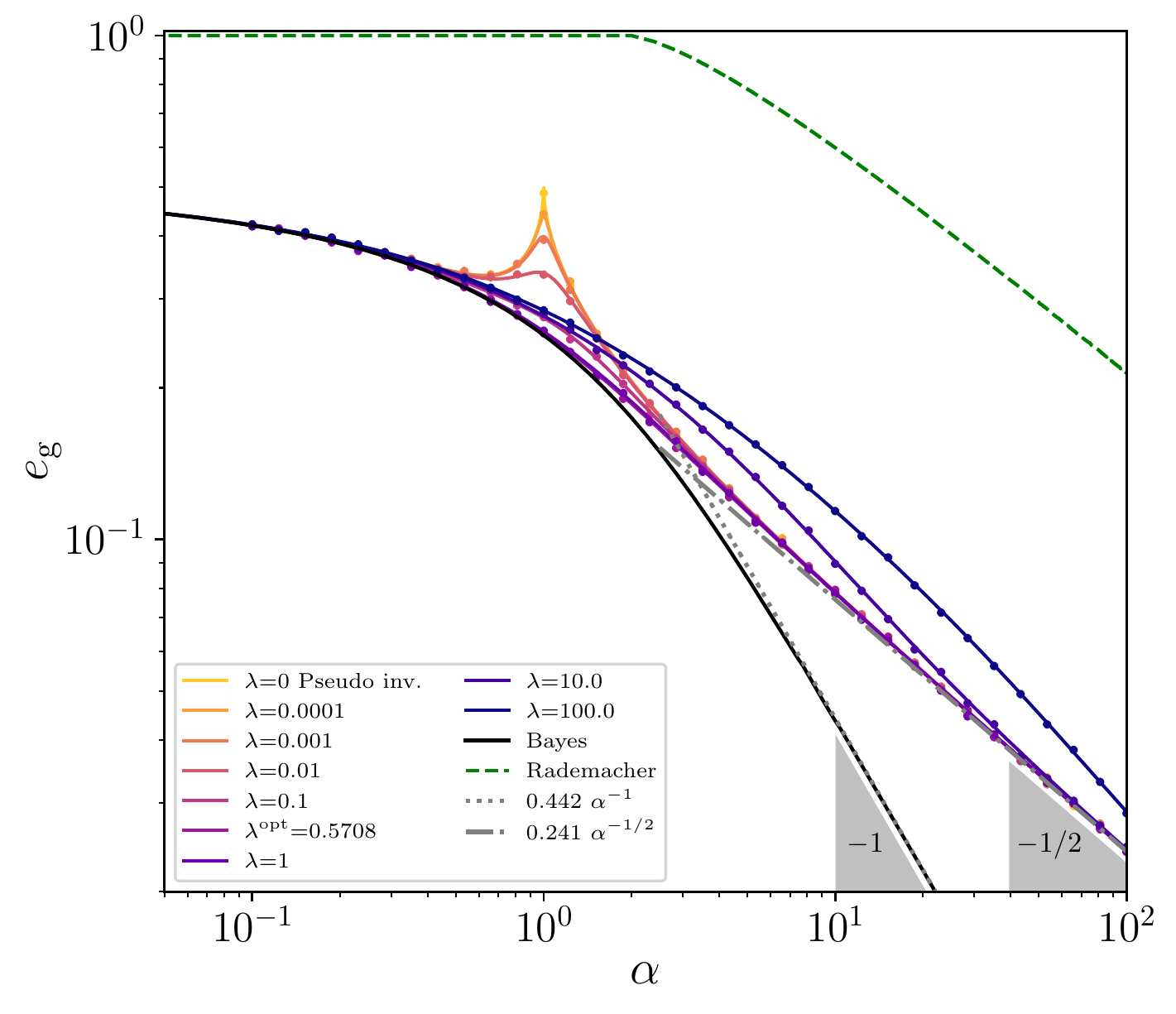}
		\caption{Ridge regression: square loss with $\rL_2$ regularization. Interpolation-peak, at $\alpha=1$, is maximal  for the pseudo-inverse estimator $\lambda=0$ (yellow line) that reaches $e_{\textrm{g}}=0.5$.
		}
		\label{fig:gen_error_ridge}	
    \end{subfigure}
    \vskip\baselineskip
    \begin{subfigure}[b]{\textwidth}
		\centering
	\includegraphics[width=0.49\linewidth]{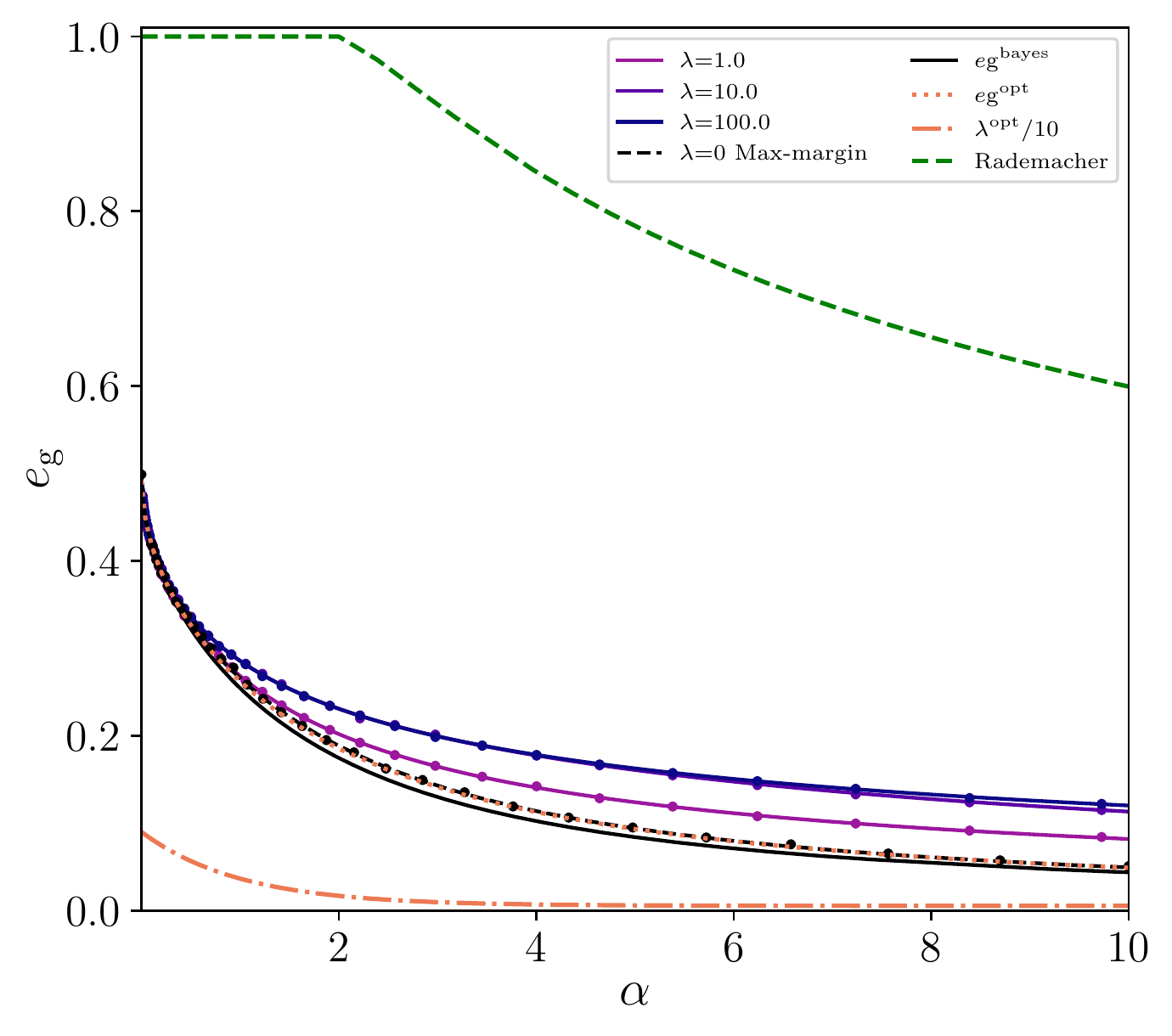}
	\hfill
	\includegraphics[width=0.49\linewidth]{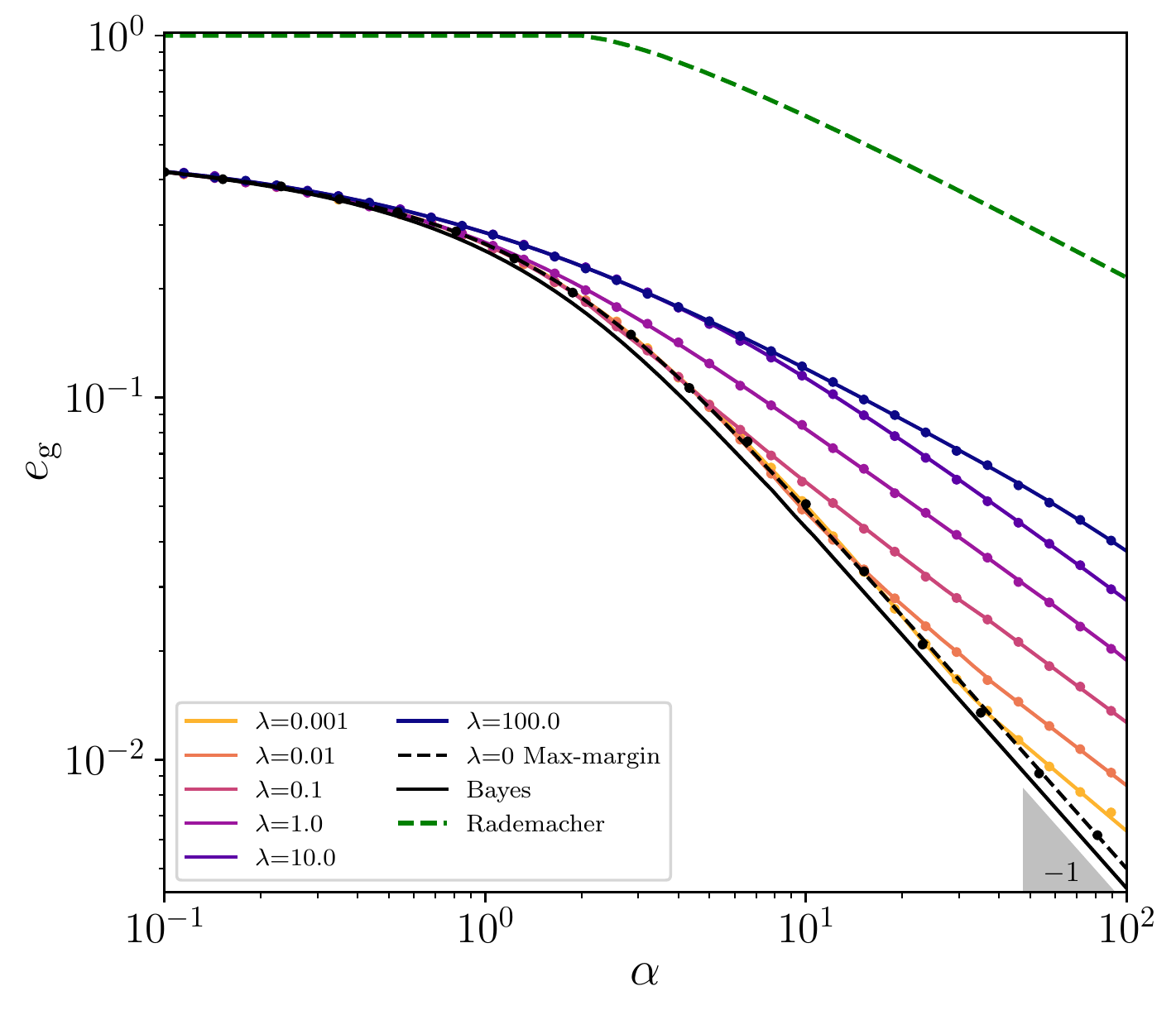}
		\caption{Hinge regression: hinge loss with $\rL_2$ regularization. For clarity the rescaled value of $\lambda^{\textrm{opt}}/10$ (dotted-dashed orange) is shown as well as its generalization error $e_{\textrm{g}}^{\textrm{opt}}$ (dotted orange) that is slightly below and almost indistinguishable of the max-margin performances (dashed black).}
		\label{fig:gen_error_hinge}	
	\end{subfigure}
	\phantomcaption
\end{figure}

\begin{figure}
    \centering
    \ContinuedFloat
    \begin{subfigure}[b]{\textwidth}
		\centering
		\includegraphics[width=0.49\linewidth]{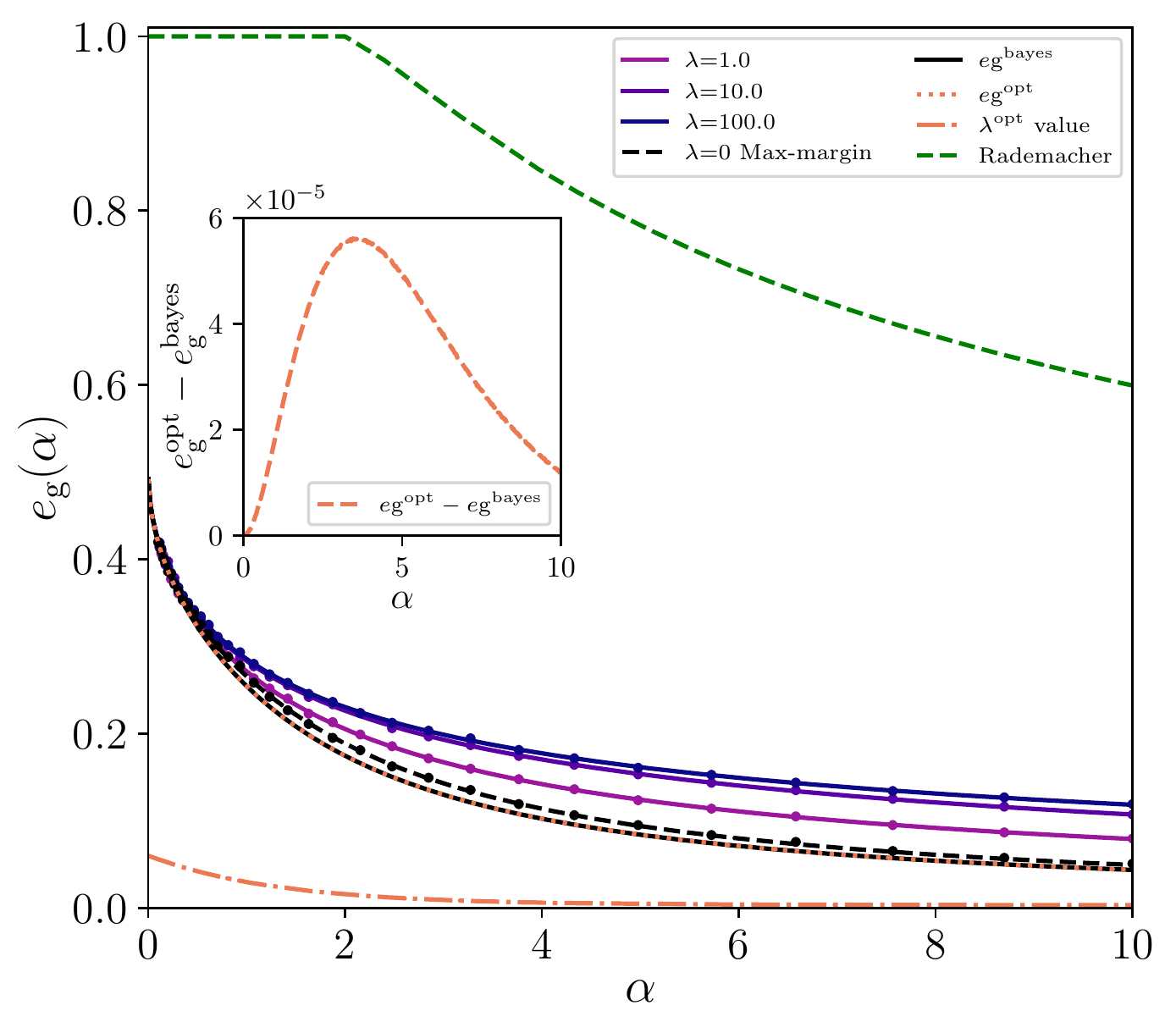}
		\hfill
		\includegraphics[width=0.49\linewidth]{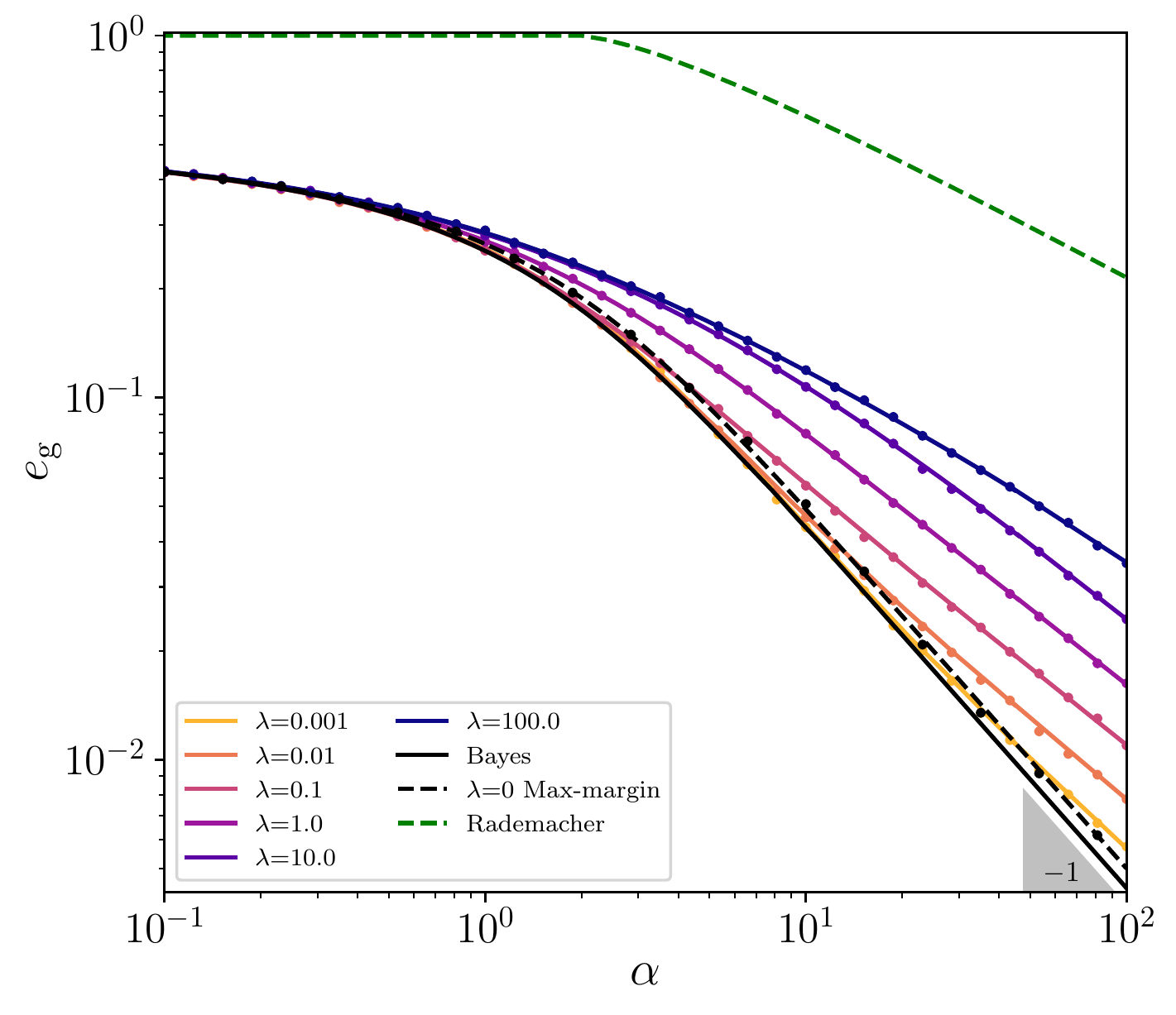}
	 \caption{Logistic regression: logistic loss with $\rL_2$ regularization - The value of $\lambda^{\textrm{opt}}$ (dotted-dashed orange) is shown as well as its generalization error $e_{\textrm{g}}^{\textrm{opt}}$ (dotted orange). Visually indistinguishable from the Bayes-optimal line, their difference $e_{\textrm{g}}^{\textrm{opt}}-e_{\textrm{g}}^{\textrm{bayes}}$ is shown as an inset (dashed orange).}
	 \label{fig:gen_error_logistic}	
	 \end{subfigure}
     \caption{Asymptotic generalization error for $\rL_2$ regularization ($\ndim \to \infty$) as a function of $\alpha$ for different regularizations strengths $\lambda$, compared to numerical simulation (points) of ridge regression for $\ndim=10^{3}$ and averaged over $n_s=20$ samples. Numerics has been performed with the default methods \textit{Ridge}, \textit{LinearSVC},  \textit{LogisticRegression} of \textsf{scikit-learn} package \cite{scikit-learn}. Bayes optimal performances are shown with a black line and goes as $\Theta\(\alpha^{-1}\)$, while the Rademacher complexity (dashed green) decrease as $\Theta\(\alpha^{-1/2}\)$. Both hinge and logistic converge to max-margin estimator (limit $\lambda=0$) which is shown in dashed black and deceases as $\Theta(\alpha^{-1})$, while Ridge decreases as $\Theta(\alpha^{-1/2})$.
     }
     \label{fig:gen_error_global}
\end{figure}

\subsection{Generalization rates at large $\alpha$} 
Finally, we turn to the very instructive behavior at large values of $\alpha$ when a large amount of data is available. First, we notice that the Bayes-optimal generalization error, whose large $\alpha$ analysis is performed in SM.~V.1 of \cite{aubin2020generalization}, decreases as $e_{\textrm{g}}^{\textrm{bayes}} \underset{\alpha \to \infty}{\sim} 0.4417 \alpha^{-1}$. Compared to this optimal value, ridge regression gives poor performances in this regime. For any value of the regularization $\lambda$ --- and in particular for both the pseudo-inverse case at $\lambda=0$ and the optimal estimator $\lambda^{\textrm{opt}}$ --- its generalization performances decrease much slower than the Bayes rate, and goes only as $e_{\textrm{g}}^{\textrm{ridge}} \!\underset{\alpha \to \infty}{\sim} \!  0.2405 \alpha^{-1/2}$, see SM.~V.3 of \cite{aubin2020generalization} for the derivation.
Hinge and logistic regressions present a radically different, and more favorable, behavior. Fig.~\ref{fig:gen_error_hinge}-\ref{fig:gen_error_logistic} show that keeping $\lambda$ finite when $\alpha$ goes to $\infty$, does not yield the Bayes-optimal rates. However the max-margin solution (that corresponds to the $\lambda \to 0$ limit of these estimators) gives extremely good performances $e_{\textrm{g}}^{\textrm{logistic,hinge}} \underset{\lambda \to 0}{\sim} e_{\textrm{g}}^{\textrm{max-margin}} \!\underset{\alpha \to \infty}{\sim} \! 0.500 \alpha^{-1}$ see derivation in SM.~V.4 of \cite{aubin2020generalization}. This is the same rate as the Bayes one, only that the constant is slightly higher.

\subsection{Comparison with VC and Rademacher statistical bounds} 
Given the fact that both the max-margin estimator and the optimized logistic achieve optimal generalization rates going as $\Theta\(\alpha^{-1}\)$, it is of interest to compare those rates to the prediction of statistical learning theory bounds. Statistical learning analysis (see e.g.~\cite{vapnik2006estimation,Bartlett98,shalev2014understanding}) relies to a large extent on the \aclink{VC} dimension analysis and on the so-called \emph{Rademacher complexity}. The uniform convergence result states that if the Rademacher complexity or the \aclink{VC} dimension $d_{\vc}$ is finite, then for a large enough number of samples the generalization gap will vanish uniformly over all possible values of parameters. Informally, uniform convergence tells us that with high probability, for any value of the weights $\vec{w}$, the generalization gap satisfies $ {\mR}_{\textrm{population}}(\vec{w}) - {\cal R}_{\textrm{empirical}}^{\nsamples} (\vec{w}) = \Theta\(\sqrt{d_{\vc} / \nsamples}\)$
where $d_{\vc} = \ndim-1$ for our \aclink{GLM} hypothesis class. Therefore, given that the empirical risk can go to \emph{zero} (since our data are separable), this provides a generalization error upper-bound $ e_{\textrm{g}} \!\leq\!  \Theta(\alpha^{-1/2})$.
This is much worse that what we observe in practice, where we reach the Bayes rate $e_{\textrm{g}} = \Theta(\alpha^{-1})$. Tighter bounds can be obtained using the Rademacher complexity, and this was studied recently, using the aforementioned \emph{replica method} \cite{abbara2020rademacher} for the very same problem as presented in \Chap\ref{chap:rademacher}. We reproduced their results and plotted the Rademacher complexity generalization bound in Fig.\ref{fig:gen_error_global} (dashed-green) that decreases as $\Theta\(\alpha^{-1/2}\)$ for the binary classification task eq.~\eqref{main:teacher_sign}.

One may wonder if this could be somehow improved. Another statistical-physics heuristic computation, however, suggests that, unfortunately, uniform bound are plagued to a slow rate $\Theta\(\alpha^{-1/2}\)$. Indeed, the authors of \cite{engel1993statistical} showed with a replica method-style computation that \emph{there exists} some set of weights, in the binary classification task.~\eqref{main:teacher_sign}, that lead to $\Theta\(\alpha^{-1/2}\)$ rates: the uniform bound is thus tight. The gap observed between the uniform bound and the almost Bayes-optimal results observed in practice in this case is therefore not a paradox, but an illustration that the price to pay for uniform convergence is the inability to describe the optimal rates one can sometimes get in practice. Therefore, we believe, that the fact this phenomena can be observed in a such simple problem sheds an interesting light on the current debate in understanding generalization in deep learning \cite{zhang2016understanding}.

Remarking our synthetic dataset is linearly separable, we may try to take this fact into consideration to improve the generalization rate. In particular, it can be done using the max-margin  based generalization error for separable data:
\begin{theorem}[Hard-margin generalization bound \cite{vapnik2006estimation,Bartlett98,shalev2014understanding}]
	Given a set $S=\{\vec{x}_1,\cdots, \vec{x}_\nsamples\}$ such that $\forall \mu \in \lb \nsamples \rb, \|\vec{x}_\mu\| \leq r$. Let $\hat{\vec{w}}$ the hard-margin \aclink{SVM} estimator on $S$ drawn with distribution $D$. With probability $1-\delta$, the generalization error is bounded by
	\begin{align}
		e_{\textrm{g}}(\alpha) \underset{\alpha \to \infty}{\leq} \(4 r \|\hat{\vec{w}}\|  + \sqrt{\log\(4/\delta\) \log_2 \|\hat{\vec{w}}\|} \)/\sqrt{\nsamples}\,.
	\end{align}
\end{theorem}
In our case one has $r^2 \simeq \frac{1}{\ndim} \EE_{\vec{x}} \| \vec{x} \|_2^2 = \frac{1}{\ndim}\sum_{i=1}^\ndim \EE ~x_i^2 \underlim{\ndim}{\infty} 1$. On the other hand, in the large size limit, the norm of the estimator $\|\hat{\vec{w}}\|_2 / \sqrt{\ndim} \underlim{\ndim}{\infty} \sqrt{q}$, that yields $e_{\textrm{g}}(\alpha)\leq 4 \sqrt{\frac{q}{\alpha}}$. 
We now need to plug the values of the norm $q$ obtained by our max-margin solution to finally obtain the results. Unfortunately, this bound turns out to be even worse than the previous one. Indeed the norm of the hard margin estimator $q$ is found to grow with $\alpha$ in the solution of the fixed point equation, and therefore the margin decay rather fast, rendering the bound vacuous. For small values of $\alpha$, one finds that $q\sim \alpha$ that provides a vacuous constant generalization bound $e_{\textrm{g}}\leq \Theta\(1\)$, while for large $\alpha$, $q\sim \alpha^2$ that yields an even worse bound $e_{\textrm{g}}\leq \Theta\(\sqrt{\alpha}\)$. Clearly, max-margin based bounds do not perform well in this high-dimensional example.

\section{Reaching Bayes optimality}
\label{sec:optimality}
Given the fact that logistic and hinge losses reach values extremely close to Bayes optimal generalization performances, one may wonder if by somehow slightly altering these losses one could actually reach the Bayesian values with a plug-in estimator obtained by \aclink{ERM}. This is what we achieve in this section, by constructing a non-convex optimization problem with a specially tuned loss and regularization, whose solution yields Bayes-optimal generalization. Recent insights have shown that indeed one can sometime re-interpret Bayesian estimation as an optimization program in inverse
problems \cite{gribonval2011should,NIPS2013_4868,gribonval2018characterization,gribonval2019bayesian}. In
particular, \cite{Advani2016} showed explicitly, on the basis of the non-rigorous replica method of statistical mechanics, that some Bayes-optimal reconstruction problems could be turned into convex M-estimation. 

Matching \aclink{ERM} and Bayes-optimal generalization errors eqs.~\eqref{main:generalization_errors}-\eqref{main:generalization_error_bayes} with overlaps respectively solutions of eq.~\eqref{main:fixed_point_equations_replicas}-\eqref{main:fixed_point_equations_bayes} and assuming that $\mZ_{\w^\star}\(\gamma,\Lambda\)$ and $\mZ_{\out^\star}\(y,\omega,V\)$, defined in \App\ref{appendix:definitions:distributions:committee}, are log-concave in $\gamma$ and $\omega$, we define the optimal loss and regularizer $l^{\textrm{opt}}$, $r^{\textrm{opt}}$:
\begin{align}
	\begin{aligned}
		l^{\textrm{opt}}\(y,z\) &=- \min_\omega \( \frac{(z-\omega)^2}{2 (\rho_{\w^\star}-q_\bayes)} + \log \mZ_{\out^\star} \(y,\omega,\rho_{\w^\star}-q_\bayes\) \)\,, \spacecase
		r^{\textrm{opt}}\(w\) &= - \min_{\gamma} \( \frac{1}{2} \hat{q}_\bayes w^2 - \gamma w + \log \mZ_{\w^\star} \(\gamma, \hat{q}_\bayes\) \)\,,  
		\label{main:opt_loss_reg}
	\end{aligned}
\end{align}
with $ (q_\bayes,\hat{q}_\bayes)$ solution of eq.~\eqref{main:fixed_point_equations_bayes}. Following these considerations, we provide the following theorem:
\begin{theorem}
	The result of empirical risk minimization eq.~(\ref{main:training_loss}) with $l^{\textrm{opt}}$ and $r^{\textrm{opt}}$ in eq.~\eqref{main:opt_loss_reg}, leads to Bayes optimal generalization error in the high-dimensional regime.
\end{theorem}
\begin{proof}
	The derivation is largely inspired by 
	\cite{Opper1991,Kinouchi1996,gribonval2011should, Bean2013,Advani2016,Advani2016b,Donoho2016,NIPS2013_4868,gribonval2018characterization,gribonval2019bayesian}.
	First we note that the so called Bayes-optimal \aclink{AMP} algorithm \cite{rangan2011generalized}, presented in \Alg\ref{alg:AMP} in \Chap\ref{chap:committee_machine}, for $K=1$ in the context of the \aclink{GLM}, is provably convergent. With Bayes-optimal update functions $f_{\out}^{\textrm{bayes}} (y,\omega,V) = \partial_\omega \log \( \mZ_{\out^\star} \)$ and $f_{\w}^{\textrm{bayes}}(\gamma ,\Lambda) = \partial_\gamma \log\(\mZ_{\w^\star}\)$, it indeed reaches Bayes-optimal performances, see \cite{barbier2017phase}.
	Instead performing Bayes-optimal (\aclink{MMSE}) estimation, we can simply use the \aclink{AMP} algorithm and change the denoising functions to perform \aclink{ERM} (\aclink{MAP}) estimation with 
		\begin{align*}
			f_{\out}^{{\textrm{erm}}, l} (y, \omega, V) &=  - \partial_{\omega} \mM_{V}[l(y,.)](\omega)\,, \spacecase
			f_{\w}^{{\textrm{erm}}, r}(\gamma, \Lambda) &=  \Lambda^{-1}\gamma - \Lambda^{-1} \partial_{\Lambda^{-1}\gamma}\mM_{\Lambda^{-1}}\[ r(.) \] (\Lambda^{-1}\gamma)\,,
		\end{align*}
	detailed in \App\ref{appendix:update_functions:map_updates}.
	The corresponding \aclink{GAMP} algorithms achieve potentially different fixed points and performances. 
	As \aclink{GAMP} provably converges to the optimal generalization error with Bayes-optimal updates, it is sufficient to enforce that at each time step $t$ the Bayes-optimal and \aclink{ERM} denoising functions are equal $f^{\textrm{bayes}}=f^{\textrm{erm}}$, to insure that \aclink{GAMP} algorithm for \aclink{ERM} estimation matches the same performances. 
	Enforcing the constraint $f_{\out}^{\textrm{bayes}}\(y, \omega, V\) = f_{\out}^{{\textrm{erm}}, l}\(y, \omega, V\)$ yields
	 \begin{align*}
	\partial_\omega \log \( \mZ_{\out^\star} \)\(y, \omega, V\) &= - \partial_\omega \mM_V\[l^{\textrm{opt}}\(y,.\)\] (\omega)\,.
	\end{align*}
	Integrating, leaving aside the constant that will not influence the final result, and taking the Moreau-Yosida regularization on both sides, we obtain:
	\begin{align*}
		\mM_V\[\log  \mZ_{\out^\star} \(y, ., V\)\]\(\omega\) = \mM_V\[- \mM_V\[l^{\textrm{opt}}\(y,.\)\] (\omega)\] = - l^{\textrm{opt}}\(y,\omega\)\,,
	\end{align*}
	where we invert the Moreau-Yosida regularization in the last equality that is valid as long as $\mZ_{\out^\star}(y,\omega,V)$ is assumed to be log-concave in $\omega$, see \cite{Advani2016} for a derivation.
	We finally obtain the \emph{optimal loss} $l^{\textrm{opt}}$
	\begin{align}
	\begin{aligned}
			l^{\textrm{opt}}\(y,z\) &= - \mM_V\[\log \( \mZ_{\out^\star} \)\(y, ., V\)\]\(z\)\\
			 &= - \min_\omega \( \frac{(z-\omega)^2}{2 V} + \log \mZ_{\out^\star} \(y,\omega,V\) \) \,.
	\end{aligned}
	\label{main:optimality:loss}
	\end{align}
	Introducing a rescaled prior denoising distribution
	\begin{align*}
	\begin{aligned}
			&\td{Q}_{\w^\star}(w; \gamma,\Lambda) \equiv \displaystyle \frac{1}{\td{\mZ}_{\w^\star} (\gamma,\Lambda)} \rP_{\w^\star}(w) e^{ - \frac{1}{2} \Lambda\(w - \Lambda^{-1} \gamma  \)^2  }\,, \\ 
			&\log\(\td{\mZ}_{\w^\star} (\gamma,\Lambda)\) = \log\(\mZ_{\w^\star} (\gamma,\Lambda)\) - \frac{1}{2}\Lambda^{-1}\gamma^2 \,,
	\end{aligned}
	\end{align*}
	so that the the prior updates read
	\begin{align*}
	\begin{aligned}
	 	f_{\w}^{\textrm{bayes}}\( \gamma, \Lambda \) &= \Lambda^{-1}\gamma + \Lambda^{-1} \partial_{\Lambda^{-1}\gamma} \log\(\td{\mZ}_{\w^\star}\)\,, \spacecase
	 	f_{\w}^{{\textrm{erm}},r}\( \gamma, \Lambda \) &= \Lambda^{-1}\gamma - \Lambda^{-1} \partial_{\Lambda^{-1}\gamma}\mM_{\Lambda^{-1}}\[ r \] (\Lambda^{-1}\gamma) \,.
	\end{aligned}
	\end{align*}
	Imposing the equivalence $f_{\w}^{\textrm{bayes}}\( \gamma, \Lambda \) =  f_{\w}^{{\textrm{erm}}, r}\( \gamma, \Lambda \)$ yields
	\begin{align*}
			\partial_{\Lambda^{-1}\gamma} \log\(\td{\mZ}_{\w^\star}\) &= -\partial_{\Lambda^{-1}\gamma}\mM_{\Lambda^{-1}}\[ r^{\textrm{opt}} \] (\Lambda^{-1}\gamma) \,,
	\end{align*}
	and assuming that $\mZ_{\w}(\gamma, \Lambda)$ is log-concave in $\gamma$, we may invert the Moreau-Yosida regularization, that leads to the expression for the optimal regularizer $r^{\textrm{opt}}$
	\begin{align}
	\begin{aligned}
		r^{\textrm{opt}}\(\Lambda^{-1} \gamma\) &= - \mM_{\Lambda^{-1}}\[ \log\(\td{\mZ}_{\w^\star}\)\(.,\Lambda^{-1}\) \]\(w\)  \\
		&= - \min_{\gamma} \( \frac{1}{2}\Lambda w^2 - \gamma w + \log \mZ_{\w^\star}\(\gamma, \Lambda\) \) \,. 
	\end{aligned}
	\label{main:optimality:reg}	
	\end{align}
	Finally, enforcing the equivalence between the \aclink{AMP} algorithm for the minimization of the \aclink{ERM} and the Bayes-optimal \aclink{AMP} lead to the expressions for the optimal loss $l^{\textrm{opt}}$ and regularizer $r^{\textrm{opt}}$ in \eqref{main:opt_loss_reg}.
	The last step is to characterize the undetermined variances $V$ and $\Lambda$ involved in \eqref{main:optimality:loss} and \eqref{main:optimality:reg}. To achieve the Bayes-optimal performances, we therefore use the variances $V$ and $\Lambda$ solutions of the Bayes-optimal \aclink{GAMP} algorithm. In the large size limit $\ndim \to \infty$, taking the expectation over the ground truth $\vec{w}^\star$ and the input data $\mat{X}$ the parameters $V$ and $\Lambda$ concentrate and are given by the \aclink{SE} of the \aclink{GAMP} algorithm \cite{barbier2017phase}
	\begin{align}
	\lim_{\ndim \to \infty} \EE_{\vec{w}^\star, \mat{X}}\[ V \] &= \rho_{\w^\star} - q_\bayes\,, && \lim_{\ndim \to \infty}\EE_{\vec{w}^\star, \mat{X}}\[ \Lambda \] = \hat{q}_\bayes\,,
	\end{align}
	where $q_\bayes$ and $\hat{q}_\bayes$ are solutions of the Bayes-optimal set of fixed point equations eq.~\eqref{main:fixed_point_equations_bayes}.
	 This shows that \aclink{AMP} applied to the \aclink{ERM} problem corresponding to \eqref{main:opt_loss_reg} both converge to its fixed point and reach Bayes-optimal performances. The theorem finally follows by noting, see \cite{montanari2012graphical,gerbelot2020asymptotic}, that the \aclink{AMP} fixed point corresponds to the extremization conditions of the loss.
\end{proof}

\begin{figure}[!htb]
	\vspace{-0.3cm}
 	\centering
        \includegraphics[width=0.49\linewidth]{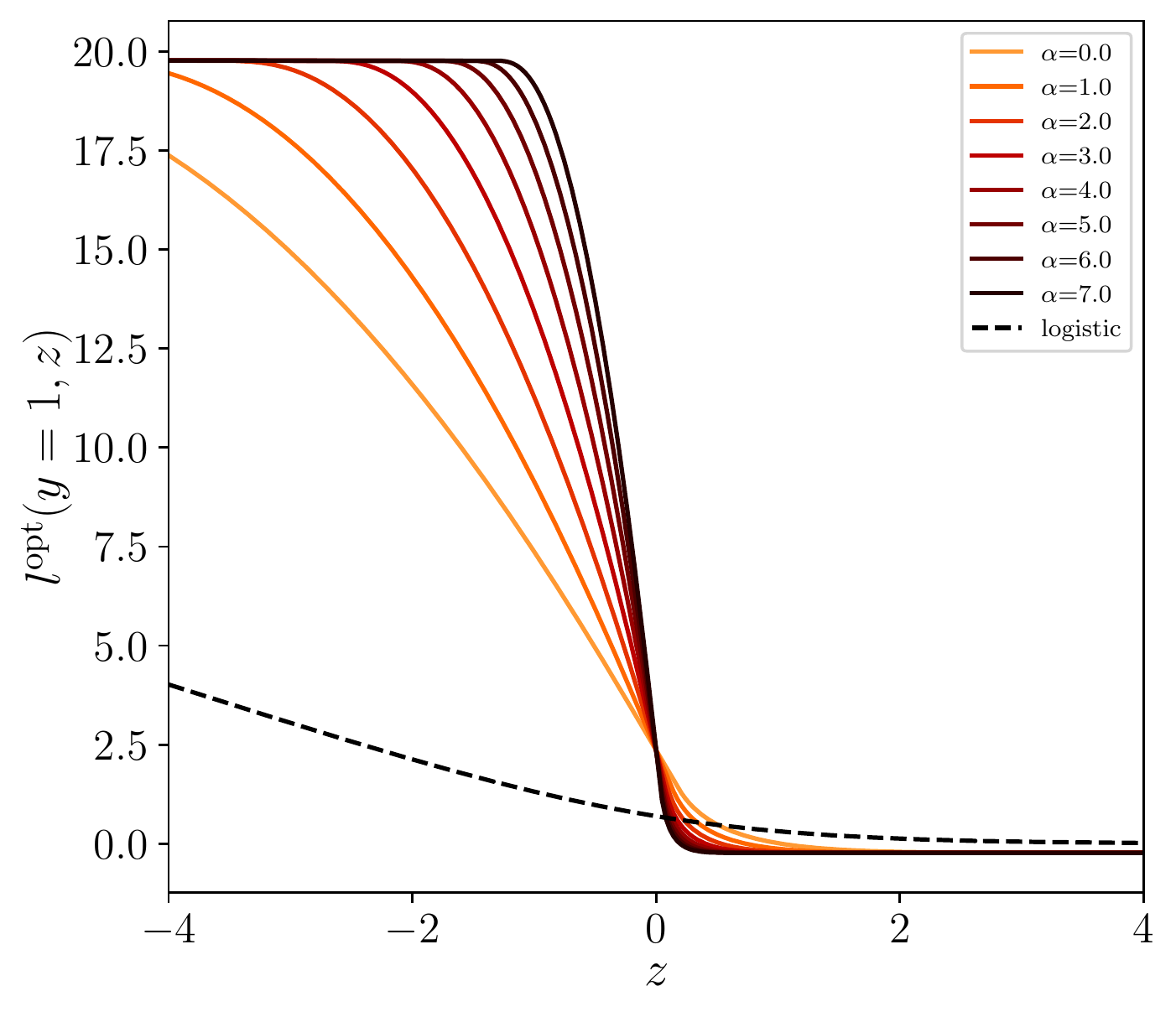}
        \includegraphics[width=0.48\linewidth]{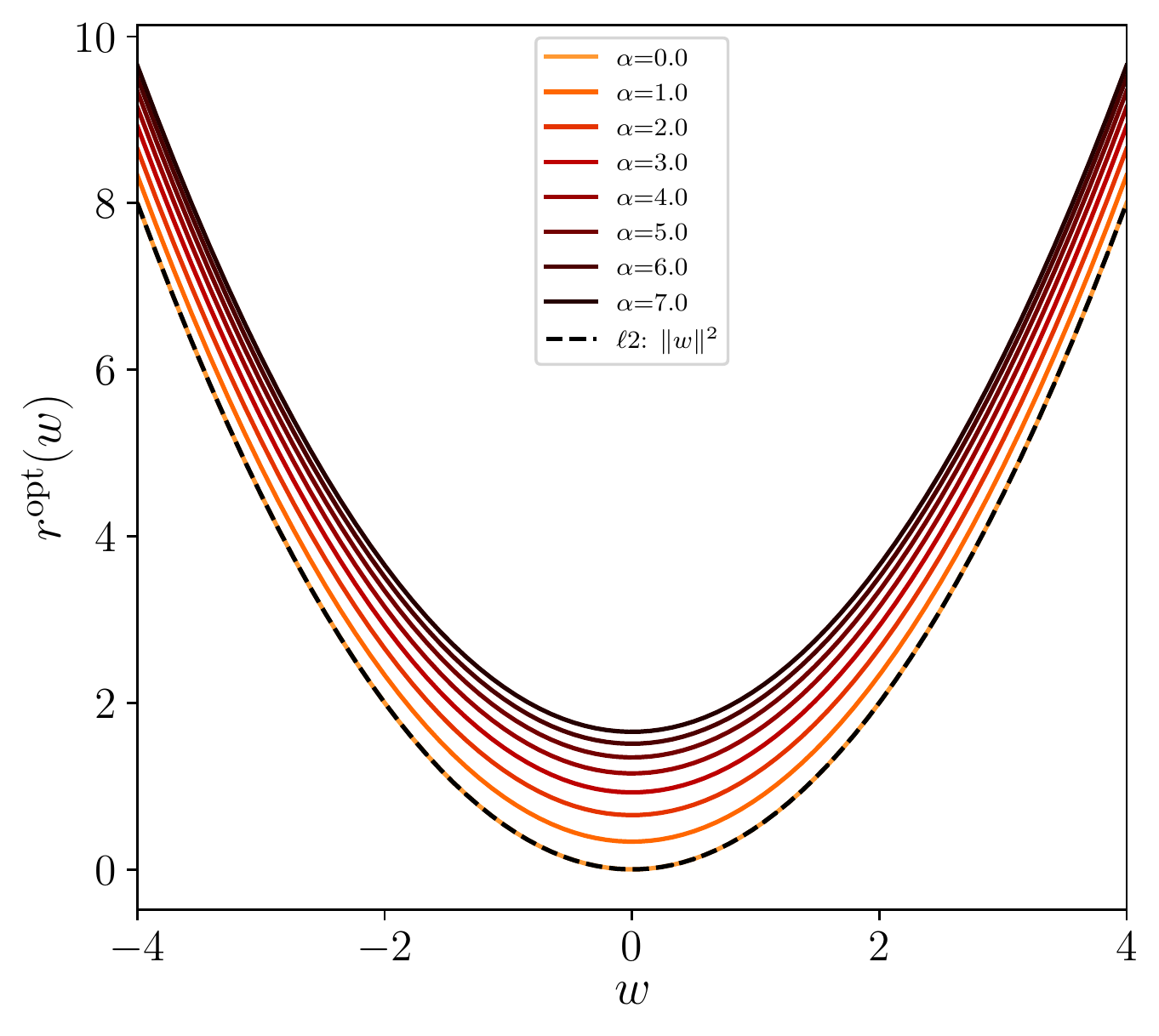}
        \caption{Optimal loss $l^{\textrm{opt}}\(y=1,z\)$ and regularizer $r^{\textrm{opt}}\(w\)$ for model eq.~\eqref{main:teacher_sign} as a function of $\alpha$. The logistic loss and the $\ell_2$ regularizer are plotted in dashed black for comparison.
     			}
        \label{fig:opt_loss_reg_sign_gaussian}
 \end{figure}
\begin{figure}[!htb]
\centering
        \includegraphics[scale=0.5]{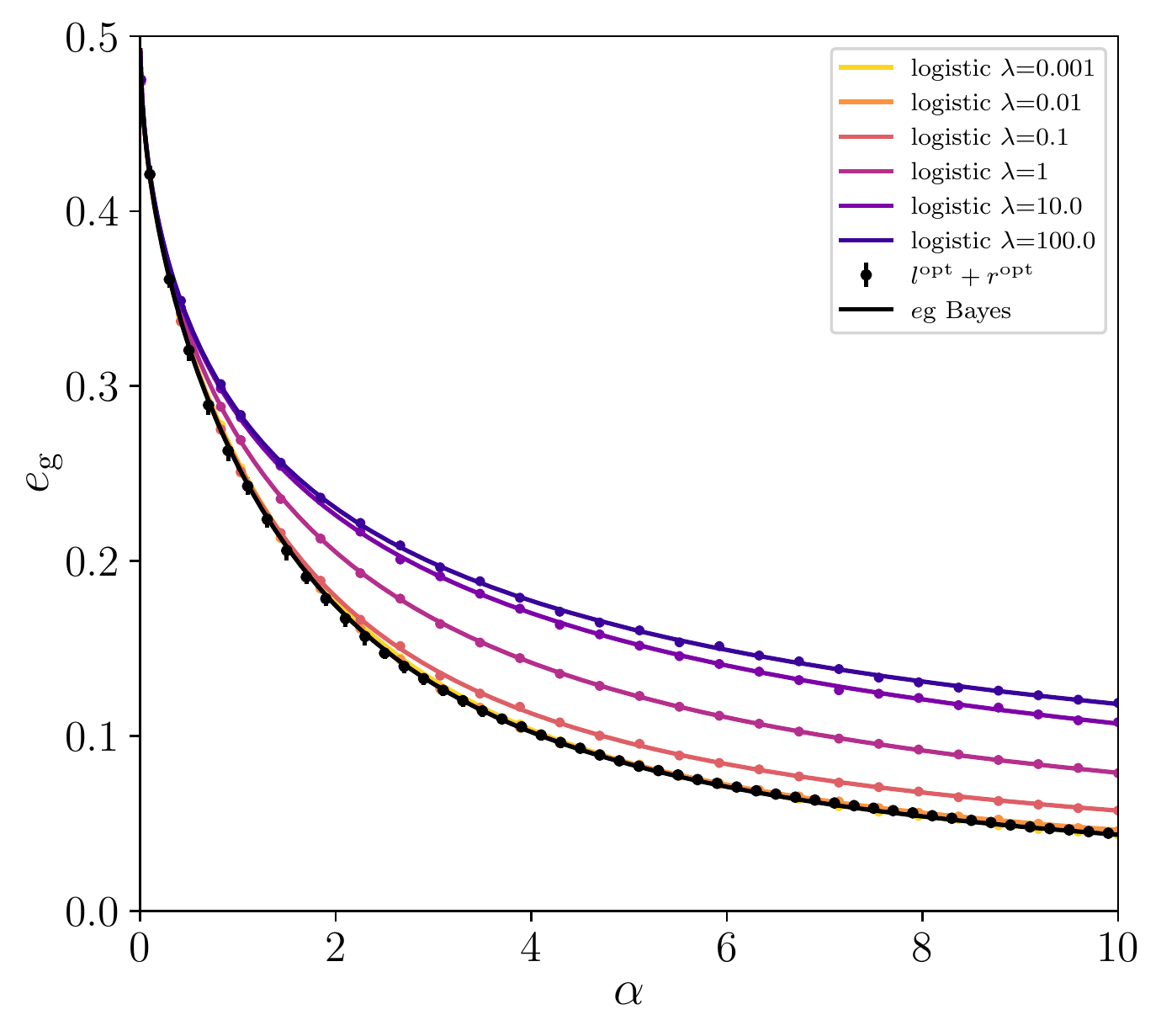}
        \caption{Generalization error obtained by optimization of the optimal loss $l^{\textrm{opt}}$ and $r^{\textrm{opt}}$ for the model \eqref{main:teacher_sign}, compared to $\rL_2$ logistic regression and Bayes-optimal performances. Numerics has been performed with \textsf{scipy.optimize.minimize} with the \textsf{L-BFGS-B} solver for $\ndim=10^3$ and averaged over $n_s=10$ instances. The error bars are barely visible.}
        \label{fig:opt_loss_reg_sign_gaussian_numerics}
\end{figure}

The optimal loss and regularizer $\lambda^{\textrm{opt}}$ and $r^{\textrm{opt}}$ for the model \eqref{main:teacher_sign} are illustrated in Fig.~\ref{fig:opt_loss_reg_sign_gaussian}. 
Notice in particular that even though the loss $l^{\textrm{opt}}$ is not convex (but seems quasi-convex), numerical simulations of \aclink{ERM} (black dots) presented in Fig.~\ref{fig:opt_loss_reg_sign_gaussian_numerics} show that \aclink{ERM} achieves indeed the Bayes-optimal performances (black line) even at finite dimension. 

	\ifthenelse{\equal{\format}{oneside}}
	{
	\clearpage\null\thispagestyle{empty}\newpage
	\clearpage\null\thispagestyle{empty}\newpage
	}
	{
	\clearpage\null\thispagestyle{empty}\newpage
	\clearpage\null\thispagestyle{empty}\newpage
	\cleardoublepage}
\subpartpage{II~ B}{Theory for the statistical estimation with random multi-layer neural networks generative priors}
	
	\chapter*{Outline and motivations}
\chaptermark{Outline and motivations}
	Another recent ongoing direction of research aims to extend the mean-field methods to the combination of known and already analyzed elementary models such as the \aclink{GLM} \cite{barbier2017phase} or the low-rank matrix factorization \cite{lesieur2017constrained}. Combining the corresponding graphical models leads naturally to the description of more complex \aclink{JPD}. However, understanding how and when this \emph{plug-in} of different models is justified is a promising research direction.
	In particular, this approach was successfully applied to the inference in multi-layer \aclink{GLM} estimation \cite{manoel2017multi}, for \aclink{i.i.d} weight matrices. It was later generalized to orthogonally invariant weight matrices with the corresponding \aclink{VAMP} algorithm \cite{fletcher2018inference}.\\
	
	Within this general plug-in approach, we consider estimation problems of the form $\vec{y} = \Gamma(\vec{x}^\star)$ where the operator $\Gamma$ represents different noisy channels such as linear inverse problems, spiked matrix estimation or phase retrieval.
	The ground truth signal $\vec{x}^\star$ must be estimated from the noisy observations $\vec{y}$ and the knowledge of the operator $\Gamma$.
	To perform this statistical reconstruction, in signal processing we often use a low-dimensional parametrization of the signal $\vec{x}^\star$ with for instance a \emph{sparse} dimensionality reduction technique. 
	Naturally, exploiting the structure of the signal drastically helps to achieve better accuracy for larger signal-to-noise ratio. 
	Recently, this sparsity structure has been challenged and successfully replaced by generative priors based on neural networks, such as \aclink{GAN} or \aclink{VAE}, that demonstrated to be particularly performant in various estimation applications. 
	
	In \Chap\ref{chap:generative_spiked} and \Chap\ref{chap:generative_phase} we respectively investigate the low-rank matrix factorization and the phase retrieval and compressed sensing estimation problems with a multi-layer feed-forward \aclink{DNN} generative prior with \aclink{i.i.d} random weights. 
	In this series of works, we investigate and provide a theory of estimation with random generative priors. Especially, we derive sharp asymptotics for the information-theoretically optimal performances and also for the algorithmic performances of a structured polynomial \aclink{AMP} algorithm naturally built from the \aclink{AMP} algorithms on the sub-models. 
	In the analyzed cases, we observed that generative priors have smaller statistical-to-algorithmic gaps than sparse priors, giving theoretical support to previous experimental observations that generative priors might be advantageous in terms of algorithmic performance compared to classical sparse separable priors.
	
	Additionally, in the context of the low-rank matrix factorization, we also take advantage of the structured model to design a new enhanced spectral algorithm \aclink{LAMP} based on the \emph{linearization} of the \aclink{AMP} algorithm and that beats \aclink{PCA} on synthetic and real data.

	Finally, in this general \emph{plug-in} approach, instead deriving and implementing from scratch the corresponding structured \aclink{AMP} algorithms, we developed the \textsf{tramp} python package, standing for \emph{TRee Approximate Message Passing}. The package provides an implementation of \aclink{EP} for modular compositional inference in high-dimensional tree-structured models. 
	We do not reprint the corresponding paper \cite{baker2020tramp} but the source code is publicly available at \href{https://github.com/sphinxteam/tramp}{\url{https://github.com/sphinxteam/tramp}}. 
	Nevertheless, similarly to previous works \cite{tramel2016inferring, bora2017compressed, fletcher2018inference}, in \Sec\ref{chap:generative_phase:applications_real}, we empirically explore the reconstruction of \textsf{tramp} on common estimation tasks on real datasets by making use of \aclink{VAE} generative priors learned on the MNIST dataset \cite{mnist10}.

	\ifthenelse{\equal{\format}{oneside}}
	{
	\clearpage\null\thispagestyle{empty}\newpage
	}
	{
	\clearpage\null\thispagestyle{empty}\newpage
	\clearpage\null\thispagestyle{empty}\newpage
	\cleardoublepage}
	
	\chapter{The spiked matrix matrix model with generative priors}	
	\chaptermark{The spiked matrix matrix model with generative priors}
	\label{chap:generative_spiked}
	
Exploiting structure for efficient signal reconstruction is a central endeavor in modern signal processing. Notable technological advances - such as JPEG and MP3 compression for example - stem from the fact that images and sound admit a sparse representation in wavelet and Fourier bases. In a seminal work, Donoho, Cand\`es and Tao have shown that underparametrized linear systems can be inverted if the signal is assumed to be sparse. This result opened the door for novel sub-Nyquist sampling strategies leveraged by sparsity which are at the heart of \aclink{CS} \cite{donoho2006compressed}. But interest in sparse representations reaches far beyond \aclink{CS}, and similar results have been derived for other signal processing tasks, such as sparse coding and sparse \aclink{PCA}. Despite the remarkable success of these results, they broadly assume the latent sparse representation is given, thus relying on expert knowledge for signal pre-processing.

Recent progress in deep learning has witnessed a surge of interest in neural network-based generative models. Opposed to sparsity, generative networks are trained to learn a latent representation of the structured signal. The expressiveness of neural networks allied with the capacity to capture hierarchical representations led to impressive results in signal modelling, the most notable perhaps being \aclink{GAN} or \aclink{VAE}, which can be trained to generate realistic images of human faces \cite{goodfellow2014generative}. An important and natural question to ask is whether signals from generative models enjoy the same aforementioned interesting properties as sparse signals in reconstruction tasks. Early results in regression-related problems suggest that the latent structure in generative models can be leveraged to improve signal reconstruction \cite{tramel2016inferring,bora2017compressed, manoel2017multi,hand2017global,
  fletcher2018inference,hand2018phase,mixon2018sunlayer}, indeed suggesting that \cite{BlogSoledad}:
\begin{equation}
{\textrm Generative \, \, models \, \, are \, \, the \, \, new \, \, sparsity.} \nonumber
\end{equation}
In this chapter we give a further step in this direction by analyzing a class of random-neural generative priors in an unsupervised task: rank-one (a.k.a. spiked) matrix factorization. Given a "data" matrix $\mat{Y}\in\mathbb{R}^{n\times p}$, the problem consists in finding two vectors, also called the \emph{spikes}, $\vec{u}\in\mathbb{R}^{n}$ and $\vec{v}\in\mathbb{R}^{p}$ such that $\mat{Y}$ can be factorized as $\mat{Y}=\vec{u}\vec{v}^{\intercal}+\sqrt{\Delta}\bxi$, where $\bxi$ is an \aclink{i.i.d} noise matrix of unit variance. This model is widely studied as a prototype for \aclink{PCA}, since for small noise  ($\Delta < 1$) and Gaussian spikes $\vec{u},\vec{v}$, the optimal estimator is given by the leading principal component of $\mat{Y}$ \cite{baik2005phase}. Optimality relies on the assumption of unstructured spikes, and no longer hold if one of the spikes is sparse. In a similar spirit to \aclink{CS}, the investigation of sparse spikes in this model resulted into bespoke algorithms widely studied under the umbrella of sparse-\aclink{PCA}, e.g. \cite{jenatton2010structured}. 

An important conclusion of the aforementioned works is the existence of an algorithmic gap for sparse signal reconstruction. In other words, even if signal reconstruction is \emph{a priori} possible, no polynomial-time algorithm is known. For spiked-matrix factorization, this means that even though the best known sparse-\aclink{PCA} algorithm perform better than vanilla \aclink{PCA}, it doesn't reach the optimal threshold set by the theoretical, and practically intractable, Bayesian estimator. As we will show, this is in sharp contract to the class of neural generative models we study, for which we provide a polynomial time algorithm reaching the optimal theoretical performance, suggesting instead that:
\begin{equation}
{\textrm Generative \, \, models \, \, are \, \, better \, \, than \, \, sparsity.} \nonumber
\end{equation}
Before moving to the bulk of the technical analysis, we give a detailed introduction of the model and regime will study, followed by an account of our main contributions.

\section{Model and studied regime}
\label{sec:model}
We will focus on the following two widely studied models in the sparse-\aclink{PCA} literature \cite{rangan2012iterative,deshpande2014information,lesieur2015phase,barbier2016mutual,perry2016optimality,lelarge2019fundamental,miolane2017fundamental}:
\paragraph{Spiked Wigner model ($\vec{v} \vec{v}^\intercal$)}
 Consider an unknown vector, the spike, $\vec{v}^\star \in \bbR^{p}$ drawn
 from a distribution $\rP_\v$;  we observe a matrix $\mat{Y} \in \bbR^{ p
   \times p}$ with a symmetric noise term $\bxi \in \bbR^{p \times p}$ and $\Delta > 0$:
\begin{align}
	\mat{Y} = \frac{1}{\sqrt{p}} {\vec{v}^\star} {\vec{v}^\star}^\intercal + \sqrt{\Delta} \bxi \, ,\label{Wigner}
\end{align}
where
$\xi_{ij} {\sim} \mN\left(0,1\right)$ \aclink{i.i.d}. The aim is to recover the hidden spike ${\vec{v}^\star}$ from the knowledge of $\mat{Y}$, up to a global sign.
\paragraph{Spiked Wishart (or spiked covariance) model ($\vec{u} \vec{v}^\intercal$)}
Consider two unknown vectors ${\vec{u}}^{\star}\in \bbR^{n}$  and $\vec{v}^{\star} \in \bbR^{p}$ drawn from distributions $\rP_\u$ and $\rP_\v$
and let $\bxi \in \bbR^{n \times p}$ with $\xi_{\mu i}
{\sim} \mN\left(0,1\right)$ \aclink{i.i.d} and $\Delta > 0$, we observe
\begin{align}
	\mat{Y} = \frac{1}{\sqrt{p}} {\vec{u}^\star} {\vec{v}^\star}^\intercal + \sqrt{\Delta} \bxi \,, \label{Wishart}
\end{align}
the goal is to find back the hidden spikes ${\vec{u}}^{\star}$ and ${\vec{v}}^\star$ from $\mat{Y} \in \bbR^{n \times p}$.

The noisy high-dimensional limit that we consider in this work, also called the
\emph{ thermodynamic limit}, is $p,n \! \to\!  \infty$ while $\beta\!
\equiv\! n/p \!=\! \Theta(1)$, and
the noise $\bxi$ has a variance $\Delta\!=\!\Theta(1)$.
The prior $\rP_\v$ is representing the spike $\vec{v}$ via a $k$-dimensional
parametrization with $\alpha\!\equiv\! p/k \!=\! \Theta(1)$. In the sparse case,
$k$ is the number of non-zeros components of $\vec{v}^{\star}$, while in
generative models $k$ is the number of latent variables.

\subsection{Considered generative models}
\label{sec:generative_models}
The simplest non-separable prior $\rP_\v$ that we consider is the
Gaussian model with a covariance matrix $\bSigma$, that is
$\rP_\v(\vec{v})={\cal N}_\vec{v}(\bzero,\bSigma)$. This prior is not compressive,
yet it captures some structure and can be simply estimated
from data via the empirical covariance. We use this prior later to produce
Fig.~\ref{main:experiement_mnist}.

To exploit the
practically observed power of generative models, it would be desirable to consider
models (e.g. \aclink{GAN}, \aclink{VAE}, restricted Boltzmann
machines, or others) trained on datasets of examples of possible spikes. Such
training, however, leads to correlations between the weights of the
underlying neural networks for which
the theoretical part of the present work does not apply readily. To
keep tractability in a closed form, and subsequent theoretical insights, we focus on multi-layer generative models where all the
weight matrices $\mat{W}^{(l)}$, $l=1,\dots,L$, are fixed, layer-wise independent,
\aclink{i.i.d} Gaussian with zero mean and unit variance. Let $\vec{v} \in \bbR^{p}$ be the output of such a generative model
\begin{equation}
    \vec{v} = \varphi^{(L)} \( \frac{1}{\sqrt{k_{L}}} \mat{W}^{(L)} \dots \varphi^{(1)} \(\frac{1}{\sqrt{k}} \mat{W}^{(1)} \bz \)\,.
    \dots \)\,,
    \label{main:eqMLmodel}
  \end{equation}
with $\bz\in \bbR^{k}$ a latent variable drawn from separable distribution
$\rP_\z$, with $\rho_z = \EE_{\rP_\z} \left[z^2\right]$. $\forall l \in \lb L \rb, ~\varphi^{(l)}$ are the element-wise activation functions that can be either deterministic or stochastic. It will be useful to define the hidden variables $\bh^{(l)}\in\mathbb{R}^{k_{l}}$ obtained from the output of layer $l-1$. The hidden variable $\bh^{(l+1)}\in\mathbb{R}^{k_{l+1}}$ is then given by
\begin{align*}
    \bh^{(l+1)} = \varphi^{(l)}\left(\frac{1}{\sqrt{k_{l}}}\mat{W}^{(l)}\bh^{(l)}\right) \hhspace \Leftrightarrow \hhspace \bh^{(l+1)}\sim \rP_{\out}^{(l)}\left(~\cdot~\Big|\frac{1}{\sqrt{k_{l}}}\mat{W}^{(l)}\bh^{(l)}\right)
\end{align*}
with $\vec{h}^{(0)}=\vec{z}$ and $\vec{h}^{(L+1)}=\vec{v}$.
The densities $\rP_{\out}^{(l)}$ over $\mathbb{R}^{k_{l+1}}$ parametrize the input/output relationship at each layer of the generative network. Note that since $\varphi^{(l)}$ act component-wise $\rP_{\out}^{(l)}$ is a separable distribution, and factorize in a product of identical $k_{l+1}$ scalar distributions over $\mathbb{R}$ which, abusing notation, we we will denote by $\rP_{\out}^{(l)}$. For instance, a deterministic layer $l$ with non-linearity $\varphi^{(l)}$ is fully characterized by the scalar density $\rP_{\out}^{(l)}(x|z)=\delta(x-\varphi(z))$.
  
In the setting considered in this work the ground-truth spike $\vec{v}^\star$ is
generated using a ground-truth value of the latent variable
$\bz^\star$. The spike is then estimated from the knowledge of the
data matrix $\mat{Y}$, and the known form of the spiked-matrix and of the
generative model. In particular the matrices $\mat{W}^{(l)}\in\mathbb{R}^{k_{l+1}\times k_{l}}$ are known, as are the
parameters $\beta, \Delta$, $\rP_\z$, $\rP_\u$, $\rP_\v$
and $\varphi^{(l)}$. Only the spikes $\vec{v}^\star$, $\vec{u}^\star$ and the latent vector
$\bz^\star$ are unknown, and are to be inferred.

For concreteness and simplicity, the generative model that will be
  analyzed in most examples given in the present work is
  the single-layer case of (\ref{main:eqMLmodel}) with $L=1$. We define the total compression ratio $\alpha \equiv p/k$. In what follows we will illustrate our results for
  $\varphi$ being linear, sign and \aclink{ReLU} functions.

\subsection{Summary of main contributions}
First, we provide an information-theoretical analysis for the performance of the optimal estimator for the spiked-matrix models (\ref{Wigner}) and (\ref{Wishart}). This analysis is based on a rigorous expression for the mutual information between the matrix $\mat{Y}$ and a general spike $\vec{v}^{\star}$ from a non-separable distribution $\rP_\v$ in $\mathbb{R}^{p}$, and holds in the afore defined thermodynamic limit. Evaluating this expression on the generative priors discussed in \Sec\ref{sec:generative_models}, we obtain the optimal statistical threshold $\Delta_c$ below which the spike $\vec{v}^{\star}$ can be reconstructed. On a second moment, we derive an \aclink{AMP} algorithm for the models (\ref{Wigner}) and (\ref{Wishart}), and show that, for the all the generative architectures analysed, they attain the same performance previously derived for the Bayesian optimal estimator. Next, we propose a simple spectral method derived from our \aclink{AMP} algorithm reaching the same statistical threshold $\Delta_{c}$. Finally, we show that this same spectral method can be, in certain cases, rigorously derived from a Random Matrix Theory.

Our main findings are in stark contrast to the known results for sparse-\aclink{PCA}, and therefore it is useful to present them in this context. We draw two main conclusions from the present work:

{\bfseries (i) No algorithmic gap with generative-model priors:}
Sharp and detailed results are known in the thermodynamic limit (as
defined above) when
the spike $\vec{v}^\star$ is sampled from a separable distribution $\rP_\v$. A detailed account of several examples can be found in
\cite{lesieur2017constrained}. The main finding for sparse priors $\rP_\v$ is that
when the sparsity $\rho = k/p = 1/\alpha$ is large enough then there exist optimal
algorithms \cite{deshpande2014information}, while for
$\rho$ small enough there is a striking gap between statistically optimal performance and the one of best
known algorithms \cite{lesieur2015phase}. The small-$\rho$ expansion
studied in 
\cite{lesieur2017constrained} is consistent with the well-known results for exact
recovery of the support of $\vec{v}^\star$ \cite{amini2009high,berthet2013computational}, which is one of
the best-known cases in which gaps between statistical and best-known algorithmic performance were described.

Our analysis of the spiked-matrix models with generative priors reveals that in
this case known algorithms are able to obtain (asymptotically) optimal
performance even when the dimension is greatly reduced, i.e. $\alpha \gg 1$. Analogous
conclusion about the lack of algorithmic gaps was reached for the
problem of phase retrieval under a generative prior in
\cite{hand2018phase}. This result suggests that plausibly generative
priors are better than sparsity as they lead to algorithmically easier
problems.

{\bfseries (ii) Spectral algorithms reaching statistical threshold:}
Arguably the most basic algorithm used to solve the spiked-matrix
model is based on the leading singular vectors of the matrix $\mat{Y}$. We
will refer to this as \aclink{PCA}. 
Previous work on spiked-matrix models \cite{perry2016optimality,lesieur2017constrained} established that in the thermodynamic limit and for separable priors of zero mean \aclink{PCA}
reaches the best performance of all known efficient algorithms in terms of the value of noise
$\Delta$ below which it is able to provide positive correlation between
its estimator and the ground-truth spike. 
While for sparse priors
positive correlation is statistically reachable even for larger values of
$\Delta$ \cite{perry2016optimality,lesieur2017constrained}, no efficient algorithm beating the \aclink{PCA} threshold is
known. Notice that this result holds only for sparsity $\rho=\Theta(1)$. A
line of works shows that when sparsity $k$ scales slower than linearly
with $p$, algorithms more performant than \aclink{PCA} exist
\cite{amini2009high,deshpande2014sparse}.

In the case of generative priors we find in this contribution that other spectral
methods improve on the canonical \aclink{PCA}. We design a spectral method,
called \aclink{LAMP}, that under certain assumptions, e.g. zero mean of the spikes, reach the statistically optimal threshold,
meaning that for larger values of noise variance no other (even exponential) algorithm is
able to reach positive correlation with the spike. Again this is a
striking difference with the sparse separable prior, making the generative
priors algorithmically more attractive. We demonstrate the performance
of \aclink{LAMP} on the spiked-matrix model when
the spike is taken to be one of the fashion-MNIST images \cite{fashionmnist2017} showing
considerable improvement over canonical \aclink{PCA}.
Each of the following sections is dedicated to one of the results above.

\section{Analysis of information theoretically optimal estimation}
\label{sec:information_theory}
In this section, we derive a set of fixed point equations, known as \aclink{SE} equations, that fully characterize the performance of the optimal estimator for the spike $\bv^{\star}$. For the sake of concreteness, the results in this section are given for the Wigner model, and can be fully generalized to the Wishart case presented in \App B.2 of \cite{aubin2019spiked}. 

\subsection{Rigorous mutual information}
From an optimization perspective, the problem we want to solve is to find the estimator $\bv^{\star}$ that minimizes the \aclink{MSE}
\begin{align} 
\text{mse}(\Delta) = \mathbb{E} ||\hat{\bv}-\bv^{\star}||_{2}^{2}.\label{eq:mse}
\end{align}
Since the information about the generative model $\rP_\v$ of the spike is given, we know that the estimator minimizing eq.~\eqref{eq:mse} is given by the mean of the posterior distribution of the spike, i.e. $\hat{\bv}^{\text{opt}} = \mathbb{E}_{\rP(\bv^{\star}|\mat{Y})}\bv$, where $\rP(\bv^{\star}|\mat{Y})$ is written from Bayes rule as
\begin{align}
    \rP(\bv^{\star}|\mat{Y}) =\frac{\rP_\v(\bv^{\star})}{\rP(\mat{Y})}\prod\limits_{1\leq i<j\leq p}\frac{1}{\sqrt{2\pi\Delta}}\exp\(-\frac{1}{2\Delta}\left(y_{ij}-\frac{v_{i}^{\star}v^{\star}_{j}}{\sqrt{p}}\right)^2\)
    \label{eq:app:defs:posterioruu}\,.
\end{align}
The expression above is written in full generality, and for the time being we have not assumed anything about $\rP_\v$. The naive approach of estimating $\hat{\bv}^{\text{opt}}$ from exact sampling of the posterior is intractable numerically, specially in the large-dimensional limit $p\to\infty$ of interest. However, it is still possible to track the performance of the optimal estimator without direct sampling through the I-MMSE theorem connecting the \aclink{MMSE} to a derivative of the mutual information between the signal and the data \cite{GuoShamaiVerdu_IMMSE}. Following this rationale, our first main result is a rigorous expression for the mutual information between the ground-truth spike
$\bv^\star$ and the observation $\mat{Y}$, defined as
$ \mI(\mat{Y} ; \bv^\star)=\mD_{\textrm{KL}}
(\rP_{(\v^\star,\textrm{Y})}\| \rP_{\v^\star} \rP_{\textrm{Y}})$, valid in the thermodynamic limit defined in \Sec\ref{sec:model}. 
\begin{thm}[Mutual information for the
  spiked Wigner model with structured spike]\label{theorem_uu} Informally, assume the
  spike $\bv^\star$ come from a sequence (of growing dimension $p$) of
  a generic structured prior $\rP_\v$ on $\bbR^p$, then
  \begin{align} \lim_{p \to \infty} i_p &\equiv \lim_{p\to \infty}
    \frac {\mI(\mat{Y};\bv^\star)}p = \inf_{\rho_v \ge q_v \ge 0} {i}_{\rs}(\Delta,q_v),\\
{\text with}~~~
    i_{\rs}(\Delta,q_v) &~\equiv
    \frac{(\rho_v-q_v)^2}{4\Delta} + \lim_{p \to \infty}
    \frac{\mI\left(\bv;\bv+\sqrt{\frac{\Delta}{q_v}} \bxi \right)}p
    \, \label{eq:information_theory:limip}\end{align}
and $\bxi$ being a
  Gaussian vector with zero mean, unit diagonal variance and
  $\rho_v=\lim\limits_{p \to \infty} \E_{\rP_\v}[\bv^\intercal\bv]/p$.
\end{thm}
The proof for this theorem is left in \App C of \cite{aubin2019spiked}, and instead we draw its consequences. Our theorem connects the asymptotic mutual information of the spiked model with generative prior $\rP_\v$ to the mutual information between $\bv$ taken from $\rP_\v$ and its noisy version, $\mI(\bv;\bv+\sqrt{{\Delta}/{q_v}}\bxi)$. As mentioned before, the mutual information is intimately connected to the performance of the optimal estimator, and one can prove in particular that for the spiked-matrix model
\cite{AlaouiKrzakala} the \aclink{MMSE} on the spike $\bv^{\star}$ is
asymptotically given by:
\begin{align}
	{\textrm{MMSE}}_v = \rho_v-q_v^\star  \,, \label{eq:MMSE}
\end{align}
where $q_v^\star$ is the optimizer of the function
$i_{\rs}\(\Delta , q_v\)$. Computing this later mutual information is itself a high-dimensional task, hard in full generality, but it can be done for a range of non-trivial $\rP_\v$. The simplest tractable case is when the prior $\rP_\v$ is
separable, then it yields back exactly the formula previously known from
\cite{krzakala_mutual_2016,barbier2016mutual,lelarge2019fundamental}. It
can also be computed for the correlated Gaussian generative model,
$\rP_\v(\bv)={\cal N}_\bv(\vec{0},\bSigma)$, for which 
$\mI(\bv;\bv+\sqrt{{\Delta}/{q_v}}\bxi) = \Tr{\log{(\rI_p + q_v
    \bSigma/\Delta)}}/2$ is readily known. 

More interestingly, the mutual information associated to the
multi-layer generative prior with random weights from eq.~\eqref{main:eqMLmodel}, explicitly written as
\begin{align}
\begin{aligned}
    \rP_\v(\bv) &= \int \prod\limits_{l=1}^{L}\prod\limits_{\nu_{l}=1}^{k_{l}}\dd h^{(l)}_{\nu_{l}} ~ \rP_{\out}^{(l-1)}\left(h^{(l)}_{\nu_{l}}\Big|\frac{1}{\sqrt{k_{l-1}}}\sum\limits_{\nu_{l-1}=1}^{k_{l-1}}w^{(l-1)}_{\nu_{l}\nu_{l-1}}h_{\nu_{l-1}}\right)\\
    & \qquad \qquad \qquad \qquad \times  \prod\limits_{i=1}^{p}\rP^{(L)}_{\out}\left(v_{i}\Big|\frac{1}{\sqrt{k_{L}}}\sum\limits_{\nu_{L}=1}^{k_{L}}w_{i \nu_{L}}^{(L)}h_{L}\right)\,,
   \end{aligned}
    \label{eq:app:intro:mlgenerative_prior}
\end{align}
\noindent can also be asymptotically computed. Indeed, the corresponding single-layer formula for this mutual information
has been derived and proven in \cite{barbier2017phase}. For the multi-layer case the mutual information formula has been derived in
\cite{manoel2017multi,reeves2017additivity} and proven for the case of two layers in
\cite{gabrie2018entropy}. Theorem \ref{theorem_uu} together with the
results from \cite{barbier2017phase, manoel2017multi,
  reeves2017additivity, gabrie2018entropy}
yields the following formula for the spiked Wigner model (\ref{Wigner}) with multi-layer
generative prior (\ref{main:eqMLmodel}):
\begin{align}
    &i_{\rs} (\Delta,q_v) = \frac{\rho_{v}^2}{4\Delta}+\frac{1}{4\Delta}q_{v}^2 \nonumber \\
    	& \qquad \qquad +\frac{1}{\alpha}\underset{\{\hat{q}_{l}, q_{l}\}_l}{\extr}\left[\frac{1}{2}\sum\limits_{l=1}^{L}\alpha_{l}\hat{q}_{l}q_{l}-\sum\limits_{l=2}^{L}\alpha_{l}\Psi^{(l)}_{\out}\left(\hat{q}_{l},q_{l-1}\right) \right. \label{main:free_entropy_uu} \\
    	& \qquad \qquad \qquad \qquad \qquad \qquad \left.   -\alpha\Psi^{(L+1)}_{\out}\left(\frac{q_{v}}{\Delta},q_{L}\right)-\Psi_{z}\left(\hat{q}_{z}\right)\right]. \nonumber
\end{align}
where $\alpha_{l} = k_{l}/k$ (note that in particular $\alpha_1=1$) and the functions $\Psi_z, \Psi_\out$ are defined by
\begin{align}
	\Psi_z (x) & \equiv \EE_{\xi} \[ \mZ_z\( x^{1/2} \xi ,x  \)
          \log \( \mZ_z\( x^{1/2} \xi ,x  \) \) \] \,
                     ,  \label{main:definition_Psi_z} 
        \spacecase
	\Psi^{(l)}_\out (x,y) & \equiv \EE_{\xi, \eta} \[\mZ^{(l)}_{\out}\( x^{1/2} \xi , x , y^{1/2} \eta  , \rho_{l} - y \) \right. \nonumber \\
         & \left. \hspace{2cm} \log\( \mZ^{(l)}_\out\( x^{1/2} \xi , x , y^{1/2} \eta  ,
          \rho_{l} - y \) \) \]\, ,
                \label{main:definition_Psi_out}
\end{align}
with $\xi, \eta \sim \mN\(0,1\)$ \aclink{i.i.d}, $\rho_{l}$ is the second moment of the hidden variable $h_{l}$ and $\mZ_z$, $\mZ^{(l)}_\out$ are the normalizations of the following denoising scalar distributions:
\begin{align}
\begin{aligned}
	\rQ_\z\(z;\gamma, \Lambda\) &\equiv \displaystyle \frac{\rP_\z(z)}{\mZ_z(\gamma, \Lambda)} ~ e^{ - \frac{\Lambda }{2} z^2  + \gamma z  }\,, \\
	\rQ^{(l)}_\out\(v, x; B,A,\omega,V\) &\equiv \displaystyle \frac{\rP^{(l)}_\out(v |x)}{\mZ^{(l)}_\out(B, A,\omega, V)} ~ e^{ -\frac{A}{2} v^2 + B v } ~ \frac{e^{ -\frac{\(x - \omega\)^2}{2V}  }}{\sqrt{2\pi V}} \,.
\end{aligned}
\label{main:definition_Z}
\end{align}
Result (\ref{main:free_entropy_uu}) is remarkable in that it
connects the asymptotic mutual
information of a high-dimensional model with a simple
scalar formula that can be easily evaluated. Moreover, it fully characterize the statistical performance of the optimal estimator, allowing us to readily identify the statistical thresholds separating the region between possible and impossible inference of the spike.
We now draw the consequences of eq.~\eqref{main:free_entropy_uu} for the most common choices of activation.

\subsection{Optimal performance and statistical thresholds: phase diagrams}
\label{sec:examples}
Taking the extremization over $q_{v}$ and $(\hat{q}_l, q_{l})_{1\leq l \leq L}$ in eq.~(\ref{main:free_entropy_uu}), we obtain the following system of coupled fixed point equations:
\begin{align}
\begin{cases}
	q_{v} = \Lambda_{x}\left(\frac{q_{v}}{\Delta}, q_{L}\right)\\
    q_{L} = \Lambda_{x}\left(\hat{q}_{L}, q_{L-1}\right)\\
\hspace{1cm}\vdots\\
{q}_{l} = \Lambda_{x}\left(\hat{q}_{l},q_{l-1}\right) \\
\hspace{1cm}\vdots\\
q_{z} =\Lambda_{z}\left(\hat{q}_{z}\right)
\end{cases}
\,,
&&
\begin{cases}
\hat{q}_{L} =\tilde{\alpha}_{L}\Lambda_{\out}\left(\frac{q_{v}}{\Delta}, q_{L}\right)	\\
\hat{q}_{L-1} =\tilde{\alpha}_{L-1}\Lambda_{\out}\left(\hat{q}_{L}, q_{L-1}\right)\\
\hspace{1cm}\vdots\\
\hat{q}_{l} =\tilde{\alpha}_{l} \Lambda_{\out}\left(\hat{q}_{l+1}, q_{l}\right)\\
\hspace{1cm}\vdots\\
\hat{q}_{z} =\tilde{\alpha}_{1} \Lambda_{\out}\left(\hat{q}_{2},q_{z}\right)
\end{cases}
\,,
\label{main:SE_uu}
\end{align}
\noindent where we have defined the update functions
\begin{align*}
	\Lambda_{x}(x,y) &\equiv 2 \partial_{x}\Psi_{\out}(x,y)\,, &&\Lambda_{\out}(x,y) \equiv 2\partial_{y}\Psi_{\out}(x,y)\,,\\
	\Lambda_{z}(x) &\equiv 2\partial_{x}\Psi_{z}(x)\,,
\end{align*}
and the layer-wise aspect ratios $\tilde{\alpha}_{l} = k_{l+1}/k_{l} = \alpha_{l+1}/\alpha_{l}$. As previously discussed, the fixed point of these equations provide all the information about the performance of the Bayes-optimal estimator through eq.~\eqref{eq:MMSE}.

An important first question that can be answered from eqs.~\eqref{main:SE_uu} is when does the Bayes-optimal estimator performs better than a random guess from the prior distribution $\rP_\v$. For instance, we intuitively expect that when the prior is not biased towards a particular direction in $\mathbb{R}^{p}$ and for very high noise $\Delta\gg 1$ better-than-random estimation is not possible. In terms of fixed points of eqs.~\eqref{main:SE_uu}, this situation corresponds to
the existence of the \emph{non-informative} fixed point $q^{\star}_{v} =
0$ (i.e. maximum ${\textrm{MSE}}_v =
\rho_v$, or zero overlap with the spike).
Evaluating the right-hand side of eqs.~\eqref{main:SE_uu} at $q_{v} =
0$, we can see that $q_{v}^{\star} = 0$ is a fixed point if
\begin{align}
  \mathbb{E}_{\rP_\z}\[z\] = 0 \andcase \mathbb{E}_{Q^{(l), 0}_{\out}}\[v\]
  = 0\, , \label{trivial_fixed}
\end{align}
where $Q_{\out}^{(l), 0}(v,x) \equiv Q^{(l)}_{\out}(v,x; 0,0,0,\rho_{l})$
from eq.~(\ref{main:definition_Z}). Note that for multi-layer network with  deterministic
channels and $\varphi^{(l)}\equiv \varphi$ for all $l$, the second condition is equivalent to $\varphi$ being an odd
function.

When the condition (\ref{trivial_fixed}) holds,
$(q_{v},q_{L}, \hat{q}_{L},\dots,\hat{q}_{z}, q_{z}) = (0,0,0, \dots, 0, 0)$ is a fixed point of
eq.~\eqref{main:SE_uu}. The numerical stability of this fixed point is determined by whether it is an attractor of the dynamics, and therefore
determines a phase transition point $\Delta_{c}$, defined as the noise below which the fixed point $\vec{0}\in\mathbb{R}^{L+1}$ becomes a repeller. The character of the fixed point can be determined by a standard linear stability analysis of the fixed point equations. The transition will then correspond to the value of $\Delta$ for which the largest eigenvalue of the Jacobian of the eqs.~\eqref{main:SE_uu} at $0$ becomes greater than one. This Jacobian is given explicitly by
\begin{align}
\scalemath{0.58}{
\begin{blockarray}{cccccccccccccc}
q_{v} & \hat{q}_{L} & q_{L} & \hat{q}_{L-1} & q_{L-1} & \cdots & \hat{q}_{l+1} & q_{l+1} & \hat{q}_{l} & q_{l}& \cdots & \hat{q}_{z} & q_{z} & \\
\begin{block}{(ccccccccccccc)c}
\frac{1}{\Delta}m^{(L)}_{vv} & 0 & \frac{1}{\rho_{L}^2} m^{(L)}_{vx} & 0 & 0 & \cdots  & 0 & 0 & 0 & 0 & \cdots & 0 & 0 & q_{v}\\
\frac{\tilde{\alpha}_{L}}{\Delta}m^{(L)}_{vx} & 0 & \frac{\tilde{\alpha}_{L}}{\rho_{L}^2}m^{(L)}_{xx} & 0 & 0 & \cdots & 0 & 0 & 0 & 0 & \cdots  & 0 & 0 & \hat{q}_{L}\\
0 & m^{(L-1)}_{vv} & 0 & 0 & \frac{1}{\rho_{L-1}^2} m^{(L-1)}_{vx} & \cdots & 0 & 0 & 0 & 0 & \cdots & 0 & 0 & q_{L}\\
0 & \tilde{\alpha}_{L-1}m^{(L-1)}_{vx} & 0 & 0 & \frac{\tilde{\alpha}_{L-1}}{\rho_{L-1}^2}m^{(L-1)}_{xx} & \cdots  & 0 & 0 & 0 & 0 & \cdots & 0 & 0 & \hat{q}_{L-1}\\
0 & 0 & 0 & m^{(L-2)}_{vv} & 0 & \cdots  & 0 & 0 & 0 & 0 & \cdots  & 0 & 0 & q_{L-1}\\
0 & 0 & 0 & \tilde{\alpha}_{L-2}m^{(L-2)}_{vx} & 0 & \cdots & 0 & 0 & 0 & 0 & \cdots  & 0 & 0 & \hat{q}_{L-2}\\
 &  & \vdots &  &  &  \vdots &  &  & \vdots &  &  & \vdots  &  & \vdots\\
0 & 0 & 0 & 0 & 0 & \cdots & m^{(l)}_{vv} & 0 & 0 & \frac{1}{\rho_{l}^2}m^{(l)}_{vx} &  \cdots & 0 & 0 & q_{l+1}\\
0 & 0 & 0 & 0 & 0 & \cdots & \tilde{\alpha}_{l}m^{(l)}_{vx} & 0 & 0 & \frac{\tilde{\alpha}_{l}}{\rho_{l}^2}m^{(l)}_{xx} & \cdots & 0 & 0 & \hat{q}_{l}\\
 &  & \vdots &  &  & \vdots &  &  & \vdots &  &  & \vdots  &  & \vdots\\
0 & 0 & 0 & 0 & 0 & \cdots & 0 & 0 & 0 & 0 & \cdots & m_{zz} & 0 & q_{z}\\
\end{block}
\end{blockarray}
}
 \label{eq:phasetransition:jacobian}
\end{align}
where we have defined the following shorthand for the second moments of $Q_{\out}^{(l), 0}(v,x)$:
\begin{align}
\begin{aligned}
    m^{(l)}_{vv} &= \left(\mathbb{E}_{Q_{\out}^{(l), 0}}v^2\right)^2, && m^{(l)}_{vx} = \left(\mathbb{E}_{Q_{\out}^{(l), 0}}vx\right)^2,\\ 
    m^{(l)}_{xx} &= \left(\mathbb{E}_{Q_{\out}^{(l), 0}}x^2-\rho_{l}\right)^2, && m_{zz} = \left(\mathbb{E}_{\rP_\z}z^2\right)^2\,.
\end{aligned}
\end{align}
\begin{figure}[tb!]
	\centering
		\includegraphics[width=1.0\linewidth]{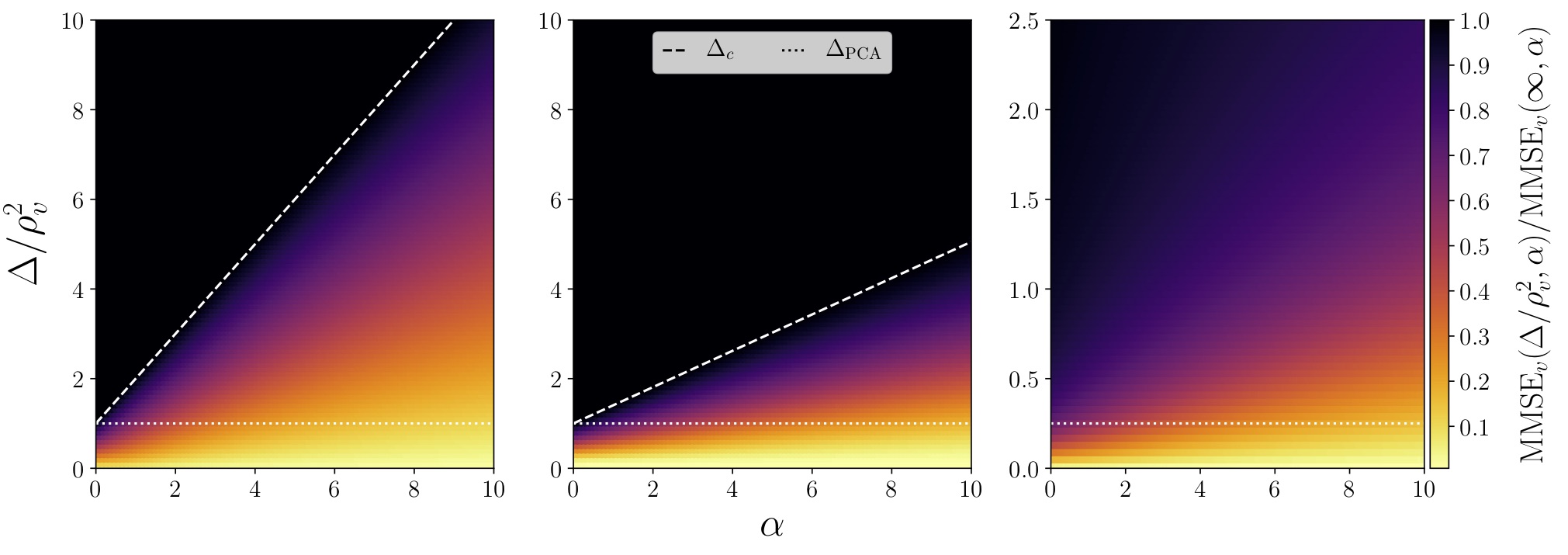}
	\caption{Spiked Wigner model: ${\textrm{MMSE}}_v$ on the spike as a function of noise to signal ratio $\Delta/\rho_v^2$, and single-layer generative prior 
	      with
          compression ratio $\alpha$ for \Left linear  $\rho_v=1$, \Center sign
          $\rho_v=1$, and \Right relu $\rho_v=1/2$ activations. Dashed white lines
          mark the phase transitions $\Delta_c$, matched by both the
          AMP and LAMP algorithms. Dotted white line
          marks the phase transition of canonical PCA.}
	\label{main:fig_map_mse_delta_alpha}
\end{figure}
\begin{figure}[tb!]
	\centering
		\includegraphics[width=1.0\linewidth]{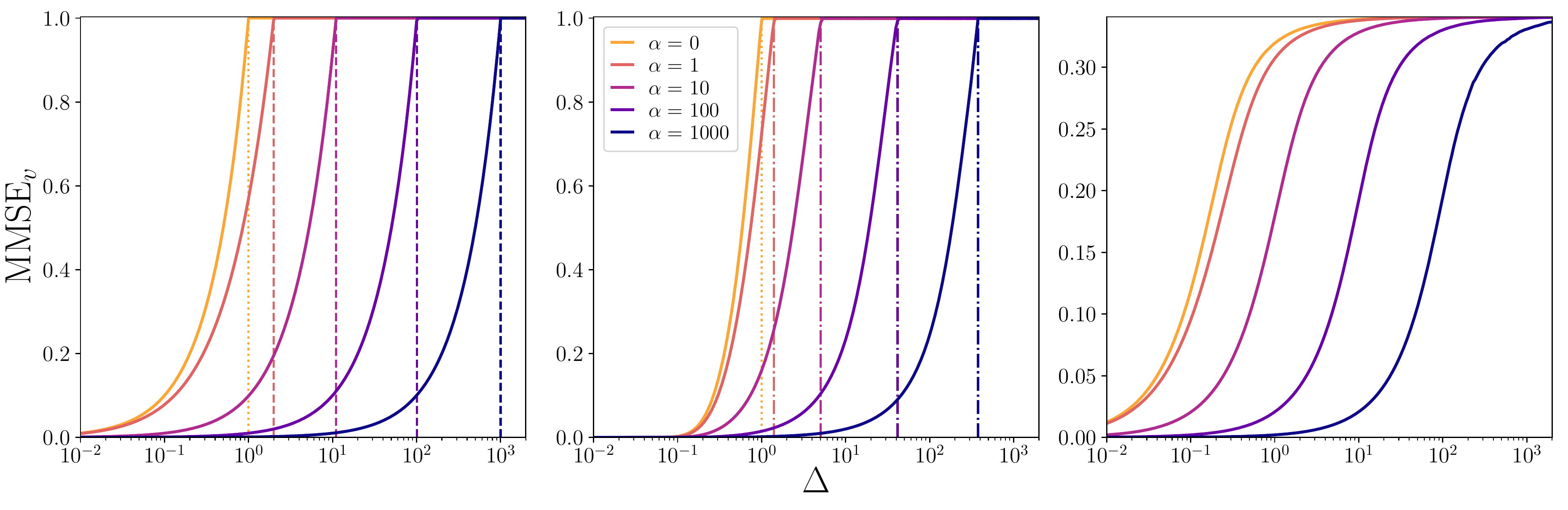}
	\caption{Spiked Wigner model: ${\textrm{MMSE}}_v$ as a function of
          noise $\Delta$ for $L=1$ a wide range of compression ratios
          $\alpha=0,1,10,100,1000$, for \Left linear, \Center sign, and \Right relu activations. Unique stable fixed
          point of (\ref{main:SE_uu}) is found for all these cases.}
	\label{main:fig_mse_u_SE}
\end{figure}
This result is given in full generality, and it is instructive to compute $\Delta_{c}$ in specific cases. 

First, consider the case of a single-layer generative prior $L=1$. Fix 
$\rP_\z(z) = \mathcal{N}_z(0,1)$ and $\rP^{(1)}_{\out}(v|x)=
\delta(v-\varphi(x))$, for $\varphi\in\{\text{linear}, \text{sign}, \text{relu}\}$. The first two choices of non-linearities are odd, and therefore in these cases we expect a transition as discussed above. It can be readily computed from the Jacobian eq.~\eqref{eq:phasetransition:jacobian} and yield $\Delta_{c} = 1+\alpha$ for linear activation and $\Delta_{c} = 1+\frac{4}{\pi^2}\alpha$ for sign activation. In both cases, since $\alpha>0$, it is clear that knowledge of the generative prior improve reconstruction in the sense that the spike can be better reconstructed for larger amplitude of noise $\Delta$. Moreover, the larger $\alpha$ (i.e. the smaller the latent dimension with respect to the signal dimension), the better the reconstruction. 

\Fig\ref{main:fig_map_mse_delta_alpha} summarizes this discussion. We numerically solve the fixed point eqs.~\eqref{main:SE_uu} and plot the \aclink{MMSE} obtained from the fixed point in a heat map, for the linear, sign and relu
activations. The white dashed line marks the threshold
$\Delta_c$ obtained analytically from the Jacobian in eq.~\eqref{eq:phasetransition:jacobian}. The property that we find the most
striking is that in these three evaluated cases, for all values of
$\Delta$ and $\alpha$ that we analyzed, we always found that
eq.~(\ref{main:SE_uu}) has a unique stable fixed point. Thus we have
not identified, in the physics terminology, any first order phase transition. \Fig\ref{main:fig_mse_u_SE} shows some examples of numerical \aclink{MMSE} curves for three nonlinearities discussed and different values of $\alpha$. The fixed point equations were solved iteratively from
uncorrelated initial condition, and from initial condition
corresponding to the ground truth signal, and found that both lead to the same fixed point. 
\begin{figure}[tb!]
	\centering
		\includegraphics[width=1.0\linewidth]{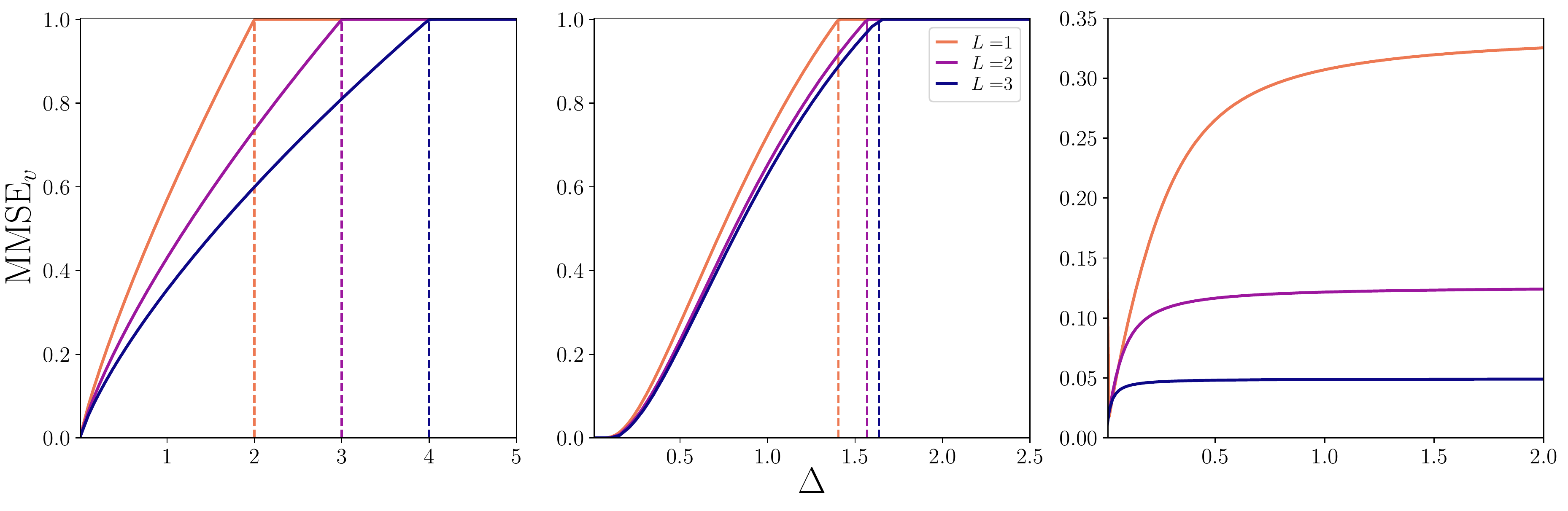}
	\caption{Spiked Wigner model: ${\textrm{MMSE}}_v$ as a function of
          noise $\Delta$ for $L=1,2,3$ with constant compressive ratio $\alpha_1 = \alpha_2 = \alpha_3 = 1$, for \Left linear, \Center sign, and \Right relu activations. The second moments of the variable $v$ for $L=1,2,3$ are for linear and sign activations $\rho_v^{(L)}=1$, while for relu $\rho_v^{(L)}=1/2^L$.}
	\label{main:fig_mse_u_SE_ML}
\end{figure}  
This observation generalizes to deeper $L>1$ generative priors. Consider $\rP_\z(z) = \mathcal{N}_z(0,1)$ and layer-wise constant activation $\rP^{(l)}_{\out}(v|x)=
\delta(v-\varphi(x))$. For the previous odd activation functions discussed, we find that
\begin{itemize}
\item[] {\bfseries Linear activation:} For $\varphi(x)=x$ the leading
      eigenvalue of the Jacobian becomes one at 
      \begin{align}
          \Delta_{c} =\displaystyle
      1+ \sum_{l=1}^{L} \frac{\alpha}{\alpha_{l}}\,.
      \end{align}
      Note in particular that for $L=1$ and in the limit $\alpha = 0$ we recover the phase
    transition
$\Delta_{c}=1$ known from the case with separable prior \cite{lesieur2017constrained}. For
$\alpha>0$, we have $\Delta_{c}>1$ meaning the spike can be estimated
more efficiently when its structure is accounted for. In particular, the deeper the generative network for the spike, the easier estimation becomes.

\item[] {\bfseries Sign activation:} For $\varphi(x)=\sign(x)$ the
      leading eigenvalue of the Jacobian becomes one at 
      \begin{align}
      \Delta_{c} =\displaystyle
      1+\sum_{l=1}^{L}\left(\frac{4}{\pi^2}\right)^{l} \frac{\alpha}{\alpha_{l}}. 
      \end{align}
      For $L=1$ and $\alpha=0$, $\rP_\v =
      \text{Bern(1/2)}$, and the transition $\Delta_{c}=1$ agrees with
      the one found for a separable prior distribution \cite{lesieur2017constrained}. As in the linear case, for
      $\alpha>0$, we can estimate the spike for
      larger values of noise than in
      the separable case, and depth also improves estimation.
\end{itemize}
Note that we also did not observe first order transitions for deeper networks, at least in the first-to-come-in-mind cases that we have investigated, i.e. deterministic deep networks with $\varphi^{(l)} \equiv \varphi \in\{\text{linear}, \text{sign}, \text{relu}\}$. However, we do not expect this behavior to be completely general neither. One can engineer a situation, for instance with a very shifted relu on the last layer, and a very large intermediate layer, so that the spike $\textbf{v}$ becomes effectively sparse with weakly correlated, almost independent, components, thus recovering the classical algorithmic gap \cite{lesieur2017constrained}.

So far we have only discussed the performance of the information theoretic optimal estimator - averting the question of estimating the signal itself. In the next section we close this gap by introducing an \aclink{AMP} algorithm for signal reconstruction. Our algorithm has the advantaged that its performance can tracked down exactly in the thermodynamic limit, and we will show that in the cases we analyzed it exactly follows the same fixed point equations \eqref{main:SE_uu} as the ones derived for optimal estimator.

\section{Approximate message passing with generative priors}
\label{main:sec_amp}
Naive sampling from the high-dimensional posterior distribution is exponentially costly, ruling this approach out from an algorithmic perspective. One should therefore appeal to algorithmically tractable approximations. \aclink{AMP} algorithms have proven to be particularly useful for problems defined on random graphs, and successful examples abound in the literature 

In this section we derive and analyze an \aclink{AMP} algorithm tailored for spiked estimation with generative priors. Next, we show that the \aclink{MSE} of our algorithm can be tracked exactly in the thermodynamic limit, and that moreover it coincides with the optimal performance discussed in \Sec\ref{sec:information_theory} even for large $\alpha$. This result is particularly interesting when compared to the known performance of message passing algorithms for sparse-\aclink{PCA}, for which \aclink{AMP} is not able to reach optimal statistical performance in the small sparsity regime \cite{lesieur_statistical_2017}.

\aclink{AMP} algorithms for spiked matrix estimation with separable priors are well known \cite{metzler2016denoising,manoel2017multi,berthier2017state}. Our derivation draw on previous works on extending \aclink{AMP} to non-separable priors \cite{metzler2016denoising,manoel2017multi,berthier2017state} and we first focus on the more general Wishart model ($\bu\bv^\intercal$). After, we discuss how to get the corresponding result for the Wigner model ($\bv\bv^\intercal$) with a simple change.

\subsection{Derivation of the Approximate Message Passing algorithm}
\label{sec:spiked:amp_derivation}
\aclink{AMP} algorithms can be derived systematically for problems that can be written in terms of an acyclic factor graph. The standard idea is to simplify the corresponding \aclink{BP} equations in the limit of a large number of variables. Together with a Gaussian Ansatz for the distribution of the \aclink{BP} messages, the expansion of the \aclink{BP} yield a set of $\Theta\left(k^2\right)$ simplified equations known as \aclink{rBP} equations. The last step to get the corresponding \aclink{AMP} algorithm is to remove the target dependency of the messages that further reduces the number of iterative equations to $\Theta\left(k\right)$. 

Our derivation is closely related to the derivation of \aclink{AMP} for a series of statistical inference problems with factorized priors, see for example \cite{lesieur2017constrained} and references therein. In the interest of the reader, instead of repeating the cumbersome steps described above, we rather describe how two known and simple \aclink{AMP} algorithms for independent inference problems can be combined into one for the corresponding structured problem. In particular, this is illustrated for the spiked-matrix estimation with single-layer generative model prior, which can be seen as the combination of a \emph{rank-one matrix factorization problem} (MF) \cite{lesieur2017constrained} with a \aclink{GLM} \cite{barbier2017phase}. Note that the multi-layer case follows by iterating this procedure.

\subsubsection{Combining factor graphs}
Consider the factor graphs for the MF and the \aclink{GLM} problems with separable priors, drawn in \Fig\ref{appendix:factor_graph_vu}. The key idea is to replace the separable prior $\rP_\v$ for the \emph{structured} variable $\bv$ in the MF model (in green) by a factorized connection channel $\rP_\out$ (see definition \cite{barbier2017phase}) linking the input $\bv$ with the output factors of the GLM (in red). The resulting factor graph for the structured  Wishart model is given in \Fig\ref{appendix:factor_graph_vu}, with the same color code.
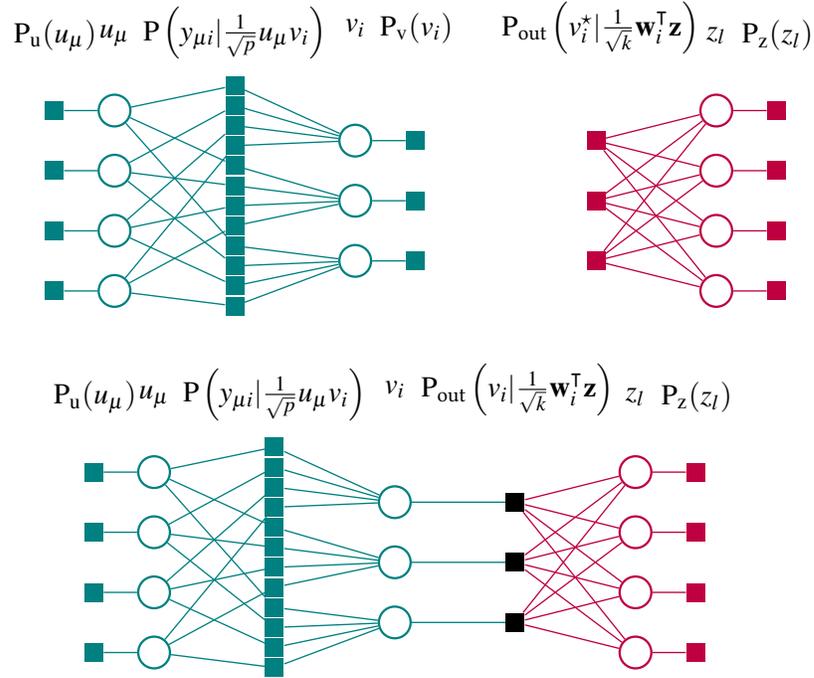
\begin{figure}[!ht]
\centering
    \begin{tikzpicture}[scale=0.8]
	\tikzstyle{factor}=[rectangle,fill=black,minimum size=7pt,inner sep=1pt]
    \tikzstyle{factorMF}=[rectangle,fill=teal,minimum size=7pt,inner sep=1pt]
    \tikzstyle{factorGLM}=[rectangle,fill=purple,minimum size=7pt,inner sep=1pt]
    \tikzstyle{latentMF}=[circle,draw=teal,line width=0.3mm,minimum size=12pt]
    \tikzstyle{latentGLM}=[circle,draw=purple,line width=0.3mm,minimum size=12pt]
    \tikzstyle{latent}=[circle,draw=black,line width=0.3mm,minimum size=12pt]
    \tikzstyle{annot} = [text width=2.5cm, text centered]
    \tikzstyle{edgeMF} = [draw=teal,line width=0.5]
    \tikzstyle{edgeGLM} = [draw=purple,line width=0.5]
    \tikzstyle{edge} = [draw=black,line width=0.5]
	\def\Nv{4}
	\def\Nu{3}
	\def\Nf{12}
		\foreach \name / \y in {1,...,\Nv}
        	\node[latentMF] (V-\name) at (-4,-\y) {};
        \foreach \name / \y in {1,...,\Nv}
        	\node[factorMF] (PV-\name) at (-5,-\y) {};
       	\foreach \name / \y in {1,...,\Nf}
        	\node[factorMF] (F-\name) at (-2,-0.25-\y/3) {};
        \foreach \name / \y in {1,...,\Nu}
        	\node[latentMF] (U-\name) at (0,-0.5-\y) {};
        \foreach \name / \y in {1,...,\Nu}
        	\node[factorMF] (FMF-\name) at (1,-0.5-\y) {};
        	
         \foreach \name / \y in {1,...,\Nu}
        	\node[factorGLM] (F2-\name) at (4,-0.5-\y) {};
        \foreach \name / \y in {1,...,\Nv}
        	\node[latentGLM] (W-\name) at (6,-\y) {};
        \foreach \name / \y in {1,...,\Nv}
        	\node[factorGLM] (PW-\name) at (7,-\y) {};

	\edge [-,edgeMF]{V-1}{F-1};
	\edge [-,edgeMF]{V-2}{F-2};
	\edge [-,edgeMF]{V-3}{F-3};
	\edge [-,edgeMF]{V-4}{F-4};
	\edge [-,edgeMF]{V-1}{F-5};
	\edge [-,edgeMF]{V-2}{F-6};
	\edge [-,edgeMF]{V-3}{F-7};
	\edge [-,edgeMF]{V-4}{F-8};
	\edge [-,edgeMF]{V-1}{F-9};
	\edge [-,edgeMF]{V-2}{F-10};
	\edge [-,edgeMF]{V-3}{F-11};
	\edge [-,edgeMF]{V-4}{F-12};

    \foreach \j in {1,...,4}
        	\edge[-,edgeMF]{F-\j}{U-1} ;
    \foreach \j in {5,...,8}
        	\edge[-,edgeMF]{F-\j}{U-2};
    \foreach \j in {9,...,12}
        	\edge[-,edgeMF]{F-\j}{U-3};
    \foreach \j in {1,...,\Nu}
        	\edge[-,edgeMF]{U-\j}{FMF-\j};

    \foreach \i in {1,...,\Nu}
    	 \foreach \j in {1,...,\Nv}
        	\edge[-,edgeGLM]{F2-\i}{W-\j};
    \foreach \i in {1,...,\Nv}
        \edge[-,edgeMF]{PV-\i}{V-\i};

    \foreach \j in {1,...,\Nv}
    	\edge[-,edgeGLM]{PW-\j}{W-\j};

	\node[annot,above of=PV-1, node distance=1cm] {$\rP_\u(u_\mu)$};
	\node[annot,above of=V-1, node distance=1cm] {$u_\mu $};
	\node[annot,above of=F-1, node distance=0.7cm] {$\rP\(y_{\mu i }|\frac{1}{\sqrt{p}} u_\mu v_{i}\)$};
	\node[annot,above of=U-1, node distance=1.5cm] {$v_{i} $};
	\node[annot,above of=FMF-1, node distance=1.5cm] {$\rP_\v(v_i)$};
	\node[annot,above of=F2-1, node distance=1.5cm] {$\rP_\out\(v_{i}^\star|\frac{1}{\sqrt{k}}\vec{w}_{i}^\intercal \bz\)$};
	\node[annot,above of=W-1, node distance=1cm] (hl) {$z_l$};
	\node[annot,above of=PW-1, node distance=1cm] {$\rP_\z(z_l)$};
    \end{tikzpicture}\\
    
    \hspace{3cm}
	\begin{tikzpicture}[scale=0.8]
	\tikzstyle{factor}=[rectangle,fill=black,minimum size=7pt,inner sep=1pt]
    \tikzstyle{factorMF}=[rectangle,fill=teal,minimum size=7pt,inner sep=1pt]
    \tikzstyle{factorGLM}=[rectangle,fill=purple,minimum size=7pt,inner sep=1pt]
    \tikzstyle{latentMF}=[circle,draw=teal,line width=0.3mm,minimum size=12pt]
    \tikzstyle{latentGLM}=[circle,draw=purple,line width=0.3mm,minimum size=12pt]
    \tikzstyle{latent}=[circle,draw=black,line width=0.3mm,minimum size=12pt]
    \tikzstyle{annot} = [text width=2.5cm, text centered]
    \tikzstyle{edgeMF} = [draw=teal,line width=0.5]
    \tikzstyle{edgeGLM} = [draw=purple,line width=0.5]
    \tikzstyle{edge} = [draw=black,line width=0.5]
	\def\Nv{4}
	\def\Nu{3}
	\def\Nf{12}
		\foreach \name / \y in {1,...,\Nv}
        	\node[latentMF] (V-\name) at (0,-\y) {};
        \foreach \name / \y in {1,...,\Nv}
        	\node[factorMF] (PV-\name) at (-1,-\y) {};
       	\foreach \name / \y in {1,...,\Nf}
        	\node[factorMF] (F-\name) at (2,-0.25-\y/3) {};
        \foreach \name / \y in {1,...,\Nu}
        	\node[latentMF] (U-\name) at (4,-0.5-\y) {};
         \foreach \name / \y in {1,...,\Nu}
        	\node[factor] (F2-\name) at (6,-0.5-\y) {};
        \foreach \name / \y in {1,...,\Nv}
        	\node[latentGLM] (W-\name) at (8,-\y) {};
        \foreach \name / \y in {1,...,\Nv}
        	\node[factorGLM] (PW-\name) at (9,-\y) {};

	\edge [-,edgeMF]{V-1}{F-1};
	\edge [-,edgeMF]{V-2}{F-2};
	\edge [-,edgeMF]{V-3}{F-3};
	\edge [-,edgeMF]{V-4}{F-4};
	\edge [-,edgeMF]{V-1}{F-5};
	\edge [-,edgeMF]{V-2}{F-6};
	\edge [-,edgeMF]{V-3}{F-7};
	\edge [-,edgeMF]{V-4}{F-8};
	\edge [-,edgeMF]{V-1}{F-9};
	\edge [-,edgeMF]{V-2}{F-10};
	\edge [-,edgeMF]{V-3}{F-11};
	\edge [-,edgeMF]{V-4}{F-12};

    \foreach \j in {1,...,4}
        	\edge[-,edgeMF]{F-\j}{U-1} ;
    \foreach \j in {5,...,8}
        	\edge[-,edgeMF]{F-\j}{U-2};
    \foreach \j in {9,...,12}
        	\edge[-,edgeMF]{F-\j}{U-3};
    \foreach \j in {1,...,\Nu}
        	\edge[-,edgeMF]{U-\j}{F2-\j};
    \foreach \i in {1,...,\Nu}
    	 \foreach \j in {1,...,\Nv}
        	\edge[-,edgeGLM]{F2-\i}{W-\j};
    \foreach \i in {1,...,\Nv}
        \edge[-,edgeMF]{PV-\i}{V-\i};

    \foreach \j in {1,...,\Nv}
    	\edge[-,edgeGLM]{PW-\j}{W-\j};

	\node[annot,above of=PV-1, node distance=1cm] {$\rP_\u(u_\mu)$};
	\node[annot,above of=V-1, node distance=1cm] {$u_\mu $};
	\node[annot,above of=F-1, node distance=0.7cm] {$\rP\(y_{\mu i }|\frac{1}{\sqrt{p}} u_\mu v_{i}\)$};
	\node[annot,above of=U-1, node distance=1.5cm] {$v_{i} $};
	\node[annot,above of=F2-1, node distance=1.5cm] {$\rP_\out\( v_{i}|\frac{1}{\sqrt{k}}\vec{w}_{i}^\intercal \bz\)$};
	\node[annot,above of=W-1, node distance=1cm] (hl) {$z_l$};
	\node[annot,above of=PW-1, node distance=1cm] {$\rP_\z(z_l)$};
\end{tikzpicture}\\
\caption{Factor graphs corresponding to a \textbf{(upper left)} low-rank matrix factorization model with separable priors $\rP_\u, \rP_\v$ on $\bu,\bv$ , \textbf{(uper right)} a generalized linear model with observations $\bv^\star$ and prior $\rP_\z$ on $\bz$, and finally to \textbf{(bottom)} a low-rank matrix factorization layer (green) with a GLM prior (red) where the separable prior $\rP_\v(v_i)$ is replaced by correlated factor $\rP_\out(v_i | . )$.}
\label{appendix:factor_graph_vu}
\end{figure}

\subsubsection{Combining AMP algorithms}
As for the factor graphs, we start by recalling the \aclink{AMP} update equations in the Bayes-optimal case for the two problems in question with separable priors.\\

\paragraph{AMP equations for the Wishart MF layer (variables $\bv$ and $\bu$)}
Consider the low-rank matrix factorization model $\mat{Y} = \frac{1}{\sqrt{p}} {\bu^\star} {\bv^\star}^\intercal + \sqrt{\Delta} \bxi$ with separable priors $\rP_\u$ and $\rP_\v$ for the variables $\bu$ and $\bv$, illustrated in \Fig\ref{appendix:factor_graph_vu} \textbf{(upper left)}. The corresponding Bayes-optimal \aclink{AMP} equations, given in \cite{lesieur2017constrained}, read:
\begin{equation}
	\begin{cases}
	\hat{\bu}^{t+1} = \vec{f}_\u(\vec{b}_u^t , \mat{A}_u^t)   \, , \\
	\hat{\mat{C}}_u^{t+1} = \partial_{\vec{b}}\vec{f}_\u(\vec{b}_u^t , \mat{A}_u^t)  \, , \\
	\hat{\bv}^{t+1} = \vec{f}_\v(\vec{b}_v^t , \mat{A}_v^t)   \, , \\
	\hat{\mat{C}}_v^{t+1} = \partial_{\vec{b}}\vec{f}_v(\vec{b}_v^t , \mat{A}_v^t)  \, , \\
	\end{cases}
\andcase
\begin{cases}
	\vec{b}_v^{t} &= \frac{1}{\Delta} \frac{\mat{Y}^\intercal}{\sqrt{p}} \hat{\bu}^t  - \frac{1}{\Delta} \frac{\vec{1}_n^\intercal  \hat{\mat{C}}_u^t}{p} \hat{\bv}^{t-1} \,, \\
	\mat{A}_v^t &= \frac{1}{\Delta} \frac{\|\hat {\bu}^t \|_2^2}{p} \mat{I}_p \, , \\
	\vec{b}_u^{t} &= \frac{1}{\Delta} \frac{\mat{Y}}{\sqrt{p}} \hat{\bv}^t  - \frac{1}{\Delta} \frac{\vec{1}_p^\intercal  \hat{\mat{C}}_v^t}{p} \hat{\bu}^{t-1}  \, , \\
	\mat{A}_u^t &=   \frac{1}{\Delta} \frac{\|\hat {\bv}^t \|_2^2}{p} \mat{I}_n \, , \\
\end{cases}
\label{appendix:amp:mf}
\end{equation}

where 
the update functions $f_u$ and $f_v$ are respectively the means of the distributions $Q_\u$ and $Q_\v$, defined similarly to eq.~\eqref{main:definition_Z} as 
\begin{align}
    \rQ_\u(u; b,A) &\equiv \displaystyle \frac{\rP_\u(u)}{\mZ_{u} (b,A)}  e^{ -\frac{1}{2} A  u^2  + b u}\,, \hhspace \rQ_\v(v; b, A) \equiv \displaystyle \frac{\rP_\v( v)}{\mZ_v (b, A)}  e^{ -\frac{1}{2} A   v^2  + b v }\,.
    \label{appendix:definition_Zuv}
\end{align}

\paragraph{AMP equations for the GLM layer (variable $\bz$)}
On the other hand, the Bayes-optimal \aclink{AMP} equations for the \aclink{GLM} model $\bv^{\star} = \varphi\left(\frac{1}{\sqrt{k}}\mat{W}\bz^{\star}\right)$ with $z^{\star}_{l}\underset{\iid}{\sim} \rP_\z,$ given in \cite{barbier2017phase} and illustrated in \Fig\ref{appendix:factor_graph_vu} read
\begin{equation}
\begin{cases}
	\hat{\bz}^{t+1} = \vec{f}_\z({\boldsymbol \bgamma}^t , \bLambda^t) \,,  \\
	\hat{\mat{C}}_z^{t+1} = \partial_\bgamma \vec{f}_\z({\boldsymbol \bgamma}^t , \bLambda^t)\,, \\
	\bg^{t} = \vec{f}_\out\(\bv^\star, \bomega^{t} , \mat{V}^{t} \) \,,
	\end{cases}
	\andcase
	\begin{cases}
	\bgamma^t =  \frac{1}{\sqrt{k}} \mat{W}^\intercal \bg^{t} + \bLambda^t \hat{\bz}^t \,, \\
	\bLambda^t =   \frac{1}{k} (\mat{W}^2)^\intercal (\bg^t)^2 \mat{I}_k \,, \\
	\bomega^t =  \frac{1}{\sqrt{k}} \mat{W} \hat{\bz}^{t} - \mat{V}^t \bg^{t-1} \,, \\
	\mat{V}^t =\frac{1}{k} (\mat{W}^2)\hat{\mat{C}}_z^{t} \mat{I}_p  \,,
	\end{cases}
	\label{appendix:amp:glm}
\end{equation}
where the operation $\left(\cdot\right)^2$ is taken component-wise. $f_\z$ is the mean of $\rQ_\z$ defined in eq.~\eqref{main:definition_Z} and $f_{\out}$ is the mean of $V^{-1}(x-\omega)$ with respect to 
\begin{align}
   \rQ_{\out}\(x;v^\star,\omega, V \) =\frac{\rP_\out\(v^\star | x \)}{\mZ_{\out}(v^\star, \omega,V)}  \frac{e^{ -\frac{1}{2}V^{-1}  \(x - \omega\)^2  }}{\sqrt{2\pi V}} 
\end{align}

\paragraph{Module composition}
In principle, composing the \aclink{AMP} equations for the inference problems above is non-trivial and requires a full-blown derivation from the \aclink{BP} equations on the composed factor graph in \Fig\ref{appendix:factor_graph_vu}. Surprisingly, the upshot of this cumbersome computation is rather simple: the \aclink{AMP} equations for the composed model are equivalent to coupling the MF eqs.~\eqref{appendix:amp:mf} and the \aclink{GLM} eqs.~\eqref{appendix:amp:glm} by replacing $\rQ_\v(v;b,A)$ and $\rQ_{\out}(x; v^\star, \omega, V)$ with the following joint distribution:
\begin{align}
	\rQ_\out(v,x;b,A,\omega, V) &\equiv  \displaystyle\frac{\rP_\out\(v | x \)}{\mZ_{\out}(b, A, \omega, v)} ~ e^{ -\frac{1}{2} A v^2 + b v }  ~\frac{e^{ -\frac{1}{2}V^{-1}  \(x - \omega\)^2  }}{\sqrt{2\pi V}} \,.
	\label{appendix:Qout_joint}
\end{align}
The associated update functions $f_\v$, $f_{\out}$ are thus replaced by the means of $v$ and $V^{-1}(x-\omega)$ with respect to this new joint distribution $\rQ_\out$. Replacing the separable distributions $\rQ_\u$ and $\rQ_\out$ by the joint distribution eq.~\eqref{appendix:Qout_joint} and corresponding update functions as described above in eq.~\eqref{appendix:amp:mf}-\eqref{appendix:amp:glm}, we obtain the \aclink{AMP} algorithm for structured model. Additionally, we note that the \aclink{AMP} equations above are also valid for arbitrary weight matrix $\mat{W}\in\mathbb{R}^{p\times k}$. In the case of interest where $w_{il}\underset{\iid}{\sim}\mathcal{N}(0,1)$, using $\EE\[w_{il}^2\]=1$ we can further simplify that leads to the following algorithm \Alg\ref{main:AMP_uv_bayes}.

The \aclink{AMP} algorithm for the Wigner model is very similar and can be readily obtained by imposing at each time step $\(\hat{\bu}^t,\hat{\mat{C}}^{t}_u \)=\(\hat{\bv}^t, \hat{\mat{C}}^{t}_v\)$ and removing the redundant equations in \Alg\ref{main:AMP_uv_bayes}.

\begin{algorithm}[!htb]
\caption{Bayes-optimal AMP algorithm for the spiked Wishart model with single-layer generative prior.}
\label{main:AMP_uv_bayes}
\begin{algorithmic}
    \STATE {\bfseries Input:} vector $\mat{Y} \in \vec{b}R^{n \times  p}$ and matrix $\mat{W}\in \vec{b}R^{p \times k}$:
    \STATE \emph{Initialize to zero:} $( \bg, \hat{\bu}, \hat{\bv}, \vec{b}_{v} , \mat{A}_{v}, \vec{b}_{u} , \mat{A}_{u})^{t=0}$
	\STATE \emph{Initialize with:} $\hat{\bu}^{t=1}=\mN(\vec{0} ,\bsigma^2)$, $\hat{\bv}^{t=1}=\mN(\vec{0},\bsigma^2)$, $\hat{\bz}^{t=1}=\mN(\vec{0},\bsigma^2)$, $\hat{\mat{C}}^{t=1}_u = \mat{I}_n$, $\hat{\mat{C}}^{t=1}_v = \mat{I}_p$, $\hat{\mat{C}}^{t=1}_z=\mat{I}_k$.
    \REPEAT
    \STATE \emph{Spiked layer:}
    \STATE  $ \vec{b}_u^{t} = \frac{1}{\Delta} \frac{\mat{Y}}{\sqrt{p}} \hat{\bv}^t  - \frac{1}{\Delta}  \frac{\vec{1}_p^\intercal \hat{\mat{C}}_v^t}{p} \mat{I}_n \hat{\bu}^{t-1}$ \andcase $\mat{A}_u^t =  \frac{1}{\Delta} \frac{\|\hat{\bv}^t\|_2^2}{p}  \mat{I}_n$
    \STATE  $ \vec{b}_v^{t} = \frac{1}{\Delta} \frac{\mat{Y}^\intercal}{\sqrt{p}} \hat{\bu}^t  - \frac{1}{\Delta}\frac{\vec{1}_n^\intercal \hat{\mat{C}}_u^t}{p}  \mat{I}_p \hat{\bv}^{t-1}$ \andcase $\mat{A}_v^t =  \frac{1}{\Delta} \frac{\|\hat{\bu}^t\|_2^2}{p}  \mat{I}_p$
    \STATE \emph{Generative layer:}
    \STATE $\mat{V}^t =\frac{1}{k} \(\vec{1}_k^\intercal\hat{\mat{C}}_z^{t} \) \mat{I}_p  \andcase  \bomega^t =  \frac{1}{\sqrt{k}} \mat{W} \hat{\bz}^{t} - \mat{V}^t \bg^{t-1}$  \andcase ${\bg}^{t} = \vec{f}_{\out}\({ \vec{b}}^{t}_v, \mat{A}^{t}_v, \bomega^{t}, \mat{V}^{t} \) $ 
    \STATE $\bLambda^t =  \frac{1}{k} \|\bg^t\|_2^2 \mat{I}_k  \andcase \bgamma^t =  \frac{1}{\sqrt{k}} \mat{W}^\intercal \bg^{t} + \bLambda^t \hat{\bz}^t $
    \STATE \emph{Update of the estimated marginals:}
    \STATE $\hat{\bu}^{t+1} = \vec{f}_\u( \vec{b}_u^t , \mat{A}_u^t) \hhspace \andcase  \hhspace \hat{\mat{C}}_u^{t+1} = \partial_{\vec{b}}\vec{f}_\u( \vec{b}_u^t , \mat{A}_u^t) $
    \STATE $\hat{\bv}^{t+1} = \vec{f}_\v( \vec{b}_v^t , \mat{A}_v^t,\bomega^{t} , \mat{V}^{t} ) \hhspace  \andcase \hhspace  \hat{\mat{C}}_v^{t+1} = \partial_{\vec{b}}\vec{f}_\v( \vec{b}_v^t , \mat{A}_v^t,\bomega^{t} , \mat{V}^{t}) $
    \STATE $\hat{\bz}^{t+1} = \vec{f}_\z(\bgamma^t , \bLambda^t) \andcase \hat{\mat{C}}_z^{t+1} = \partial_\bgamma \vec{f}_\z(\bgamma^t , \bLambda^t)$ 
    \STATE ${t} = {t} + 1$
    \UNTIL{Convergence}
    \STATE {\bfseries Output: $\hat{\bu}, \hat{\bv}, \hat{\bz}$}
\end{algorithmic}
\end{algorithm}

\subsection{State evolution equations}
\label{sec:amp:se}
Perhaps the most important virtue of \aclink{AMP}-type algorithms is that their asymptotic performance can be tracked exactly via a set of scalar equations called \emph{state evolution}. The order parameters involved are the average overlap between the estimated signals and the ground truth, and are closely related to the mean square error obtained by the algorithm. This fact has been proven for a range of models including the spiked matrix models with separable priors in \cite{javanmard2013state}, and with non-separable priors in \cite{berthier2017state}. Adapting the steps of these works, we now derive the state evolution equations for our structured model. As before, we focus on the derivation for the general Wishart model $\bu\bv^{\intercal}$, from which the Bayes-optimal \aclink{SE} equations for the symmetric $\bv\bv^{\intercal}$ can be readily obtained.

\subsubsection{Relaxed-Belief Propagation equations}
Note that the standard derivation starts from the \aclink{rBP} equations, which are roughly equivalent to \aclink{AMP} updates up to the Onsager terms containing messages with delayed time indices $(\cdot)^{t-1}$. We briefly recall them below where we introduced the parameters $s_{j \mu} \equiv \frac{y_{j \mu}}{\Delta}$ and $r_{j \mu} \equiv -\frac{1}{\Delta} + s_{j \mu}^2$, $\forall j\in \lb n \rb  \mu \in \lb p \rb$.
\begingroup

\begin{align}
\begin{aligned}
	\textbf{Variable u}&\\
	\hat{u}_{j \to j \mu}^{t+1} &= f_\u\(b^{u,t}_{j \to j \mu}, A^{u,t}_{j\to j \mu}\)\,, \hhspace \hat{C}_{j \to j \mu}^{u,t+1} = \partial_{b} f_\u \( b^{u,t}_{j \to j \mu}, A^{u,t}_{j\to j \mu} \) \,, \\
	b^{u,t}_{j \to j \mu} &= \frac{1}{\sqrt{p}} \displaystyle \sum_{\nu \ne \mu}^p s_{j \nu}\hat{v}_{\nu \to j \nu}^t\,, \\  
	a^{u,t}_{j\to j \mu} &=  \frac{1}{p} \displaystyle \sum_{\nu \ne \mu}^p s_{j \nu}^2 (\hat{v}_{\nu \to j \nu}^t )^2 - r_{j \nu} ( \hat{C}^{v,t}_{\nu  \to j \nu} + (\hat{v}_{\nu \to j \nu}^t)^2 ) \,,  \\
	\textbf{Variable v}&\\
	\hat{v}_{\mu \to j \mu}^{t+1} &= f_\v (b^{v,t}_{\mu \to j \mu}, A^{v,t}_{\mu \to j \mu} , \omega_{\mu}^t , V_{\mu}^t )\,,\\
	\hat{C}^{v,t+1}_{\mu \to j \mu} &= \partial_b f_\v (b^{v,t}_{\mu \to j \mu},A^{v,t}_{\mu \to j \mu} , \omega_{\mu}^t , V_{\mu}^t ) \,,  \\
	b^{v,t}_{\mu \to j \mu} &= \frac{1}{\sqrt{p}} \displaystyle \sum_{l \ne j }^n s_{l \mu} \hat{u}_{l \to l \mu}^t\,, \\
	A^{v,t}_{\mu \to j \mu} &=  \frac{1}{p} \displaystyle \sum_{l \ne j}^n s_{l \mu}^2 (\hat{u}_{l \to l \mu}^t)^2 -r_{l \mu} ( \hat{C}^{u,t}_{l  \to l \mu} + (\hat{u}_{l \to l \mu}^t)^2 )  \,, \\
	\omega_{\mu}^t &= \frac{1}{\sqrt{k}}
		\displaystyle \sum_{i=1}^{k} w_{\mu i} \hat{z}_{i \to \mu }^t\,, \hhspace V_{\mu}^t =  \frac{1}{k} \displaystyle   \sum_{i=1}^{k} w_{\mu i}^2\hat{C}^{z,t}_{i \to \mu}\,,\\
\end{aligned}
\label{main:rbp_equations_1}
\end{align}
\begin{align}
\begin{aligned}
	\textbf{Variable z}&\\
	\hat{z}_{i \to \mu}^{t+1} &= f_\z\(\gamma_{i \to \mu}^t,\Lambda_{i \to \mu}^t \)\,, \hhspace  \hat{C}_{i \to \mu}^{z,t+1} = \partial_\gamma f_\z \( \gamma_{i \to \mu}^t,\Lambda_{i \to \mu}^t \) \,,  \\
	\gamma_{i \to \mu}^t &=  \displaystyle  \sum_{\nu \ne \mu }^p  b_{\nu \to i}^{z,t} \,, \hhspace 
	\Lambda_{i \to \mu}^t =  \displaystyle \sum_{\nu \ne \mu }^p  A_{\nu \to i}^{z,t} \,,  \\
	b^{z,t}_{\nu \to i} &=  \frac{w_{\nu i}}{\sqrt{k}} f_{\out}(b^{v,t}_{\nu},A^{v,t}_{\nu},\omega_{\nu \to i}^t,V_{\nu \to i}^t)\,,\\ 
	A^{z,t}_{\nu \to i} &=   -\frac{w_{\nu i}^2}{k} \partial_\omega f_{\out}(b^{v,t}_{\nu},A^{v,t}_{\nu},\omega_{\nu \to i}^t, V_{\nu \to i}^t)  \,, \\
	b^{v,t}_{\nu} &= \frac{1}{\sqrt{p}} \displaystyle \sum_{j=1}^n s_{j \nu} \hat{u}_{j \to j \nu}^t\,,\\
	A^{v,t}_{\nu} &=  \frac{1}{p} \displaystyle \sum_{j =1}^n s_{j \nu}^2 (\hat{u}_{j \to j \nu}^t)^2 -r_{j \nu} ( \hat{C}^{u,t}_{j  \to j \nu} + (\hat{u}_{j \to j \nu}^t)^2 ) \,,  \\
	\omega_{\nu \to i }^t &=  \frac{1}{\sqrt{k}}\displaystyle \sum_{l\ne i}^{k} w_{\nu l} \hat{z}_{l \to \nu }^t \,, \hhspace V_{\nu \to i}^t =   \frac{1}{k}\displaystyle \sum_{l\ne i}^{k}  w_{\nu l}^2 \hat{C}^{z,t}_{l \to \nu}\,.
\end{aligned}
\label{main:rbp_equations_2}
\end{align}

We take this as our starting point and refer the curious reader to \cite{lesieur2017constrained} for more details. The first step is to define the overlap parameters that measure the reconstruction of our inference problem:
\begin{align}
\label{appendix:amp:overlap_definition}
	q_u^t &\equiv \EE_{\bu^\star} \lim_{n\to \infty} \frac{(\hat{\bu}^t)^\intercal \hat{\bu}^t}{n}  = \EE_{\bu^\star} \lim_{n\to \infty} \frac{(\hat{\bu}^t)^\intercal \bu^\star}{n}\equiv m_u^t\,,\\ 
	q_v^t &\equiv   \EE_{\bv^\star} \lim_{p\to \infty} \frac{(\hat{\bv}^t)^\intercal \hat{\bv}^t}{p} = \EE_{\bv^\star} \lim_{p\to \infty} \frac{(\hat{\bv}^t)^\intercal \bv^\star}{p}  \equiv m_v^t\,, \spacecase 
	q_z^t &\equiv \EE_{\bz^\star}\lim_{k\to \infty} \frac{(\hat{\bz}^t)^\intercal \hat{\bz}^t}{k} = \EE_{\bz^\star} \lim_{k\to \infty} \frac{(\hat{\bz}^t)^\intercal \bz^\star}{k}  \equiv m_z^t\,, \nonumber
\end{align}
where we used the Nishimori identity see \cite{lesieur2017constrained} or \App\ref{appendix:replica_computation:nishimori} to obtain the equality between order parameters $q_x^t = m_x^t$ for $x\in \{v,u,z\}$.

\subsubsection{Average distributions}
Next, to see how these order parameters come into play, we compute the distribution of the \aclink{rBP} messages in eqs.~(\ref{main:rbp_equations_1}-\ref{main:rbp_equations_2}), taking the average over the random variables $\mat{W}$, $\bxi$, the planted solutions $\bv^\star, \bu^\star, \bz^\star$ and taking the limit $p \to \infty$. Note that using the \aclink{BP} independence assumption over the messages and keeping only dominant terms in the $1/p$ expansion, the dependency in the target node disappears and yields:

\paragraph*{$\bullet$ Average over $\vec{b}_u,\bA_u$}
\begin{align}
	\EE\[ \vec{b}_u^t \] &= \frac{1}{\sqrt{p}\Delta} \EE\[ \mat{Y} \hat{\bv}^t \] = \frac{1}{\sqrt{p}\Delta} \EE\[ \(\frac{\bu^\star(\bv^\star)^\intercal}{\sqrt{p}}  + \sqrt{\Delta}\bxi \) \hat{\bv}^t \]\nonumber \\ 
	& \underlim{p}{\infty}  \frac{q_v^t}{\Delta} \bu^\star \,, \nonumber \spacecase
	\EE\[ \vec{b}_u^t (\vec{b}_u^t)^\intercal \] &= \frac{1}{p\Delta^2} \EE\[ \mat{Y} \hat{\bv}^t(\hat{\bv}^t)^\intercal \mat{Y}^\intercal \]  = \frac{1}{\Delta}  \frac{1}{p} \EE\[ \xi \hat{\bv}^t(\hat{\bv}^t)^\intercal \xi^\intercal \]  + o\(1/p\) \nonumber \\
	&\underlim{p}{\infty} \frac{q_v^t}{\Delta} \rI_n \,,\spacecase
	\EE \[ \mat{A}_u^t \] &= \EE\[ \frac{1}{\Delta} \frac{\|\hat{\bv}^t\|_2^2}{p}  \rI_n \]  \underlim{p}{\infty} \frac{q_v^t}{\Delta} \rI_n \,.\nonumber 
\end{align}

\paragraph*{$\bullet$ Average over $\vec{b}_v,\bA_v$}
\begin{align}
	\EE\[ \vec{b}_v^t \] & \underlim{p}{\infty}  \beta\frac{q_u^t}{\Delta} \bv^\star \,, \hhspace \EE\[ \vec{b}_v^t (\vec{b}_v^t)^\intercal \] \underlim{p}{\infty} \beta \frac{q_u^t}{\Delta} \rI_p \,, \hhspace \EE \[ \mat{A}_v^t \] \underlim{p}{\infty} \beta \frac{q_u^t}{\Delta}\rI_p \,.  
\end{align}
\paragraph*{$\bullet$ Average over $\bomega, \mat{V}$}
\begin{align}
\begin{aligned}
	\EE\[\bomega^t\] &= \vec{0}_p \,,  \hhspace \EE\[\bomega^t (\bomega^t)^\intercal \] = \EE\[ \frac{1}{k} \mat{W} \hat{\bz}^{t} (\hat{\bz}^{t})^\intercal \mat{W}^\intercal \] \underlim{n}{\infty} q_z^t \rI_p \,,\\ 
	\EE \[ \mat{V}^t \] &\underlim{k}{\infty} (\rho_z - q_z^t) \rI_p \,. 
\end{aligned}
\end{align}

Wrapping the above equations together, we obtained the distributions of means and variances $\vec{b}_u, \mat{A}_u$, $\vec{b}_v, \mat{A}_v$ and $\bomega, \mat{V}$:
\begin{align}
\begin{aligned}
	\vec{b}_u &\sim  \frac{q_v^t}{\Delta} \bu^\star + \sqrt{\frac{q_v^t}{\Delta}} \bxi_u \,,&& \mat{A}_u^t \sim \frac{q_v^t}{\Delta} \rI_n  \,,\\
	\vec{b}_v &\sim  \beta\frac{q_u^t}{\Delta} \bv^\star + \sqrt{\beta\frac{q_u^t}{\Delta}} \bxi_v \,,&& \mat{A}_v^t \sim  \beta  \frac{q_u^t}{\Delta}\rI_p \,,\\ 
	\bomega &\sim  \sqrt{q_z^t} \boldeta \,,&& \mat{V} \sim (\rho_z - q_z^t) \rI_p \,,
\end{aligned}
	 \label{appendix:amp:se_from_amp:distributions}
\end{align}
with $\bxi_u \sim \mN\( \vec{0}_n, \rI_n \), \bxi_v\sim \mN\( \vec{0}_p, \rI_p \), \boldeta \sim \mN\( \vec{0}_p, \rI_p  \)$.\\

\subsubsection{State evolution equations in the Wishart model}
With the averaged limiting distributions of all the messages, we can now compute the state evolution of the overlaps. Using the definition of the overlaps eq.~\eqref{appendix:amp:overlap_definition} and distributions in eq.~\eqref{appendix:amp:se_from_amp:distributions}, we obtain:
\paragraph*{Variable $\bu$}
\begin{align}
	q_u^{t+1} & = \EE_{\bu^\star} \lim_{n\to\infty} \frac{1}{n} \vec{f}_\u(\vec{b}_u^t , \mat{A}_u^t)^\intercal	\vec{f}_u(\vec{b}_u^t , \mat{A}_u^t) \nonumber \\ 
	&= \EE_{u^\star,\xi}\[ f_\u\(\frac{q_v^t}{\Delta} u^\star + \sqrt{\frac{q_v^t}{\Delta}} \xi , \frac{q_v^t}{\Delta} \)^2 \] \label{appendix:amp:se_from_amp:se_nb_qu} 
\end{align}
where $u^\star \sim \rP_{\u}$, $\xi \sim \mN(0,1)$.
\paragraph*{Variable $\bv$}
\begin{align}
	q_v^{t+1} &= \EE_{\bv^\star} \lim_{p\to\infty} \frac{1}{p} \vec{f}_\v(\vec{b}_v^t , \vec{b}v^t,\bomega^{t} , \mat{V}^{t} )^\intercal	f_v(\vec{b}_v^t , \vec{b}v^t,\bomega^{t} , \mat{V}^{t} ) \nonumber \\
	&= \EE_{v^\star,\xi,\eta}\[ f_\v\(\frac{\beta q_u^t}{\Delta} v^\star  + \sqrt{\frac{ \beta q_u^t}{\Delta}} \xi , \beta  \frac{q_u^t}{\Delta}, \sqrt{q_z^t}\eta ,\rho_z-q_z^t\)^2 \] \label{appendix:amp:se_from_amp:se_nb_qv}
\end{align}
where $v^\star \sim \rP_{\v}$, $\xi \sim \mN(0,1)$.

\paragraph*{Variable $\hat{\bz}$ and $\bz$}
Even if the \emph{hat} overlap does not have as much physical meaning as the standard overlaps that quantify the reconstruction performances, we define it as
\begin{align}
	\hat{q}_z^t &\equiv \alpha \EE_{v^\star,\xi,\eta}\[ f_{\out}\(\frac{\beta q_u^t}{\Delta} v^\star  + \sqrt{\frac{ \beta q_u^t}{\Delta}} \xi , \beta  \frac{q_u^t}{\Delta} \, \sqrt{q_z^t}\eta , \rho_z-q_z^t \)^2 \] \label{appendix:amp:se_from_amp:se_nb_qhat}\,, 
\end{align}
with $v^\star \sim \rP_{\v}$, $\xi,\eta \sim \mN(0,1)$.
Averages of the messages of the variable $\bz$ are explicitly expressed as a function of the \emph{hat} overlaps introduced just above: 
\begin{align}
	\EE\[\bgamma^t\] & \underlim{k}{\infty} \hat{q}_z^t  \bz^\star,\hhspace \EE\[\bgamma^t (\bgamma^t)^\intercal \] \underlim{k}{\infty}   \hat{q}_z^t \rI_k \andcase \EE\[ \bLambda^t \] \underlim{k}{\infty} \hat{q}_z^t \rI_k\,.
\end{align}
At the leading order, we obtain
\begin{align}
	\bgamma^t &\sim \hat{q}_z^t  \bz^\star + \sqrt{\hat{q}_z^t} \bxi \,,\hhspace  \bLambda^t \sim \hat{q}_z^t \rI_k\, \hhspace \textrm{with} \hhspace \bxi \sim \mN\( \vec{0}_k, \rI_k \)
\end{align}
and finally
\begin{align}
	q_z^{t+1} &\equiv \EE_{\bz^\star} \lim_{k\to\infty} \frac{1}{k} \vec{f}_z(\bgamma^t, \bLambda^t)^\intercal \vec{f}_z(\bgamma^t, \bLambda^t)\nonumber \\
	&= \EE_{z^\star,\xi}\[ f_\z\( \hat{q}_z^t  z^\star + \sqrt{\hat{q}_z^t} \xi , \hat{q}_z^t \)^2 \]\,.\label{appendix:amp:se_from_amp:se_nb_qz}
\end{align}
As a conclusion, equations (\ref{appendix:amp:se_from_amp:se_nb_qu}-
	\ref{appendix:amp:se_from_amp:se_nb_qhat}, 
	\ref{appendix:amp:se_from_amp:se_nb_qz}) constitute the closed set of \aclink{SE} equations of the Bayes-optimal \aclink{AMP} algorithm for the Wishart model.
	
\subsubsection{State evolution equations in the Wigner model}
Finally, similarly to the derivation of the \aclink{AMP} algorithm, the \aclink{SE} equations for the Wigner model ($\bv\bv^{\intercal}$) are obtained as a particular case of the above by simply restricting $q_u^t = q_v^t$ and $\beta=1$. In the end, performing a change of variable, this leaves us with only three coupled equations:
\begin{align}
\label{main:SE_AMP_uu}
	q_z^{t+1} &= \displaystyle \EE_{\xi}\[ \mZ_z \times f_z^2 \(\sqrt{\hat{q}_z^t} \xi , \hat{q}_z^t \)\] = \displaystyle  2\partial_{\hat{q}_{z}}\Psi_{z}\left(\hat{q}_{z}^t\right) \,, \notag \spacecase
	\hat{q}_z^t &= \displaystyle\alpha \EE_{\xi,\eta}\[ \mZ_{\out} \times f_{\out}^2 \( \sqrt{\frac{q_v^t}{\Delta}} \xi , \frac{q_v^t}{\Delta} , \sqrt{q_z^t}\eta , \rho_z - q_z^t \) \] \notag \\
	 &= \displaystyle 2\alpha\partial_{q_{z}}\Psi_{\out}\left(\frac{q_{v}^t}{\Delta},q_{z}^t\right) \,,  \spacecase
	q_v^{t+1} &= \displaystyle\EE_{\xi,\eta}\[ \mZ_{\out} \times f_v^2\(\sqrt{\frac{q_v^t}{\Delta}} \xi ,\frac{q_v^t}{\Delta}, \sqrt{q_z^t}\eta , \rho_z - q_z^t\) \] \notag\\
	&= \displaystyle 2\partial_{q_{v}}\Psi_{\out}\left(\frac{q_{v}^t}{\Delta},q_{z}^t\right) \,,\notag
\end{align}
\noindent with initialization $q_v^{t=0} = \varepsilon $, $q_z^{t=0}=\varepsilon$ and a small $\varepsilon>0$.
We notice immediately that \eqref{main:SE_AMP_uu} are the same equations as the fixed point equations related to the Bayes-optimal estimation \eqref{main:SE_uu} with specific time-indices and initialization, but crucially the same fixed points.
Thus the analysis of fixed points in \Sec\ref{sec:examples} applies straightforwardly here. In particular, since in all cases analyzed we found the stable fixed point of (\ref{main:SE_uu}) to be unique, we conclude that our \aclink{AMP} algorithm reaches asymptotically optimal performance in these cases.

We can further check this result by numerically comparing the runs of \aclink{AMP} on finite size instances with the state evolution curves already presented in \Fig\ref{main:fig_mse_u_SE}, also giving an idea of the amplitude of the finite size effects. This experiment is illustrated in \Fig\ref{main:bbp_lamp_amp_se}, together with a curve for \aclink{PCA} and for \aclink{LAMP}, a spectral method we derive from \aclink{AMP} in the next section. A code for reproducing this experiment is provided in \href{https://github.com/sphinxteam/StructuredPrior_demo}{GitHub repository} \cite{StructuredPrior_demo_repo}.

\begin{figure}[tb!]
\centering
	\includegraphics[width=1.0\linewidth]{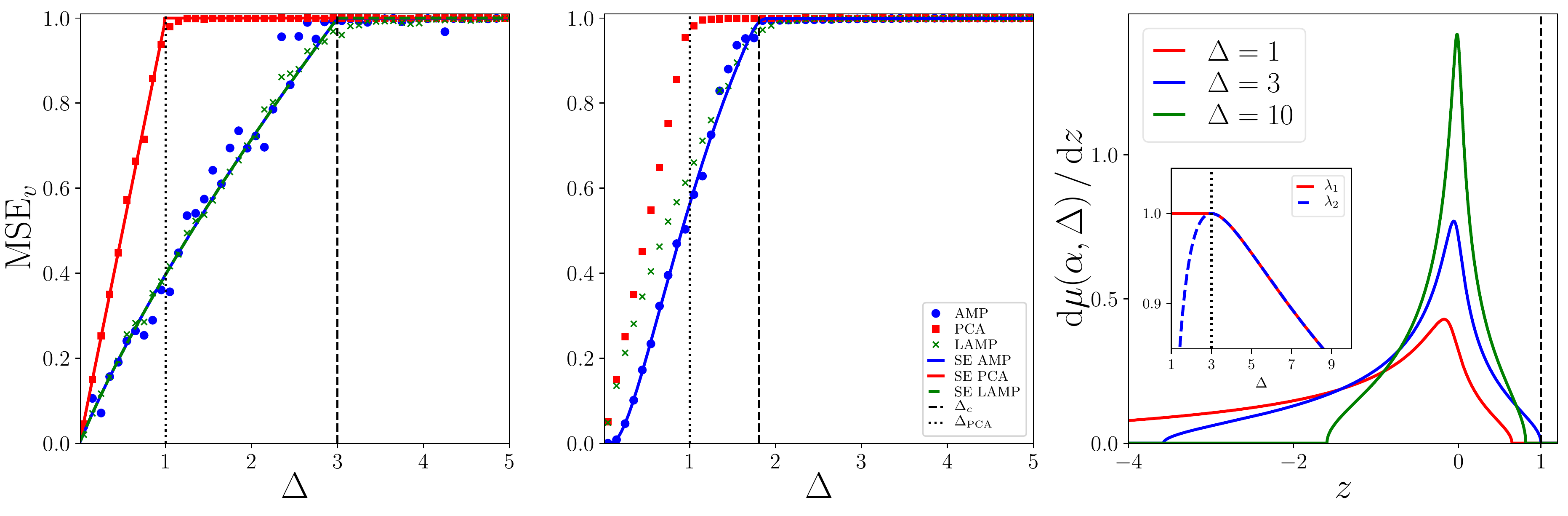}
	\caption{Comparison between PCA, LAMP and AMP for {(left)}
          the linear, {(center)} and sign activations, at compression ratio
          $\alpha=2$. Lines correspond to the theoretical asymptotic
          performance of PCA (red line), LAMP (green line) and AMP
          (blue line). Dots correspond to simulations
          of PCA (red squares), LAMP (green crosses) for $k=10^4$ and AMP (blue
          points) for $k=5.10^3$, $\sigma^2=1$. Notice that the spectral estimators have been rescaled by a factor $(q_{v,\textrm{AMP}}^\star)^{1/2}$ to fairly compare AMP with PCA and LAMP.
          {(Right)} Illustration of the spectral phase transition in the
          matrix $\bGamma_p^{vv}$ eq.~(\ref{eq:matrix_symmetric}) at $\alpha=2$ with
          an informative leading eigenvector with eigenvalue equal to
          $1$ out of the bulk  for   $\Delta\le 1+\alpha$.  We show
          the bulk spectral density $\mu(\alpha, \Delta)$. The inset
          shows the two leading eigenvalues.}
	\label{main:bbp_lamp_amp_se}
\end{figure}

\section{LAMP: a spectral algorithm for generative priors}
\label{sec:spectralmethods}
Spectral algorithms are the most popular and simplest methods for solving the spiked
matrix estimation problem. For instance, canonical \aclink{PCA} estimates the spike from the leading
eigenvector of the matrix $\mat{Y}$. A classical result from \aclink{BBP} \cite{baik2005phase} shows that this eigenvector
is correlated with the signal if and only if the signal-to-noise ratio
$\rho_v^2/\Delta >1$. For sparse separable priors with
$\rho_v^2=\Theta(1)$, $\Delta_{\textrm{PCA}} = \rho_v^2$ is also the threshold
for \aclink{AMP} and it is conjectured that no polynomial algorithm can improve
upon it \cite{lesieur2017constrained}. In the previous section we have shown
that our structured \aclink{AMP} algorithm has a consistently better performance than \aclink{PCA}, and in particular achieve the optimal threshold for better-than-random recovery. This is not a surprise, since different from \aclink{AMP}, vanilla \aclink{PCA} doesn't take into account the information available from the prior.

Despite all its virtues, \aclink{AMP} is unarguably a convoluted algorithm. It would be desirable to have a simpler spectral algorithm taking into account the structured nature of the prior. In this section we design a spectral algorithm, hereafter named \aclink{LAMP}, matching the \aclink{AMP} recovery threshold. Our derivation follows the strategy pioneered in \cite{krzakala_spectral_2013}, consisting on the linearization of the \aclink{AMP} equations around the non-informative fixed point. 
In this section, the discussion is framed on the Wigner model, the Wishart case being a straightforward generalization.

In order for the $q_{v} =0 $ expansion to be well-defined, we first need to insure that this is indeed a fixed point. Indeed, this condition was already discussed in \Sec\ref{sec:information_theory} for the fixed point equations for the Bayes-optimal estimator. Not surprisingly, the same conditions can be obtained independently from the \aclink{AMP} equations by analyzing when $\hat{\vec{v}} = \vec{0}$ is a fixed point, and are repeated below for convenience.
\begin{align}
		(\hat{\vec{v}}, \hat{\bz}) = ( \vec{0}, \vec{0} ) \hhspace \textrm{ if } \hhspace  \mC \equiv \left\{ \hhspace
			\EE_{\rQ_\out^0} \[ v \] = 0 \andcase
			\EE_{\P_\v} \[ z \] = 0 \right \} \,. 
\end{align}
That these conditions agree exactly to the ones in eq.~\eqref{trivial_fixed}
is just a rephrasing of the fact that the \aclink{AMP} \aclink{SE} equations in eqs.~\eqref{main:SE_AMP_uu} have the same fixed points as the Bayes-optimal estimator. 
\subsection{Linearizing the AMP equations}
To lighten notations, we denote with $|_{\star}$ quantities that are evaluated at $(\vec{b}_v, \mat{A}_v, \bomega, \mat{V}, \bgamma, \bLambda) = (\vec{0}, \mat{0}, \vec{0}, \rho_z\mat{I}_p, \vec{0} ,\mat{0})$, and we linearize the \aclink{AMP} equations in \Alg\eqref{main:AMP_uv_bayes} around the uninformative fixed point
\begin{align}
\begin{aligned}
	(\hat{\vec{v}}, \hat{\mat{C}}_v) &= (\vec{0}, \rho_v \mat{I}_p  ),\hhspace (\hat{\bz}, \hat{\mat{C}}_z) = ( \vec{0}, \rho_z \mat{I}_k), \hhspace (\vec{b}_v, \mat{A}_v) = (\vec{0} ,\mat{0}),\\
	(\bgamma,\bLambda) &= (\vec{0}, \mat{0}), \hhspace (\bomega, \mat{V}, \bg) = (\vec{0}, \rho_z \mat{I}_p, \vec{0} )\,.
\end{aligned}
\end{align}
In a scalar formulation, this yields
\begin{align}
\begin{aligned}
		\delta \hat{v}_i^{t+1} &= \partial_b f_\v|_{\star} \delta b^{v,t}_{i}   + \partial_A f_\v|_{\star}  \delta A^{v,t}_{i}  + \partial_\omega f_\v|_{\star} \delta \omega^{t}_{i}   + \partial_V f_\v|_{\star}  \delta V^{t}_{i} \,, \spacecase
		\delta \hat{c}_{i}^{v,t+1} &= \partial_{b,b}^2 f_\v|_{\star}  \delta b^{v,t}_{i}  + \partial_{A,b}^2 f_\v|_{\star}  \delta A^{v,t}_{i}  + \partial_{\omega,b}^2 f_\v|_{\star}  \delta \omega^{t}_{i}   + \partial_{V,b}^2 f_\v|_{\star}  \delta V^{t}_{i} \,,
			\spacecase
		\delta \hat{z}_{l}^{t+1} &= \partial_\gamma f_\z|_{\star}   \delta \gamma_l^t  + \partial_\Lambda f_\z|_{\star} \delta \Lambda_l^t \,, \spacecase
		\delta \hat{c}_{i}^{z,t+1} &=  \partial_{\gamma,\gamma}^2 f_\z|_{\star}  \delta \gamma_l^t  + \partial_{\Lambda,\gamma}^2 f_\z|_{\star}  \delta \Lambda_l^t \,, \spacecase
		\delta g_i^{t} &=  \partial_b f_{\out}|_{\star}   \delta b^{v,t}_{i}  + \partial_A f_{\out}|_{\star} \delta A^{v,t}_{i}  + \partial_\omega f_{\out}|_{\star} \delta \omega^{t}_{i}  + \partial_V f_{\out}|_{\star}  \delta V^{t}_{i} \,, 
\end{aligned}	
\label{eq:lamp_uu:v_z_g}
\end{align}
with
\begin{align}
\begin{aligned}
		\delta b^{v,t}_{i} &= \frac{1}{\Delta}	 \sum_{j=1}^p \frac{y_{j i}}{\sqrt{p}}  \delta \hat{v}_{j}^{t}  -  \frac{1}{\Delta} \( \displaystyle \sum_{j=1}^p  \frac{\hat{c}_{j}^{v,t}|_\star}{p}  \) \delta\hat{v}_{i}^{t-1} -  \frac{1}{\Delta} \( \displaystyle \sum_{j=1}^p  \frac{ \delta\hat{c}_{j}^{v,t}}{p}  \) \hat{v}_{i}^{t-1}|_\star \,,  \\ 
		\delta A^{v,t}_{i} &=  \frac{2}{\Delta} \displaystyle \sum_{j=1}^p \frac{\hat{v}_{j}^t|_\star \delta \hat{v}_{j}^t}{p} = 0  \spacecase
		\delta \omega_{i}^t &= \frac{1}{\sqrt{k}}\displaystyle \sum_{l=1}^k w_{i l} \delta \hat{z}_{l}^{t} - \delta V_{i}^t g_i^{t-1}|_\star - V_{i}^t|_\star \delta g_i^{t-1} \,, \hhspace \delta V_{i}^t  = \frac{1}{k} \displaystyle   \sum_{l=1}^k \delta \hat{c}^{z,t}_{l}\,, 
			\spacecase
		\delta \gamma_{l}^t &= \frac{1}{\sqrt{k}} \sum_{i=1}^p  W_{i l} \delta \bg_i^{t} + \delta\Lambda_{l}^t \hat{\bz}_{l}^t|_\star + \Lambda_{l}^t|_\star \delta\hat{z}_{l}^t\,, \hhspace
		\delta \Lambda^t_{l} =  \frac{2}{k} \sum_{i=1}^p g_i^{t}|_\star \delta g_i^{t} = 0 \,.
\end{aligned}
\label{eq:lamp_uu:b_A_gamma_Lambda}
\end{align}
These equations can be simplified and closed over three vectors $\hat{\vec{v}} \in \bbR^{p}$, $\hat{\bz} \in \bbR^{k}$ and $\bomega \in \bbR^{p}$, where we used the existence condition $\mC$ that leads to $\partial_\omega f_{\out}|_\star = \partial_V f_{\out}|_\star = 0$. Finally, inserting eq.~\eqref{eq:lamp_uu:b_A_gamma_Lambda} in \eqref{eq:lamp_uu:v_z_g}, rewriting
the partial derivatives of $f_v$, $f_z$ and $f_{\out}$ at the fixed point $|_{\star}$ as moments of the distributions $\rP_\z$ and $\rQ_\out$ and simplifying the expression using the condition $\mC$, we finally obtain
\begin{align}
	\delta \hat{\vec{v}}^{t+1} =& \displaystyle  \frac{1}{\Delta} \rho_v \( \displaystyle 	 \frac{\mat{Y}}{\sqrt{p}}  \delta \hat{\vec{v}}^{t}  - \rho_v \mat{I}_{p} \delta\hat{\vec{v}}^{t-1} \)  + \rho_z^{-1} \EE_{\rQ_\out^0}[v x] \mat{I}_p \delta \bomega^{t} \nonumber \\
	& \hspace{2cm} + \frac{\EE_{\rQ_\out^0}[v x^2] \EE_{\rP_\z} \[z^3\] }{2\rho_z^3}  \frac{\vec{1}_p \vec{1}_k^\intercal }{k} \delta \hat{\bz}^{t} \label{eq:lamp_uu:v_cv_final} \,, \spacecase
    \delta \hat{\bz}^{t+1} =& \displaystyle \frac{1}{\Delta} \EE_{\rQ_\out^0}[v x]   \frac{\mat{W}^\intercal}{\sqrt{k}}     \[	  \frac{\mat{Y}}{\sqrt{p}}  \delta \hat{\vec{v}}^{t}  - \rho_v \mat{I}_p \delta \hat{\vec{v}}^{t-1} \]  \,,
	\label{eq:lamp_uu:z_cz_final}
		\spacecase
	\delta \bomega^{t+1} =& \displaystyle  \frac{1}{\Delta}  \(\displaystyle  \EE_{\rQ_\out^0}[v x]   \frac{\mat{W}\mat{W}^\intercal }{k}     \[	  \frac{\mat{Y}}{\sqrt{p}}  \delta \hat{\vec{v}}^{t}  - \rho_v \mat{I}_p \delta\hat{\vec{v}}^{t-1} \] \) \nonumber \\
	& \hspace{2cm} - \EE_{\rQ_\out^0}[v x] \[ \displaystyle	 \frac{\mat{Y}}{\sqrt{p}}  \delta \hat{\vec{v}}^{t-1}  -  \rho_v \mat{I}_p \delta\hat{\vec{v}}^{t-2} \] \,. \label{eq:lamp_uu:omega_V_final}
\end{align}

Inserting eq.~\eqref{eq:lamp_uu:z_cz_final}-\eqref{eq:lamp_uu:omega_V_final} in \eqref{eq:lamp_uu:v_cv_final} and dropping heuristically the time indices, we finally obtain the closed linear equation $\hat{\vec{v}}= \bGamma_{p}^{vv}\hat{\vec{v}}$, where the \aclink{LAMP} operator $\bGamma_{p}^{vv}$ is given by
\begin{align}
 \bGamma_p^{vv} = \frac{1}{\Delta}  \( (a-b) \mat{I}_p  +  b  \frac{\mat{W} \mat{W}^\intercal}{k} + c  \frac{\vec{1}_p \vec{1}_k^\intercal}{k} \frac{\mat{W}^\intercal}{\sqrt{k} }  \) \times \(\frac{\mat{Y}}{\sqrt{p}}  -  a \mat{I}_p  \) \, , \label{eq:spectral:gammas}
\end{align}
\noindent where the parameters are simply the moments of distributions $\rP_\z$ and $\rQ_\out^0$
\begin{align}
\begin{aligned}
		a &\equiv \EE_{\rQ_\out^0} [v^2] = \rho_v\,,\hhspace 
		b \equiv \rho_z^{-1} \EE_{\rQ_\out^0} [v x]^2 \,,\\
		c &\equiv \frac{1}{2} \rho_z^{-3} \EE_{\rP_\z} \[z^3\]  \EE_{\rQ_\out^0}[v x^2] \EE_{\rQ_\out^0}[v x]\,.
\end{aligned}
\end{align}
Note that in most of the cases we studied, the parameter $c$, taking into account the skewness of the variable $\bz$, is zero, simplifying considerably the structured matrix. 
Moreover, for the specific examples already discussed in \Sec\ref{sec:information_theory}, the \aclink{LAMP} operator $\bGamma_p^{vv}$ is very simple. For instance, for Gaussian $z$ and $\rP_{\out}(v|x) = \delta(v-\text{sign}(x))$, we
have $(a,b,c)=(1,2 / \pi,0)$. Instead, for linear activation we get $(a,b,c)=(1,1,0)$. Note that in this last case, the LAMP operator can be written as 
\begin{align}\label{eq:matrix_symmetric}
  \bGamma_p^{vv} &= \frac{1}{\Delta} \mat{K}_p \, \left[\frac{\mat{Y}}{\sqrt{p}}
                  - \mat{I}_p\right]~\text{with}~~\mat{K}_p=\frac{ \mat{W}
                  \mat{W}^\intercal}k = \bSigma  \approx \frac{1}{n}\sum_\alpha
                  \vec{v}^\alpha (\vec{v}^\alpha)^\intercal \, ,
\end{align}
\noindent or, in other words, $\mat{K}_p$ is the covariance matrix of the structured spike $\vec{v}$. The same observation holds for the sign activation function. Interestingly, the covariance matrix $\bSigma$ can be empirically estimated directly from samples of spikes, without the knowledge of the generative model $(\varphi, \mat{W})$ itself, suggesting a simple practical implementation of \aclink{LAMP}. 
Therefore we finally use a more generic definition for \aclink{LAMP} as expressed in \Alg\ref{main:lamp_algo}.
\begin{algorithm}[!htb]
\begin{algorithmic}
   \STATE {\bfseries Input:} Observed matrix $\mat{Y}\in \bbR^{p\times p}$, prior $\rP_\v$ on $\vec{v} \in \bbR^{p}$
   \STATE Take the leading eigenvector $\hat{\vec{v}} \in \bbR^{p}$ of $ \bGamma_p^{vv} \equiv  \mat{K}_p \, \left[\frac{\mat{Y}}{\sqrt{p}}
                  - \mat{I}_p\right]~\text{with}~~\mat{K}_p=\EE_{\rP_\v} \[ \vec{v} \vec{v}^\intercal \] \,.$
\end{algorithmic}
\caption{LAMP spectral algorithm for the Wigner model.}
\label{main:lamp_algo}
\end{algorithm}
From this perspective, \aclink{LAMP} in \Alg\ref{main:lamp_algo} can be interpreted as a \aclink{PCA} that takes into account the structure of the prior by incorporating the non-trivial correlations through $\mat{K}_p$ into the spectral estimation. In particular taking $\rP_\v(\vec{v}) =\mN_{\vec{v}}\(\vec{0}, \mat{I}_p\)$, we obtain $\bGamma_p^{vv} = \frac{1}{\Delta} \, \left[\frac{\mat{Y}}{\sqrt{p}}
                  - \mat{I}_p\right]$ and recognize the \aclink{PCA} operator that has been shifted.
Analogously to the state evolution for \aclink{AMP}, the asymptotic performance of both 
\aclink{PCA} and \aclink{LAMP} can be evaluated in a closed-form for the spiked Wigner model
with single-layer generative prior with linear activation. The
corresponding expressions are derived in the next section and plotted in
Fig.~\ref{main:bbp_lamp_amp_se} for the three considered algorithms.

\subsection{State evolution for LAMP and PCA in the linear case}
\label{appendix:subsec:SE_lamp}

As we have already mentioned in \Sec\ref{sec:amp:se}, one of the greatest virtues of \aclink{AMP} is being able to track its asymptotic performance through a set of simple scalar state evolution equations. Interestingly, we can also derive state evolution equations for the \aclink{LAMP} algorithm in the linear case. This allows a direct comparison between the performance of \aclink{LAMP} and the performance of \aclink{PCA}.

For the noiseless linear channel $\rP_{\out} (v|x) = \delta\(v - x\)$, the set of eqs.~(\ref{eq:lamp_uu:v_cv_final}-\ref{eq:lamp_uu:omega_V_final}) are already linear. Hence the \aclink{LAMP} spectral method flows directly from the \aclink{AMP} \Alg\ref{main:AMP_uv_bayes}. As a consequence, this means that the state evolution equations associated to the spectral method are simply dictated by the set of \aclink{AMP} \aclink{SE} equations eq.~\eqref{main:SE_AMP_uu}. 

However, it is worth stressing that as the \aclink{LAMP} returns a normalized estimator, the \aclink{LAMP} \aclink{MSE} is not given by the \aclink{AMP} mean squared error. We now compute the overlaps and mean squared error performed by this spectral algorithm.

Recall that $m_v$ and $q_v$ are the parameters defined in eq.~\eqref{appendix:amp:overlap_definition}, respectively measuring the overlaps between the ground truth $\vec{v}^\star$ and the estimator $\hat{\vec{v}}$, and the norm of the estimator. In the general case, the \aclink{MMSE} eq.~\eqref{eq:MMSE} becomes:
\begin{align*}
     {\textrm{MMSE}}_v &= \rho_v + \EE_{\vec{v}^\star} \lim_{p\to \infty} \frac{1}{p} \|\hat{\vec{v}}\|_2^2 - 2 \EE_{\vec{v}^\star} \lim_{p\to \infty} \frac{1}{p} \hat{\vec{v}}^\intercal \vec{v}^\star =\rho_v + q_v - 2 m_v \,,
\end{align*} 

However the \aclink{LAMP} spectral method computes the normalized leading eigenvector of the structured matrix $\bGamma_p^{vv}$. Hence the norm of the \aclink{LAMP} estimator is $\|\hat{\vec{v}}\|_{\textrm{LAMP}}^2 = q_{v,\textrm{LAMP}} = 1$, while the Bayes-optimal \aclink{AMP} estimator is not normalized with $\|\hat{\vec{v}}\|_{\textrm{AMP}}^2=q_{v,\textrm{AMP}}^\star=m_{v,\textrm{AMP}}^\star \ne 1$, solutions of eq.~\eqref{main:SE_AMP_uu}. As the non-normalized \aclink{LAMP} estimator follows \aclink{AMP} state evolutions in the \emph{linear case}, the overlap with the ground truth is thus given by:
\begin{align*}
\begin{aligned}
	m_{v,\textrm{LAMP}} &\equiv  \EE_{\vec{v}^\star} \lim_{p\to \infty} \frac{1}{p} \hat{\vec{v}}_{\textrm{LAMP}}^\intercal \vec{v}^\star  = \EE_{\vec{v}^\star} \lim_{p\to \infty} \frac{1}{p} \(\frac{\hat{\vec{v}}_{\textrm{AMP}}}{\|\hat{\vec{v}}\|_{\textrm{AMP}}}\)^\intercal \vec{v}^\star \\
	&= \frac{m_{v,\textrm{AMP}}^\star}{\(q_{v,\textrm{AMP}}^\star\)^{1/2}} = \(m_{v,\textrm{AMP}}^\star\)^{1/2}\,.
\end{aligned}
\end{align*}
Finally the mean squared error performed by the \aclink{LAMP} method is easily obtained from the optimal overlap reached by the \aclink{AMP} algorithm and yields
\begin{align*}
     {\textrm{MSE}}_{v,\textrm{LAMP}} = \rho_v + 1 - 2 \(q_{v,\textrm{AMP}}^\star\)^{1/2} \,.
\end{align*}

The respective result for \aclink{PCA} can be obtained from the observation that for the linear case, the $\alpha=0$ \aclink{LAMP} operator reduces exactly to the matrix $\mat{Y}$. In other words, in this case \aclink{LAMP} reduces to \aclink{PCA}. In terms of the prior, this is clear since $\alpha=0$ is equivalent to a separable Gaussian prior, for which the spectral algorithm derived from \aclink{AMP} is exactly given by \aclink{PCA} \cite{lesieur2017constrained}. Therefore we can simply state that the mean squared error performed by \aclink{PCA} is computed using the optimal overlap reached by \aclink{AMP} at $\alpha=0$:
\begin{align*}
     {\textrm{MSE}}_{v,\textrm{PCA}} = \rho_v + 1 - 2 \(q_{v,\textrm{AMP}}^\star|_{\alpha=0}\)^{1/2} \,.
\end{align*}
In order to fairly compare \aclink{PCA}, \aclink{LAMP} and \aclink{AMP} in Fig.~\ref{main:bbp_lamp_amp_se}, instead of showing the \aclink{MSE} corresponding to the \emph{normalized} \aclink{PCA} and \aclink{LAMP} estimators, we rescale these spectral estimators by the optimal normalisation $\(q_{v,\textrm{AMP}}^\star\)^{1/2}$ (obtained from \aclink{AMP} for instance) so that the renormalized \aclink{MSE} are given by
\begin{align*}
     {\textrm{MSE}}_{v,\textrm{LAMP}} &= \rho_v - m_{v,\textrm{LAMP}}^\star \,,  && {\textrm{MSE}}_{v,\textrm{PCA}} = \rho_v - m_{v,\textrm{PCA}}^\star \,. 
\end{align*}
Therefore in the \emph{linear case} we simply obtain that \aclink{LAMP} is strictly equivalent to \aclink{AMP}, while \aclink{PCA} is sub-optimal:
\begin{align*}
     {\textrm{MSE}}_{v,\textrm{LAMP}} &= \rho_v - q_{v,\textrm{AMP}}^\star \,,  && {\textrm{MSE}}_{v,\textrm{PCA}} = \rho_v - q_{v,\textrm{AMP}}^\star|_{\alpha=0} \,.
\end{align*}
\Fig~\ref{main:bbp_lamp_amp_se} shows good agreement between the state evolution for \aclink{LAMP} and \aclink{PCA} with linear activation (solid lines) and the respective finite instance numerical simulations (points). 

\subsection{A random matrix perspective on the recovery threshold}
Remarkably, the performance of the spectral method based on matrix (\ref{eq:matrix_symmetric}) can be
investigated independently of \aclink{AMP} using random matrix theory.
An analysis of the random matrix (\ref{eq:matrix_symmetric}) shows that a spectral phase transition for
generative prior with linear activation appears at 
$\Delta_c=1+\alpha$ (as for \aclink{AMP}).
This transition is analogous to the well-known \aclink{BBP}
transition \cite{baik2005phase}, but for a non-GOE random matrix
(\ref{eq:matrix_symmetric}). 
For the spiked Wigner models with linear generative
prior we prove two detailed theorems
describing the behavior of the supremum of the bulk spectral density, the transition of the largest eigenvalue 
and the correlation of the corresponding eigenvector. {\color{black}The theorems counterparts for the linear Wishart model are very similar, and are presented in appendix.}
We assume in the following that $\rho_v = 1$ to simplify the analysis (without any loss of generality).
Recall that we have 
\begin{align}\label{eq_app:gammap_vv}
	\bGamma^{vv}_p &\equiv \left[\frac{1}{k} \mat{W} \mat{W}^\intercal \right] \, \left[\frac{1}{\sqrt{\Delta p}} \bxi + \frac{1}{\Delta} \frac{\vec{v} \vec{v}^\intercal}{p} - \frac{1}{\Delta}\mat{I}_p\right].
\end{align}
Here $\bxi / \sqrt{p}$ is a matrix from the Gaussian Orthogonal Ensemble, i.e. $\bxi$ is a real symmetric matrix with entries drawn independently from a Gaussian distribution with zero mean and variance
$\EE \, \xi_{ij}^2 = (1+\delta_{ij})$.
\begin{thm}[Bulk of the spectral density, spiked Wigner, linear activation]
	\label{thm:main_lambdamax_uu_case}
	For any $\alpha, \Delta > 0$, as $p \to +\infty$, the spectral measure of $\bGamma_p^{vv}$ converges almost surely and in the weak sense 
	to a well-defined and compactly supported probability measure $\mu(\alpha, \Delta)$, and we denote $\mathrm{supp}\, \mu$ its support.	
	We separate two cases:
	\begin{itemize}
		\item[$(i)$] If $\Delta \leq \frac{1}{4}$, then $\mathrm{supp}\, \mu \subseteq \bbR_-$.
		\item[$(ii)$] Assume now  $\Delta > \frac{1}{4}$ and denote $z_1(\Delta) \equiv -\Delta^{-1} + 2 \Delta^{-1/2} > 0$.
		Let $\rho_\Delta$ be the probability measure on $\bbR$ with density
		\begin{align}\label{eq_app:def_rho_Delta}
			\rho_\Delta(\mathrm{d}t) &= \frac{\sqrt{\Delta}}{2\pi} \sqrt{4 - \Delta \left(t+\frac{1}{\Delta} \right)^2} \mathds{1}\left\{\left|t+\frac{1}{\Delta}\right| \leq \frac{2}{\sqrt{\Delta}}\right\} \, \mathrm{d}t.
		\end{align}
		Note that the supremum of the support of $\rho_\Delta$ is $z_1(\Delta)$.
		The following equation admits a unique solution for $s \in (-z_1(\Delta)^{-1},0)$:
		\begin{align}\label{eq_app:se_vv}
			\alpha \int \rho_\Delta(\mathrm{d}t) \left(\frac{st}{1+ st}\right)^2 &= 1.
		\end{align}
		We denote this solution as $s_{\textrm{edge}}(\alpha, \Delta)$ (or simply $s_{\textrm{edge}}$).
		The supremum of the support of $\mu(\alpha,\Delta)$ is denoted $\lambda_{\textrm{max}}(\alpha, \Delta)$ (or simply $\lambda_{\textrm{max}}$).
		It is given by:
		\begin{align}
			\lambda_{\textrm{max}} &=
			\begin{dcases}
				-\frac{1}{s_{\textrm{edge}}} + \alpha \int \rho_\Delta(\mathrm{d}t) \frac{t}{1 + s_{\textrm{edge}}t} \quad & \text{ if } \alpha \leq 1,\\
				\max\left(0,-\frac{1}{s_{\textrm{edge}}} + \alpha \int \rho_\Delta(\mathrm{d}t) \frac{t}{1 + s_{\textrm{edge}}t} \right)\quad & \text{ if } \alpha > 1.
			\end{dcases}
		\end{align}
	\end{itemize}
	As a function of $\Delta$, $\lambda_{\textrm{max}}$ has a unique global maximum, reached exactly at the point $\Delta = \Delta_c(\alpha) = 1 + \alpha$.
	Moreover, $\lambda_{\textrm{max}}(\alpha, \Delta_c(\alpha)) = 1$.
\end{thm}

\begin{thm}[Transition of the largest eigenvalue and eigenvector,
  spiked Wigner, linear activation]
	\label{thm:main_transition_uu_case}
	Let $\alpha > 0$.
	We denote $\lambda_1 \geq \lambda_2$ the first and second eigenvalues of $\bGamma^{vv}_p$.
	\begin{itemize}
		\item If $\Delta \geq \Delta_c(\alpha)$, then as $p \to \infty$ we have a.s.
$\lambda_1 {\to} \lambda_{\textrm{max}}$ and $\lambda_2
{\to} \lambda_{\textrm{max}}$.
		\item If $\Delta \leq
\Delta_c(\alpha)$, then as $p \to \infty$ we have a.s. $\lambda_1
{\to} 1$ and $\lambda_2 {\to}
\lambda_{\textrm{max}}$.
	\end{itemize}
        Further, denoting $\tilde{\vec{v}}$ a normalized
($\norm{\tilde{\vec{v}}}_2^2 = p$ ) eigenvector of $\bGamma^{vv}_p$ with
eigenvalue $\lambda_1$, then $|\tilde{\vec{v}}^\intercal \vec{v}^\star |^2
/p^2 {\to} \epsilon(\Delta)$ a.s., where 
$\epsilon(\Delta) = 0$ for all $\Delta \geq \Delta_c(\alpha)$,
		$\epsilon(\Delta) > 0$ for all $\Delta < \Delta_c(\alpha)$ and $\lim_{\Delta \to 0}\epsilon(\Delta) =1$.
\end{thm}

Thm.~\ref{thm:main_lambdamax_uu_case} and
Thm.~\ref{thm:main_transition_uu_case} are illustrated in
Fig.~\ref{main:bbp_lamp_amp_se}. 
The proof gives the value of $\epsilon(\Delta)$, which turns out to
lead to the same \aclink{MSE} as in \Fig\ref{main:bbp_lamp_amp_se} in the linear case. The proofs of theorems~\ref{thm:main_lambdamax_uu_case} and \ref{thm:main_transition_uu_case} are left in \cite{aubin2019spiked}, along with the precise arguments used to derive the 
eigenvalue density, the transition of $\lambda_1$ and the computation of $\epsilon(\Delta)$. These arguments are solely based on random matrix theory. The method of proof of Theorem~\ref{thm:main_transition_uu_case} is very much inspired by \cite{benaych2011eigenvalues}, and allows us to 
compute numerically the squared correlation $\epsilon(\Delta)$.
Note that while all the calculations are justified, refinements would be needed in order to be completely rigorous.
These refinements would follow exactly some proofs of \cite{silverstein1995empirical} and \cite{benaych2011eigenvalues}, 
so we will refer to them when necessary.
A Mathematica demonstration notebook is provided in the \href{https://github.com/sphinxteam/StructuredPrior_demo}{GitHub repository \cite{StructuredPrior_demo_repo}}.

In the non-linear case the random matrix analysis is harder to perform. In the matrix $\bGamma_p^{vv}$, the Wishart matrix $\mat{W}\mat{W}^\intercal/k$ is replaced by $a\mat{I} + b \mat{W}\mat{W}^\intercal$ with $a,b\geq 0$. It is thus not possible to relate the spectrum of $\bGamma_p^{vv}$ to the one of a symmetric matrix of the type $\mat{W} \mat{Z} \mat{W}^\intercal$ with $\mat{W}$ a gaussian \aclink{i.i.d} matrix. Some techniques from free probability could make the computation nevertheless possible, but we leave this analysis for future work. 

\subsection{Applying LAMP to real data}
As we have already remarked, the \aclink{LAMP} operator in eq.~\eqref{eq:matrix_symmetric} only depend on the generative prior through its covariance. An interesting exercise is to apply \aclink{LAMP} for real data by simply using the empirical covariance for $n$ samples of the spikes, $\vec{v}^\alpha$, $\alpha=1,\dots,n$. 

For illustration, we perform the following experiment: the spikes $\vec{v}^{\star}$ are drawn from the standard Fashion-MNIST dataset \cite{fashionmnist2017}, and are used to generate the spiked matrix $\mat{Y}$ according to eq.~\eqref{Wigner}. We then apply our \aclink{LAMP} algorithm to reconstruct the spikes, repeating this experiment for different values of noise $\Delta$. In \Fig\ref{main:experiement_mnist} we compare the reconstruction by \aclink{LAMP} with standard \aclink{PCA} over $\mat{Y}$. In principle, we have no theoretical guarantees about the performance of \aclink{LAMP}, since the Fashion-MNIST images are not drawn from the generative class studied above. Nevertheless, it is striking to observe that \aclink{LAMP} outperforms \aclink{PCA}.

A demonstration notebook illustrating this experiment is provided in the  \href{https://github.com/sphinxteam/StructuredPrior_demo}{GitHub repository} \cite{StructuredPrior_demo_repo}.

\begin{figure}[tb!]
\centering
	\includegraphics[width=1.0\linewidth]{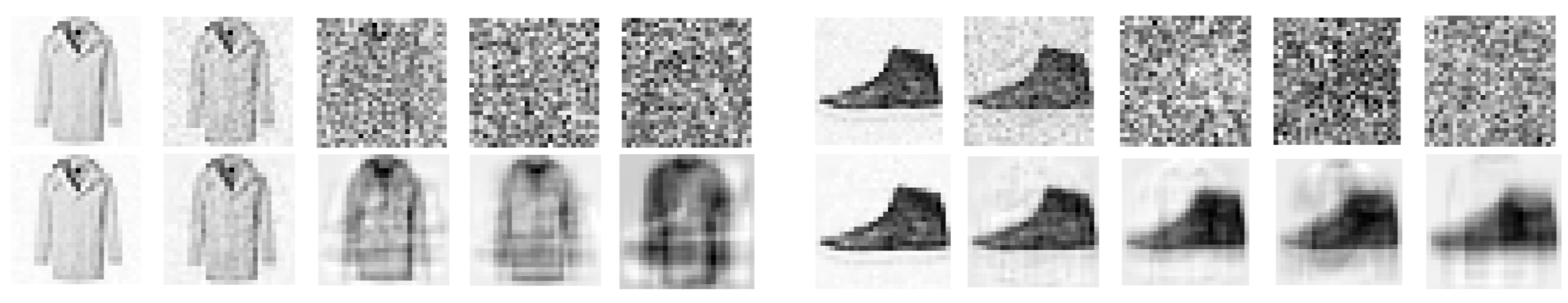}
	\caption{Illustration of canonical PCA (top line) and the LAMP
          (bottom line)
          spectral methods (\ref{eq:matrix_symmetric}) on the spiked
          Wigner model. The covariance $\bSigma$ is estimated
          empirically from the FashionMNIST database \cite{fashionmnist2017}. The
          estimation of the spike is shown for two images from
          FashionMNIST, with           (from left to right), noise variance $\Delta=0.01,0.1,1,2,10$.}
	\label{main:experiement_mnist}
\end{figure}

	\ifthenelse{\equal{\format}{oneside}}
	{
	\clearpage\null\thispagestyle{empty}\newpage
	}
	{
	\clearpage\null\thispagestyle{empty}\newpage
	\cleardoublepage}
	
	\chapter{Exact asymptotics for phase retrieval and compressed sensing with random generative priors}
	\chaptermark{Phase retrieval with random generative priors}
	\label{chap:generative_phase}
	Over the past decade the study of compressed sensing has lead to  significant developments in the field of signal processing, with novel sub-Nyquist sampling strategies and a veritable explosion of work in sparse representation. A central observation is that sparsity allows one to measure the signal with fewer observations than its dimension \cite{donoho2006compressed,candes2006near}.
The success of neural networks in the recent years suggests another powerful and generic way of representing signals with multi-layer generative priors, such as those used in generative
adversarial networks \aclink{GAN} \cite{goodfellow2014generative} and \aclink{VAE}. It is therefore natural to replace sparsity by generative neural network models in compressed sensing and other inverse problems, a strategy that was successfully explored in a number of papers, e.g. \cite{tramel2016approximate,tramel2016inferring,bora2017compressed,manoel2017multi,hand2017global,fletcher2018inference,hand2018phase,mixon2018sunlayer,aubin2019spiked}.
While this direction of research seems to have many promising applications, a systematic theory of what can be efficiently achieved still falls short of the one developed over the past decade for sparse signal processing. 
Our aim is therefore to dialogue with the broad program of studying how generative models can help solving inverse problems using the toolbox of statistical physics.
In this chapter, we build on a line of work allowing for theoretical analysis in the case the measurement and the weight matrices of the prior are random \cite{manoel2017multi,reeves2017additivity,fletcher2018inference,gabrie2018entropy,aubin2019spiked} similarly to \Chap\ref{chap:generative_spiked}.

We employ tools originally developed in the context of statistical physics to derive precise asymptotics for the information-theoretically optimal thresholds for signal recovery and for the performance of the best known polynomial algorithm in two such inverse problems: (real-valued) phase retrieval and compressed sensing. These two problems of interest can be framed as a \emph{generalized linear estimation}. Given a set of observations $\vec{y}\in\mathbb{R}^{n}$ generated
from a fixed (but unknown) signal $\vec{x}^{\star}\in\mathbb{R}^{d}$ as 
\begin{equation}
\vec{y}=\varphi\left(\mat{A}\vec{x}^{\star}\right),
\end{equation}
the goal is to reconstruct $\vec{x}^{\star}$ from the knowledge of $\vec{y}$, $\varphi$ and $\mat{A}\in\bbR^{n \times d}$. \aclink{CS} and \aclink{PR} are particular instances of this problem, corresponding to $\varphi(x)= x$ and $\varphi(x) = |x|$ respectively. Two key questions in these inverse problems are a) how many observations $n$ are required for theoretically reconstructing the signal $\vec{x}^{\star}$, and b) how this can be done in practice - \ie. to find an efficient algorithm for reconstruction. Signal structure plays an important role in the answer to both these questions, and have been the subject of intense investigation in the literature. A typical situation is to consider signals admitting a low-dimensional representation, such as sparse signals, for which $k-d$ of the $d$ components of $\vec{x}^{*}$ are exactly zero, see e.g. \cite{candes2015phase, netrapalli2013phase}.

In this work, we consider instead structured signals drawn from a generative model $\vec{x}^{\star} = G(\vec{z})$, where $\vec{z}\in\mathbb{R}^{k}$ is a low-dimensional latent representation of $\vec{x}^{\star}$. In particular, we will focus in generative multi-layer neural networks, and in order to provide a sharp asymptotic theory, we will restrict the analysis to an ensamble of random networks with known random weights:
\begin{equation}
\vec{x}^{\star} = G\left(\vec{z}\right) = \sigma^{(L)}\left(\mat{W}^{(L)}\sigma^{(L-1)}\left(\mat{W}^{(L-1)}\cdots\sigma^{(1)}\left(\mat{W}^{(1)}\vec{z}\right)\cdots\right)\right), 
\label{eq:generative_model}
\end{equation}
\noindent where $\sigma^{(l)}:\mathbb{R}\to\mathbb{R}$, $1\leq l\leq L$ are component-wise non-linearities. As aforementioned, we take $\mat{A}\in\mathbb{R}^{n\times d}$ and ${\mat{W}}^{(l)}\in\mathbb{R}^{k_{l}\times k_{l-1}}$ to have \aclink{i.i.d} Gaussian entries with zero means and variances $1/d$ and $1/k_{l-1}$ 
respectively, and focus on the high-dimensional regime defined by taking $n, d, k_{l} \to \infty$ while keeping the measurement rate $\alpha = n/d$ and the layer-wise aspect ratios $\beta_{l} = k_{l+1}/k_{l}$ constant. We stress that in this regime the depth $L$ is of order one when compared to the width of the generative network, which scales with the input dimension $d$. With this observation in mind, we adopt the standard terminology in machine learning of denoting networks with $L>1$ as \emph{deep}.
To provide a comparison with previous results for sparse signals, it is useful to define the total compression factor $\rho = k/d$. We note, however, that the comparison between generative and sparse priors herein is not based on a quantitative comparison between the reconstruction estimation errors. Indeed, since data is generated differently in both cases, such a comparison would make little sense. Instead, we compare qualitative properties of the phase diagrams, taking as a surrogate for algorithmic hardness the size of the statistical-to-algorithmic gap in these two different reconstruction problems.
Our results hold for latent variables drawn from an arbitrary separable distribution $\vec{z}\sim \rP_\z$, and for arbitrary activations $\sigma^{(l)}$, although for concreteness we present results for $\vec{z}\sim \mathcal{N}(\vec{0},\mat{I}_{k})$ and $\sigma^{(l)}\in\{\text{linear}, \text{ReLU}\}$, as it is commonly the case in practice with \aclink{GAN} or \aclink{VAE}. 

\paragraph{Previous results on sparsity:} Sparsity is probably the most widely studied type of signal structure in linear estimation and phase retrieval. It is thus instructive to recall the main results for sparse signal reconstruction in these inverse problems in the high-dimensional regime with random measurement matrices studied in this manuscript. Optimal statistical and algorithmic thresholds have been established non-rigorously using the replica-method in a series of works \cite{wu2012optimal,krzakala2012probabilistic,reeves2012compressed,zdeborova2016statistical}. Later the information theoretic results, as well as the corresponding \aclink{MMSE}, has been rigorously proven in \cite{barbier2016mutual,Reeves,barbier2017phase}. So far, the best known polynomial time algorithm in this context is the \aclink{AMP} algorithm, the new avatar of the mean-field approach pioneered in statistical mechanics \cite{mezard1987spin}, that has been introduced in \cite{donoho2009message,rangan2011generalized, krzakala_statistical-physics-based_2012,schniter2014compressive,metzler2017coherent} for these problems, and can be rigorously analyzed \cite{bayati2011dynamics}. For both (noiseless) compressed sensing and phase retrieval, the information theoretic limit for a perfect signal recovery is given by $\alpha>\alpha_{\textrm{IT}}=\rho_s$, with $\rho_s$ being the fraction of non-zero components of the signal~$\vec{x}^{\star}$.

The ability of \aclink{AMP} to exactly reconstruct the signal, however, is different. A non-trivial line $\alpha^{\textrm{sparse}}_{\textrm{alg}}(\rho_s)>\alpha_{\textrm{IT}}$ appears below which \aclink{AMP} fails. No polynomial algorithm achieving better performance for these problems is known. Strikingly, as discussed in \cite{barbier2017phase}, the behaviour of the sparse linear estimation and phase retrieval is drastically different: while $\alpha^{\textrm{sparse}}_{\textrm{alg}}(\rho_s)$ is going to zero as $\rho_s\to 0$ for sparse linear estimation hence allowing for compressed sensing, it is not the case for the phase retrieval, for which  $\alpha^{\textrm{sparse}}_{\textrm{alg}} \to 1/2$ as  $\rho_s\to 0$. As a consequence, \emph{no efficient approach to real-valued compressed phase retrieval with small but order one $\rho_s$ in the high-dimensional limit is known}. 

\paragraph{Summary of results:}
In this work, we replace the sparse prior by the multi-layer generative model introduced in eq.~\eqref{eq:generative_model}. Our main contribution is specifying the interplay between the number of measurements needed for exact reconstruction of the signal, parametrized by $\alpha$, and its latent dimension~$k$. Of particular interest is the comparison between a sparse and separable signal (having a fraction $\rho_s$ of non-zero components) and the structured generative model above, parametrized by $\rho=k/d$. While the number of unknown latent variables is the same in both cases if $\rho=\rho_s$, the upshot is that generative models offer algorithmic advantages over sparsity. More precisely:
\begin{enumerate}
	\item We analyze the \aclink{MMSE} of the optimal Bayesian estimator for the compressed sensing and phase retrieval problems with generative priors of arbitrary depth, choice of activation and prior distribution for the latent variable. We derive sufficient conditions for the existence of an \emph{undetectable phase} in which better-than-random estimation of $\vec{x}^{\star}$ is impossible, and characterize in full generality the threshold $\alpha_{c}$ beyond which partial signal recovery becomes statistically possible. 
	\item Fixing our attention on the natural choices of activations $\sigma\in\{\text{linear}$, \aclink{ReLU}$\}$, we establish the threshold $\alpha_{\IT}$ above which perfect signal reconstruction is theoretically possible. This threshold can be intuitively understood with a simple counting argument.
	\item We analyze the performance of the associated \aclink{AMP} algorithm \cite{manoel2017multi}, conjectured to be the best known polynomial time algorithm in this setting. This allows us to establish the algorithmic threshold $\alpha_{\alg}$ below which no known algorithm is able to perfectly reconstruct $\vec{x}^{\star}$. 	 
\end{enumerate}
As expected, the thresholds $\{\alpha_{c}, \alpha_{\IT}, \alpha_{\alg}\}$ are functions of the compression factor $\rho$, the number of layers $L$, the aspect ratios $\{\beta_{l}\}_{l=1}^L$ and the activation functions. In particular, for a fixed architecture we find that the algorithmic gap $\Delta_{\alg} = \alpha_{\alg} - \alpha_{\IT}$ is drastically reduced with the depth $L$ of the generative model, beating the algorithmic hindrance identified in \cite{barbier2017phase} for compressive phase retrieval with sparse encoding.

\section{Information theoretical analysis}
\label{sec:it_analysis}

\subsection{Performance of the Bayes-optimal estimator}
In our analysis we assume that the model generating the observations $\vec{y}\in\mathbb{R}^{n}$ is known. Therefore, the optimal estimator minimizing the mean-squared-error in our setting is given by the Bayesian estimator
\begin{align}
\hat{\vec{x}}^{\text{opt}} = \underset{\hat{\vec{x}}}{\text{argmin}}||\hat{\vec{x}}-\vec{x}^{\star}||^2_2 = \mathbb{E}_{\rP(\vec{x}|\vec{y})}\[ \vec{x}\]\,.
\label{eq:optimal_estimator}
\end{align}
The posterior distribution of the signal given the observations is in general given by:
\begin{align}
	\rP(\vec{x}|\vec{y}) &= \frac{1}{\mZ_\ndim(\vec{y})}\rP_{\x}(\vec{x})\prod\limits_{\mu=1}^{n}\delta\left(y^{\mu}-\varphi\left(\sum\limits_{j=1}^{d} a^{\mu}_{j}x_j\right)\right),
\end{align}
\noindent where the normalization $\mZ_\ndim(\vec{y})$ is known as the \emph{partition function}, and $\varphi$ is the nonlinearity defining the estimation problem, e.g. $\varphi(x) = |x|$ for phase retrieval and $\varphi(x) = x$ for linear estimation. We note that the presented approach generalizes straightforwardly to account for the presence of noise, but we focus in this work on the analysis of the noiseless case. For the generative model in eq.~\eqref{eq:generative_model}, the prior distribution $\rP_\x$ reads
\begin{align}
\rP_\x(\vec{x}) = \int_{\mathbb{R}^{k}}\dd \vec{z}~ \rP_\z(\vec{z})\prod\limits_{l=1}^{L}\int_{\mathbb{R}^{k_{l}}}\dd \vec{h}^{(l)}~\rP^{(l)}_{\out}\left(\vec{h}^{(l+1)}\Big|\mat{W}^{(l)}\vec{h}^{(l)}\right)\,,
\label{eq:Px}
\end{align}
\noindent where for notational convenience we denoted $\vec{x} \equiv \vec{h}^{(L+1)}$, $\vec{z} \equiv \vec{h}^{(1)}$ and defined the likelihoods $\rP^{(l)}_{\out}$ parametrising the output distribution of each layer given its input. As before, this Bayesian treatment also accounts for stochastic activation functions, even though we focus here on deterministic ones.

Although exact sampling from the posterior is intractable in the high-dimensional regime, it is still possible to track the behavior of the minimum-mean-squared-error estimator as a function of the model parameters. Our main results are based on the line of works comparing, on one hand, the information-theoretically best possible reconstruction, analyzing the ideal Bayesian inference decoder, regardless of the computation cost, and on the other, the best reconstruction using the most efficient known polynomial algorithm - the approximate message passing.

Our analysis builds upon the statistical physics inspired multi-layer formalism introduced in \cite{manoel2017multi}, who showed using the cavity and replica methods that the minimum mean-squared-error achieved by the Bayes-optimal estimator defined in eq.~\eqref{eq:optimal_estimator} can be written, in the limit of $n,d \to \infty$ and $\alpha=n/d=\Theta(1)$ for a generic prior distribution $\rP_\x$ as
\begin{align}
    \text{mmse}(\alpha) = \lim\limits_{d\to\infty}\frac{1}{d}\mathbb{E}||\hat{\vec{x}}^{\text{opt}}-\vec{x}^{\star}||^{2}_{2} = \rho_{x} - q_x^{\star}
    \label{eq:mmse}
\end{align}
\noindent where $\rho_{x}$ is the second moment of $\rP_\x$ and the scalar parameter $q_{x}^{\star}\in[0,\rho_{x}]$ is the solution of the following \emph{free energy} extremisation problem
\begin{align}
   \Phi = -\lim\limits_{d\to\infty}\frac{1}{d}\mathbb{E}_{y}\log\mZ_\ndim(\vec{y}) =   \underset{q_{x},\hat{q}_{x}}{\extr}\left\{\frac{1}{2}\hat{q}_{x}q_{x}-\alpha\Psi_{y}\left(q_{x}\right)-\Psi_{x}(\hat{q}_{x})\right\}\,,\label{eq:freeen}
\end{align}
\noindent with the so-called potentials $(\Psi_{y}, \Psi_{x})$ given by
\begin{align}
\begin{aligned}
\Psi_{y}(t) &= \mathbb{E}_{\xi}\left[\int_\bbR \dd y~\mathcal{Z}_{y}\left(y;\sqrt{t}\xi, t\right)\log\mathcal{Z}_{y}\left(y;\sqrt{t}\xi, t\right)\right]\, ,\\
\Psi_{x}(r) &= \lim\limits_{d\to\infty}\frac{1}{d} ~\mathbb{E}_{\xi}\left[\mathcal{Z}_{x}(\sqrt{r}\xi,r)\log{\mathcal{Z}_{x}(\sqrt{r}\xi,r)}\right]\, ,
\end{aligned}
\end{align}
\noindent where $\xi\sim\mathcal{N}(0,1)$ and $\mathcal{Z}_{y}$, $\mathcal{Z}_{x}$ are the normalizations of the auxiliary distributions
\begin{align}
\rQ_\y\left(x; y,\omega,V\right) &= \frac{1}{\mathcal{Z}_{y}(y;\omega, V)}\frac{e^{-\frac{1}{2V}(x-\omega)^2}}{\sqrt{2\pi V}}\delta\left(y-\varphi(x)\right)\,, \\
\rQ_\x\left(\vec{x};b,A\right) &=\frac{\rP_\x(\vec{x})}{\mathcal{Z}_{x}(b,A)} e^{-\frac{A}{2}x_{j}^2+b x_{j}}\,. \nonumber
\end{align}
Note that this expression is valid for arbitrary distribution $\rP_\x$, as long as the limit in $\Psi_{x}$ is well-defined. In particular, it reduces to the known result in \cite{krzakala2012probabilistic,barbier2017phase} when $\rP_\x$ factorizes. In principle, for correlated $\rP_x$ such as in the generative model of eq.~\eqref{eq:Px} computing $\Psi_{x}$ is itself a hard problem. However, we can see eq.~\eqref{eq:Px} as a chain of generalized linear models. In the limit where $k_{l}\to\infty$ with $\rho = k/d = \Theta(1)$, $L= \Theta(1)$ and $\beta_{l}=k_{l+1}/k_{l} = \Theta(1)$ we can apply the observation above iteratively, layer-wise, up to the input layer for which $\rP_\z$ factorizes - and is easy to compute. This yields \cite{manoel2017multi}
\begin{multline}
\Phi = \underset{q_{x},\hat{q}_{x}, \{q_{l},\hat{q}_{l}\}}{\extr}\left\{-\frac{1}{2}\hat{q}_{x}q_{x}-\frac{\rho}{2}\sum\limits_{l=1}^{L}\beta_{l}q_{l}\hat{q}_{l}+\alpha\Psi_{y}\left(q_{x}\right) \right. \\ \left. + \rho\sum\limits_{l=2}^{L}\beta_{l}\Psi^{(l)}_{\out}\left(\hat{q}_{l}, q_{l-1}\right)+\Psi^{(L+1)}_{\out}(\hat{q}_{x}, q_{L})+\rho\Psi_{z}\left(\hat{q}_{z}\right)\right\}\label{eq:freeen:multilayer},
\end{multline}
\noindent where we have introduced the additional potentials $(\Psi_{\out}, \Psi_{z})$
\begin{align}
\begin{aligned}
\Psi^{(l)}_{\out}(r,s) &= \mathbb{E}_{\xi,\eta}\left[\mathcal{Z}^{(l)}_{\out}(\sqrt{r}\xi,r,\sqrt{s}\xi, \rho_{l-1}-s) \right. \\
&\left. \hspace{2cm} \log{\mathcal{Z}^{(l)}_{\out}(\sqrt{r}\xi,r,\sqrt{s}\xi, \rho_{l-1}-s)}\right]\,, \\
\Psi_{z}(t) &= \mathbb{E}_{\xi}\left[\mathcal{Z}_{z}(\sqrt{t}\xi,t)\log{\mathcal{Z}_{z}(\sqrt{t}\xi,t)}\right]\,,
\end{aligned}
\end{align}
\noindent defined in terms of the following auxiliary distributions
\begin{align}
\begin{aligned}
\rQ^{(l)}_{\out}(x,z;b,A,\omega,V) &= \frac{e^{-\frac{A}{2}x^2+b x}}{\mathcal{Z}_{\out}(b,A,\omega,V)} \frac{e^{-\frac{1}{2V}\left(z-\omega\right)^2}}{\sqrt{2\pi V}}\rP^{(l)}_{\out}(x|z)\,,\\
Q_{\z}\left(z; b,A\right) &= \frac{e^{-\frac{A}{2}z^2+b z}}{\mathcal{Z}_{z}(b,A)}~\rP_\z(z)\,,
\end{aligned}	
\end{align}
\noindent and with $\rho_{l}$ the second moment of the hidden variable $\vec{h}^{(l)}$.

These predictions, that have also been derived with different heuristics in \cite{reeves2017additivity}, were rigorously proven for two-layers in \cite{gabrie2018entropy}, while deeper architectures requires additional assumptions on the concentration of the free energies to be under a rigorous control. Eq.~\eqref{eq:freeen:multilayer} thus reduces the asymptotics of the high-dimensional estimation problem to a low-dimensional extremisation problem over the $2(L+1)$ variables $(q_{x}, \hat{q}_{x}, \{q_{l},\hat{q}_{l}\}_{l=1}^L)$, allowing for a mathematically sound and rigorous investigation. These parameters are also known as the \emph{overlaps}, since they parametrize the overlap between the Bayes-optimal estimator and ground-truth signal at each layer. Solving eq.~\eqref{eq:freeen} provides two important statistical thresholds: the \emph{weak recovery} threshold $\alpha_{c}$ above which better-than-random (i.e. ${\textrm{mmse}} < \rho_{x}$) reconstruction becomes theoretically possible and the \emph{perfect reconstruction} threshold, above which perfect signal recovery (i.e. when ${\textrm{mmse}} = 0$) becomes possible.

Interestingly, the free energy eq.~\eqref{eq:freeen:multilayer} also provides information about the algorithmic hardness of the problem. The above extremisation problem is closely related the state evolution of the \aclink{AMP} algorithm for this problem, as derived in \cite{manoel2017multi}, and generalized in \cite{fletcher2018inference}. It is conjectured to provide the best polynomial time algorithm for the estimation of $\vec{x}^{\star}$ in our considered setting. Specifically, the algorithm reaches a mean-squared error that corresponds to the local extremiser reached by gradient descent in the function (\ref{eq:freeen:multilayer}) starting with uninformative initial conditions.  

While so far we summarized results that follow from previous works, these results were up to our knowledge not systematically evaluated and analyzed for the linear estimation and phase retrieval with generative priors. This analysis and its consequences is the object of the rest of this work and constitutes the original contributions of this work. 

\subsection{Weak recovery threshold}
Solutions for the extremisation in eq.~\eqref{eq:freeen:multilayer} can be found by solving the fixed point equations, obtained by taking the gradient of eq.~\eqref{eq:freeen:multilayer} with respect of the parameters $(q_{x},\hat{q}_{x}, \{q_{l}, \hat{q}_{l}\}_{l=1}^{L})$:
\begin{align}
\begin{cases}
\hat{q}_{x} =\alpha\Lambda_{y}\left(q_{x}\right)	\\
\hat{q}_{L} =\beta_{L}\Lambda_{\out}\left(\hat{q}_{x}, q_{L}\right)	\\
\hat{q}_{L-1} =\beta_{L-1}\Lambda_{\out}\left(\hat{q}_{L}, q_{L-1}\right)\\
\hspace{1cm}\vdots\\
\hat{q}_{l} =\beta_{l} \Lambda_{\out}\left(\hat{q}_{l+1}, q_{l}\right)\\
\hspace{1cm}\vdots\\
\hat{q}_{z} =\beta_{1} \Lambda_{\out}\left(\hat{q}_{2},q_{z}\right)
\end{cases}\,, &&
\begin{cases}
	q_{x} = \Lambda_{x}\left(\hat{q}_{x}, q_{L}\right)\\
    q_{L} = \Lambda_{x}\left(\hat{q}_{L}, q_{L-1}\right)\\
\hspace{1cm}\vdots\\
{q}_{l} = \Lambda_{x}\left(\hat{q}_{l},q_{l-1}\right) \\
\hspace{1cm}\vdots\\
q_{z} =\Lambda_{z}\left(\hat{q}_{z}\right)
\end{cases}
\,,
	\label{eq:SE_uu}
\end{align}
\noindent where $\Lambda_{y}(t) = 2~\partial_{t}\Psi_{y}(t)$, $\Lambda_{z}(t) = 2~\partial_{t}\Psi_{z}(t)$, $\Lambda_{x}(t) = 2~\partial_{r}\Psi_{\out}(r,s)$, $\Lambda_{\out}(t)$ $= 2~\partial_{s}\Psi_{\out}(r,s)$.
The weak recovery threshold $\alpha_{c}$ is defined as the value above which one can estimate $\vec{x}^{\star}$ better than a random draw from the prior $\rP_\x$. In terms of the \aclink{MMSE} it is defined as
\begin{align}
\alpha_{c} = \underset{\alpha\geq 0}{\textrm{argmax}	}\{{\textrm{mmse}}(\alpha)=\rho_{x}\}.
\end{align}
From eq.~\eqref{eq:mmse}, it is clear that an uninformative solution ${\textrm{mmse}}=\rho_{x}$ of eq.~\eqref{eq:freeen:multilayer} corresponds to a fixed point $q_{x} = 0$. For both the phase retrieval and linear estimation, evaluating the right-hand side of eqs.~\eqref{eq:SE_uu} at $q_{x}=0$ we can see that $\hat{q}^{\star}_{x}=0$ is a fixed point if $\sigma$ is an odd function and if
\begin{align}
\mathbb{E}_{\rP_\z} \[ z \] = 0, \hspace{ 1.5cm }\textrm{ and }\hspace{ 1.5cm } \mathbb{E}_{\rQ^{(l), 0}_{\out}}\[ x \] = 0 \,,
\label{eq:stability_conditions} 	
\end{align}
\noindent where $\rQ^{(l), 0}_{\out}(x,z) = \rQ^{(l)}_{\out}(x,z;0,0,0,\rho_{l-1})$. These conditions reflect the intuition that if the prior~$\rP_\z$ or the likelihoods $\rP_{\out}^{(l)}$ are biased towards certain values, this knowledge helps the statistician estimating better than a random guess. If these conditions are satisfied, then $\alpha_{c}$ can be obtained as the point for which the fixed point $q_{x}=0$ becomes unstable. The stability condition is determined by the eigenvalues of the Jacobian of eqs.~\eqref{eq:SE_uu} around the fixed point $(q^{\star}_{x}, \hat{q}^{\star}_{x}, \{q^{\star}_{l}, \hat{q}^{\star}_{l}\}_{l=1}^L) = 0$. More precisely, the fixed point becomes unstable as soon as one eigenvalue of the Jacobian is bigger than one. Expanding the update functions around the fixed point and using the conditions in eq.~\eqref{eq:stability_conditions},
\begin{align}
\Lambda_{y}(t) &\underset{t\ll 1}{=} \frac{1}{\rho^2_{x}}\int\dd y~\mathcal{Z}_{y}(y;0,\rho_{x})\left(\mathbb{E}_{\rQ^{0}_{y}}[\rho_{x}-x^2]\right)^2 t+\Theta \left(t^{3/2}\right) \notag \,,\\
\Lambda^{(l)}_{x}(r,s) &\underset{r,s\ll 1}{=} \left(\mathbb{E}_{\rQ^{(l), 0}_{\out}}[x^2]\right)^2~r+\frac{1}{\rho_{l-1}^2}\left(\mathbb{E}_{\rQ^{(l), 0}_{\out}}[xz]\right)^2 s+\Theta \left(r^{3/2},s^{3/2}\right)\,, \\
\Lambda^{(l)}_{\out}(r,s) &\underset{r,s\ll 1}{=}\left(\mathbb{E}_{\rQ^{(l), 0}_{\out}}[xz]\right)^2  r+\frac{1}{\rho_{l-1}^2}	\left(\mathbb{E}_{\rQ^{(l), 0}_{\out}}[z^2]-\rho_{l-1}\right)^2 s \notag\\
& \hspace{4cm} + \Theta\left(r^{3/2},s^{3/2}\right)\notag \,, \\
\Lambda_{z}(t) &\underset{t\ll 1}{=}\left(\mathbb{E}_{\rP_z}[z^2]\right)^2 t+\Theta\left(t^{3/2}\right)\,. \notag
\end{align}
For a generative prior with depth $L$, the Jacobian is a cumbersome sparse $(L+1) \times (L+1)$ matrix, with all the entries given by the six partial derivatives above. For the sake of conciseness we only write it here for $L=1$:
\begin{align}
\scalemath{0.8}{
\begin{pmatrix}
0 & \left(\mathbb{E}_{\rQ^{0}_{\out}}\[x^2\]\right)^2 &\frac{1}{\rho_{z}^{2}}\left(\mathbb{E}_{\rQ^{0}_{\out}}\[xz\] \right)^2 & 0\\ 	
\frac{\alpha}{\rho_{x}^2}\int\dd y~\mathcal{Z}^{0}_{y} \left(\mathbb{E}_{\rQ^{0}_{y}}\[\rho_{x}-x^2\]\right)^2 & 0 & 0 & 0\\
0 & 0 & 0 & \left(\mathbb{E}_{\rP_\z}\[z^2\]\right)^2\\
0 & \beta\left(\mathbb{E}_{\rQ^{0}_{\out}}\[xz\]\right)^2 &\frac{\beta}{\rho_{z}^2}\left(\mathbb{E}_{\rQ^{0}_{\out}}\[z^2\]-\rho_{z}\right)^2 & 0
\end{pmatrix}}.
\end{align}
Note that this holds for any choice of $\rP_{\out}^{(l)}$ and latent space distribution $\rP_\z$, as long as conditions eq.~\eqref{eq:stability_conditions} hold. For the phase retrieval with a linear generative model for instance $\rP^{(l)}(x|z) = \delta(x-z)$, we find $\alpha_{c} = \frac{1}{2}\frac{1}{1+\rho^{-1}}$. For a linear network of depth $L$ this generalizes to
\begin{align}
\alpha_{c} = \frac{1}{2}\left(1+\sum\limits_{l=1}^{L}\prod\limits_{k=0}^{l-1}\beta_{L-k}\right)^{-1}.
\label{eq:weak_recovery}
\end{align}
The linear estimation problem has exactly the same threshold, but without the global $1/2$ factor. Since $\rho, \beta_{l}\geq 0$, it is clear that $\alpha_{c}$ is decreasing in the depth $L$ of the network. This analytical formula is verified by numerically solving eqs.~\eqref{eq:SE_uu}, see Figs.~\ref{main:phase_diagramm_CS} and \ref{main:phase_diagramm_PR}. For other choices of activation satisfying condition \eqref{eq:stability_conditions} (e.g. the sign function), we always find that depth helps in the weak recovery of the signal.

\subsection{Perfect recovery threshold}
We now turn our attention to the perfect recovery threshold, above which perfect signal reconstruction becomes statistically possible. Formally, it can be defined as
\begin{align}
\alpha_{\IT} &= \underset{\alpha\geq 0}{\text{argmin}}\{\text{mmse}(\alpha) = 0\},
\end{align}
\noindent and corresponds to the global minimum of the free energy in eq.~\eqref{eq:freeen:multilayer}. Numerically, the perfect recovery threshold is found by solving the fixed point equations \eqref{eq:SE_uu} from an informed initialization $q_{x}\approx \rho_{x}$, corresponding to $\text{mmse} \approx 0$ according to eq.~\eqref{eq:mmse}. The resulting fixed point is then checked to be a minimizer of the free energy eq.~\eqref{eq:freeen:multilayer}. Different from $\alpha_{c}$, it cannot be computed analytically for an arbitrary architecture. However, for the compressed sensing and phase retrieval problems with $\sigma\in\{\text{linear}, \text{ReLU}\}$ generative priors, $\alpha_{\IT}$ can be analytically computed by generalizing a simple argument based on the invertibility of the linear system of equations at each layer, originally used in the usual compressive sensing \cite{candes2006near, tao2009}. 

First, consider the linear estimation problem with a deep linear generative prior, i.e. $\vec{y} = \mat{A} \vec{x}^{\star}\in\mathbb{R}^{n}$ with $\vec{x}^{\star} = \mat{W}^{(L)}\dots \mat{W}^{(1)}\vec{z}\in\mathbb{R}^{d}$ and $\mat{A}, \{\mat{W}^{(l)}\}_{l=1}^L$ \aclink{i.i.d} Gaussian matrices, that are full rank with high probability. For $n > d$, the system $\vec{y} = \mat{A} \vec{x}^{\star}$ is overdetermined as there are more equations than unknowns. Hence the information theoretical threshold has to verify $\alpha_{\IT} = \frac{n_{\IT}}{d} \leq 1$. For $L=0$ (i.e. $\vec{x}^{\star}$ is Gaussian \aclink{i.i.d}), we have exactly $\alpha_{\IT}^{(0)}=1$ as the prior does not give any additional information for solving the linear system. 
For $L\geq 1$ though, at each level $l \in \lb L \rb$, we need to solve successively $\vec{h}^{(l)} \in \bbR^{k_l}$ in the linear system $\vec{y} = \mat{A} \mat{W}^{(L)} \cdots \mat{W}^{(l)} \vec{h}^{(l)}$. Again as $\mat{A} \mat{W}^{(L)} \cdots \mat{W}^{(l)} \in\bbR^{ n \times k_l}$, if $n > k_l$ the system is over-constrained. Hence the information theoretical threshold for this equation is such that $\forall l \in \lb L \rb,  n_{\IT}^{(l)} \leq k_l  \Leftrightarrow \alpha_{\IT}^{(l)} \leq \prod\limits_{k=1}^{l} \frac{1}{\beta_{L-k+1}} $. And note that $\rho \equiv \prod\limits_{k=1}^{L} \frac{1}{\beta_{L-k+1}}$.
Hence, the information theoretical threshold is obtained by taking the smallest of the above values $\alpha_{\IT}^{(l)}$:
			\begin{align}
			 	\alpha_{\IT} = \min_{l\in [0:L]} \alpha_{\IT}^{(l)} = \min \left(1, \left\{ \prod\limits_{k=1}^{l} \frac{1}{\beta_{L-k+1}}  \right\}_{l=1}^{L-1}, \rho \right)\,.
			 	\label{eq:alphaIT}
			\end{align}

This result generalizes to the real-valued phase retrieval problem. First, we note that by the data processing inequality taking $\vec{y} = |\mat{A}\vec{x}^{\star}|$ cannot increase the information about $\vec{x}^{\star}$, and therefore the transition in phase retrieval cannot be \emph{better} than for compressed sensing. Secondly, an inefficient algorithm exists that achieve the same performance as compressed sensing for the real valued phase retrieval: one just needs to try all the possible $2^n$ assignments for the sign, and then solve the corresponding compressed sensing problem. This strategy that will work as soon as the compressed sensing problem is solvable. Eq.~(\ref{eq:alphaIT}) is thus valid for the real phase retrieval problem as well.

One can finally generalize this analysis for a non-linear generative prior with \aclink{ReLU} activation at each layer, i.e. $\vec{x}^{\star} = \text{relu}\left(\mat{W}^{(L)}\text{relu}\left(\cdots \mat{W}^{(1)}\vec{z}\right)\cdots\right)$. Noting that on average $\vec{x}$ has half of zero entries and half of \aclink{i.i.d} Gaussian entries, the system can be reorganized and simplified $\vec{y} = \td{\mat{A}} \td{\vec{x}}$, with $\td{\vec{x}}\in \bbR^{d/2}$ the extracted vector of $\vec{x}$ with on average $d/2$ strictly positive entries and the corresponding reduced matrix $\td{\mat{A}} \in \bbR^{n \times d/2}$, is over-constrained for $n > d/2 $ and hence the information theoretical threshold verifies $\alpha_{\IT} = \frac{n_{\IT}}{d} \leq \frac 12 $. 
Noting that this observation remains valid for generative layers, we will have on average at each layer an input vector $\vec{h}^{(l)}$ with half of zero entries and half of Gaussian distributed entries - except at the very first layer for which the input $\vec{z}\in\mathbb{R}^{k}$ is dense. Repeating the above arguments yields the following perfect recovery threshold
			\begin{align}
			 	\alpha_{\IT} =  \text{min}\left(\frac{1}{2}, \left\{ \frac{1}{2} \prod\limits_{k=1}^{l} \frac{1}{\beta_{L-k+1}}  \right\}_{l=1}^{L-1}, \rho \right)\,.
			 	\label{eq:alphaIT2}
			\end{align}
	for both the linear estimation and phase retrieval problems. Both these results are consistent  with the solution of the saddle-point eqs.~\eqref{eq:SE_uu} with a informed initialisation, see Figs.~\ref{main:phase_diagramm_PR}-\ref{fig:multilayer:compression}.

\subsection{Algorithmic threshold}
\label{sec:algo}
The discussion so far focused on the statistical limitations for signal recovery, regardless of the cost of the reconstruction procedure. In practice, however, one is concerned with the algorithmic costs for reconstruction. In the high-dimensional regime we are interested, where the number of observations scale with the number of parameters in the model, only (low)-polynomial time algorithms are manageable in practice. Remarkably, the formula in eq.~\eqref{eq:freeen:multilayer} also provides useful information about the algorithmic hindrances for the inverse problems under consideration. Indeed, with a corresponding choice of iteration schedule and initialization, the fixed point equations eq.~\eqref{eq:freeen:multilayer} are identical to the state evolution describing the asymptotic performance of an associated \aclink{AMP} algorithm \cite{manoel2017multi,fletcher2018inference}. Moreover, the \aclink{AMP} aforementioned is the \emph{best known} polynomial time algorithm for the estimation problem under consideration, and it is conjectured to be the optimal polynomial algorithm in this setting.

The \aclink{AMP} \emph{state evolution} corresponds to initializing the overlap parameters $(q_{x},q_{l}) \approx 0$ and updating, at each time step $t$ the hat variables $\hat{q}_{x}^{t} = \alpha\Lambda_{y}(q_{x}^{t})$ before the overlaps $q_{x}^{t+1} = \Lambda_{x}(\hat{q}_{x}^{t}, q_{L}^{t})$, etc. In Fig.~\ref{fig:mses} we illustrate this equivalence by comparing the \aclink{MSE} obtained by iterating eqs.~\eqref{eq:SE_uu} with the averaged \aclink{MSE} obtained by actually running the \aclink{AMP} algorithm from \cite{manoel2017multi} for a specific architecture and implemented with the \textsf{tramp} python package \cite{baker2020tramp}. In particular even though the \aclink{AMP} state evolution is not yet rigorously proven, we see a strong agreement of our analytical results with \aclink{AMP} simulations.

Note that, by construction, the performance of the Bayes-optimal estimator corresponds to the global minimum of the scalar potential in eq.~\eqref{eq:freeen:multilayer}. If this potential is convex, eqs.~\eqref{eq:SE_uu} will converge to the global minimum, and the asymptotic performance of the associated \aclink{AMP} algorithm will be optimal. However, if the potential has also a local minimum, initializing the fixed point equations will converge to the different minima depending on the initialization. In this case, the \aclink{MSE} associated to the \aclink{AMP} algorithm (corresponding to the local minimum) differs from the Bayes-optimal one (by construction the global minimum). In the later setting, we define the \emph{algorithmic threshold} as the threshold above which \aclink{AMP} is able to perfectly reconstruct the signal - or equivalently for which $\text{mmse} = 0$ when eqs.~\eqref{eq:SE_uu} are iterated from $q_{x}^{t=0}=q_{l}^{t=0}=\epsilon\ll 1$. Note that by definition $\alpha_{\IT} < \alpha_{\alg}$, and we refer to $\Delta_{\alg} = \alpha_{\alg} - \alpha_{\IT}$ as the algorithmic gap. See Fig.~\ref{fig:landscape} for an illustration of the evolution of the free energy landscape for increasing $\alpha$.

Studying the existence of an algorithmic gap for the linear estimation and phase retrieval problems, and how it depends on the architecture and depth of the generative prior, is the subject of the next section.

\begin{figure}[htb!]
	\centering
		\includegraphics[width=0.49\linewidth]{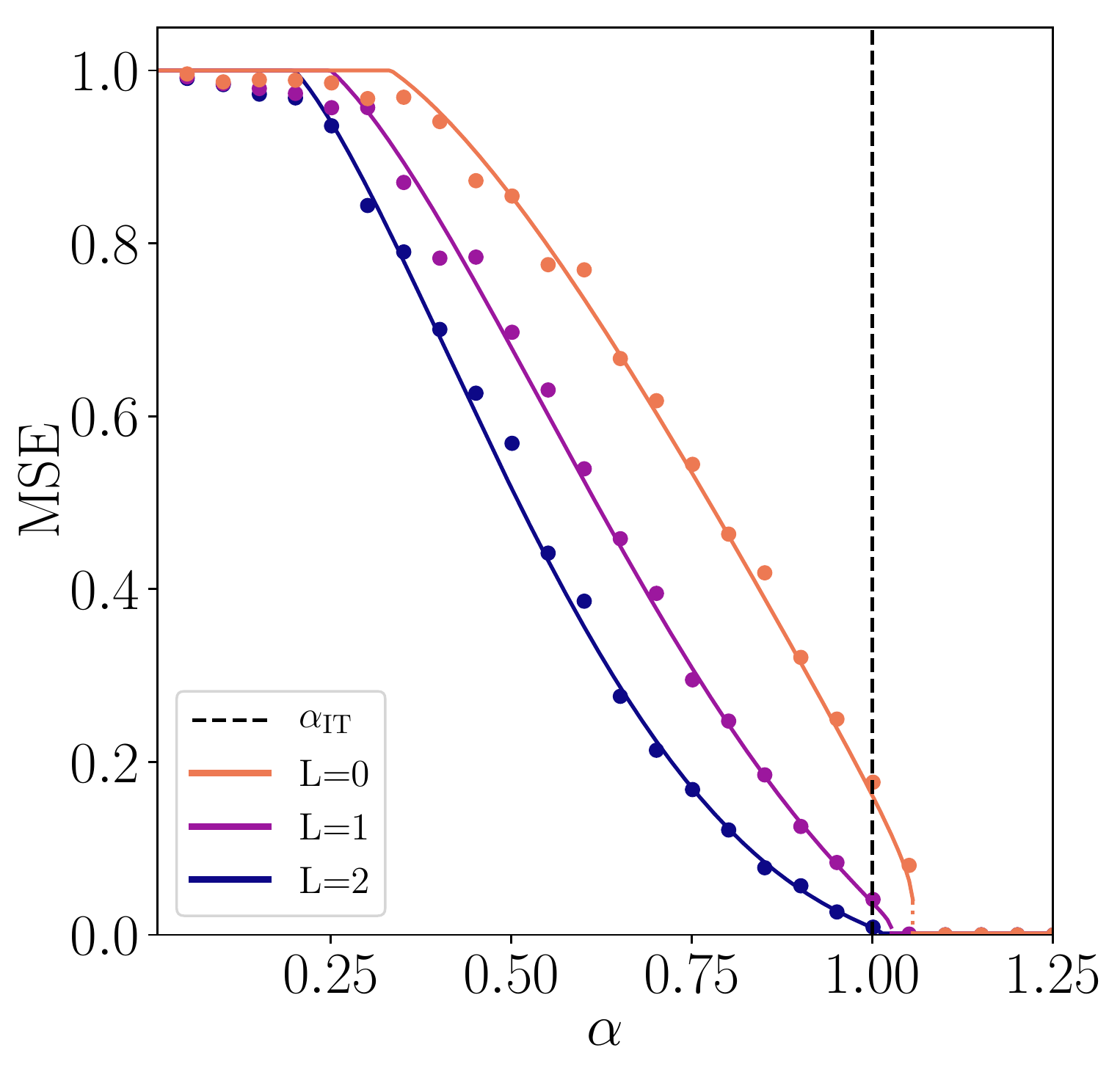}%
		\includegraphics[width=0.48\linewidth]{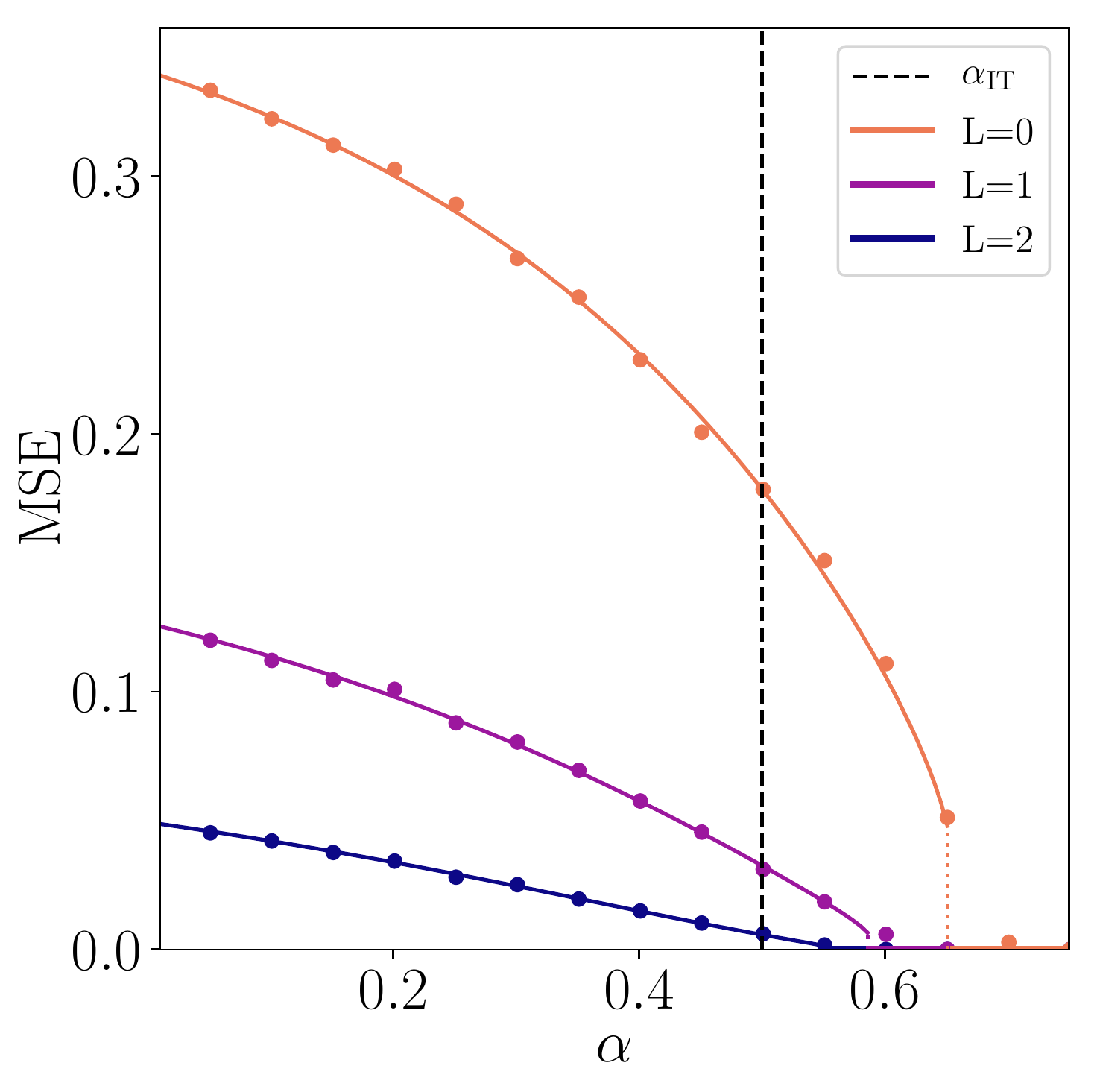}
	\caption{Mean squared error obtained by running the AMP algorithm (dots) from \cite{manoel2017multi} and implemented with the \textsf{tramp} package \cite{baker2020tramp}, for $d=2.10^3$ averaged on $10$ samples, compared to the MSE obtained from the state evolution eqs.~\eqref{eq:SE_uu} with uninformative initialization $q_{x}=q_{l}\approx 0$ (solid line) for the phase retrieval problem with linear \Left and relu \Right generative prior networks. Different curves correspond to different depths $L$, with fixed $\rho = 2$ and layer-wise aspect ratios $\beta_{l} = 1$. The dashed vertical line corresponds to $\alpha_{\IT}$. To illustrate for instance in the linear case \Leftn, $(\alpha_c^{L=0}, \alpha_c^{L=1},\alpha_c^{L=2}) = (1/3, 1/4, 1/5)$, $\alpha_{\IT}=1$ and  $(\alpha_{\alg}^{L=0}, \alpha_{\alg}^{L=1}, \alpha_{\alg}^{L=2}) = (1.056, 1.026 , 1.011)$.
	\label{fig:mses}
	\vspace{-0.2cm}
	}
\end{figure}

\section{Phase diagrams}
\label{sec:phase_diagrams} 
In this section we summarize the previous discussions in plots in the $(\rho,\alpha)$-plane, hereafter named \emph{phase diagrams}. Phase diagrams quantify the quality of signal reconstruction for a fixed architecture $(\beta_1, \dots, \beta_{L-1})$ \footnote{Note that $\beta_{L}$ is fixed from the knowledge of $(\rho, \beta_1, \dots, \beta_{L-1})$.} as a function of the compression $\rho$. Moreover, it allows a direct visual comparison between the phase diagram for a sparse Gaussian prior and the multi-layer generative prior. For both the phase retrieval and compressed sensing problems we distinguish the following regions of parameters limited by the thresholds of \Sec\ref{sec:it_analysis}: 
\begin{itemize}
    \item 
    \emph{Undetectable} region where the best achievable error is as bad as a random guess from the prior as if no measurement $\vec{y}$ were available. Corresponds to $\alpha < \alpha_{c}$.
    \item 
    \emph{Weak recovery} region where the optimal reconstruction error is better than the one of a random guess from the prior, but exact reconstruction cannot be achieved. Corresponds to $\alpha_c < \alpha < \alpha_{\IT}$.
    \item 
    \emph{Hard} region where exact reconstruction can be achieved information-theoretically, but no efficient algorithm achieving it is known. Corresponds to $\alpha_{\IT}<\alpha<\alpha_{\alg}$ 
   \item 
    The so-called \emph{easy} region where the aforementioned \aclink{AMP} algorithm for this problem achieves exact reconstruction of the signal. Corresponds to $\alpha > \alpha_{\alg}$.
\end{itemize}
As already explained, we locate the corresponding phase transitions in the following manner:
for the weak recovery threshold $\alpha_c$, we notice that the fixed point corresponding to an error as bad as a random guess corresponds to the values of the order parameters $q_{x}, q_{l} = 0$. This is an extremiser of the free energy (\ref{eq:freeen}) when the prior $\rP_\z$ has zero mean and the non-linearity $\varphi$ is an even function. This condition is satisfied for both the linear estimation and the phase retrieval problem with linear generative priors that leads to zero-mean distributions on the components of the signal, but is not achieved for a generative prior with \aclink{ReLU} activation, since it biases estimation.
In case this uninformative fixed point exists, we investigate its stability under the state evolution of the \aclink{AMP} algorithm, thus defining the threshold $\alpha_c$. For $\alpha<\alpha_c$ the fixed point is stable, implying the algorithm is not able to find an estimator better than random guess. In contrast, for $\alpha>\alpha_c$ the \aclink{AMP} algorithm provides an estimator better than random guess. For phase retrieval with linear generative model in the setting of the present paper, this analysis leads to the threshold derived in \eqref{eq:weak_recovery}.
If there exists a region where the performance of the \aclink{AMP} algorithm and the information-theoretic one do not agree we call it the \emph{hard} region. The hard region is delimited by threshold $\alpha_{\IT}$ and $\alpha_{\alg}$. 

The statistical and algorithmic thresholds defined above admit an alternative and instructive description in terms of free energy landscape, see Fig. ~\ref{fig:landscape}. Consider a fixed $\rho$: for small $\alpha$ the free energy \eqref{eq:freeen:multilayer} has a single global minimum with small overlap (high \aclink{MSE}) with the ground truth solution $\vec{x}^\star$, referred as the \emph{uninformative} fixed point. 
At a value $\alpha_{\textrm{sp}}$, known as the \emph{first spinodal transition}, a second local minimum appears with higher overlap (smaller \aclink{MSE}) with the ground truth, referred as \emph{informative} fixed point. The later fixed point becomes a global minimum of the free energy at $\alpha_{\IT} > \alpha_{\textrm{sp}}$, while the uninformative fixed point becomes a local minimum. A second spinodal transition occurs at $\alpha_{\alg}$ when the informed fixed point becomes unstable.  
Numerically, the informed and uninformative fixed points can be reached by iterating the saddle-point equations from different initial conditions. When the two are present, the informed fixed point can be reached by iterating from $q_{x}\approx \rho_{x}$, which corresponds to a minimum overlap with the ground truth $\bx^{\star}$, and the uninformative fixed point from $q_{x}\approx 0$, corresponding to no initial overlap with the signal. In the noiseless linear estimation and phase retrieval studied here we observe $\alpha_{\IT} = \alpha_{\textrm{sp}}$. 

\begin{figure}[t]
	\centering
	\begin{tikzpicture}[scale=0.16]
    \tikzstyle{dot}=[circle,minimum size=10pt, scale=0.4]
    \tikzstyle{green}=[fill=green!70!black]
    \tikzstyle{orange}=[fill=orange!90!black]
    \tikzstyle{yellow}=[fill=yellow!95!black]
    \tikzstyle{red}=[fill=red!90!black]
    \tikzstyle{annot}=[text width=3cm, text centered, font=\footnotesize]
    \tikzstyle{line}=[-,thick, black]
	\draw [line] (0,0) to [out=-65,in=180] (3.5,-2) to [out=0,in=200] (9,1) ;
	\node[dot, green] at (3.5,-2) {}; 
	\draw [line] (11,0) to [out=-45,in=180] (13.5,-2) to [out=0,in=-180] (15.5,0) to [out=0,in=-180] (16,0) to [out=10,in=220] (19,2)  ;
	\node[dot, green] at (13.5,-2) {}; 
	\node[dot, orange] at (16,0) {}; 
	\node[annot, below = 0.25cm] at (15, -1) {$\alpha_{\textrm{sp}}$}  ;
	\draw [line] (15, -6.5) to (15, -5.5) ;
	\draw [line] (21,0) to [out=-45,in=180] (23.5,-2) to [out=0,in=180] (25,0) to [out=0,in=-180] (26.5,-1) to [out=0,in=180] (29,0);
	\node[dot, green] (BO) at (23.5,-2) {}; 
	\node[dot, orange] (BO) at (26.5,-1) {}; 
	\draw[densely dotted] (33.5,-2) to (36.5,-2);
	\draw [line] (31,0) to [out=-45,in=180] (33.5,-2) to [out=0,in=180] (35,0) to [out=0,in=-180] (36.5,-2) to [out=0,in=180] (39,0);
	\node[dot, red]  at (33.5,-2) {}; 
	\node[dot, green] at (36.5,-2) {}; 
	\node[annot, below = 0.25cm] at (35, -1) {$\alpha_{\textrm{IT}}$}  ;
	\draw [line] (35, -6.5) to (35, -5.5) ;
	\draw [line] (41,0) to [out=-45,in=180] (43.5,-1) to [out=0,in=180] (45,0) to [out=0,in=-180] (46.5,-2) to [out=0,in=180] (49,0);
	\node[dot, red] at (43.5,-1) {}; 
	\node[dot, green] at (46.5,-2) {}; 
	\draw [line] (51,1) to [out=-70,in=180] (53.5,0) to [out=0,in=-180] (56.5,-2) to [out=0,in=180] (59,0);
	\node[dot, red] at (53, 0) {}; 
	\node[dot, green] at (56.5,-2) {}; 
	\node[annot, below = 0.25cm] at (55, -1) {$\alpha_{\textrm{alg}}$} ;
	\draw [line] (55, -6.5) to (55, -5.5) ;
	\draw [line] (61,1) to [out=0,in=-180] (66.5,-2) to [out=0,in=180] (69,0);
	\node[dot, green] at (66.5,-2) {};
	\draw[draw=none, orange] (0,-6) to (0,-6.5) to (35,-6.5) to (35, -6) to (0, -6);
	\draw[draw=none, yellow] (35,-6) to (35,-6.5) to (55,-6.5) to (55, -6) to (35, -6);
	\draw[draw=none, green] (55,-6) to (55,-6.5) to (69,-6.5) to (69, -6) to (55, -6);
	\draw [->,thick, black] (0,-6) to (70,-6);
	\node[text centered, font=\footnotesize, below = 0.25cm] at (17.5, -6.2) {\textbf{Weak recovery}}  ;
	\node[annot, below = 0.25cm] at (45, -6.2) {\textbf{Hard}}  ;
	\node[annot, below = 0.25cm] at (62.5, -6.2) {\textbf{Easy}}  ;
	\end{tikzpicture}
\caption{Illustration of the free energy landscape as a function of the overlap with the ground truth solution, when one increases $\alpha$. For small $\alpha<\alpha_{\textrm{sp}}$, there exists a unique global minimum, whose overlap with the solution is small  (high MSE). 
	At $\alpha=\alpha_{\textrm{sp}}$, a \emph{local} minimum (orange dot) with higher overlap (small MSE) appears. By definition, the global minimum corresponds to the MMSE of the problem, which is the MSE attained by the Bayes-optimal estimator (green dot). For $\alpha<\alpha_{\textrm{IT}}$ the accessible solution, i.e the global minimum (green dot) has a high MSE while a better solution exists but has a higher free energy (weak recovery phase). At $\alpha=\alpha_{\textrm{IT}}$ the two minima are global and have the same free energy.  Between $\alpha_{\textrm{IT}} < \alpha <\alpha_{\textrm{alg}}$ (hard phase), the local minimum with higher MSE corresponds to the performance of the AMP estimator (red dot). Above $\alpha_{\textrm{alg}}$ only the small MSE minima survive and the AMP estimator is able to achieve the Bayes-optimal performance (easy phase). 
}
\label{fig:landscape}
\end{figure}
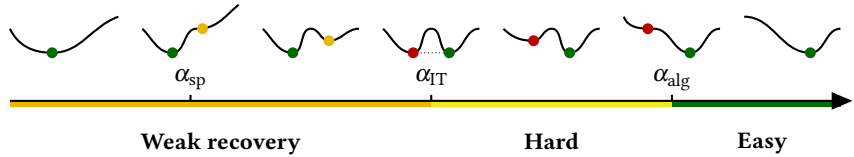

\subsection{Single-layer generative prior}
\label{sec:phase_diagrams:single_layer}
First, we consider the case where the signal is generated from a single-layer generative prior, $\vec{x}^{\star} = \sigma(\mat{W}\vec{z})$ with $\vec{z}\sim\mathcal{N}(\vec{0},\mat{I}_{k})$. We analyze both the compressed sensing and the phase retrieval problem, for $\sigma\in\{\text{linear}, \text{ReLU}\}$. In this case the only free parameters of the model are $(\rho, \alpha)$, and therefore the phase diagram fully characterizes the recovery in these inverse problems. The aim is to compare with the phase diagram of a sparse prior with density $\rho_{s} = \rho$ of nonzero components.

\Fig\ref{main:phase_diagramm_CS} depicts the compressed sensing problem with linear \Left and \aclink{ReLU} \Right generative priors. We depict the phase transitions defined above. On the left hand side we compare to the algorithmic phase transition known from \cite{krzakala_statistical-physics-based_2012} for sparse separable prior with fraction $1-\rho$ of zero entries and $\rho$ of Gaussian entries of zero mean presenting an algorithmically hard phase for $\rho < \alpha < \alpha_{\textrm{alg}}^{\textrm{sp}arse}(\rho)$. 

In the case of compressed sensing with linear generative prior we do not observe any hard phase and exact recovery is possible for $\alpha \ge \min (\rho,1)$ due to invertibility (or the lack of there-of) of the matrix product~$\mat{A}\mat{W}$. With \aclink{ReLU} generative prior we have $\alpha_{\IT} = \min(\rho,1/2)$ and the hard phase exists and has interesting properties: The $\rho \to \infty$ limit corresponds to the separable prior, and thus in this limit $\alpha_{\textrm{alg}}(\rho \to \infty) = \alpha_{\textrm{alg}}^{\textrm{sp}arse}(\rho_s=1/2)$. Curiously we observe $\alpha_{\textrm{alg}} > \alpha_{\IT}$ for all $\rho \in (0,\infty)$ except at $\rho=1/2$. Moreover the size of the hard phase is very small for $\rho < 1/2$ when compared to the one for compressed sensing with separable priors, suggesting that exploring structure in terms of generative models might be algorithmically advantageous over sparsity.     

\Fig\ref{main:phase_diagramm_PR} depicts the phase diagram for the phase retrieval problem with linear \Left and \aclink{ReLU} \Right generative priors. The information-theoretic transition is the same as the one for compressed sensing, while numerical inspection shows that $\alpha_{\textrm{alg}}^{\textrm{PR}} > \alpha_{\textrm{alg}}^{\textrm{CS}}$ for all $\rho \neq 0, 1/2,1$. In the left hand side we depict also the algorithmic transition corresponding to the sparse separable prior with non-zero components being Gaussian of zero mean, $\alpha_{\textrm{alg}}^{\textrm{sp}arse}(\rho_s)$, as taken from \cite{barbier2017phase}. 
Crucially, in that case the algorithmic transition to exact recovery does not fall bellow $\alpha= 1/2$ even for very small (yet finite) $\rho_s$, thus effectively disabling the possibility to sense compressively. In contrast, with both the linear and \aclink{ReLU} generative priors we observe $\alpha_{\textrm{alg}}(\rho \to 0) \to 0$. More specifically, the theory for the linear prior implies that $\alpha_{\textrm{alg}}/\rho (\rho \to 0) \to \alpha_{\textrm{alg}}^{\textrm{sp}arse}(\rho_s=1) \approx 1.128$ with the hard phase being largely reduced. 
Again the hard phase disappears entirely for $\rho=1$ for the linear model and $\rho=1/2$ for \aclink{ReLU}. 

\begin{figure}[htb!]
	\centering
		\includegraphics[width=0.47\linewidth]{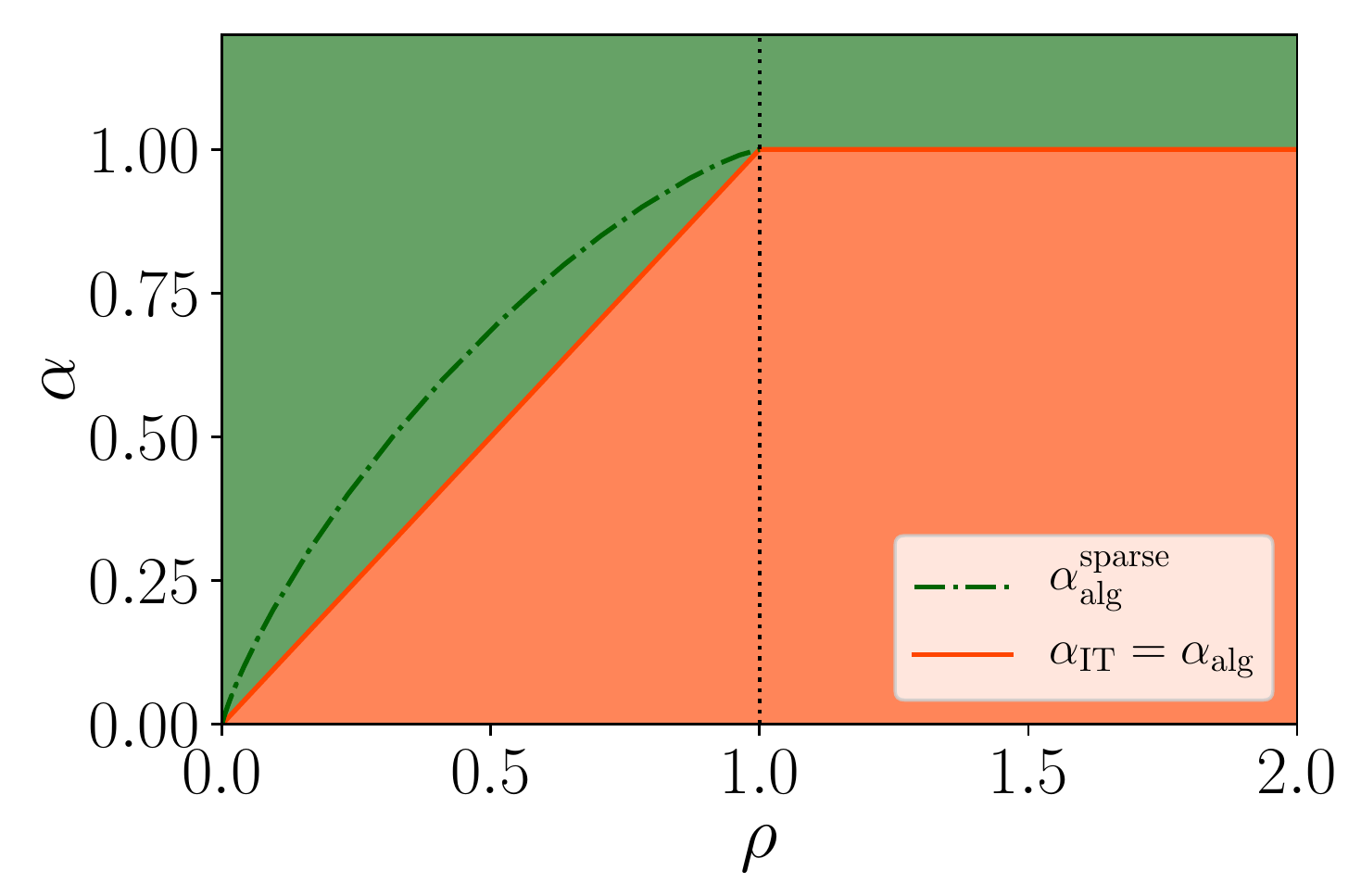}%
		\includegraphics[width=0.49\linewidth]{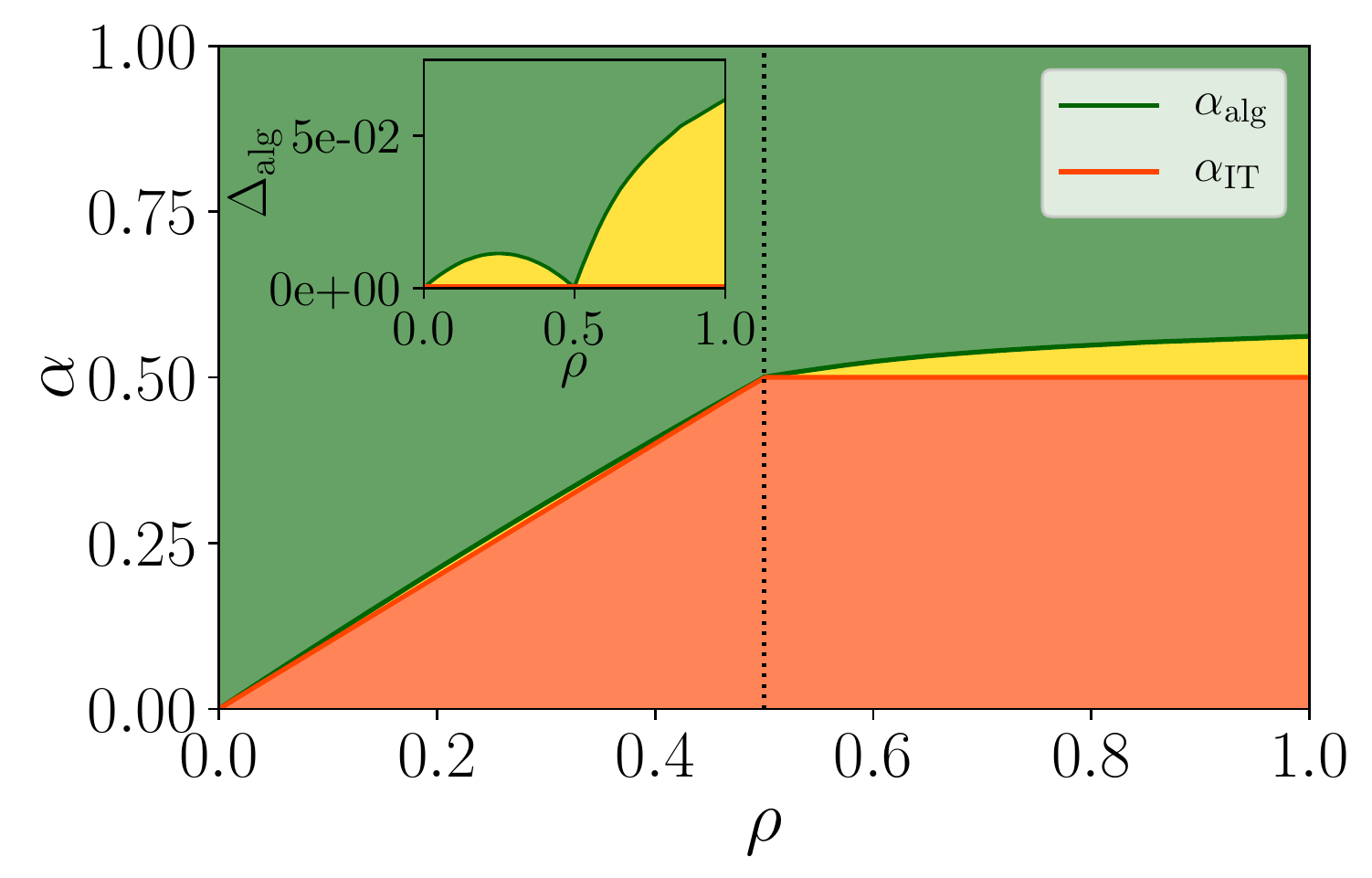}
	\caption{Phase diagrams for the compressed sensing problem with \Left linear generative prior and \Right ReLU generative prior, in the plane $(\rho,\alpha)$. The $\alpha_{\IT}$ (red line) represents the information theoretic transition for perfect reconstruction and $\alpha_{\textrm{alg}}$ (green line)  the algorithmic transition to perfect reconstruction. In the left part we depict for comparison the algorithmic phase transition for sparse separable prior $\alpha_{\textrm{alg}}^{\textrm{sp}arse}$ (dashed-dotted green line). The inset in the right part depicts the difference $\Delta_{\textrm{alg}} = \alpha_{\textrm{alg}} - \alpha_{\IT}$. Colored areas correspond respectively to the \textit{weak recovery} (orange), \textit{hard} (yellow) and \textit{easy} (green) phases. The behavior of the free energy landscape for increasing $\alpha$ and fixed $\rho$ is illustrated in \Fig\ref{fig:landscape}. 
	\label{main:phase_diagramm_CS}
	\vspace{-0.2cm}
	}
\end{figure}

\begin{figure}[htb!]
	\centering
		\includegraphics[width=0.49\linewidth]{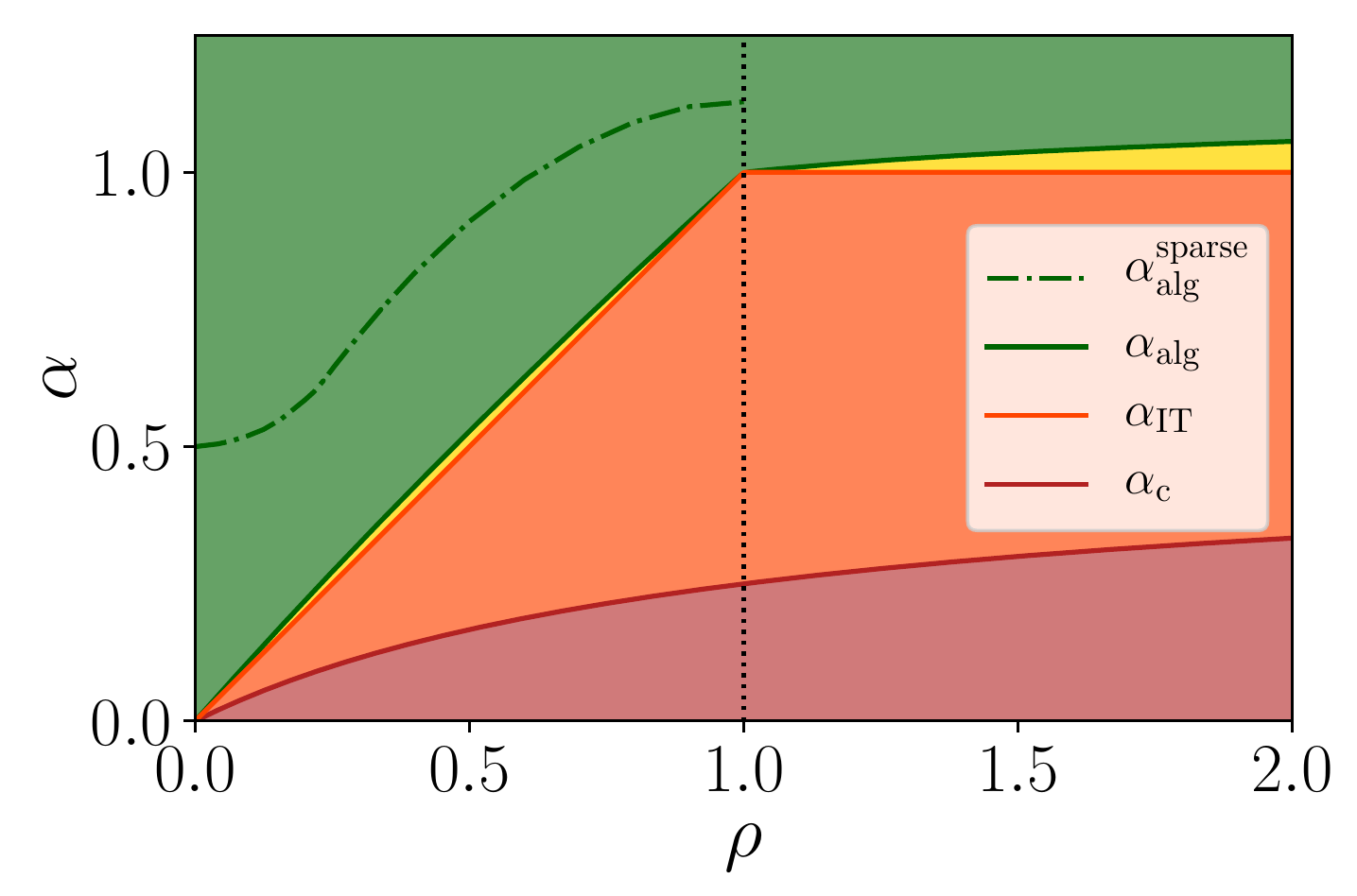}
		\includegraphics[width=0.49\linewidth]{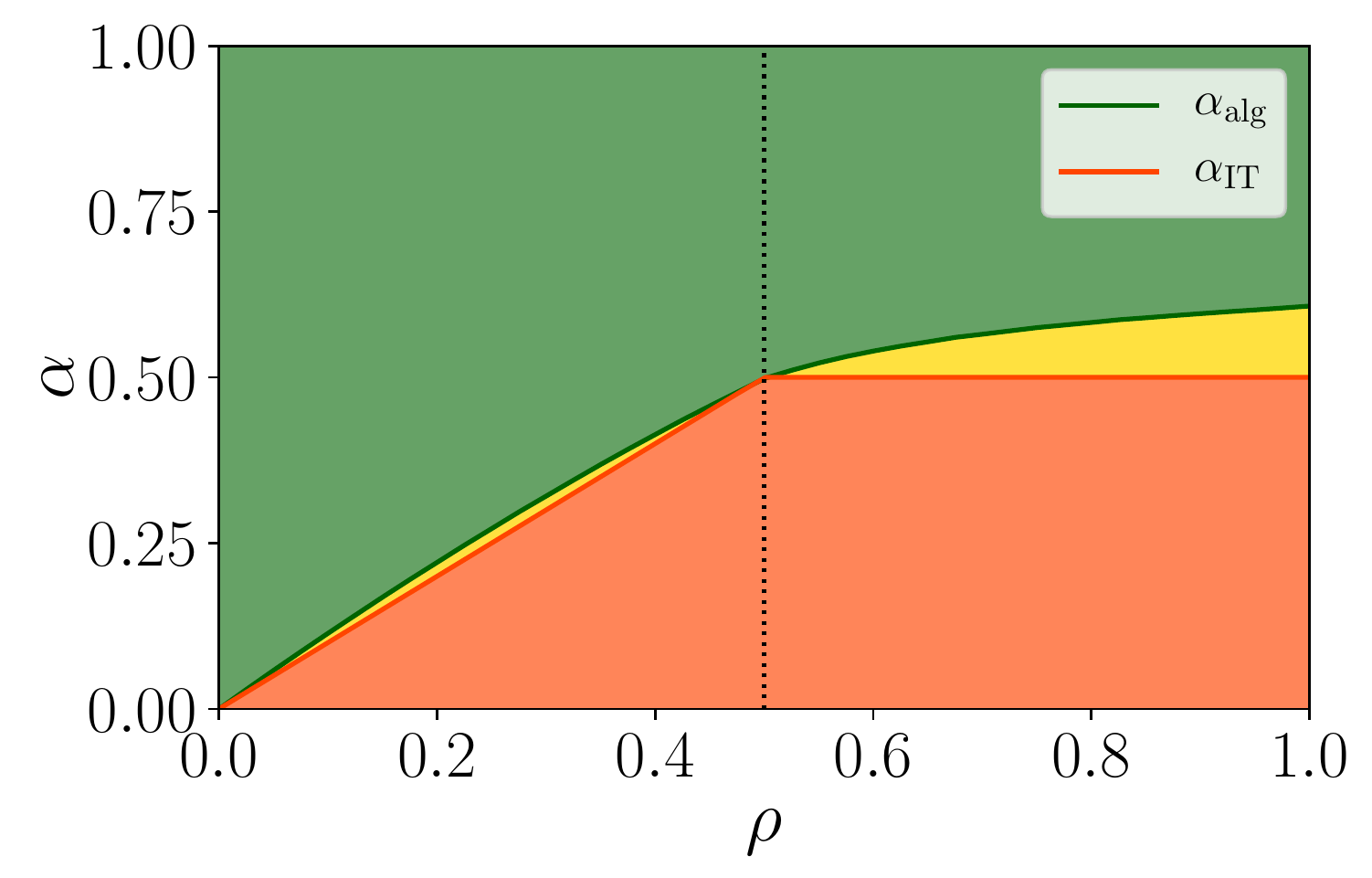}
	\caption{The same as \Fig\ref{main:phase_diagramm_CS} for the phase retrieval problem with \Left linear generative prior and \Right ReLU generative prior. A major result is that while with sparse separable priors (green dashed-dotted line) compressed phase retrieval is algorithmically hard for $\alpha < 1/2$, with generative priors compressed phase retrieval is tractable down to vanishing $\alpha$ (green line). 
	In the left part we depict additionally the \textit{weak recovery} transition $\alpha_c = \rho/ [2(1+\rho)]$ (dark red line). 
	It splits the \textit{no-exact-recovery} phase  into the \emph{undetectable} (dark red) and the \emph{weak-recovery} region (orange). 
	}
	\label{main:phase_diagramm_PR}
\end{figure}

\subsection{Multi-layer generative prior}
From the discussion above, we conclude that generative priors are algorithmically advantageous over sparse priors, allowing compressive sensing for the phase retrieval problem. We now investigate how the role of depth of the prior in this discussion. As before, we analyze both the linear estimation and phase retrieval problems, fixing $\sigma^{(l)} \equiv \sigma \in \{\text{linear}, \text{ReLU}\}$ at every layer $1\leq l \leq L$. Different from the $L=1$ case discussed above, for $L>1$ we have other $L-1$ free parameters characterizing the layer-wise compression factors $\left(\beta_1, \dots, \beta_{L-1}\right)$.

First, we fix $\beta_{l}$ and investigate the role played by depth. \Fig\ref{fig:multilayer:depth} depicts the phase diagrams for compressed sensing \Left
and phase retrieval \Right with \aclink{ReLU} activation with varying depth, and a fixed architecture $\beta_{l}=3$ for $1\leq l \leq L$ and note that all these curves share the same $\alpha_{\IT} = \text{min}(0.5, \rho)$. It is clear that depth improves even more the small gap already observed for a single-layer generative prior. The algorithmic advantage of multi-layer generative priors in the phase retrieval problem has been previously observed in a similar setting in \cite{hand2018phase}.

Next, we investigate the role played by the layer-wise compression factor $\beta_{l}$. \Fig\ref{fig:multilayer:compression} depicts the phase diagrams for the compressed sensing \Left and phase retrieval \Right with \aclink{ReLU} activation for fixed depth $L=2$, and varying $\beta\equiv \beta_{1}$. According to the result in \eqref{eq:alphaIT}, we have $\alpha_{\IT} = \text{min}\left(1/2,\rho, 1/2\beta\right)$. It is interesting to note that there is a trade-off between compression $\beta<2$ and the algorithmic gap, in the following sense. For $\rho<0.5$ fixed, $\alpha_{\IT}$ decreases with decreasing $\beta\ll 1$: compression helps perfect recovery. However, the algorithmic gap $\Delta_{\alg}$ becomes wider for fixed $\rho<0.5$ and decreasing $\beta\ll 1$.

These observations also hold for a linear generative model. In \Fig\ref{fig:multilayer:linear} we have a closer look by plotting the algorithmic gap $\Delta_{\alg} \equiv \alpha_{\alg}-\alpha_{\IT}$ in the phase retrieval problem. On the left, we fix $L=4$ and plot the gap for increasing values of $\beta \equiv \beta_{l}$, leading to increasing $\Delta_{\alg}$. On the right, we fix $\beta = 2$ and vary the depth, observing a monotonically decreasing $\Delta_{\alg}$.

\begin{figure}[htb!]
	\centering
		\includegraphics[width=0.49\linewidth]{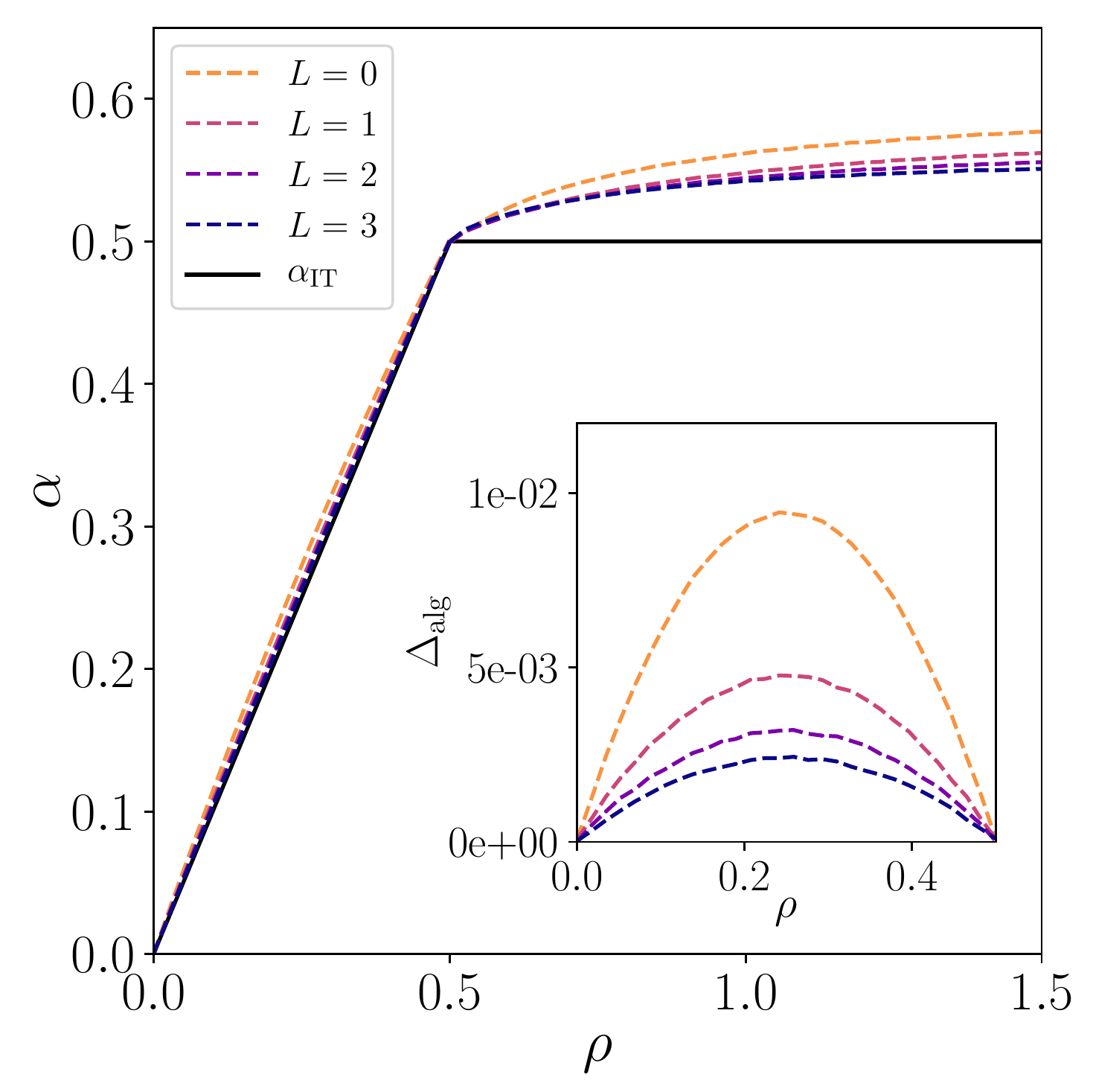}
		\includegraphics[width=0.49\linewidth]{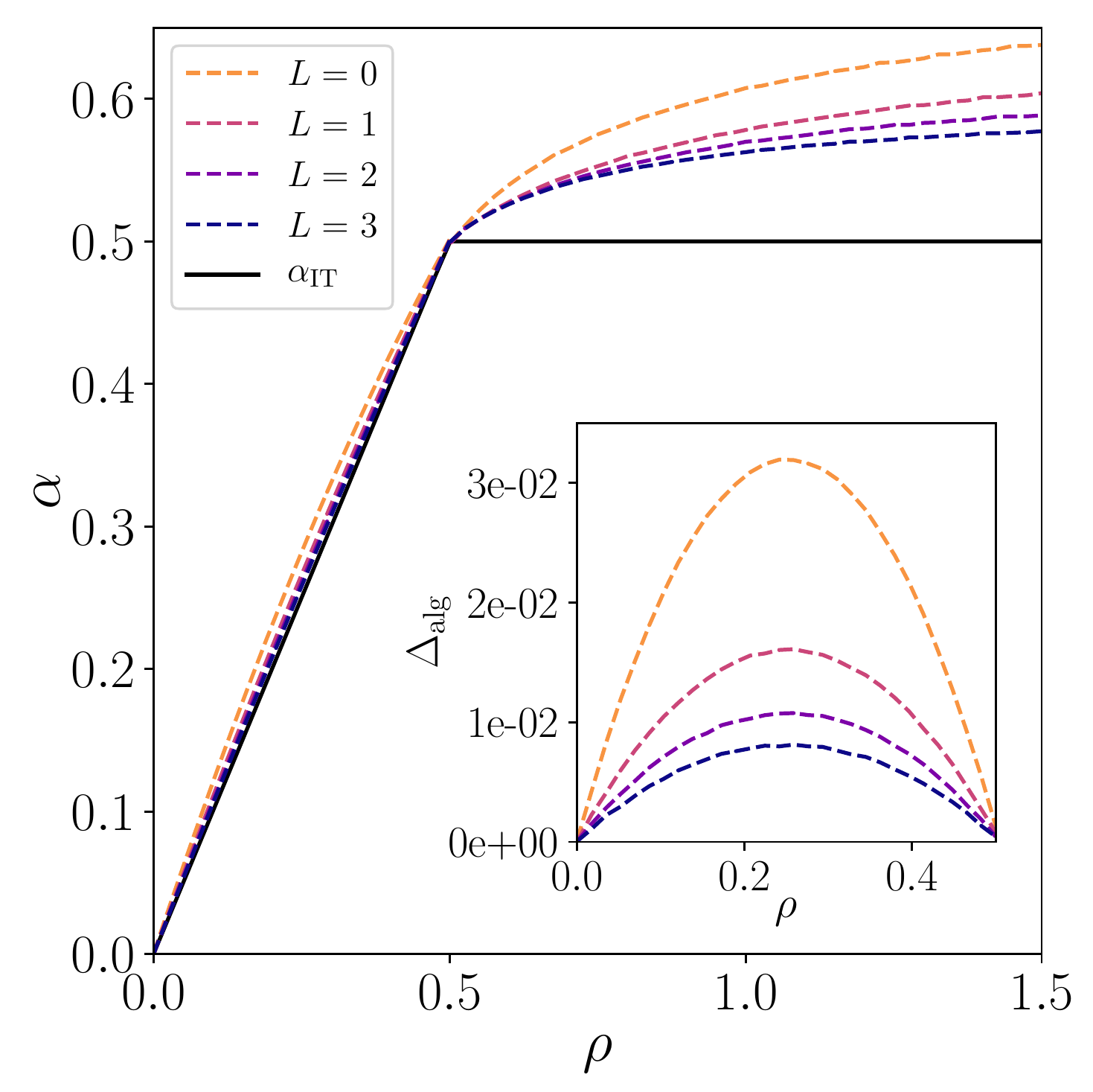}
	\caption{Phase diagrams for the compressed sensing \Left and phase retrieval \Right problems for different depths of the prior, with ReLU activation and fixed layer-wise compression $\beta_{l}=3$. Dashed lines represent the algorithmic threshold $\alpha_{\alg}$ and solid lines the perfect recovery threshold $\alpha_{\IT}$. We note that the algorithmic gap $\Delta_{\textrm{alg}}$ (shown in insets) decreases with the network depth $L$.}
	\label{fig:multilayer:depth}
\end{figure}

\begin{figure}[htb!]
	\centering
		\includegraphics[width=0.49\linewidth]{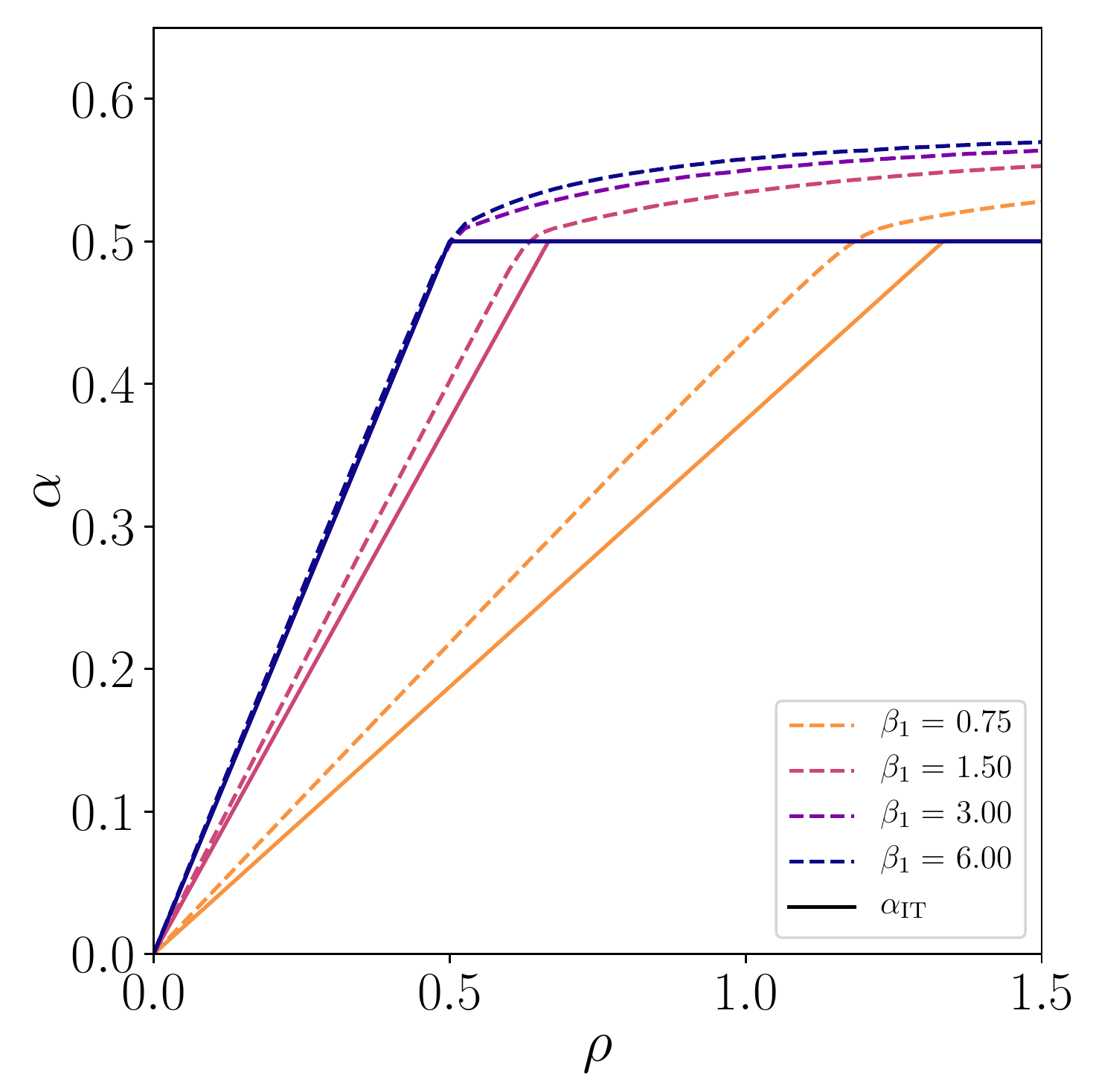}%
		\includegraphics[width=0.49\linewidth]{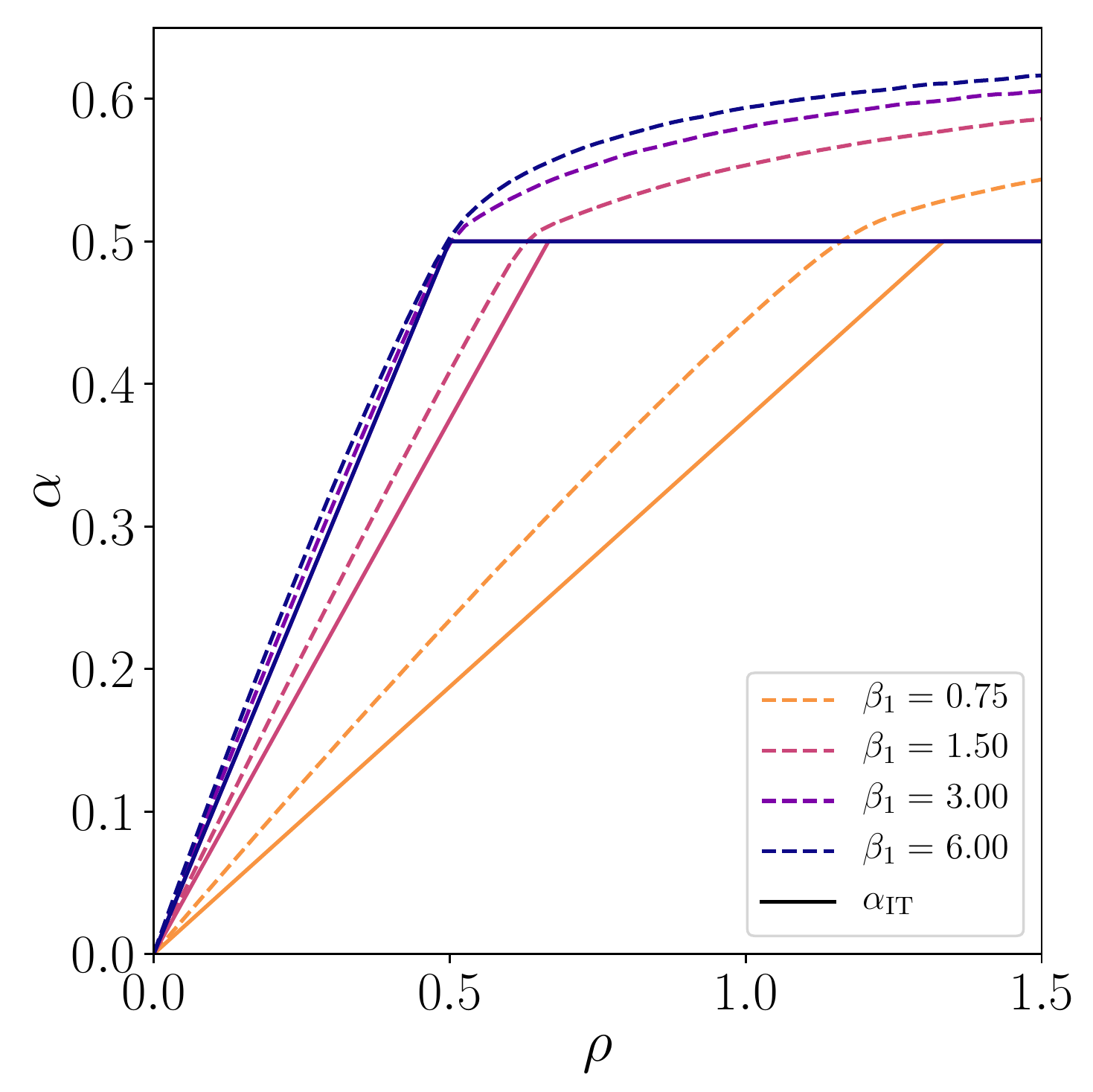}
	\caption{Phase diagrams for the compressed sensing \Left and phase retrieval \Right problems with $L=2$ and ReLU activation for different values of the layer-wise compression factor $\beta_{1}$. Dashed lines represent the algorithmic threshold $\alpha_{\alg}$ and solid lines the perfect recovery threshold $\alpha_{\IT}$. We note that for a given $\rho < 0.5$, $\alpha_{\IT}$ is decreasing with $\beta\ll 1$. However, the algorithmic gap $\Delta_{\alg}$ (shown in the inset) grows for decreasing $\beta$. Note that for $\beta_1\geq 2$ the hard phase is hardly visible at $\rho=0.5$, even though it disappears only in the large width limit, for both compressed sensing and phase retrieval settings.}	\label{fig:multilayer:compression}
	\vspace{-0.2cm}
\end{figure}

\begin{figure}[htb!]
	\centering
		\includegraphics[width=0.49\linewidth]{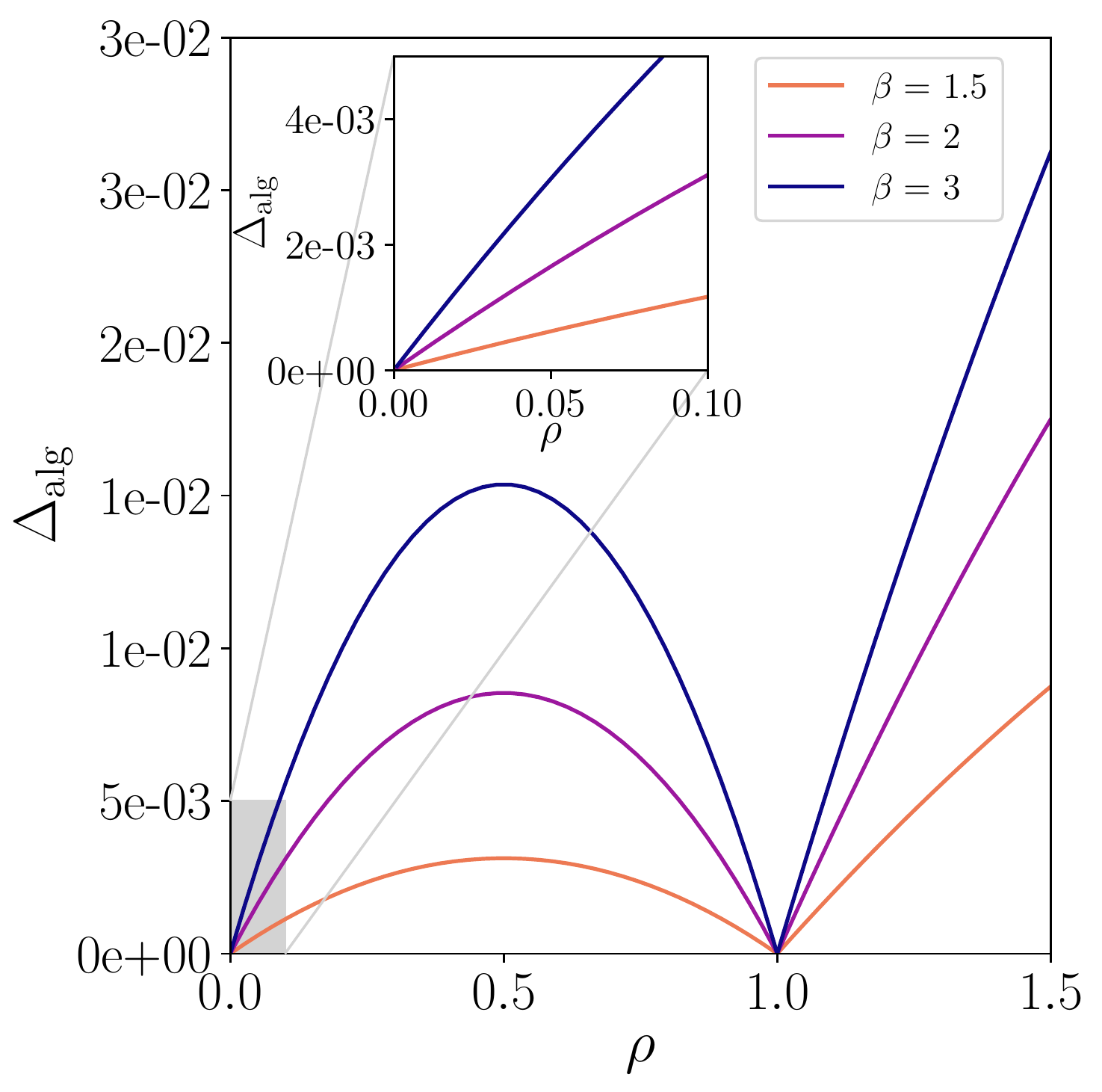}
		\includegraphics[width=0.49\linewidth]{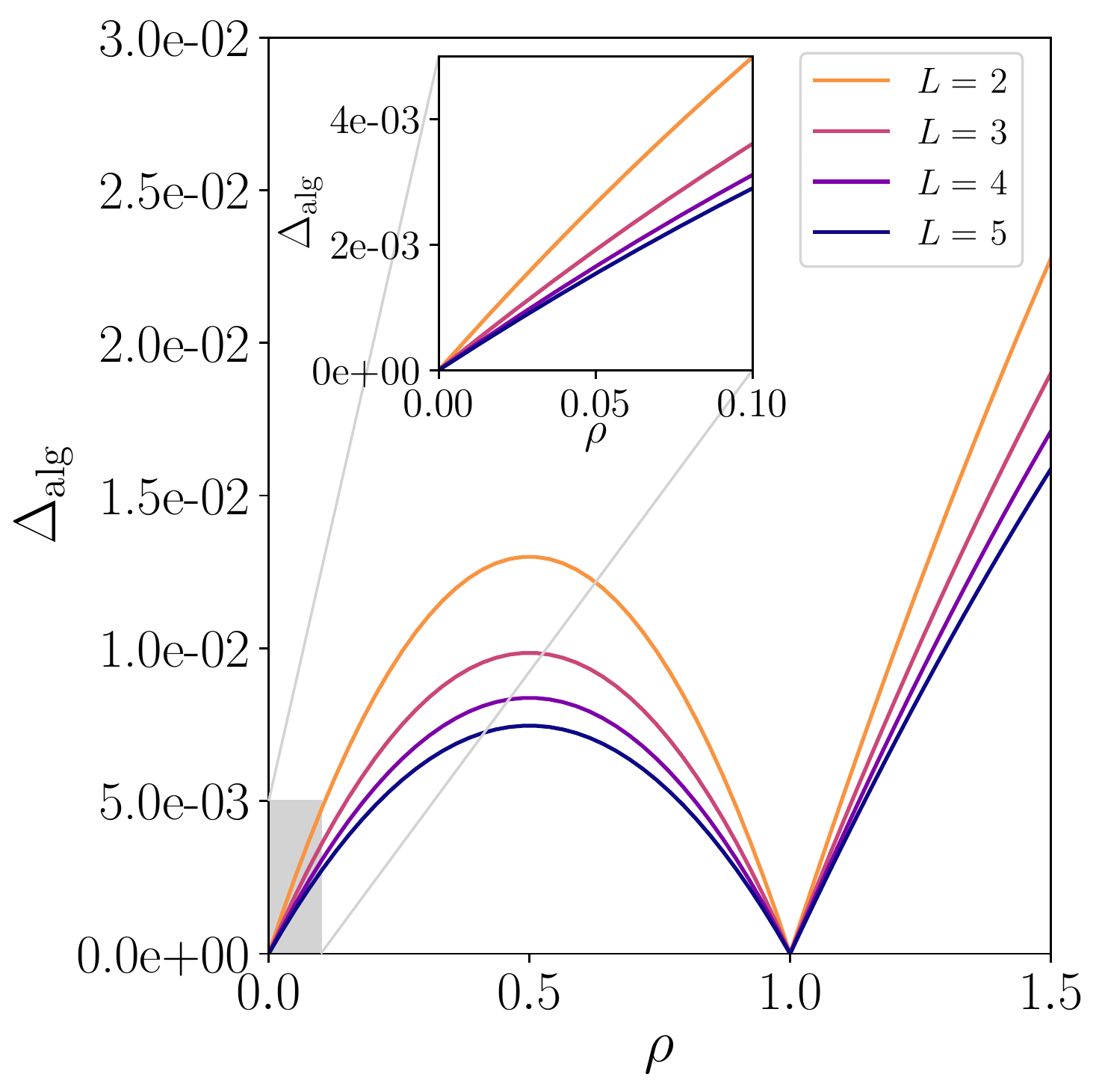}
	\caption{Algorithmic gap $\Delta_{\textrm{alg}} = \alpha_{\textrm{alg}} - \alpha_{\IT}$ for small $\rho$ and linear activation, as a function of \Left the compression $\beta\equiv \beta_{l}$ for fixed depth $L=4$ and of \Right depth for a fixed compression $\beta=2$.}
	\label{fig:multilayer:linear}
\end{figure}

\section*{Conclusion and perspectives}
In this chapter, we analyzed how generative priors from an ensemble of random multi-layer neural networks impact signal reconstruction in the high-dimensional limit of two important inverse problems: real-valued phase retrieval and linear estimation. More specifically, we characterized the phase diagrams describing the interplay between number of measurements needed at a given signal compression $\rho$, for a range of shallow and multi-layer architectures for the generative prior. We observed that although present, the algorithmic gap significantly decreases with depth in the  studied architectures. This is particularly striking when compared with sparse priors at $\rho\ll 1$, for which the algorithmic gap is considerably wider. In practice, this means generative models given by random multi-layer neural networks allow for efficient compressive sensing in these problems.

In this work we have only considered independent random weight matrices for both the estimation layer and for the generative model. Ideally, one would like to introduce correlations in a setting closer to reality 
to show that the smaller computation-to-statistical gap also appears in  real-life tasks.
The hurdle is that in those cases one does not know what is the theoretically optimal performance nor what are the optimal polynomial algorithms, so that one cannot evaluate the computation-to-statistical empirically in those cases. 
Yet another tractable case is the study of random rotationally invariant or unitary sensing matrices, as in \cite{kabashima2008inference,fletcher2018inference,barbier2018mutual,dudeja2019information}. In a different direction, it would be interesting to observe the phenomenology from this work in an experimental setting, for instance using a generative model, such as \aclink{GAN} or \aclink{VAE}, trained on a real dataset to improve the performance of \aclink{AMP} algorithms in a practical task. This is the purpose of the next section.

\section{Estimation with non i.i.d generative priors}
\label{chap:generative_phase:applications_real}
Instead reproducing the \emph{plug-in} approach illustrated in the context of the spiked matrix model with generative prior in \Chap\ref{chap:generative_spiked} \Sec\ref{sec:spiked:amp_derivation} to derive the \aclink{AMP} algorithm for each structured model, we developed a python package \tramp, standing for \emph{TRee Approximate Message Passing}, that automatically build the corresponding \aclink{AMP} algorithm from the sub-models.

Moreover, the package provides an implementation of \aclink{EP} for modular compositional inference in high-dimensional tree-structured models, which is more robust than the classical \aclink{AMP}.
In particular, while the classical \aclink{AMP}, discussed in the previous section, is restricted to \aclink{i.i.d} weights, \aclink{EP} implemented in \tramp is able to handle non-\aclink{i.i.d} weights such as the ones obtained after training of a \aclink{GAN} or \aclink{VAE}.
More details on the implementation can be found in \cite{baker2020tramp} and the source code publicly available at \href{https://github.com/sphinxteam/tramp}{\url{https://github.com/sphinxteam/tramp}}. 

To illustrate the performances of \aclink{EP} on structured models with correlated weights, we consider a signal $\vec{x}\in \bbR^{\ndim}$ (with $\ndim=784$) drawn from the MNIST data set \cite{mnist10}. We want to reconstruct the original image from a corrupted observation $\vec{y} = \varphi(\vec{x}) \in\bbR^{\ndim}$, where $\varphi:\bbR^{\ndim}\to\bbR^{\ndim}$ represents a noisy channel. In the following the noisy channel represents either a Gaussian additive channel or an inpainting channel, that erases some pixels of the input image. 
In order to reconstruct correctly the MNIST image, we investigate the possibility of using a generative prior such as a \aclink{VAE} along the lines of \cite{bora2017compressed,fletcher2018inference}. Note that information theoretical and approximate message passing properties of reconstruction of a low rank or \aclink{GLM} channel, using a dense feed-forward neural network generative prior with \aclink{i.i.d} weights has been studied in particular in \citep{aubin2019spiked, aubin2019exact}. However, neither information theoretical or algorithmic perspective was investigated to handle a \emph{trained} generative prior with non-\aclink{i.i.d} weights, such as the ones we consider in this section.
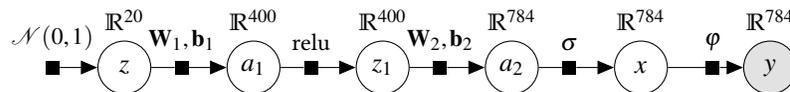
\begin{figure}[H]
\centering
  \begin{tikzpicture}[scale=0.9]
    \node[latent, label=$\bbR^{20}$]  (z)  {$z$} ;
    \node[latent, right=1 of z, label=$\bbR^{400}$]  (a1)  {$a_1$} ;
    \node[latent, right=1 of a1, label=$\bbR^{400}$] (z1) {$z_1$} ;
    \node[latent, right=1 of z1, label=$\bbR^{784}$] (a2) {$a_2$} ;
    \node[latent, right=1 of a2, label=$\bbR^{784}$] (x) {$x$} ;
    \node[obs, right=1 of x, label=$\bbR^{784}$] (y) {$y$} ;
    \factor[left=0.5 of z] {p} {$\mN(0,1)$} {} {z} ;
    \factor[left=0.5 of a1] {p} {$\mat{W}_1, \vec{b}_1$} {z} {a1} ;
    \factor[left=0.5 of z1] {p} {$\text{relu}$} {a1} {z1} ;
    \factor[left=0.5 of a2] {p} {$\mat{W}_2, \vec{b}_2$} {z1} {a2} ;
    \factor[left=0.5 of x] {p} {$\sigma$} {a2} {x} ;
    \factor[right=0.5 of x] {p} {$\varphi$} {x} {y} ;
  \end{tikzpicture}
 \caption{Denoising/inpainting of a MNIST image with a VAE prior. The weights $\mat{W}_1, \mat{W}_2$ and biases $\vec{b}_1, \vec{b}_2$ are learned beforehand on the MNIST data set and fixed during the reconstruction.}
 \label{fig:factorgraph_vae}
\end{figure}
Following \cite{fletcher2018inference}, we use a structured prior coming from a \aclink{VAE} trained itself on the MNIST data set beforehand. The \aclink{VAE} architecture is summarized in \Fig\ref{fig:factorgraph_vae} and the training procedure follows closely the canonical one detailed in \citep{keras_vae}. We consider two common inference tasks: denoising and inpainting, which are simpler than the one considered in the previous section.
\paragraph{Denoising:}
In that case, the corrupted channel $\varphi_{\textrm{den}, \Delta}$ adds a Gaussian noise and corresponds to the noisy component-wise channel
$$\varphi_{\textrm{den}, \Delta}(\vec{x}) = \vec{x} + \bxi \textrm{ with } \xi_i \sim \mN(0,\Delta)\,.$$ 
\paragraph{Inpaiting:}
The corrupted channel erases a few pixels of the input image and corresponds formally to
$$\varphi_{\textrm{inp}, \textrm{I}_\alpha}(\vec{x}) = \vec{x} - m(\vec{x})\,,$$
where $m$ represents a mask applied component-wise. Let $\alpha \in [0;1]$, $\textrm{I}_{\alpha}$ denotes the set of erased indexes of size $\left\lfloor \alpha \ndim \right\rfloor$ and the masks acts according to $m(x_i)= \id\[ x_i \in \textrm{I}_{\alpha}\]$. As an illustration, we consider two different manner of generating the erased interval $I_\alpha$:
\begin{enumerate}
    \item A central horizontal band of width $\lfloor\alpha \ndim\rfloor$: $\textrm{I}_{\alpha}^{\textrm{band}}=[\lfloor\frac{\ndim}{2}(1 -\alpha)\rfloor ; \lfloor\frac{\ndim}{2}(1 + \alpha)\rfloor]$
    \item  $\lfloor\alpha \ndim\rfloor$ indices drawn uniformly at random: $\textrm{I}_{\alpha}^{\textrm{uni}} \sim \mathcal{U}([1,\ndim];\lfloor\alpha \ndim \rfloor)$
\end{enumerate}

Solving these inference tasks in \tramp is straightforward: first declare the structured model \Fig\ref{fig:factorgraph_vae} and then run \aclink{EP}. A few MNIST samples $\vec{x}^*\in\bbR^{784}$ in the test set, which were not used to train the \aclink{VAE}, compared to the noisy observations $\vec{y}\in\bbR^{784}$ and \tramp reconstructions $\hat{\vec{x}}$ are presented in \Fig\ref{fig:vae_denoising}. It suggest that the \aclink{EP} implementation of \tramp is able to use the trained \aclink{VAE} prior information to either denoise very noisy observations or reconstruct missing pixels of the MNIST images.
\begin{figure}[H]
\centering
\begin{minipage}[c]{0.45\linewidth}
  \includegraphics[width=\linewidth]{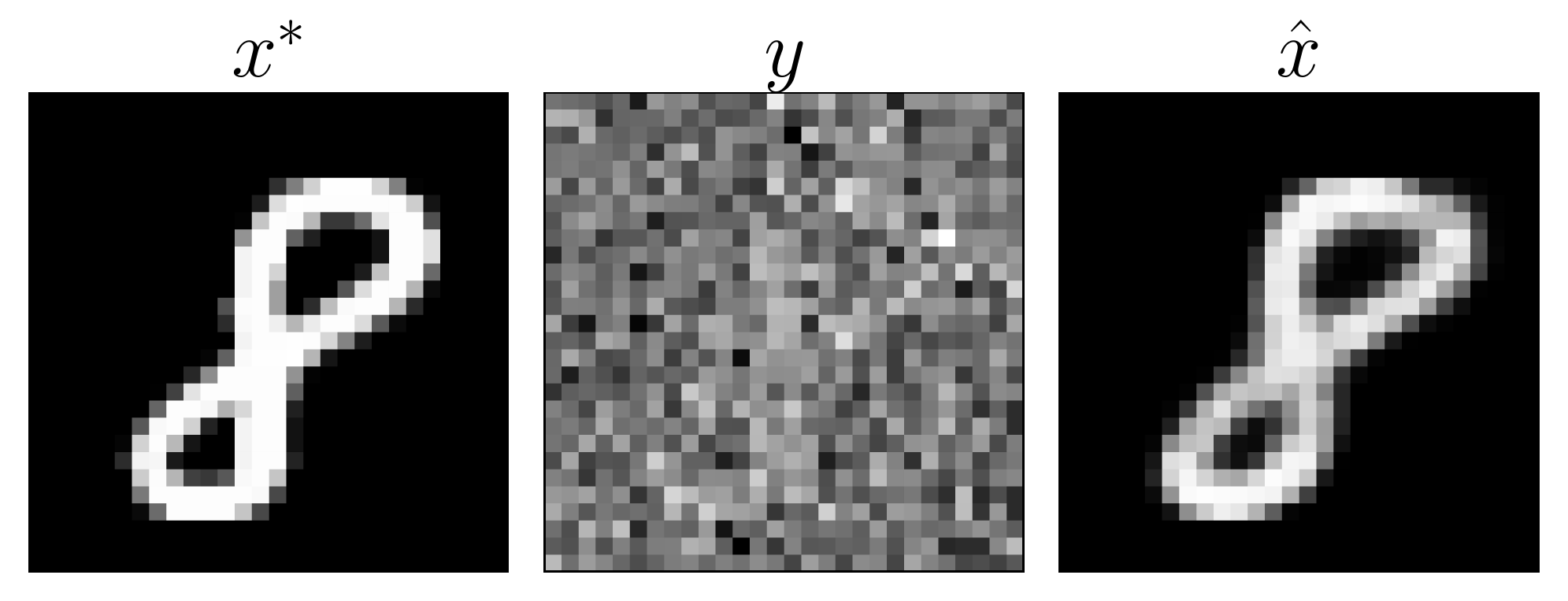}
  \includegraphics[width=\linewidth]{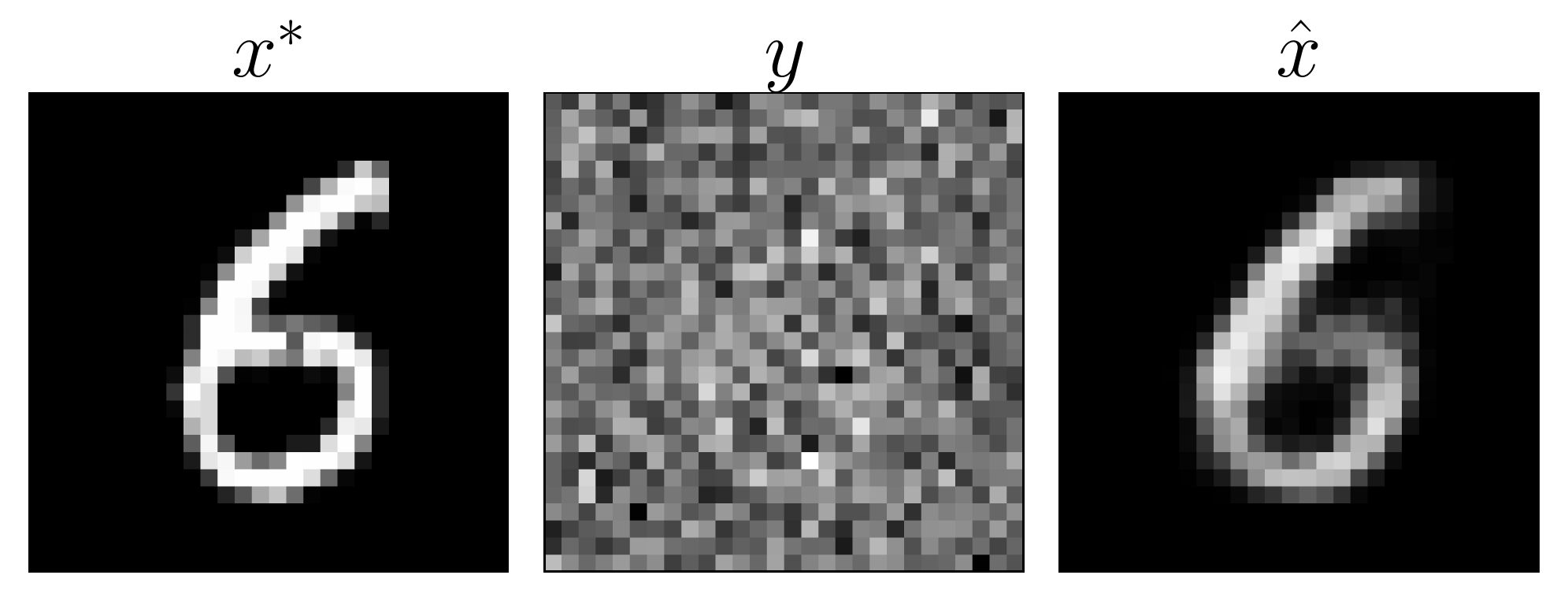}
\end{minipage}
\begin{minipage}[c]{0.45\linewidth}
  \includegraphics[width=\linewidth]{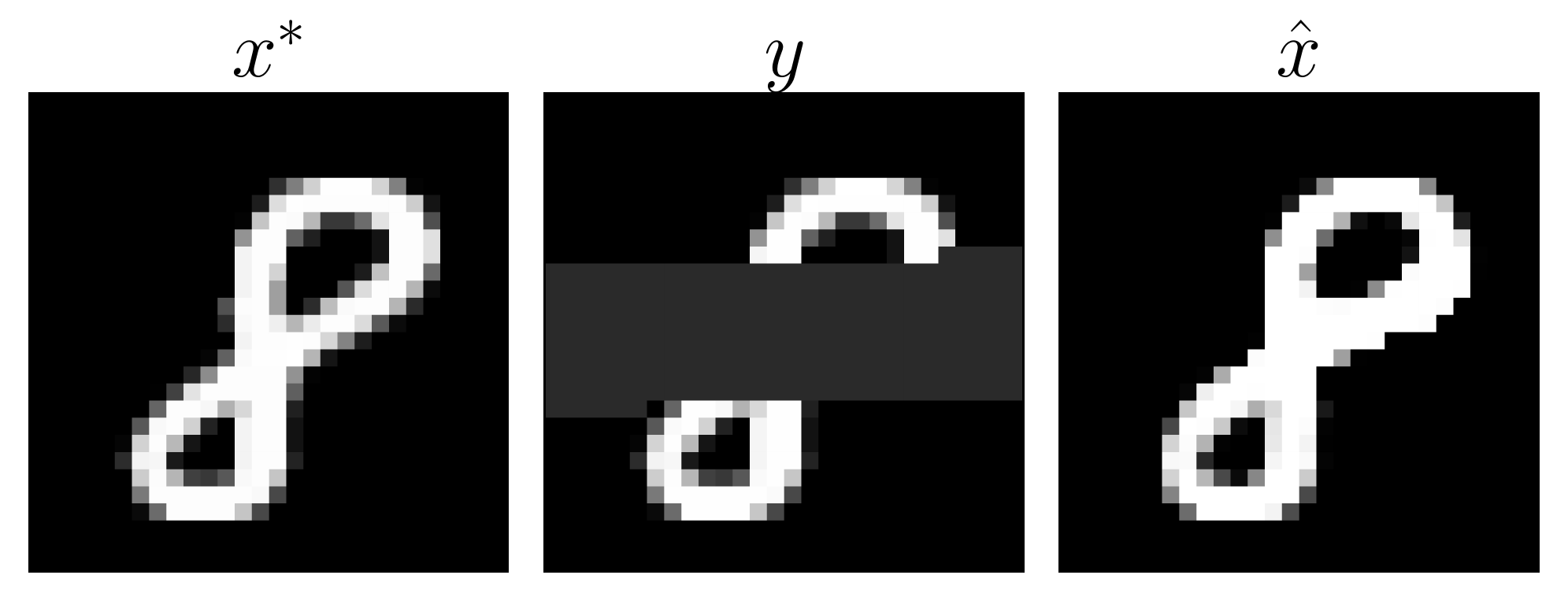}
  \includegraphics[width=\linewidth]{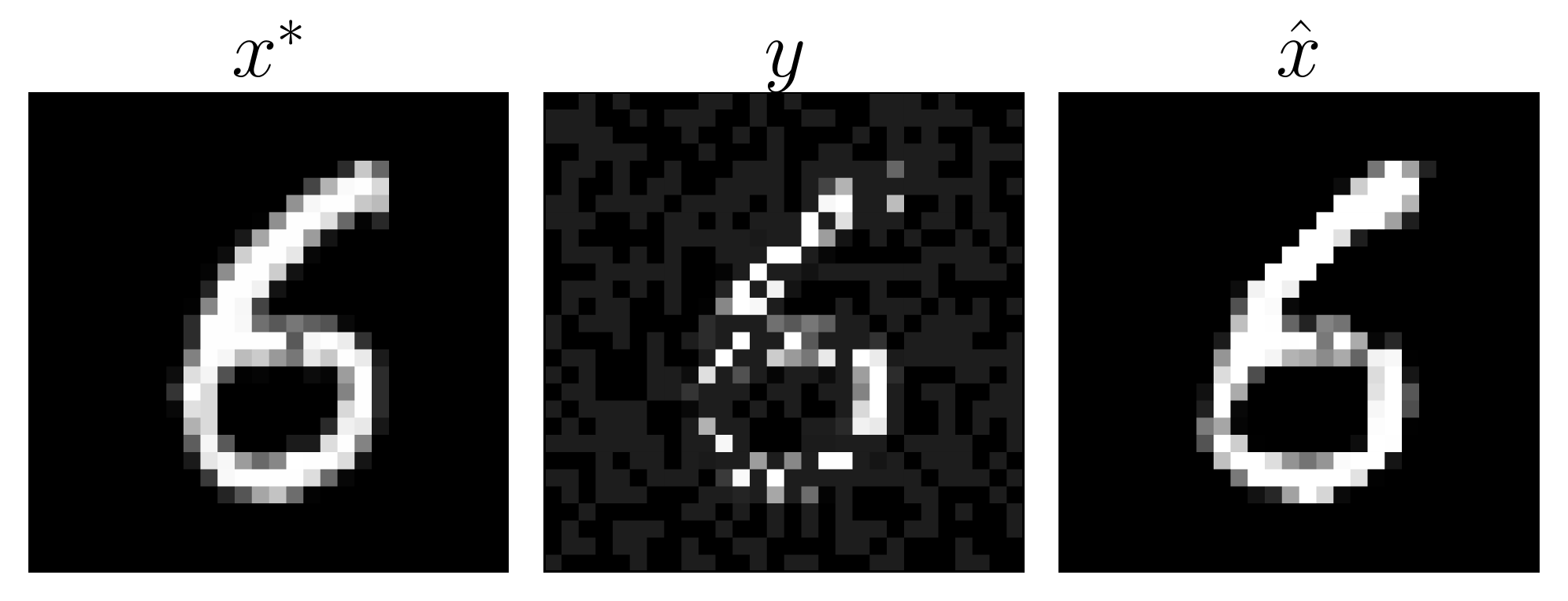}
\end{minipage}
    \caption{Illustration of the \tramp prediction $\hat{x}$ using a VAE prior from observation $y = \varphi(x^*)$ with $x^*$ a MNIST sample.
    \textbf{(Left)} Denoising $\varphi = \varphi_{\textrm{den}, \Delta}$ with $\Delta =4$.   
    \textbf{(Right-upper)} Band-inpainting $\varphi_{\textrm{inp}, I_\alpha^{\textrm{band}}}$ with $\alpha=0.3$
    \textbf{(Right-lower)} Uniform-inpainting $\varphi_{\textrm{inp}, I_\alpha^{\textrm{uni}}}$ with $\alpha=0.5$.}
    \label{fig:vae_denoising}
\end{figure}
However, analyzing the \aclink{SE} of \aclink{EP} or \aclink{VAMP} for such complex prior distribution encoded in the correlated weights of the \aclink{VAE} is still an ongoing line of research. As a conclusion, this direction shall be pushed further to provide a more accurate theoretical comparison between generative priors and separable sparse priors, and finally conclude on their respective performances on real data.

\cleardoublepage
\ifthenelse{\equal{\format}{oneside}}
	{
	\clearpage\null\thispagestyle{empty}\newpage
	}
	{\cleardoublepage}
	
\appendix
\ctparttext{}
\part{Appendices}
\ifthenelse{\equal{\format}{oneside}}
	{\clearpage\null\thispagestyle{empty}}
	{\cleardoublepage}
	
\chapter{Definitions and mathematical identities}

\section{Gaussian distribution and multivariate central limit theorem}	
\label{appendix:probability:clt}
		Consider $\vec{m}\in\bbR^{\ndim}$ and $\bSigma \in \bbR^{\ndim \times \ndim} $ a symmetric positive definite matrix. For $\vec{x}\in\bbR^{\ndim}$, the Gaussian probability distribution is defined by
		\begin{align}
			\mN_{\vec{x}}\( \vec{m}, \Sigma \) \equiv e^{-\frac{(\vec{x}-\vec{m})^\intercal \bSigma^{-1} (\vec{x}-\vec{m}) }{2}}/ \sqrt{\det{2\pi\bSigma}}\,.
		\end{align}
		The Gaussian vector $\vec{x}$ has mean $\EE\[ \vec{x} \] = \vec{m}$ and variance $\Var{\vec{x}} = \bSigma$. The Gaussian distribution is crucial as it turns out to be the fixed point distribution of the sum of \aclink{i.i.d} random variables as stated by the \aclink{CLT}:
		
		\begin{proposition}[ Multivariate Central Limit Theorem]
		Let $\{\vec{x}_1, \cdots, \vec{x}_n\}$ a sequence of \aclink{i.i.d} random vectors in $\bbR^{d}$ such that $\EE\[\vec{x}\] = \vec{m}$ and covariance matrix $\bSigma $. Defining $\vec{s}_\nsamples = \frac{1}{\nsamples} \sum_{\mu=1}^\nsamples \vec{x}_{\mu}$, then as $\nsamples$ approaches infinity, the sum converges in distribution to a Gaussian law:
			\begin{align}
				\sqrt{\nsamples}\(\vec{s}_\nsamples - \vec{m}\) \xrightarrow[\nsamples \to \infty ]{d} \mN_{\vec{x}}\(\vec{0},\bSigma\)\,.
			\end{align}
		\end{proposition}
		
\section{Hubbard-Stratonovich transformation}
\label{appendix:replica_computation:hubbard}

The Hubbard-Stratonovich transformation is a simple Gaussian identity based on the fact that:

\begin{proposition}[Hubbard-Stratonovich transformation]
	 For $\bxi \sim \mN\(\vec{0},\rI_\ndim\)$ and a symmetric positive definite matrix $\mat{A} \in \bbR^{\ndim \times \ndim}$, for all $\vec{x} \in \bbR^{\ndim}$
	 \begin{align}
	\EE_{\bxi}\exp\(  \bxi^\intercal \mat{A}^{1/2} \vec{x} \) = \(2\pi\)^{-\ndim/2} \int_{\bbR^\ndim} \d \bxi e^{-\frac{1}{2} \vec{x}^\intercal \vec{x} + \bxi^\intercal \mat{A}^{1/2} \vec{x} } = e^{\frac{1}{2} \vec{x}^\intercal \mat{A} \vec{x} }\,.
	\end{align}
\end{proposition}

\section{Nishimori identity}
\label{appendix:replica_computation:nishimori}
We recall the Nishimori identity from \cite{nishimori1980exact, nishimori1981internal, nishimori2001statistical,zdeborova2016statistical,lesieur2017constrained}:
\begin{proposition}[Nishimori identity] \label{prop:nishimori}
	Let $(\rX,\rY)$ a couple of random variables. Let $\{\vec{x}_{\mu}\}_{\mu=1}^{\nsamples}$ $\nsamples \geq 1$ samples drawn \aclink{i.i.d} from $\rP(\rX | \rY)$. Let us denote $ \langle . \rangle$ the expectation over the posterior distribution $\rP(\rX | \rY)$ and $\EE$ the expectation with respect to $(\rX, \rY)$. For all continuous bounded function $f$:
	\begin{align*}
		\EE\[ \left\langle  f(\rY, \vec{x}_{1}, \cdots , \vec{x}_{\nsamples -1}, \vec{x}_{\nsamples} \right\rangle\] = \EE\[ \left\langle  f(\rY, \vec{x}_{1}, \cdots , \vec{x}_{\nsamples -1}, \rX \right\rangle\]\,.
	\end{align*}
\end{proposition}
\begin{proof}
	This is a simple consequence of the Bayes formula.
	It is equivalent to sample the couple $(\rX, \rY)$ according to its joint distribution $\rP(\rX, \rY)$ or to sample first $\rY$ according to its marginal distribution $\rP(\rY)$ 
	and then to sample $\rX$ conditionally to $\rY$ from its conditional distribution $\rP(\rX|\rY)$. Thus the $(\nsamples +1)$-tuple $(\rY, \vec{x}_{1}, \cdots , \vec{x}_{\nsamples -1}, \vec{x}_{\nsamples})$ is equal 
	in law to $(\rY,\vec{x}_{1}, \cdots , \vec{x}_{\nsamples -1}, \rX)$.
	\end{proof}

\section{Denoising distributions, updates and free entropy terms}
\label{appendix:update_functions}

In this section, we introduce the $K$-dimensional probability distributions involved in the replica free entropies and from which the \aclink{AMP} update equations are derived in the context of \emph{committee machines}. The multivariate formulation can be simplify to scalar expressions for $K=1$ in the case of \aclink{GLM}.

\subsection{MMSE estimation with committee machines}
\label{appendix:update_functions:mmse_updates}

Analyzing the joint distribution $\rP\(\vec{y},\mat{X}\)$ for \aclink{MMSE} estimation in the high-dimensional regime boils down to introducing the denoising distributions $\rQ_{\w}, \rQ_{\out}$ on $\vec{w}\in\bbR^K$ and $\vec{z}\in\bbR^K$ and their respective normalizations $\mZ_{\w}$, $\mZ_{\out}$ in \Sec\ref{appendix:definitions:distributions:committee}. We define as well the denoising functions $\vec{f}_\w, \partial_\bgamma \vec{f}_\w$, $\vec{f}_\out, \partial_\bomega \vec{f}_\out$ in \Sec\ref{appendix:definitions:updates:committee}, that play a central role in Bayesian inference. Note in particular that they correspond to the \emph{updates} of the \aclink{GAMP} algorithm in \cite{rangan2011generalized} that we recall in \Alg\ref{alg:appendix:amp:committee_machine} for the committee machine hypothesis class. They are simply defined as the derivatives of $\log \mZ_{\w}$ and $\log \mZ_{\out}$. Finally the free entropy can be expressed as a function of simple free entropy terms $\Psi_\w, \Psi_\out$ defined in \Sec\ref{appendix:definitions:free_entropy_terms:committee}.\\

Consider $y\in \bbR$, $\bgamma, \bomega \in \bbR^{K}$, $\bLambda,\bV \in \mS_K^+$, the ensemble of symmetric positive matrices of size $K \times K$, and vectors to infer $\vec{w}, \vec{z} \in \bbR^K$, with prior distributions $\rP_\w$, $\rP_\out$.
				
	\subsubsection{Denoising distributions}
	\label{appendix:definitions:distributions:committee}

		\begin{align}
			\rQ_{\w}(\vec{w}; \bgamma, \bLambda) & \equiv \displaystyle \frac{\rP_{\w}(\vec{w}) }{\mZ_{\w} (\bgamma,\bLambda)} e^{ - \frac{1}{2} \vec{w}^\intercal \bLambda \vec{w}  + \bgamma^\intercal \vec{w}  }\,, \label{appendix:definitions:update_generic:committee:prior:Qw} \\
			\mZ_{\w} (\bgamma,\bLambda) &\equiv \EE_{\vec{w}\sim \rP_\w} \[ e^{ - \frac{1}{2} \vec{w}^\intercal \bLambda \vec{w}  + \bgamma^\intercal \vec{w}  } \] \label{appendix:definitions:update_generic:committee:prior:Zw}\\
			&= \int_{\bbR^K} \d \vec{w} ~ \rp_{\w}(\vec{w})  e^{ - \frac{1}{2} \vec{w}^\intercal \bLambda \vec{w}  + \bgamma^\intercal \vec{w}  }\,, \notag \Spacecase
			\rQ_{\out} (\vec{z}; y, \bomega, \bV ) & \equiv  \displaystyle\frac{\rP_{\out}\( y | \vec{z}\) }{\mZ_{\out}(y, \bomega, \bV)}  \frac{e^{ -\frac{1}{2}  \(\vec{z} - \bomega\)^\intercal \bV^{-1} \(\vec{z} - \bomega\)  }}{\sqrt{\det{2\pi \bV}}}\,, \label{appendix:definitions:update_generic:committee:channel:Qout}  \\
			\mZ_{\out}(y, \bomega, \bV) &\equiv \EE_{\vec{z} \sim \mN(\vec{0},\rI_K)}\[ \rP_{\out}\(y | \bV^{1/2} \vec{z} + \bomega \) \]   \label{appendix:definitions:update_generic:committee:channel:Zout} \\
			&= \int_{\bbR^K} \d \vec{z} ~ \rp_{\out}\(y | \vec{z} \) \frac{e^{ -\frac{1}{2}  \(\vec{z} - \bomega\)^\intercal \bV^{-1} \(\vec{z} - \bomega\)  }}{\sqrt{\det{2\pi \bV}}}\,. \notag
		\end{align}

	\subsubsection{Denoising updates}
	\label{appendix:definitions:updates:committee}
	
		\begin{align}
		\vec{f}_{\w}(\bgamma ,\bLambda) &\equiv  \partial_\bgamma \log\(\mZ_{\w}(\bgamma,\bLambda)\) = \EE_{\rQ_{\w}} \[ \vec{w} \]\,, \label{appendix:definitions:update_generic:committee:prior:fw} \Spacecase 
		\partial_\bgamma \vec{f}_{\w} (\bgamma ,\bLambda) &\equiv  \EE_{\rQ_{\w}} \[\vec{w} \vec{w}^\intercal \] - \vec{f}_{\w}(\bgamma ,\bLambda) ^{\otimes 2} \label{appendix:definitions:update_generic:committee:prior:dfw}\,,\Spacecase
		\vec{f}_{\out} (y,\bomega, \bV) &\equiv \partial_\bomega \log \( \mZ_{\out}(y, \bomega, \bV) \) = \bV^{-1} \EE_{\rQ_{\out}} \[ \vec{z} - \bomega\]\,, \label{appendix:definitions:update_generic:committee:channel:fout} \Spacecase	
		\partial_{\bomega} \vec{f}_{\out} (y, \bomega, \bV) &\equiv \displaystyle \frac{\partial \vec{f}_{\out}(y,\bomega, \bV)}{\partial \bomega}\label{appendix:definitions:update_generic:committee:channel:dfout} \\
			&= \bV^{-1} \EE_{\rQ_{\out}} \[ \(\vec{z} - \bomega\)^{\otimes2} \]\bV^{-1} - \bV^{-1} - \vec{f}_{\out} (y,\bomega, \bV)^{\otimes2} \,.\notag
		\end{align}
			
	\subsubsection{Free entropy terms}
	\label{appendix:definitions:free_entropy_terms:committee}
		For overlap matrices $\mat{Q}^\star, \mat{Q} \in \bbR^{K\times K}$, and second moments $\brho^\star,\brho\in \bbR^{K\times K}$,
		\begin{align}
			\Psi_{\w} (\mat{Q}^\star, \mat{Q}) & \equiv \EE_{\bxi} \[ \mZ_\w \( (\mat{Q}^\star)^{1/2} \bxi , \mat{Q}^\star  \) \log \( \mZ_\w \( \mat{Q}^{1/2} \bxi ,\mat{Q}  \) \) \] \,, 
			\label{appendix:definitions:free_entropy:committee:prior}\spacecase
			\Psi_{\out} (\mat{Q}^\star, \mat{Q}, \brho^\star, \brho) & \equiv \EE_{\bxi} \[\mZ_{\out}\( (\mat{Q}^\star)^{1/2} \bxi , \brho^\star - \mat{Q}^\star\) \right. \notag \\
			& \qquad \qquad  \left. \times \log\( \mZ_{\out}\( \mat{Q}^{1/2} \bxi  , \brho - \mat{Q} \) \) \] \label{appendix:definitions:free_entropy:committee:channel}
		\end{align}

\subsection{MAP estimation with GLM}
\label{appendix:update_functions:map_updates}
Before defining similar denoising functions to analyze \aclink{MAP} estimation, we first recall the definition of the Moreau-Yosida regularization in the scalar case $K=1$.

\subsubsection{Moreau-Yosida regularization and proximal}
\label{appendix:definitions:moreau-yosida}
Let $\Sigma>0$, $f(,z)$ a convex function in $z \in \bbR$, the Moreau-Yosida regularization $\mM_\bSigma$ and the proximal map $\mP_\bSigma$ are defined by
\begin{align}
	\mP_{\Sigma}[f(,.)](x) &= \argmin_z \[f(,z) + \frac{1}{2\Sigma}\(z - x\)^2\] \,, \label{appendix:definitions:proximal}\\
	\mM_{\Sigma}[f(,.)](x) &= \min_z \[f(,z) + \frac{1}{2\Sigma}\(z - x\)^2\] \,, 
	\label{appendix:definitions:moreau}
\end{align}
where $(,.)$ denotes all the arguments of the function $f$. 

\subsubsection{MAP denoising functions}
\label{appendix:definitions:update_functions_map}

The \aclink{MAP} denoising functions for any convex loss $l(,.)$ and convex separable regularizer $r(.)$ can be written in terms of the Moreau-Yosida regularization or the proximal map as follows
\begin{align}
	f_{\w}^{{\map}, r}(\gamma, \Lambda) &\equiv  \mP_{\Lambda^{-1}}\[ r(.) \](\Lambda^{-1}\gamma) \label{eq:appendix:definitions:update_functions_map_prior} \\
	&= \Lambda^{-1}\gamma - \Lambda^{-1} \partial_{\Lambda^{-1}\gamma}\mM_{\Lambda^{-1}}\[ r(.) \] (\Lambda^{-1}\gamma)\,, \nonumber \spacecase 
	f_{\out}^{{\map}, l} (y, \omega, V) &\equiv  - \partial_{\omega} \mM_{V}[l(y,.)](\omega) \label{eq:appendix:definitions:update_functions_map_channel} \\
	&= V^{-1} \( \mP_{V}[l(y,.)](\omega) - \omega\)\,. \nonumber
\end{align}
The derivation and the applications are detailed in \App I.3 of \cite{aubin2020generalization}.

\ifthenelse{\equal{\format}{oneside}}
	{\clearpage\null\thispagestyle{empty}
	\clearpage\null\thispagestyle{empty}}
	{\cleardoublepage}
	
\chapter{Replica computations}
\label{appendix:replica_computation}
	
	\section{Teacher-student - Committee machine with i.i.d data} 
	\label{appendix:replica_computation:committee}
	
In this section, we present the heuristic derivation of the replica formula of Theorem~\ref{main-thm} using the replica method, presented in \Sec\ref{main:sec:mean_field:replica_method}, in the context of  of the \emph{committee machine}. This computation is necessary to properly guess the formula that we then prove using the adaptive interpolation method. The reader interested in the replica approach to neural networks and the committee machine is invited to look as well to some of the classical papers \cite{gardner1988optimal,mezard1989space,schwarze1992generalization,schwarze1993generalization,schwarze1993learning,monasson1995learning}.
In the \emph{teacher-student} setting, the committee machine estimation problem consists of trying to estimate a \emph{teacher} signal $\mat{W}^\star \in \bbR^{\ndim \times K}$ from a set of $\nsamples$ input-output observations $\{\mat{X}, \vec{y} \}\in \bbR^{\nsamples \times \ndim} \times \bbR^{\nsamples}$ generated according to
\begin{align*}
	\vec{y} = \varphi_\out \(\frac{1}{\sqrt{\ndim}} \mat{X} \mat{W}^\star \) = \varphi_\out \( \left\{ \frac{1}{\sqrt{\ndim}} \mat{X} \vec{w}_k^\star \right \}_{k=1}^K \) \,.
\end{align*}
The student, within the same committee machine hypothesis class, with parameters $\mat{W}\in \bbR^{\ndim \times K}$, tries to learn the teacher rule generated by the ground truth weights $\mat{W}^\star$.
Committee machines are a simple vectorized generalization of \aclink{GLM}, defined in \Sec\ref{main:introduction:glm_class}, whose estimation is performed simultaneously with $K \geq 1$ \aclink{GLM}. Therefore, the replica computation is shown only in the general committee machine case and final expressions for \aclink{GLM} will be derived as a particular case for $K=1$.

We will assume that the matrix of data inputs $\mat{X} \in \bbR^{\nsamples \times \ndim }$ is drawn \aclink{i.i.d} with density $\rp_{\x}$. We will consider them to be \aclink{i.i.d} Gaussian with zero mean and unit variance: $\forall \mu \in \lb \nsamples \rb, \vec{x}_{\mu} \sim \mN_{\vec{x}}\(\vec{0}, \mat{I}_{\ndim}\)$. The function $\varphi_\out : \bbR^{K} \mapsto \bbR$ represents a deterministic, or stochastic function associated to a probability distribution $\rP_\out$, applied component-wise to each sample. Notice that the factor $\frac{1}{\sqrt{\ndim}}$ is present to insure that the variance of the input data is normalized to the unit.

\subsection{Replica calculation}
\label{appendix:replica_computation:committee}

	\subsubsection{On statistical estimation}
	\label{appendix:replica_computation:committee:estimation}
		Both \aclink{MMSE} and \aclink{MAP} estimations boil down to the analysis of the posterior distribution $\rP\(\mat{W} | \vec{y}, \mat{X}\)$ expressed by the Bayes rule
			\begin{align}
				\rP\(\mat{W} | \vec{y}, \mat{X}\)  &= \frac{\bbP\(\vec{y} | \mat{W}, \mat{X} \) \bbP\(\mat{W}\) }{\rP\(\vec{y},\mat{X}\)} = \frac{\rP_{\out} \(\vec{y} | \mat{W}, \mat{X} \) \rP_{\w}\(\mat{W}\)}{\mZ_\ndim\(\{\vec{y}, \mat{X}\} \)}   \,.
				\label{appendix:replica:committee:bayes_formula}
			\end{align}
			The joint distribution is also called the \emph{partition function} $\rP\(\vec{y},\mat{X}\) \equiv \mZ_\ndim\(\{\vec{y}, \mat{X}\} \)$. To connect with the statistical physics formalism, we introduce the corresponding Hamiltonian, for separable distributions $\rP_{\out}, \rP_{\w}$ along one dimension, by
		\begin{align*}
			\mH_\ndim\(\mat{W},\{\vec{y},\mat{X}\}\) &=- \log \rP_{\out} \(\vec{y} | \mat{W}, \mat{X} \) -  \log \rP_{\w}\(\mat{W}\)\,,\\
			&= - \sum_{\mu=1}^\nsamples \log \rP_{\out} \(y_\mu | \mat{W}, \vec{x}_{\mu} \) - \sum_{i=1}^\ndim \rP_{\w}\(\vec{w}_i\)\,.
		\end{align*}
		The spin variables classically denoted $\bsigma$ are replaced by the weights of the model $\mat{W} \in \bbR^{\ndim \times K}$ and they interact through the random dataset $\{\vec{y}, \mat{X}\}$ that plays the role of the quenched exchange interactions $\mat{J}$. However here, the interactions are not \emph{pairwise}, as it is often the case in the Ising-like models, but instead \emph{fully connected}, meaning that each variable $\vec{w}_i \in \bbR^{K}$ is connected to every other spin $\{\vec{w}_j\}_{j \in \partial j \setminus i }$ as represented in the factor graph in \Fig\ref{fig:appendix:factor_graph_committee}.
			\begin{figure}[htb!]
			\centering
			\begin{tikzpicture}[scale=0.8, auto, swap]
			    \foreach \i in {1,...,6}
			        \node[var] (X\i) at (1.5*\i,0) {};
			    \node at (11, 0) {$ \vec{w}_i \in \bbR^K $};
			
			    \foreach \mu in {1,...,4}
			        \node[inter] (Y\mu) at (1.5+1.5*\mu,-2) {};
			    \foreach \i in {1,...,6}
			        \foreach \mu in {1,...,4}
			            \path[edge] (X\i) -- (Y\mu);
			    \node at (10, -2) {};
			    \node (F) at (11, -2) {$ \rp_{\out}\(y_\mu | \mat{W}, \vec{x}_{\mu} \) $};			
			    \foreach \i in {1,...,6} {
			        \node[field] (P\i) at (1.5*\i,1) {};
			        \path[edge] (X\i) -- (P\i);
			    }
			    \node at (11, 1) {$ \rp_\w(\vec{w}_i) $};
			\end{tikzpicture}
			\caption{Factor graph corresponding to the committee machines hypothesis class. The vectorial variables to infer $\vec{w}_i$ are fully connected through the quenched disorder $\vec{y} \sim \rP_{\out^\star} (.)$ and each variable follow a one-body interaction with a separable prior distribution $\rp_\w(\vec{w}_i)$.}
			\label{fig:appendix:factor_graph_committee}
			\end{figure}
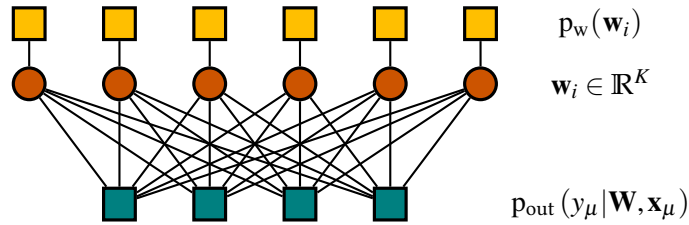
		The partition function at inverse temperature $\beta$ is therefore defined by 
			\begin{align}
			\begin{aligned}
			\mZ_\ndim\(\{\vec{y}, \mat{X}\}; \beta\) &\equiv \rP\(\vec{y},\mat{X}\) = \int_{\bbR^{\ndim\times K}} \d\mat{W} ~ e^{-\beta \mH_\ndim \(\mat{W},\{\vec{y},\mat{X}\}\)}  \\
			&= \int_{\bbR^{\ndim\times K}} \d\vec{w} ~ e^{\beta\( \log \rp_{\out} \(\vec{y} | \mat{W}, \mat{X} \) + \log P_{\w}\(\mat{W}\) \)}\\
			&= \int_{\bbR^{\ndim\times K}} \d\vec{w} ~ \rp_{\out} \(\vec{y} | \mat{W}, \mat{X} \) \rp_{\w}\(\mat{W}\)\,,
			\end{aligned}
			\end{align}
			and can be exactly mapped to Bayesian estimation for $\beta=1$. In the context of \aclink{ERM}, \aclink{MAP} estimation can be analyzed  by taking the limit $\beta\to \infty$ as detailed in \Chap\ref{chap:erm}. 
			In the considered modern high-dimensional regime with $\ndim \to \infty$, $\nsamples \to\infty$, $\alpha = \nsamples/\ndim = \Theta(1)$ and $K=\Theta(1)$, we are interested  in computing the \emph{free entropy} $\Phi$ \eqref{main:intro:averaged_free_entropy_free_energy}, \emph{averaged} over the input data $\mat{X}$ and teacher weights $\mat{W}^\star$, or equivalently over the output labels $\vec{y}$ generated from it, defined as
			\begin{align}
				\Phi(\alpha) \equiv \lim_{\ndim \to \infty}  \frac{1}{\ndim}  \EE_{\vec{y},\mat{X}} \[\log  \mZ_\ndim\(\vec{y}, \mat{X}\) \]\,.
				\label{appendix:replica:committee:free_entropy}
			\end{align}
			The heuristic replica method described in \Sec\ref{main:sec:mean_field:replica_method} allows to compute the above average over the random dataset $\{\vec{y}, \mat{X}\}$, that plays the role of the quenched disorder in usual spin glasses.
			We show the computation for the more involved committee machine model class and generalization of the \aclink{GLM} class, only for \aclink{i.i.d} data. The cumbersome computation for non \aclink{i.i.d} data can be performed as well and lead to more complex expressions and has been performed in particular in \cite{kabashima2008inference} in the case of the \aclink{GLM}. 
										
		\subsubsection{Replica computation}
		\label{appendix:replica_computation:committee:replica_derivation}

			We present here the replica computation of the averaged free entropy $\Phi(\alpha)$ in eq.~\eqref{appendix:replica:committee:free_entropy} for arbitrary \emph{student} prior and channel distributions $\rP_{\w},\rP_{\w^\star}$ and $\rP_{\out},\rP_{\out^\star}$, so that the computation remains valid for both the Bayes-optimal and mismatched settings.
			The average in eq.~\eqref{appendix:replica:committee:free_entropy} is intractable in general, and the computation relies on the so called \emph{replica trick}, see \Sec\ref{main:intro:replicas:replica_trick}, that consists in applying the identity  
			\begin{align}
				\EE_{\vec{y},\mat{X}} \[ \lim_{\ndim \to \infty} \frac{1}{\ndim} \log  \mZ_\ndim\(\vec{y}, \mat{X}\) \] =  \lim_{r \to 0} \[ \lim_{\ndim \to \infty}  \frac{1}{\ndim}  \frac{\partial \log \EE_{\vec{y},\mat{X}} \[  \mZ_\ndim\(\vec{y}, \mat{X}\)^r\] }{\partial r} \]\,.
				\label{appendix:replica:committee:replica_trick}
			\end{align}
			The replica trick has been used in a series of previous works to compute the free energy density of \aclink{GLM} for separable distributions \cite{krzakala2012probabilistic} and has been rigorous proved in this case by \cite{barbier2017phase}. 
			Eq.~\eqref{appendix:replica:committee:replica_trick} is interesting in the sense that it reduces the intractable average to the computation of the moments of the averaged partition function, which are easier quantities to compute. Note that for $r \in \bbN$, $\mZ_\ndim\(\vec{y}, \mat{X}\)^r = \prod_{a=1}^r \mZ_\ndim\(\vec{y}, \mat{X}\)$ represents the partition function of $r$ identical non-interacting copies of the initial system, called \emph{replicas}. Taking the quenched average over the disorder will then correlate the replicas, before taking the number of replicas $r\to 0$.
			Therefore, we assume there exists an analytical continuation so that $r\in \bbR$ and the limit is well defined. Finally, notice that we exchanged the order of the limits $r \to 0$ and $\ndim \to \infty$. These technicalities are crucial points but are not rigorously justified and we will ignore them in the rest of the computation.
			First, in order to decouple the contributions of the channel $\rP_\out$ and the prior $\rP_\w$, we introduce the variable $\mat{Z} = \frac{1}{\sqrt{\ndim}} \mat{X} \mat{W}$ and a Dirac-delta integral:
			\begin{align*}
				\mZ_\ndim\(\vec{y}, \mat{X}\) &= \int_{\bbR^{\nsamples\times K}} \d\vec{z} ~ \rp_{\out}\(\vec{y} | \mat{Z} \) \int_{\bbR^{\ndim\times K}} \d\vec{w} ~ \rp_{\w}\(\mat{W}\) \delta\(\mat{Z} - \frac{1}{\sqrt{\ndim}} \mat{X} \mat{W}\)\,.
			\end{align*} 
			Thus the replicated partition function for an integer $r\in \bbN$ in eq.~\eqref{appendix:replica:committee:replica_trick} can be  written as
			\begin{align}
			&\EE_{\vec{y},\mat{X}} \[  \mZ_\ndim\(\vec{y}, \mat{X}\)^r \] \nonumber \\
			&=  \EE_{\vec{y},\mat{X}} \[  \prod_{a=1}^r \int_{\bbR^\nsamples} \d\mat{Z}^a ~ \rp_{\out^a}\(\vec{y} | \mat{Z}^a \) \right. \nonumber \\
			& \qquad \qquad \left. \times  \int_{\bbR^{\ndim \times K}} \d\mat{W}^a ~ \rp_{\w^a}\(\mat{W}^a\) \delta\(\mat{Z}^a - \frac{1}{\sqrt{\ndim}}\mat{X} \mat{W}^a\)\] \nonumber \\ 
			&= \EE_{\mat{X}} \int_{\bbR^\nsamples} \d \vec{y} ~ \int_{\bbR^{\nsamples \times K}} \d\mat{Z}^\star ~ \rp_{\out^\star}\(\vec{y} | \mat{Z}^\star \)  \\
			& \qquad \qquad \times  \int_{\bbR^{\ndim \times K}} \d\mat{W}^\star ~ \rp_{\w^\star}\(\mat{W}^\star\) \delta\(\mat{Z}^\star - \frac{1}{\sqrt{\ndim}}\mat{X} \mat{W}^\star\)\nonumber \\
			& \times \[  \prod_{a=1}^r \int_{\bbR^{\nsamples \times K}} \d\mat{Z}^a ~ \rp_{\out^a}\(\vec{y} | \mat{Z}^a \) \int_{\bbR^{\ndim\times K}} \d\mat{W}^a ~ \rp_{\w^a}\(\mat{W}^a\) \delta\(\mat{Z}^a - \frac{1}{\sqrt{\ndim}}\mat{X} \mat{W}^a\)\]  \nonumber \\
			&= \int_{\bbR^\nsamples} \d \vec{y} ~ \prod_{a=0}^r \int_{\bbR^{\nsamples \times K}} \d\mat{Z}^a ~ \rp_{\out^a}\(\vec{y} | \mat{Z}^a \) \int_{\bbR^{\ndim \times K}} \d\mat{W}^a ~ \rp_{\w^a}\(\mat{W}^a\) \nonumber \\
			& \qquad \qquad  \times \underbrace{\EE_{\mat{X}} \prod_{a=0}^r \delta\(\mat{Z}^a - \frac{1}{\sqrt{\ndim}}\mat{X} \mat{W}^a\)}_{(I)}\,. \nonumber
			\end{align}
			Note that the average over $\vec{y}$ is equivalent to the one over the ground truth vector $\mat{W}^\star$ in the case of a \emph{teacher-student}, which can be conveniently grouped with the other terms by just extending the replica indices and
			considering it as a new replica $\mat{W}^0$ with index $a=0$, leading to a total of $r+1$ replicas. 
			
		\paragraph{Average over the i.i.d input data $\mat{X}$}	
		\label{appendix:replica_computation:committee:average}	
			Remains to compute the average over $\mat{X}$ in the term $(I)$.
			We suppose that inputs are drawn from an \aclink{i.i.d} distribution, for example a Gaussian $\rp_{\textrm{x}}(\vec{x}) = \mN_{\vec{x}}\(\vec{0},\mat{I}_\ndim\)$. More precisely, for $(i,j) \in \lb \ndim\rb^2$, $(\mu,\nu) \in \lb \nsamples \rb^2$, $\EE_\mat{X} \[ x_{\mu i} x_{\nu j} \] =  \delta_{\mu\nu} \delta_{ij}$. 
			By definition, the average in $(I)$ defines the probability density $\rp_{\z^a} (\mat{Z}^a)$ and as $\forall k \in \lb K \rb, \forall \mu \in \lb\nsamples \rb$, $z_{\mu k}^a =\frac{1}{\sqrt{\ndim}} \sum_{i=1}^\ndim x_{\mu i} w_{ik}^a$ is the sum of \aclink{i.i.d} random variables, the \aclink{CLT} insures that in the thermodynamic limit $\ndim \to \infty$, $z_{\mu k}^a$ follows a Gaussian multivariate distribution, with first moments given by:
			\begin{align*}
					\EE_{\mat{X}}[z_{\mu k}^a] &= \frac{1}{\sqrt{\ndim}} \sum_{i=1}^\ndim \EE_{\mat{X}}\[x_{\mu i}\] w_{ik}^a = 0 \spacecase
					\EE_{\mat{X}}[z_{\mu k}^a z_{\nu k'}^b] &= \frac{1}{\ndim} \sum_{ij} \EE_{\mat{X}}\[x_{\mu i} x_{\mu j}\] w_{ik}^a w_{jk'}^b  = \frac{1}{\ndim} \sum_{ij}  \delta_{ij} w_{ik}^a w_{jk'}^b \delta_{\mu \nu}\\
					& \equiv \delta_{\mu \nu} Q^{a k}_{b k'}.
			\end{align*}
			Notice that averaging over the quenched disorder introduced correlations between replicas, which were initially independent, described by the symmetric \emph{overlap} matrix $\{ Q^{a k}_{b k'}\}_{kk'}$ of size $(r+1)K \times (r+1)K$.
			This matrix order parameter measures the correlations between the replicated matrices $\{\mat{W}^a\}_{a=0}^r$ and is formally defined by
			$$\mat{Q}(\{\mat{W}^a\}_{a=0}^r)\equiv\(\frac{1}{\ndim} \sum_{i=1}^{\ndim} w_{ik}^a w_{ik'}^b \)^{a,b=0..r}_{k,k'=1..K}\,,$$ 
			such that $\forall (a,b) \in \lb 0 : r \rb^2 $, $\mat{Q}^{ab} \in \bbR^{K \times K}$. 
			Therefore, again by the \aclink{CLT}, in the limit $d\to\infty$, the hidden variable $\mat{Z}^a \in \bbR^{\nsamples \times K}$ converges in distribution to the multivariate distribution
			$$\rp_{\z^a}\(\mat{Z}^a|\mat{Q}\) = \exp \left[- \frac{1}{2} \sum_{\mu =1}^\nsamples \sum_{a,b=0}^r \sum_{k,k'=1}^K z^{a}_{\mu k} z^b_{\mu k'}(\mat{Q}^{-1})_{k k'}^{ab} \right] / \( \det{2\pi \mat{Q}} \)^{\frac{\nsamples}{2}}\,.$$ 	
			Inserting this back in the replicated partition function finally writes
			\begin{align*}
			&\EE_{\vec{y},\mat{X}} \[  \mZ_\ndim\(\vec{y}, \mat{X}\)^r \] = \\
			& \qquad \int_{\bbR^\nsamples} \d \vec{y} ~ \prod_{a=0}^r \int_{\bbR^{\nsamples \times K}} \d\mat{Z}^a ~ \rp_{\out^a}\(\vec{y} | \mat{Z}^a \) \rp_{\z^a}\(\mat{Z}^a|\mat{Q}\) \int_{\bbR^{\ndim \times K}} \d\mat{W}^a ~ \rp_{\w^a}\(\mat{W}^a\)
			\end{align*}
						
			\paragraph{Fourier representation}
			Next we introduce the change of variable for the new order parameter $\mat{Q}^{ab}$ with a Dirac-$\delta$ distribution and its Fourier representation. For a variable $x \in \bbR$, the distribution $\delta(x)$ can be written as an integral over a purely imaginary parameter $\hat{x}$:
			\begin{align*}
				\delta\(x\) = \frac{1}{2 i\pi} \int_{i \bbR} \d \hat{x} e^{-\hat{x} x }\,.
			\end{align*}
			 Applying the above identity to the change of variable, we obtain
			 \begin{align*}
			 	&1 = \int_{\bbR^{(K \times r+1)^2 }} \d \mat{Q} \prod_{0 \leq a \leq b \leq r, 1 \leq k,k'\leq K} \delta \(\ndim Q^{ab}_{kk'}-\sum_{i=1}^\ndim w_{ik}^a w_{ik'}^b \)\\
				&\propto \int \int_{\bbR^{(K \times r+1)^2}} \d \mat{Q} \d\hat{\mat{Q}} ~ \exp \( - \ndim \sum_{a=0}^r \sum_{k, k'}^K Q^{aa}_{k k'} \hat{Q}^{aa}_{k k'} - \frac{\ndim}{2} \sum_{a \neq b}^r \sum_{k, k'}^K Q^{a b}_{k k'} \hat{Q}^{a b}_{k k'}  \) \\
				&  \qquad \qquad \times \exp \( \sum_{a=0}^r \sum_{k, k'}^K  \hat{Q}^{a a}_{k k'} w_k^a w_{k'}^a  + \frac{1}{2} \sum_{a \neq b}^r \sum_{k,k'}^K \hat{Q}^{a b}_{k k'} w_k^a w_{k'}^b\) \,,
				\end{align*} 
			that involves a new ad-hoc purely imaginary matrix parameter $\hat{\mat{Q}} \in i\bbR^{(K \times (r+1))^2 }$.
			Finally, multiplying the replicated partition function by $1$, using the Cauchy theorem and rotating the integration, it becomes an integral over the symmetric matrices $\mat{Q} \in \bbR^{(K \times r+1)^2}$ and $\hat{\mat{Q}} \in \bbR^{(K \times r+1)^2}$
			 \begin{align*}       
			  &\EE_{\vec{y},\mat{X}} \[  \mZ_\ndim\(\vec{y}, \mat{X}\)^r \] \\
			  &= \int \int_{\bbR^{(K \times r+1)^2}} \d \mat{Q} \d\hat{\mat{Q}} ~ \exp \( - \ndim \sum_{a=0}^r \sum_{k,k'}^K Q^{aa}_{kk'} \hat{Q}^{aa}_{kk'} - \frac{\ndim}{2} \sum_{a \neq b}^r \sum_{k, k'}^K Q^{a b}_{k k'} \hat{Q}^{a b}_{k k'}  \) \\
			  & \qquad \qquad \times \exp \( \sum_{a=0}^r \sum_{k, k'}  \hat{Q}^{a a}_{k k'} w_k^a w_{k'}^a  + \frac{1}{2} \sum_{a \neq b} \sum_{k, k'} \hat{Q}^{a b}_{k k'} w_k^a w_{k'}^b\)\\
			  & \int_{\bbR^\nsamples} \d \vec{y}  \prod_{a=0}^r \int_{\bbR^{\nsamples \times K}} \d\mat{Z}^a ~ \rp_{\out^a}\(\vec{y} | \mat{Z}^a \) \rp_{\z^a}\(\mat{Z}^a | \mat{Q} \) \int_{\bbR^{\ndim \times K}} \d\mat{W}^a ~ \rp_{\w^a}\(\mat{W}^a\) \\
			  &= \int \int_{\bbR^{(K \times r+1)^2}} \d \mat{Q} \d\hat{\mat{Q}} ~ \exp \( - \ndim \sum_{a=0}^r \sum_{k, k'}^K Q^{aa}_{kk'} \hat{Q}^{aa}_{kk'} - \frac{\ndim}{2} \sum_{a \neq b}^r \sum_{k, k'}^K Q^{a b}_{k k'} \hat{Q}^{a b}_{k k'}  \) \\
			  & \qquad \qquad \times \exp \( \sum_{a=0}^r \sum_{k, k'}^K  \hat{Q}^{a a}_{k k'} w_k^a w_{k'}^a  + \frac{1}{2} \sum_{a \neq b}^r \sum_{k,k'}^K \hat{Q}^{a b}_{k k'} w_k^a w_{k'}^b\)\\
			  &  \[\int_\bbR \d y  \prod_{a=0}^r \int_{\bbR^{K}} \d\vec{z}^a ~ \rp_{\out^a}\(y | \vec{z}^a \) \rp_{\z^a}\(\vec{z}^a | \mat{Q} \)\]^\nsamples \[\prod_{a=0}^r \int_{\bbR^{K}} \d\vec{w}^a ~ \rp_{\w^a}\(\vec{w}^a\)\]^{\ndim} \\
			  &\simeq  \int \int_{\bbR^{(K \times r+1)^2}} \d \mat{Q} \d\hat{\mat{Q}} ~ e^{\ndim \Phi^{(r)} (\mat{Q},\hat{\mat{Q}} )} \,,
			\end{align*}    
			where in the last step, we used a Laplace method \cite{Rong89} and omitted the sub-leading factors in the thermodynamic limit $\ndim \to \infty$ to evaluate it as a function of the free entropy potential defined by
			\begin{align}
			\begin{aligned}
				\Phi^{(r)} (\mat{Q},\hat{\mat{Q}}) &=  - \sum_{a=0}^r \sum_{k, k'}^K Q^{a a}_{k k'} \hat{Q}^{a a}_{k k'} -\frac{1}{2}\sum_{a \neq b}^r \sum_{k, k'}^K Q^{a b}_{k k'} \hat{Q}^{a b}_{k k'} \\
				&+ \log \Psi_{\w}^{(r)} (\hat{\mat{Q}})  + \alpha \log \Psi_{\out}^{(r)}(\mat{Q})\,,
			      \spacecase
			      \Psi_{\w}^{(r)} (\hat{\mat{Q}}) &= \displaystyle \prod_{a=0}^r \int_{\mathbb{R}^{K}} \d \vec{w}^a ~ 
			      \rp_{\w^a}\(\vec{w}^a\) \\
			      & \hhspace \times \exp \( \sum_{a=0}^r \sum_{k k'}  \hat{Q}^{a a}_{k k'} w_k^a w_{k'}^a  + \frac{1}{2} \sum_{a \neq b} \sum_{k,k'} \hat{Q}^{a b}_{k k'} w_k^a w_{k'}^b\)\,, \spacecase
			      \Psi_{\out}^{(r)}(\mat{Q}) &= \displaystyle \prod_{a=0}^r  \int_\bbR \d y \int_{\mathbb{R}^{K}}  \d \vec{z}^a ~ \rp_{\out^a}\(y | \vec{z}^a\) \rp_{\z^a}\(\vec{z}^a | \mat{Q}\)\,,
			\end{aligned}
			    \label{appendix:intro:replicas:Phi_r}
			\end{align}
			and where we decoupled the variable $\Z^a \in \bbR^{\nsamples \times K}$ and $\mW^a \in \bbR^{\ndim \times K}$ along the rows
			\begin{align*}
				\rp_{\out^a}\(\vec{y} | \mat{Z}^a \) &= \displaystyle \prod_{\mu=1}^\nsamples \rp_{\out^a}\(y_\mu | \vec{z}^a_\mu \)\,, \text{ with } \vec{z}^a_\mu \in \bbR^{K}\,, \\
				\rp_{\z^a} (\mat{Z}^a| \mat{Q}) &= \displaystyle \prod_{\mu=1}^\nsamples \rp(\vec{z}^a_\mu | \mat{Q})\,, \\
				\rp_{\w^a}\(\mat{W}^a\) &= \displaystyle \prod_{i=1}^\ndim \rp_{\w}\( \vec{w}^a_i \)\,, \text{ with } \vec{w}^a_i \in \bbR^{K}\,,\\
				\rp_{\z^a}\(\vec{z}^a | \mat{Q}\) &= \exp \left[- \frac{1}{2} \sum_{a,b=0}^r \sum_{k,k'=1}^K z^{a}_{k} z^b_{k'}(\mat{Q}^{-1})_{k k'}^{ab} \right] / \( \det{2\pi \mat{Q}} \)^{\frac{1}{2}}\,.
			\end{align*}
			Note that the averaged replicated partition function of this fully connected model can be expressed as a saddle point equation only because distributions $\rP_{\out},\rP_{\out^\star}$ and $\rP_\w,\rP_{\w^\star}$ are separable so that a pre-factor scaling with the system size $\ndim$ dominates the exponential distribution.
			Finally, switching the two limits $r\to 0$ and $\ndim \to \infty$, the quenched free entropy $\Phi$ simplifies as a saddle point equation
			\begin{equation}
				\Phi (\alpha) = \extr_{ \mat{Q}, \hat{\mat{Q}} } \left\{\lim_{r\rightarrow 0} \frac{\partial \Phi^{(r)}(\mat{Q},\hat{\mat{Q}})}{\partial  r} \right\}\,,
			\end{equation}
			over symmetric matrices $\mat{Q}\in \bbR^{(K \times r+1)^2}$ and $\hat{\mat{Q}} \in \bbR^{(K \times r+1)^2}$. 
			To summarize, we managed to get rid of the original high-dimensional integrals and replace them by an optimization in the space of matrices, which, in this form, is still intractable. We not only have to search in the space of $(r+1)\times (r+1)$ matrices to find the extremiser of $\Phi^{(r)}$, but we also need to compute the limit $r\to 0^{+}$.
			In the following we will assume a simple Ansatz for these matrices in order to first obtain an analytic expression in $r$ before taking the derivative with respect to $r$.

		\subsubsection{Replica Symmetric free entropy}
		\label{appendix:replicas:committee:rs_free_entropy}			
			Our goal is to express the functional $\Phi^{(r)}(\mat{Q},\hat{\mat{Q}})$ appearing in the free entropy as an analytical function of $r$, in order to perform the replica trick. 
			
			\paragraph{Replica symmetric ansatz}
			To do so, we will assume that the extremum of $\Phi^{(r)}$ is attained at a point in $\mat{Q},\hat{\mat{Q}}$ space such that a \emph{replica symmetry} property is verified. More concretely, we assume: 
			\begin{align}
			\begin{aligned}
			\exists \mat{Q} \in \mathbb{R}^{K \times K} \text{ s.t } \quad \forall a \in \lb 0 : r \rb \quad \forall (k,k') \in \lb K \rb^2 \quad Q^{a a}_{k k'} &= Q_{k k'}\,,\\
			\exists \mat{Q}^\star \in \mathbb{R}^{K \times K} \text{ s.t } \quad \forall (k,k') \in \lb K \rb^2 \quad Q^{0 0}_{k k'} &= Q^\star_{k k'}\,,\\
			\exists \mat{q} \in \mathbb{R}^{K \times K} \text{ s.t } \quad \forall (a < b) \in \lb 0 : r \rb^2 \quad \forall (k,k') \in \lb K \rb^2 \quad Q^{a b}_{k k'} &= q_{k k'}\,,\\
			\exists \mat{m} \in \mathbb{R}^{K \times K} \text{ s.t } \quad \forall a \in \lb 0 : r\rb \quad \forall (k,k') \in \lb K \rb^2 \quad Q^{0 a}_{k k'} &= m_{k k'}\,,
			\end{aligned}
			\end{align}
			and similarly for the ad-hoc parameter
			\begin{align}
			\begin{aligned}
			\exists \hat{\mat{Q}} \in \mathbb{R}^{K \times K} \text{ s.t } \quad \forall a \in \lb 0 : r \rb \quad \forall (k,k') \in \lb K \rb^2 \quad \hat{Q}^{a a}_{k k'} &= -\frac{1}{2} \hat{Q}_{k k'}\,,\\
			\exists \hat{\mat{Q}}^\star \in \mathbb{R}^{K \times K} \text{ s.t } \quad \forall (k,k') \in \lb K \rb^2 \quad \hat{Q}^{0 0}_{k k'} &= \hat{Q}^\star_{k k'}\,, \\
			\exists \hat{\mat{q}} \in \mathbb{R}^{K \times K} \text{ s.t } \quad \forall (a < b) \in \lb 0 : r \rb^2 \quad \forall (k,k') \in \lb K \rb^2 \quad \hat{Q}^{a b}_{k k'} &= \hat{q}_{k k'}\,, \\
			\exists \hat{\mat{m}} \in \mathbb{R}^{K \times K} \text{ s.t } \quad \forall a \in \lb 0 : r \rb \quad \forall (k,k') \in \lb K \rb^2 \quad \hat{Q}^{0 a}_{k k'} &= \hat{m}_{k k'}\,.
			\end{aligned}\
			\end{align} 
			The factor $-\frac{1}{2}$ is not necessary bu useful to recover commonly used formulations.
			This Ansatz can be represented by symmetric \aclink{RS} matrices $\mat{Q}^{(\rs)} \in \bbR^{(K \times r+1)^2}$ and $\hat{\mat{Q}}^{(\rs)} \in \bbR^{(K \times r+1)^2}$
			\begin{equation}
			\begin{aligned}[c]
			\mat{Q}^{(\rs)} = \scalemath{0.95}{\begin{pmatrix} 
			\mat{Q}^\star & \mat{m} & \cdots & \mat{m} \\
			\mat{m}  & \mat{Q}  & \mat{q} & ...  \\
			\vdots & \mat{q} & \ddots & \mat{q}   \\
			\mat{m}  &... & \mat{q}  & \mat{Q}     \\
			\end{pmatrix}}
			\end{aligned}
			\hspace{0.2cm}
			\textrm{and} 
			\hspace{0.2cm}
			\begin{aligned}[c]
			\hat{\mat{Q}}^{(\rs)}=\scalemath{0.9}{\begin{pmatrix} 
			 \hat{\mat{Q}}^\star & \hat{\mat{m}} & ... & \hat{\mat{m}}\\
			\hat{\mat{m}} &-\frac{1}{2}\hat{\mat{Q}} & \hat{\mat{q}} & ...  \\
			 \vdots & \hat{\mat{q}} & \ddots & \hat{\mat{q}}  \\
			\hat{\mat{m}} &... & \hat{\mat{q}} & -\frac{1}{2}\hat{\mat{Q}}\\  
			\end{pmatrix}}\,,
			\end{aligned}
			\end{equation}
			where the \emph{overlap} parameters may be reinterpreted as the scalar product between the replicas
			\begin{align*}
				\forall (a,b) \in \lb r \rb^2,~ \mat{q} = \frac{1}{\ndim} \mat{W}^{a \intercal} \mat{W}^{b}\,,
			\end{align*}
			the self-overlap of each replica
			\begin{align*}
				\forall a \in \lb r \rb,~ \mat{Q} = \frac{1}{\ndim} \mat{W}^{a \intercal} \mat{W}^{a}\,,
			\end{align*}
			the scalar product with the ground truth
			\begin{align*}
				\forall a \in \lb r \rb,~ \vec{m} = \frac{1}{\ndim} \mat{W}^{\star \intercal} \mat{W}^{a}\,,
			\end{align*}
			and the second moment of the ground truth distribution
			\begin{align*}
				\mat{Q}^\star = \frac{1}{\ndim} \mat{W}^{\star \intercal} \mat{W}^{\star}\,.
			\end{align*}
			The above Ansatz simplifies in the scalar \aclink{GLM} case with $K=1$ to
			$q = \frac{1}{\ndim} \vec{w}^a \cdot \vec{w}^b$ for $a \ne b$, a norm $Q= \frac{1}{\ndim} \|\vec{w}^a\|_2^2$, an overlap with the ground truth $m =\frac{1}{\ndim} \vec{w}^a \cdot \vec{w}^\star$ and a second moment $Q^\star= \frac{1}{\ndim} \|\vec{w}^\star\|_2^2$.
			
			Let's compute separately the terms involved in the functional $\Phi^{(r)}(\mat{Q},\hat{\mat{Q}})$ in \eqref{appendix:intro:replicas:Phi_r} by applying this Ansatz: the first is a trace term, the second term $\Psi_{\w}^{(r)}$ depends on the prior distributions $\rP_\w$, $\rP_{\w^\star}$ and finally the third term $\Psi_{\out}^{(r)}$ depends on the channel distributions $\rP_{\out^\star}$, $\rP_\out$.
						
			\paragraph{Trace term} 
				The trace term in \eqref{appendix:intro:replicas:Phi_r} can be easily computed at the \aclink{RS} fixed point and takes the following form
				\begin{align*}
					&\left. - \sum_{a=0}^r \sum_{k, k'}^K Q^{a a}_{k k'} \hat{Q}^{a a}_{k k'} -\frac{1}{2}\sum_{a \neq b}^r \sum_{k, k'}^K Q^{a b}_{k k'} \hat{Q}^{a b}_{k k'} \right|_{\rs} \\
					&= \frac{1}{2} \tr{\mat{Q}^\star \hat{\mat{Q}}^\star} + \frac{1}{2} r \tr{\mat{Q} \hat{\mat{Q}}} -  r \tr{\mat{m} \hat{\mat{m}}}  - \frac{r(r-1)}{2} \tr{\mat{q} \hat{\mat{q}}}\,,
				\end{align*}
				and taking the derivative and the limit $r\to 0$ we obtain
				\begin{align}
				\begin{aligned}
					&\lim_{r \to 0} \partial_r \( \left. - \sum_{a=0}^r \sum_{k, k'}^K Q^{a a}_{k k'} \hat{Q}^{a a}_{k k'} -\frac{1}{2}\sum_{a \neq b}^r \sum_{k, k'}^K Q^{a b}_{k k'} \hat{Q}^{a b}_{k k'} \)\right|_{\rs} \\
					&  \qquad \qquad = \frac{1}{2} \tr{\mat{Q} \hat{\mat{Q}}} -  \tr{\mat{m} \hat{\mat{m}}}  + \frac{1}{2} \tr{\mat{q} \hat{\mat{q}}}
					\label{appendix:replicas:committee:trace} 
				\end{aligned}
				\end{align}
			
			\paragraph{Prior integral $\Psi_{\w}^{(r)}$} Evaluated at the \aclink{RS} fixed point the quadratic form reads			      
				\begin{align*}
					&\sum_{a=0}^r \sum_{k k'}  \hat{Q}^{a a}_{k k'} w_k^a w_{k}^a  + \frac{1}{2} \sum_{a \neq b} \sum_{k,k'} \hat{Q}^{a b}_{k k'} w_k^a w_{k'}^b\\
					&= \vec{w}^{\star \intercal} \hat{\mat{Q}}^\star \vec{w}^{\star} + \sum_{a=1}^r \vec{w}^{\star \intercal} \hat{\mat{m}} \vec{w}^a - \frac{1}{2} \sum_{a=1}^r  \vec{w}^{a \intercal} \hat{\mat{Q}} \vec{w}^{a} + \sum_{1 \leq a < b \leq r}  \vec{w}^{a \intercal} \hat{\mat{q}} \vec{w}^{b} \\
					&= \vec{w}^{\star \intercal} \hat{\mat{Q}}^\star \vec{w}^{\star} + \sum_{a=1}^r \vec{w}^{\star \intercal} \hat{\mat{m}} \vec{w}^a - \frac{1}{2} \sum_{a=1}^r  \vec{w}^{a \intercal} \(\hat{\mat{Q}} + \hat{\mat{q}} \) \vec{w}^{a}\\
					& + \frac{1}{2} \(\sum_{a=1}^r  \vec{w}^{a} \)^\intercal \hat{\mat{q}} \(\sum_{a=1}^r  \vec{w}^{a} \)\,.
				\end{align*}
				Using a Hubbard-Stratonovich transformation presented in \App\ref{appendix:replica_computation:hubbard}, the prior integral can be further simplified 
				\begin{align}
				\begin{aligned}
					&\left.\Psi_{\w}^{(r)} (\hat{\mat{Q}})\right|_{\rs} = \displaystyle \int_{\mathbb{R}^{(r+1) \times K}} \d \mat{W} 
			        \rp_{\td{\w}}\(\mat{W}\) e^{\sum_{a=0}^r \sum_{k, k'} ^K \hat{Q}^{a a}_{k k'} w_k^a w_{k}^a  + \frac{1}{2} \sum_{a \neq b} \sum_{k,k'}^K \hat{Q}^{a b}_{k k'} w_k^a w_{k'}^b} \\
					&= \EE_{\bxi, \vec{w}^\star \sim \rP_{\w^\star}} \[ e^{\vec{w}^{\star \intercal} \hat{\mat{Q}}^\star \vec{w}^{\star} } \EE_{\vec{w} \sim \rP_{\w}} \[ e^{\( \vec{w}^\intercal \hat{\vec{m}} \vec{w}^\star  - \frac{1}{2}\vec{w}^\intercal  (\hat{\mat{Q}} + \hat{\mat{q}}) \vec{w} + \vec{w}^\intercal \hat{\mat{q}}^{1/2} \bxi \) } \]^r   \]\,.
				\end{aligned}
				\label{appendix:replicas:committee:Psi_w_rs}
				\end{align}
					
			\paragraph{Channel integral $\Psi_{\out}^{(r)}$}
				Let us focus on the matrix $\mat{Q}^{(\rs)}$ involved in the expression of $\Psi_{\out}^{(r)}$ in \eqref{appendix:intro:replicas:Phi_r}. The elements of its inverse block matrix
				\begin{equation}
				   \(\mat{Q}^{(\rs)}\)^{-1}=\begin{bmatrix}
				   \td{\mat{Q}}^\star & \td{\mat{m}} & \cdots & \td{\mat{m}}  \\
				   \td{\mat{m}} & \td{\mat{Q}} & \td{\mat{q}} & \cdots \\
				   \vdots & \td{\mat{q}} & \ddots & \td{\mat{q}} \\
				   \td{\mat{m}} & \cdots & \td{\mat{q}} & \td{\mat{Q}} \\
				  \end{bmatrix}
				\end{equation}
				can be computed and given by
				\begin{align*}
					&\td{\mat{Q}}^\star = \(\mat{Q}^\star - r \mat{m} (\mat{Q} + (r-1) \mat{q})^{-1} \mat{m}^\intercal \)^{-1}   \spacecase
					&\td{\mat{m}} = -\(\mat{Q}^\star - r \mat{m} (\mat{Q} + (r-1) \mat{q})^{-1} \mat{m}^\intercal \)^{-1} \mat{m} ( \mat{Q} + (r-1) \mat{q})^{-1}  \spacecase
					&\td{\mat{Q}} = (\mat{Q}-\mat{q})^{-1} - (\mat{Q} +(r-1) \mat{q})^{-1} \mat{q} (\mat{Q}-\mat{q})^{-1} \\
					& \qquad  + ( \mat{Q} + (r-1) \mat{q} )^{-1} \mat{m}^\intercal \\ 
					& \qquad \qquad \times \(\mat{Q}^\star - r \mat{m} ( \mat{Q} + (r-1) \mat{q})^{-1} \mat{m}^\intercal \)^{-1} \mat{m} ( \mat{Q} + (r-1)\mat{q})^{-1}\spacecase
					&\td{\mat{q}} = \td{\mat{Q}} - (\mat{Q}-\mat{q})^{-1}
				\end{align*}
				and its determinant by
				\begin{align}
					\det{\mat{Q}^{(\rs)}} &= \det{\mat{Q}-\mat{q}}^{r-1} \det{\mat{Q} + (r-1)\mat{q}} \nonumber\\
					& \qquad \qquad \times \det{\mat{Q}^\star - r\mat{m}( \mat{Q} + (r-1)\mat{q} )^{-1} \mat{m}^\intercal } \,. 
					\label{appendix:replicas:committee:det_rs}
				\end{align}
				Therefore the quadratic form in $\rp_{\z^a}(\mat{z}^a | \mat{Q}^{(\rs)}$ reads				
				\begin{align*}
					&- \frac{1}{2} \sum_{a,b} \sum_{k,k'} z^{a}_{k} z^b_{k'}(\mat{Q}^{-1})_{k k'}^{ab} \\
					&= -\frac{1}{2} \vec{z}^{\star \intercal} \td{\mat{Q}}^\star \vec{z} 
					-\sum\limits_{a=1}^{r} \vec{z}^{\star \intercal} \td{\mat{m}} \vec{z}^{a}\\
					&
					\qquad \qquad  -\frac{1}{2}\sum\limits_{a=1}^{r} \vec{z}^{a \intercal} \( \td{\mat{Q}} - \td{\mat{q}} \) \vec{z}^{a}
					-\frac{1}{2} \(\sum\limits_{a}^{r} \vec{z}^{a}\)^\intercal \td{\mat{q}} \(\sum\limits_{a}^{r} \vec{z}^{a}\) \,,
				\end{align*}
				and using another Gaussian transformation, see \App\ref{appendix:replica_computation:hubbard}, we finally obtain
				\begin{align}
				&\left. \Psi_{\out}^{(r)}(\mat{Q}) \right|_{\rs} =  \displaystyle \int \d y \int_{\mathbb{R}^{(r+1) \times K}}  \d \mat{Z} ~ 
			      \rp_{\out}\(y | \mat{Z}\) \rp\(\mat{Z} | \mat{Q}\) \nonumber \\
			      &= \displaystyle \int \d y \int_{\mathbb{R}^{(r+1) \times K}}  \d \mat{Z} 
			      \rp_{\out}\(y | \mat{Z}\) e^{- \frac{1}{2} \sum_{a,b=0}^r \sum_{k,k'=1}^K z^{a}_{k} z^b_{k'}(\mat{Q}^{-1})_{k k'}^{ab} } \nonumber\\
			      & \qquad \qquad \qquad \qquad \qquad \qquad  \qquad \qquad / \( \det{2\pi \mat{Q}}^{(\rs)} \)^{\frac{1}{2}}
				\nonumber \\
				&= \int \d y ~ \EE_{\bxi} \e^{- \frac{1}{2}\log(\det{2\pi \mat{Q}^{(\rs)}} )} \times \int \d \vec{z}^\star ~ \rp_{\out^\star}\(y | \vec{z}^\star \) e^{ -\frac{1}{2} \vec{z}^{\star \intercal} \td{\mat{Q}}^\star \vec{z}^\star} \label{appendix:replicas:committee:Psi_out_rs} \\ 
				& \times\[ \int \d \vec{z} ~ \rp_{\out}\(y | \vec{z}\) \exp\( - \vec{z}^{\star \intercal} \td{\mat{m}} \vec{z} - \frac{1}{2} \vec{z}^\intercal \( \td{\mat{Q}} - \td{\mat{q}} \) \vec{z}  + \vec{z}^\intercal (-\td{\mat{q}})^{1/2} \bxi\) \]^r\,, \nonumber
				\end{align}
				with $\det{\mat{Q}^{(\rs)}}$ given by \eqref{appendix:replicas:committee:det_rs}.
			
		\subsubsection{Consistency conditions $r\to 0$: $\Theta(1)$ terms}
		\label{appendix:replicas:committee:rs_consistency}
			It remains to take the limit $r \to 0^+$ of the expressions for $\Psi_{\w}^{(r)}$ and $\Psi_{\out}^{(r)}$ that are now analytical in $r$.
			First, our assumptions must be consistent and thus we need to check the consistency conditions in the limit $r \to 0$. Indeed, if $\Phi^{(r)}$ is finite we could obtain divergence taking the limit $\lim_{r \to 0} \frac{1}{r} \Phi^{(r)} = \infty$. Therefore to avoid such divergence, we must at least impose that $\lim_{r \to 0} \Phi^{(r)} = 0$:			  
			\begin{align*}
				\lim_{r\to 0} \Phi^{(r)} (\mat{Q},\hat{\mat{Q}}) &= - \tr{\mat{Q}^\star \hat{\mat{Q}}^\star} +\log \Psi_{\w}^{0} (\hat{\mat{Q}}^\star)+\alpha \log \Psi_{\out}^{0}(\mat{Q}^\star)
			\end{align*}	
			with 
			\begin{align*}
				\Psi_{\w}^{0} (\hat{\mat{Q}}^\star) &\equiv \EE_{\vec{w}^\star} \exp\( \vec{w}^{\star \intercal}  \hat{\mat{Q}}^\star \vec{w}^{\star}\)\,, \spacecase 
				\Psi_{\out}^{0}(\mat{Q}^\star) &\equiv \int_\bbR \d y ~  \int \d \vec{z}^\star ~ \rp_{\out^\star}\(y | \vec{z}^\star \) \mN_{\vec{z}^\star}\(\vec{0}, \mat{Q}^\star \)  = 1 \,.
			\end{align*}		
			Taking the saddle point equations over $\mat{Q}^\star$ and $\hat{\mat{Q}}^\star$, imposing the consistency condition $\lim_{r\to 0} \Phi^{(r)} (\mat{Q},\hat{\mat{Q}}) = 0$, we finally obtain
			\begin{align}
				\hat{\mat{Q}}^\star = \mat{0} \andcase \mat{Q}^\star = \EE_{\vec{w}^\star}\[ \vec{w}^{\star \intercal} \vec{w}^{\star} \]\,.
				\label{appendix:replicas:committee:consistency}
			\end{align}
			
		\subsubsection{Replica trick $r \to 0$ limit: $\Theta(r)$ terms}
		\label{appendix:replicas:committee:rs_limit_r_0}  
			Imposing the conditions \eqref{appendix:replicas:committee:consistency} avoids divergence in the replica trick, and we can therefore proceed with the $\Theta(r)$ terms.  
				
		\paragraph{Prior integral $\Psi_{\w}^{(r)}$}
			The limit $r\to 0$ and the derivative of the logarithm of the prior integral \eqref{appendix:replicas:committee:Psi_w_rs} can be trivially computed
			\begin{align*}
			&\lim_{r \to 0} \partial_r \left. \log \Psi_{\w}^{(r)} (\hat{\mat{Q}})\right|_{\rs} \\
			&= \EE_{\bxi, \vec{w}^\star} \\
			& \times \log \[ \EE_{\vec{w}}  \exp \(\[  \vec{w}^{\star \intercal} \hat{\mat{m}} \vec{w} - \frac{1}{2} \vec{w}^\intercal (\hat{\mat{Q}} + \hat{\mat{q}}) \vec{w} + \bxi^\intercal \hat{\mat{q}}^{1/2}  \vec{w} \] \) \]\,,
			\end{align*}	
			with $\bxi \sim \mN(\vec{0},\mat{I}_K)$ and $\vec{w}^\star \sim \rP_{\w^\star}$. 
			To conclude, we can symmetrize and decouple the \emph{teacher} and \emph{student} expectations $\EE_{\vec{w}^\star},\EE_{\vec{w}}$. By performing the change of variable $\bxi \leftarrow \bxi + \hat{\mat{q}}^{-1/2} \hat{\mat{m}} \vec{w}^\star$, we finally obtain		
			\begin{align}
			&\lim_{r \to 0} \partial_r \left. \log \Psi_{\w}^{(r)} (\hat{\mat{Q}})\right|_{\rs} \nonumber \\ 
			&=  \EE_{\bxi, \vec{w}^\star}\exp\(-\frac{1}{2} \vec{w}^{\star \intercal} \hat{\mat{m}}^\intercal \hat{\mat{q}}^{-1}\hat{\mat{m}} \vec{w}^\star + \bxi^\intercal \hat{\mat{q}}^{-1/2} \hat{\mat{m}} \vec{w}^\star  \)  \label{appendix:replicas:committee:log_Psi_w} \\
			& \hspace{1cm} \times \log \[ \EE_{\vec{w}}  \exp \(\[ - \frac{1}{2}\vec{w}^{\intercal} (\hat{\mat{Q}} + \hat{\mat{q}})\vec{w}  + \bxi^\intercal \hat{\mat{q}}^{1/2} \vec{w} \] \) \] \nonumber \\
			&\equiv \EE_{\bxi, \vec{w}^\star} \mZ_{\w^\star}\(\hat{\mat{m}}\hat{\mat{q}}^{-1/2} \bxi, \hat{\mat{m}}^\intercal \hat{\mat{q}}^{-1}\hat{\mat{m}} \) \log \mZ_\w\(\hat{\mat{q}}^{1/2} \bxi , \hat{\mat{Q}} + \hat{\mat{q}}\)\,, \nonumber
			\end{align}
			with the corresponding denoising distribution $\rQ_\w$ and functions $\mZ_{\w^\star},\mZ_{\w}$ defined in \Sec\ref{appendix:definitions:distributions:committee} respectively with distribution $\rP_{\w^\star}$ and $\rP_\w$.

		\paragraph{Prior integral $\Psi_{\out}^{(r)}$}
			The limit $r\to 0$ and the derivative of the logarithm of the channel integral \eqref{appendix:replicas:committee:Psi_out_rs} is more tricky. First, the limit of the determinant simplifies easily and yields
			\begin{align*}
				\det{\mat{Q}^{(\rs)}} \underlim{r}{0} \det{\mat{Q}^\star}
			\end{align*}
			and the matrix elements of $\(\mat{Q}^{(\rs)}\)^{-1}$ in this limit become
			\begin{align*}
				\td{\mat{Q}}^\star & \underlim{r}{0} \(\mat{Q}^\star\)^{-1}\,,  \spacecase
				\td{\mat{m}} & \underlim{r}{0} -\(\mat{Q}^\star \)^{-1} \mat{m} ( \mat{Q} - \mat{q})^{-1}\,,  \spacecase
				\td{\mat{q}} & \underlim{r}{0} - (\mat{Q} - \mat{q})^{-1}\(\mat{q} -  \mat{m} \(\mat{Q}^\star \)^{-1} \mat{m} \)( \mat{Q}-\mat{q})^{-1}\,, \spacecase
				\td{\mat{Q}} & \underlim{r}{0}  \td{\mat{q}} + (\mat{Q}-\mat{q})^{-1}\,.
			\end{align*}
			By taking properly the $r \to 0$ limit and performing several change of variables 
			, we finally obtain
			\begin{align}
			&\lim_{r\to 0} \partial_r \left. \log  \Psi_{\out}^{(r)} (\mat{Q})\right|_{\rs} \nonumber \\
				&= \int_\bbR \d y ~ \EE_{\bxi}  \int_{\bbR^K} \d \vec{z}^\star ~ \rp_{\out^\star}\(y | \vec{z}^\star \) \nonumber \\ 
				& \times \exp\(-\frac{1}{2} \(\vec{z}^\star - \mat{m}\mat{q}^{-1/2}\bxi\)^\intercal \(\mat{Q}^\star - \mat{m}^\intercal \mat{q}^{-1} \mat{m} \) \(\vec{z}^\star - \mat{m}\mat{q}^{-1/2}\bxi\) \)\nonumber \\
				&\times \log \[ \int_{\bbR^K} \d \vec{z} ~ \rp_{\out}\(y | \vec{z} \) e^{ - \frac{1}{2} \( \vec{z} - \mat{q}^{-1/2} \bxi \)^\intercal \( \mat{Q} - \mat{q} \) \( \vec{z} - \mat{q}^{-1/2} \bxi \)}  \] \label{appendix:replicas:committee:log_Psi_out} \\
				&= \int_\bbR \d y ~ \EE_{\bxi} \mZ_{\out^\star}\( \mat{m}\mat{q}^{-1/2}\bxi, \mat{Q}^\star - \mat{m}^\intercal \mat{q}^{-1} \mat{m} \)  \log \mZ_{\out}\(\mat{q}^{1/2} \bxi, \mat{Q}-\mat{q}\)\,, \nonumber
			\end{align}
			where the denoising distribution $\rQ_{\out}$ and functions $\mZ_{\out^\star},\mZ_{\out}$ are defined in \Sec\ref{appendix:definitions:distributions:committee} for the distributions $\rP_{\out^\star}$, $\rP_{\out}$.

\subsubsection{Summary - Mismatched case}
In the mismatched case, where the teacher and the student have not the same prior distributions, we finally obtain the replica symmetric free entropy $\Phi_{\rs}$ for the committee machine hypothesis class:
\begin{align}
	&\Phi_{\rs}(\alpha) \equiv \EE_{\vec{y},\mat{X}} \[ \lim_{\ndim \to \infty} \frac{1}{\ndim} \log\( \mZ_\ndim\(\vec{y}, \mat{X}\) \) \] \label{appendix:replicas:free_entropy_non_bayes} \nonumber \\
	&=  \extr_{\mat{Q},\hat{\mat{Q}}, \mat{q}, \hat{\mat{q}}, \mat{m}, \hat{\mat{m}}} \left \{  - \tr{ \mat{m} \hat{\mat{m}}} + \frac{1}{2} \tr{ \mat{Q} \hat{\mat{Q}}} + \frac{1}{2} \tr{  \mat{q} \hat{\mat{q}}}  \right. \\
	& \left. \qquad \qquad \qquad \qquad  \qquad \qquad + \Psi_\w\(\hat{\mat{Q}},\hat{\mat{m}},\hat{\mat{q}}  \) + \alpha \Psi_\out\( \mat{Q}, \mat{m}, \mat{q};  \brho_{\w^\star}\) \right\} \nonumber \,,
\end{align}
where $\brho_{\w^\star} \equiv \lim_{\ndim \to \infty} \mat{Q}^\star = \lim_{\ndim \to \infty} \EE_{\vec{w}^\star} \frac{1}{\ndim} \mat{W}^{\star\intercal} \mat{W}^{\star} $ and the channel and prior integrals are defined by
\begin{align}
	\Psi_\w\(\hat{\mat{Q}},\hat{\mat{m}},\hat{\mat{q}}  \) &\equiv \EE_{\bxi} \[ \mZ_{\w^\star}\( \hat{\mat{m}}  \hat{\mat{q}}^{-1/2}  \bxi , \hat{\mat{m}} \hat{\mat{q}}^{-1} \hat{\mat{m}}  \) \right. \label{appendix:replicas:committee:free_entropy_terms_non_bayes} \\
	& \left. \qquad \qquad \qquad \qquad \times  \log \mZ_{\w} \(  \hat{\mat{q}}^{1/2} \bxi, \hat{\mat{Q}} + \hat{\mat{q}} \)   \]\,, \nonumber \spacecase
	\Psi_\out\(\mat{Q},\mat{m}, \mat{q}; \brho_{\w^\star}\) &\equiv \EE_{y, \bxi } \[ \mZ_{\out^\star} \( y,  \mat{m} \mat{q}^{-1/2} \bxi, \brho_{\w^\star} - \mat{m}^\intercal \mat{q}^{-1} \mat{m}  \)\right. \nonumber \\
	& \left. \qquad \qquad \qquad \qquad \times \log \mZ_{\out} \( y,  \mat{q}^{1/2} \bxi, \mat{Q} - \mat{q} \)  \]\,,\nonumber
\end{align}
where again $\mZ_{\out^\star},\mZ_{\w^\star}$ and $\mZ_{\out},\mZ_{\w}$ are defined in \Sec\ref{appendix:definitions:distributions:committee} and depend respectively on channel and prior distributions of the \emph{teacher} and \emph{student}.

\subsubsection{Summary - Bayes optimal MMSE estimation}
For \aclink{MMSE} estimation in the Bayes-optimal setting, the student has access to the ground truth distributions of the teacher $\rP_\out\(\vec{y} | \mat{Z} \) = \rP_{\out^\star}\(\vec{y} | \mat{Z}\) $ and $\rP_\w\(\mat{W}\) = \rP_{\w^\star}(\mat{W})$, and therefore $\mZ_{\out}=\mZ_{\out^\star}$, $\mZ_{\w}=\mZ_{\w^\star}$. 
In this idealistic setting, the Nishimori conditions, recalled in \App\ref{appendix:replica_computation:nishimori}, imply that 
\begin{align}
\begin{aligned}
	\mat{Q} &= \mat{Q}_{\w^\star}\,, && \hat{\mat{Q}}=\mat{0}\,, && \mat{m}= \mat{q} \equiv \mat{q}_\bayes \,, && \hat{\mat{m}}=\hat{\mat{q}} \equiv  \hat{\mat{q}}_\bayes \,.
\end{aligned}
	\label{appendix:replicas:committee:nishimori}
\end{align}
Therefore the free entropy in \eq\eqref{appendix:replicas:committee:free_entropy_terms_non_bayes} simplifies as an optimization problem over the \emph{overlaps} $\mat{q}_\bayes,\hat{\mat{q}}_\bayes \in \bbR^{K\times K}$
\begin{align}
	\Phi_{\rs}^\bayes (\alpha) &=  \extr_{\mat{q}_\bayes,\hat{\mat{q}}_\bayes} \left \{ - \frac{1}{2} \tr{\mat{q}_\bayes \hat{\mat{q}}_\bayes}  + \Psi_{\w}^\bayes\(\hat{\mat{q}}_\bayes  \) + \alpha \Psi_{\out}^\bayes\(\mat{q}_\bayes; \brho_{\w^\star}\) \right\} \,,
	\label{appendix:free_entropy_bayes}
\end{align}
with free entropy terms $\Psi_{\w}^\bayes$ and $\Psi_{\out}^\bayes$ given by
\begin{align}
		\Psi_{\w}^\bayes\(\hat{\mat{q}}\) &= \EE_{\bxi} \[ \mZ_{\w^\star} \(  \hat{\mat{q}}^{1/2}\bxi,   \hat{\mat{q}} \)  \log \mZ_{\w^\star} \(  \hat{\mat{q}}^{1/2}\bxi,   \hat{\mat{q}} \)   \] \,, \nonumber \spacecase 
		\Psi_{\out}^\bayes \(\mat{q}; \brho_{\w^\star}\) &= \EE_{y, \bxi } \[ \mZ_{\out} \( y,  \mat{q}^{1/2} \bxi, \brho_{\w^\star} - \mat{q} \) \right. \\
		& \left. \qquad \qquad \qquad \qquad  \times  \log \mZ_{\out} \( y,  \mat{q}^{1/2}\bxi, \brho_{\w^\star} - \mat{q} \)  \]\,. \nonumber
\end{align} 

\paragraph{Application to the GLM}
For the \aclink{GLM} hypothesis class, the same equations are valid if we take $K=1$ for both the teacher and the student. As a result, we recover the replica symmetric free entropy in the Bayes-optimal setting rigorously proven in \cite{barbier2017phase}.

\subsection{Fixed point equations}
\label{appendix:replica_computation:committee:fixed_point}

The overlaps parameters, such as $\mat{m}, \mat{q}$, play a crucial role since they measure the performances of the statistical estimation. Their behaviours are respectively characterized by the extremization of the free entropy \eqref{appendix:replicas:free_entropy_non_bayes} in the mismatched setting and \eqref{appendix:free_entropy_bayes} in the Bayes-optimal case.
In this section, we give the expressions of the corresponding fixed point equations, whose derivations can be found in \cite{aubin2020generalization} \App IV.4-5 for $K=1$ which can be extended to $K \ge 1$.

\subsubsection{Mismatched setting}
\label{appendix:fixed_point_erm}
Extremizing the free entropy eq.~\eqref{appendix:replicas:free_entropy_non_bayes}, we easily obtain the set of six fixed point equations
\begin{align}
\begin{aligned}
	\hat{\mat{Q}} &= -2 \alpha \partial_{\mat{Q}} \Psi_\out\(\mat{Q},\mat{m}, \mat{q}; \brho_{\w^\star}\) \,, \qquad 
			&& \mat{Q} = - 2  \partial_{\hat{\mat{Q}}} \Psi_{\w}\(\hat{\mat{Q}},\hat{\mat{m}},\hat{\mat{q}}  \)\spacecase
	\hat{\mat{q}} &= -2 \alpha \partial_{\mat{q}} \Psi_\out\(\mat{Q},\mat{m}, \mat{q}; \brho_{\w^\star}\)\,, 
			&& \mat{q} =   -2\partial_{\hat{\mat{q}}} \Psi_\w\(\hat{\mat{Q}},\hat{\mat{m}},\hat{\mat{q}}  \) \,,\spacecase
	\hat{\mat{m}} &= \alpha \partial_{\mat{m}} \Psi_\out\(\mat{Q},\mat{m}, \mat{q}; \brho_{\w^\star}\)\,,
			&& \mat{m} = \partial_{\hat{\mat{m}}} \Psi_\w\(\hat{\mat{Q}},\hat{\mat{m}},\hat{\mat{q}}  \) \,.
\end{aligned}
\label{appendix:se_equations_generic_not_simplified}
\end{align}
Interestingly, these equations can be reformulated as functions of $\mZ_{\out^\star}$, $\mZ_{\w^\star}$ and the denoising functions $f_{\out^\star}, f_{\w^\star}, f_\out, f_{\w}$ defined in \eqref{appendix:definitions:update_generic:committee:prior:fw}-\eqref{appendix:definitions:update_generic:committee:channel:fout} in \Sec\ref{appendix:definitions:updates:committee}.
Defining the natural variables $\bSigma = \mat{Q} - \mat{q}$ and $\hat{\bSigma}= \hat{\mat{Q}}+\hat{\mat{q}}$ they can reformulated as
\begin{align}
\begin{aligned}
	\hat{\mat{m}} &= \alpha \EE_{y, \bxi } \[ \mZ_{\out^\star} \times  \vec{f}_{\out^\star} \( y,  \mat{m} \mat{q}^{-1/2} \bxi, \brho_{\w^\star} - \mat{m}^\intercal \mat{q}^{-1}\mat{m}  \)  \right. \\
	& \left. \qquad \qquad \qquad \qquad \qquad \qquad \qquad \qquad \times \vec{f}_{\out} \( y,  \mat{q}^{1/2}\bxi, \bSigma \)^\intercal    \]\,, \\
	\hat{\mat{q}} &= \alpha \EE_{y, \bxi } \[ \mZ_{\out^\star} \(y,  \mat{m} \mat{q}^{-1/2} \bxi, \brho_{\w^\star} - \mat{m}^\intercal \mat{q}^{-1}\mat{m}    \)   \vec{f}_{\out} \(  y,  \mat{q}^{1/2}\bxi, \bSigma\)^{\otimes 2}    \]\,, \\
	\hat{\bSigma} &= - \alpha \EE_{y, \bxi } \[ \mZ_{\out^\star} \( y,  \mat{m} \mat{q}^{-1/2} \bxi, \brho_{\w^\star} - \mat{m}^\intercal \mat{q}^{-1}\mat{m}   \)  \right. \\
	& \left. \qquad \qquad\qquad \qquad \qquad \qquad \qquad \qquad \times \partial_\bomega \vec{f}_{\out} \(  y,  \mat{q}^{1/2}\bxi, \bSigma \)    \]\,,  \\
	\mat{m} &= \EE_{\bxi} \[ \mZ_{\w^\star} \times f_{\w^\star}\( \hat{\mat{m}}  \hat{\mat{q}}^{-1/2}  \bxi , \hat{\mat{m}}^\intercal \hat{\mat{q}}^{-1} \hat{\mat{m}}  \) \vec{f}_{\w} \(  \hat{\mat{q}}^{1/2}\bxi  , \hat{\bSigma} \)   \]\,, \\
	\mat{q} &= \EE_{\bxi} \[ \mZ_{\w^\star}\( \hat{\mat{m}}  \hat{\mat{q}}^{-1/2}  \bxi , \hat{\mat{m}}^\intercal \hat{\mat{q}}^{-1} \hat{\mat{m}}   \) \vec{f}_{\w} \( \hat{\mat{q}}^{1/2}\bxi  , \hat{\bSigma}\)^2   \]\,, \\
	\bSigma &= \EE_{\bxi} \[ \mZ_{\w^\star}\(\hat{\mat{m}}  \hat{\mat{q}}^{-1/2}  \bxi , \hat{\mat{m}}^\intercal \hat{\mat{q}}^{-1} \hat{\mat{m}}   \)  \partial_\bgamma \vec{f}_{\w} \(  \hat{\mat{q}}^{1/2}\bxi  , \hat{\bSigma}\) \] \,,
\end{aligned}
\label{appendix:se_equations_generic}
\end{align}
where we use the abusive notation $\EE_{y} = \int_\bbR \d y$.

\subsubsection{Bayes-optimal estimation}
Extremizing the Bayes-optimal free entropy eq.~\eqref{appendix:free_entropy_bayes}, we easily obtain the set of fixed point equations over the scalar parameters $\mat{q}_\bayes, \hat{\mat{q}}_\bayes$. It can be deduced from eq.~\eqref{appendix:se_equations_generic} using the Nishimori conditions $\vec{f}_{\w} = \vec{f}_{\w^\star}$, $\vec{f}_{\out}=\vec{f}_{\out^\star}$, $\mat{m}=\mat{q}, \bSigma = \brho_{\w^\star} - \mat{q}, \hat{\mat{m}}=\hat{\mat{q}}$ and $\hat{\bSigma}=\hat{\mat{q}}$ that lead to
\begin{align}
	\hat{\mat{q}}_\bayes &= \alpha \EE_{y, \bxi } \[ \mZ_{\out^\star} \( y,  \mat{q}_\bayes^{1/2}\bxi, \brho_{\w^\star} - \mat{q}_\bayes \)   \vec{f}_{\out^\star} \( y,  \mat{q}_\bayes^{1/2}\bxi, \brho_{\w^\star} - \mat{q}_\bayes \)^{\otimes 2}    \] \,, \nonumber \spacecase
	\mat{q}_\bayes &= \EE_{\bxi} \[ \mZ_{\w^\star}\( \hat{\mat{q}}_\bayes^{1/2}  \bxi , \hat{\mat{q}}_\bayes  \) \vec{f}_{\w^\star} \(  \hat{\mat{q}}_\bayes^{1/2}\bxi, \hat{\mat{q}}_\bayes\)^{\otimes 2}   \] \,.
\label{appendix:se_equations_generic:bayes}
\end{align}

	\newpage
	\section{Random labels - GLM with i.i.d data}
	\label{appendix:replica_computation:random_labels:iid} 
	In this section, we present the replica computation of \aclink{GLM} corresponding to the hypothesis class $\mF_{\varphi}$ in eq.~\eqref{main:glm_hypothesis_class}. We focus on data ${\{\vec{x}_1^\intercal, \dots, \vec{x}_\nsamples^\intercal \} = \mat{X}\in \bbR^{\nsamples \times \ndim}}$, with $\alpha = \nsamples / \ndim$, drawn \aclink{i.i.d} from a distribution $\P_\x(\vec{x}) =\mN_{\vec{x}}(\bzero,\mat{I}_\ndim)$, and labels $\vec{y}$ drawn randomly from $ \rP_\y(.)$.
We consider for the moment a generic prior distribution $\vec{w} \sim \rP_\w(.)$ that factorizes, and a component-wise activation function $\varphi(.)$. 
Defining the linear transformation applied by the model $z_\mu \equiv \frac{1}{\sqrt{d}} \vec{w}^{\intercal} \vec{x}_\mu$, we introduce the corresponding cost function of a given sample $\(\vec{x}_\mu, y_\mu \)$ according to $V(y_\mu|z_\mu) = \id\[ y_\mu \ne \varphi (z_\mu) \]$ which is $0$ if the the estimator classifies the example correctly (\ie when $ y_\mu = \varphi (z_\mu)$ ) and 1 otherwise. 
Finally we define the \emph{constraint function} $\mC$ at inverse temperature $\beta$
\begin{equation}
    \mC(\vec{y} | \vec{z}, \beta) \equiv \prod_{\mu=1}^\nsamples e^{- \beta V(y_\mu|z_\mu)} = e^{-\beta \mH_\ndim \(\{\vec{y},\mat{X}\}, \vec{w}\) }\,,
    \label{appendix:replica_constraint}
\end{equation}
which, denoting the output of the estimator $f_{\vec{w}}\( \vec{x}_{\mu}\)=\varphi\( z_\mu \)$, depends explicitly on the Hamiltonian
\begin{align}
	 \mH_\ndim \(\{\vec{y},\mat{X}\}, \vec{w}\) \equiv 	\sum_{\mu=1}^\nsamples \id \[ y_{\mu} \ne 
    f_{\vec{w}}\( \vec{x}_{\mu}\) \]\,.
\end{align}
Notice that at zero temperature the \emph{soft} constraint function $\mC$ converges to a \emph{hard} constraint function $ \mC (\vec{y} | \vec{z}, \beta) \underset{\beta \to \infty}{\longrightarrow} \prod_{\mu=1}^\nsamples \id\[ V(y_\mu|z_\mu) = 0\]$, which tolerates only configurations that satisfy simultaneously all the constraints.
In this context, the partition function simply reads 
\begin{align}
	\mZ_\ndim( \{\vec{y}, \mat{X}\}, \alpha, \beta) = \int_{\bbR^d} \d \rP_\w\( \vec{w} \)  \mC(\vec{y} | \vec{z}, \beta)  \,.
\end{align}

In order to compute the quenched free entropy average, we use the replica trick, see \Sec\ref{main:intro:replicas:replica_trick}, and consider the partition function of $r \in \mathbb{N}$ identical copies of the initial system. Assuming there exists an analytical continuation for $r \rightarrow 0^+$ and we can revert limits, the averaged free entropy $\Phi(\alpha, \beta) \equiv \lim_{d \to \infty} \frac{1}{d } \EE_{\vec{y}, \mat{X}}\log \mZ_\ndim( \{\vec{y}, \mat{X}\}, \alpha, \beta)$ of the initial system becomes 
\begin{align}
    \Phi(\alpha, \beta) =  \lim_{r \to 0} \[ \lim_{d \to \infty}  \frac{1}{\ndim}  \frac{\partial  \log \EE_{\vec{y}, \mat{X}} \[ \mZ_\ndim(\{\vec{y}, \mat{X}\}, \alpha, \beta)^r\] }{\partial r} \] \,,
    \label{appendix:free_entropy}
\end{align}
where the replicated partition function writes
\begin{align}
&\EE_{\vec{y}, \mat{X}} \[ \mZ_\ndim(\{\vec{y}, \mat{X}\}, \alpha, \beta)^r\] =  \int_{\bbR^\nsamples} \d \rP_\y\(\vec{y}\) ~ \int_{\bbR^{\nsamples \times \ndim}} \d\rP_\x(\mat{X}) ~ \mZ_\ndim(\{\vec{y}, \mat{X}\}, \alpha, \beta)^r \nonumber \\
&= \int_{\bbR^\nsamples} \d \rP_\y\(\vec{y}\) ~ \int_{\bbR^{\nsamples \times \ndim}} \d\rP_\x(\mat{X})  \label{appendix:average_Zn} \\
& \qquad \times \prod_{a=1}^r \int_{\bbR^\ndim} \d \rP_\w\( \vec{w}^a \) ~ \prod_{\mu=1}^\nsamples \int \d z_\mu^a ~ \mC (y_\mu|z_\mu^a,\beta)\delta\(z_\mu^a-\ \frac{1}{\sqrt{d}} \vec{w}^{a \intercal} \vec{x}_\mu \)  \,. \nonumber
\end{align}

\subsection{Average over iid inputs}
As the data matrix is taken (Gaussian) \aclink{i.i.d}, for $(i,j)\in \llbracket \ndim \rrbracket^2$, $(\mu,\nu) \in \llbracket \nsamples \rrbracket^2$, $\EE_\mat{X} [x_{\mu i} x_{\nu j}] = \delta_{\mu\nu} \delta_{ij}$. Hence $z_{\mu}^a =\frac{1}{\sqrt{d}} \sum_{i=1}^\ndim x_{\mu i} w_i^a$ is the sum of \aclink{i.i.d} random variables.
The \aclink{CLT} guarantees that in the large size limit $ \ndim \to \infty$, $z_{\mu}^a\sim$ \newline
$\mN\(\EE_{\mat{X}}[z_\mu^a]  ,\EE_{\mat{X}}[z_\mu^a z_\mu^b] \)$, with the two first moments given by
\begin{align}
\begin{aligned}
		\EE_{\mat{X}}[z_\mu^a] &= \frac{1}{\sqrt{d}} \sum_{i=1}^\ndim \EE_{\mat{X}}[x_{\mu i}] w_i^a =0 \,,\spacecase
		\EE_{\mat{X}}[z_\mu^a z_\nu^b] &= \frac{1}{d} \sum_{ij} \EE_{\mat{X}}[x_{\mu i}x_{\nu j}] w_i^a w_j^b  = \(\frac{1}{d} \sum_{i=1}^\ndim w_i^a w_i^b \) \delta_{\mu \nu}  \, .
\end{aligned}
\end{align}

In the following, we introduce the overlap matrix $\mat{Q}\equiv\(\frac{1}{d}\vec{w}^a \cdot \vec{w}^b\)_{a,b=1..r} \in \bbR^{r \times r}$ and we define the replicated vectors $\td{\vec{z}}_{\mu} \in \bbR^{r} \equiv (z^a_{\mu})_{a=1..r}$, $\td{\vec{w}}_i \equiv (w_i^a)_{a=1..r}$ $\in \bbR^{r}$. From the above calculation $\td{\vec{z}}_{\mu}$ follows a multivariate Gaussian distribution $\td{\vec{z}}_{\mu} \sim \rP_\z(\td{\vec{z}},\mat{Q}) \triangleq \mathcal{N}_{\td{\vec{z}}}( \vec{0}_r, \mat{Q})$ and $
        \rP_\w(\td{\vec{w}}_i) = \prod_{a=1}^r
        \rP_\w(\td{w}^a_i)$. Introducing the change of variable and the Fourier representation of the Dirac-$\delta$ distribution that involves a new ad-hoc matrix order parameter $\hat{\mat{Q}}\in \bbR^{r \times r}$:
\begin{align}
    1 &= \int_{\bbR^{r \times r}} \d \mat{Q} ~ \prod_{a \leq b} \delta \(d Q_{ab}-\sum_{i=1}^\ndim w_i^a w_i^b \) \\
    &\propto \int_{\bbR^{r \times r}} \d \mat{Q} ~ \int_{\bbR^{r \times r}} d\hat{\mat{Q}} ~ \exp \(-\frac{d}{2}\tr{\mat{Q} \hat{\mat{Q}}} \)  \exp\( \frac{1}{2}\sum_{i=1}^\ndim \td{\vec{w}}_i^{\intercal} \hat{\mat{Q}} \td{\vec{w}}_i\), \notag
\end{align}
the replicated partition function factorizes and becomes an integral over the matrix order parameters $\mat{Q}$ and $\hat{\mat{Q}}$, that can be evaluated using a Laplace method in the $d \to \infty$ limit,
\begin{align}
	\EE_{\vec{y}, \mat{X}} \[ \mZ_\ndim(\{\vec{y}, \mat{X}\}, \alpha, \beta)^r\] &\propto \int \d \mat{Q} ~ \d \hat{\mat{Q}} ~ e^{d \Phi^{(r)}\(\mat{Q}, \hat{\mat{Q}}, \alpha, \beta \) } \label{appendix:expectation_Zn} \\
	&\underset{d \to \infty}{\simeq} \exp \( d \cdot  \extr_{\mat{Q}, \hat{\mat{Q}}} \left\{ \Phi^{(r)}\(\mat{Q}, \hat{\mat{Q}}, \alpha, \beta \) \right\} \), \nonumber
\end{align}
where the replica potential is defined by
\begin{align}
     \Phi^{(r)}\(\mat{Q}, \hat{\mat{Q}}, \alpha, \beta \) &\equiv  -\frac{1}{2}\tr{\mat{Q}\hat{\mat{Q}}} +\log \Psi_{\w}^{(r)} (\hat{\mat{Q}})+\alpha\log \Psi_{\out}^{(r)}(\mat{Q},\beta)\,, \nonumber
      \spacecase
      \Psi_{\w}^{(r)} (\hat{\mat{Q}}) &= \displaystyle \int_{\mathbb{R}^r} \d \rP_\w(\td{\vec{w}}) ~ e^{ \frac{1}{2}\td{\vec{w}}^{\intercal} \hat{\mat{Q}} \td{\vec{w}} }  \,, \label{appendix:S_r} \spacecase
     \Psi_{\out}^{(r)}(\mat{Q},\beta) &=  \displaystyle \int \d \rP_\y\(y\) ~ \int_{\mathbb{R}^r}  \d \rP_{\z}(\td{\vec{z}},\mat{Q}) ~ \mC (y|\td{\vec{z}},\beta). \nonumber
\end{align}
Finally, using eq.~\eqref{appendix:free_entropy} and switching the two limits $r \to 0$ and $d \to \infty$, the quenched free entropy $\Phi$ simplifies as an extremization problem
\begin{equation}
\Phi (\alpha, \beta) = \extr_{\mat{Q}, \hat{\mat{Q}}} \left\{\lim_{r\rightarrow 0} \frac{\partial \Phi^{(r)} (\mat{Q},\hat{\mat{Q}}, \alpha, \beta)}{\partial  r} \right\} ,
\label{appendix:free_entropy2}
\end{equation}
over general symmetric matrices $\mat{Q}$ and $\hat{\mat{Q}}$. In the following we assume simple Ansätze for these matrices that allow to obtain analytic expressions in $r$ in order to take the derivative and the limit $r\to 0^+$.

\subsection{Annealed computation}
\label{appendix:annealed_calculation}
We can use the replica calculation \eqref{appendix:S_r} to compute the \emph{annealed} free entropy $\Phi^a (\alpha) =  \log \EE_{\vec{y}, \mat{X}} \[ \mZ_\ndim(\{\vec{y}, \mat{X}\}, \alpha)\]$, see \Sec\ref{main:intro:disordered:quenched_spin_glass:annealed}, by assuming there exists a single replica with $r=1$, $\mat{Q} = q$ and $\hat{\mat{Q}}=\hat{q}$ \cite{krauth1989storage}.
\begin{equation}
	\Phi^a(\alpha, \beta) = \underset{q ,\hat{q}}{\extr} \left\{  -\frac{1}{2}q \hat{q} +  \log  \Psi_\w^a (\hat{q})  + \alpha \log  \Psi_\out^a (q, \beta) \right\} \,,
	\label{eq:appendix:annealed_calculation:annealed_free_entropy}
\end{equation}
with
\begin{align}
\begin{aligned}
		\Psi_\w^a (\hat{q}) &\equiv  \int_\bbR \d\rP_\w(w) ~ \exp \( \frac{1}{2} \hat{q} w^2 \) \,, \\
		\Psi_\out^a (q, \beta) &\equiv  \int_\bbR \d \rP_y ~ \int_\bbR \d z ~ \frac{e^{-\frac{z^2}{2 q}}}{\sqrt{2\pi q}} \mC (y|z, \beta)\,.
\end{aligned}
\end{align}
Finally, in the case of binary weights $\rP(w) = \(\delta(w-1) + \delta(w+1) \) $ we obtain $\Psi_\w^a (\hat{q}) =2 \exp\( \frac{1}{2} \hat{q} \) $. Taking the derivative of \eqref{eq:appendix:annealed_calculation:annealed_free_entropy} with respect to $\hat{q}$ we obtain $q=1$ so that the annealed free entropy writes
\begin{align}
\Phi^a(\alpha, \beta) = \log(2) + \alpha \log\(  \int_\bbR  \d \rP_\y(y) ~ \int_\bbR  \D z ~ \mC (y|z, \beta) \)\,.
\label{appendix:annealed_entropy}
\end{align}
We can compute therefore the annealed capacity $\alpha_a$ at zero temperature $\beta \to \infty$, such that the annealed entropy $\Phi^a(\alpha, \beta \to \infty)$ vanishes:
\begin{equation}
	\alpha_a = \frac{-\log(2)}{\log\(   \int_\bbR  \d \rP_\y(y) ~ \int_\bbR   \D z ~ \mC (y|z) \)}\,.
	\label{appendix:annealed_capacity}
\end{equation}

\subsection{Choosing an Ansatz}
Back to the quenched average computation in \eqref{appendix:free_entropy2}, optimizing over the space of matrices is intractable. Therefore, one needs to assume simple Ansätze about the matrices structure to push the computation further, see \Sec\ref{main:sec:mean_field:replica_method:replica_Ansatz}, such as the so-called
\begin{itemize}
    \item Replica Symmetry (\aclink{RS}) Ansatz: $\mat{Q}^{(\rs)} = (Q-q_0) \mat{I}_r + q_0 \mat{J}_r$
    \item 1-Step Replica Symmetry Breaking (\aclink{1RSB}) Ansatz: $\mat{Q}^{(\textrm{1rsb})}= (Q-q_1) \mat{I}_r  + (q_1-q_0) \mat{I}_{r/x_0} \otimes \mat{J}_{x_0} + q_0 \mat{J}_r $\,,
    \item 2-Step Replica Symmetry Breaking (\aclink{2RSB}) Ansatz: 
    $$\mat{Q}^{(\textrm{2rsb})} = \( Q - q_2 \) \mat{I}_r + \( q_2 - q_1 \) \mat{I}_{r/x_1} \otimes \mat{J}_{x_1} +   \( q_1 - q_0 \) \mat{I}_{r/x_0} \otimes \mat{J}_{x_0} +q_0 \mat{J}_r  $$
\end{itemize}
where $\mat{I}_k$ is the identity matrix of size $k$, and $\mat{J}_k$ is the matrix of size $k$ full of ones.
Plugging these Ansätze, taking the derivative and the $r\to 0^+$ limit, extremizing over the space of matrices boils down to much simpler optimization problems over a few scalar order parameters, as illustrated in the next sections.

\subsection{RS free entropy for i.i.d data}
\label{appendix:replicas_iid:rs}

Let us compute the free entropy potential $\Phi^{(r)} (\mat{Q},\hat{\mat{Q}}, \alpha, \beta)$ in \eqref{appendix:S_r} in the \aclink{RS} Ansatz. The latter assumes that all replicas remain equivalent with a common overlap $q_0 = \frac{1}{d} \sum_{i=1}^\ndim w_i^a w_i^b$ for $a \ne b$ and a norm $Q= \frac{1}{d} \sum_{i=1}^\ndim w_i^a w_i^a$, leading to the following expressions for matrices $\mat{Q}$ and $\hat{\mat{Q}} \in \mathbb{R}^{r\times r}$:
\begin{equation}
\begin{aligned}[c]
\mat{Q}^{(\rs)} =
\scalemath{0.9}{\begin{pmatrix} 
 Q & q_0 & \cdots & q_0 \\ 
 q_0 & Q & \ddots & \vdots  \\
\vdots & \ddots & \ddots & q_0  \\
 q_0 & \cdots & q_0 & Q   
\end{pmatrix}}
\end{aligned}
\qquad
\textrm{and} 
\qquad
\begin{aligned}[c]
\hat{\mat{Q}}^{(\rs)}=
\scalemath{0.9}{\begin{pmatrix} 
  \hat{Q} & \hat{q}_0 & \cdots & \hat{q}_0 \\ 
\hat{q}_0 &  \hat{Q} & \ddots & \vdots  \\
\vdots & \ddots & \ddots & \hat{q}_0  \\
\hat{q}_0 & \cdots &\hat{q}_0 &  \hat{Q}   
\end{pmatrix}}
.
\end{aligned}
\end{equation}
Let us compute separately the terms involved in the functional $\Phi^{(r)} (\mat{Q},\hat{\mat{Q}}, \alpha, \beta)$ \eq\eqref{appendix:S_r}: the first is a trace term, the second a term  $\Psi_{\w}^{(r)}$ depends on the prior distribution $\rP_\w$ and finally the third $\Psi_{\out}^{(r)}$ on the constraint $\mC(y|z)$ in \eqref{appendix:replica_constraint}. 
\paragraph{Trace} 
The trace term can be easily computed as
\begin{equation}
	\left.\frac{1}{2}\tr{\mat{Q}\hat{\mat{Q}}} \right|_{\rs} =\frac{1}{2} \( r Q\hat{Q} + r(r-1) q_0\hat{q}_0 \)\,.
\end{equation}

\paragraph{Prior integral} Evaluated at the \aclink{RS} fixed point, and using a Hubbard-Stratonovich transformation, see \ref{appendix:replica_computation:hubbard}, the prior integral can be further simplified 
\begin{align}
	&\left.\Psi_{\w}^{(r)} (\hat{\mat{Q}})\right|_{\rs} = \int \d \rP_\w(\td{\vec{w}})  ~ e^{ \frac{1}{2}\td{\vec{w}}^{\intercal} \hat{\mat{Q}}^{(\rs)} \td{\vec{w}} } =   \int \d \rP_\w(\td{\vec{w}}) \\
	&\qquad  \times \exp\( {\frac{(\hat{Q}- \hat{q}_0  )}{2}\sum_{a=1}^r (w^a)^2}\) \exp\(\hat{q}_0 \(  \sum_{a=1}^r  w^a  \)^2 \) \nonumber \\
	&\qquad  = \int \D \xi_0 ~ \[ \int \d \rP_\w(w) ~ \exp\( \( {\frac{(\hat{Q} - \hat{q}_0  )}{2}  w^2}+ \xi_0 \sqrt{\hat{q}_0} w \) \) \]^r \,. \nonumber  
\end{align}

\paragraph{Constraint integral} 
Recall the vector $\td{\vec{z}}\sim \rP_\z \triangleq \mathcal{N}(\vec{0},\mat{Q})$ follows a Gaussian distribution with zero mean and covariance matrix $\mat{Q}$. In the \aclink{RS} Ansatz, the covariance can be rewritten as a linear combination of the identity $\mat{I}_r$ and $\mat{J}_r$: $\left. \mat{Q} \right|_{\rs} = (Q-q_0) \mat{I}_r + q_0 \mat{J}_r$, that allows to split the variable $z^a = \sqrt{Q-q_0} u^a + \sqrt{q_0} \xi_0   $ with  $\xi_0 \sim \mathcal{N}(0,1)$ and $ \forall a \in \lb r \rb,~ u_a \sim \mathcal{N}(0,1)$. The constraint integral finally reads
\begin{align}
& \left. \Psi_{\out}^{(r)}(\mat{Q},\beta) \right|_{\rs} = \displaystyle \int \d \rP_{y}\(y\) ~ \int_{\mathbb{R}^r}  \d \rP_\z(\td{\vec{z}}) ~ \mC (y|\td{\vec{z}},\beta) \nonumber\\
& = \int \d \rP_{y}\(y\) \int \d \xi_0 ~ \int  \prod_{a=1}^r \d u^a ~  \mC \(y| \sqrt{Q-q_0} u^a + \sqrt{q_0} \xi_0 , \beta \)\\
	 &= \int \d \rP_{y}\(y\) ~ \int \d \xi_0  ~ \left[ \int  \d z ~ \mC \(y|  \sqrt{Q-q_0} z +  \sqrt{q_0} \xi_0, \beta \) \right]^r \,. \nonumber
\end{align}
Finally, putting pieces together, the functional $\Phi^{(r)} (\mat{Q},\hat{\mat{Q}}, \alpha, \beta)$ taken at the \aclink{RS} fixed point has an explicit formula and dependency in $r$:
\begin{align}
	&\left.\Phi^{(r)} (\mat{Q},\hat{\mat{Q}}, \alpha, \beta) \right|_{\rs} \underset{r\to 0}{\simeq} -\frac{1}{2} \( r Q\hat{Q} + r(r-1) q_0\hat{q}_0 \)  \nonumber\\
	& + r \int \d \xi_0 ~ \log\( \int \d \rP_\w(w) ~ \exp{ \( {\frac{(\hat{Q}- \hat{q}_0  )}{2}  w^2}+ \xi_0\sqrt{\hat{q}_0} w \)}   \) \\
	& + r\alpha  \int \d \rP_{y}\(y\) \int \d \xi_0 ~  \log\(   \int  \d z ~ \mC \(y | \sqrt{Q-q_0} z + \sqrt{q_0} \xi_0 , \beta \)  \).\nonumber
\end{align}

\subsubsection{Summary of RS free entropy - general case}
Taking the derivative with respect to $r$ and the $r\to 0^+$ limit, the  \aclink{RS} free entropy has a simple expression
\begin{align}
    \Phi^{(\rs)}(\alpha, \beta) &= \underset{ q_0, \hat{q}_0}{\textbf{extr}}     \left\{   -\frac{1}{2}Q\hat{Q} + \frac{1}{2}q_0\hat{q}_0+  \Psi_{\w}^{(\rs)}(\hat{q}_0)   +\alpha  \Psi_{\out}^{(\rs)}(q_0, \beta)    \right\},  \nonumber \\
    \Psi_{\w}^{\rs}(\hat{q}_0) &\equiv \EE_{\xi_0}  \log   \EE_{w} \[ \exp \( {\frac{(\hat{Q}- \hat{q}_0  )}{2}  w^2}+ \xi_0 \sqrt{\hat{q}_0} w \)  \]  \,, \label{appendix:replicas:free_energy_rs_out_w} \\
	\Psi_{\out}^{\rs}(q_0, \beta) &\equiv  \EE_y ~ \EE_{\xi_0}  \log  \EE_{z} \[\mC \(y \big | \sqrt{Q - q_0} z + \sqrt{q_0}\xi_0, \beta \)  \]\,, \nonumber
\end{align} 
where $\xi_0, z \sim \mN(0,1)$, $w\sim \rP_\w(.)$, $y\sim \rP_{y}(.) $ and $Q=\hat{Q}=1$.\\

\subsubsection{Summary of RS free entropy - spherical case}
In the spherical (or equivalently in the Gaussian case with a correctly defined variance) such that the weights verify $\|\td{\vec{w}}\|_2^2=d$, $\Psi_{\w}^{(r)} (\hat{\mat{Q}})$ in eq.~\eqref{appendix:S_r} can be directly integrated
\begin{align}
	\Psi_{\w}^{(r)} (\hat{\mat{Q}}) = \displaystyle \int_{\|\td{\vec{w}}\|_2^2=d} \d \td{\vec{w}} ~ \exp\( \frac{1}{2}\td{\vec{w}}^{\intercal} \hat{\mat{Q}} \td{\vec{w}} \) = -\frac{1}{2} \log \det{2\pi (\mat{I}_r + \hat{\mat{Q}})}\,.
\end{align}
Besides, taking the derivative of \eq\eqref{appendix:S_r} with respect to $\hat{\mat{Q}}$ we obtain $\mat{Q}^{-1} = (\mat{I}_r + \hat{\mat{Q}})$. Injecting it, we can get rid of $\hat{\mat{Q}}$ and obtain
\begin{align}
    \Phi^{(r)}\(\mat{Q}, \alpha, \beta \) &\equiv  \frac{1}{2} \log \det{2\pi \mat{Q}} +\alpha\log \Psi_{\out}^{(r)}(\mat{Q},\beta)\,.
	\label{app:replicas:spherical_simplification}	
\end{align}
\paragraph{Determinant}
The above determinant reads in the \aclink{RS} Ansatz
\begin{equation}
	\left.\frac{1}{2}\det{\mat{Q}} \right|_{\rs} \simeq \frac{r}{2} \( \log(1-q_0) + \frac{q_0}{1 + (r-1)q_0} + \dots \)\,,
	\label{app:replicas:rs:determinant}
\end{equation}
so that it leads to the \aclink{RS} free entropy 
\begin{align}
\label{app:spherical:rs_free_energy_simplified}
	\Phi^{(\rs)}(\alpha,\beta) &= \extr_{q_0} \left\{ \frac{1}{2(1-q_0)} +\frac{1}{2}\log(2\pi) +\frac{1}{2} \log(1-q_0) \right. \nonumber \\
	& \qquad \qquad \qquad \qquad \qquad \qquad \left. + \alpha \Psi_{\out}^{\rs}(q_0, \beta) \right\}\,, 
\end{align}
with $\Psi_{\out}^{\rs}$ defined in eq.~\eqref{appendix:replicas:free_energy_rs_out_w}.

\subsection{RS Stability}
\label{appendix:AT_stability}

\subsubsection{De Almeida Thouless RS stability}
The stability of a given saddle point Ansatz is related to the positivity the Hessian of the functional $-\Phi^{(r)}$. Following \cite{Almeida1978, gardner1988optimal,engel2001statistical}, the stability analysis leads to computing the first unstable eigenvalues of the Hessian, the so-called \emph{replicons eigenvalues}. 
In the context of the \aclink{RS} Ansatz $\lambda_3^A$ and $\lambda_3^B$ can be expressed as functions of $\{g_i^w,f_i^z\}_{i=0}^2$ defined in \Chap\ref{chap:binary_perceptron} - \eq\eqref{main:f_z_g_w_rs}:
\begin{align}
\begin{aligned}
		\lambda_3^A(q_0) &= \frac{1}{(Q-q_0)^2} \EE_{\xi_0}\[\frac{\(f_0^{z}(f_0^{z}-f_2^{z}) + (f_1^{z})^2 \)^2}{(f_0^{z})^4}(\xi_0,q_0)\] \,,\\ 
		\lambda_3^B (\hat{q}_0) &= \EE_{\xi_0}\[ \frac{ \(g_0^{w}g_2^{w} -(g_1^{w})^2  \)^2 }{(g_0^{w})^4}(\xi_0,\hat{q}_0) \]
		\,,
\end{aligned}		
\end{align}
for $\xi_0\sim\mN(0,1)$. The instability \aclink{dAT}-line is defined when the determinant of the Hessian vanishes, \ie when the first negative eigenvalues appear. This translates as an implicit equation over $\alpha$, where $q_0,\hat{q}_0$ are solution of the saddle point equations eq.~\eqref{main:phi_RS} at $\alpha=\alpha_\textrm{at}$: 
\begin{align}
	\frac{1}{\alpha_\textrm{at}} &=   \lambda_{3}^A \(q_0(\alpha_\textrm{at}),\beta \) \times \lambda_{3}^B \(\hat{q}_0(\alpha_\textrm{at}) \) \,.
\end{align}  
However for $\alpha < \alpha_\textrm{at}$, $(q_0,\hat{q}_0)=(0,0)$ is the only solution. 
Defining for $z \sim \mN(0,1)$ and $w\sim \rP_\w$
\begin{align}
\td{f}_i^{z} &\equiv \EE_z\[ z^i \varphi(z) \] \,, && \td{g}_i^{w} \equiv \EE_w \[ w^i \exp\( w^2 / 2 \) \]\,,
\label{appendix:fgtd}
\end{align},
this expression can be simplified in the case where the prior distribution $\rP_\w$ and the activation $\varphi$ are symmetric. In fact the symmetry imposes $\td{f}_1^{z}=0$ and $\td{g}_1^{w} = 0$ and the condition simplifies to
\begin{align}
	\frac{1}{\alpha_\textrm{at}} &= \(\frac{\td{f}_2^{z}-\td{f}_0^{z}}{\td{f}_0^{z}}\)^2 \(\frac{\td{g}_2^{w} }{\td{g}_0^{w}}\)^2 \, .
	\label{appendix:AT_line}
\end{align}  

\subsubsection{Existence and stability of the RS fixed point}
 We provide an alternative approach to get the instability condition of the \aclink{RS} solution for symmetric prior $\rP_\w$ and activation $\varphi$. In this symmetric case, the stability can be derived from the existence and stability of the symmetric fixed point $(q_0,\hat{q}_0)=(0,0)$. Let us define
\begin{align}
\begin{aligned}
	F(q_0) &\equiv \alpha  \EE_{\xi_0}\[ \frac{(f_1^{z})^2-2 \xi_0 \sqrt{q_0}f_0^{z} f_1^{z} + q_0 \xi_0^2 (f_0^{z})^2}{(1-q_0)^2  (f_0^{z})^2}(\xi_0,q_0) \] \,, \\
	G( \hat{q}_0) &\equiv  \EE_{\xi_0} \[ \frac{g_2^{w} - \xi_0 \hat{q}_0^{-1/2}g_1^{w}}{g_0^{w}}(\xi_0,\hat{q}_0) \] \,.
\end{aligned}
\end{align}
In fact the saddle point equations at the \aclink{RS} fixed point eq.~\eqref{appendix:replicas:free_energy_rs_out_w} can be written using the functions $F, G$, and can be reduced to a single fixed point equation over $q_0$
\begin{align}
	\begin{cases}
         q_0=G(\hat{q}_0)\,, \\
         \hat{q}_0=F(q_0)\,,
     \end{cases} \qquad
     \Rightarrow \qquad
         q_0=G \circ F (q_0) \equiv H(q_0)\,.
\label{appendix:H_q0}
\end{align}
The \aclink{RS} stability of the fixed point $(q_0,\hat{q}_0)=(0,0)$ can be analyzed from the above fixed point equation eq.~\eqref{appendix:H_q0}. Computing $F, F', G, G'$ in the limit $(q_0,\hat{q}_0) \to (0,0)$, expanding $\{f_i^z$,$g_i^w\}_i$ as functions of $\{\td{f}^z_i,\td{g}^w_i \}_i$ and finally using the symmetry conditions $\td{f}_1^{z}=0$ and $\td{g}_1^{w} = 0$, we finally obtain
\begin{align*}
		F(q_0) &\underset{q_0 \to 0}{=} \alpha    \left [ \(\frac{\td{f}_1^{z}}{\td{f}_0^{z}}\)^2 +q_0\( \frac{(\td{f}_2^{z}-\td{f}_0^{z})^2}{(\td{f}_0^{z})^2} +3\frac{(\td{f}_1^{z})^4}{(\td{f}_0^{z})^4} \right.\right. \\ 
		& \left.\left. -4 \frac{(\td{f}_1^{z})^2(\td{f}_2^{z}-\td{f}_0^{z})}{(\td{f}_0^{z})^3}   \)  + \Theta(q_0^2) \right]{\sim} \alpha q_0    \(\frac{\td{f}_2^{z}-\td{f}_0^{z}}{\td{f}_0^{z}}\)^2 \underset{q_0 \to 0}{\longrightarrow} 0 \,, \vspace{0.2cm} \\
		\partial_{q_0} F (q_0) &\underset{q_0 \to 0}{=} \alpha  \left[ \(\frac{\td{f}_2^{z}-\td{f}_0^{z}}{\td{f}_0^{z}}\)^2 + \(\frac{\td{f}_1^{z}}{\td{f}_0^{z}}\)^2 \left (3\frac{(\td{f}_1^{z})^2}{(\td{f}_0^{z})^2} -4 \frac{(\td{f}_2^{z}-\td{f}_0^{z})}{\td{f}_0^{z}}\) \right. \\
		  &\left. \qquad \qquad  + \Theta(q_0) \right] \underset{q_0 \to 0}{\longrightarrow}  \alpha    \(\frac{\td{f}_2^{z}-\td{f}_0^{z}}{\td{f}_0^{z}}\)^2 \,, \vspace{0.2cm} \\
		G(\hat{q}_0) &\underset{\hat{q}_0 \to 0}{=} \(\frac{\td{g}_1^{w} }{\td{g}_0^{w}}\)^2  + \hat{q}_0\( \(\frac{\td{g}_2^{w} }{\td{g}_0^{w}}\)^2 + \frac{\td{g}_1^{w}}{\td{g}_0^{w}} \(3 \(\frac{\td{g}_1^{w} }{\td{g}_0^{w}}\)^3-4\frac{\td{g}_1^{w}\td{g}_2^{w}}{(\td{g}_0^{w})^2} \) \) \\
		 & \qquad + \Theta( \hat{q}_0^{3/2}) \underset{ \hat{q}_0 \to 0}{\longrightarrow} 0  \,, \vspace{0.2cm} \\
		\partial_{\hat{q}_0} G(\hat{q}_0) &\underset{\hat{q}_0 \to 0}{=}  \(\frac{\td{g}_2^{w} }{\td{g}_0^{w}}\)^2 + \frac{\td{g}_1^{w}}{\td{g}_0^{w}} \(3 \(\frac{\td{g}_1^{w} }{\td{g}_0^{w}}\)^3-4\frac{\td{g}_1^{w}\td{g}_2^{w}}{(\td{g}_0^{w})^2} \) + \Theta(\sqrt{ \hat{q}_0}) \\
		 &\qquad \underset{ \hat{q}_0 \to 0}{\longrightarrow} \(\frac{\td{g}_2^{w} }{\td{g}_0^{w}}\)^2 \,.
\end{align*}
Finally, the existence and stability conditions of the fixed point $(q_0,\hat{q}_0)=(0,0)$  
translate as an explicit condition over $\alpha$ that implicitly defines $\alpha_\textrm{at}$
\begin{equation}
	\begin{cases}
		H(q_0) = G \circ F(q_0) \underset{q_0 \to 0}{\to} 0 \vspace{0.3cm} \\
		\left. \frac{\partial H}{\partial q_0}\right|_{q_0=0} =  \left. \frac{\partial G}{\partial \hat{q}_0} \times \frac{\partial F}{\partial q_0}\right|_{q_0=0} \leq 1 \,, 
	\end{cases}
	\Rightarrow 
	\alpha \leq  \left[   \(\frac{\td{f}_2^{z}-\td{f}_0^{z}}{\td{f}_0^{z}}\)^2 \(\frac{\td{g}_2^{w} }{\td{g}_0^{w}}\)^2 \right]^{-1} \equiv \alpha_\textrm{at}\,.
	\label{stability}
\end{equation}

\subsection{1RSB free entropy for i.i.d data}
\label{appendix:replica_1RSB}

The free entropy potential \eq\eqref{appendix:S_r} can also be evaluated at the simplest non trivial fixed point: the one-step Replica Symmetry Breaking Ansatz (\aclink{1RSB}), see \Sec\ref{main:sec:mean_field:replica_method:replica_Ansatz}. 
Instead of assuming that replicas are equivalent, it states that the symmetry between the replicas is broken and that the replicas are clustered in different \emph{states}, with inner-overlap $q_1$ and outer-overlap $q_0$. Translating this analytically, the matrices can be expressed as function of the Parisi parameter $x_0$, which controls the size of the clusters:
\begin{align}
\begin{aligned}
		\mat{Q}^{(\textrm{1rsb})} &= q_0 \mat{J}_r + \( q_1 - q_0 \) \mat{I}_{\frac{r}{x_0}} \otimes \mat{J}_{x_0} +  \( Q - q_1 \) \mat{I}_r \,, \\
		\hat{\mat{Q}}^{(\textrm{1rsb})} &= \hat{q}_0 \mat{J}_r + \( \hat{q}_1 - \hat{q}_0 \) \mat{I}_{\frac{r}{x_0}} \otimes \mat{J}_{x_0} +  \( \hat{Q} - \hat{q}_1 \) \mat{I}_r \,.
\end{aligned}
\end{align}

\paragraph{Trace term} 
Again, the trace term can be easily computed
\begin{equation}
	\left.\frac{1}{2}\tr{\mat{Q}\hat{\mat{Q}}} \right|_{\textrm{1rsb}} =\frac{1}{2} \( r Q\hat{Q} + r(x_0-1)q_1\hat{q}_1 + r(r-x_0)q_0\hat{q}_0 \).
	\label{appendix:1RSB_Tr}
\end{equation}

\paragraph{Prior integral}
To decouple the replicas with different overlaps $q_0, q_1$, and using Hubbard-Stratonovich transformations in \App\ref{appendix:replica_computation:hubbard}, the prior integral can be written 
\begin{align}
	&\left.\Psi_{\w}^{(r)} (\hat{\mat{Q}})\right|_{\textrm{1rsb}} =  \int_{\bbR^r} \d \rP_\w(\td{\vec{w}}) ~ \exp\(  \frac{(\hat{Q}- \hat{q}_1  )}{2}\sum_{a=1}^r (w^a)^2 \right. \nonumber \\
	& \quad \quad \quad \left. + \frac{(\hat{q}_1-\hat{q}_0)}{2} \sum_{k=1}^{\frac{r}{x_0}} \sum_{a,b=(k-1)x_0 + 1 }^{kx_0} w^a w^b + \frac{\hat{q}_0}{2} \(\sum_{a=1}^r w^a \)^2 \) \nonumber\\
	&= \int \D \xi_0 ~ \left[\int \D \xi_1 ~ \right. 	\label{appendix:1RSB_Iw} \\
	& \left. \left[ \int \d \rP_\w(w) ~ \exp\( \frac{(\hat{Q}- \hat{q}_1  )}{2} w^2 + \(\sqrt{\hat{q}_0}\xi_0+\sqrt{\hat{q}_1-\hat{q}_0}\xi_1 \)w \) \right ]^{x_0} \right]^{\frac{r}{x_0}}\,, \nonumber
\end{align}
with $\xi_0, \xi_1 \sim \mN(0,1)$.

\paragraph{Constraint integral} The replicated vector $\td{\vec{z}} \sim \rP_\z(.) \triangleq \mN_{\vec{z}}\(\vec{0} ,\mat{Q}^{(\textrm{1rsb})}\)$ follows a Gaussian vector with zero mean and covariance matrix $\mat{Q}^{(\textrm{1rsb})}$ that can be decomposed in a sum of normal Gaussian vectors $\xi_0 \sim \mN(0,1)$, $\forall k \in \lb 1 ; \frac{r}{x_0} \rb$, $\xi_k \sim \mN(0,1)$ and $ \forall a \in \lb (k-1) x +1 ; k x \rb$, $u_a \sim \mN(0,1)$:
\begin{equation*}
z^a = \sqrt{q_0} \xi_0 + \sqrt{q_1-q_0} \xi_k  + \sqrt{Q-q_1} u_{a}\,.
\end{equation*}
Finally, the constraint integral reads
\begin{align}
&\left. \Psi_{\out}^{(r)}(Q,\beta) \right|_{\textrm{1rsb}} \nonumber\\ 
& = \int \d \rP_{y}\(y\) \int \D \xi_0 ~ \int \prod_{k=1}^{\frac{r}{x_0}}  \D \xi_k \nonumber \\
& \qquad \times \int \prod_{a=(k-1)x+1}^{kx}  \D u_a ~ \mC \(y|\sqrt{q_0} \xi_0 + \sqrt{q_1-q_0} \xi_k  + \sqrt{Q-q_1} u_{a}, \beta\) \nonumber \\
&= \int \d \rP_{y}\(y\) \int \D \xi_0  \label{appendix:1RSB_Iz} \\ 
& \quad \times \left[ \int \D \xi_1  \left[ \int \D z ~ \mC \(y|\sqrt{q_0} \xi_0 + \sqrt{q_1-q_0} \xi_1  + \sqrt{Q-q_1} z, \beta \)\right]^x \right]^{\frac{r}{x_0}}. \nonumber 
\end{align}
 
 \subsubsection{Summary of the 1RSB free entropy - general case}
Gathering the previous computations \eq(\ref{appendix:1RSB_Tr}, \ref{appendix:1RSB_Iw}, \ref{appendix:1RSB_Iz}), the functional $\Phi^{(r)}$ evaluated at the \aclink{1RSB} fixed point reads:
\begin{align}
\begin{aligned}
	&	\left.\Phi^{(r)} (\mat{Q},\hat{\mat{Q}}, \alpha, \beta) \right|_{\textrm{1rsb}} \nonumber\\
	&\underset{r\to 0}{\simeq}
	- \frac{1}{2} \( r Q\hat{Q} + r(x_0-1)q_1\hat{q}_1 + r(r-x_0)q_0\hat{q}_0 \) \\
	& \qquad \qquad \qquad \qquad \qquad \qquad + r \Psi_{\w}^{(\textrm{1rsb})}(\hat{\vec{q}}) + r \alpha  \Psi_{\out}^{(\textrm{1rsb})}(\vec{q},\beta)
\end{aligned}
\end{align}
with 
\begin{align}
	&\Psi_{\w}^{(\textrm{1rsb})}(\hat{\vec{q}}, x_0) \equiv \frac{1}{x_0}
                \EE_{\xi_0} \nonumber  \\
    & \log\( \EE_{\xi_1} \EE_w \[\exp\(
                  \frac{(\hat{Q}- \hat{q}_1  )}{2} w^2 +
                  \(\sqrt{\hat{q}_0}\xi_0+\sqrt{\hat{q}_1-\hat{q}_0}\xi_1
                  \)w \)  \]^{x_0}\) \,,  \nonumber  \\
    &\Psi_{\out}^{(\textrm{1rsb})}(\vec{q}, x_0, \beta) \equiv \frac{1}{x_0}
                \EE_y \EE_{\xi_0}  \\
                & \qquad  \log\(  \EE_{\xi_1} \EE_{z} \[\mC( y \big | \sqrt{q_0} \xi_0 +
                \sqrt{q_1-q_0} \xi_1  + \sqrt{Q-q_1} z,\beta) \]^x_0 \)\,,   \nonumber    
\end{align}
where $\vec{q} =(q_0, q_1)$, $\hat{\vec{q}} =(\hat{q}_0, \hat{q}_1)$, $\xi_0, \xi_1,z \sim \mN(0,1)$, $w \sim \rP_\w(.)$, $y\sim \rP_{y}(.)$ and $Q=\hat{Q}=1$.
Finally taking the derivative with respect to $r$ and the limit $r\to 0^+$, we obtain the \aclink{1RSB} free entropy 
\begin{multline}
    \label{appendix:free_energy_1rsb}
	\Phi^{(\textrm{1rsb})}(\alpha, \beta) = \underset{ \vec{q}, \hat{\vec{q}}, x_0}{\textbf{extr}} \left\{  \frac{1}{2} \(  q_1\hat{q}_1 - Q\hat{Q} \) + \frac{x_0}{2} \(q_0\hat{q}_0 - q_1\hat{q}_1 \)   \right. \\
	\left. + \Psi_{\w}^{(\textrm{1rsb})}(\hat{\vec{q}},x_0)   +\alpha \Psi_{\out}^{(\textrm{1rsb})}(\vec{q},x_0,\beta)    \right\} \,.
\end{multline}

\subsubsection{Summary of the 1RSB free entropy - spherical case}
In the \aclink{1RSB}, the simplification \eq\eqref{app:replicas:spherical_simplification} remains valid. Therefore, we can simply compute the determinant in the \aclink{1RSB} Ansatz.
\paragraph{Determinant}
\begin{align*}
\begin{aligned}
	\left. \det{\mat{Q}} \right|_{\textrm{1rsb}} &= \( r q_0 + x_0 (q_1-q_0) +(1-q_1)\) \\
	 & \times \( 1 -q_1 \)^{r - r/x_0} \times \( x_0 (q_1-q_0) + (1-q_1) \)^{r/x_0-1}\,,
\end{aligned}
\end{align*}
so that
\begin{align*}
\begin{aligned}
	&\left. \log \det{\mat{Q}} \right|_{\textrm{1rsb}} \simeq r \( \frac{x_0-1}{x_0} \log (1- q_1) + \right. \\
	& \qquad \left. \frac{1}{x_0} \log\( x_0 (q_1-q_0) + (1-q_1) \) + \frac{q_0}{x_0(q_1-q_0) + (1-q_1)}  \)\,.
\end{aligned}
\label{app:det_1RSB}
\end{align*}
Using the above expression for the determinant and the simplified replica potential in eq.~\eqref{app:replicas:spherical_simplification} we obtain
\begin{align}
	&\Phi^{(\textrm{1rsb})}(\alpha, \beta) = \extr_{q_0, q_1, x} \left\{  \frac{1}{2}\log(2\pi) + \frac{x-1}{2x} \log (1- q_1) + \right. \nonumber \\
	& \qquad \left. + \frac{1}{2x} \log\( x(q_1-q_0) + (1-q_1) \)  + \frac{q_0}{2 \( x(q_1-q_0) + (1-q_1) \) } \right. \\
	& \qquad \left.  +  \alpha   \Psi_{\out}^{(\textrm{1rsb})}(\vec{q},\beta) \right\} \,. \nonumber
	\label{app:spherical:1rsb_free_energy_simplified}
\end{align}


\subsection{Ground state energies}
\label{appendix:replicas_ground_state_spherical}
We focus on the particular case of the spherical perceptron with parameters $\vec{w} \in \bbR^{d}$ lying on the sphere and verifying $\|\vec{w}\|_2^2=d$. 

\paragraph{RS capacity}
In the case of the step-perceptron activation function $\varphi(z) = \theta(z- \kappa)$ for $\kappa \geq 0$, we can compute the capacity $\alpha_c$ taking the extremization over $q_0$ in eq.~\eqref{app:spherical:rs_free_energy_simplified}:
\begin{align}
	q_0 = -2\alpha(1-q_0)^2  \partial_{q_0}\Psi_{\out}^{(\rs)} (q_0) \simeq 2\alpha(1-q_0)^2  \int_{-\infty}^{\kappa} \d t \frac{(\kappa -t)^2}{(1-q_0)^2} \,.
\end{align}
At the critical capacity, we have $q_0 \to 1$, which leads to the expression
\begin{align}
	\alpha_c = \(\int_{-\infty}^{\kappa} \d t (\kappa - t)^2\)^{-1} \,.
\end{align}
Notice that for $\kappa = 0$, this approach performed in \cite{gardner1988optimal} naturally leads to Cover's result $\alpha_c = 2$ \cite{cover1965geometrical}. Above this capacity $\alpha_c$, the constraints cannot be satisfied simultaneously and the ground state energy is necessarily positive.  

\subsubsection{RS ground state energy} 
To compute the ground state energy, we first need to take both limits $q_0 \to 1$ and $\beta \to \infty$, keeping the product $\chi = \beta (Q-q_0)$ finite \cite{Majer1993a, Erichsen1992, Whyte1996}.
Recall eq.~\eqref{appendix:replicas:free_energy_rs_out_w}, we obtain using the definition of $\mC$ in \eqref{appendix:replica_constraint}
\begin{align}
	&\Psi_{\out}^{\rs}(q_0, \beta) \equiv  \EE_y ~ \EE_{\xi_0}  \log  \EE_{z} \[\mC \(y \big | \sqrt{Q - q_0} z + \sqrt{q_0}\xi_0, \beta \)  \] \nonumber\\
	& = \int \d \rP_\y(y) ~ \int \D \xi_0 ~  \log\(    \int \d z ~ \mN_z\( \sqrt{q}_0 \xi_0 , Q - q_0\)  e^{-\beta V(y|z)  } \) \nonumber \\
	& \underset{(q_0,\beta) \to (1,\infty)}{\simeq}    -\frac{1}{2} \log(2\pi (Q-q_0))  \\
	& \qquad \qquad - \beta \int \d \rP_\y(y) \int \D \xi_0 ~ \min_{z} \[ V(y|z) +\frac{\(z - \xi_0 \)^2}{2 \chi}  \] \nonumber\,.
	\label{app:spherical:gs:Psi_out}
\end{align}
Taking the limits $q_0 \to 1$, $\beta \to \infty$ in eq.~\eqref{app:spherical:rs_free_energy_simplified},  we obtain to the \aclink{RS} ground state energy of the spherical perceptron
\begin{align}
    e_{\gs}^{\rs} &=  \extr_{\chi} \left\{ -\frac{1}{2 \chi} + \alpha  \EE_{y, \xi_0} \min_{z} \[ V(y|z) + \frac{\(z - \xi_0\)^2}{2 \chi}  \]   \right\}
\end{align}

\paragraph{Application to the step-perceptron}
For the step function $V(y|z) = \theta(\kappa-z)$ with $\rP_\y(y)=\delta(y-1)$ and $\kappa \geq 0$, it leads to the Gardner expression \cite{gardner1988optimal}
\begin{align}
	 e_{\gs}^{\rs} &=  \extr_{\chi} \left\{ -\frac{1}{2\chi} + \alpha  \( \int_{-\infty}^{\kappa- \sqrt{2\chi}} \D \xi_0   + \int_{\kappa-\sqrt{2\chi}}^{\kappa} \D \xi_0 ~ \frac{(\xi_0 - \kappa)^2}{2\chi} \)   \right\}
\end{align}

\subsubsection{1RSB ground state energy}
\label{app:spherical:gs_1rsb}
To compute the ground state energy in the \aclink{1RSB} Ansatz, we take similarly the limits $q_1 \to 1$, $\beta \to \infty$ and $x_0\to 0$, keeping the products $ \chi \equiv \beta (Q-q_1)$ and $\omega_0 \equiv  x_0 \beta$ finite \cite{Whyte1996}, with $\Delta q = 1 - q_0$
\begin{align}
\begin{aligned}
 &\Psi_{\out}^{(\textrm{1rsb})}(\vec{q},\beta) \equiv \\
 &\frac{1}{x_0} \EE_y ~ \EE_{\xi_0}  \log\(  \EE_{\xi_1} \EE_{z} \[\mC( y \big | \sqrt{q_0} \xi_0 +
                \sqrt{q_1-q_0} \xi_1  + \sqrt{Q-q_1} z,\beta) \]^{x_0} \) \nonumber\\
     &= \frac{1}{x_0} \int \d \rP_\y(y)
                \int \D \xi_0 ~  \log \\  
     & \qquad  \int \D \xi_1 ~ \( \int \d z ~ \mN_z \(\sqrt{q_0} \xi_0 +
                \sqrt{q_1-q_0} \xi_1 , 1-q_1  \) e^{-\beta V(y|z)} \)^{x_0} \\
     &\simeq \frac{1}{x_0}\int \d \rP_\y(y)
                \int \D \xi_0 ~ \log \\
     & \qquad \int \D \xi_1 ~ e^{-x_0 \beta \min_z \[V(y|z) + \frac{1}{2 \beta (1-q_1)} \(z - \sqrt{q_0} \xi_0 -
                \sqrt{q_1-q_0} \xi_1 \)^2 \]}      \nonumber    \,.
\end{aligned}
\end{align}
Finally, taking $q_1 \to 1$ with $\beta \to \infty$ and $x\to 0$ in eq.~\eqref{app:spherical:1rsb_free_energy_simplified}, defining  $\Omega_0 \equiv \frac{\omega_0}{\chi}$, we obtain the \aclink{1RSB} ground state energy
\begin{align}
e_{\gs}^{(\textrm{1rsb})} &= \extr_{\chi, \Omega_0, q_0} \left\{  \frac{1}{2\Omega_0 \chi} \log\( 1 + \Omega_0 \Delta q \) + \frac{q_0}{2\chi \( 1 + \Omega_0 \Delta q \) }   \right. \\
& \left. \qquad + \frac{\alpha}{\chi \Omega_0}  \EE_{\xi_0}  \log \EE_{\xi_1} e^{-\Omega_0 \chi \min_z \[V(y | z) + \frac{1}{2 \chi } \(z - \sqrt{q_0} \xi_0 -
                \sqrt{\Delta q} \xi_1 \)^2 \]}   \right\}\,. \nonumber
\end{align}

\subsubsection{2RSB ground state energy $e_\gs$}
\label{app:spherical:gs_2rsb}
Similarly taking $q_2 \to 1$ with $\beta \to \infty$, we define $\Omega_0  \equiv  \frac{x_0 \beta}{\chi}, \Omega_1 \equiv  \frac{x_1 \beta}{\chi}$ and we obtain similarly the \aclink{2RSB} ground state energy of the spherical perceptron \cite{Whyte1996}
\begin{align}
	&e_{\gs}^{(\textrm{2rsb})} = \extr_{\chi,\Omega_1, \Omega_0, q_1,q_0,} \left\{ \frac{q_0}{2\chi(1+\Omega_1 (1-q_1) + \Omega_0 (q_1-q_0)}\right.  \nonumber \\
	&\left. +  \frac{1}{2 \Omega_1 \chi} \log(1+\Omega_1 (1-q_1))  \right. \nonumber\\
	&\left. + \frac{1}{2\Omega_0\chi} \log\(1+\frac{\Omega_0(q_1-q_0)}{1+\Omega_1 (1-q_1)}\) + \frac{\alpha}{\chi \Omega_0} \EE_{\xi_0} \right. \\
	&\left. + \log \EE_{\xi_1} \[\EE_{\xi_2} \exp \(-\Omega_1 \chi \min_z \[V(y | z) \right.\right.\right.\right. \nonumber\\ 
	&\left.\left.\left.\left. + \frac{1}{2 \chi } \(z - \sqrt{q_0} \xi_0 -
                \sqrt{q_1-q_0} \xi_1 - \sqrt{1-q_1} \xi_2 \)^2 \]\) \]^{\Omega_0 / \Omega_1} \right\} \nonumber
\end{align}
and notice that taking $q_1=q_0, x_0=x_1$ we recover the \aclink{1RSB} expression.

\ifthenelse{\equal{\format}{oneside}}
	{\clearpage\null\thispagestyle{empty}}
	{\cleardoublepage}
\chapter{AMP derivation - Committee machine}
	\label{appendix:amp:committee} 
	The \aclink{AMP} algorithm can be seen as Taylor expansion of the loopy \aclink{BP} approach 
\cite{mezard1987spin,mezard2009information,wainwright2008graphical},
similar to the so-called \aclink{TAP} equation in spin
glass theory \cite{thouless1977solution}. While the behavior of \aclink{AMP}
can be rigorously studied
\cite{bayati2011dynamics,javanmard2013state,bayati2015universality},
it is useful and instructive to see how the derivation can be
performed in the framework of \aclink{BP} and the cavity
method, as was pioneered in \cite{mezard1989space} for the single
layer problem. The derivation uses the \aclink{GAMP} notations of
\cite{rangan2011generalized} and follows closely the one of \cite{zdeborova2016statistical}.
The computation is presented for the committee machine hypothesis class, which is the vectorized version of the  \aclink{GLM}, with $K\geq 1$ vectorial parameters $\mat{W} = \{\vec{w}_k\}_{k=1}^K \in\bbR^{\ndim \times K}$.

\section{Factor graph and BP equations}
\label{appendix:amp:derivation:bp_eqs}
		As a central illustration, we present the instructive derivation of the \aclink{rBP} equations starting with the \aclink{BP} equations in the context of committee machines, already discussed in \App\ref{appendix:replica_computation:committee}. We recall the \aclink{JPD}
			\begin{align}
				\rP_\ndim\(\mat{W} | \vec{y}, \mat{X}\)  &= \frac{\rP_\out(\vec{y} | \mat{Z}) \rP_\w(\mat{W}) }{\mZ_\ndim(\vec{y},\mat{X})} = \frac{\prod_{\mu=1}^\nsamples \rP_\out(y_\mu | \vec{z}_\mu) \prod_{i=1}^\ndim \rP_{\w}(\vec{w}_i)  }{\mZ_\ndim(\vec{y},\mat{X})} \,,
			\end{align}
			where we defined $\mat{Z} =\frac{1}{\sqrt{\ndim}} \mat{X} \mat{Z} \in \bbR^{\nsamples \times K}$ and we assume that the channel and prior distributions factorize over factors $\rP_{\out}(y_\mu | \vec{z}_\mu)$ and variables $\rP_{\w}(\vec{w}_i)$. 	
			
	\subsection{Factor graph}
			The posterior distribution may be represented by the following bipartite factor graph in \Fig\ref{appendix:fig:amp:factor_graph_committee}.
			\begin{figure}[htb!]
				\centering
			\begin{tikzpicture}[scale=0.85, auto, swap]
			    \foreach \i in {1,...,6}
			        \node[var] (X\i) at (1.5*\i,0) {};
			    \node at (12, 0) {$ \vec{w}_i \in \bbR^K $};
			
			    \foreach \mu in {1,...,4}
			        \node[inter] (Y\mu) at (1.5+1.5*\mu,-2) {};
			    \foreach \i in {1,...,6}
			        \foreach \mu in {1,...,4}
			            \path[edge] (X\i) -- (Y\mu);
			    \node at (10, -2) {};
			    \node (F) at (12, -2) {$ \rP_{\out}\(y_\mu | \frac{1}{\sqrt{\ndim}} \vec{x}_\mu^\intercal \mat{W} \) $};			
			    \foreach \i in {1,...,6} {
			        \node[field] (P\i) at (1.5*\i,1) {};
			        \path[edge] (X\i) -- (P\i);
			    }
			    \node at (12, 1) {$ \rP_\w(\vec{w}_i) $};
			    \path[-latex, teal, very thick] (Y1) edge node[left]{$\td{m}_{\mu \to i}(\vec{w}_i)$} (X1);
			    \path[-latex, burntorange, very thick] (X6) edge node[right]{$m_{i \to \mu}(\vec{w}_i)$} (Y4);
			\end{tikzpicture}
			\caption{Factor graph representation of the joint distribution for committee machines.}
			\label{appendix:fig:amp:factor_graph_committee}
			\end{figure}
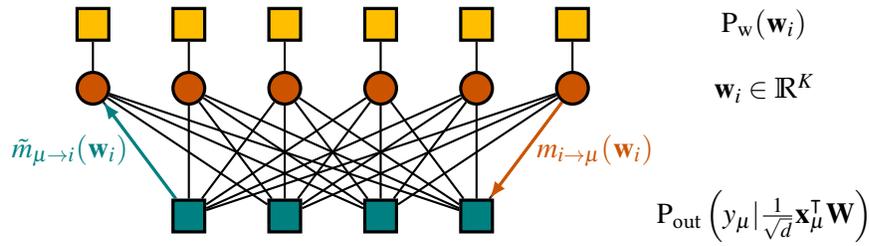
		In the following, we attach a set of \emph{messages} $\{m_{i\to \mu},\tilde{m}_{\mu \to i}\}_{i=1..n}^{\mu=1..m}$ to the edges of this bipartite factor graph. These messages correspond to the marginal probabilities of $\vec{w}_i \in \bbR^K$ if we remove the edges $(i \to \mu)$ or $(\mu \to i)$. 
			We define the auxiliary variable $\vec{z}_\indsamples = \frac{1}{\sqrt{\ndim}}\vec{x}_\indsamples^\intercal \mat{W} \in \bbR^K $ which is $\Theta(1)$ thanks to the pre-factor rescaling $1/\sqrt{d}$. This scaling is crucial as it allows the \aclink{BP} equations to hold true even though the factor graph is not tree-like and is instead fully connected with short loops.
				
	\subsection{BP equations}		
		The \aclink{BP} equations (also called the sum-product equations) for $\vec{w}_i=(w_{ik})_{k=1..K} \in \bbR^K$ on the factor graph \Fig\ref{appendix:fig:amp:factor_graph_committee} can be formulated, see \Sec\ref{main:intro:mean_field:bp}, as:
			\begin{align}
					m_{i\to \mu}^{t+1} (\vec{w}_i) &= \displaystyle \frac{1}{\mZ_{i\to \mu}} \rP_\w (\vec{w}_i) \prod\limits_{\nu \neq \mu}^\nsamples \tilde{m}_{\nu \to i}^t (\vec{w}_i) \label{appendix:amp:bp_equations_committee} \\
					\tilde{m}_{\mu \to i}^t (\vec{w}_i) &=  \displaystyle \frac{1}{\mZ_{\mu \to i}} \int_{\bbR^K} \prod\limits_{j\neq i}^\ndim \d \vec{w}_j ~ \rP_\out \(y_{\mu} |  \frac{1}{\sqrt{\ndim}} \sum_{j=1}^\ndim  x_{\mu j}\vec{w}_{j} \)  m_{j \to \mu}^t (\vec{w}_j )\,, \nonumber
			\end{align}
										
\section{Relaxed BP equations}
\label{appendix:amp:derivation:rbp_eqs}
		The idea of the relaxed \aclink{BP} equations is to simply expand in the limit $\ndim \to \infty$ the set of $\Theta(\ndim^2)$ messages $\td{m}$ of the \aclink{BP} equations in \eqref{appendix:amp:bp_equations_committee} before plugging them in $m$. Truncating the expansion and keeping only terms of order $\Theta\(1/\ndim\)$, messages become \textit{Gaussian}.
			Hence messages are therefore parametrized only by the mean $\hat{\vec{w}}_{i\to \mu}^t$ and the covariance matrix $\hat{\mat{C}}_{i\to \mu}^t$ of the marginal distribution at time $t$:
			\begin{align}
			\begin{aligned}
				\hat{\vec{w}}_{i\to \mu}^t &\equiv \displaystyle \int_{\bbR^K} \d \vec{w}_i ~
					 m_{i \to \mu}^t (\vec{w}_i) ~ \vec{w}_i \spacecase
					 \hat{\mat{C}}_{i \to \mu}^t &\equiv \displaystyle \int_{\bbR^K} \d \vec{w}_i ~
					 m_{i \to \mu}^t (\vec{w}_i) ~ \vec{w}_i \vec{w}_i^\intercal - \hat{\vec{w}}_{i \to \mu}^t(\hat{\vec{w}}_{i \to \mu}^t)^\intercal
			\end{aligned}
			\label{appendix:amp:committee:what_chat}
			\end{align}
		To decouple the argument of $\rP_{\out}$, we first by introducing its Fourier transform $\hat{\rP}_{\out}$ according to
			\begin{align*}
			&\rP_\out \(y_{\mu} |  \frac{1}{\sqrt{\ndim}} \sum_{j=1}^\ndim  x_{\mu j}\vec{w}_{j} \) = \frac{1}{(2\pi)^{K/2}}\\
			& \qquad \qquad  \times \int_{\bbR^K} \d \bxi \exp\( i \bxi^\intercal \( \displaystyle \frac{1}{\sqrt{\ndim}} \sum_{j=1}^\ndim  x_{\mu j}\vec{w}_{j}\) \hat{\rP}_{\out}(y_{\mu} , \bxi )    \).	
			\end{align*}
			Injecting this representation in the \aclink{BP} equations, \eqref{appendix:amp:bp_equations_committee} becomes:
			\begin{align}
			&\tilde{m}_{\mu \to i}^t (\vec{w}_i ) = 
			\frac{1}{(2\pi)^{K/2}\mZ_{\mu\to i} }
			\int_{\bbR^K} \d \bxi ~ \hat{\rP}_{\out}(y_{\mu} , \bxi) ~ 
			\exp\(i  \bxi^\intercal \frac{1}{\sqrt{\ndim}} x_{\mu i} \vec{w}_i \)\nonumber\\
			 &\qquad \times\prod\limits_{j\neq i}^\ndim \underbrace{\int_{\bbR^K} \d \vec{w}_j ~
					 m_{j \to \mu}^t (\vec{w}_j ) ~ \exp\( i  \bxi^\intercal \frac{1}{\sqrt{\ndim}} x_{\mu j} \vec{w}_j ) \)}_{\equiv I_j}\label{appendix:amp:bp_equations_committe:mtilde}
			\end{align}
			In the limit $\ndim \to \infty$ the term $I_j$ can be easily expanded and expressed using $\hat{\vec{w}}$ and $\hat{\mat{C}}$ in \eqref{appendix:amp:committee:what_chat}:
			\begin{align*}
			 I_j &= \int_{\bbR^K} \d \vec{w}_j ~
					 m_{j \to \mu}^t (\vec{w}_j ) ~ \exp\( i \frac{x_{\mu j}}{\sqrt{\ndim}} \bxi^\intercal \vec{w}_j ) \) \\
					 & \simeq  \exp\( i \frac{x_{\mu j}}{\sqrt{\ndim}} \bxi^\intercal  \hat{\vec{w}}_{j\to \mu}^t -  \frac{1}{2} \frac{x_{\mu j}^2}{\ndim}  \bxi^\intercal \hat{\mat{C}}_{j\to \mu}^t  \bxi \).
			\end{align*} 
			Finally using the inverse Fourier transform:
			\begin{align*}
			&\tilde{m}_{\mu \to i}^t (\vec{w}_i ) = 
			\frac{1}{(2\pi)^{K/2} \mZ_{\mu \to i}}
			\int_{\bbR^K} \d \vec{z} \rP_{\out}(y_\mu | \vec{z} ) 
			\int_{\bbR^K} \d \bxi  
			e^{-i \bxi^\intercal \vec{z}}
			e^{ i x_{\mu i} \bxi^\intercal \vec{w}_i} \\
			&\qquad\qquad\qquad\times\prod\limits_{j\neq i}^\ndim \exp\( i \frac{x_{\mu j}}{\sqrt{\ndim}} \bxi^\intercal \hat{\vec{w}}_{j\to \mu}^t -  \frac{1}{2}\frac{x_{\mu j}^2}{\ndim} \bxi^\intercal \hat{\mat{C}}_{j\to \mu}^t \bxi \) \\
			&= \frac{1}{(2\pi)^{K} \mZ_{\mu\to i}}
			\int_{\bbR} \d \vec{z} ~ \rP_{\out}(y_\mu | \vec{z} )\\
			& \qquad\qquad \int_{\bbR^K} \d \bxi ~ e^{-i \bxi^\intercal \vec{z}}
			e^{ i x_{\mu i} \bxi^\intercal \vec{w}_i} e^{i \sum\limits_{j\neq i}^\ndim\frac{x_{\mu j}}{\sqrt{\ndim}}  \bxi^\intercal \hat{\vec{w}}_{j\to \mu}^t } e^{-  \frac{1}{2} \sum\limits_{j\neq i}^\ndim\frac{x_{\mu j}^2}{\ndim}  \bxi^\intercal  \hat{\mat{C}}_{j \to \mu }^t \bxi} \\
			&= \frac{1}{(2\pi)^K \mZ_{\mu\to i}} \int_{\bbR^K} \d \vec{z} ~ \rP_{\out}(y_\mu | \vec{z}) \\ 
			& \qquad \quad \times \sqrt{\frac{(2\pi)^K}{\det{V_{\mu \to i}^t}}} \underbrace{e^{-\frac{1}{2} \( \vec{z} -\frac{x_{\mu i}}{\sqrt{\ndim}} \vec{w}_i -\bomega_{\mu \to i}^t \)^\intercal (\mat{V}_{\mu \to i}^t)^{-1} \( \vec{z} -\frac{x_{\mu i}}{\sqrt{\ndim}} \vec{w}_i -\bomega_{\mu \to i}^t \)}}_{\equiv H_{\mu \to i}}\,,
			\end{align*}
			where we defined the mean and variance, depending on the node $i$:
			\begin{align*}
				\bomega_{\mu \to i}^t &\equiv  \frac{1}{\sqrt{\ndim}} \sum\limits_{j\neq i}^\ndim x_{\mu j}  \hat{\vec{w}}_{j\to \mu}^t \,, &&
				\mat{V}_{\mu \to i}^t \equiv  \frac{1}{\ndim} \sum\limits_{j\neq i}^\ndim x_{\mu j}^2  \hat{\mat{C}}_{j \to \mu}^t\,.
			\end{align*}
			Again, in the limit $\ndim\to \infty$, the term $H_{\mu \to i}$ can be expanded as
			\begin{align*}
				H_{\mu \to i} &\simeq  e^{-\frac{1}{2} \( \vec{z} -\bomega_{\mu \to i}^t \)^\intercal (\mat{V}_{\mu \to i}^t)^{-1} \( \vec{z} -\bomega_{\mu \to i}^t \) } \\
				& \times 
				\( 1 + \frac{x_{\mu i}}{\sqrt{\ndim}} \vec{w}_i^\intercal (\mat{V}_{\mu \to i}^t)^{-1} (\vec{z} -\bomega_{\mu \to i}^t) -\frac{1}{2}\frac{x_{\mu i}^2}{\ndim} \vec{w}_i^\intercal (\mat{V}_{\mu \to i}^t)^{-1} \vec{w}_i \right.\\
			& \left. + \frac{1}{2} \frac{x_{\mu i}^2}{\ndim} \vec{w}_i^\intercal (\mat{V}_{\mu \to i}^t)^{-1} (\vec{z} -\bomega_{\mu \to i}^t) (\vec{z} - \bomega_{\mu \to i}^t)^\intercal  (\mat{V}_{\mu \to i}^t)^{-1} \vec{w}_i \).
			\end{align*}
			Putting all pieces together, the message $\tilde{m}_{\mu \to i}$ can be expressed using definitions of $\vec{f}_\out$ and $\partial_{\omega} \vec{f}_\out$ in \App\ref{appendix:definitions:updates:committee}. We finally obtain
			\begin{align*}
			&\tilde{m}_{\mu  \to i}^t (\vec{w}_i ) \sim \frac{1}{\mZ_{\mu \to i}} \left \{1 +  \frac{x_{\mu i}}{\sqrt{\ndim}} \vec{w}_{i}^\intercal  \vec{f}_\out (y_{\mu}, \bomega_{\mu \to i}^t, \mat{V}_{\mu \to i}^t) \right. \\
			&\left. \qquad\qquad \qquad\qquad + \frac{1}{2} \frac{x_{\mu i}^2}{\ndim} \vec{w}_{i}^\intercal \vec{f}_\out  \vec{f}_\out^\intercal (y_{\mu}, \bomega_{\mu \to i}^t, \mat{V}_{\mu \to i}^t) \vec{w}_{i} \right. \\
			& \left.\qquad\qquad \qquad\qquad  + \frac{1}{2} \frac{x_{\mu i}^2}{\ndim} \vec{w}_{i}^\intercal  \partial_\bomega \vec{f}_\out(y_{\mu}, \bomega_{\mu \to i}^t, \mat{V}_{\mu \to i}^t)  \vec{w}_{i}
			\right\}\\
			&= \frac{1}{\mZ_{\mu \to i}} \left\{ 1 + \vec{w}_{i}^\intercal  \vec{b}_{\mu \to i}^t +\frac{1}{2}  \vec{w}_{i}^\intercal  \vec{b}_{\mu \to i}^t (\vec{b}_{\mu \to i}^t)^\intercal  (\vec{w}_{i}) -\frac{1}{2} \vec{w}_{i}^\intercal  \mat{A}_{\mu \to i}^t w_{i} \right\} \\
			&=\sqrt{\frac{\det{\mat{A}_{\mu \to i}^t}}{(2\pi)^K}} e^{-\frac{1}{2}\(\vec{w}_{i}^\intercal  - (\mat{A}_{\mu \to i}^t)^{-1}\vec{b}_{\mu \to i}^t \)^\intercal  \mat{A}_{\mu \to i}^t\(\vec{w}_{i}^\intercal  - (\mat{A}_{\mu \to i}^t)^{-1}\vec{b}_{\mu \to i}^t \) }
			\end{align*}
			with the following definitions of $\mat{A}_{\mu \to i}$ and $\vec{b}_{\mu \to i}$
			\begin{align*}
					\vec{b}_{\mu \to i}^t &\equiv  \frac{x_{\mu i}}{\sqrt{\ndim}} \vec{f}_\out (y_{\mu}, \bomega_{\mu \to i}^t, \mat{V}_{\mu \to i}^t) \,,\\
					\mat{A}_{\mu \to i}^t &\equiv - \frac{x_{\mu i}^2}{\ndim}  \partial_\bomega \vec{f}_\out(y_{\mu}, \bomega_{\mu \to i}^t, \mat{V}_{\mu \to i}^t)\,.
			\end{align*}
			The set of \aclink{BP} equations can finally be closed over the Gaussian messages $\{m_{i\to \mu}\}_{i=1..\ndim}^{\mu=1..\nsamples}$ according to
			\begin{align*}
				 m_{i\to \mu}^{t+1} (\vec{w}_i) &= \frac{1}{\mZ_{i\to \mu}} \rP_\w (\vec{w}_i) \prod\limits_{\nu \neq \mu}^\nsamples \sqrt{\frac{\det{\mat{A}_{\nu \to i}^t}}{(2\pi)^K}} \\
				 & \qquad \qquad \qquad \times e^{-\frac{1}{2}\(\vec{w}_{i} - (\mat{A}_{\nu \to i}^t)^{-1}\vec{b}_{\nu \to i}^t \)^\intercal  \mat{A}_{\nu \to i}^t\(w_{i} - (\mat{A}_{\nu \to i}^t)^{-1}\vec{b}_{\nu \to i}^t \) }.
			\end{align*}
			In the end, computing the mean and variance of the product of Gaussians, the messages are updated using $\vec{f}_\w$ and $\partial_\bgamma \vec{f}_\w$, defined in \App\ref{appendix:definitions:updates:committee}, according to
			\begin{align*}
				\hat{\vec{w}}_{i\to \mu}^{t+1} &= \vec{f}_\w( \bgamma_{\mu \to i}^t, \bLambda_{\mu \to i}^t  )\,,
				&& \hat{\mat{C}}_{i \to \mu}^{t+1}=\partial_\bgamma \vec{f}_\w( \bgamma_{\mu \to i}^t, \bLambda_{\mu \to i}^t )\,,
			\end{align*}
			with
			\begin{align*}
				\bgamma_{\mu \to i}^t &= \sum\limits_{\nu \ne \mu}^\nsamples  \vec{b}_{\nu \to i}^t \,, 
				&& \bLambda_{\mu \to i}^t = \sum\limits_{\nu \ne \mu}^\nsamples  \mat{A}_{\nu \to i}^t \,.
			\end{align*}
			
		\paragraph{Summary of the rBP equations}
			In the end, the \aclink{rBP} equations are simply the following set of equations:
			\begin{align}
			\label{appendix:amp:committee:relaxed_bp_summary}
			\begin{aligned}
				\hat{\vec{w}}_{i\to \mu}^{t+1} &= \vec{f}_\w(\bgamma_{\mu \to i}^t,\bLambda_{\mu \to i}^t )\,, 
				&&\hat{\mat{C}}_{i \to \mu}^{t+1} = \partial_\bgamma \vec{f}_\w(\bgamma_{\mu \to i}^t, \bLambda_{\mu \to i}^t)  \\
				\bgamma_{\mu \to i}^t &=  \sum\limits_{\nu \ne \mu}^\nsamples  \vec{b}_{\nu \to i}^t \,, 
				&& \bLambda_{\mu \to i}^t =  \sum\limits_{\nu \ne \mu}^\nsamples  \mat{A}_{\nu \to i}^t   \\
			\vec{b}_{\mu \to i}^t &=  \frac{x_{\mu i}}{\sqrt{\ndim}} \vec{f}_\out (y_{\mu}, \bomega_{\mu \to i}^t, \mat{V}_{\mu \to i}^t) \,, \\ 
			\mat{A}_{\mu \to i}^t &= - \frac{x_{\mu i}^2}{\ndim}  \partial_\bomega \vec{f}_\out(y_{\mu}, \bomega_{\mu \to i}^t, \mat{V}_{\mu \to i}^t) \\
				\bomega_{\mu \to i}^t &= \sum\limits_{j\neq i}^\ndim\frac{x_{\mu j}}{\sqrt{\ndim}}   \hat{\vec{w}}_{j\to \mu}^t\,,
				&& \mat{V}_{\mu \to i}^t = \sum\limits_{j\neq i}^\ndim\frac{x_{\mu j}^2}{\ndim} \hat{\mat{C}}_{j\to \mu}^t\,.
			\end{aligned}
			\end{align}
			
\section{AMP algorithm}
\label{appendix:amp:derivation:amp_eqs}
	The \aclink{rBP} equations \eq\eqref{appendix:amp:committee:relaxed_bp_summary} contains $\Theta(\ndim^2)$ messages. However all the messages depend weakly on the target node. The missing message is negligible in the limit $\ndim \to \infty$, that allows us to expand the \aclink{rBP} around the \emph{full} messages:
			\begin{align}
			\begin{aligned}
				\bomega_{\mu}^t &\equiv \sum\limits_{j = 1}^\ndim\frac{x_{\mu j}}{\sqrt{\ndim}}   \hat{\vec{w}}_{j\to \mu}^t\,, 
				&& \mat{V}_{\mu}^t \equiv \sum\limits_{j=1}^\ndim  \frac{x_{\mu j}^2}{\ndim}  \hat{\mat{C}}_{j\to \mu}^t \\
				\bgamma_{i}^t & \equiv  \sum\limits_{\mu =1}^\nsamples  \vec{b}_{\mu \to i}^t \,,
				&& \bLambda_{i}^t \equiv  \sum\limits_{\mu =1}^\nsamples  \mat{A}_{\mu \to i}^t \,.
			\end{aligned}		
			\end{align}
			By completing the sum, we naturally remove the target node dependence and reduce the set of messages to $\Theta(\ndim)$. Let us now perform the expansion of the \aclink{rBP} messages.
	
	\paragraph{Partial covariance $\vec{f}_\w$: $\bLambda_{\mu \to i}^t$}
			\begin{align*}
				&\bLambda_{\mu \to i}^t =  \sum\limits_{\nu \ne \mu}^\nsamples  \mat{A}_{\nu \to i}^t 
				=  \sum\limits_{\nu =1 }^\nsamples  \mat{A}_{\nu \to i}^t - \mat{A}_{\mu \to i}^t  \\
				&= \bLambda_{i}^t - \mat{A}_{\mu \to i}^t = \bLambda_{i  }^t + \Theta\(\frac{1}{\ndim} \) \,.
			\end{align*}
			
	\paragraph{Partial mean $\vec{f}_\w$: $\bgamma_{\mu \to i}^t$}
			\begin{align*}
				\bgamma_{\mu \to i}^t &= \sum\limits_{\nu \ne \mu}^\nsamples  \vec{b}_{\nu \to i}^t  = \sum\limits_{\nu =1}^\nsamples  \vec{b}_{\nu \to i}^t -  \vec{b}_{\mu \to i}^t = \bgamma_{i}^t - \vec{b}_{\mu \to i}^t + \Theta\(\frac{1}{\ndim}\)\,.
			\end{align*}
			
	\paragraph{Mean $\hat{\vec{w}}_{i\to \mu}^{t+1}$ update}
			\begin{align*}
				\hat{\vec{w}}_{i\to \mu}^{t+1} &= \vec{f}_\w(\bgamma_{\mu \to i}^t , \bLambda_{\mu \to i}^t ) = \vec{f}_\w\(\bgamma_{i}^t - \vec{b}_{\mu \to i}^t,  \bLambda_{i}^t \) + \Theta\(\frac{1}{\ndim}\)\\
				&= \vec{f}_\w\(\bgamma_{i}^t, \bLambda_{i}^t \) -  \partial_\bgamma \vec{f}_\w \(\bgamma_{i}^t, \bLambda_{i}^t\) \vec{b}_{\mu \to i}^t  + \Theta\( \frac{1}{\ndim} \) \\
				&= \hat{\vec{w}}_{i}^{t+1} - \hat{\mat{C}}_{i}^{t+1} \vec{b}_{\mu \to i}^t  + \Theta\( \frac{1}{\ndim} \)\\
				&= \hat{\vec{w}}_{i}^{t+1} - \frac{x_{\mu i}}{\sqrt{\ndim}} \hat{\mat{C}}_{i}^{t+1} \vec{f}_\out (y_{\mu}, \bomega_{\mu}^t, \mat{V}_{\mu}^t) + \Theta\( \frac{1}{\ndim} \)\,.
			\end{align*}
			where we defined the prior updates
			\begin{align*}
				\hat{\vec{w}}_{i}^{t+1} &\equiv \vec{f}_\w\( \bgamma_{i}^t, \bLambda_{i}^t \)\,, && \hat{\mat{C}}_{i}^{t+1}\equiv \partial_\bgamma  \vec{f}_\w\(\bgamma_{i}^t, \bLambda_{i}^t \)\,,
			\end{align*}
			and used the fact that $\vec{b}_{\mu \to i}^t \simeq \frac{x_{\mu i}}{\sqrt{\ndim}} \hat{\mat{C}}_{i}^{t+1} \vec{f}_\out (y_{\mu}, \bomega_{\mu}^t, \mat{V}_{\mu}^t) $ by expanding the equation over $\vec{b}_{\mu \to i}^t$ in \eqref{appendix:amp:committee:relaxed_bp_summary}.
			
		\paragraph{Covariance $\hat{\mat{C}}_{i\to \mu}^{t+1}$ update}
			\begin{align*}
				\hat{\mat{C}}_{i\to \mu}^{t+1} &= \partial_\bgamma \vec{f}_\w(\bgamma_{\mu \to i}^t, \bLambda_{\mu \to i}^t) \\
				&\simeq \partial_\bgamma \vec{f}_\w(\bgamma_{i}^t, \bLambda_{i}^t)  + \Theta\( \frac{1}{\sqrt{\ndim}} \) = \hat{\mat{C}}_{i}^{t+1} + \Theta\( \frac{1}{\sqrt{\ndim}} \) \,.
			\end{align*}
		
		\paragraph{Channel update function $\vec{f}_\out(y_{\mu}, \bomega_{\mu \to i}^t, \mat{V}_{\mu \to i}^t)$}
	
			\begin{align*}
				&\vec{f}_\out(y_{\mu}, \bomega_{\mu \to i}^t, \mat{V}_{\mu \to i}^t) = \vec{f}_\out \(y_{\mu}, \bomega_{\mu}^t - \frac{x_{\mu i}}{\sqrt{\ndim}}   \hat{\vec{w}}_{i\to \mu}^t, \mat{V}_{\mu}^t - \frac{x_{\mu i}^2}{\ndim}   \hat{\mat{C}}_{i \to l}^t \)\\
				&= \vec{f}_\out \( y_{\mu}, \bomega_{\mu}^t, \mat{V}_{\mu}^t \) - \frac{x_{\mu i}}{\sqrt{\ndim}} \partial_\bomega \vec{f}_\out\( y_{\mu}, \bomega_{\mu}^t, \mat{V}_{\mu}^t \)   \underbrace{\hat{\vec{w}}_{i\to \mu}^t}_{=\hat{\vec{w}}_{i}^t + \Theta\( \frac{1}{\sqrt{\ndim}}\)} + \Theta\( \frac{1}{\ndim}\)\\
				&= \vec{f}_\out \( y_{\mu}, \bomega_{\mu}^t, \mat{V}_{\mu}^t \)-\frac{x_{\mu i}}{\sqrt{\ndim}} \partial_\bomega \vec{f}_\out\( y_{\mu}, \bomega_{\mu}^t, \mat{V}_{\mu}^t \)   \hat{\vec{w}}_{i}^t + \Theta\( \frac{1}{\ndim}\)\,.
			\end{align*}
			
		\paragraph{Covariance $\vec{f}_\out$: $\mat{V}_{\mu}^t$}
			\begin{align*}
			 \mat{V}_{\mu}^t &\equiv \sum\limits_{j=1}^\ndim  \frac{x_{\mu j}^2}{\ndim}  \hat{\mat{C}}_{j\to \mu}^t = \sum\limits_{j=1}^\ndim  \frac{x_{\mu j}^2}{\ndim}  \hat{\mat{C}}_{j\to \mu}^t + \Theta \( \frac{1}{\ndim^{3/2}}\) \,.
			\end{align*}
			
		\paragraph{Mean $\vec{f}_\out$: $\bomega_{\mu}^t$}
			\begin{align*}
				\bomega_{\mu}^t &= \sum\limits_{i = 1}^{\ndim} \frac{x_{\mu i}}{\sqrt{\ndim}}   \hat{\vec{w}}_{i\to \mu}^t \\
				&= \sum\limits_{i = 1}^{\ndim} \frac{x_{\mu i}}{\sqrt{\ndim}} \(\hat{\vec{w}}_{i}^t - x_{\mu i} \hat{\mat{C}}_{i}^t \vec{f}_\out (y_{\mu}, \bomega_{\mu}^{t-1}, \mat{V}_{\mu}^{t-1}) + \Theta\( \frac{1}{\ndim} \)  \) \\
				&= \sum\limits_{i = 1}^{\ndim} \frac{x_{\mu i}}{\sqrt{\ndim}} \hat{\vec{w}}_{i}^t -   \sum\limits_{i = 1}^{\ndim} \frac{x_{\mu i}^2}{\ndim}\hat{\mat{C}}_{i}^t \vec{f}_\out ( y_{\mu}, \bomega_{\mu}^{t-1}, \mat{V}_{\mu}^{t-1}) + \Theta \( \frac{1}{\ndim^{3/2}}\)\,.
			\end{align*}
			
		\paragraph{Covariance $\vec{f}_\w$: $\bLambda_{i}^t$}
			\begin{align*}
			\bLambda_{i}^t &\equiv  \sum\limits_{\mu =1}^\nsamples  \mat{A}_{\mu \to i}^t  =   \sum\limits_{\nu =1}^\nsamples - \frac{x_{\mu i}^2}{\ndim}  \partial_\bomega \vec{f}_\out(y_{\mu}, \bomega_{\mu \to i}^t, \mat{V}_{\mu \to i}^t)   \\
			&=   \sum\limits_{\mu =1}^\nsamples - \frac{x_{\mu i}^2}{\ndim}  \partial_\bomega \vec{f}_\out(y_{\mu}, \bomega_{\mu}^t, \mat{V}_{\mu}^t)  + \Theta\( \frac{1}{\ndim^{3/2}}\) \,.
			\end{align*}
			
		\paragraph{Mean $\vec{f}_\w$: $\bgamma_{i}^t$}
			\begin{align*}
			\bgamma_{i}^t &=  \sum\limits_{\mu =1}^\nsamples  \vec{b}_{\mu \to i}^t =  \sum\limits_{\mu =1}^\nsamples   \frac{x_{\mu i}}{\sqrt{\ndim}} \vec{f}_\out (y_{\mu}, \bomega_{\mu \to i}^t, \mat{V}_{\mu \to i}^t) \\
			&= \sum\limits_{\mu =1}^\nsamples    \frac{x_{\mu i}}{\sqrt{\ndim}} \vec{f}_\out (y_{\mu}, \bomega_{\mu}^t, \mat{V}_{\mu}^t) \\
			& \qquad -\frac{x_{\mu i}^2}{\ndim} \partial_\bomega \vec{f}_\out (y_{\mu}, \bomega_{\mu}^t, \mat{V}_{\mu}^t) \hat{\vec{w}}_{i}^t  + \Theta\(\frac{1}{\ndim^{3/2}} \)\,.
			\end{align*}
			
		\subsubsection*{Summary - AMP algorithm}
		We finally obtain the \aclink{AMP} algorithm as a reduced set of $\Theta(\ndim)$ messages in \Alg\ref{alg:appendix:amp:committee_machine}.
			\begin{algorithm} 
			\begin{algorithmic}
			    \STATE {\bfseries Input:} vector $\vec{y} \in \bbR^\nsamples$ and matrix $\mat{X}\in \bbR^{\nsamples \times \ndim}$:
			    \STATE \emph{Initialize}: $\hat{\vec{w}}_i$, $\vec{f}_{\out,\mu} \in \bbR^K$ and $\hat{\mat{V}}_i$, $\partial_{\bomega} \vec{f}_{\out, \mu} \in \bbR^{K\times K}$ for $ 1 \leq i \leq \ndim $ and $ 1 \leq \mu \leq \nsamples $ at $t=0$.
			    \REPEAT   
			    \STATE \noindent Channel: Update the mean $\omega_{\mu} \in \bbR^K$ and variance $V_{\mu}\in \bbR^{K\times K}$: \spacecase
			    \indent $\mat{V}_{\mu}^t = \sum\limits_{i=1}^\ndim  \frac{x_{\mu j}^2}{\ndim}  \hat{\mat{C}}_{i}^t $\\ 
			    \indent $\bomega_{\mu}^t = \sum\limits_{i = 1}^{\ndim} \frac{x_{\mu i}}{\sqrt{\ndim}} \hat{\vec{w}}_{i}^t -   \mat{V}_{\mu}^t \vec{f}_{\out,\mu}^{t-1}$\,, \\
			    \STATE \noindent Update $\vec{f}_{\out, \mu}$ and $\partial_\bomega \vec{f}_{\out,\mu}$: \spacecase
			    $\vec{f}_{\out,\mu}^t = \vec{f}_\out \( y_{\mu}, \bomega_{\mu}^t, \mat{V}_{\mu}^t \)$\,, $ \partial_\bomega \vec{f}_{\out^,\mu}^t = \partial_\bomega\vec{f}_\out \( y_{\mu}, \bomega_{\mu}^t, \mat{V}_{\mu}^t \)$ \spacecase
			    \STATE \noindent Prior: Update the mean $\bgamma_i \in \bbR^K$ and variance $\bLambda_i \in \bbR^{K\times K}$:\spacecase
			    $ \bLambda_{i}^t =  \sum\limits_{\mu =1}^\nsamples - \frac{x_{\mu i}^2}{\ndim}  \partial_\bomega \vec{f}_{\out, \mu} $\spacecase
			    $\bgamma_i^t = \sum\limits_{\mu =1}^\nsamples    \frac{x_{\mu i}}{\sqrt{\ndim}} \vec{f}_{\out,\mu} + \bLambda_{i}^t \hat{\vec{w}}_{i}^t $\,,
			    \STATE Update the estimated marginals $\hat{\vec{w}}_i \in \bbR$ and $\hat{\mat{C}}_i \in \bbR^+$: \spacecase
			   $\hat{\vec{w}}_{i}^{t+1}= \vec{f}_\w\( \bgamma_{i}^t, \bLambda_{i}^t \)$\,, $\hat{\mat{C}}_{i}^{t+1}= \partial_\bgamma  \vec{f}_\w\(\bgamma_{i}^t, \bLambda_{i}^t \)$\spacecase
			    \STATE ${t} \leftarrow {t} + 1$ 
			    \UNTIL{Convergence on
			    $\hat{\vec{w}}_i$, $\hat{\mat{C}}_i$.} 
			    \STATE {\bfseries Output:}
			    $\{\hat{\vec{w}_i}\}_{i=1}^\ndim$ and $\{\hat{\mat{C}}_i\}_{i=1}^\ndim$.
			\end{algorithmic}
			\caption{Approximate Message Passing algorithm for committee machines.}
  			\label{alg:appendix:amp:committee_machine}
			\end{algorithm}
			
\section{State evolution equations of AMP}
\label{appendix:amp:derivation:se_eqs}
In this section we derive the behavior of the \aclink{AMP} algorithm in  \Alg\ref{alg:appendix:amp:committee_machine} in the thermodynamic limit $\ndim \to \infty$. This average asymptotic behavior can be tracked with some overlap parameters at time $t$, $\mat{m}^t$, $\mat{q}^t$, $\bSigma^t$, that respectively measure the correlation of the \aclink{AMP} estimator with the ground truth, the norms of student and teacher weights, the estimator variance and the second moment of the teacher network $ \brho_{\w^\star}$, defined by
\begin{align}
\begin{aligned}
	\mat{m}^t &\equiv \displaystyle \EE \lim_{\ndim \to \infty} \frac{1}{\ndim}\hat{\mat{W}}^{t \intercal} \hat{\mat{W}}^{\star} \,, 
	&& \mat{q}^t  \equiv \displaystyle \EE \lim_{\ndim \to \infty} \frac{1}{\ndim} \hat{\mat{W}}^{t \intercal} \hat{\mat{W}}^{t}  \,, \\ 
	\bSigma^t & \equiv \displaystyle \EE \lim_{\ndim \to \infty} \frac{1}{\ndim} \sum_{i=1}^\ndim \hat{\mat{C}}_{i}^{t} \,, &&
	\brho_{\w^\star} \equiv \displaystyle \EE \lim_{\ndim \to \infty} \frac{1}{\ndim} \mat{W}^{\star \intercal} \mat{W}^{\star} \,,
\end{aligned}
\label{appendix:amp:derivation:se_eqs:overlaps}
\end{align}
where the expectation is over ground truth signals $\mat{W}^\star$ and input data $\mat{X}$. The aim is to derive the asymptotic behavior of these overlap parameters, called \aclink{SE}. The idea is simply to compute the overlap distributions starting with the set of \aclink{rBP} equations in \eqref{appendix:amp:committee:relaxed_bp_summary}.

\subsection{Messages distribution}
In order to get the asymptotic behavior of the overlap parameters, we first need to compute the distribution of $\mat{W}^{t+1}$ and, as a result, of the mean $\bgamma_{\mu \to i}^t$ and covariance $\bLambda_{\mu \to i}^t$. 
Recalling that under the \aclink{BP} assumption incoming messages are independent, the messages $\bomega_{\mu \to i}^t$ and $\vec{z}_{\mu}$ are the sum of independent variables and follow Gaussian distributions. 
However, these two variables are correlated and we need to compute correctly the covariance matrix.

To compute it, we will make use of different ingredients. 
First, we recall that in the \aclink{T-S} scenario, the output has been generated by a teacher such that $\forall \mu \in \lb \nsamples \rb,~ y_\mu = \varphi_{\out^\star} \(\frac{1}{\sqrt{\ndim}} \vec{x}_\mu^\intercal \mat{W}^\star \)$. By convenience, we define $\vec{z}_\mu \equiv \frac{1}{\sqrt{\ndim}} \vec{x}_\mu^\intercal \mat{W}^\star = \frac{1}{\sqrt{\ndim}} \sum_{i=1}^\ndim x_{\mu i} \vec{w}_i^\star $ and $z_{\mu \to i} \equiv  \frac{1}{\sqrt{\ndim}} \sum_{j \ne i}^\ndim x_{\mu j} \vec{w}_j^\star$. 
Second, in the case the input data are \aclink{i.i.d} Gaussian, we have $\EE_{\mat{X}}[x_{\mu i}] = 0 $ and $\EE_{\mat{X}} [x_{\mu i}^2] = 1$.

\paragraph{Partial mean $\vec{f}_\out$: $\bomega_{\mu \to i}^t$}
Let's compute the first two moments, using expansions of the \aclink{rBP} equations \eqref{appendix:amp:committee:relaxed_bp_summary}:
\begin{align*}
	&\EE\[ \bomega_{\mu \to i}^t\]  = \frac{1}{\sqrt{\ndim}} \sum\limits_{j \neq i}^\ndim  \EE_{\mat{X}} \[ x_{\mu j}\]   \EE \[ \hat{\vec{w}}_{j\to \mu}^t \] = \vec{0}\,, \\
	&\EE \[ \bomega_{\mu \to i}^t \( \bomega_{\mu \to i}^t\)^\intercal \]  = \frac{1}{\ndim} \sum\limits_{j \neq i}^\ndim \EE_{\mat{X}} \[ x_{\mu j}^2\]  \EE\[ \hat{\vec{w}}_{j\to \mu}^t \(\hat{\vec{w}}_{j\to \mu}^t\)^\intercal\]\\
	&= \frac{1}{\ndim} \sum\limits_{i=1}^\ndim \EE_{\mat{X}} \[ x_{\mu j}^2\]  \EE\[ \hat{\vec{w}}_{i}^t \(\hat{\vec{w}}_{i}\)^\intercal\] + \Theta\(\ndim^{-3/2}\)  \underlim{\ndim}{\infty} \mat{q}^t\,.
\end{align*}	

\paragraph{Hidden variable $\vec{z}_{\mu}$}
Let us compute the first moments of the hidden variable $\vec{z}_{\mu}$:
\begin{align*}
	\EE \[ \vec{z}_\mu \]  &= \frac{1}{\sqrt{\ndim}} \sum\limits_{i=1}^\ndim \EE_{\mat{X}}\[x_{\mu i}\] \EE_{\mat{W}^\star}\[\vec{w}_i^\star\]  = \vec{0} \,, \\
	\EE \[  \vec{z}_\mu  \vec{z}_\mu^\intercal \]  &= \frac{1}{\ndim} \sum\limits_{i=1}^\ndim \EE_{\mat{X}}\[x_{\mu i}^2\] \EE_{\mat{W}^\star}\[ \vec{w}_i^\star (\vec{w}_i^\star)^\intercal\] \underlim{\ndim}{\infty}\brho_{\w^\star} \,.
\end{align*}

\paragraph{Correlation between $\vec{z}_{\mu}$ and $\bomega_{\mu \to i}^t$}
The cross correlation is given by
\begin{align*}
	&\EE \[ \bomega_{\mu \to i}^t  \vec{z}_\mu^\intercal \] =  \frac{1}{\ndim} \sum\limits_{j\neq i, k=1 }^\ndim \EE_{\mat{X}} \[x_{\mu j} x_{\mu k} \]  \EE_{\mat{W}^\star} \[ \hat{\vec{w}}_{j\to \mu}^t (\vec{w}_{k}^\star)^\intercal \]  \\
	&=  \frac{1}{\ndim} \sum\limits_{j\neq i}^\ndim  \EE_{\mat{W}^\star} \[ \hat{\vec{w}}_{j\to \mu}^t (\vec{w}_{j}^\star)^\intercal \] = \frac{1}{\ndim} \sum\limits_{i}^\ndim  \EE_{\mat{W}^\star} \[ \hat{\vec{w}}_{i}^t (\vec{w}_{i}^\star)^\intercal \] + \Theta\(\ndim^{-3/2}\)\\
	& \qquad \underlim{\ndim}{\infty} \mat{m}^t \,.
\end{align*}
Hence asymptotically the random vector ($\vec{z}_\mu$, $\bomega_{\mu \to i}^t$) follow a multivariate Gaussian distribution with covariance matrix 
$ \mat{Q}^t = 
\begin{bmatrix}
    \brho_{\w^\star} & \mat{m}^t \\
    \mat{m}^t & \mat{q}^t  \\
\end{bmatrix} \in \bbR^{(2K) \times (2K)}$.
 
\paragraph{Partial variance $\vec{f}_\out$: $\mat{V}_{\mu \to i}$} 
$\mat{V}_{\mu \to i}$ concentrates around its mean:
\begin{align*}
	\EE \[ \mat{V}_{\mu \to i}^t \] &= \frac{1}{\ndim} \sum\limits_{j\neq i}^\ndim \EE_{\mat{X}} \[ x_{\mu j}^2 \] \hat{\mat{C}}_{j\to \mu}^t = \frac{1}{\ndim} \sum\limits_{i}^\ndim \hat{\mat{C}}_{i}^t + \Theta(\ndim^{-3/2}) \underlim{\ndim}{\infty} \bSigma^t \,.
\end{align*}

\paragraph{Ad-hoc overlaps}
Let us define some other ad-hoc order parameters, that will appear in the following:
\begin{align}
\begin{aligned}
		\hat{\mat{q}}^t & \equiv \alpha \EE_{\bomega,\vec{z}} \[ \vec{f}_{\out} (\varphi_{\out^\star}(\vec{z}), \bomega, \bSigma^t )^{\otimes 2}  \]  \,, \\
		 \hat{\mat{m}}^t &\equiv \alpha \EE_{\bomega, \vec{z}} \[ \partial_\vec{z} \vec{f}_{\out}(\varphi_{\out^\star}(\vec{z}), \bomega, \bSigma^t ) \]  \,, \\
		  \hat{\bchi}^t &\equiv \alpha \EE_{\bomega, \vec{z}} \[ - \partial_\bomega \vec{f}_{\out}(\varphi_{\out^\star}(\vec{z}), \bomega, \bSigma^t ) \] \,. \\
\end{aligned}
\end{align}

\paragraph{Partial mean $\vec{f}_\w$: $\bgamma_{\mu \to i}^t$} 
Using the expression $y_\nu = \varphi_{\out^\star}\( \vec{z}_{\nu \to i} + \frac{1}{\sqrt{\ndim}} x_{\nu i} \vec{w}_i^\star \)$ and expanding $\bgamma_{\mu \to i}^t$, we obtain
\begin{align*}
	&\bgamma_{\mu \to i}^t =  \sum\limits_{\nu \ne \mu}^\nsamples  \vec{b}_{\nu \to i}^t  =  \sum\limits_{\nu \ne \mu}^\nsamples  \frac{x_{\nu i}}{\sqrt{\ndim}} \vec{f}_\out \( y_\nu , \bomega_{\nu \to i}^t, \mat{V}_{\nu \to i}^t\) \\
	&=   \frac{1}{\sqrt{\ndim}} \sum\limits_{\nu \ne \mu}^\nsamples  x_{\nu i} \vec{f}_\out \( \varphi_{\out^\star}\( \vec{z}_{\nu \to i} \), \bomega_{\nu \to i}^t, \mat{V}_{\nu \to i}^t\) \\
	& + \frac{1}{\ndim} \sum\limits_{\nu \ne \mu}^\nsamples x_{\nu i}^2 \partial_\vec{z}\vec{f}_\out \( \varphi_{\out^\star}\( \vec{z}_{\nu \to i}\), \bomega_{\nu \to i}^t, \mat{V}_{\nu \to i}^t\) \vec{w}_i^\star \,.
\end{align*}
Thus, taking the average
\begin{align*}
	\EE\[ \bgamma_{\mu \to i}^t \] &= \vec{0} + \frac{1}{\ndim} \sum\limits_{\nu \ne \mu}^\nsamples \EE_{\vec{z}, \bomega} \[ \partial_\vec{z}\vec{f}_\out \( \varphi_{\out^\star}\( \vec{z}_{\nu \to i}\), \bomega_{\nu \to i}^t, \mat{V}_{\nu \to i}^t\) \] \vec{w}_i^\star \\
	& \underlim{\ndim}{\infty} \hat{\mat{m}}^t  \vec{w}_i^\star \,, \\
	\EE\[ (\bgamma_{\mu \to i}^t)^{\otimes 2} \] &= \frac{1}{\ndim} \sum\limits_{\nu \ne \mu}^\nsamples  \EE_{\vec{z}, \bomega}\[ \vec{f}_\out \( \varphi_{\out^\star}\( \vec{z}_{\nu \to i} \), \bomega_{\nu \to i}^t, \mat{V}_{\nu \to i}^t\)^{\otimes 2} \] \\
	& \underlim{\ndim}{\infty} \hat{\mat{q}}^t\,.
\end{align*}
Hence $ \bgamma_{\mu \to i}^t \sim  \hat{\mat{m}}^t \vec{w}_i^\star  + ( \hat{\mat{q}}^t)^{1/2}\bxi $ with $\bxi \sim \mN(\vec{0},\rI_K)$. 

\paragraph{Partial covariance $\vec{f}_\w$: $\bLambda_{\mu \to i}^t$}
\begin{align*}
	\bLambda_{\mu \to i}^t &= \sum\limits_{\nu \ne \mu}^\nsamples  \mat{A}_{\nu \to i}^t =  - \frac{1}{\ndim}  \sum\limits_{\nu \ne \mu}^\nsamples x_{\mu i}^2 \partial_\bomega \vec{f}_\out(y_{\nu}, \bomega_{\nu \to i}^t, \mat{V}_{\nu \to i}^t) \\
	& = - \frac{1}{\ndim}  \sum\limits_{\nu \ne \mu}^\nsamples x_{\mu i}^2 \partial_\bomega \vec{f}_\out( \varphi_{\out^\star}(\vec{z}_{\nu \to i}) , \bomega_{\nu \to i}^t, \mat{V}_{\nu \to i}^t)  + \Theta\(\ndim^{-3/2} \)
\end{align*}
and taking the average 
\begin{align*}
	\EE \[ \bLambda_{\mu \to i}^t\] &= - \frac{1}{\ndim}  \sum\limits_{\nu \ne \mu}^\nsamples \EE_{\vec{z}, \bomega}\[ \partial_\bomega \vec{f}_\out( \varphi_{\out^\star}(\vec{z}_{\nu \to i}) , \bomega_{\nu \to i}^t, \mat{V}_{\nu \to i}^t)\]\\
	& \qquad \underlim{\ndim}{\infty}  \hat{\bchi}^t\,,
\end{align*} 
so that in the thermodynamic limit $\bLambda_{\mu \to i}^t \sim \hat{\bchi}^t$.

\subsection{Summary of the SE - mismatched setting}
Using the definition of the overlaps in \eqref{appendix:amp:derivation:se_eqs:overlaps} at time $t+1$ and the message distributions, we finally obtain the set of \aclink{SE} equations of the \aclink{AMP} algorithm in \Alg\ref{alg:appendix:amp:committee_machine} in the mismatched setting:
\begin{align}
\begin{aligned}
	\mat{m}^{t+1} & \equiv \displaystyle \EE \lim_{\ndim \to \infty} \frac{1}{\ndim}\hat{\mat{W}}^{t+1 \intercal} \hat{\mat{W}}^{\star} = \EE_{\vec{w}^\star, \bxi} \[  \vec{f}_\w\( \hat{\mat{m}}^t \vec{w}^\star  + ( \hat{\mat{q}}^t)^{1/2}\bxi, \hat{\bchi}^t \) \vec{w}^{\star \intercal} \]  \,, \\
	\mat{q}^{t+1} &\equiv \displaystyle \EE \lim_{\ndim \to \infty} \frac{1}{\ndim} \hat{\mat{W}}^{t+1 \intercal} \hat{\mat{W}}^{t+1}= \EE_{\vec{w}^\star, \bxi} \[  \vec{f}_\w\( \hat{\mat{m}}^t \vec{w}^\star  + ( \hat{\mat{q}}^t)^{1/2}\bxi, \hat{\bchi}^t \)^{\otimes 2} \] \,, \\ 
	\bSigma^{t+1} & \equiv \displaystyle \EE \lim_{\ndim \to \infty} \frac{1}{\ndim} \sum_{i=1}^\ndim \hat{\mat{C}}_{i}^{t+1} = \EE_{\vec{w}^\star, \bxi} \[  \partial_\bgamma \vec{f}_\w\( \hat{\mat{m}}^t \vec{w}^\star  + ( \hat{\mat{q}}^t)^{1/2}\bxi, \hat{\bchi}^t \) \] \,,
\end{aligned}
\end{align}
and
\begin{align}
\begin{aligned}
	\hat{\mat{q}}^t &= \alpha\int_{\bbR^K} \int_{\bbR^K} \d \bomega ~ \d \vec{z} ~ \mN_{(\vec{z}, \bomega)}\(\vec{0}_{2K}, \mat{Q}^t \) \vec{f}_{\out} (\varphi_{\out^\star}(\vec{z}), \bomega, \bSigma^t )^{\otimes 2} \\
	\hat{\mat{m}}^t &= \alpha\int_{\bbR^K} \int_{\bbR^K} \d \bomega ~ \d \vec{z} ~ \mN_{(\vec{z}, \bomega)}\(\vec{0}_{2K}, \mat{Q}^t \)) \partial_\vec{z} \vec{f}_{\out}(\varphi_{\out^\star}(\vec{z}), \bomega, \bSigma^t ) \,, \\
	\hat{\bchi}^t &= -\alpha \int_{\bbR^K} \int_{\bbR^K} \d \bomega ~ \d \vec{z} ~ \mN_{(\vec{z}, \bomega)}\(\vec{0}_{2K}, \mat{Q}^t \)  \partial_\bomega \vec{f}_{\out}(\varphi_{\out^\star}(\vec{z}), \bomega, \bSigma^t ) \,. 
\end{aligned}
\end{align}
with $ \mat{Q}^t = 
\begin{bmatrix}
    \brho_{\w^\star} & \mat{m}^t \\
    \mat{m}^t & \mat{q}^t  \\
\end{bmatrix} \in \bbR^{(2K) \times (2K)}$.

\subsection{Summary of the SE - Bayes-optimal setting}
In the Bayes-optimal setting, the student $\rP_\w = \rP_{\w^\star}$ and $\rP_\out = \rP_{\out^\star}$, so that we have $\vec{f}_\w = \vec{f}_{\w^\star}$ and $\vec{f}_\out = \vec{f}_{\out^\star}$. Moreover, the Nishimori conditions, recalled in \App\ref{appendix:replica_computation:nishimori}, imply that
\begin{align*}
	\mat{m}^t &= \mat{q}^t \equiv \mat{q}^t_\bayes  \,, && \hat{\mat{q}}^t  = \hat{\mat{m}}^t = \hat{\bchi}^t \equiv \hat{\mat{q}}^t_\bayes \,, && \bSigma^{t} =  \brho_{\w^\star} - \mat{q}^t\,.
\end{align*}
Therefore the set of \aclink{SE} equations simplify and reduce to
\begin{align}
	\mat{q}^{t+1}_\bayes &= \EE_{\vec{w}^\star, \bxi} \[  \vec{f}_{\w^\star}\( \hat{\mat{q}}^t_\bayes \vec{w}^\star  + ( \hat{\mat{q}}^t_\bayes)^{1/2}\bxi, \hat{\mat{q}}^t_\bayes \)^{\otimes 2} \] \label{appendix:amp:se:bayes} \\
	\hat{\mat{q}}^t_\bayes &= \alpha\int_{\bbR^K} \int_{\bbR^K} \d \bomega ~ \d \vec{z} ~ \mN_{(\vec{z}, \bomega)}\(\vec{0}_{2K} , \mat{Q}^t_\bayes \) \vec{f}_{\out^\star} (\varphi_{\out^\star}(\vec{z}), \bomega,  \brho_{\w^\star} - \mat{q}^t_\bayes )^{\otimes 2} \nonumber
\end{align}
with the simplified covariance matrix $\mat{Q}^t_\bayes=\begin{bmatrix}
     \brho_{\w^\star} & \mat{q}^t_\bayes \\
    \mat{q}^t_\bayes & \mat{q}^t_\bayes  \\
  \end{bmatrix} $.

\subsection{Consistence with the replica computation}
Very surprisingly, the \aclink{SE} of the \aclink{AMP} algorithm can be obtained in a convoluted and more rapid way. It turns out that in the Bayes-optimal setting, \aclink{AMP} performs a gradient ascent on the \aclink{RS} free entropy in \eqref{appendix:free_entropy_bayes}. Meaning that at convergence, and under good initialization, the \aclink{AMP} overlaps are given by the saddle point equations of the \aclink{RS} free entropy $\Phi^{(\rs)}$. 
To see this, we shall start performing the change of variable $\bxi \leftarrow \bxi + \(\hat{\mat{q}}^t_\bayes\)^{1/2} \vec{w}^\star$ in \eqref{appendix:amp:se:bayes} so that we directly obtain the first equation of \eqref{appendix:se_equations_generic:bayes} with the corresponding time indices
\begin{align}
	\mat{q}^{t+1}_\bayes &= \EE_{\vec{w}^\star, \bxi} \[ \mZ_{\w^\star}\(( \hat{\mat{q}}^t_\bayes)^{1/2}\bxi, \hat{\mat{q}}^t_\bayes\)  \vec{f}_{\w^\star}\( ( \hat{\mat{q}}^t_\bayes)^{1/2}\bxi, \hat{\mat{q}}^t_\bayes \)^{\otimes 2} \]\,.
\end{align}
Moreover in this setting, we notice that variables $\bomega_{\mu \to i}^t$ and $\vec{z}_\mu - \bomega_{\mu \to i}^t$ become independent since
\begin{align*}
	\EE \[ \bomega_{\mu \to i}^t\(  \vec{z}_\mu - \bomega_{\mu \to i}^t \)^\intercal \] & \underlim{\ndim}{\infty} \mat{m}^t - \mat{q}^t = \mat{q}^t_\bayes - \mat{q}^t_\bayes = \vec{0}\,,\\
	\EE \[ \bomega_{\mu \to i}^t (\bomega_{\mu \to i}^t)^\intercal \]& \underlim{\ndim}{\infty} \mat{q}^t_\bayes\,, \\
	\EE \[ \(  \vec{z}_\mu - \bomega_{\mu \to i}^t \) \(  \vec{z}_\mu - \bomega_{\mu \to i}^t \)^\intercal \] &\underlim{\ndim}{\infty} \brho_{\w^\star} - \mat{q}^t_\bayes \,,
\end{align*}
so that the multivariate Gaussian distribution factorize to
$$\mN_{(\vec{z}, \bomega)}\(\vec{0}, \mat{Q}^t_\bayes \) = \mN_{\bomega}\(\vec{0}_K, \mat{q}^t_\bayes \) \mN_{\vec{z}}\(\bomega,  \brho_{\w^\star} - \mat{q}^t_\bayes \).$$ 
Using $\rP_{\out^\star}(y|\vec{z}) = \delta\( y - \varphi_{\out^\star}(\vec{z})\)$ the second equation of \eqref{appendix:amp:se:bayes} becomes
\begin{align*}
	\hat{\mat{q}}^t &= \alpha\int_{\bbR^K} \int_{\bbR^K} \d \bomega ~ \d \vec{z} ~ \mN_{(\vec{z}, \bomega)}\(\vec{0}_{2K}, \mat{Q}^t_\bayes \) \vec{f}_{\out^\star} (\varphi_{\out^\star}(\vec{z}), \bomega,  \brho_{\w^\star} - \mat{q}^t_\bayes )^{\otimes 2} \\
	&= \alpha \int_\bbR \d y ~ \int_{\bbR^K} \d \bomega ~ \mN_{\bomega}\(\vec{0}_K, \mat{q}^t_\bayes \)  \\
	&\qquad \times \int_{\bbR^K} \d \vec{z} ~ \rp_{\out^\star}(y | \vec{z}) \mN_{\vec{z}}\( \bomega ;  \brho_{\w^\star} - \mat{q}^t_\bayes \)\vec{f}_{\out^\star} (y, \bomega,  \brho_{\w^\star} - \mat{q}^t_\bayes )^{\otimes 2}\\
	&= \alpha \int_\bbR  \d y ~ \int_{\bbR^K} \d \bxi ~ \mN_{\bxi}\(\vec{0} ; \rI_K \)  \int_{\bbR^K} \d \vec{z} ~ \rp_{\out^\star}(y | \vec{z}) \\
	& \qquad \qquad \times  \mN_{\vec{z}}\( (\mat{q}^t_\bayes)^{1/2}\bxi ;  \brho_{\w^\star} - \mat{q}^t_\bayes \) \vec{f}_{\out^\star} (y, (\mat{q}^t_\bayes)^{1/2}\bxi ,  \brho_{\w^\star} - \mat{q}^t_\bayes)^{\otimes 2} \tag{Change of variable $\bxi \leftarrow (\mat{q}^t_\bayes)^{-1/2} \bomega^t$}\\
	&= \alpha \int_\bbR  \d y ~ \EE_{\bxi} \mZ_{\out^\star}\(y, (\mat{q}^t_\bayes)^{1/2}\bxi ,  \brho_{\w^\star} - \mat{q}^t_\bayes\)  \\
	& \qquad \qquad \qquad \qquad  \times \vec{f}_{\out^\star}\(y,  (\mat{q}^t_\bayes)^{1/2}\bxi ,  \brho_{\w^\star} - \mat{q}^t_\bayes  \)\,,
\end{align*}
which is exactly the second fixed point equation of the \aclink{RS} free entropy \eqref{appendix:amp:se:bayes}.


\ifthenelse{\equal{\format}{oneside}}
	{\clearpage\null\thispagestyle{empty}}
	{\cleardoublepage}
\ctparttext{}
\part{Bibliography}
\label{app:bibliography} 

\manualmark 
\markboth{\spacedlowsmallcaps{\bibname}}{\spacedlowsmallcaps{\bibname}} 
\refstepcounter{dummy}
\addcontentsline{toc}{section}{}	


\renewcommand*{\bibfont}{\small} 
\setstretch{0.8}
\setlength\bibitemsep{0pt}
\printbibliography

\ifthenelse{\equal{\format}{oneside}}
	{}
	{\clearpage\null\thispagestyle{empty}\newpage}


\thispagestyle{empty}

\hfill

\vfill
\vspace{12cm}

\noindent Ph.D dissertation of \myName \\ 
\textit{\myTitlesmall}\textcopyright\ \myTime

\bigskip

\noindent\spacedlowsmallcaps{Supervisors}: \\
Lenka Zdeborov{\'a} and Florent Krzakala 

\medskip

\noindent\spacedlowsmallcaps{Location}: \\
\myDepartment\\
\myUni, \myLocation \\

\medskip

\noindent\spacedlowsmallcaps{Time Frame}: \\
October 1, 2017 --- December 16, 2020

\vfill 
\ifthenelse{\equal{\format}{oneside}}
	{}
	{\newpage\null\thispagestyle{empty}}
	
\ifthenelse{\equal{\cover}{true}}
	{\newcommand{\paragraphBack}[1]{\paragraph*{\fontfamily{fvs}\fontseries{b}\fontsize{10}{10}\selectfont #1}}

\thispagestyle{empty}
\newgeometry{top=1.25cm, bottom=1.25cm, left=2cm, right=2cm}

\noindent 
\includegraphics[height=2cm]{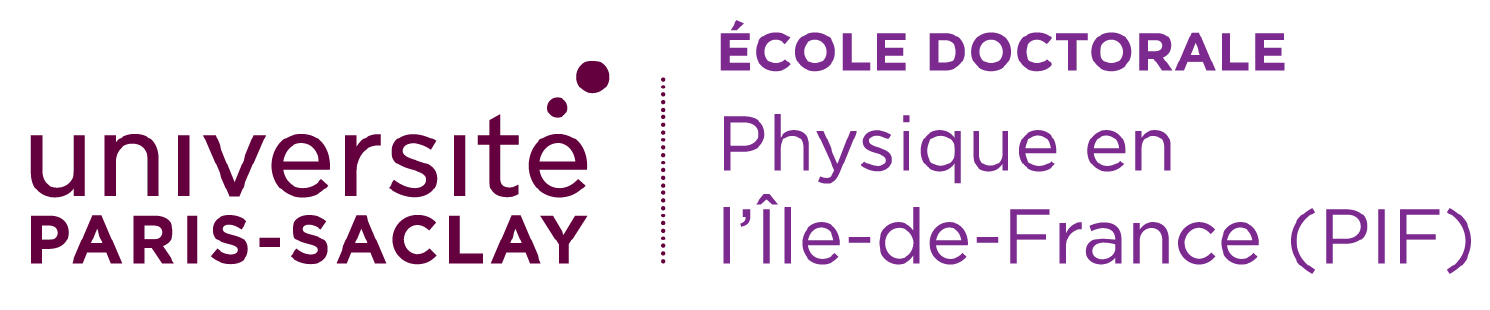}
\vspace{0.25cm}

\fontfamily{fvs}\fontseries{m}\fontsize{10}{10}\selectfont

\begin{mdframed}[linecolor=Prune,linewidth=1]
\begin{footnotesize}
\vspace{-.25cm}
\paragraphBack{Titre:} Méthodes à champ moyen et perspectives algorithmiques pour l'apprentissage automatique en haute dimension.
\vspace{-.25cm}
\paragraphBack{Mots clés:} Physique statistique, apprentissage automatique, réseaux de neurones, estimation statistique, algorithmes de passage de messages, méthode des répliques.
\end{footnotesize}

\begin{footnotesize}
\vspace{-.5cm}
\begin{multicols}{2}
\paragraphBack{Résumé:} 
À une époque où l'utilisation des données a atteint un niveau sans précédent, l'apprentissage machine, et plus particulièrement l'apprentissage profond basé sur des réseaux de neurones artificiels, a été responsable de très importants progrès pratiques. 
Leur utilisation est désormais omniprésente dans de nombreux domaines d'application, de la classification d'images à la reconnaissance vocale en passant par la prédiction de séries temporelles et l'analyse de texte. 
Pourtant, la compréhension de nombreux algorithmes utilisés en pratique est principalement empirique et leur comportement reste difficile à analyser. 
Ces lacunes théoriques soulèvent de nombreuses questions sur leur efficacité et leurs potentiels risques. 
Etablir des fondements théoriques sur lesquels asseoir les observations numériques est devenu l'un des défis majeurs de la communauté scientifique.
La principale difficulté qui se pose lors de l’analyse de la plupart des algorithmes d'apprentissage automatique est de traiter analytiquement et numériquement un grand nombre de variables aléatoires en interaction. 
Dans ce manuscrit, nous revisitons une approche basée sur les outils de la physique statistique des systèmes désordonnés. Développés au long d’une riche littérature, ils ont été précisément conçus pour décrire le comportement macroscopique d'un grand nombre de particules, à partir de leurs interactions microscopiques. 
Au cœur de ce travail, nous mettons fortement à profit le lien profond entre la méthode des répliques et les algorithmes de passage de messages pour mettre en lumière les diagrammes de phase de divers modèles théoriques, en portant l’accent sur les potentiels écarts entre seuils statistiques et algorithmiques.
Nous nous concentrons essentiellement sur des tâches et données synthétiques générées dans le paradigme enseignant-élève. En particulier, nous appliquons ces méthodes à champ moyen à l'analyse Bayes-optimale des machines à comité, à l'analyse des bornes de généralisation de Rademacher pour les perceptrons, et à la minimisation du risque empirique dans le contexte des modèles linéaires généralisés. 
Enfin, nous développons un cadre pour analyser des modèles d'estimation avec des informations \emph{à priori} structurées, produites par exemple par des réseaux de neurones génératifs avec des poids aléatoires.
\end{multicols}
\end{footnotesize}
\end{mdframed}

\vspace{0.75cm}

\begin{mdframed}[linecolor=Prune,linewidth=1]
\begin{footnotesize}
\vspace{-.25cm}
\paragraphBack{Title:} Mean-field methods and algorithmic perspectives for high-dimensional machine learning

\vspace{-.25cm}
\paragraphBack{Keywords:} Statistical physics, machine learning, neural networks, statistical estimation, message passing algorithms, replica method
\end{footnotesize}

\begin{footnotesize}
\vspace{-.5cm}
\begin{multicols}{2}
\paragraphBack{Abstract:}
At a time when the use of data has reached an unprecedented level, machine learning, and more specifically deep learning based on artificial neural networks, has been responsible for very important practical advances. 
Their use is now ubiquitous in many fields of application, from image classification, text mining to speech recognition, including time series prediction and text analysis. 
However, the understanding of many algorithms used in practice is mainly empirical and their behavior remains difficult to analyze. 
These theoretical gaps raise many questions about their effectiveness and potential risks. 
Establishing theoretical foundations on which to base numerical observations has become one of the fundamental challenges of the scientific community.
The main difficulty that arises in the analysis of most machine learning algorithms is to handle, analytically and numerically, a large number of interacting random variables.
In this manuscript, we revisit an approach based on the tools of statistical physics of disordered systems. 
Developed through a rich literature, they have been precisely designed to infer the macroscopic behavior of a large number of particles from their microscopic interactions. 
At the heart of this work, we strongly capitalize on the deep connection between the replica method and message passing algorithms in order to shed light on the phase diagrams of various theoretical models, with an emphasis on the potential differences between statistical and algorithmic thresholds. 
We essentially focus on synthetic tasks and data generated in the teacher-student paradigm. In particular, we apply these mean-field methods to the Bayes-optimal analysis of committee machines, to the worst-case analysis of Rademacher generalization bounds for perceptrons, and to empirical risk minimization in the context of generalized linear models. Finally, we develop a framework to analyze estimation models with structured prior informations, produced for instance by deep neural networks based generative models with random weights.
\end{multicols}
\end{footnotesize}
\end{mdframed}

\vspace{1cm}

\fontfamily{fvs}\fontseries{m}\selectfont
{\color{Prune} \noindent \textbf{Université Paris-Saclay}\\
Espace Technologique / Immeuble Discovery\\
Route de l’Orme aux Merisiers RD 128 / 91190 Saint-Aubin, France}
} 
	{\newpage\null\thispagestyle{empty}}
	
\end{document}